\RequirePackage{fix-cm}
\documentclass[natbib,smallextended]{svjour3}       % onecolumn (second format)
\smartqed  % flush right qed marks, e.g. at end of proof
\usepackage{epsf}
\usepackage{graphicx}
\usepackage{longtable}
\usepackage[colorlinks=true,citecolor=blue,urlcolor=blue,breaklinks]{hyperref}
\usepackage[svgnames]{xcolor}
\newcommand{\figwidth}{0.7\textwidth}
\newcommand{\figwidthb}{0.9\textwidth}
\newcommand{\fig}{.}

\newcommand{\rv}{}

\newcommand{\citecl}[1]
{\citeauthor{#1} \citeyear{#1}}

\newcommand{\noprintlabel}{}
\newcommand{\be}{\begin{equation}}
\newcommand{\bea}{\begin{eqnarray}}
\newcommand{\ee}
{\end{equation}}
\newcommand{\eea}
{\end{eqnarray}}
\newcommand{\eel}[1]
{\label{#1}
\end{equation}
\ifx\noprintlabel\undefined{\small\bf [#1]}\par\noindent\fi}
\newcommand{\eenl}
{\end{equation}}
\newcommand{\eeal}[1]
{\label{#1}
\end{eqnarray}
\ifx\noprintlabel\undefined{\bf [#1]}\par\noindent\fi}
\newcommand{\eeanl}[1]
{\end{eqnarray}}
\newcommand{\figlab}[1]
{\label{#1}\ifx\noprintlabel\undefined{\bf [#1]}\fi}

\newcommand{\clabel}[1]
{\label{#1}\ifx\noprintlabel\undefined{\bf [#1]}\fi}
\newcommand{\biblab}[1]
{\ifx\noprintlabel\undefined{\bf [#1]}\fi}

\newcommand{\Eq}[1]{(\ref{#1})}

\newcommand{\cf}{{cf.}}
\newcommand{\eg}{{e.g.}}
\newcommand{\ie}{{i.e.}}
\newcommand{\bolddelta}{\delta\kern-0.45em\delta\kern-0.45em\delta}
\newcommand{\boldr}{\mbox{\boldmath$r$}}
\newcommand{\bolddelr}{\bolddelta \boldr}

\newcommand{\cm}{\,{\rm cm}}
\newcommand{\nm}{\,{\rm nm}}
\newcommand{\m}{\,{\rm m}}
\newcommand{\dyn}{\,{\rm dyn}}
\newcommand{\erg}{\,{\rm erg}}
\newcommand{\g}{\,{\rm g}}
\newcommand{\s}{\,{\rm s}}
\newcommand{\yr}{\,{\rm yr}}
\newcommand{\Gyr}{\,{\rm Gyr}}
\newcommand{\K}{\,{\rm K}}

\newcommand{\MeV}{\,{\rm MeV}}
\newcommand{\SNU}{\,{\rm SNU}}
\newcommand{\Zs}{Z_{\rm s}}
\newcommand{\Ys}{Y_{\rm s}}
\newcommand{\Xs}{X_{\rm s}}
\newcommand{\Rs}{R_{\rm s}}
\newcommand{\Ls}{L_{\rm s}}
\newcommand{\Lsun}{\,{\rm L}_\odot}
\newcommand{\Rsun}{\,{\rm R}_\odot}
\newcommand{\Msun}{\,{\rm M}_\odot}
\newcommand{\Teff}{T_{\rm eff}}
\newcommand{\dd}{{\rm d}}
\newcommand{\amu}{m_{\rm u}}
\newcommand{\kB}{k_{\rm B}}
\newcommand{\nablaad}{\nabla_{\rm ad}}
\newcommand{\nablarad}{\nabla_{\rm rad}}
\newcommand{\Fcon}{F_{\rm con}}
\newcommand{\pt}{p_{\rm t}}
\newcommand{\Frad}{F_{\rm rad}}
\newcommand{\alphamlt}{\alpha_{\rm ML}}
\newcommand{\eplus}{{\rm e^+}}
\newcommand{\eminus}{{\rm e^-}}
\newcommand{\hyd}{{{}^{1}\rm H}}
\newcommand{\deut}{{{}^{2}\rm D}}
\newcommand{\neut}{{\rm n}}
\newcommand{\helthree}{{{}^{3}\rm He}}
\newcommand{\helfour}{{{}^{4}\rm He}}
\newcommand{\berseven}{{{}^{7}\rm Be}}

\newcommand{\litseven}{{{}^{7}\rm Li}}
\newcommand{\bereight}{{{}^{8}\rm Be}}
\newcommand{\boreight}{{{}^{8}\rm B}}
\newcommand{\cartwelve}{{{}^{12}\rm C}}
\newcommand{\carthirteen}{{{}^{13}\rm C}}
\newcommand{\nitthirteen}{{{}^{13}\rm N}}
\newcommand{\nitfourteen}{{{}^{14}\rm N}}
\newcommand{\nitfifteen}{{{}^{15}\rm N}}
\newcommand{\oxyfifteen}{{{}^{15}\rm O}}

\newcommand{\oxyseventeen}{{{}^{17}\rm O}}
\newcommand{\floseventeen}{{{}^{17}\rm F}}

\newcommand{\PPII}{\hbox{\rm PP-II}}
\newcommand{\PPIII}{\hbox{\rm PP-III}}
\newcommand{\rt}{r_{\rm t}}
\newcommand{\kh}{k_{\rm h}}
\newcommand{\clight}{\tilde c}
\newcommand{\nHz}{\,{\rm nHz}}

\newcommand{\CF}{{\cal F}}

\newcommand{\modname}[2]{[{#1}]}
\newcommand{\MS}{\modname{S}{d.02c}}

\newcommand{\Mage}{\modname{Age}{d.06c}}
\newcommand{\MRs}{\modname{$R_{\rm s}$}{d.05c}}
\newcommand{\MLs}{\modname{$L_{\rm s}$}{d.41c}}
\newcommand{\Mopa}{\modname{Opc.~7.0}{d.15c}}
\newcommand{\Mopb}{\modname{Opc.~6.5}{d.16c}}
\newcommand{\Mopalninesix}{\modname{OPAL96}{d.07c}}
\newcommand{\Mopzerofive}{\modname{OP05}{d.36c}}
\newcommand{\Mgsnineeight}{\modname{GS98}{d.37c}}
\newcommand{\Meoszerofive}{\modname{Liv05}{d.40c}}
\newcommand{\Msurfop}{\modname{Surf.~opac.}{d.35c}}
\newcommand{\MCM}{\modname{CM}{d.24c}}
\newcommand{\Madelb}{\modname{Adelb11}{d.34c}}
\newcommand{\Mnacre}{\modname{NACRE}{d.39c}}
\newcommand{\Mhethreeeq}{\modname{$\helthree$ eql.}{d.02c\_eq}}

\newcommand{\Mnoelscrn}{\modname{No el.scrn}{d.20c}}
\newcommand{\Mdifa}{\modname{Dc}{d.17c}}
\newcommand{\Mdifb}{\modname{DVc}{d.18c}}
\newcommand{\Mnodif}{\modname{No diff.}{03c}}
\newcommand{\Magszerofive}{\modname{AGS05}{d.21c}}
\newcommand{\Magsszeronine}{\modname{AGSS09}{d.22c}}

\begin{document}

\title{Solar structure and evolution}

\author{J{\o}rgen Christensen-Dalsgaard}

\institute{J. Christensen-Dalsgaard \at
	Stellar Astrophysics Centre, and \\
Department of Physics and Astronomy, Aarhus University, \\
Ny Munkegade, 8000 Aarhus C, Denmark \\
	\email{jcd@phys.au.dk \\
	Orchid: 0000-0001-5137-0966}
}

\date{Received: date / Accepted: date}

\maketitle

\begin{abstract}
The Sun provides a critical benchmark for the general study of
stellar structure and evolution.
Also, knowledge about the internal properties of the Sun is important
for the understanding of solar atmospheric phenomena, including the
solar magnetic cycle.
Here I provide a brief overview of the theory of stellar structure and
evolution, including the physical processes and parameters that are
involved.
This is followed by a discussion of solar evolution, extending from the
birth to the latest stages.
As a background for the interpretation of observations
related to the solar interior I provide a rather extensive analysis
of the sensitivity of solar models to the assumptions underlying their
calculation.
I then discuss the detailed information about the solar interior that
has become available through helioseismic investigations and the detection of 
solar neutrinos, with further constraints provided by the observed
abundances of the lightest elements.
Revisions in the determination of the solar surface abundances have
led to increased discrepancies, discussed in some detail,
between the observational inferences and solar models.
I finally briefly address the relation of the Sun to other similar stars and
the prospects for asteroseismic investigations of stellar structure
and evolution.
\end{abstract}

\vfill\eject
\noindent
\rule{10cm}{0.2mm}

Dedicated to the memory of the late Professor Michael J. Thompson, 
who tragically passed away in October 2018.
His great understanding of the Sun and the helioseismic techniques 
used to probe its properties, shared through our collaboration over two decades,
is an essential basis for the work presented here.

\noindent
\rule{10cm}{0.2mm}

\setcounter{tocdepth}{3}
\tableofcontents

%===========================================================================

\section{Introduction}
\clabel{sec:introduction}

%\notecd [Basic role of stellar evolution in astrophysics.]
%
The study of stellar properties and stellar evolution plays a central
role in astrophysics. 
Observations of stars determine the chemical composition, age and
distance of the varied components of the Milky Way Galaxy and hence form
the basis for studies of Galactic evolution.
Stellar abundances and their evolution, particularly for lithium, are
also a crucial component of the study of Big-Bang nucleosynthesis.
Understanding of the pulsational properties of Cepheids underlies their use
as distance indicators and hence the basic unit of distance measurement in
the Universe.
The detailed properties of supernovae are important for the study of
element nucleosynthesis, while supernovae of Type Ia are crucial for
determining the large-scale properties of the Universe, including the
evidence for a dominant component of `dark energy'.
In all these cases an accurate understanding, and modelling, of
stellar interiors and their evolution is required for reliable results.

%\notecd [Emphasis of connection with physics, potential tests of physics.]

Modelling stellar evolution depends on a detailed treatment of the physics
of stellar interiors.
Insofar as the star is regarded as nearly spherically symmetric the
basic equations of stellar equilibrium are relatively straightforward
(see Sect.~\ref{sec:basicmod}),
but the detailed properties, often referred to as microphysics, of
matter in a star are extremely complex, yet of major importance to
the modelling. 
This includes the thermodynamical properties, as specified by the
equation of state, the interaction between matter and radiation
described by the opacity, the nuclear processes generating energy
and causing the evolution of the element composition, and the 
diffusion and settling of elements.
Equally important are potential hydrodynamical processes caused by
various instabilities which may contribute to the transport of energy
and material, hence causing partial or full mixing of given regions in a star.
It is obvious that sufficiently detailed observations of stellar properties,
and comparison with models, may provide a possibility for testing the
physics used in the model calculation, hence allowing investigations of
physical processes far beyond the conditions that can be reached in a
terrestrial laboratory.

%\notecd [Special role of Sun, as being extremely well studied by helioseismology;
%hence essential to understand it as a prerequisite for understanding 
%other stars.]

Amongst stars, the Sun obviously plays a very special role, both to
our daily life and as an astrophysical object.
Its proximity allows very precise, and probably accurate, determination
of its global parameters, as well as extremely detailed investigations
of phenomena in the solar atmosphere, compared with other stars.
%\notecd [Worth checking Gustafsson on normality of Sun].
Indications are that the Sun is typical for its mass and age
%\notecd [Check further references and coordinate with Sect.~\ref{sec:stars}].
\citep[{\eg},][]{Gustaf1998, Robles2008, Struga2017},%
\footnote{However, subtle and potentially very interesting differences
have been found between the solar surface composition and
the composition of similar stars; see Sect.~\ref{sec:twins},
in particular Fig.~\ref{fig:solarabund}.}
although a detailed analysis by \citet{Reinho2020} 
of photometric variability observed with
the {\it Kepler} spacecraft indicated that solar magnetic activity
may be rather low compared with solar-like stars.
Also, conditions in the solar interior are relatively benign,
providing some hope that reasonably realistic modelling can be carried out.
Thus it is an ideal case for investigations of stellar structure
and evolution.
Interestingly, there still remain very significant discrepancies between
the observed properties of the Sun and solar models.

%\end{document}

%\notecd [A note or two on the history, including the neutrino problem, interplay
%between helioseismology and improvements in solar modelling.
%Possibly already here hint at problems with new composition.]

A good overview of the development of the study
of stellar structure and evolution was provided by \citet{Tassou2004}.
Also, \citet{Shaviv2009} gave an excellent wide-ranging and
deep description of the evolution of the field, including an extensive
discussion of the relevant observational basis, the underlying physics, and
related aspects, such as the early tension between estimates of the age of the
Earth and the Sun.
The application of physics to the understanding of stellar 
interiors developed from the middle of the 19th century.
The first derivation of stellar models based on mechanical equilibrium
was carried out by \citet{Lane1870}.%
\footnote{It is interesting to note that the lengthy title of that paper
explicitly refers to the use of `the laws of gases as known to
terrestrial experiments'; the application of terrestrial physics
to the modelling of stars of course remains a key aspect 
to the study of stellar interiors.}
Further development of the theory of such models, summarized in
an extensive bibliographical note by \citet{Chandr1939}, was carried out
by Ritter, Lord Kelvin and others, culminating in the monograph
by \citet{Emden1907}.
These models were based on the condition of hydrostatic equilibrium,
combined with a simplified, so-called polytropic, equation of state.
Major advances came with the application of the theory of radiative
transfer, and quantum-mechanical calculations of atomic absorption
coefficients, to the energy transport in stellar interiors.
This allowed theoretical estimates to be made of the relation between
the stellar mass and luminosity, even without detailed knowledge about
the stellar energy sources
(for a masterly discussion of these developments, see \citecl{Edding1926}).
Further investigations of the properties of stellar opacity
led to the conclusion that stellar matter was dominated by hydrogen
\citep{Stromg1932, Stromg1933}, in agreement with the
detailed determination of the composition of the solar photosphere by
\citet{Russel1929},
as well as with the analysis of a broad range of stars by \citet{Unsold1931}.
%\notecd [Should do a little about Str\"omgren here].
Although stellar modelling had proceeded up to this point without
any definite information about the sources of stellar energy,
this issue was evidently of very great interest.
As early as 1920 \citet{Edding1920} and others
noted that the fusion of hydrogen into helium might produce the required
energy, over the solar lifetime,
but a mechanism making the fusion possible, given the strong Coulomb
repulsion between the nuclei, was lacking.
This mechanism was provided by Gamow's development of the treatment
of quantum-mechanical barrier penetration between reacting nuclei,
resulting in the identification of the 
dominant reactions in hydrogen fusion through the PP chains
and the CNO cycle ({\cf} Sect.~\ref{sec:engenr}) by
\citet{vonWei1937, vonWei1938}, \citet{Bethe1938} and \citet{Bethe1939}.
With this,
%the identification by \notecd [somebody and sombody, including Bethe]
%the source of stellar energy in hydrogen fusion,
the major ingredients required for the modelling of the solar interior
and evolution had been established.

%\notecd [A few words about early convection treatment].
An important aspect of solar structure is the presence of an outer
convection zone.
Following the introduction by Karl Schwarzschild \citep{Schwar1906}
of the criterion for
convective instability in stellar atmospheres,
\citet{Unsold1930} noted that such instability would be expected 
in the lower photosphere of the Sun.
As a very important result, \citet{Bierma1932} noted that 
the temperature gradient resulting from the consequent convective
energy transport would in general be close to adiabatic;
as a result, the structure of the convection zone depends little
on the details of the convective energy transport.
Also, he found that the resulting convective region in the Sun extended
to very substantial depths, reaching a temperature of $10^7 \K$.
Further calculations by, for example, \citet{Bierma1942} and
\citet{Rudkjo1942}, taking into account more detailed models of the
solar atmosphere, generally confirmed these results.
In an interesting short paper \citet{Stromg1950} summarized these early results.
He noted that the presence of ${}^7{\rm Li}$ in the solar atmosphere
clearly showed that convective mixing could extend at most to a temperature
of $3.5 \times 10^6 \K$,%
\footnote{A modern value for this limit is $2.5 \times 10^6 \K$; see
Sect.~\ref{sec:lightcomp}.}
beyond which lithium would be destroyed by nuclear reactions.
He also pointed out that a revision of determinations of the composition
of the solar atmosphere, relative to the one assumed by Biermann,
had reduced the heavy-element abundance and that this would reduce the
temperature at the base of the convection zone to the acceptable value
of $2.5 \times 10^6 \K$.
Although these models are highly simplified, the use of the lithium 
abundance as a constraint on the extent of convective mixing, and
the effect of a composition adjustment on the convection-zone depth,
remain highly relevant, as discussed below.

%\notecd [Need to check for early models specifically of the Sun;
%see Gough \& Weiss.  See \citet{Schwar1958}.]
Specific computations of solar models must satisfy the known observational
constraints for the Sun, namely that solar radius and luminosity
be reached at solar age, for a $1 M_\odot$ model.
As discussed by Martin Schwarzschild \citep{Schwar1957}
this can be achieved by adjusting
the composition and the characteristics of the convection zone.
They noted that no independent determination of the initial hydrogen and
helium abundances $X_0$ and $Y_0$
was possible and consequently determined models
for specified initial values of the hydrogen abundance.
The convection zone was assumed to have an adiabatic stratification
and to consist of fully ionized material, such that it was
characterized by the adiabatic constant $K$ in the relation 
$p = K \rho^{5/3}$ between pressure $p$ and density $\rho$.
Given $X_0$ the values of $Y_0$ and $K$ were then determined to
obtain a model with the correct luminosity and radius.
Although since substantially refined, this remains the basic principle
for the calibration of solar models (see Sect.~\ref{sec:calib}).
A detailed discussion of the calibration of the properties of the
convection zone was provided by \citet{Gough1976}.

Given the calibration, the observed mass, radius and luminosity 
clearly provide no test of the validity of the solar model.
An important potential for testing solar models became evident with
the realization \citep{Fowler1958}
that nuclear reactions in the solar core produce huge numbers 
of neutrinos which in principle may be measured, given a suitable
detector \citep{Davis1964, Bahcal1964}.
%\notecd [Early precursor experiment?]
The first results of a large-scale experiment \citep{Davis1968} 
%(\notecd Davis [1968?])
surprisingly showed an upper limit to the neutrino flux substantially
below the predictions of the then current solar models.
Further experiments using a variety of techniques, and additional
computations, did not eliminate this discrepancy, the predictions being
higher by a factor 2--3 than the experiment, until the beginning of
the present millennium.

An independent way of testing solar models, with potentially much
higher selectivity, became available with the detection of solar
oscillations \citep[see][for further details on the history
of the field]{Christ2004}.
Oscillations with periods near 5 minutes were discovered by 
\citet{Leight1962}.
Their character as standing acoustic waves was proposed independently by 
\citet{Ulrich1970} and \citet{Leibac1971},
leading also to the expectation
that their frequencies could be used to probe the outer parts of the Sun.
This identification was confirmed observationally by \citet{Deubne1975},
whose data clearly showed the modal character of the oscillations.
The observed modes had short horizontal wavelength
and extended only a few per cent into the Sun.
Indications of global oscillations in the solar diameter were presented by
\citet{Hill1976}, immediately suggesting that detailed information about
the whole solar interior could be obtained from analysis of their frequencies
\citep[{\eg},][]{Christ1976}.
Although Hill's data have not been confirmed by later studies, they served
as important inspirations for such studies,
now known as \emph{helioseismology}.%
\footnote{This term was apparently introduced in the scientific
literature by \citet{Severn1979}.}

%\notecd [perhaps a little more history here; Hill??
%reference to C-D Yale for historical details]
Early analyses of the short-wavelength five-minute oscillations
\citep{Gough1977a, Ulrich1977} showed that the solar convection zone was 
substantially deeper than in the models of the epoch.
A major breakthrough was the detection of global five-minute oscillations
by \citet{Claver1979} and \citet{Grec1980}
and the subsequent identification of modes in the
five-minute band over a broad range of horizontal wavelengths
\citep{Duvall1983}.
Observations of these modes have formed the basis for the dramatic 
development of helioseismology over the last three decades.
With the increasing precision and detail of the observed oscillation 
frequencies, increasing sophistication was applied to solar modelling,
generally leading to improved agreement between models and observations.
Important examples were the realization
that the opacity of the solar interior should be increased to match the
inferred sound-speed profile \citep{Christ1985},
that sophisticated equations of
state were required to match the observed frequencies \citep{ChristDL1988},
and that the inclusion of diffusion and settling substantially 
improved the agreement between the models and the Sun \citep{Christ1993}.
%(\notecd [possibly earlier Cox et al.]; \citep{Christ1993}).
Remarkably, these developments in the model physics, motivated by
but not directly fitted to, the steadily improving observations,
led to models in good overall agreement with the inferred solar
structure \citep[{\eg},][]{Christ1996, Goughetal1996, Bahcal1997, Brun1998}.
The remaining discrepancies were highly significant and clearly required
changes to the physics of the solar interior, however.
Interestingly, later revisions of the measured solar surface abundance
now result in rather larger discrepancies between models and observations,
indicating that more basic modifications to the modelling may be required.

In the present review I provide an overview of these issues, covering both the 
modelling and the sensitivity of solar models to the physical assumptions
and the inferences drawn from various observations and their interpretation.
Chapter~\ref{sec:solarmod} presents the tools required to model the Sun
and its evolution,
including some emphasis on the underlying physical properties of solar matter.
In Chapter~\ref{sec:evol} I present a brief overview of 
the evolution of a solar-mass star.
A detailed discussion of the sensitivity of solar models to changes in the
model parameters or physics is provided in Chapter~\ref{sec:standard},
using as reference
case the widely used so-called Model~S \citep{Christ1996}.
Chapter~\ref{sec:test} discusses the observations available
to test our understanding of solar structure and evolution, i.e.,
helioseismology, solar neutrinos
and the details of the solar surface composition;
in discussing the helioseismic results a brief presentation of 
results on solar internal rotation is also provided.
In Chapter~\ref{sec:abundprob} the serious issues raised 
by the revised determinations of
the solar composition after 2000 are discussed in detail, 
including the revisions to solar modelling which have attempted to
obtain agreement with the helioseismically inferred structure under
the constraints of these revised abundances.
Finally Chapter~\ref{sec:stars} gives a very brief presentation
of studies of other stars,
including the place of the Sun in relation to solar-like stars,
and Chapter~\ref{sec:concluding} provides a few concluding remarks.
In support of the numerical results provided here,
Appendix~\ref{sec:numacc} briefly addresses
the important issue of the numerical accuracy of the computed models.

%\end{document}
%% \newpage

\section{Modelling the Sun}
\clabel{sec:solarmod}

\subsection{Basics of stellar modelling}
\clabel{sec:basicmod}

%\notecd [Usual assumptions of `standard' stellar modelling.]
%
Stellar models are generally calculated under a number of simplifying 
approximations, of varying justification.
In most cases rotation and other effects causing departures from
spherical symmetry are neglected and hence the star is regarded as spherically
symmetric. 
Also, with the exception of convection, hydrodynamical instabilities are
neglected, while convection is treated in a highly simplified manner.
The mass of the star is assumed to be constant, so that no significant
mass loss is included.
In contrast to these simplifications of the `macrophysics'
the microphysics is included with considerable, although
certainly inadequate, detail.
In recent calculations effects of diffusion and settling are typically
included, at least in computations of solar models.
The result of these approximations is what is often called a 
`standard solar model', although still obviously depending on the
assumptions made in the details of the calculation.%
\footnote{The notion of `standard model' develops over time;
for example, until around 1995 diffusion and settling
would not generally be regarded as part of `standard' solar modelling.}
Even so, such models computed independently, with recent formulations
of the microphysics, give rather similar results.
In this paper I generally restrict the discussion to standard models,
although discussing the effects of some of the generalizations.
It might be noted that the present Sun is in fact one case where the
standard assumptions may have some validity:
at least the Sun rotates sufficiently slowly that direct dynamical
effects of rotation are likely to be negligible. 
On the other hand, rotation was probably faster in the past
and the loss and redistribution of angular momentum may well have led
to instabilities and hence mixing affecting the present composition profile.

%\notecd [Basic equations (with inspiration from Leiden paper). Macrophysics and
%microphysics. Obviously include diffusion and settling here.]

%\notecd [ ****** START OF THEFT FROM LEIDEN]
With the assumption of spherical symmetry the model is
characterized by the distance $r$ to the centre.
Hydrostatic equilibrium requires a balance between the pressure gradient
and gravity which may then be written as
\be
{\dd p \over \dd r}  = - {G m \rho \over r^2} \; , 
\eel{eq:hydrostat}
where $p$ is pressure, $\rho$ is density,
$m$ is the mass of the sphere contained within $r$, 
and $G$ is the gravitational constant.
Also, obviously,
\be
{\dd m \over \dd r}  =  4 \pi r^2 \rho \; . 
\eel{eq:mass}
The energy equation relates the energy generation to the energy flow
and the change in the internal energy of the gas:
\be
{\dd L \over \dd r}  =  4 \pi r^2 \left[ \rho \epsilon -
\rho {\dd \over \dd t }\left( {e \over \rho }\right)
 + {p \over \rho }{\dd \rho \over \dd t }\right] \; ;
\eel{eq:lum}
here 
$L$ is the energy flow through the surface of the sphere of radius $r$,
$\epsilon$ is the rate of nuclear energy generation%
\footnote{reduced for the emission of neutrinos which escape the star
and hence do not contribute to the energy budget.}
per unit mass and unit time,
$e$ is the internal energy per unit volume and $t$ is time.%
\footnote{For a star evolving in near thermal equilibrium,
such as is the case during main-sequence evolution,
the terms in the time derivatives are small.}
The gradient of temperature $T$ is determined by the requirements of
energy transport, from the central regions where nuclear reactions take
place to the surface where the energy is radiated.
The temperature gradient is conventionally written in terms of
$\nabla = \dd \ln T / \dd \ln p$ as
\be
{\dd T \over \dd r}  =  \nabla {T \over p} {\dd p \over \dd r} \; .
\eel{eq:nabla}
The form of $\nabla$ depends on the mode of energy transport;
for radiative transport in the diffusion approximation
\be
\nabla = \nablarad
\equiv {3 \over 16 \pi a \tilde c  G} {\kappa p \over T^4}{L(r) \over m (r)} 
\; ,
\eel{eq:nablarad}
where $\kappa$ is the opacity,
$a$ is the radiation energy density constant and $\tilde c$ is
the speed of light.
Finally, we need to consider the rate of change of the composition,
which controls stellar evolution.
In a main-sequence star such as the Sun the dominant effect is the
fusion of hydrogen to helium;
however, we must also take into account the changes in composition resulting
from diffusion and settling.
The rate of change of the abundance $X_i$ by mass of element $i$ is
therefore given by
\be
{\partial X_i \over \partial t}
= {\cal R}_i + {1 \over r^2 \rho} {\partial \over \partial r}
\left [ r^2 \rho \left(D_i {\partial X_i \over \partial r} 
+ V_i X_i \right) \right] \; ,
\eel{eq:diffusion}
%\notecd [might need more general equation]
where ${\cal R}_i$ is the rate of change resulting from nuclear
reactions, $D_i$ is the diffusion coefficient and $V_i$
is the settling velocity.
%with similar equations for the abundance of other relevant elements.

To these basic equations we must add the treatment of the microphysics.
This is discussed in Sect.~\ref{sec:microphys} below.

I have so far ignored the convective instability.
This sets in if the density decreases more slowly with position than
for an adiabatic change, {\ie},
\be
{\dd \ln \rho \over \dd \ln p} < {1 \over \Gamma_1} \; ,
\eel{eq:convstab}
where $\Gamma_1 = (\partial \ln p / \partial \ln \rho)_{\rm ad}$,
the derivative being taken for an adiabatic change.
In stellar modelling this condition is often replaced by
\be
{\dd \ln T \over \dd \ln p} \equiv \nabla > \nabla_{\rm ad} 
\equiv \left(\dd \ln T \over \dd \ln p \right)_{\rm ad}  \; ,
\eel{eq:schwarz}
which is equivalent in the case of a uniform composition.%
\footnote{For the complications arising when composition is not uniform,
see for example \citet{Kippen2012}.}
%\notecd [UPDATE: But note that these complications may have to be discussed if
%we consider semiconvection later, as found by \citet{Bahcal2001}].
Thus a layer is convectively unstable if the radiative gradient
$\nablarad$ ({\cf} Eq.~\ref{eq:nablarad}) exceeds $\nabla_{\rm ad}$.
In this case convective motion sets in, with hotter gas rising and cooler
gas sinking, both contributing to the energy transport towards the surface.
The structure of the convective flow should clearly be such that
the combined radiative and convective energy transport
at any point in the convection zone
%through a surface of radius $r$
match the luminosity.
The conditions in stellar interiors are such that complex, possibly
turbulent, flows are expected over a broad range of scales
\citep[e.g.,][]{Schuma2020}.
Also, the convective flux at a given location obviously
represents conditions over a range of positions in the star,
sampled by a moving convective eddy,
so that convective transport is intrinsically non-local.
As a related issue,
motion is inevitably induced outside the immediate unstable region,
also potentially affecting the energy transport and structure, although
this is often ignored.
However, in computations of stellar evolution these complexities 
are almost always reduced to a grossly simplified local description which
allows the computation of the average temperature gradient in terms
of local conditions, as
\be
\nabla = \nabla_{\rm conv}(\rho, T, L, \ldots) \; ,
\eel{eq:nablaconv}
applied in regions of convective instability (see Sect.~\ref{sec:convection}).
%\notecd [This might deserve a reference or two, either here or later
%Check \citet{Nordlu2009a} for review of local treatments.]

%\notecd [ ****** END OF THEFT FROM LEIDEN]

%\notecd [Surface boundary conditions to be discussed in near-surface section].

The equations are supplemented by boundary conditions.
The centre, which is a regular singular point, can be treated through
a series expansion in $r$.
For example, it follows from Eq.~\Eq{eq:mass} for the mass
and Eq.~\Eq{eq:hydrostat} of hydrostatic support that 
\be
m = {4 \over 3} \pi \rho_{\rm c} r^3 + .... \; , \qquad
p = p_{\rm c} - {2 \over 3} \pi \rho_{\rm c}^2 r^2 + .... \; , 
\eel{eq:centexp}
where $\rho_{\rm c}$ and $p_{\rm c}$ are the central density and pressure.
A discussion of the expansions to second significant order in $r$,
and techniques for incorporating them in the central boundary
conditions, was given by \citet{Christ1982}.
At the surface, the model must include the stellar atmosphere.
Since this requires a more complex description of radiative transfer
than provided by the diffusion approximation (Eq.~\ref{eq:frad}),
separately calculated detailed
atmospheric models are often matched to the interior solution,
thus effectively providing the surface boundary condition.
Simpler alternatives are discussed in Sect.~\ref{sec:surface}.

%\notecd [Briefly on solution techniques.
%Might be worth emphasizing precision issues here, and test them in appendix.]

The equations and boundary conditions are most often solved using
finite-difference methods, by what in 
the stellar-evolution community is known as the Henyey technique
\citep[{\eg},][]{Henyey1959, Henyey1964}.%
\footnote{
%\notecd [Something about Newton-Raphson-Kantorovich, in
%more general numerical community.]
The general numerical techniques were presented by \citet{Richtm1957}.
The resulting nonlinear difference equations are solved using the
Newton-Raphson technique, the convergence of which
was demonstrated in the present context 
by Kantorovich \citep[see][]{Henric1962}.
The basic package used in the Aarhus evolution code,
developed by D.~O. Gough,
\citep{Christ1982} consequently goes under the name {\tt nrk},
for Newton--Raphson--Kantorovich 
%\notecd [really needs a suitable Gough reference]
\citep[see][]{Toomre1977}.}
This was discussed in some detail by \citet{Clayto1968} and \citet{Kippen2012}.
%\notecd [In any case these general references, and others,
%should be given early].
The presence of the time dependence, in the energy equation and the
description of the composition evolution, is an additional complication.
The detailed implementation in the Aarhus STellar Evolution Code (ASTEC),
used in the following to compute examples of solar models,
was discussed in some detail by \citet{Christ2008a}.

An important issue is the question of numerical accuracy, in the sense
of providing an accurate solution to the problem, given the assumptions
about micro- and macrophysics.
It is evident that the accuracy must be substantially higher than the
effects of, for example, those potential errors in the physics which
are investigated through comparisons between the models and observations.
{\it Ab initio} analyses of the computational errors are unlikely to
be useful, given the complexity of the equations.
As discussed in Appendix \ref{sec:numacc}, computations with
differing spatial and temporal resolution provide estimates of the
intrinsic precision of the calculation.
Additional tests, which may also uncover errors in programming,
are provided by comparisons between independently computed models,
with carefully controlled identical physics
\citep[\eg,][]{Gabrie1991, ChristReit1995, Lebret2008, Montei2008}.
%(\notecd [usual references, also Lebreton et al., ESTA volume]).

\subsection{Basic properties of the Sun}
\clabel{sec:basicpar}

%\notecd [Discuss values of mass, radius, luminosity, age to be used (with
%a brief discussion of where they come from; age in particular).]

The Sun is unique amongst stars 
in that its global parameters can be determined with high precision.
From planetary motion the product $G M_\odot$ of the gravitational
constant and the solar mass is know with very high accuracy,
as $1.32712438 \times 10^{26} \cm^3 \s^{-2}$.
Even though $G$ is the least precisely determined of the fundamental constants 
this still allows the solar mass to be determined with a precision
far exceeding the precision of the determination of other stellar masses.
The 2014 recommendations of CODATA%
\footnote{Committee on Data for Science and Technology}
\citep{Mohr2016} 
give a value 
$G = 6.67408 \pm 0.00031 \times 10^{-8} \cm^3 \g^{-1} \s^{-2}$,
corresponding to $\Msun = 1.98848 \times 10^{33} \g$.
However, the solar mass has traditionally been taken to be 
$\Msun = 1.989 \times 10^{33} \g$, corresponding to 
$G = 6.672320 \times 10^{-8} \cm^3 \g^{-1} \s^{-2}$;
in the calculations reported in the present paper I use the latter
values of $\Msun$ and $G$, even though these are not entirely consistent
with the CODATA 2014 recommendations.
I note that \citet{Christ2005} found that variations to $G$ and $\Msun$,
keeping their product fixed, had very small effects on the resulting 
solar models.

The angular diameter of the Sun can be determined with very substantial
precision, although the level in the solar atmosphere to which the
value refers obviously has to be carefully specified.
%\notecd [Possibly, but certainly not certainly, Haberreiter ref.]
From such measurements, and the known mean distance between the Earth 
and the Sun, the solar photospheric radius, referring to the point
where the temperature equals the effective temperature, has been
determined as $6.95508 \pm 0.00026 \times 10^{10} \cm$ by \citet{Brown1998};
this was adopted by \citet{Cox2000}.
\citet{Haberr2008} obtained the value 
$6.95658 \pm 0.00014 \times 10^{10} \cm$,
which within errors is consistent with the value of \citet{Brown1998}.
However, most solar modelling has used the older value
$\Rsun = 6.9599 \times 10^{10} \cm$ \citep{Auwers1891},
as quoted, for example, by \citet{Allen1973};
thus, for most of the models presented here I use this value.
%\notecd [And we need to decide what to use for the model(s) to present here!]

From bolometric measurements of the solar `constant' from space the
total solar luminosity can be determined, given the Sun-Earth distance,
if it is assumed that the solar flux is independent of latitude;
although no evidence
has been found to question this assumption, it is perhaps of some concern
that measurements of the solar irradiance have only been made close to
the ecliptic plane.
An additional complication is provided by the variation in solar irradiance
with phase in the solar cycle of around 0.1\%, peak to peak
\citep[for a review, see][]{Frohli2004};
since the cause of this variation is uncertain it is difficult to estimate the 
appropriate luminosity corresponding to equilibrium conditions.
The value $\Lsun = 3.846 \times 10^{33} \erg \s^{-1}$ 
\citep[obtained from the average irradiance quoted by][]{Willso1997}
has often been used and will generally be applied here.
However, recently \citet{Kopp2016} has obtained a revised irradiance,
as an average over solar cycle 23,
leading to $\Lsun = 3.828 \times 10^{33} \erg \s^{-1}$.

The solar radius and luminosity are often used as units in characterizing
other stars, although with some uncertainty about the precise values that
are used.
In 2015 this led to Resolution B3 of the International Astronomical Union%
\footnote{See \url{https://www.iau.org/static/resolutions/IAU2015_English.pdf}}
\citep[see][]{Mamaje2015, Prsa2016},
defining the nominal solar radius ${\cal R}_\odot^N = 6.957 \times 10^8 \m$,
suitably rounded from the value obtained by \citet{Haberr2008},
and the nominal solar luminosity 
${\cal L}_\odot^N = 3.828 \times 10^{33} \erg \s^{-1}$ from \citet{Kopp2016}.

The solar age {\rv $t_\odot$} can be estimated from radioactive dating 
of meteorites combined with a model of the evolution of the solar system,
relating the formation of the meteorites to the arrival of the Sun on
the main sequence.
{\rv Detailed} discussions of meteoritic dating were provided by
Wasserburg, in \citet{Bahcal1995}, {\rv and by \citet{Connel2012}.
Wasserburg found $t_\odot = 4.570 \pm 0.006 \times 10^9$\, years,
with very similar although more accurate values obtained by Connelly {\etal}}
Uncertainties in the modelling of the early solar system obviously
affect {\rv how this relates to solar age.
For simplicity, in the following I simply identify this age with the time
since the arrival of the Sun on the main sequence.%
\footnote{{\rv In fact, according to \citet{Mamaje2009, Connel2012}
the formation of the meteoritic components probably
occur within 3\,Myr after the Sun passes the `birthline' 
(see Fig.~\ref{fig:pms} below) and hence roughly 40\,Myr before it reaches
the zero-age main sequence.
Further constraints on the formation of the meteoritic components,
relative to early solar evolution, were obtained by \citet{Brenne2020}
based on the distribution of molybdenum isotopes in calcium-aluminium-rich
inclusions.}}
Despite the remaining uncertainty this}
still provides an independent measure of a
stellar age of far better accuracy than is available for any other star.

%\notecd [Somewhere also need to discuss, in an overall sense, the determination
%of solar abundances (with later perhaps a separate LR paper!)]

The solar surface abundance can be determined from spectroscopic analysis
\citep[for reviews, see][]{Asplun2005, Asplun2009}.
Additional information about the primordial composition of the solar system,
and hence likely the Sun, is obtained from analysis of meteorites.
A major difficulty is the lack of a reliable determination from spectroscopy
of the solar helium abundance.
Lines of helium, an element then not known from the laboratory,
were first detected in the solar spectrum;%
\footnote{For a brief description of the discovery of helium,
see \citet{Ashbro1968}.}
however, these lines are formed under rather uncertain, and very complex,
conditions in the upper solar atmosphere, making an accurate 
abundance determination from the observed line strengths infeasible;
the same is true of other noble gases, with neon being a particularly important
example.
For those elements with lines formed in deeper parts of
the atmosphere the spectroscopic analysis yields reasonably precise
abundance determinations \citep[\eg,][]{Allend2016};
however, given that the helium abundance is
unknown these are only relative, typically specified as a fraction
of the hydrogen abundance.
Detailed analyses were provided by \citet{Anders1989} 
and \citet{Greves1993}, the latter
leading to a commonly used present ratio $\Zs/\Xs = 0.0245$
between the surface abundances $\Xs$ and $\Zs$ by mass of hydrogen and
elements heavier than helium, respectively.
Also, for most refractory elements there is good agreement between
the solar abundances and those inferred from primitive meteorites.
A striking exception is the abundance of lithium which has been reduced
in the solar photosphere by a factor of around {\rv 150, relative to
the meteoritic abundance \citep{Asplun2009}.}
This is presumably the result of lithium destruction by nuclear reaction,
which would take place to the observed extent over the solar lifetime
at a temperature of around $2.5 \times 10^6 \K$,
indicating that matter currently at the solar surface has been 
mixed down to this temperature.
On the other hand, the abundance of beryllium, which would be destroyed
at temperatures above around $3.5 \times 10^6 \K$, has apparently not been 
significantly reduced relative to the primordial value 
\citep{Balach1998, Asplun2004a},
so that significant mixing has not reached this temperature.
These abundance determinations obviously provide interesting constraints
on mixing processes in the solar interior during solar evolution
(see Sect.~\ref{sec:lightcomp}).

Since 2000 major revisions of solar abundance determinations have been
carried out, through the use of three-dimensional (3D)
hydrodynamical simulations of the solar atmosphere 
\citep[][see also Sect.~\ref{sec:convection}]{Nordlu2009a}.
This resulted in a substantial decrease in the inferred abundances of,
in particular, oxygen, carbon and nitrogen
\citep[for a summary, see][]{Asplun2009},
%\notecd [suitable selection of references; Asplund review; also new ARAA review
%which should be imminent],
resulting in $\Zs/\Xs = 0.0181$.
The resulting decrease in the opacity in the radiative interior has
substantial consequences for solar models and their comparison 
with helioseismic results; I return to this in Sect.~\ref{sec:abundprob}.
%\notecd [Also reference to von Steiger and Zurbuchen 2016.]

Observations of the solar surface show that the Sun is rotating
differentially, with an angular velocity that is highest at the equator.
This was evident already quite early from measurements of the apparent
motion of sunspots across the solar disk \citep{Carrin1863},
and has been observed also in the Doppler velocity of the solar atmosphere.
In an analysis of an extended series of Doppler measurements,
\citet{Ulrich1988} obtained the surface angular velocity $\Omega$ as
\be
{\Omega \over 2 \pi} = (415.5 - 65.3 \cos^2 \theta - 
66.7 \cos^4 \theta ) \, {\rm nHz} \; 
\eel{eq:surfrot}
as a function of co-latitude $\theta$,
corresponding to rotation periods of 25.6\,d at the equator and
31.7\,d at a latitude of $60^\circ$.

As discussed in Sect.~\ref{sec:helio}, helioseismology has provided
very detailed information about the properties of the solar interior.
Here I note that the depth of the solar convection zone has been
determined as $0.287 R$, with errors as small as $0.001 R$
\citep[e.g.,][]{Christ1991b, Basu1997b}.
Also, the effect of helium ionization on the sound speed in the outer parts
of the solar convection zone allows a determination of the solar envelope helium
abundance $Y_{\rm s}$, although with some sensitivity to the equation of state;
the results are close to $Y_{\rm s} = 0.25$
\citep[e.g.,][]{Voront1991b, Basu1998}.

\subsection{Microphysics}
\clabel{sec:microphys}

%\notecd [Stress that this is where the difficulties are, in the present simple
%formulation.
%Illustrate some aspects, using Model~S (likely to be discussed in more
%detail later).]
%
Within the framework of `standard solar models' most of the complexity
in the calculation lies in the determination of the microphysics, and
hence very considerable effort has gone into calculations of the relevant
physics.
In comparing the resulting models with observations,
particularly helioseismic inferences, to test the validity of these
physical results one must, however, obviously keep in mind potential errors in
the approximations defining the standard models.

In this section I provide a relatively brief discussion of the various
formulations that have been used for the physics.
To illustrate some of the effects comparisons are made based on the
structure of the present Sun discussed in more detail in
Sect.~\ref{sec:standard} below.
A detailed discussion of the physics of stellar interiors was provided
by \citet{Cox1968} and updated by \citet{Weiss2004}; 
for a concise review of the treatment of the equation of state and opacity,
see \citet{Dappen2000}.

%most naturally, given the form of the equations, in terms of
%$(p, T, \{X_i\})$, where $X_i$ are the abundances of the relevant elements.
%The determination of $\rho$, $u$ and other required thermodynamical variables
%follows from the equation of state, extensively discussed elsewhere in this
%volume.
%The opacity (describing the absorption and scattering of radiation by the
%constituents of the gas) is most naturally obtained as
%$\kappa = \kappa(\rho, T, \{X_i\})$.
%Evidently the calculation of the opacity requires knowledge of the relevant
%atomic parameters;
%however, the opacity also depends crucially on the thermodynamic state
%of the gas, through the occupation of atomic ionization and excitation
%levels and the perturbation of atomic states by neighbouring particles.
%Similarly, the rates of nuclear reactions depend not only on the nuclear
%parameters but also, as discussed by Shaviv in this volume,
%on the interactions between the particles in the gas which leads to 
%partial screening of the repulsive Coulomb potential between the nuclei.
%Also, calculation of the diffusion and settling coefficients depends
%on the thermodynamic state.

\subsubsection{Equation of state}
\clabel{sec:eos}

%\notecd [Usual discussion (but briefly, and with ample references) to chemical
%and physical picture. Probably show some examples of results, effects,
%already here.]
%
The thermodynamic properties of stellar matter, defined by the equation of
state, play a crucial role in stellar modelling.
This directly involves the relation between pressure, density, temperature
and composition.
In addition, the adiabatic compressibility $\Gamma_1$ affects the
adiabatic sound speed ({\cf} Eq.~\ref{eq:soundspeed}) and hence the
oscillation frequencies of the star, whereas other thermodynamic
derivatives are important in the treatment of convective energy transport.

The treatment of the equation of state involves the determination of all
relevant thermodynamic quantities, for example defined as functions of
$(\rho, T,\{X_i\})$, where $X_i$ are the abundances of the relevant elements;
the composition is often characterized by the abundances $X$, $Y$ and $Z$
by mass of hydrogen, helium and heavier elements with, obviously,
$X + Y + Z = 1$.
This should take into account the interaction between the different 
constituents of the gas, including partial ionization.
Also, pressure and internal energy from radiation must be included,
although they play a comparatively minor role in the Sun.
An important constraint on the treatment is that it be thermodynamically
consistent such that all thermodynamic relations are satisfied between
the computed quantities \citep[\eg,][]{Dappen1993}.
Thus it would not, for example, be consistent to add the contribution of
Coulomb effects to pressure and internal energy without making corresponding
corrections to other quantities, including the thermodynamical potentials
that control the ionization.
%\notecd [Here probably a few D\"appen references!]

A particular problem concerns ionization in the solar core.
As pointed out by, \eg, \citet{Christ1992} %\notecd [not ideal]
straightforward
application of the Saha equation would predict a substantial degree
of recombination of hydrogen at the centre of the Sun, yet the volume
available to each hydrogen nuclei does not allow this.
In fact, ionization must be largely controlled by interactions
between the constituents of the gas, not included in the Saha equation,
and often somewhat misleadingly denoted pressure ionization.
These effects are taken into account in formulations of the equation
of state at various levels of detail,
generally showing that ionization is almost complete in the solar core.
%\notecd [Although one might want to question that?]
The simplest approach, which is certainly not thermodynamically consistent,
is to enforce full ionization above a certain density or pressure.

\begin{figure}[htp]
%\def\epsfsize#1#2{0.5#1}
%\centerline{\epsfbox{\fig/ionzones.eps}}
\centerline{\includegraphics[width=\figwidth]{\fig/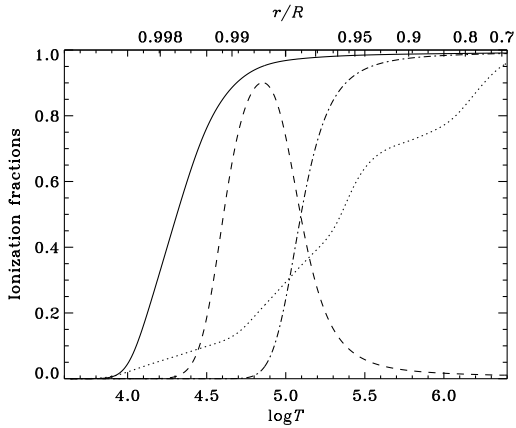}}
\caption{Fractional ionization in a model of the present Sun
(Model~S; see Sect.~\ref{sec:models}), as a function
of the logarithm of the temperature (in K; bottom) and of
fractional radius (top).	
The ionization was calculated with the CEFF equation of state (see below).
The solid curve shows the fraction of ionized hydrogen, the dashed and 
dot-dashed curves the fraction of singly and fully ionized helium,
respectively, and the dotted curve shows the average degree of ionization
of the heavy elements.
}
\clabel{fig:ionization}
\end{figure}

A simple approximation to the solar equation of state is that of a
fully ionized ideal gas, according to which 
\be
p \simeq {\kB \rho T \over \mu \amu} \;, \qquad \nablaad \simeq 2/5 \; , \qquad
\Gamma_1 \simeq 5/3 \; ; 
\eel{eq:idealg}
here $\kB$ is Boltzmann's constant, $\amu$ is the atomic mass unit and
$\mu$ is the mean molecular weight which can be approximated by
\be
\mu = {4 \over 3 + 5 X - Z} \; .
\eel{eq:meanmol}
However, departures from this simple relation must obviously be taken into
account in solar modelling.
The most important of this is
partial ionization, particularly relatively near the surface
where hydrogen and helium ionize.
Figure~\ref{fig:ionization} shows the fractional ionization in a model of
the present Sun. 
As discussed in Section\,\ref{sec:heliostruc} the effects of
the ionization of helium on $\Gamma_1$ provides {\rv a} strong diagnostics
of the solar envelope helium abundance.

Other effects are smaller but highly significant, particularly given
the high precision with which the solar interior can be probed with
helioseismology.
Radiation pressure, $p_{\rm rad} = 1/3 a T^4$, 
and other effects of radiation are small but not entirely negligible.
Coulomb interactions between particles in the gas need to be taken into 
account; a measure of their importance is given by 
\be
\Gamma_{\rm e} = {e^2 \over d_{\rm e} \kB T } \; , \qquad
\hbox{with} \quad
d_{\rm e} = \left({3 \over 4 \pi n_{\rm e}} \right)^{1/3} \; ,
\eel{eq:coulomb}
which determines the ratio between the average Coulomb and thermal energy
of an electron; here $e$ is the charge of an electron,
and $d_{\rm e}$ is the average distance between the electrons,
$n_{\rm e}$ being the electron density per unit volume.
Also, in the core effects of partial electron degeneracy must be included;
the importance of degeneracy is measured by
%\notecd [reference??]
\be
\zeta_{\rm e} = \lambda_{\rm e}^3 n_{\rm e} 
= {4 \over \sqrt{\pi}} F_{1/2} (\psi) \simeq 2 e^\psi \; ,
\eel{eq:deg}
where
\be
\lambda_{\rm e} = {h \over (2 \pi m_{\rm e} \kB T)^{1/2}}
\eel{eq:debroigle}
is the de Broigle wavelength of an electron, $h$ being Planck's 
constant and $m_{\rm e}$ the mass of an electron.
In Eq.~\Eq{eq:deg} $\psi$ is the electron degeneracy parameter
and $F_\nu(\psi)$ is the Fermi integral,
\be
F_\nu(y) = \int_0^\infty {x^\nu \over 1 + \exp(y+x)} \dd x \; .
\eel{eq:fermi}
The last approximation in Eq.~\Eq{eq:deg} is valid for small
degeneracy, $\psi \ll -1$; in this case the correction to the electron
pressure $p_{\rm e}$, relative to the value for an ideal non-degenerate
electron gas, is
\be
{p_{\rm e} \over n_{\rm e} \kB T} -1 \simeq 2^{-5/2} e^\psi
\simeq 2^{-7/2} \zeta_e 
\eel{eq:dpedeg}
\citep[see also][]{Chandr1939}.
Finally, the mean thermal energy of an electron is not negligible compared with
the rest-mass energy of the electron near the solar centre, so relativistic
effects should be taken into account;
their importance is measured by 
\be
x_{\rm e} = {\kB T \over m_{\rm e} \clight^2} \; ;
\eel{eq:elrel}
at the centre of the present Sun $x_{\rm e} \simeq 0.0026$.
As an important example, the relativistic effects cause a change 
\be
{\delta \Gamma_1 \over \Gamma_1}
\simeq - {2 + 2X \over 3 +5 X} x_{\rm e}
\eel{eq:elrelgam}
in $\Gamma_1$,
which is readily detectable from helioseismic analyses \citep{Elliot1998a}.

\begin{figure}[htp]
\centerline{\includegraphics[width=\figwidth]{\fig/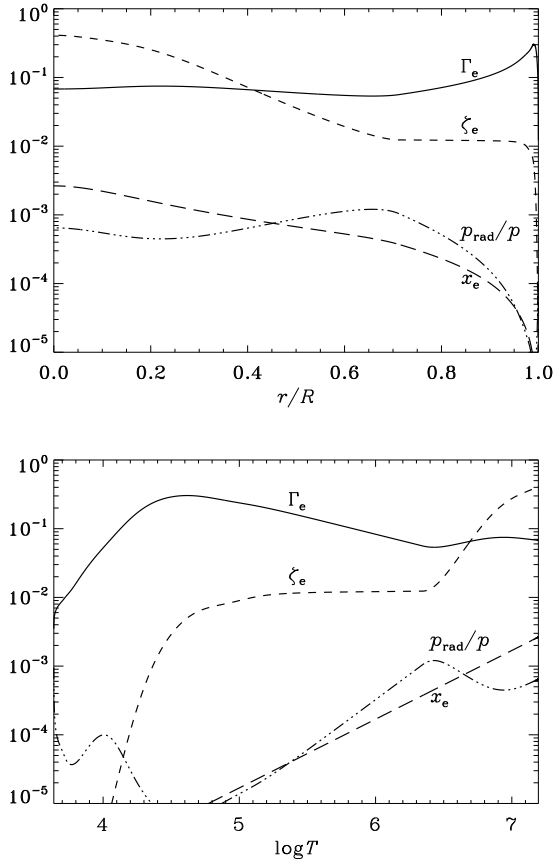}}
\caption{Measures of non-ideal effects in the equation of state in
a model of the present Sun (Model~S; see Sect.~\ref{sec:models}), 
as a function of fractional radius (top panel)
and temperature (bottom panel).
The solid line shows $\Gamma_{\rm e}$ ({\cf} Eq.~\ref{eq:coulomb})
which measures the importance of Coulomb effects.
The short-dashed line shows $\zeta_{\rm e}$
({\cf} Eq.~\ref{eq:deg}) which measures effects of electron degeneracy.
(Note that in $\Gamma_{\rm e}$ and $\zeta_{\rm e}$ the electron number density
was obtained with the CEFF equation of state; see below.)
The long-dashed line shows $x_{\rm e}$ 
({\cf} Eq.~\ref{eq:elrel}),
the ratio between the thermal energy and rest-mass energy of electrons.
Finally, the triple-dot-dashed line shows $p_{\rm rad}/p$,
the ratio between radiation and total pressure.
%\notecd [Might label lines in figure.]
}
\clabel{fig:eos}
\end{figure}

The magnitude of these departures from a simple ideal gas are summarized 
in Fig.~\ref{fig:eos}, for a standard solar model.
Given the precision of helioseismic inferences, none of the effects 
can be ignored.
Coulomb effects are relatively substantial throughout the model, although
peaking near the surface.
Inclusion of these effects, in the so-called MHD equation of state
(see below) was shown by \citet{ChristDL1988} to lead to a substantial
improvement in the agreement between the observed and computed frequencies.
Electron degeneracy has a significant effect in the core of the model
while, as already noted, relativistic effects for the electrons have
been detected in helioseismic inversion \citep{Elliot1998a}.

The computation of the equation of state has been reviewed by
\citet{Dappen1993, Dappen2004, Dappen2007, Dappen2010, Christ1992, Baturi2013}.
%\notecd [Dappen; \citecl{Christ1992}; \citecl{Dappen2004, Dappen2007}.]
Extensive discussions of issues related to the equation of state in
astrophysical systems were provided by \citet{Celebo2004}.
The procedures can be divided into what has been called the chemical picture
and the physical picture.
In the former, the gas is treated as a mixture of different components
(molecules, atoms, ions, nuclei and electrons) each contributing to the
thermodynamical quantities.
Approximations to the contributions from these components are used to
determine the free energy of the system, and the equilibrium state is
determined by minimizing the free energy at given temperature and density,
say, under the relevant stoichiometric constraints.
The level of complexity and, one may hope, realism of the formulation
depends on the treatment of the different contributions to the free energy.
In the physical picture, the basic constituents are taken to be nuclei and
electrons, and the state of the gas, including the formation of ions and
atoms, derives from the interaction between these constituents.
In practice, this is dealt with in terms of activity expansions
\citep{Rogers1981},
the level of complexity depending on the number of terms included.
%\notecd [Likely more to say here (or below); references????].

A simple form of the chemical picture is the so-called EFF equation of state
\citep{Egglet1973}.
%\notecd [Eggleton, Faulkner \& Flannery].
This treats ionization with the basic Saha equation, although adding
a contribution to the free energy which ensures full ionization at
high electron densities.
Partial degeneracy and relativistic effects are covered with 
an approximate expansion.
Because of its simplicity it can be included directly in a stellar
evolution code and hence it has found fairly widespread use;
however, it is certainly not sufficiently accurate to be used for
computation of realistic solar models.
An extension of this treatment, the CEFF equation of state including
in addition Coulomb effects treated in the Debye-H\"uckel approximation,
was introduced by \citet{Christ1992}.
A comprehensive equation of state based on the chemical treatment
has been provided in the so-called MHD%
\footnote{for Mihalas, Hummer and D\"appen}
equation of state \citep{Mihala1988, Mihala1990, Dappen1988, Nayfon1999}.
%\notecd [probably need more complete set of relevant references].
This includes a probabilistic treatment of the occupation of states
in atoms and ions \citep{Hummer1988}, based on the perturbations caused
by surrounding neutral and charged constituents of the gas, and including
excluded-volume effects.
Also, Coulomb effects and effects of partial degeneracy are taken into account.
The MHD treatment and other physically realistic equations of state
are too complex (so far) to be included directly into
stellar evolution codes.
Instead, they are used to set up tables which are then interpolated to
obtain the quantities required in the evolution calculation.
Thus both the table properties and the interpolation procedures become
important for the accuracy of the representation of the physics.
Issues of interpolation were addressed by \citet{Baturi2019}.

The physical treatment of the equation of state, for realistic 
stellar mixtures, has been developed by the OPAL group
at the Lawrence Livermore National Laboratory,
in what they call the ACTEX equation of state (for ACTivity EXpansion),
in connection with the calculation of opacities.
For this purpose it has obviously been necessary to extend the treatment
to include also a determination of atomic energy levels and their
perturbations from the surrounding medium.
The result is often referred to as the OPAL tables.
Extensive tables, in the following OPAL\,1996, were initially provided by 
\citet{Rogers1996}, with later updates presented by \citet{Rogers2002}.
%\notecd [UPDATE: Need input on possible problems with more recent tables,
%inconsistency in $c_V$.]

Interestingly, relativistic effects were ignored in the original
formulations of both the MHD and the OPAL tables,
while they were included, in approximate form, in the simple formulation
of \citet{Egglet1973}.
Following the realization by \citet{Elliot1998a}, based on helioseismology,
that this was inadequate, updated tables taking these effects into
account have been produced by \citet{Gong2001a} and \citet{Rogers2002}.
The latter tables, with additional updates,
are known as the OPAL\,2005 equation-of-state tables%
\footnote{see \href{http://opalopacity.llnl.gov/EOS_2005}{http://opalopacity.llnl.gov/EOS\underline{\hphantom{a}}2005}}
and are seeing widespread use.

\begin{figure}[htp]
\centerline{\includegraphics[width=\figwidth]{\fig/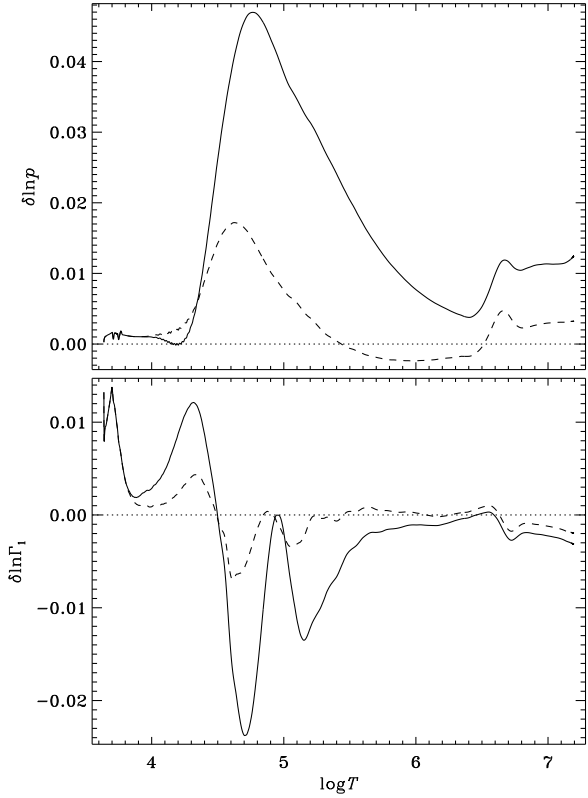}}
\caption{Comparison of equations of state at fixed $(\rho, T)$ and
composition corresponding to the structure of the present Sun
(specifically Model~S of \citet{Christ1996}), in the sense
(modified equation of state) -- (model), plotted against the
logarithm of the temperature in the model;
the model used the original \citep[OPAL\,1996;][]{Rogers1996}
equation of state.
The top panel shows the difference in pressure and the bottom panel the
difference in $\Gamma_1$
Solid lines show the EFF equation of state \citep{Egglet1973},
and dashed lines the CEFF equation of state \citep{Christ1992}.
For the comparison the same relative composition of the heavy elements
was chosen for the EFF and CEFF calculations as in the OPAL tables.
% Uses same composition as OPAL tables:
% C, N, O, Ne = 0.190661 0.055849 0.542979 0.210511
}
\clabel{fig:compeos1}
\end{figure}

\begin{figure}[htp]
\centerline{\includegraphics[width=\figwidth]{\fig/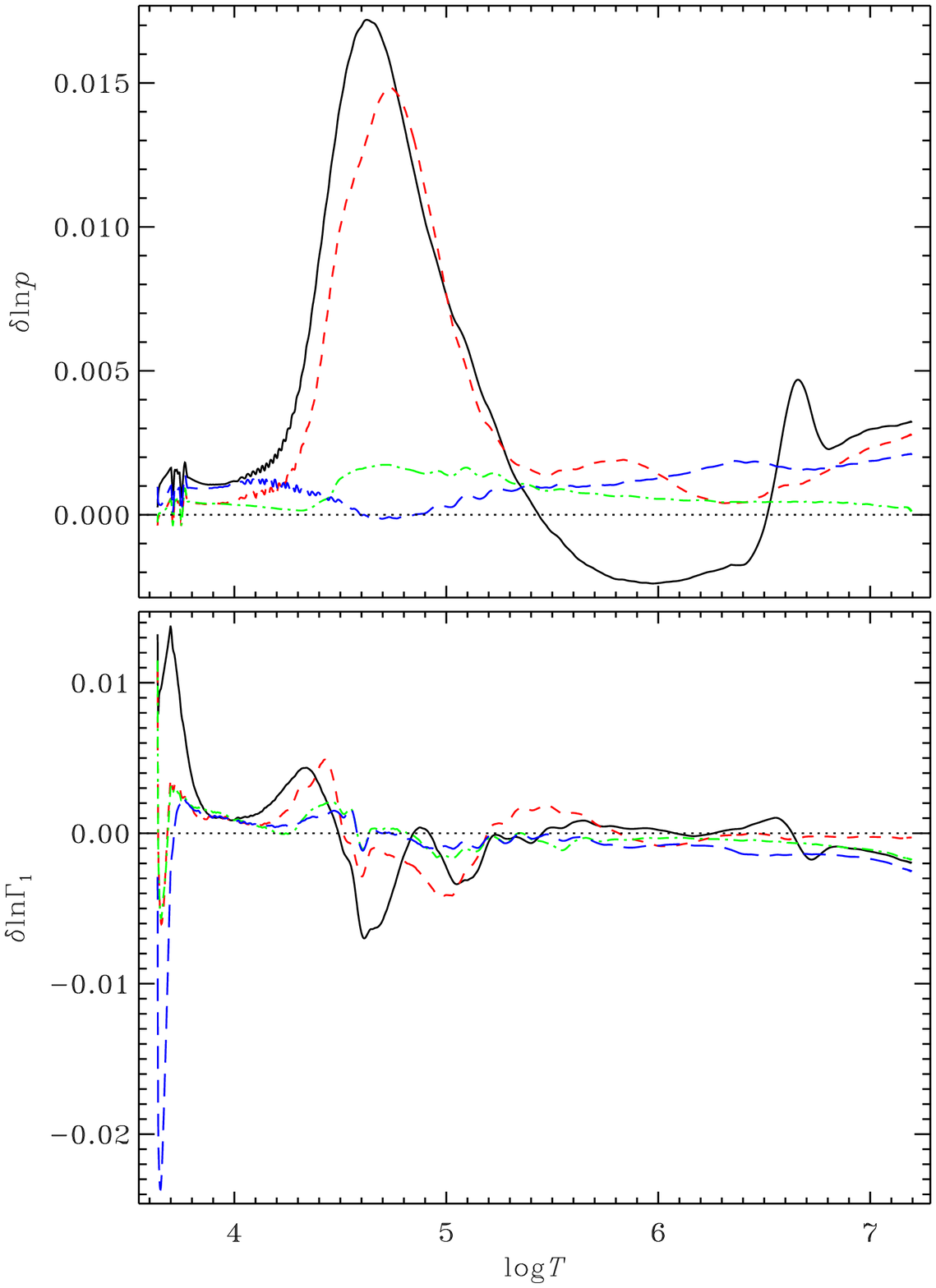}}
\caption{As Fig.~\ref{fig:compeos1},
but showing CEFF (black solid lines), the MHD equation of state
\citep[][red dashed lines]{Mihala1990}
the OPAL\,2005 equation of state \citep[][green dot-dashed lines]{Rogers2002},
and the SAHA-S equation of state \citep[][blue long-dashed lines]{Gryazn2004}.
%\notecd [UPDATE: Need to check MHD compositions, and/or possible updates].
Note that the relative composition of the heavy elements may differ between
the different implementations.
}
\clabel{fig:compeos2}
\end{figure}

To illustrate the effects of using the different formulations,
Figs~\ref{fig:compeos1} and \ref{fig:compeos2}%
\footnote{The analysis of the equation of state and opacity (see below)
used interpolation routines developed by Werner D\"appen and G\"unter Houdek.}
show relative differences in
$p$ and $\Gamma_1$ for various equations of state at the conditions in
a model of the present Sun, using the OPAL\,1996 equation of state as
reference.
It is clear that the inclusion of Coulomb effects in CEFF captures a substantial
part of the inadequacies of the simple EFF formulation, although the remaining
differences are certainly very significant.
In the bottom panel of Fig.~\ref{fig:compeos2} it should be noticed that
the MHD and OPAL\,1996 formulations share the lack of proper treatment
of relativistic effects and hence have very similar behaviour of $\Gamma_1$ 
at the highest temperatures.
This is corrected in both CEFF and OPAL\,2005 which therefore show very
similar departures from OPAL\,1996 at high temperature.
%\notecd [perhaps here or before a note on the importance of $\Gamma_1$
%for helioseismology?]
A detailed comparison between the MHD and OPAL formulations was 
carried out by \citet{Trampe2006}.

%\notecd [Probably also need to discuss briefly other formulations; Irwin etc.
%Check MESA papers for suitable references.]
Further developments of the MHD equation of state have been undertaken to
emulate aspects of the OPAL equation of state in a flexible manner, allowing
the calculation of extensive consistent and physically more realistic 
tables \citep{Liang2004, Dappen2009},
or developing a similar emulation in the simpler CEFF equation of state, 
which might enable bypassing the table calculations \citep{LinDap2010}.
A comprehensive update of the MHD equation of state is being prepared
by R. Trampedach.
The implementation of these developments in solar and stellar model 
calculations will be very interesting. 

An independent development of an equation of state in the chemical picture
has been carried out in the so-called SAHA-S formulation
\citep{Gryazn2004, Baturi2013, Baturi2017}.%
%\footnote{see \url{http://crydee.sai.msu.ru/SAHA-S_EOS/}}
\footnote{\href{http://crydee.sai.msu.ru/SAHA-S_EOS}{http://crydee.sai.msu.ru/SAHA-S\underline{\hphantom{a}}EOS}}. 
Results for this equation of state are shown in Fig.~\ref{fig:compeos2}
with the blue long-dashed curve.
Apart from a rather stronger variation in $\Gamma_1$ in the atmosphere 
{\rv due to the wide variety of molecular species included,}
the SAHA-S formulation is clearly quite similar to OPAL\,2005.
%\notecd [Check new tables for comparisons; 
%link to tables???;
%see Baturin {\etal} (2017; CESAM2k etc.), 
%including more recent Gryaznov references].
Also, Alan W. Irwin has developed the FreeEOS formulation,%
\footnote{available as open source at \url{http://freeeos.sourceforge.net}}
based on free-energy minimization \citep[see][]{Cassis2003a},
which allows rapid calculation of an equation of state
that closely matches the OPAL equation of state.

\subsubsection{Opacity}
\clabel{sec:opac}

%\notecd [UPDATE: Could be interesting to get new plot of dominant opacity sources!
%(With Sylvain Turcotte's help, perhaps??. Or Maelle Lepennec?)]

In stellar interiors, the diffusion approximation for radiative
transfer, implied by Eq.~\Eq{eq:nablarad}, is adequate, and
the opacity is determined as the Rosseland mean opacity,
\be
\kappa^{-1} \equiv \kappa_{\rm R}^{-1} = {\pi \over a \clight T^3}
\int_0^\infty \kappa_\nu^{-1} {\dd B_\nu \over \dd T} \dd \nu 
\eel{eq:rosseland}
\citep{Rossel1924},
where $\kappa_\nu$ is the monochromatic opacity at (radiation) frequency
$\nu$ and $B_\nu$ is the Planck function.
The computation of stellar opacities is generally so complicated %
%\footnote{\notecd [although one could comment on the approximations
%used in Los Alamos solar modelling, at least at some point]}
that opacities have to be obtained in stellar modelling through
interpolation in tables.
The computation of the tables includes contributions of transitions
between the different levels of the atoms and ions in the gas,
including as far as possible the effects of level perturbations;
an extensive review of opacity calculations was provided by
\citet{Pain2017}.
The thermodynamic state of the gas, including the degrees of
ionization and the distribution amongst the levels, is an important
ingredient in the calculation;
indeed, both the MHD and the OPAL equations of state were
developed as bases for new opacity calculations.
Within the convection zone, solar structure is essentially independent
of opacity, since the temperature gradient is nearly adiabatic.
Below the convection zone the opacity is dominated by heavy elements;
hence it is sensitive not only to the total heavy-element abundance $Z$
but also to the relative distribution of the individual elements.
This is illustrated in Fig.~\ref{fig:kapder} showing the sensitivity
of the opacity to variations in the dominant contributions to the
heavy elements.
Evidently iron is an important contribution to the opacity,
particularly in the solar core, but other elements such
as oxygen, neon and silicon also play major roles.
Modelling the solar atmosphere requires low-temperature opacities,
including effects of molecules;
in the calculation of the structure of calibrated solar models the
resulting uncertainties are largely suppressed by changes in the
treatment of convection ({\cf} Fig.~\ref{fig:changesurfopac}).

\begin{figure}[htp]
\centerline{\includegraphics[width=\figwidth]{\fig/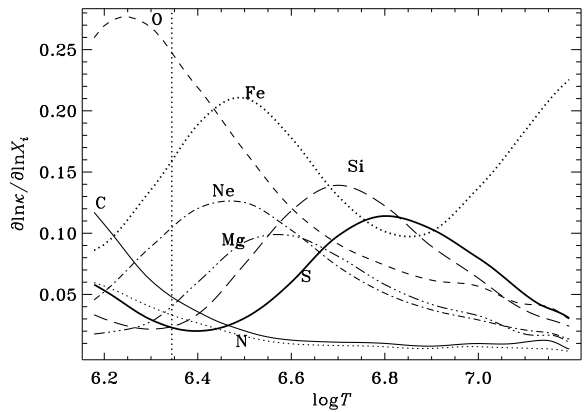}}
\caption{Logarithmic derivatives of the opacity with respect
to contributions to the total heavy-element abundance
of the different elements indicated, evaluated for OPAL opacities
\citep{Iglesi1996} in the radiative part of a standard solar model.
The vertical dotted line marks the temperature at the base of the convection
zone in the present Sun.
Courtesy of H.~M. Antia.
%\notecd [Could be more to say, here or later]
}
\clabel{fig:kapder}
\end{figure}

Early models used for helioseismic analysis generally used the
\citet{Cox1970} and \citet{Cox1976} tables.
An early inference of the solar internal sound speed
\citep{Christ1985} showed that the solar sound speed was higher 
below the convection zone than the sound speed of a model
using the \citet{Cox1976} tables,
prompting the suggestion that the opacity had to be increased
by around 20 per cent at temperatures higher than $2 \times 10^6 \K$.
This followed an earlier plea by \citet{Simon1982} for a reexamination
of the opacity calculations in connection with problems in the
interpretation of double-mode Cepheids and in the understanding
of the excitation of oscillations in $\beta$ Cephei stars;
%\notecd [check the point about stability]
it was subsequently demonstrated by \citet{Andrea1988}
that agreement between observed and computed period ratios for
double-mode $\delta$~Scuti stars and Cepheids could be obtained 
by a substantial opacity increase, by a factor of 2.7, in the range
$\log T = 5.2 - 5.9$.

These results motivated a reanalysis of the opacities by the
Livermore group, who pointed out \citep{Iglesi1987} that the contribution
from line absorption in metals had been seriously underestimated in 
earlier opacity calculations.
This work resulted in the OPAL tables
\citep[{\eg},][in the following OPAL92]{Iglesi1991, Iglesi1992, Rogers1992,
Rogers1994}.
%\notecd [could do with more references here].
Owing to the inclusion of numerous transitions in iron-group elements
and a better treatment of the level perturbations and associated
line broadening these new calculations did indeed show very substantial
opacity increases, qualitatively matching the requirements from the helioseismic
sound-speed inference; also, this led largely to agreement with 
evolution models of the period ratios for
RR Lyrae and Cepheid double-mode pulsators
%\notecd [Cepheids?? See also Cox 1991; ApJ 381, 71, on RR Lyrae stars.]
\citep[\eg,][]{Cox1991, Moskal1992a, Kanbur1994}
%\notecd [a few references]
and to opacity-driven instability in the $\beta$ Cephei models
\citep[\eg,][]{Cox1992, Kiriak1992, Moskal1992b}.
%\notecd [further references].
These results are excellent examples of stellar pulsations, and in
particular helioseismology, providing input to the understanding
of basic physical processes.

\begin{figure}[htp]
\centerline{\includegraphics[width=\figwidth]{\fig/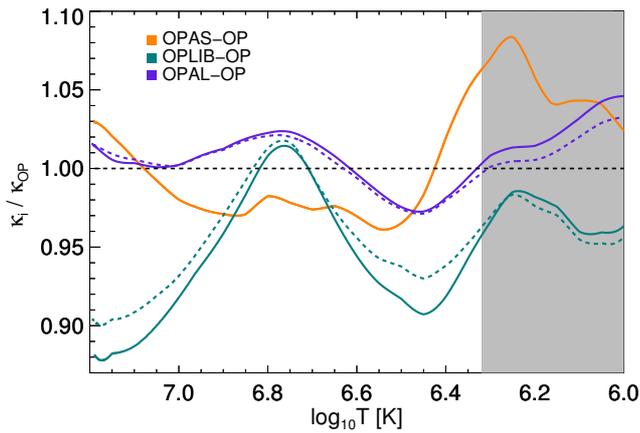}}
\caption{Comparison of the OPAL, OPLIB and OPAS opacities (see text)
relative to the OP opacities.
The dashed curves are for the \citet{Greves1998} composition,
while the solid curves are for the \citet{Asplun2009} composition
(see also Sect.~\ref{sec:newcomp}).
From Villante, Serenelli and Vinyoles (in preparation).
Figure courtesy of Aldo Serenelli.
}
\clabel{fig:opcomp}
\end{figure}

The OPAL tables, with further developments 
\citep[{\eg},][in the following OPAL96]{Iglesi1996},%
\footnote{The tables are available at \url{https://opalopacity.llnl.gov/}}
have seen widespread use in solar and stellar modelling.
In parallel with the OPAL calculations, independent calculations
were carried out within the Opacity Project (OP) \citep{Seaton1994},
with results in good agreement with those of OPAL96 
at relatively low density and temperature,
although larger discrepancies were found under conditions 
relevant to the solar radiative interior \citep{Iglesi1995}.
More recent updates to the OP opacities, in the following OP05,
have decreased these discrepancies
substantially, to a level of $5 - 10$ per cent \citep{Seaton2004, Badnel2005}.%
\footnote{The OP tables are available at
\url{http://cdsweb.u-strasbg.fr/topbase/OpacityTables.html}}
A recent effort is under way at the CEA, France, resulting in the so-called
OPAS tables%
\footnote{available at 
\url{http://cdsarc.u-strasbg.fr/viz-bin/Cat?J/ApJS/220/2\#sRM2.1}}
\citep{Blanca2012, Mondet2015}.
Also, the Los Alamos group has updated their calculations, in the 
OPLIB tables \citep{Colgan2016}.%
\footnote{ available at \url{http://aphysics2.lanl.gov/opacity/lanl/}}
A review of these recent opacity results was provided by \citet{Turck2016},
while Fig.~\ref{fig:opcomp} shows a comparison of the opacity values
in a model of the present Sun.
%I return to opacity issues in connection with updates to the inferred solar
%surface composition in Sect.~\ref{sec:modcorcomp}.

The opacity tables discussed so far typically include few or no molecular lines.
Thus the opacity at low temperature (often taken to be below $10^4 \K$)
must be obtained from separate tables, suitably matched to the opacity
at higher temperature.
Tables provided by \citet{Kurucz1991} and \citet{Alexan1994}
have often been used.
A set of tables with a more complete equation of state
and improved treatment of grains was provided by \citet{Fergus2005}.
%\notecd [Update?].

I note that the potential uncertainties in the opacity calculations have
gained renewed interest in connection with the apparent discrepancies 
between helioseismic inferences and solar models computed with revised
inferences of solar surface composition.
I return to this in Sect.~\ref{sec:modcorcomp}.

\subsubsection{Energy generation}
\clabel{sec:engenr}

The basic energy generation in the Sun takes place through hydrogen fusion
to helium which may be schematically written as
\be
4 \hyd \rightarrow \helfour + 2 {\rm e}^+ + 2 \nu_{\rm e} \; .
\eel{eq:hydfus}
Here the emission of the two positrons results from the required conversion
of two protons to neutrons, as also implied by conservation of charge in
the process, and the two electron neutrinos ensure conservation of lepton
number.
Evidently the positrons are immediately annihilated by two electrons,
resulting in further release of energy.
Thus the net reaction can formally be regarded as the fusion of 
four hydrogen \emph{atoms} into a helium atom; this is convenient from
the point of view of calculating the energy release based on tables 
of atomic masses. 
The result is that each reaction in Eq.~\Eq{eq:hydfus}
generates $26.73 \MeV$.
However,
the neutrinos have a negligible probability for interaction with matter
in the Sun, and hence the energy contributed to the neutrinos must be
subtracted to obtain the energy generation rate $\epsilon$
actually available to the Sun.
Thus $\epsilon$ depends on the energy of the emitted neutrinos
and hence on the details of the reactions resulting in the net reaction
in Eq.~\Eq{eq:hydfus}.
As discussed in Sect.~\ref{sec:neutr} 
%\notecd[already discussion in the introduction?? Probably not.]
detection of the emitted neutrinos provides 
a crucial confirmation of the presence of nuclear reactions in the
solar core and a probe of the properties of the neutrinos.

%\notecd [Neutrino losses, point to later section on neutrino measurements
%(which should be brief, in anticipation of full paper).]

The detailed properties of nuclear reactions in stellar interiors have
been discussed by, for example, \citet{Clayto1968}.
Reactions require tunneling through the potential barrier resulting
from the Coulomb repulsion between the two nuclei.
Thus to a first approximation reactions between more highly charged
nuclei are expected to have a lower probability.
Also, the temperature dependence of the reactions depends strongly
on the charges of the reacting nuclei.
The dependence on temperature of the reaction rate $r_{12}$ between two nuclei
1 and 2 is often approximated as $r_{12} \propto T^n$, where 
\be
n = {\eta -2 \over 3} \; , \qquad 
\eta = 42.487 ({\cal Z}_1 {\cal Z}_2 {\cal A})^{1/3} T_6^{-1/3} \; ;
\eel{eq:tdep}
here ${\cal Z}_1 e$ and ${\cal Z}_2 e$ are the charges of the two nuclei,
${\cal A} = {\cal A}_1 {\cal A}_2/( {\cal A}_1 + {\cal A}_2)$ is the
reduced mass of the nuclei in atomic mass units, 
${\cal A}_1$ and ${\cal A}_2$ being the masses of the nuclei,
and $T_6 = T/(10^6 \K)$.%
\footnote{The numerical constant is based on the CODATA 1986
recommendations \citep{Cohen1987}.}
However, the specific properties of the interacting nuclei also play
a major role for the reaction rate.
Furthermore, 
the conversion of protons into neutrons and the production of neutrinos
involve the \emph{weak interaction} which takes place with comparatively low
probability.
This has a strong effect on the rates of reactions where this conversion
takes place.

The net reaction in Eq.~\Eq{eq:hydfus} obviously has to take place through
a number of intermediate steps.
%As \notecd [quite likely] discussed in \notecd [the introduction]
The dominant series of reactions starts directly with the fusion of
two hydrogen nuclei; 
the full sequence of reactions is%
\footnote{using the notation $A(a,b)B$ for the reaction
$A + a \rightarrow B + b$, including fairly obvious generalizations}
\be
\hyd(\hyd, \eplus \nu_{\rm e})\,\deut(\hyd, \gamma)\,\helthree
(\helthree,2 \hyd)\, \helfour \; .
\eel{eq:PPI}
This sequence of reactions is known as the \emph{PP-I chain} and
clearly corresponds to Eq.~\Eq{eq:hydfus}.
The average energy of the neutrinos lost in the first reaction in the chain
is $0.263 \MeV$. 
Thus the effective energy production for each resulting $\helfour$ is
$26.21 \MeV$.

Two alternative chains, PP-II and PP-III, continue with the fusion of
$\helthree$ and $\helfour$ after the production of $\helthree$:
\bea
\label{eq:PPII}
&& \helthree(\helfour, \gamma)\,\berseven(\eminus, \nu_{\rm e})\,\litseven
(\hyd, \helfour)\,\helfour \phantom{{\rm He})\,\helfour}\qquad
(\PPII) \nonumber \\
&& \phantom{\helthree(\helfour, \gamma)}
 \quad \Downarrow \\
&& \phantom{\helthree(\helfour, \gamma)\, }
\berseven(\hyd, \gamma)\, \boreight(,\eplus \nu_{\rm e})\,\bereight
(, \helfour)\,\helfour \qquad (\PPIII) \nonumber
\eeanl{eq:PPII}
Here the total average neutrino losses per produced $\helfour$ are
$1.06 \MeV$ and $7.46 \MeV$, respectively. 
At the centre of the present Sun the contributions of the PP-I, PP-II and
PP-III reactions to the total energy generation by the PP chains,
excluding neutrinos,
are 23, 77 and 0.2 per cent, respectively;
owing to a much higher temperature sensitivity of the PP-II and PP-III chains
the corresponding contributions to the solar luminosity are
77, 23 and 0.02 per cent.
However, even though insignificant for the energy generation, 
the PP-III chain is very important for the study of neutrino emission
from the Sun due to the high energies of the neutrinos emitted in
the decay of $\boreight$.

Of the reactions in the PP chains the initial reaction, fusing
two hydrogen nuclei, has by far the lowest rate per pair of reacting nuclei.
This is a result of the effect of the weak interaction in the conversion
of a proton into a neutron, coupled with the penetration of the
Coulomb barrier.%
\footnote{The other two reactions in the PP chains 
requiring the weak interaction 
involve positron emission or electron capture and hence take place
rapidly, compared with the nuclear reactions.}
Thus the overall rate of the chains is controlled by this reaction;
since the charges of the interacting nuclei is relatively low,
it has a modest temperature sensitivity, approximately as $T^4$
({\cf} Eq.~\Eq{eq:tdep}).
The distribution of the reactions between the different branches
depends on the branching ratios at the reactions destroying
$\helthree$ and $\berseven$; as a result PP-II and in particular
PP-III become more important with increasing temperature, 
with important consequences for the neutrino spectrum of the Sun.

In principle, the full reaction network should be considered as a function
of time, to follow the changing abundances resulting from the nuclear
reactions.
In practice the relevant reaction timescales for the reactions involving
$\deut$, $\berseven$ and $\litseven$ are so short that the reactions
can be assumed to be in equilibrium under solar conditions
({\eg}, \citecl{Clayto1968}); the resulting equilibrium abundances
are minute.%
\footnote{However, the details of these reactions, including deuterium
burning, are important during pre-main-sequence evolution.}
On the other hand, the timescales for the reactions involving $\helthree$
are comparable with the timescale of solar evolution, at least in the
outer parts of the core;
thus the calculation should follow the detailed evolution with time
of the $\helthree$ abundance.
The resulting abundance profile in a model of the present Sun
is illustrated in Fig.~\ref{fig:he3};
below the maximum $\helthree$ has reached nuclear equilibrium, 
with an abundance that increases with decreasing temperature.
The location of this maximum moves further out with increasing age.
It was found by \citet{Christ1974} that the establishment of this $\helthree$
profile caused instability to a few low-degree g modes early in the
evolution of the Sun, {\rv although the analysis neglected potentially important
nonadiabatic effects in the convective envelope.
In a more reliable calculation
\citet{Sonoi2012} found a similar instability in low-mass very low-metallicity
stars, where the low metallicity reduces the effect of the convection zone
on the mode instability.
They noted that the oscillation-induced mixing originally proposed for
the Sun by \citet{Dilke1972} could have a significant effect on 
the evolution of these stars.}

\begin{figure}[htp]
\centerline{\includegraphics[width=\figwidth]{\fig/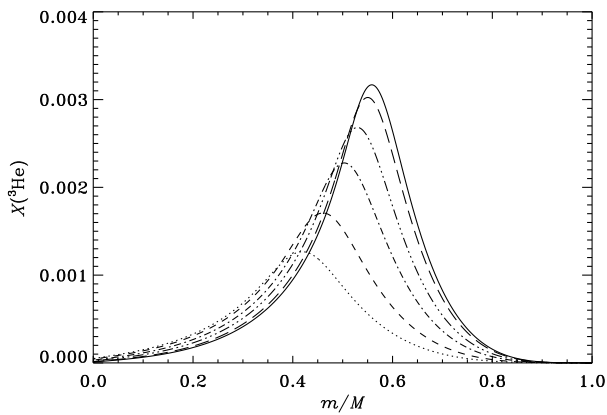}}
\caption{
Evolution of the abundance of ${}^3 {\rm He}$.
The solid curve shows the abundance in a model of the present Sun,
while the dotted, dashed, dot-dashed, triple-dot-dashed and 
long-dashed curves show the abundances at ages
0.5, 1.0, 2.0, 2.9 and 3.9 Gyr, respectively.
The initial abundance was assumed to be zero.
%\notecd [Note, unless changed, that this started with zero initial abundance.]
}
\clabel{fig:he3}
\end{figure}

The primordial abundances of light elements, as inferred from
solar-system abundances, are crucial constraints on models of the
Big Bang \citep[{\eg}][]{Geiss2007}.
This includes the abundances of $\deut$ and $\helthree$,
with $\deut$ burning ({\cf} Eq.~\ref{eq:PPI}) taking place at 
sufficiently low temperature that the primordial $\deut$ has largely
been converted to $\helthree$.
The $\helthree/\helfour$ ratio can be determined from the solar wind;
the resulting value can probably be taken as representative for
matter in the solar convection zone and hence provides a constraint on
the extent to which the convection zone has been enriched by 
$\helthree$ resulting from hydrogen burning.
This was used by, for example, \citet{Schatz1981}, \citet{Lebret1987} and
\citet{Vaucla1998} to constrain the extent of turbulent mixing
beneath the convection zone.
\citet{Heber2003} investigated the time variation in
the $\helthree/\helfour$ ratio from analysis of lunar regolith samples.
After correction for secondary processes, using the presumed constant
${}^{20}{\rm Ne}/{}^{22}{\rm Ne}$ as reference, they deduced that the
$\helthree/\helfour$ ratio has been approximately constant over the
past around 4\,Gyr, with an average value for the ratio of number densities
of $(4.47 \pm 0.13) \times 10^{-4}$.
This provides a further valuable constraint on the mixing history below the
solar convection zone.%
\footnote{The production of $\helthree$ in hydrogen burning, with
subsequent dredge-up in the red-giant phase, could be expected to lead to
general enrichment which is not observed in the Galaxy.
It was pointed out by \citet{Egglet2006} and \citet{Charbl2007} that 
thermohaline instability resulting from molecular-weight inversion
caused by $\helthree$ burning, as noted earlier by \citet{Ulrich1972},
could cause mixing in the radiative interior of red giants and hence
additional destruction of $\helthree$, accounting for this discrepancy
\citep[see also][]{Angelo2011}.
{\rv I note, however, that \citet{Deniss2011} and \citet{Maeder2013} 
questioned whether the efficiency of the thermohaline mixing would be
sufficient.}}

%\notecd [Needs a few comments on He3 equilibrium and final profile, probably.
%Somewhere this also needs to be related to observed abundances, in solar
%wind or otherwise; Maeder \& Lebreton? More recent papers, surely].

\begin{figure}[htp]
\centerline{\includegraphics[width=\figwidth]{\fig/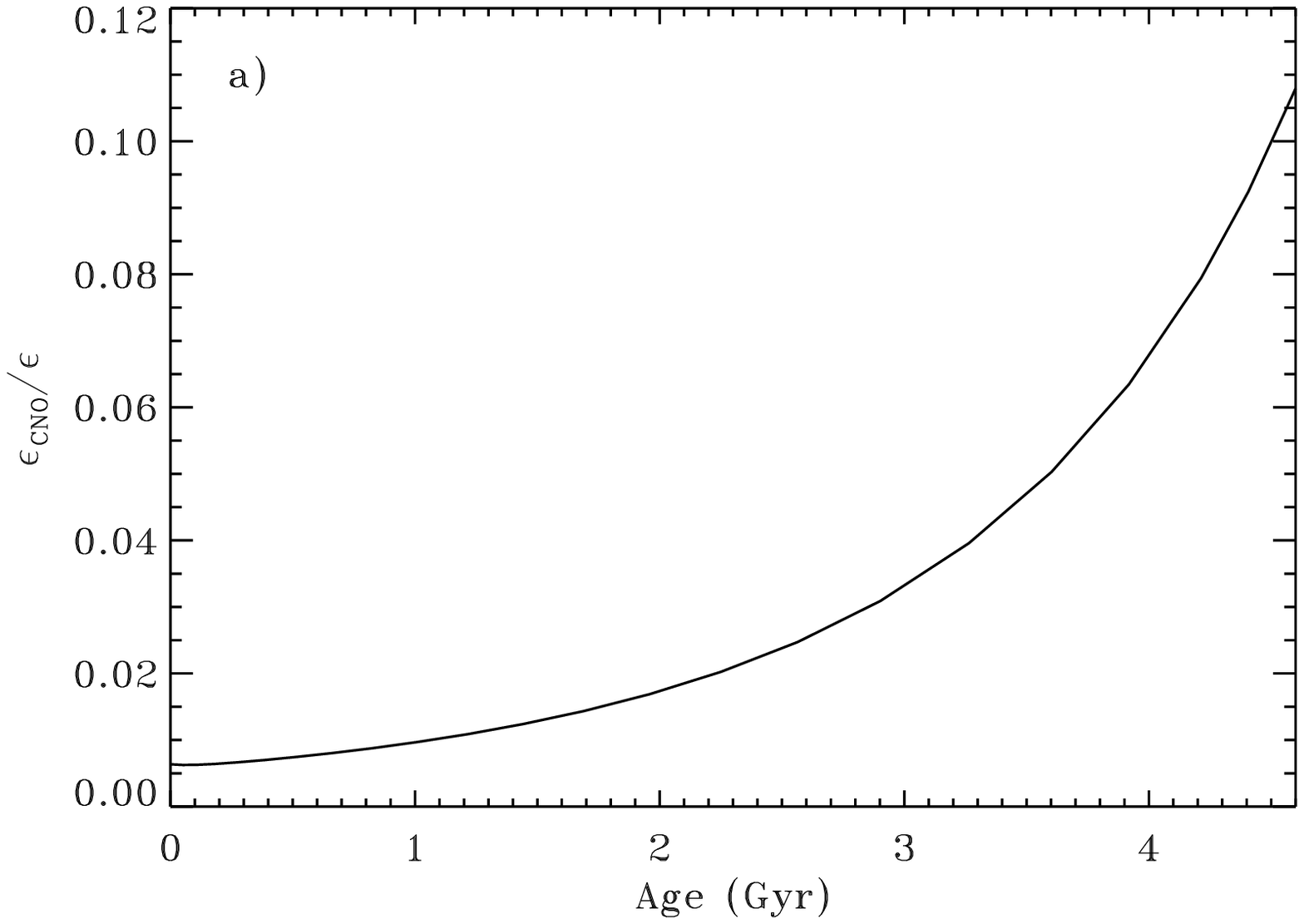}}
\centerline{\includegraphics[width=\figwidth]{\fig/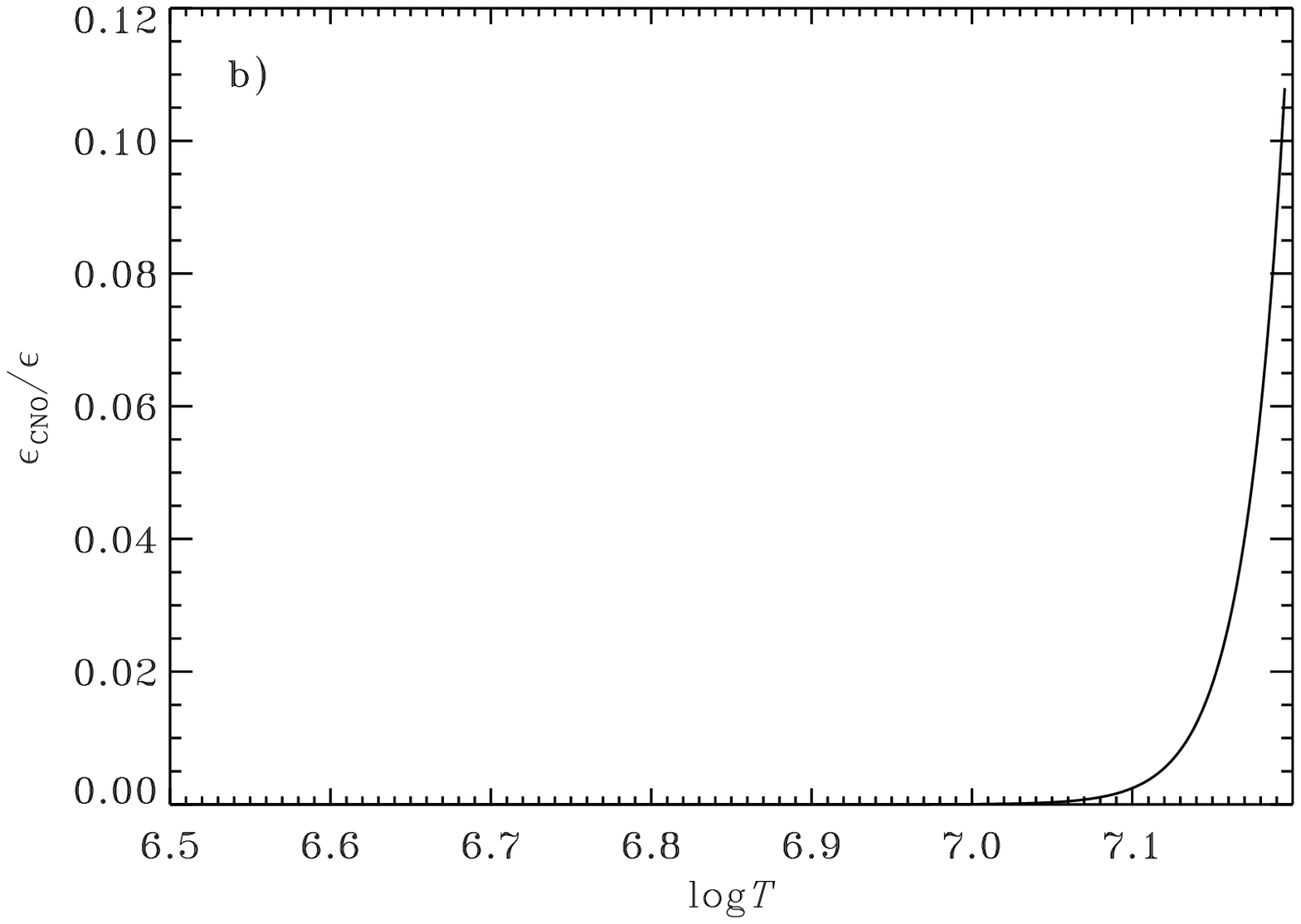}}
\caption{
Contributions of the CNO cycle to the energy generation in a solar model. 
Top panel: the ratio of $\epsilon_{\rm CNO}$ to the total $\epsilon$ at 
the centre of the model, as a function of age.
Bottom panel: the fractional contribution $\epsilon_{\rm CNO}/\epsilon$ as a
function of position in a model of present Sun.
}
\clabel{fig:cno}
\end{figure}

%\notecd [As probably also discussed in the introduction]
A second set of processes resulting in the net reaction 
in Eq.~\Eq{eq:hydfus} involves successive reactions with
isotopes of carbon, nitrogen and oxygen:

\vbox{
\be
\cartwelve(\hyd, \gamma)\,\nitthirteen(\eplus \nu_{\rm e})\,\carthirteen
(\hyd, \gamma)\,\nitfourteen(\hyd, \gamma)\,\oxyfifteen
(\eplus \nu_{\rm e})\,\nitfifteen(\hyd, \helfour)\,\cartwelve \; .
\eel{eq:CNO}
\setlength{\unitlength}{1mm}
\begin{picture}(15,0)
\thicklines
\put(21,0){\vector(0,4){5}}
\put(20.8,0){\line(4,0){104.2}}
\put(125,0){\line(0,4){5}}
\end{picture}
}
\noindent
This \emph{CNO cycle} is obviously a catalytic process, with the net result of 
converting hydrogen into helium.
The reaction with the lowest rate in this cycle is proton capture
on $\nitfourteen$ which therefore controls the overall rate of the cycle;
this leads to a temperature dependence of roughly $T^{20}$ under
solar conditions, owing to the high nuclear charge of nitrogen
({\cf} Eq.~\Eq{eq:tdep}).
As a result, the CNO cycle is significant mainly very near the solar
centre, and its importance increases rapidly with increasing age of the
model, due to the increase in core temperature
({\cf} Fig.~\ref{fig:cno}a).
Owing to the strong temperature dependence it is strongly 
concentrated near the centre, as illustrated in Fig.~\ref{fig:cno}b.
Thus, although in the present Sun the central contribution
to the energy-generation rate is 11 per cent,
the CNO cycle only contributes 1.3 per cent to the luminosity.
As a consequence of the ${}^{14}{\rm N}$ bottleneck in the CN cycles
almost all the initial carbon is converted into nitrogen by the reactions.
An additional side branch mainly serves to convert oxygen into nitrogen;
under the conditions leading up to the present Sun
this is relatively unimportant, causing an increase in the central abundance
of $\nitfourteen$ by around 12 per cent in the present Sun, relative to
the initial abundance.

%\notecd [Make plots of CNO contribution at centre, as function of age,
%and as function of position in the present Sun].

%\notecd [Discuss different compilations of data (a comparison would be good, but
%is perhaps too much, for now?)]

The computation of nuclear reaction rates requires nuclear parameters,
determined from experiments or, in the case of the $\hyd + \hyd$ reaction,
from theoretical considerations.
In addition to affecting the energy-generation rate the details of
the reactions have a substantial effect on the branching ratios
in the PP chains and hence on the production rate of the high-energy
$\boreight$ neutrinos.
The reaction rate, averaged over the thermal energy distribution of the
nuclei, is typically expressed as a function of temperature in terms
of a factor describing the penetration of the Coulomb barrier%
\footnote{This gives rise to the dominant temperature dependence
described by Eq.~\Eq{eq:tdep}.}
and a correction factor provided as an expansion in temperature.
A substantial number of compilations of data for nuclear reactions
have been made, starting with the classical, and much used,
sets by \citet{Fowler1967, Fowler1975}.
\citet{Bahcal1995} provided an updated set of parameters
specifically for the computation of solar models.
Two extensive and commonly used compilations of parameters have been
provided by \citet{Adelbe1998} and \citet{Angulo1999}.
Revised parameters for the important reaction
$\nitfourteen(\hyd, \gamma)\,\oxyfifteen$, which controls the overall rate
of the CNO cycle, have been obtained \citep{Formic2004, Angulo2005},
reducing the rate by a factor of almost 2.
An updated set of nuclear parameters specifically for solar modelling was
provided by \citet{Adelbe2011},
including also the revised rates for $\nitfourteen(\hyd, \gamma)\,\oxyfifteen$.
%\notecd [Somewhere comment on small effect on solar models, based on
%model ...july05/l8bi.d.13 computed for S.B. This may still need a check.]
%\notecd [Possibly further updates.]
%\notecd [Could plot total energy generation rate and $\boreight$ rate for
%B\&P95, Adelberger and Angulo?
%Note however that what we call B\&P95 (for {\tt ivreng = 6}) might be
%a little questionable!]

%\notecd [Try to make sense of electron screening, with reference also to recent
%Shaviv results, claims.]
The nuclear reactions take place in a plasma, with charged particles
that modify the interaction between the nuclei.
A classical and widely used treatment of this effect was developed by
\citet{Salpet1954}, with a mean-field treatment of the plasma in
the Debye-H\"uckel approximation;
this shows that the nuclei are surrounded by clouds of electrons which 
partly screen the Coulomb repulsion between the nuclei and hence
increase the reaction rate.
Following criticism of Salpeter's result by \citet{Shaviv1996},
\citet{Brugge1997, Brugge2000} made a more careful analysis of the
thermodynamical assumptions underlying the derivation, 
confirming Salpeter's result and in the second paper
extending it to take into account quantum-mechanical exclusion and
polarization of the screening cloud;
in the solar case, however, such effects are largely insignificant.
On the other hand, the mean-field approximation may be questionable in cases,
such as the solar core, where the average number of electrons within the
radius of the screening cloud is very small.
This has given rise to extensive discussions of dynamic effects in
the screening \citep[{\eg}][]{Shaviv2001}.
\citet{Bahcal2002} argued that such effects, and other claims of
problems with the Salpeter formulation, were irrelevant.
However, molecular-dynamics simulations of stellar plasma
strongly suggest that dynamical effects may in fact substantially
influence the screening \citep{Shaviv2004a, Shaviv2004b}.
Further investigations along these lines are clearly needed.
Thus it is encouraging that \citet{Mussac2007} started 
independent molecular-dynamics simulations.
Initial results by the group \citep{Mao2009}
confirmed the earlier conclusions by Shaviv;
a more detailed analysis by \citet{Mussac2011} found evidence for a
slight \emph{reduction} in the reaction rate as a result of plasma effects.
%\notecd [Check for updates. Certainly need Mussack's 2011 papers].
Interestingly, \citet{Weiss2001} noted that the solar structure as inferred
from helioseismology ({\cf} Sect.~\ref{sec:heliostruc}) can be used
to constrain the departures from the simple Salpeter formulation;
in particular, they found that a model computed assuming no screening
was inconsistent with the helioseismically inferred sound speed.
These issues clearly need further investigations.

\subsubsection{Diffusion and settling}
\clabel{sec:diffus}

%\notecd [Usual Michaud \& Proffitt story. Discuss also other formulations, but
%briefly.]
%
As indicated in Eq.~\Eq{eq:diffusion} the temporal evolution of
stellar internal abundances must take into account effects of
diffusion and settling.
Crudely speaking, settling due to gravity and thermal effects tends
to establish composition gradients; diffusion, described by
the diffusion coefficient $D_i$, tends to smooth out such gradients,
including those that are established through nuclear reactions.
A brief review of these processes was provided by \citet{Michau1993}.
They were discussed in some detail already by \citet{Edding1926};
he concluded that they might lead to unacceptable changes in surface
composition unless suppressed by processes that redistributed the
composition, such as circulation.
%\notecd [Check again for Eddington statements; referred to in M\&P].
%\notecd [V. slow processes in deep stellar interiors, hence insignificant?
%However, potentially dominant in outer parts of the star and hence
%must be counteracted].

%\notecd [Might be worth looking at Chapman \& Cowling (1970; 3rd edition).]

A brief review of diffusion was provided by \citet{Thoul2007}.
The basic equations describing the microscopic motion of matter in a star
are the Boltzmann equations for the velocity distribution of each
type of particle.
The treatment of diffusion and settling in stars has generally been
based on approximate solutions of the Boltzmann equations 
presented  by \citet{Burger1969}.
This results in a set of equations for momentum, energy and mass 
conservation for each species which can be solved numerically 
to obtain the relevant quantities such as $D_i$ and
$V_i$ in Eq.~\Eq{eq:diffusion}.
The equations depend on the collisions between particles in the gas,
greatly complicated by the long-range nature of the Coulomb force
between charged particles (electrons and ions);
these are typically described in terms of coefficients
based on the screened Debye-H\"uckel potential,
mentioned above in connection with Coulomb effects in the equation of state
and electron screening in nuclear reactions,
and depending on the ionization state of the ions.
As emphasized initially by \citet{Michau1970}
the gravitational force on the particles may be modified by radiative
effects, depending on the detailed ionization and excitation state
of the individual species and hence varying strongly between different
elements or with position in the star.%
\footnote{Interestingly, already \citet{Edding1926} pointed out the
possibility of such differential effects of radiation pressure.}
It should be noted that the typical diffusion and settling timescales,
although possibly short on a stellar evolution timescale, are generally
much longer than the timescales associated with large-scale
hydrodynamical motions.
Thus regions affected by such motion, particularly convection zones,
can generally be assumed to be fully mixed;
in the solar case microscopic diffusion and settling is only relevant
beneath the convective envelope.
Formally, hydrodynamical mixing can be incorporated by maintaining
Eq.~\Eq{eq:diffusion} but with a very large value of $D_i$
\citep[{\eg},][]{Egglet1971}.
%\notecd [is there not a relevant Eggleton reference for this?].
%\notecd [UPDATE: Also very briefly for convection in \citet{Richer2000},
%again with $D_{\rm T} \simeq \langle v \ell \rangle$;
%although hardly needs reference here.
%Something like this is used in setting {\tt difcon} in 
%{\tt s/r evolrhs.d...}, it seems,
%with $D_{\rm T} = \langle v \ell \rangle/3$.]
%\notecd [UPDATE: \citet{Richar2001} for semiconvection, in connection with
%\citet{Bahcal2001}].

%\notecd [Summarize M\&P; refer to detailed expressions in ESTA paper.]
%\notecd [Discuss also comparison with other formulations,
%including \citet{Thoul1994}].

\citet{Michau1993} presented relatively simple approximations to
the diffusion and settling coefficients for hydrogen as well
as for heavy elements regarded as trace elements
\citep[see also][]{Christ2008a}.
These were based on solutions of Burger's equations, adjusting coefficients
to obtain a reasonable fit to the numerical results.
These approximations were also compared with the results of the numerical
solutions by \citet{Thoul1994} who in addition presented simpler,
and rather less accurate, approximate expressions for the coefficients.

%\notecd [Need to check what was done in earlier papers: Noerdlinger;
%Wambsganss; Cox et al., Bahcall \& Loeb (1990).
%\citet{Proffi1991} give a fairly extensive
%comparison of these early results, in fact].

Although diffusion and settling have been considered since the early
seventies \citep[\eg,][]{Michau1970} to explain peculiar abundances in
some stars, it seems that \citet{Noerdl1977} was the first to include
these effects in solar modelling; indeed, the early estimates by
\citet{Edding1926} suggested that the effects would be fairly small.
In fact, including helium diffusion and settling
Noerdlinger found a reduction of about 0.023 in the
surface helium abundance $\Ys$, from the initial value.
Roughly similar results were obtained by \citet{Gabrie1984} and
\citet{Cox1989}, the latter authors considering a broad range of
elements, while \cite{Wambsg1988} found a much smaller reduction.
\citet{Proffi1991} provided a detailed comparison of these early results,
although without explaining the discrepant value found by Wambsganss.
\citet{Bahcal1992a, Bahcal1992b} made careful calculations of models with
helium diffusion and settling,
using the then up-to-date physics, and emphasizing
the importance of calibrating the models to yield the observed present
surface ratio $\Zs/\Xs$ between the abundances of heavy elements and
hydrogen; they found that the inclusion of diffusion and settling increased
the neutrino capture rates from the models by up to around 10 per cent.
A careful analysis of the effects of heavy-element diffusion and settling
on solar models and their neutrino fluxes was presented by 
\citet{Proffi1994}.

\citet{Gabrie1984} concluded that the inclusion of helium diffusion and
settling had little effect on the oscillation frequencies of the model,
while \citet{Cox1989}, in their more detailed treatment, actually found
that the model with diffusion and settling showed a larger difference
between observed and model frequencies than did the model that did not include
these effects.
However, \citet{Christ1993} showed that the inclusion of helium diffusion 
and settling substantially decreased the difference in sound speed between
the Sun and the model, as inferred from a helioseismic 
differential asymptotic inversion.
Further inverse analyses of observed solar oscillation frequencies have
confirmed this result, thus strongly supporting the reality of these effects
in the Sun and contributing to making diffusion and settling a part of
`the standard solar model' \citep[\eg,][]{Christ2007}.
Further evidence is the difference between the initial helium abundance
required to calibrate solar models and the helioseismically inferred
envelope helium abundance (see Sect.~\ref{sec:heliostruc}), which is
largely accounted for by the effects of helium settling.
%\notecd [e.g. C-D \& Di Mauro Porto2006; this also addresses effect of new
%abundances, although this will be discussed later].
%\notecd [At some point might need to state that including diffusion also
%helps on envelope helium abundance, compared with helioseismology,
%and on the depth of the convection zone.
%Discuss in section on helioseismology (where the He effect is already hinted
%at)].

\begin{figure}[htp]
\centerline{\includegraphics[width=\figwidth]{\fig/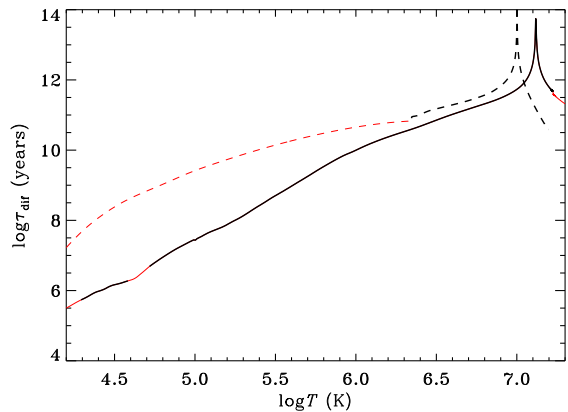}}
\caption{
Diffusion timescales for helium, defined by the term in $V_i X_i$ in
Eq.~(\ref{eq:diffusion}), for a model of the present Sun (dashed)
and a zero-age main-sequence $2 \Msun$ model (continuous).
The thinner red parts of the curves mark the fully mixed convection zones.
Adapted from \citet{Aerts2010}.
%\notecd [perhaps with suitable copyright statement].
}
\clabel{fig:diftime}
\end{figure}

%\notecd [Possibly here have a plot of diffusion and settling coefficients
%in present Sun.
%Might extend curves, with different line style or similar, to conditions
%within the convection zone to illustrate qualitatively effects in stars
%with thinner convection zones.

%\notecd [Also need to look at Dar \& Shaviv etc., probably].

%\notecd [Should check with Sylvain T. on detailed settling results.
%Probably a brief discussion of the fact that they are irrelevant for the
%Sun, indicating at what level.
%Note that Fig.~14 of \citet{Turcot1998a} very clearly illustrates
%that radiative effects are fairly small, and actually the even simpler
%approximation of having everything settle like O is in reasonable
%agreement with it!]

Detailed calculations of atomic data for the OPAL and OP opacity projects
({\cf} Sect.~\ref{sec:opac})
have allowed precise calculations of the radiative effects on settling
\citep{Richer1998}.
As mentioned above such effects are highly selective, affecting
different elements differently.
As a result, not only does the heavy-element abundance change
as a result of settling, but the relative mixture of the heavy
elements varies as a function of stellar age and position in the star.
As is evident from Fig.~\ref{fig:kapder} this has a substantial effect
on the opacities.
To take such effects consistently into account the opacities must therefore
be calculated from the appropriate mixture at each point in the model,
requiring appropriately mixing monochromatic 
contributions from individual elements
and calculating the Rosseland mean ({\cf} Eq.~\ref{eq:rosseland}).
Such calculations are feasible \citep{Turcot1998a} although obviously
very demanding on computing resources in terms of time and storage.
\cite{Turcot1998a} carried out detailed calculations of this
nature for the Sun. 
Here the relatively high temperatures and resulting ionization beneath the
convective envelope, where diffusion and settling are relevant,
result in modest effects of radiative acceleration
and little variation in the relative heavy-element abundances.
In fact, Fig.~14 of \citeauthor{Turcot1998a} shows that neglecting
radiative effects and assuming all heavy elements to settle at
the same rate, corresponding to fully ionized oxygen,
yield results somewhat closer to the full detailed treatment
than does neglecting radiative effects and taking partial ionization fully
into account.
The rather reassuring conclusion is that, as far as solar modelling is
concerned, the simple procedure of treating all heavy elements as one
is adequate \citep[see also][]{Turcot1998b}.
This simpler approach, neglecting radiative effects, is in fact what is
used for the models presented here.
%\notecd [Or with more details on where].
%\notecd [Check T \& C-D on oscillation effects!].

%\notecd [Might mention stronger settling in more massive stars,
%requiring additional mixing processes to ensure `normal' composition].

The timescale of diffusion and settling, 
defined by Eq.~(\ref{eq:diffusion}),
increases with increasing density and hence with depth beneath the
stellar surface, as illustrated in Fig.~\ref{fig:diftime}.
Since the convective envelope is fully mixed, the relevant timescale 
controlling the efficiency of diffusion is the value just below the
convective envelope.
In the solar case this is of order $10^{11}$ years, resulting in a modest
effect of diffusion over the solar lifetime.
In somewhat more massive main-sequence stars, however, 
with thinner outer convection zones, the time scale is short
compared with the evolution timescale; in the case illustrated for a 
$2 \Msun$ star, for example, it is around $5 \times 10^6$ years.
Thus settling has a dramatic effect on the surface abundance unless
counteracted by other effects \citep{Vaucla1974}.
This leads to a strong reduction in the helium abundance, 
likely eliminating instability due to helium driving in stars that
might otherwise be expected to be pulsationally unstable
\citep{Turcot2000}.
Also, differential radiative acceleration leads to a surface mixture
of the heavy elements very different from the solar mixture,
which is indeed observed in `chemically peculiar stars',
as already noted by \citet{Michau1970}.
\citet{Richer2000} pointed out that to match the observed abundances
even in these cases compensating effects had to be included
to reduce the effects of settling;
they suggested either sub-surface turbulence, increasing
the reservoir from which settling takes place, or mass loss bringing
fresh material less affected by settling to the surface.
An interesting analysis of these processes in controlling the observed
abundances of Sirius was presented by \citet{Michau2011}.
To obtain `normal' composition in such stars, 
processes of this nature reducing the effects of
settling are {\it a fortiori} required;%
\footnote{Already \citet{Edding1926} noted that `[i]t would be difficult to 
reconcile these results [on diffusion and settling]
with the observed spectra of stars where light
and heavy elements appear together'.}
since most main-sequence stars somewhat more massive than the Sun
rotate relatively rapidly, circulation
or hydrodynamical instabilities induced by rotation are likely candidates
\citep[\eg,][see also Sect.~\ref{sec:stars}]{Zahn1992}.
\citet{Deal2020} investigated the combined effects of rotation 
and radiatively affected
diffusion in main-sequence stars and found that this could
account for the observed surface abundances for stars with masses below
$1.3\,{\rm M}_\odot$.
For more massive stars additional mixing processes appeared to be required.
{\rv It should also be noted that such hydrodynamical models of the evolution
of rotation are unable to account for the rotation observed 
in the solar interior (see Section~\ref{sec:heliorot}).
A complete model of the transport of composition and angular momentum
in stellar interiors remains to be found.}

\subsection{The near-surface layer}
\clabel{sec:surface}

The treatment of the outermost layers of the model is complicated and
affected by substantial physical uncertainties.
In the atmosphere the diffusion approximation for radiative transport,
implicit in Eq.~\Eq{eq:nablarad}, is no longer valid;
here the full radiative-transfer equations need to be considered,
including the details of the frequency dependence of absorption and emission.
Such detailed stellar atmosphere models are available and can in
principle be incorporated in the full solar model
\citep[\eg,][]{Kurucz1991, Kurucz1996, Gustaf2008}.
%\notecd [Kurucz; also Nice, Liege atmosphere matching].
However, additional complications arise from the effects of convection
which induce motion in the atmosphere as well as strong lateral
inhomogeneities in the thermal structure.
Also, observations of the solar atmosphere strongly indicate the
importance of non-radiative heating processes in the upper parts of
the atmosphere, likely caused by acoustic or magnetic waves,
or other forms of magnetic energy dissipation,
for which no reliable models are available.
The thermal structure just beneath the photosphere is strongly affected
by the transition to convective energy transport, which determines the
temperature gradient $\nabla = \nabla_{\rm conv}$.
Also, in this region convective velocities are a substantial
fraction of the speed of sound, leading to significant momentum
transport by convection described as a `turbulent pressure',
but most often ignored in the model calculations.

From the point of view of the global structure of the Sun, these
near-surface problems are of lesser importance.
In most of the convection zone the temperature gradient is
very nearly adiabatic, $\nabla \simeq \nablaad$
(see also Fig.~\ref{fig:gwcon}).
Thus the structure is essentially determined by the (constant) value
of the specific entropy $s_{\rm conv}$;
in other words, the variations of the thermodynamical quantities within
this part of the convection zone lie on an adiabat.
In fact, if the further approximation of a fully ionized ideal gas
is made, such as is roughly valid except in the outer few per cent
of the solar radius, $\nablaad \simeq 2/5$, 
$\dd \ln p / \dd \ln \rho \simeq 5/3$, and the relation between pressure
and density can be approximated by
\be
p = K \rho^\gamma \; ,
\eel{eq:appcon}
with $\gamma = 5/3$.
In this case, therefore, the properties of the convection zone are
characterized by the adiabatic constant $K$.
Such an approximation was generally used in early calculations of
solar models ({\eg}, \citecl{Schwar1957}).
The structure of the convection zone determines its radial extent and
hence affects the radius of the model.
In the solar case the radius is known observationally with high precision;
thus the adiabat of the adiabatic part of the convection zone
({\ie}, the value of $K$ in the approximation in Eq.~\Eq{eq:appcon})
must therefore be chosen such that the model has the observed radius.
This is part of the calibration of solar models
(see Sect.~\ref{sec:calib}).

From this point of view the details of the treatment of the near-surface
layers serve to determine $s_{\rm conv}$ (or $K$).
This is obtained from the specific entropy at the bottom of the
atmosphere through the change in entropy resulting from
integrating $\nabla - \nablaad$ over the significantly
superadiabatic part of the convection zone.
%\notecd [UPDATE: Might deserve an equation, at least if needed below].
The treatment of convection typically involves parameters that
can be adjusted to control the adiabat and hence the radius of the model;
given such calibration to solar radius, the structure of the deeper 
parts of the model is largely insensitive to the details of the
treatment of the atmosphere and the convective gradient
(for an example, see Fig.~\ref{fig:canuto} below).

I note that although the detailed modelling of the near-surface layers has
modest effect on the internal properties of \emph{calibrated} solar models,
they have a substantial effect on the computed oscillation frequencies
which may affect the analysis of observed frequencies 
(see Sect.~\ref{sec:basichelio}).
Also, in computations of other stars no similar calibration based on the
observed properties is generally possible.
It is customary to apply solar-calibrated convection properties in these cases;
although this is clearly not {\it a priori} justified,
some support at least for only modest variations relative to the Sun 
over a substantial range of stellar parameters has
been found from hydrodynamical simulations of near-surface convection
({\cf} Fig.~\ref{fig:alphacal}).

\begin{figure}[htp]
\centerline{\includegraphics[width=\figwidth]{\fig/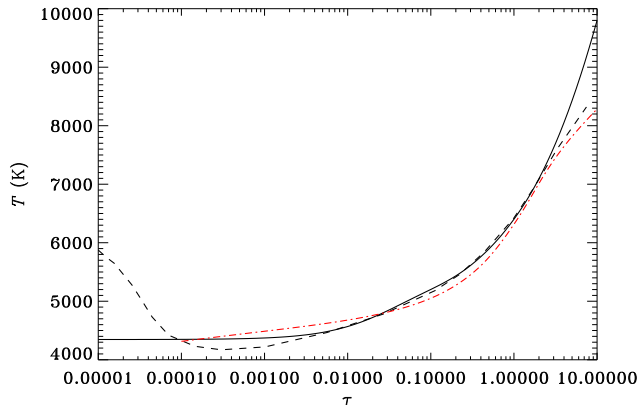}}
\caption{Comparison of the temperature structure in Model~C
of \citet{Vernaz1981} (dashed curve), 
against monochromatic optical depth $\tau$ at $500 \nm$,
and the fit given in Eq.~\Eq{eq:ttauval} (solid curve).
The red dot-dashed curve shows the $T(\tau)$ relation, against 
Rosseland mean opacity, obtained from matching a 3D hydrodynamical
simulation \citep[][see also Sect.~\ref{sec:convection}]{Trampe2014a}.
%\notecd [Could be more to say, here or later]
}
\clabel{fig:ttau}
\end{figure}

Although the atmospheric structure can be implemented in terms 
of reasonably realistic models of the solar atmosphere,
the usual procedure in modelling solar evolution is to base
the atmospheric properties on a simple relation between temperature
and optical depth $\tau$, $T = T(\tau)$;
here $\tau$ is defined by
\be
{\dd \tau \over \dd r} = - \kappa \rho \; ,
\eel{eq:tau}
with $\tau = 0$ at the top of the atmosphere.
This $T(\tau)$ relation is often expressed on the form
\be
T^4 = {3 \over 4} \Teff^4 [\tau + q(\tau)] \; , 
\eel{eq:ttauq}
defining the (generalized) Hopf function $q$.%
\footnote{This form is inspired by the exact solution for a grey 
atmosphere in radiative equilibrium; {\cf} \citet{Mark1947},
\citet{Mihala1970}, defining the so-called Hopf function $q(\tau)$.}
Given $T(\tau)$, and the equation of state and opacity as functions of 
density and temperature, 
the atmospheric structure can be obtained by integrating the equation
of hydrostatic support, which may be written as
\be
{\dd p \over \dd \tau} = {g \over \kappa} \; ,
\eel{eq:atmhyd}
where the gravitational acceleration $g$ can be taken to be constant,
at least for main-sequence stars such as the Sun.
This defines the photospheric pressure, {\eg} at the point where
$T = \Teff$, the effective temperature, and hence the outer
boundary condition for the integration of the full equations of
stellar structure.%
\footnote{A smooth transition to the diffusion approximation 
in Eq.~\Eq{eq:nablarad} can be achieved by suitably incorporating
the derivative of the assumed $T(\tau)$ relation in that equation
\citep[\eg][]{Henyey1965, Trampe2014b}.}
The $T(\tau)$ relation can be obtained from fitting to more detailed
theoretical atmospheric models, as done, for example,
by \citet{Morel1994}, who used the \citet{Kurucz1991} models.
Alternatively, a fit to a semi-empirical model of the solar atmosphere
can be used,
such as the Krishna Swamy fit \citep{Krishn1966} or the
Harvard-Smithsonian Reference Atmosphere \citep{Ginger1971}.
As an example, the \citet{Vernaz1981} Model~C $T(\tau)$ relation
is shown in Fig.~\ref{fig:ttau};
here is also shown the result of using the following approximation 
for the Hopf function in Eq.~(\ref{eq:ttauq}):
\be
q(\tau) = 1.036 -0.3134 \exp(-2.448 \tau) -0.2959 \exp(-30 \tau) \; . 
\eel{eq:ttauval}
The approximation provides a reasonable fit to the observationally
inferred temperature structure in that part of the atmosphere which 
dominates the determination of the photospheric pressure.

$T(\tau)$ relations based on a solar $q(\tau)$ are often used for 
general modelling of stars, even though the atmospheric structure may have
substantial variations with stellar properties. 
An interesting alternative is to determine $q(\tau)$, as a function of
stellar parameters, from averaged hydrodynamical simulations of the
stellar near-surface layers
\citep[\eg][]{Trampe2014a}.
%\notecd [Could be more; check for Magic]
An example based on a simulation for the present Sun is also shown in
Fig.~\ref{fig:ttau}.

\subsection{Treatment of convection}
\clabel{sec:convection}

%Even though the details of the treatment of convection do not affect the
%overall structure of solar models, after calibration
%\notecd [this badly needs a general coordination]
A detailed review of observational and theoretical aspects of solar
convection was provided by \citet{Nordlu2009a},
{\rv while \citet{Rincon2018} focused on the largest clearly observed scale of
convection on the solar surface, the supergranulation.
Further} details, including the treatment of convection in a time-dependent
environment such as a pulsating star, were reviewed by \citet{Houdek2015}.
As discussed below, extensive hydrodynamical simulations have been carried
out of the near-surface convection in the Sun and other stars.
However, direct inclusion of these simulations in stellar evolution
calculations is impractical, owing to the computational expense;
thus we must rely on simpler procedures.
It is obviously preferable to have a physically motivated description of
convection;
as discussed above (see also Sect.~\ref{sec:calib}),
solar modelling requires one or more parameters
which can be used to adjust the 
specific entropy in the adiabatic part of the convection zone and hence
the radius of the model.
In stellar modelling convection is typically treated by means of some
variant of \emph{mixing-length model} 
\citep[e.g.][]{Bierma1932, Vitens1953, Boehm1958};
a more physically-based derivation of the description was provided
by \citet{Gough1977b, Gough1977c}, in terms of the linear growth and
subsequent dissolution of unstable modes of convection.
In the commonly used physical description of this prescription%
\footnote{which hardly deserves the more impressive name of `theory'}
\citep[for further details, see][]{Kippen2012}
convection is described by the motion of blobs over a distance $\ell$,
after which the blob is dissolved in the surroundings, giving up its
excess heat.
If the temperature difference between the blob and the surroundings is 
$\Delta T$ and the typical speed of the blob is $v$, the convective flux
is of order $\Fcon \sim v c_p \rho \Delta T$, where $c_p$ is the specific heat
at constant pressure.
Assuming, for simplicity, that the motion of the blob takes place
adiabatically, $\Delta T \sim \ell T (\nabla - \nablaad) /H_p$, where
$H_p = - (\dd \ln p/\dd r)^{-1}$ is the pressure scale height.
Also, the speed of the element is determined by the work of the buoyancy
force $- \Delta \rho g$ on the element, 
where $\Delta \rho \sim - \rho \Delta T/T$ is the density difference
between the blob and the surroundings, assuming the ideal gas law
and pressure equilibrium between the blob and the surroundings.
This gives $\rho v^2 \sim - \ell g \Delta \rho 
\sim \rho \ell^2 g (\nabla - \nablaad)/H_p$.
Thus we finally obtain%
\footnote{neglecting, as is usually done, the kinetic energy flux}
\be
\Fcon \sim \rho c_p T {\ell^2 g^{1/2} \over H_p^{3/2}} 
(\nabla - \nablaad)^{3/2} \; .
\eel{eq:fcon}
To this must be added the radiative flux
\be
\Frad = {4 a \clight T^4 \over 3 \kappa \rho} {\nabla \over H_p}
\eel{eq:frad}
({\cf} Eq.~\ref{eq:nablarad});
the total flux $F = \Fcon + \Frad$ must obviously match $L/(4 \pi r^2)$,
for equilibrium.
This condition determines the temperature gradient in this description.

This description obviously depends on
the choice of $\ell$; this is typically also regarded as a measure of the
size of the convective elements.
An almost universal, if not particularly strongly physically motivated,
choice of $\ell$ is to take it as a multiple of the pressure scale height,
\be
\ell = \alphamlt H_p \; .
\eel{eq:mixlength}
From Eq.~\Eq{eq:fcon} it is obvious that $\Fcon$ then scales
as $\alphamlt^2$.
Adjusting $\alphamlt$ therefore modifies the convective efficacy
and hence the superadiabatic gradient $\nabla - \nablaad$ required
to transport the energy, thus fixing the specific entropy in the
deeper parts of the convection zone.
%\notecd [Although this may need to go closer to the discussion of the 
%calibration].
This in turn affects the structure of the convection zone, including its
radial extent, and hence the radius of the star.
As discussed in Sect.~\ref{sec:calib} the requirement that models of the
present Sun have the correct radius is typically used to determine a
value of $\alphamlt$, 
which is then often used for the modelling of other stars.

In practice, further details are added.
These involve a more complete thermodynamical description, the inclusion
of factors of order unity in the relation for the average velocity and
energy flux and expressions for the heat loss from the convective element.
Although not of particular physical significance, the choice 
made for these aspects
obviously affects the final expressions and must be taken into account in
comparisons between different calculations, particularly when it comes
to the value of $\alphamlt$ required to calibrate the model.
A detailed description of a commonly used formulation was provided by 
\citet{Boehm1958}.
%\notecd [Cite Gough \& Weiss for discussion of various early formulations,
%and pointing out that they give similar results.]
It was pointed out by \citet{Gough1976} (see also Sect.~\ref{sec:surface})
that solar models, with the appropriate calibration of the relevant
convection parameters to obtain the proper radius, are largely insensitive
to the details of the treatment of convection, although the specific values
of $\alphamlt$ may obviously differ.
It is important to keep this in mind when comparing independent solar and
stellar models.
As an additional point I note that the preceding description is entirely local:
it is assumed that $\Fcon$
is determined by conditions at a given point in the model, leading
effectively to a relation of the form \Eq{eq:nablaconv}.

%\notecd [Turbulent pressure?]
The motion of the convective elements also leads to transport of momentum
which, when averaged, appears as a contribution to hydrostatic support
in the form of a turbulent pressure of order
\be 
\pt \sim \rho v^2 
\sim {\rho \ell^2 g \over H_p} (\nabla - \nablaad) \; .
\eel{eq:pturb}
Correspondingly, hydrostatic equilibrium, Eq.~\Eq{eq:hydrostat}, is
expressed in terms of $p = p_{\rm g} + p_{\rm t}$, where $p_{\rm g}$
is the thermodynamic pressure.
On the other hand, the superadiabatic gradient $\nabla - \nabla_{\rm ad}$
in Eqs~\Eq{eq:fcon} and \Eq{eq:pturb} is essentially a thermodynamic
property and hence is determined by the gradient in $p_{\rm g}$ or,
if expressed in terms of $p$ and $p_{\rm t}$, the gradient of $p_{\rm t}$.
Consequently, including $p_{\rm t}$ consistently in Eq.~\Eq{eq:hydrostat}
increases the order of the system of differential equations within
the convection zone, leading to severe numerical difficulties at
the boundaries of the convection zone where the order changes
\citep[e.g.,][]{Stelli1976, Gough1977c}.
A detailed analysis of the resulting singular points at the convection-zone
boundaries was carried out by \citet{Gough1977b}.
As a result, although the effect of the turbulent pressure 
on the hydrostatic structure has been included in
some calculations  based on a local treatment of convection
\citep[e.g.,][]{Henyey1965, Kosovi1995} 
$\nabla - \nabla_{\rm ad}$ has generally been determined from the
total pressure, thus avoiding the difficulties at the boundaries of the
convection zone, but introducing some inconsistency \citep[{\eg}][]{Baker1979}.

It is obvious that the local treatment of convection is an approximation,
even in the simple physical picture employed here:
a convective element senses conditions over a 
range of depths in the Sun during its motion;
similarly, the convective flux at a given location must arise from an
ensemble of convective elements originating at different depths.
This indicates the need for a \emph{non-local} description of convection,
involving some averaging over the travel of a convective element and
the elements contributing to the flux.
Noting the similarity to the non-local nature of radiative transfer
\citet{Spiege1963} proposed an approximation to this averaging akin
to the Eddington approximation, leading to a set of local differential 
equations, albeit of higher order, to describe the convective properties
\citep[see also][]{Gough1977b}.
%\notecd [Need a little more detail on the approximation].
This was implemented by \citet{Balmfo1991} and \citet{Balmfo1992}.%
\footnote{They furthermore generalized the description to include the 
time-dependent case of convection in a pulsating star,
based on an earlier specific physical model by \citet{Gough1977c}
leading to the mixing-length formulation.
For reviews and applications of this aspect, see for example
\citet{Houdek1999, Houdek2000, Houdek2015}.}
An advantage of the non-local formulation is that it bypasses the
singularities caused by a consistent treatment of turbulent pressure
in a local convection formulation;
interestingly, \citet{Balmfo1992} showed that the common inconsistent 
local treatment has a non-negligible effect on the properties of the model,
compared with the local limit of the non-local treatment.
%\notecd [Also deserves a remark about Houdek, etc., although by then
%hardly as a footnote].}
%\notecd [Note on turbulent pressure, marginal convective overshoot].

%\notecd [Discuss convection simulations, compare average models with mixing-length
%treatment. Note importance of atmospheric structure ($T-\tau$ relation).
%Also discuss calibration of simple models with simulations.
%More generally about surface conditions being defined by 
%atmospheric models (no LRSP paper on solar atmospheric structure,
%it seems??)]

Alternative formulations for the convective properties have been
developed on the basis of statistical descriptions of turbulence,
thus including the full spectrum of convective eddies
\citep[e.g.,][]{Xiong1977, Xiong1989, Canuto1991, Canuto1996}
%and with parameters
%at least to some extent calibrated against laboratory turbulence experiments.
\citep[for a more detailed discussion of such Reynolds stress models, see][]
{Houdek2015}.
Even so, the descriptions typically contain an adjustable parameter,
most commonly related to a length scale,
allowing the calibration of the surface radius of solar models.

%\notecd [Need to refer to LRSP paper(s) on convection!].
A more physical description of convection is possible through
numerical simulation \citep[see][]{Nordlu2009a, Freyta2012}.
In practice this is restricted to fairly limited regions near the stellar
surface, and even then requires simplified descriptions of the behaviour
on scales smaller than the numerical grid.%
\footnote{A typical simulation of solar granulation may use a box with
a horizontal extent of $10 \times 10$\,Mm, with a horizontal cell size of
$40 \times 40$\,km, properly to resolve the granulation.
This should be compared with the viscous dissipation scale, based on
the microscopic viscosity, of order cm.}
Detailed modelling, including radiative effects in the stellar atmosphere,
has been carried out by, for example, \citet{Stein1989, Stein1998}
and \citet{Wedeme2004}.
This also includes treatments of the equation of state and opacity 
which are consistent with global stellar models and hence immediately
allow comparison with such models.
\citet{Magic2013} and \citet{Trampe2013} presented extensive grids of
simulations for a range of stellar parameters, covering the main sequence
and the lower part of the red-giant branch.

The simulations provide an alternative
to the usual simplified stellar atmosphere models, which are assumed
to be time independent and homogeneous in the horizontal direction.
A very interesting aspect is that spectral line profiles calculated from the
simulations and suitably averaged are in excellent agreement with
observations, without the conventional {\it ad hoc} inclusion of
additional line broadening through `microturbulence'
\citep[{\eg},][]{Asplun2000}.
Also, the simulations provide a very good fit to the observed solar 
limb darkening, {\ie}, the variation across the solar disk of the intensity
\citep{Pereir2013}.

The simulations of solar near-surface convection typically extend
sufficiently deeply to cover that part of the convection zone
where the temperature gradient
is substantially superadiabatic (see Fig.~\ref{fig:gwcon}).
Thus they essentially define the specific entropy of the adiabatic part
of the convection zone and hence fix the depth of the convection zone.
\citet{Rosent1999} utilized this by extending an averaged
simulation by means of a mixing-length envelope.
Interestingly, they found that the resulting convection-zone depth was
essentially consistent with the depth inferred from helioseismology
({\cf} Sect.~\ref{sec:heliostruc}), thus indicating that the simulation
had successfully matched the actual solar adiabat.

\begin{figure}[htp]
\centerline{\includegraphics[width=\figwidth]{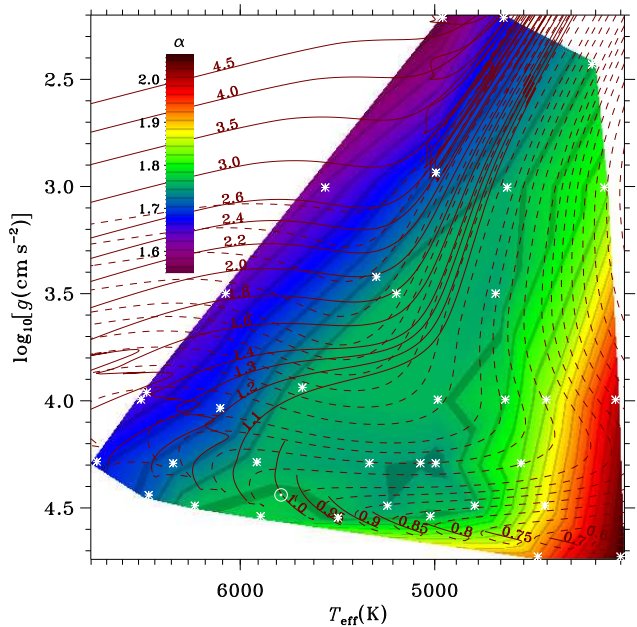}}
\caption{
Mixing-length parameter $\alphamlt$ obtained by fitting
averaged 3D radiation-hydrodynamical simulations to stellar envelope
models based on the \citet{Boehm1958} mixing-length treatment,
shown using the colour scale, against effective temperature $T_{\rm eff}$
(on a logarithmic scale) and $\log g$.
This is based on a fit to the simulations indicated by asterisks and the
solar simulation shown with $\odot$.
Stellar evolution tracks, computed with the MESA code \citep{Paxton2011},
are shown for masses between $0.65$ and $4.5 \Msun$, as indicated;
the dashed segments mark pre-main-sequence evolution.
(From \citet{Trampe2014b}.)
}
\clabel{fig:alphacal}
\end{figure}

As a generalization of these investigations, the simulations can be included
in stellar modelling through grids of atmosphere models
or suitable parameterization of simple formulations.
A convenient procedure is to determine an effective mixing-length parameter
$\alphamlt(\Teff, g)$ as a function of effective temperature and surface
gravity, such as to reproduce the entropy of the adiabatic part of the 
convection zone \citep[{\eg},][]{Ludwig1999, Trampe1999, 
Ludwig2008, Trampe2014b, Magic2015}.
It should be noted that since $\alphamlt$ determines the entropy jump
from the atmosphere to the interior of the convection zone, this calibration
is intimately tied to the assumed atmospheric structure, e.g., 
specified by a $T(\tau)$ relation also obtained from the simulations
\citep{Trampe2014a}.
As an example, Fig.~\ref{fig:alphacal} shows the calibrated $\alphamlt$
obtained by \citet{Trampe2014b}, as a function of $T_{\rm eff}$ and $\log g$.
Interestingly, the variation of $\alphamlt$ is modest in the central
part of the diagram, along the evolution tracks of stars close to solar.
Preliminary evolution calculations using these calibrations 
were carried out by \citet{Salari2015} and \citet{Mosumg2017, Mosumg2018}.
A similar analysis based on the calibration of the mixing-length parameter
was carried out by \citet{Spada2018}.
As an alternative to use the fitted mixing length,
\citet{Jorgen2017a} developed a method to include in stellar modelling
the averaged structure of the near-surface layers
obtained by interpolating in a grid of simulations. 
This was used by \citet{Jorgen2018a} to 
calculate a solar-evolution model incorporating such averaged structure
in all models along the evolution track;
similarly, \citet{Mosumg2020} calculated stellar evolution tracks
for a range of masses,
including the interpolated simulations along the evolution.

Apart from the calibration to match the solar radius 
({\cf} Sect.~\ref{sec:calib}) tests of the mixing-length parameter
and its possible dependence on stellar properties can be carried
out by comparing observations and models of red giants,
whose effective temperature depends on the assumed $\alphamlt$
\citep{Salari2002}.
A recent analysis was carried out by \citet{Tayar2017} based on APOGEE and
{\it Kepler} observations, comparing with models computed with
the YREC code \citep{vanSad2012}.
The model fits indicated a significant dependence on stellar metallicity,
with $\alphamlt$ increasing with increasing metallicity.
Interestingly, calibrations based on 3D simulations \citep{Magic2015}
did not show this trend, nor did the results obtained by Tayar {\etal}
match the values obtained
by \citet{Trampe2014b}, shown in Fig.~\ref{fig:alphacal}.
However, it should be recalled that the effect of $\alphamlt$ on
stellar structure depends on other parameters in the mixing-length treatment, 
as well as on the assumed atmospheric structure and physics of the
near-surface layers.
Thus comparison of numerical values of $\alphamlt$ or
trends with, {\eg}, metallicity requires some care;
the discrepancies may {\rv be caused by differences in
other aspects of the modelling.
In fact, in a detailed analysis \citet{Salari2018}, carefully taking into
account the other uncertainties in the modelling of the near-surface
layers, were unable to reproduce the results of \citet{Tayar2017};
on the other hand, they did find some issues when $\alpha$-enhanced stars
were included in the sample.}
%different formulations of the mixing-length treatment or differences in
%the modelling of the stellar atmospheres which, as already noted,
%is closely linked to the value of $\alphamlt$ required to recover a given
%internal structure.

\begin{figure}[htp]
\centerline{\includegraphics[width=\figwidth]{\fig/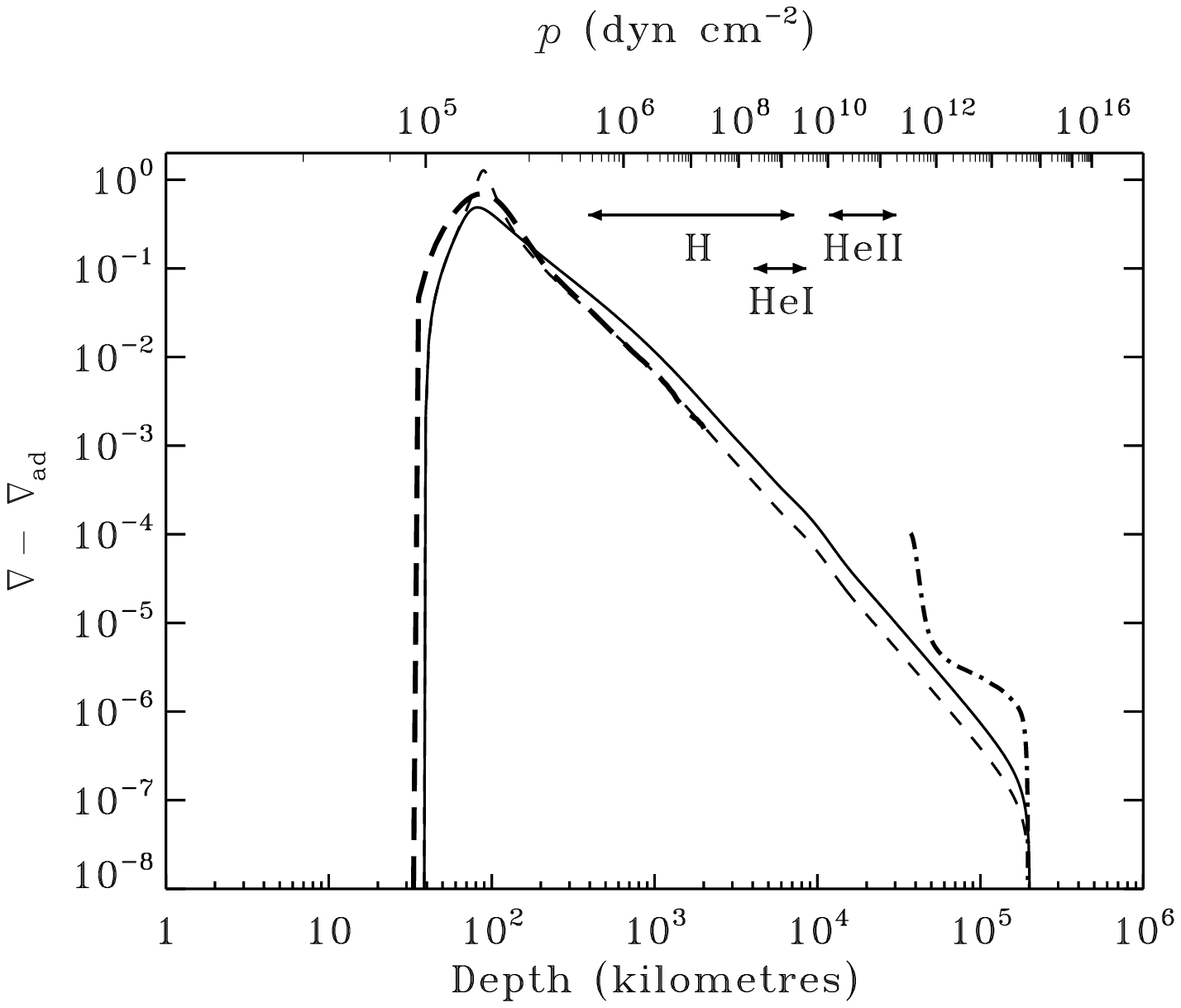}}
\caption{
Properties of the solar convection zone.
The lower abscissa is depth below the location where the temperature 
equals the effective temperature, whereas the upper abscissa is pressure $p$.
The solid curve shows the superadiabatic gradient $\nabla - \nablaad$
in Model~S of \citet{Christ1996}, using the \citet{Boehm1958} 
mixing-length treatment of convection, and the horizontal arrows indicate the
extents of the hydrogen and helium ionization zones in this model.
% Between 10 and 90 percent ionization.
Also, the short-dashed curve shows $\nabla - \nablaad$ in a model corresponding
to Model~S, including calibration to the same surface radius, but using
the \citet{Canuto1991} treatment of convection, 
and the heavy long-dashed curve shows $\nabla - \nablaad$ in an average
model resulting from hydrodynamical simulations of near-surface convection
\citep{Trampe2013}.
The heavy dot-dashed line shows the mean superadiabatic gradient in 
a hydrodynamical simulation % of the bulk of the convection zone
%\citep{Brun2002},
\citep{Feathe2016},
excluding the outer parts of the convection zone;
the initial increase in the most shallow part of the simulation is an
artifact of the imposed boundary condition.
\citep[Adapted from][]{Gough1976}.
}
\clabel{fig:gwcon}
\end{figure}

A comparison between different formulations of near-surface convection
is provided in Fig.~\ref{fig:gwcon}, in a format introduced by
\citet{Gough1976}.
The complete solar models, corresponding to 
Model~S of \citet{Christ1996}, have been calibrated to the same solar radius
({\cf} Sect.~\ref{sec:calib}) through the adjustment of suitable
parameters; 
this yields a depth of the convection zone which is essentially
consistent with the helioseismically determined value.
Evidently, regardless of the convection treatment the region of substantial
superadiabatic gradient $\nabla - \nablaad$ is confined to the
near-surface layers, as would also be predicted from the simple
analysis given above ({\cf} Eq.~\ref{eq:fcon}).
Using the \citet{Canuto1991} formulation leads to a rather higher and
sharper peak in the superadiabatic gradient than for the 
\citet{Boehm1958} mixing-length formulation.
On the other hand, it is striking that the detailed behaviour of the averaged 
superadiabatic gradient resulting from the \citet{Trampe2013} simulation
is in reasonable agreement with the results of the
calibrated mixing-length treatment.
As already noted, it also appears to lead to the correct adiabat in the
deeper parts of the convection zone.

Physically realistic simulations of near-surface convection
have been carried out extending over 96 Mm in the horizontal direction,
thus for the first time also including the scale of supergranules,
and to a depth of 20 Mm, around 10 per cent of the convection zone
\citep{Stein2006, Stein2009, Nordlu2009b}.%
\footnote{Results from these simulations can be obtained from
\url{http://steinr.pa.msu.edu/~bob/data.html}.}
%\notecd [may need update??].
Simulations have also been carried out which cover the bulk of
the convection zone, but excluding the near-surface region: it is
very difficult to include the very disparate range of temporal and
spatial scales needed to cover the entire convection zone.
Also, the microphysics of such simulations are typically somewhat
simplified.
On the other hand, the simulations take rotation into account, in
an attempt to model the transport of angular momentum and hence
the source of the surface differential rotation ({\cf} Eq.~\ref{eq:surfrot})
and the variation of rotation within the convection zone
(see also Sect.~\ref{sec:heliorot}).
A detailed review of these simulations was provided by \citet{Miesch2005}.
As an example of their relation to global solar structure,
Fig.~\ref{fig:gwcon} includes the average superadiabatic gradient
from such a simulation, appropriately located relative to the global models.
Apart from boundary effects the simulation is clearly in relatively
good agreement with the simplified treatment, in particular confirming
that this part of the model is very nearly isentropic.

An interesting issue was raised by \citet{Hanaso2012} concerning the
validity of the deeper convection simulations:
based on local helioseismology \citep[see][]{Gizon2005} using
the time distance technique they obtained estimates of the convective
velocity one or two orders of magnitude lower than obtained in the
simulations, or indeed predicted from the simple estimate in
Eq.~\Eq{eq:fcon}.
This was questioned in an analysis using the ring-diagram technique
\citep{Greer2015}, who obtained results similar to those of the simulations.
{\rv However, \citet{Hanaso2020} showed, using a helioseismic technique 
based on coupling of mode eigenfunctions, that large-scale turbulence in
the Sun is strongly suppressed compared with the results of global
numerical simulations.
Thus} there is increasing observational evidence 
for possible limitations in our understanding of the dynamics of convection 
in the Sun, {\rv in particularly at larger scales,
where there} is essentially no observational evidence for structured flows,
unlike what is seen in global simulations of the solar convection zone
\citep[for a review, see][]{Miesch2005}.
A review of the helioseismic inferences of solar convection was provided
by \citet{Hanaso2016}.
Simulations by \citet{Cosset2016} indicated that supergranules might be
the largest coherent scales of convection, with energy transport in the deeper,
essentially adiabatically stratified, parts of the convection zone being
dominated by colder compact downflowing plumes.
For a recent short review on solar convection, see \citet{Rast2020}.

%\notecd [UPDATE: Somewhere also need to say something about convective overshoot
%at base of convection zone, and tests of it].

\subsection{Calibration of solar models}
\clabel{sec:calib}

%\notecd [Discuss basics of mixing-length, composition calibration. Possibly
%illustrate how it works. 
%Could include rough analysis of the relevant sensitivity (homology
%for luminosity, simple convection zone for alpha).
%Possibly provide set of derivatives, for convenience.]

The Sun is unique amongst stars in that we have accurate determinations of its
mass, radius and luminosity
and an independent and relatively precise measure of its age from age
determinations of meteorites (see Sect.~\ref{sec:basicpar}).
It is obvious that solar models should satisfy these constraints,
as well as other observed properties of the Sun, particularly the present
ratio between the abundances of heavy elements and hydrogen.
Ideally, the constraints would provide tests of the models;
in practice, the modelling includes {\it a priori\/} three unknown parameters
which must be adjusted to match the observed properties: the initial 
hydrogen and heavy-element abundances $X_0$ and $Z_0$ and a parameter
characterizing the efficacy of convection
(see Sect.~\ref{sec:convection}).
This adjustment constitutes the calibration of solar models.

Some useful understanding of the sensitivity of the models to the parameters
can be obtained from simple homology arguments \citep[{\eg},][]{Kippen2012}.
According to these, the luminosity approximately scales with mass and 
composition as
\be
L \propto Z^{-1} (1 + X)^{-1} M^{5.5} \mu^{7.5} \; ,
\eel{eq:homlum}
assuming Kramers opacity, with $\kappa \propto Z (1 + X) \rho T^{-3.5}$,
and with $\mu$ given by Eq.~\Eq{eq:meanmol}.
Obviously, the strong sensitivity to the average mean molecular weight
means that relatively modest changes in the helium abundance can lead to
the correct luminosity.

As discussed above,
the efficacy of convection in the near-surface layers determines the 
specific entropy in the adiabatic part of the convection zone and hence
the structure of the convection zone,
thus controlling its extent and hence the radius of the model.
(When the composition is fixed by obtaining the correct luminosity the
extent of the radiative interior is largely determined.)
%This is almost entirely controlled by the details of the treatment of 
%convection in the very superficial parts of the convection zone where the
%temperature gradient is significantly superadiabatic
%(see Fig.~\ref{fig:gwcon}).
%\notecd [Somehow explain that with increasing efficacy the radius decreases!]
With increasing efficacy the superadiabatic temperature gradient 
$\nabla - \nabla_{\rm ad}$ required
to transport the flux is decreased; hence the temperature in the convection
zone is generally lower, the density (at given pressure) therefore higher,
and the mass of the convection zone occupies a smaller volume, and hence
a smaller extent in radius.
Thus the radius of the model decreases with increasing efficacy.
The actual reaction of the model is substantially more complex
but leads to the same qualitative result.
%\notecd [UPDATE: this argument is actually a little flaky, when compared with
%details of models, and hence deserves another look, in an update.
%See model results in ...gong/jan07, for model l9bi.d.02c\_dalpha, and
%in comp-gong-dalfa.idl. Of course C-D 1996 is also relevant for this.]

As discussed in Sect.~\ref{sec:convection}, the treatment of convection
and hence the properties of the superadiabatic temperature gradient
are typically obtained from the mixing-length treatment.
According to Eqs~\Eq{eq:fcon} and \Eq{eq:mixlength}, assuming
that $F_{\rm con}$ carries most of the flux and is therefore essentially fixed,
an increase in $\alphamlt$ causes an increase in the convective
efficacy and hence a decrease in $\nabla - \nabla_{\rm ad}$, corresponding,
according to the above argument, to a decrease in the model radius.
Thus by adjusting $\alphamlt$ a model with the correct radius can
be obtained.
In other simplified convection treatments, such as that of \citet{Canuto1991},
a similar efficiency parameter is typically introduced to allow radius
calibration.
When $\alphamlt$ is obtained through fitting to 3D simulations
({\cf} Fig.~\ref{fig:alphacal}) there is no {\it a priori} guarantee that
this yields the value required to obtain the correct solar radius.
In this case a correction factor can be applied to achieve the proper
solar calibration \citep{Mosumg2017}.
Of course, if the simulations provide a good representation of the outermost
layers of the Sun, as already found to be the case by \citet{Rosent1999},
this factor would be close to one, as has indeed been found in practice.
The same correction factor is then applied when the fit to the
3D simulations are used for more general stellar modelling.

The details of the calibration depend on whether or not diffusion and settling
are included.
If these effects are ignored the surface composition of the model hardly 
changes between the zero-age main sequence and the present age of the Sun.
Although the present surface abundance $\Xs$ of hydrogen is affected by
the calibration of $X_0$ the range of variation is typically so small that it
can be ignored, and the (constant, in space and time) value of the 
heavy-element abundance, and hence $Z_0$, is fixed from 
$\Zs/\Xs$ and some suitable characteristic value of $X$.
On the other hand, if diffusion and settling are included the change in
the convection-zone composition must be taken into account and the value
of $Z_0$ must be adjusted to match properly $\Zs/\Xs$.

The formal calibration problem is then, when including diffusion and 
settling, to determine the set of parameters 
$\{p_i\} = \{X_0, Z_0, \alphamlt\}$ to match the observables
$\{o_k\} = \{\Ls, \Zs/\Xs, R\}$ to the solar values
$\{o_k^\odot\} = \{L_{\rm s, \odot}, (\Zs/\Xs)_\odot, R_{\odot}\}$.
(Specifically, $R$ is here taken to be the photospheric radius,
defined at the point in the model where $T = \Teff$, the effective 
temperature.)
This is greatly simplified by the fact that variations in the parameters
generally are fairly limited.
Thus in practice 
the corrections $\{\delta p_i\}$ to the parameters can be found from the
errors in the observables, using a fixed set of derivatives, as
\be
\delta p_i = \sum_k (o_k^\odot -o_k) {\partial p_i \over \partial o_k} \; ,
\eel{eq:calcor}
where the derivatives $\{\partial p_i / \partial o_k\}$ are obtained by
varying the parameters in turn and inverting the resulting 
derivative matrix $\{\partial o_k / \partial p_i\}$.
I have found that the following values secure relatively rapid convergence
of the iteration:
\be
\begin{array}{ccc}
\rule[-0.4cm]{0cm}{1cm}\displaystyle
{\partial \ln \alphamlt \over \partial \ln \Ls} = 1.15 &
\displaystyle
{\partial \ln \alphamlt \over \partial \ln R} = -4.70 &
\displaystyle
{\partial \ln \alphamlt \over \partial \ln (\Zs/\Xs)} = 0.148 \\
\rule[-0.4cm]{0cm}{1cm}\displaystyle
{\partial \ln X_0 \over \partial \ln \Ls} = -0.137 &
\displaystyle
{\partial \ln X_0 \over \partial \ln R} = -0.087  &
\displaystyle
{\partial \ln X_0 \over \partial \ln (\Zs/\Xs)} = -0.132 \\
\rule[-0.4cm]{0cm}{1cm}\displaystyle
{\partial \ln Z_0 \over \partial \ln \Ls} = -0.111 &
\displaystyle
{\partial \ln Z_0 \over \partial \ln R} = 0.275 &
\displaystyle
{\partial \ln Z_0 \over \partial \ln (\Zs/\Xs)} = 0.864 \; .
\end{array}
\eel{eq:calfder}
%\notecd [Perhaps somewhere a description of ASTEC indicating that the 
%calibration is built-in].
These derivatives are incorporated in the ASTEC code \citep{Christ2008a}
and allow efficient and automatic calculation of calibrated solar models.
In the case where no iteration for $Z_0$ is carried out the following values
have been used:
\be
\begin{array}{cc}
\rule[-0.4cm]{0cm}{1cm}\displaystyle
{\partial \ln \alphamlt \over \partial \ln \Ls} = 1.17 &
\displaystyle
{\partial \ln \alphamlt \over \partial \ln R} = -4.75 \\
\rule[-0.4cm]{0cm}{1cm}\displaystyle
{\partial \ln X_0 \over \partial \ln \Ls} = -0.154 &
\displaystyle
{\partial \ln X_0 \over \partial \ln R} = -0.045   \; .
\end{array}
\eel{eq:calpder}
Convergence to a relative precision of $10^{-7}$ is typically obtained in
5 -- 7 iterations.

%\notecd [UPDATE: Might be something to say about the effect of the calibration 
%for stellar models, including red-giant models.
%Really needs an update beyond ARFM paper. Check recent red-giant reviews,
%possibly involving A. Weiss??]

%% \newpage

%===========================================================================

\section{The evolution of the Sun}
\clabel{sec:evol}

%\notecd [Possibly a little on pre-main-sequence evolution, star formation,
%although principally reference.]

To set the scene for this brief overview of solar evolution it is useful
to recall the characteristic timescales of stars.
Departure from hydrostatic equilibrium causes motion on a
\emph{dynamical timescale}, of order
\be
t_{\rm dyn} = \left({R^3 \over G M} \right)^{1/2}
\simeq 30 \, {\rm min} \left({R \over \Rsun}\right)^{3/2}
\left({M \over \Msun}\right)^{-1/2} \; .
\eel{eq:tdyn}
Evolution in phases where the energy is provided by release of 
gravitational energy happens on the \emph{Kelvin--Helmholz} timescale, 
of order
\be
t_{\rm KH} = {G M^2 \over L R} 
\simeq 3 \times 10^7 \yr 
\left({M \over \Msun} \right)^2
\left({R \over \Rsun}\right)^{-1}
\left({\Ls \over \Lsun}\right)^{-1} \; .
\eel{eq:tKH}
As a result of the virial theorem \citep[\eg,][]{Kippen2012}
this is also the timescale for the 
cooling of the star as a result of loss of thermal energy.
Finally, the timescale for nuclear burning on the main sequence can be
estimated as
\be
t_{\rm nuc} = {Q_{\rm H} q_{\rm c} X_0 M \over L}
\simeq 10^{10} \yr 
{M \over \Msun}
\left({\Ls \over \Lsun}\right)^{-1} \; ,
\eel{eq:tnuc}
where $Q_{\rm H}$ 
%\notecd [not the most elegant notation]
is the energy released per unit mass of consumed hydrogen and
$q_{\rm c} \simeq 0.1$ is the fraction of stellar mass that is involved
in nuclear burning on the main sequence.
Later stages of hydrogen burning typically involve smaller fractions of
the mass and take place at higher luminosity and consequently have shorter
duration; 
also, the burning of elements heavier than hydrogen release far less energy
per unit mass and the corresponding phases are therefore also relatively 
short.

\subsection{Pre-main-sequence evolution}
\clabel{sec:pmsevol}

Stars, including the Sun, are born from the collapse of gas and dust in
dense and cold molecular clouds.
Brief reviews of star formation were provided by, for example,
\citet{Lada1990} and \citet{Stahle1994};
for an extensive review, see \citet{McKee2007}. 
The collapse is triggered by gravitational instabilities, 
likely through turbulence which may have been induced by supernova explosions 
\citep{Padoan2016}.
Detailed simulations by \citet{Li2018} of star formation in
externally driven turbulence successfully reproduced the common filamentary
structure of interstellar clouds and the statistical properties of
newly formed stellar systems.
Evidence for 
the presence at the birth of the solar system of a nearby supernova,
which may have contributed to the dynamics leading to the formation of the Sun,
is provided by
decay products of short-lived radioactive nuclides found in meteorites
\citep[\eg,][]{Goswam2000, Goodso2016},
allowing a remarkably precise dating of different components of the early
solar system \citep{Connel2012}.
Further diagnostics of the early history of the solar system is provided
by the ratios of oxygen isotopes \citep{Gounel2007};
{\it in-situ} measurements of the solar wind by the {\it Genesis} 
spacecraft appear to have further complicated the picture \citep{Gaidos2009}.
%\notecd [The following bit may be out of place; check later].
%Additional information about the composition of the early solar system
%has been sought from measurements of the solar wind in the so-called
%polar coronal holes \citep{vonSte2016};
%I return to this in Sect.~\ref{sec:checkcomp}.
A detailed review of the environment of solar-system formation was
given by \citet{Adams2010}.
%\notecd [Needs update, including von Steiger and Zurbuchen 2016.]

The collapse of the cloud results in the formation of a core
which subsequently accretes matter from the surrounding cloud;
detailed simulations of these early phases of stellar evolution have been
carried out by, for example, \citet{Baraff2009}.
The angular momentum of the infalling material probably leads 
to the formation of
a disk around the star while processes likely involving magnetic fields
often result in outflow from the proto-star in highly collimated jets
along the rotation axis \citep{Shu2000},
giving rise to the so-called \emph{Herbig--Haro objects}
\citep[\eg,][]{Reipur2001}.
The gravitational energy released in the contraction of the protostar 
partly goes to heating it up and is partly released as radiation from the star;
the radiation finally stops the accretion and blows away 
the surrounding material,
such that the star becomes directly observable:
the star has reached the `birth line'.

\begin{figure}[htp]
\centerline{\includegraphics[width=\figwidth]{\fig/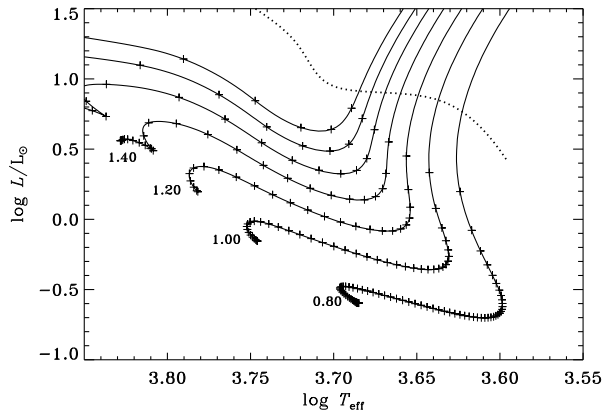}}
\caption{
Pre-main-sequence evolution of stars with masses between $0.8$ and $2 \Msun$,
as indicated,
computed with the Li\`ege stellar evolution code CL\'ES \citep{Scufla2008}.
The composition is $X = 0.7$, $Z = 0.02$.
The crosses mark the age along the tracks, in steps of $1 \, {\rm Myr}$;
the ages at the end of the tracks range from $87 \, {\rm Myr}$ at
$0.8 \Msun$ to $32\, {\rm Myr}$ at $1.4 \Msun$.
The heavy dotted line is a sketch of the so-called birth line, as shown
by \citet{Palla1993}, where the star emerges in visible light from the
material left over from its formation.
(Adapted from \citet{Aerts2010}; data courtesy of A. Miglio.)
}
\clabel{fig:pms}
\end{figure}

In these early phases matter in the protostar is relatively cool,
leading to a high opacity, and the luminosity is rather large.
Consequently, models of the star in this phase are generally fully convective,
evolving down the so-called \emph{Hayashi line} \citep{Hayash1961}
with contraction at roughly constant effective temperature,
and material in the star is fully mixed.
In this phase the temperature in the core reaches a point where deuterium
burning can take place, but since the initial deuterium content is tiny
(around $1.6 \times 10^{-5}$ of the hydrogen abundance),
the energy release has little effect on the evolution.
With further contraction the temperature in the central parts of the star
becomes so high that convection ceases and the star develops a gradually
growing central radiative region.
In this initial contraction, where energy for the luminosity and
the heating of stellar material is provided by release of 
gravitational energy, evolution takes place on the 
Kelvin-Helmholz timescale ({\cf} Eq.~\ref{eq:tKH}), along the so-called
\emph{Henyey line} \citep{Henyey1955} at increasing effective temperature
and luminosity.
With the beginning onset of substantial nuclear energy release, readjustments
of the structure of the star lead to a reduction in luminosity, and the star
settles on the \emph{zero-age main sequence} (ZAMS).
These early evolutionary phases are illustrated in Fig.~\ref{fig:pms}.
An extensive description of star formation, although possibly not completely
up to date, was given by \citet{Stahle2004}.

Interestingly, this somewhat simplistic picture has been questioned by 
more detailed
modelling of the contraction phase, starting from the initial collapsing cloud.
\citet{Wuchte2001} and
\citet{Wuchte2003} solved the spherically symmetric equations of
radiation hydrodynamics, starting from a suitable isothermal
model of the original cloud and following the formation of an
optically thick protostellar core and the accretion of further matter
on this core.
They found that deuterium burning takes place during the
accretion phase and that the model retains a substantial 
radiative core throughout the evolution; the later phases of the contraction
are parallel to the fully convective Hayashi track, but at somewhat higher
effective temperature.
These calculations were criticized by \cite{Baraff2010} on the grounds
of the assumed spherical symmetry of the infall.
However, by considering episodic infall Baraffe {\etal} also found
models with an early radiative core.
Detailed 3D modelling of collapsing molecular clouds, coupled with 
spherically symmetric modelling 
of the resulting proto-stellar and pre-main-sequence evolution
\citep{Kuffme2018, Jensen2018}
has confirmed the episodic nature of the accretion.
Also, interestingly, the results provide a plausible explanation
for the observed properties of young stellar clusters.

As discussed in Sect.~\ref{sec:twins} the detailed pre-main-sequence
evolution could have important
consequences for the interpretation of the present solar surface composition.
Given the importance of rotation and disk formation,
departures from spherical symmetry in the evolution of the
star should clearly be taken into account in the modelling.

\begin{figure}[htp]
\centerline{\includegraphics[width=\figwidth]{\fig/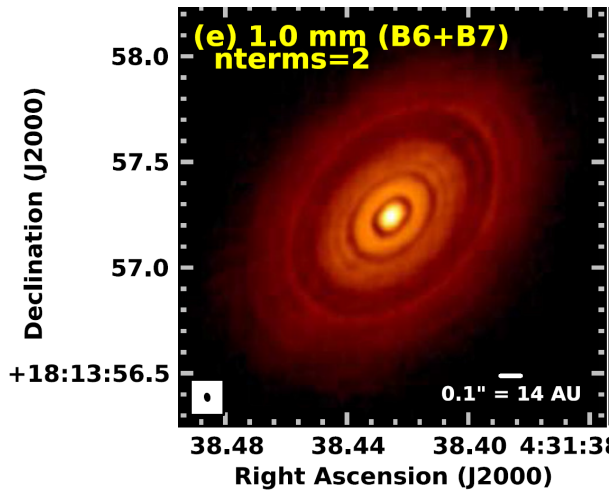}}
\caption{
ALMA observations, at a wavelength of 1\,mm,
of the planet-forming disk around the young star HL~TaU.
The lower-left inset shows the resolution.
Adopted from \citet{ALMA2015}.
}
\clabel{fig:hltau}
\end{figure}

At the end of pre-main-sequence evolution,
the temperature reaches a level where the full set of reactions
in the PP chains
(see Eqs~\ref{eq:PPI} and \ref{eq:PPII})
%\notecd [with an appropriate reference]
sets in, supplying the energy lost from the stellar surface.
At this point the contraction stops and the star enters its main-sequence
evolution, with a balance between the nuclear energy generation and the
energy loss from the surface, and hence taking place on a nuclear timescale.

%\notecd [The following may need further update].
It is likely that the early contraction, and the accretion of matter in
the disk, leads to an initial rapid rotation of the star.
In fact, it is observed that young stars generally rotate much more rapidly
than the present Sun.
However, 
in young open clusters where the stars may be assumed to share the same age 
substantial scatter in the rotation rates is found
\citep[\eg,][]{Soderb2001}.
This is a strong indication of the complex processes controlling the
evolution of angular momentum in the initial phases of proto-stellar evolution,
involving interactions between the star, the accreting disk and the
outflows, likely of magnetic origin \citep{Shu1994, Bodenh1995},
including magnetic locking between the outer layers of the star and the
inner parts of a truncated accretion disk.

Disks are commonly observed around protostars, confirming also this
part of the description \citep[e.g.,][]{Greave2005, Willia2011}.
The ubiquitous presence of planetary systems around other stars 
\citep{Batalh2014, Winn2015}
strongly suggests that the formation of planets in such
\emph{protoplanetary disks} is a common phenomenon.
This likely takes place through the formation and subsequent coalescence of
dust grains into objects of increasing size,
and finally the formation of a planetary system
\citep{Lissau1993, Aliber2005, Montme2006, Johans2017}.
Detailed discussions of the properties of such disks
and the formation of planets were provided by \citet{Armita2011, Armita2017}.
Dramatic illustrations of these planet-forming processes have been obtained
with the Atacama Large Millimeter/submillimeter Array (ALMA)
high-resolution observations 
\citep[e.g.,][]{ALMA2015, Isella2016, Harson2018}.
An example is illustrated in Fig.~\ref{fig:hltau};
modelling by \citet{Dipier2015} showed that the observed gaps are indeed
consistent with the presence of newly formed planets.
The planet-forming processes probably happen on a timescale comparable with,
or shorter than, the gravitational contraction of the star.
Thus the ages of meteorites as determined from radioactive dating likely
provide good measures of the age of the Sun since it arrived on 
the main sequence.
%Thanks to the Atacama Large Millimeter/sub-millimeter Array (ALMA) it is now
%possible to follow the evolution of the disk and signs of the formation of
%planets directly \citep[e.g.][]{Harson2018}.

\begin{figure} %[htp]
\centerline{\includegraphics[width=\figwidth]{\fig/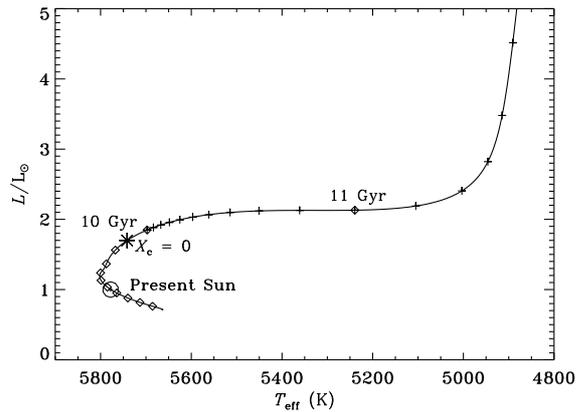}}
\caption{
Evolution track in the Herzsprung-Russell diagram of a model sequence
passing through Model~S of the present Sun 
\citep[][see also Sect.~\ref{sec:standard}]{Christ1996}.
Diamonds mark models separated by $1 \Gyr$ in age, and after an age of
$10 \Gyr$ plus symbols are at intervals of $0.1 \Gyr$.
The Sun symbol ($\odot$) indicates the location of the present Sun and
the star shows the point where hydrogen has been exhausted at the centre.
}
\clabel{fig:HR}
\end{figure}

\subsection{Main-sequence evolution}
\clabel{sec:msevol}

%\notecd [Summarize main-sequence evolution leading to the present Sun.
%Briefly mention `faint-early-Sun' problem.]
%
The evolution after the arrival on the main sequence, past the present age of
the Sun, is illustrated in Fig.~\ref{fig:HR}.
This is based on a model corresponding to Model~S of \citet{Christ1996},
discussed in more detail in Sect.~\ref{sec:models}.
Additional information about the variation with time of key quantities,
normalized to values for the present Sun,
is provided in Fig.~\ref{fig:modrat}.
The evolution is obviously driven by the gradual conversion of hydrogen
into helium in the core, leading to an increase in the mean molecular
weight of matter in the core.
This leads to a contraction of the core, an increase in the central density
and temperature and, 
in accordance with Eq.~\Eq{eq:homlum},
to an increase in the luminosity.
This evolution may be understood in simple terms by noting,
from Eq.~(\ref{eq:idealg}),
that the increase in $\mu$ would cause a decrease
in pressure inconsistent with hydrostatic balance, unless compensated for
by an increase in $\rho$ and $T$ resulting from the contraction of the core.
The increase in temperature, although partly counteracted by the
decrease in $X$, leads to an increase in the energy-generation rate and,
more importantly, to an increase in the radiative conductivity, and hence
to the increase in the luminosity.
Thus this effect is basic to the main-sequence evolution of stars;
unless non-standard effects (such as mass loss; see Sect.~\ref{sec:massloss})
are relevant there is hardly any doubt that the solar luminosity has 
undergone a fairly substantial increase since the formation of the solar
system.
A detailed analysis of this behaviour, in terms of homology scaling,
was provided by \citet{Gough1990a}.

\begin{figure}[htp]
\centerline{\includegraphics[width=\figwidth]{\fig/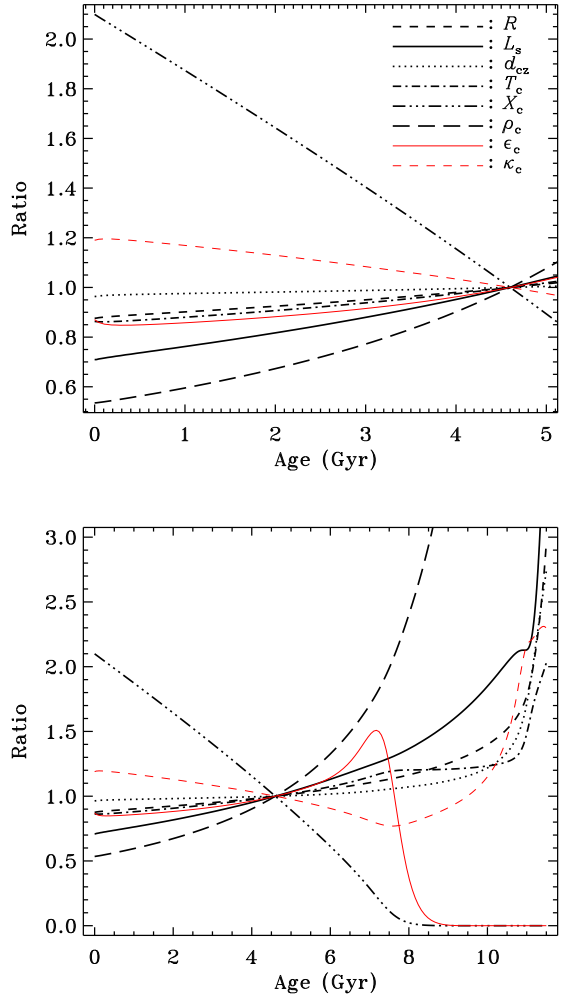}}
\caption{
%\notecd [Time variation of suitable quantities, Model~S and a little beyond].
Variation with age of quantities, normalized to the value at the present
age of the Sun, in a $1 \Msun$ evolution sequence, including Model~S
of the present Sun (see Sect.~\ref{sec:models}).
The top panel shows the evolution up to just after the present age,
whereas the bottom
panel continues the evolution beyond the exhaustion of hydrogen at the centre.
Line styles and colours are indicated in the figure.
$R$ and $\Ls$ are photospheric radius and surface luminosity,
$d_{\rm cz}$ is the depth of the convective envelope,
in units of the surface radius, and
$T_{\rm c}$, $X_{\rm c}$, $\rho_{\rm c}$, $\epsilon_{\rm c}$ and
$\kappa_{\rm c}$ are central temperature, hydrogen abundance, density,
energy-generation rate and opacity.
Values in the present Sun for most of the quantities are given in
Table~\ref{tab:modelchar};
in addition, $\epsilon_{\rm c} = 17.06 \erg \g^{-1} \s^{-1}$
and $\kappa_{\rm c} = 1.242 \cm^2 \g^{-1}$.
At the end of the illustrated part of the evolution, the
ratio $\rho_{\rm c}/\rho_{\rm c,\odot}$ is around 340, corresponding to
a central density of $5.3 \times 10^4 \g \cm^{-3}$.
}
\clabel{fig:modrat}
\end{figure}

\begin{figure}[htp]
\centerline{\includegraphics[width=\figwidth]{\fig/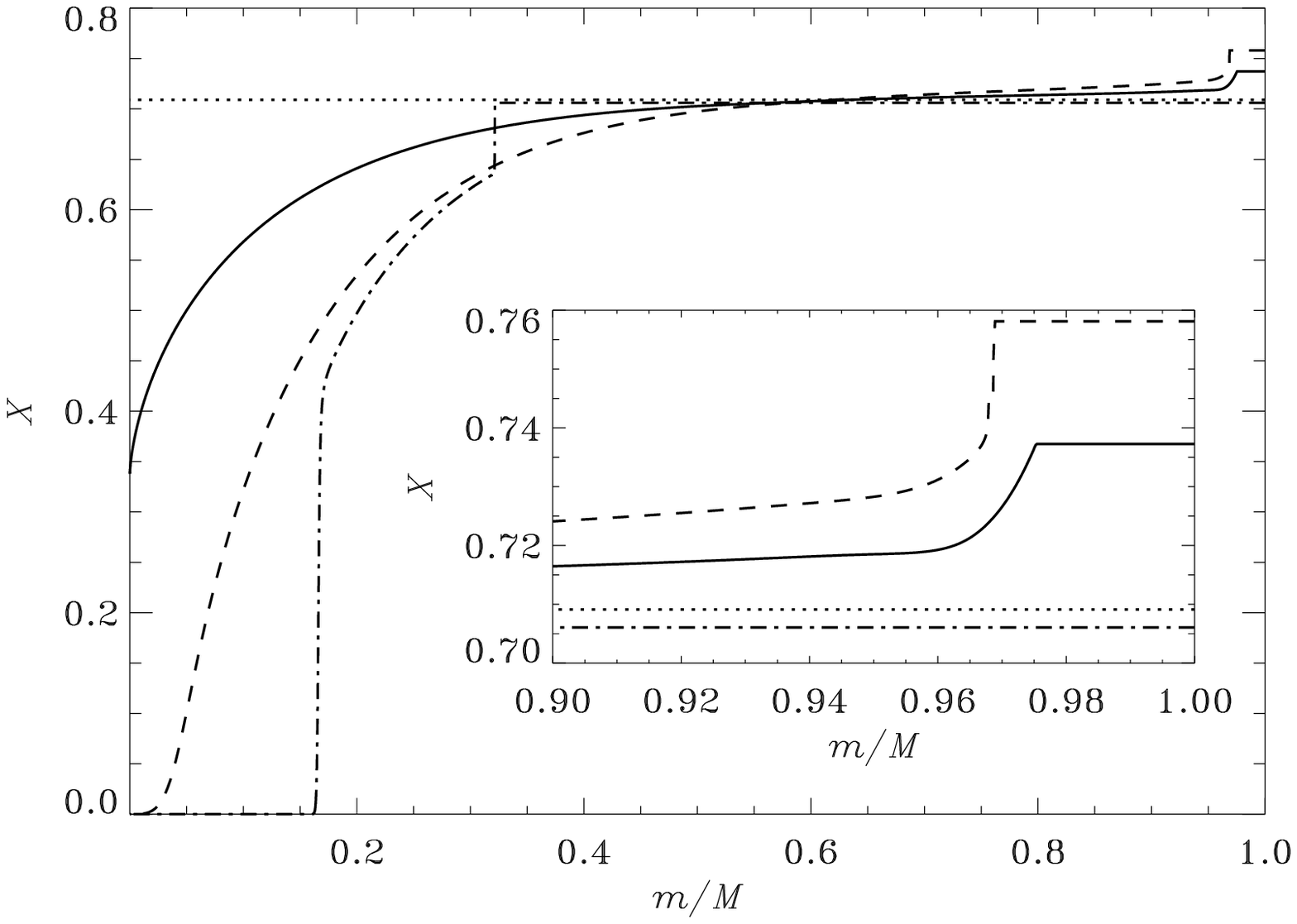}}
\caption{
Hydrogen abundance $X$ against fractional mass $m/M$ for a zero-age 
main-sequence model (dotted line), a model of age $4.6 \Gyr$
(present Sun; solid line), a model of age $9.5 \Gyr$,
where hydrogen has just been
exhausted at the centre (dashed line) and the model of age 
$11.5 \Gyr$, the final model included in Fig.~\ref{fig:HR} 
(dot-dashed line).
In the latter model the radiative core containing 32 per cent of the stellar
mass occupies only 21 per cent of the stellar radius.
The evolution sequence corresponds to Model~S of 
\citet[][see Sect.~\ref{sec:models}]{Christ1996}.
}
\clabel{fig:Xevol}
\end{figure}

Such a change in the solar energy reaching the Earth might be expected to
have climatic effects;
in fact, a naive estimate based on black-body radiative balance indicates that
the change of 30 per cent in solar luminosity shown in Fig.~\ref{fig:modrat}
would cause a change of around
7 per cent in the surface temperature of the Earth, {\ie}, around 20 K.
Thus one might expect that the Earth was very substantially colder early in
its history.
In fact, already \citet{Schwar1957} noted that, since in their
calculations the solar luminosity was about 20 per cent less than
now two billion years ago ``[t]he average temperature on the earth's surface
must then have been just about at the freezing point of water, if
we assume that it changes proportionally to the fourth root of the solar
luminosity.
Would such a low average temperature have been too cool for the algae known
to have lived at that time?''
In contrast to these models, the terrestrial surface temperature shows no
indication of dramatic changes over the past 4 Gyr, with evidence 
for liquid water in even very old geological material
\citep{Mojzsi2001, Wilde2001, Rosing2004}.
%\notecd [Could do with a check in a further revision; 
%contact Svensmark; recent LRSP paper only concerns variations on short 
%timescales].
This problem has been dubbed `the faint early Sun problem'
\citep[see also][]{Gudel2007},
and led to speculations about errors in our understanding of stellar evolution.
It seems more likely, however, that the problem lies in the simplistic
climate models used for these estimates of the temperature of the early Earth
\citep[\eg,][]{Sagan1972}.
With a substantially stronger early greenhouse effect, perhaps caused by 
a higher content of ${\rm CO_2}$, 
the present temperature could have been reached with a lower energy input.
Modelling of the early terrestrial atmosphere by \citet{vonPar2008}
suggested that the required abundances of greenhouse gasses may be 
consistent with geological evidence.
This was questioned by \citet{Rosing2010} who suggested that the dominant 
effect was a reduced cloud cover and hence lower terrestrial mean albedo
than at present,
resulting in a fainter Sun providing sufficient heating to achieve the required
surface temperature on Earth.
%\notecd [Needs update].
\citet{Shaviv2003} and \citet{Svensm2006}
noted that modulation of galactic
cosmic rays by an initially stronger solar wind could have contributed to
the warming of the early Earth, by similarly reducing the cloud cover.
Variations with time of solar activity and their possibly effects on
planetary atmospheres were also discussed by \citet{Gudel2007}.
There remains the problem of explaining the apparent stability of Earth's
temperature despite the variation in solar luminosity.
Various feedback mechanisms of a geological nature have been proposed
that may account for this \citep[{\eg},][]{Walker1981},
involving climate-dependent weathering
of rocks and ${\rm CO_2}$ outgassing from volcanoes;
a detailed review of these processes was provided by \citet{Kump2000}.%
\footnote{Alternatively, \citet{Margul1974} argued,
in the so-called \emph{Gaia hypothesis},
that the global response of the biosphere
could provide the required feedback regulation of the greenhouse effect.
Although attractive, this idea has seen little support from the available
evidence.}
%\notecd [We may return to effects of mass-loss in Sect.~\ref{sec:abundprob},
%with reference to \citet{Minton2007}].
A comprehensive review of the `faint early Sun problem' was provided by
\citet{Feulne2012}.

Beyond the present Sun the increase in luminosity continues, as is evident
from Figs~\ref{fig:HR} and \ref{fig:modrat}.
Also, the radius increases monotonically during the central hydrogen burning.
The evolution of the hydrogen-abundance profile is illustrated
in Fig.~\ref{fig:Xevol}.
The nuclear reactions cause a gradual reduction of the hydrogen in the core,
whereas helium settling, although fairly weak in the Sun, gives rise
to an increase in the hydrogen abundance in the convection zone and
the formation of a fairly sharp composition gradient at its base.
When hydrogen is exhausted at the centre there is a gradual transition
to hydrogen burning in a shell around a core consisting
predominantly of helium;
the core gradually grows in mass and contracts, leading to high central
densities and a substantial degree of degeneracy, while the 
hydrogen-burning shell becomes quite thin.
%This increase is enhanced when hydrogen is exhausted at the centre of 
%the model.
This enhances the increase in the stellar radius:
for reasons that are not entirely understood
\citep[see, however,][]{Faulkn2004}
the contraction of the core inside a burning shell leads to expansion of 
the region outside the shell.
The resulting strong expansion of the stellar surface radius leads to a 
decrease of the effective temperature and strong increase in the depth
of the convective envelope.
The evolution initially takes place at nearly constant luminosity,
on the so-called \emph{subgiant branch}.
Eventually, the star reaches a structure that, in terms of 
distance to the centre,
is nearly fully convective, apart from a radiative core of 
very small radial extent;
as a result, the star evolves towards higher luminosity with the increase in
radius, parallel and close to the Hayashi track.
At the final point illustrated in Fig.~\ref{fig:modrat} the convective envelope
extends over 68 per cent of the mass, and 79 per cent of the radius, of 
the model.
As shown in Fig.~\ref{fig:Xevol}
the resulting mixing with layers previously enriched in helium by settling
leads to a reduction in the surface hydrogen abundance.

For stars from slightly above solar mass and below
there is a systematic decrease in the rotation rate with increasing age
as the stars evolve on the main sequence;
for stars of solar mass \citep{Skuman1972, Barnes2003};
this is assumed to result from the loss of angular momentum in a magnetized
stellar wind \citep[{\eg},][]{Kawale1988, Matt2015},
presumably related to the generation of magnetic activity 
through dynamo action, as inferred in the Sun 
\citep[for a review, see][]{Charbo2020}.
Regardless of the substantial spread in early rotation rates,
these processes tend to lead to a well-defined rotation rate as a function
of age and mass, after an initial converging phase \citep[e.g.,][]{Gallet2013}.
This forms the basis for \emph{gyrochronology}, {\ie}, the determination
of ages of stars based on their rotation periods
\citep[e.g.,][]{Barnes2010, Epstei2014}.
The details of these processes, and of the subsequent redistribution of 
angular momentum in the stellar interior, are highly uncertain, however
\citep{Charbo1993, Gough1998, Talon2003, Charbl2005, Eggenb2005}.
%\notecd [Update references, check Baltimore IAU proc.]
In the solar case the result of the angular-momentum loss and redistribution,
as determined from helioseismology,
is a nearly spatially unvarying rotation in the radiative interior,
at a rate slightly below the equatorial surface rotation rate.
%\notecd [Check Howe, LRSP.]
These results, and their theoretical interpretation,
are discussed in Sect.~\ref{sec:heliorot} in the light of helioseismic
inferences of solar internal rotation.
Interestingly, by combining asteroseismic determinations of stellar ages
({\cf} Sect.~\ref{sec:astero}) with determinations of stellar rotation rates
\citet{vanSad2016} {\rv indicated} that the steady decrease of
rotation rate with increasing age slows down for stars older than a few Gyr,
indicating a weakening of the magnetic braking.
This {\rv would complicate} the use of gyrochronology for age determination of
stars older than the Sun.
{\rv However, I note that \citet{Barnes2016} questioned the
analysis by \citet{vanSad2016}.%
\footnote{{\rv A discussion that was in turn questioned by \citet{Metcal2019}}.}
Also, \citet{Lorenz2020} inferred a rotation rate 
matching the expectations for normal spin-down
for the solar twin HIP\,102152, with an age of 8\,Gyr inferred 
from isochrone fitting;
however, there may be some question about the precision of the
age and the modelling of the spin-down (van Saders, private communication).
Thus further work is clearly required to define the limits of applicability
of gyrochronology.}
%\notecd [May get back to this in Sect.~\ref{sec:stars}].
%\notecd [also possibly more to say here, with references; and need to check
%on status of other relevant LRSP papers. Miesch on rotation of the
%outer layers not obviously relevant here, but should be mentioned later.
%Internal rotation not yet there.
%Papers in Gough volume].

\subsection{Late evolutionary stages}
\clabel{sec:lateevol}

The later evolution of stars of solar mass is discussed in detail by
\citet{Kippen2012}.
The specific case of the Sun was considered by,
for example, \citet{Joerge1991} and \citet{Sackma1993}.
With continuing core contraction and expansion of the envelope the star moves
up along the Hayashi track as a \emph{red giant},
reaching a luminosity of more than $2000 \Lsun$
\citep[for a review of red-giant evolution, see][]{Salari2002};
needless to say, this is incompatible with life on Earth.
The helium core heats up, partly as a result of the contraction and partly
through heating from the hydrogen-burning shell whose temperature is 
forced to increase to match the energy required by the increasing luminosity.
When the core reaches a temperature of around $80 \times 10^6 \K$
helium burning starts, in the triple-alpha reaction producing ${}^{12}{\rm C}$.
Since the core is strongly degenerate the pressure is essentially independent
of temperature;
thus the heating associated with helium ignition initially has no effect on
the pressure and the burning takes place in a run-away process, 
a \emph{helium flash}, where
the core luminosity exceeds $10^{10} \Lsun$ for several hours.
However, the energy released is absorbed as gravitational energy in expanding
the inner parts of the star;
together with a decrease in the energy production from the
hydrogen shell-burning, this results in a drop of the surface luminosity.
Detailed calculations of the complex evolution through this phase
have been carried out by, for example, \citet{Schlat2001} 
and \citet{Cassis2003b},
and are also possible in the general-purpose MESA stellar evolution code
\citep{Paxton2011}. 
Hydrodynamical simulations in two and three dimensions of the evolution
during the flash were made by \citet{Mocak2008, Mocak2009},
confirming the importance of core convection in carrying away the energy 
generated during the flash.
Only when degeneracy is lifted by the increase in temperature 
and decrease in density does the core expand
and nuclear burning stabilizes in a phase of quiet core helium burning;
in addition to the triple-alpha reaction, ${}^{16}{\rm O}$ is produced
from ${}^4{\rm He}+{}^{12}{\rm C}$.
When helium is exhausted in the core the star again ascends
along the Hayashi track, on the \emph{asymptotic giant branch}.
Here the star enters the so-called \emph{thermally pulsing} phase where
helium repeatedly ignites in helium flashes
in a shell around the degenerate carbon-oxygen core, after which evolution
settles down again over a timescale of a few thousand years
\citep[{\eg},][]{Herwig2005}.
%\notecd [Possibly others? Check Sanya proc.]
Finally the star sheds its envelope through rapid mass loss
\citep[{\eg},][]{Willso2000, Miller2016},
leaving behind a hot and compact core
consisting predominantly of carbon and oxygen.
The Sun is expected to reach this point in its evolution at an age of around
12.4 Gyr, 7.8 Gyr from now.
The ejected material may shine due to the excitation from the ultraviolet
light emitted by the core, 
as \emph{a planetary nebula} which quickly disperses, 
with a lifetime of typically of order 10,000\,years
\citep[e.g.,][]{Gesick2018}.
The core contracts and cools over a very extended period as a white dwarf,
from its initial surface temperature of more than $10^5 \K$,
reaching a surface temperature of $4000 \K$ only after a further 10 Gyr.

The details of this evolution are still somewhat uncertain, depending in 
particular on the extent of mass loss in the red-giant phases,
and on exotic processes that may cool the core and delay helium ignition.
An uncertain issue of some practical importance is whether the solar radius
at any point reaches a size such as to engulf the Earth, taking into
account also the possible increase in the size of the Earth's orbit
resulting from mass loss from the Sun;
this depends in part on the variation of the radius during the final
thermal pulses.
In a detailed analysis of the evolutionary scenarios, 
\citet{Rybick2001} concluded that `it seems probable that the Earth will
be evaporated inside the Sun'.
This was confirmed by more recent calculations by \citet{Schrod2008},
taking into account tidal interactions between the planet and the expanding
Sun and dynamical drag in the solar atmosphere, as well as the compensating
effects of solar mass loss and their influence on the orbit of the planet.
According to their results, planets with a present distance from the Sun of
less than around 1.15 AU would be engulfed when the Sun reaches the tip of
the red-giant branch.
%\notecd [update].

It is obvious that the continued increase of solar luminosity, even on 
the main sequence, will have had catastrophic climatic consequences long
before this point is reached. 
Already \citet{Lovelo1982} noted that the increase over only 150 million
years would be larger than could be compensated for by a decreasing
greenhouse effect caused by a decrease in the atmospheric ${\rm CO_2}$ content,
to the minimum level required for photosynthesis.
In an interesting, if somewhat speculative, analysis \citet{Koryca2001}
pointed out the possibility of compensating for the increase in 
solar luminosity by increasing the size of the Earth's orbit through
engineering repeated, although infrequent, carefully controlled encounters
with a substantial asteroid.
It seems unlikely, however, that such a change could be rapid enough to
negate the effect of the increase of the solar luminosity on the red-giant
branch.
Furthermore, it is hardly necessary to point out that the Earth
may face more imminent threats to the climate as a result of the
antropogenic effects on the composition of the atmosphere
\citep[e.g.,][]{Crowle2000, Solomo2009, Cubasc2013}.

%\notecd [UPDATE: Here could also, with reference to \citet{Bahcal2001}
%and further presentation
%of result, discuss `semiconvection' that occurs beneath the 
%convective envelope at a slightly later stage.
%Probably makes more sense later, in fact, after discussing details of solar
%structure].

%% \newpage

%===========================================================================

\section{`Standard' solar models}
\clabel{sec:standard}

%\notecd [Here discuss the structure of the chosen reference case (Model~S++)
%in some detail.
%Probably for now to be done with `old' composition, so as not to
%pre-empt Pijpers {\etal}
%Of course, model should be made available electronically (and probably
%also evolution track).
%Might also make available a set of computed oscillation frequencies
%(but this is to be discussed later).]

As discussed in Sect.~\ref{sec:basicmod}, 
the concept of `standard solar model' has evolved greatly over the years;
the term goes back at least to \citet{Bahcal1969a}
who introduced it in connection with calculations of the solar neutrino flux.
It may now be taken to be a spherically symmetric model, including a relatively
simple treatment of diffusion and gravitational settling, up-to-date equation
of state, opacity and nuclear reactions, and a simple treatment of near-surface
convection.
Other potential hydrodynamical effects,
including mixing processes in the radiative interior
and the effects of rotation and its evolution, are ignored.
The evolution of the concept can be followed in several sets of solar evolution
calculations, often motivated by the solar neutrino problem
(see Sect.~\ref{sec:neutr}) and, more recently, by the availability of
detailed helioseismic constraints (see Sect.~\ref{sec:helio}).
An impressive example are the efforts of John Bahcall over an extended period.
As reviewed by \citet{Bahcal1989} early models did not include diffusion
\citep[{\eg},][]{Bahcal1968a, Bahcal1988}.
\citet{Bahcal1992b} included diffusion of helium,
whereas later models \citep[e.g.,][]{Bahcal1995, Bahcal2006}
included diffusion of both helium and heavier elements.
Other examples of standard model computations are
\citet{Turck1988},
\citet{Cox1989},
\citet{Guenth1992},
\citet{Bertho1993},
\citet{Turck1993},
\citet{Gabrie1994, GabrieM1997},
\citet{Chaboy1995},
\citet{Guenth1996},
\citet{Richar1996},
\citet{Schlat1997},
\citet{Brun1998},
\citet{Elliot1998b},
\citet{Morel1999},
\citet{Neufor2001a} and
\citet{Serene2011}.
A recent comprehensive recomputation of solar models was carried out
by \citet{Vinyol2017},
discussed in more detail in Sect.~\ref{sec:modcorcomp}.
%\notecd [Morel et al., suitable Yale, suitable Warsaw??, suitable 
%Turck-Chi\`eze, suitable Toulouse, suitable Li\`ege?].
A brief review of standard solar modelling was provided by
\citet{Serene2016a}.

As representative of standard models I here consider the so-called Model~S
of \citet{Christ1996};
details on the model calculation were provided by \citet{Christ2008a}.
Although more than two decades old, 
and to some extent based on out-dated physics, it is
still seeing substantial use for a variety of applications, including as
reference for helioseismic inversions.
Thus it provides a useful reference for discussing the effects of various
updates to the model physics.
%Similar results have been obtained in other calculations, some
%of which are discussed below.
%\notecd [Not quite logical anymore; needs a little thought].
Remarkably, as discussed in Sections~\ref{sec:heliostruc} and
\ref{sec:neutr}, such simple models are
in reasonable agreement with observations of solar oscillations and neutrinos.
%Updated versions of Model~S are discussed in Sect.~\ref{sec:newmodS}
%and will be made available in the online version of the paper.

\subsection{Model~S}
\clabel{sec:models}

Model~S was computed with the OPAL equation of state \citep{Rogers1996}
and the 1992 version of the OPAL opacities
\citep{Rogers1992}, with low-temperature opacities from
\citet{Kurucz1991}.%
\footnote{\citet{Christ1996} did not make it clear that the
earlier OPAL tables (OPAL92) were used, rather than updated OPAL96 tables
of \citet{Iglesi1996}.
A comparison between solar models computed with these two sets of tables
is provided in Fig.~\ref{fig:changeopac}.}
Nuclear reaction parameters were generally obtained from \citet{Bahcal1995},
and electron screening was treated in the weak-screening approximation of
\citet{Salpet1954}.
The computation was started from a static and chemically homogeneous
zero-age main-sequence model, and the age of the present Sun, from that state,
was assumed to be $4.6 \Gyr$.
The time evolution of the ${}^3{\rm He}$ abundance was followed, while the
other reactions in the PP chains were assumed to be in nuclear equilibrium;
to represent the pre-main-sequence evolution the initial ${}^3{\rm He}$ 
abundance was assumed to correspond to the evolution of the abundance at
constant conditions for a period of $5 \times 10^7 \yr$,
starting at zero abundance \citep[see][]{Christ1974}.
%\notecd [although might try to test effect of a more realistic initial 
%abundance.]
Similarly, the CN part of the CNO cycle ({\cf} Eq.~\ref{eq:CNO}) was assumed
to have reached nuclear equilibrium in the pre-main-sequence phase while
the conversion of ${}^{16}{\rm O}$ into ${}^{14}{\rm N}$ was followed.
The diffusion and settling of helium and heavy elements
were computed in the approximation of \citet{Michau1993};
the evolution of $Z$ was computed neglecting the effect of nuclear reactions
and representing $D_i$ and $V_i$ by
the behaviour of fully ionized ${}^{16}{\rm O}$.
%\notecd [probably with a reference to an appendix?].
Convection was treated in the \citet{Boehm1958} formalism.
The atmospheric structure was computed using the VAL $T(\tau)$ relation 
given by Eq.~\Eq{eq:ttauval} and illustrated in Fig.~\ref{fig:ttau}.
The initial composition was calibrated to obtain a present $\Zs/\Xs = 0.0245$
\citep{Greves1993},
while the surface luminosity and radius were set to 
$3.846 \times 10^{33} \erg \s^{-1}$ and $6.9599 \times 10^{10} \cm$,
respectively, to an accuracy of better than $10^{-6}$
(see Sect.~\ref{sec:basicpar}).
%\notecd [References for these quantities in any case probably given before].
%Further details of the calculation were provided by \citet{Christ2008a}.

Some basic quantities of the model of the present Sun are given
in Table~\ref{tab:modelchar} below, together with properties of other similar
models, discussed in detail in Sect.~\ref{sec:modsens}.
Also, Fig.~\ref{fig:Xmod} shows the variation of $X$ and $Z$ through the model.
It is striking that the settling of helium and heavy elements causes sharp
gradients in $X$ and $Z$ just below the convection zone.
Details of the model structure are provided at
{\tt https://github.com/jcd11/LRSP\_models}.
%\notecd [here with some link, or whatever; simple table and model in fgong
%format, perhaps defining the format also in an appendix].

\begin{figure}[htp]
\centerline{\includegraphics[width=\figwidth]{\fig/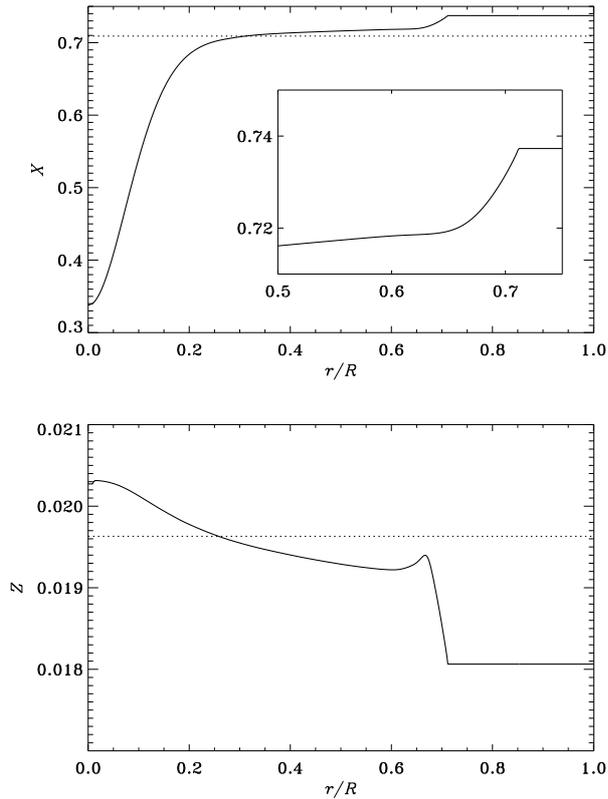}}
\caption{
Hydrogen abundance $X$ (top panel) and heavy-element abundance (lower panel)
against fractional radius, in a model (Model~S of \citecl{Christ1996})
of the present Sun.
The inset in the upper panel shows the hydrogen-abundance profile in the
vicinity of the base of the convective envelope.
The horizontal dotted lines show the initial values $X_0$ and $Z_0$.
}
\clabel{fig:Xmod}
\end{figure}

\begin{figure}[htp]
\centerline{\includegraphics[width=\figwidth]{\fig/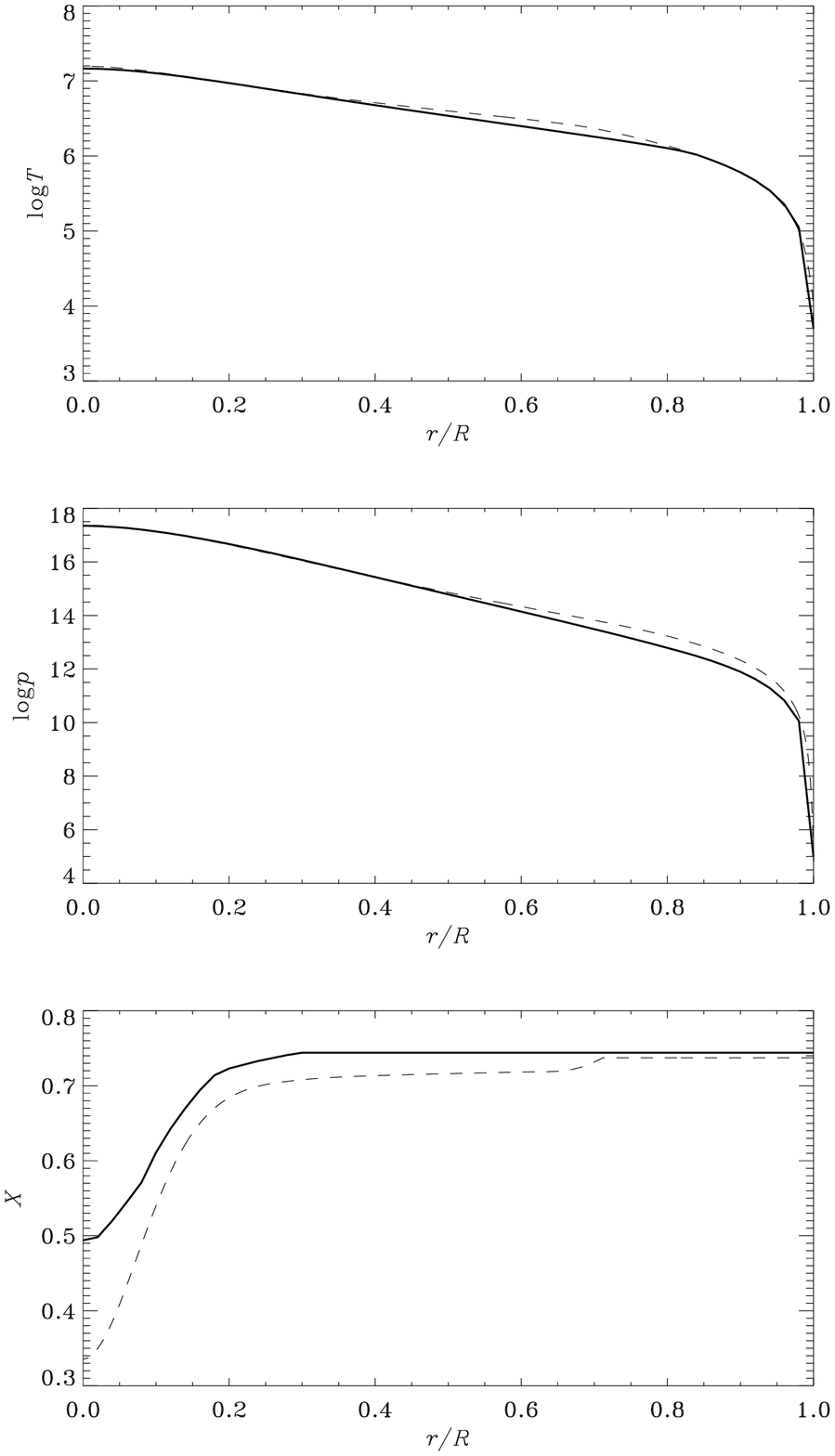}}
\caption{
Comparison of Model~S of \citet{Christ1996} (dashed curves)
with a $1 \Msun$ model computed by \citet{Weyman1957} (solid curves).
The quantities illustrated are temperature $T$, in $\K$ (top panel),
pressure $p$, in $\dyn \cm^{-2}$ (central panel) and hydrogen abundance $X$
(bottom panel).
}
\clabel{fig:weymann}
\end{figure}

It is perhaps of some interest to compare the structure of this model with
an early calibrated model of solar structure.
In Fig.~\ref{fig:weymann} Model~S is compared with a
$1 \Msun$ model computed by \cite{Weyman1957}, as quoted by
\citet{Schwar1958};
the model has solar radius and approximately solar luminosity at an age
of $4.5 \Gyr$.
It is evident that the hydrogen profile differs substantially between the 
two models, in part owing to the inclusion of settling in Model~S, but more
importantly because the Weymann model is less evolved.
On the other hand, on this scale temperature and pressure look quite similar
between the two models.
In fact, the central temperature and pressure differ by less than 10 per cent,
although there are differences of up to nearly 30 per cent in temperature 
elsewhere in the model and even larger differences in pressure.
Another significant difference is in the depth of the convective envelope
which is around $0.15 \Rsun$ in the Weymann model and 0.29 in Model~S.
Even so, given that Model~S provides a reasonable representation of 
solar structure (see Sect.~\ref{sec:heliostruc}),
it is evident that the early model succeeded in capturing important 
aspects of the structure of the Sun.
%\notecd [Might need to note that Model~S is similar to the Sun!].

\subsection{Sensitivity of the model to changes in physics or parameters}
\clabel{sec:modsens}

%\notecd [Illustrate sensitivity to some modifications to the parameters (age,
%etc.); again, could be done in more detail for online version.]
%
It is evident that the uncertainty in the input parameters, and physics,
of the calculation introduces uncertainties in the model structure.
A number of investigations have addressed aspects of these uncertainties.
An early example is provided by 
\citet{Christ1988a} who considered several different changes to the
model physics, analysing the effects on the model structure and the
resulting oscillation frequencies.
Remarkably, he found that the change to the structure was essentially
linear in the change in opacity as represented by $\log \kappa$,
even for quite substantial changes.
Such linearity in changes to $\Gamma_1$ was also found by \citet{Christ1991a}.
\citet{Boothr2003} considered a broad range of changes in the model parameters
and physics, emphasizing comparisons with the helioseismically inferred
sound speed obtained by \citet{Basu2000}.
A very ambitious investigation was carried out by \citet{Bahcal2006}
who made a Monte Carlo simulation based on 10\,000 models with random
selections of 21 parameters characterizing the models, in this way assigning
statistical properties to the computed model quantities, including 
detailed neutrino fluxes.
It was demonstrated by \citet{Jorgen2017b}
that, owing to the near linearity of the model response to changes
in parameters \citep[see also][]{Bahcal2005e},
this result could to a large extent be recovered much more economically by
computing the relevant partial derivatives with respect 
to the model parameters;
this opens the possibility for more extensive statistical analysis of this
nature.
A more systematic exploration of the linearity of the response of
solar models was carried out by \citet{Villan2010} who linearized
the equations of stellar structure in terms of various perturbations and,
consistent with the numerical experiments discussed above, demonstrated that
the resulting changes to the model closely matched the differences
between models computed with the assumed perturbations.
%\notecd [UPDATE: Could summarize early Gough efforts,
%as detailed in mail with subject
%'Re: Linearized response of solar models: BEWARE long record', from 9/3/10].
%\notecd [Possible UPDATE:
%Here discuss detailed Sackmann etc., and Bahcall analysis,
%in a fairly loose sense].

\begin{figure}[htp]
\centerline{\includegraphics[width=\figwidth]{\fig/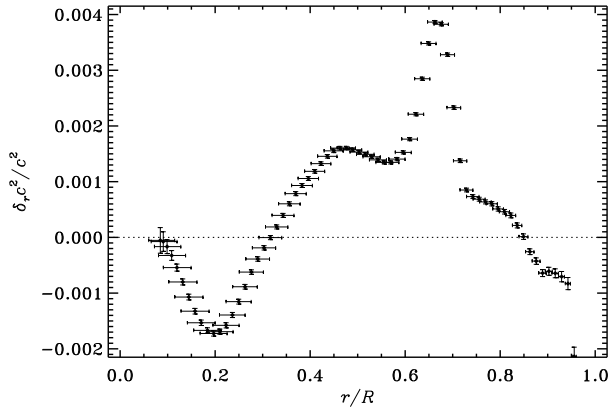}}
\caption{Results of helioseismic inversions.%
Inferred relative differences in squared sound speed
between the Sun and Model~S 
in the sense (Sun) -- (model). 
The vertical bars show $1\,\sigma$ errors in the inferred values,
based on the errors in the observed frequencies.
The horizontal bars provide a measure of the resolution of the inversion.
(Adapted from \citecl{Basu1997a}.)
}
\clabel{fig:csqinvS}
\end{figure}

Here I consider some examples of changes to the model parameters and
physics, emphasizing the updates that have taken place since the original
computation of Model~S.
When not specifically mentioned, the physical properties and parameters
of the models are the same as for Model~S (see also Table~\ref{tab:modelpar}),
which is also in most cases used as reference.
An overview of the models considered is provided by Table~\ref{tab:modelpar},
while Table~\ref{tab:modelchar} gives basic properties of the models,
and Table~\ref{tab:modeldiff} presents the differences between the modified
models and Model~S.
To put the results in context, Fig.~\ref{fig:csqinvS}
shows the helioseismically inferred difference%
\footnote{This and the subsequent helioseismic structure inversions
were carried out with an inversion code provided by Maria Pia Di Mauro
\citep[see][]{DiMaur2002, DiMaur2003}.}
in squared sound speed between the Sun and Model~S.
Note that the statistical errors in the inferences are barely visible,
compared with the size of the symbols.
The helioseismic results on solar structure are discussed in detail
in Sect.~\ref{sec:heliostruc}.

To interpret the results of such model comparisons, it is useful to note some 
simple properties of the solar convection zone 
\citep[see also][]{Gough1984b, Christ1992b, Christ1997a, Christ2005}.
%\notecd [May also need a few references to Gough, Baturin?, as appropriate].
Apart from the relatively thin ionization zones of hydrogen and helium,
pressure and density in the convection zone are approximately related by
Eq.~\Eq{eq:appcon}, with $\gamma = 5/3$; 
also, since the mass of the convection zone is only around $0.025 \Msun$
we can, as a first approximation, assume that $m \simeq M$ in the convection
zone.
In this case it is easy to show that%
\footnote{more precisely, $1/R$ should be replaced by $1 / R_*$,
such that $c^2$ within the convection zone extrapolates to 0 at $r = R_*$
\citep{Houdek2007a, Gough2013a}.}
\be
c^2 = {\gamma p \over \rho} \simeq (\gamma - 1) G M 
\left( {1 \over r} - {1 \over R} \right)  \; .
\eel{eq:approxcsq}
It follows that $c$ is unchanged at fixed $r$ between models with the same
mass and surface radius.
Also, 
\be
{\delta_r p \over p} = {\delta_r \rho \over \rho} \simeq
- {1 \over \gamma - 1} {\delta K \over K} \; ,
\eel{eq:appdelp}
where $\delta_r$ denotes the difference between two models at fixed $r$,
and $\delta K$ is the difference in $K$ between the models.
Finally, assuming the ideal gas law, Eq.~\Eq{eq:idealg},
\be
{\delta_r T \over T} \simeq {\delta_r \mu \over \mu} \; ,
\eel{eq:appdelt}
which is obviously constant.

\begin{table}
\caption{\clabel{tab:modelpar}
%\notecd [Model parameters]
Parameters of solar models. Age, $R$ and $L$ are for the model 
of the present Sun.
OPAL92, OPAL96 and OP05 refer to the opacity tables by
\citet{Rogers1992}, \citet{Iglesi1996} and \citet{Badnel2005}, respectively,
while Kur91, Alex94 and Fer05 indicate low-temperature opacities from
\citet{Kurucz1991}, \citet{Alexan1994} and \citet{Fergus2005}.
The heavy-element abundance used in the opacities are GN93 \citep{Greves1993}
or GS98 \citep{Greves1998}.
The default equation of state is the \citet{Rogers1996} implementation of the
OPAL formulation, while Model~{\Meoszerofive} used the \citet{Rogers2002} 
version.
In Models~{\Mopa} and {\Mopb} localized increases in opacity
({\cf} Eq. \ref{eq:opmod}) were included, at respectively
$\log T_\kappa = 7.0$ and $6.5$.
Model~{\MCM} replaced the mixing-length treatment of convection 
\citep{Boehm1958} by an emulation of the \citet{Canuto1991} formulation.
The default set of nuclear-reaction parameters is based on 
\citet{Bahcal1995}, while Models~{\Madelb} and {\Mnacre} used, respectively,
the set from \citet{Adelbe2011} and the NACRE set \citep{Angulo1999} with
an updated $\nitfourteen + \hyd$ reaction \citep{Formic2004}.
In Model~{\Mhethreeeq} the $\helthree$ abundance was assumed to be always
in nuclear equilibrium, while Model~{\Mnoelscrn} neglected electron
screening.
In Model~{\Mdifa} the diffusion coefficient $D_i$ was increased by a
factor 1.2 ({\cf} Eq.~\ref{eq:diffusion}), while in Model~{\Mdifb}
both $D_i$ and $V_i$ were increased by this factor.
Finally diffusion and settling were neglected in Model~{\Mnodif}.
For further details on the model physics, see Sect.~\ref{sec:microphys}.
Values or other aspects differing from Model~S are shown as {\bf bold}.
%\notecd [Needs to be implemented fully].
The detailed structure of the models is
provided at {\tt https://github.com/jcd11/LRSP\_models}.
}
\vskip 4mm
\centering
{\small
\begin{tabular}{l|ccccccc}
\hline
Model & Age & $R$  & $L$ & Opacity & Surface  & Surface & Other changes \\
      & (Gyr) & $(10^{10} \cm)$ & $(10^{33} \erg \s^{-1})$ & tables &  opacity & comp. & (see caption) \\
\hline
{\MS}     & 4.60 & 6.9599 & 3.846 & OPAL92  & Kur91 & GN93  &   -   \\
%{\MSprime}     & 4.60 & 6.9599 & 3.846 & {\bf OPAL96}  & {\bf Alex94} & GN93  &  Updated Model S   \\
{\Mage}    & {\bf 4.57} & 6.9599 & 3.846 & OPAL92  & Kur91 & GN93  &   -  \\
\MRs    & 4.60 & {\bf 6.95508} & 3.846 & OPAL92  & Kur91 & GN93  &   -  \\
\MLs    & 4.60 & 6.9599 & {\bf 3.828} & OPAL92  & Kur91 & GN93  &   -   \\
{\Meoszerofive} & 4.60 & 6.9599 & 3.846 & OPAL92  & Kur91 & GN93  & {\bf EOS Liv05}   \\
\Mopa   & 4.60 & 6.9599  & 3.846 & OPAL92  & Kur91 & GN93  &   {\bf Local $\delta \log \kappa$} \\
\Mopb   & 4.60 & 6.9599  & 3.846 & OPAL92  & Kur91 & GN93  &   {\bf Local $\delta \log \kappa$}  \\
\Mopalninesix   & 4.60 & 6.9599  & 3.846 & {\bf OPAL96}  & Kur91 & GN93  &  - \\
{\Msurfop} & 4.60 & 6.9599 & 3.846 & {\bf OPAL96}  & {\bf Fer05} & GN93  & Surf. opac.   \\
\Mgsnineeight  & 4.60 & 6.9599  & 3.846 & {\bf OPAL96}  & {\bf Fer05} &{\bf GS98}  &   -     \\
\Mopzerofive  & 4.60 & 6.9599  & 3.846 & {\bf OP05}  & {\bf Fer05} &{\bf GS98}  &   -    \\
{\MCM} & 4.60 & 6.9599 & 3.846 & OPAL92  & Kur91 & GN93  & {\bf CM conv.}   \\
{\Madelb} & 4.60 & 6.9599 & 3.846 & OPAL92  & Kur91 & GN93  & {\bf EnGen. Adelberger}   \\
{\Mnacre} & 4.60 & 6.9599 & 3.846 & OPAL92  & Kur91 & GN93  & {\bf EnGen. NACRE}   \\
{\Mhethreeeq} & 4.60 & 6.9599 & 3.846 & OPAL92  & Kur91 & GN93  & {\bf $\helthree$ nucl. eql.}   \\
{\Mnoelscrn} & 4.60 & 6.9599 & 3.846 & OPAL92  & Kur91 & GN93  & {\bf no electr. screen.} \\
\Mdifa     & 4.60 & 6.9599  & 3.846 & OPAL92  & Kur91 & GN93  & {\bf Change diff.} \\
\Mdifb    & 4.60 & 6.9599  & 3.846 & OPAL92  & Kur91 & GN93  &  {\bf Change diff., settl.} \\
\Mnodif    & 4.60 & 6.9599  & 3.846 & OPAL96  & Alex94 & GN93  &  {\bf No diffusion} \\
\hline
\end{tabular}
}
\end{table}

%\begin{table}
%\caption{\clabel{tab:modelchar}
%\notecd [Model characteristics]
%}
\vskip 4mm
%\centering
%\begin{tabular}{l|ccccccccc}
%\hline
%Model   & $X_0$ & $Z_0$ & $T_{\rm c}$ & $\rho_{\rm c}$ & $X_{\rm c}$ & $Y_{\rm s}$ & $\Zs/\Xs$ & $d_{\rm cz}/R$ \\
%        &       &       & $(10^6 \K)$ & $(\g \cm^{-3})$  &             &             &           &       &        \\
%\hline
%%l9bi.d.02c 
%S    & 0.70909 & 0.019630 & 15.669 & 153.87 & 0.33725 & 0.24466 & 0.02450 & 0.28846  \\
%%l9bi.d.06 
%\Mage   & 0.70886 & 0.019617 & 15.661 & 153.38 & 0.33888 & 0.24497 & 0.02450 & 0.28806  \\
%%l9bi.d.05 
%\MRs   & 0.70912 & 0.019628 & 15.668 & 153.86 & 0.33730 & 0.24467 & 0.02450 & 0.28866  \\
%%l9bi.d.15 
%\Mopa  & 0.70850 & 0.019628 & 15.665 & 153.71 & 0.33755 & 0.24508 & 0.02450 & 0.28794  \\
%%l9bi.d.16 
%\Mopb  & 0.70904 & 0.019623 & 15.669 & 153.91 & 0.33716 & 0.24478 & 0.02450 & 0.28868  \\
%%l9bi.d.07 
%\Mopalninesix  & 0.70815 & 0.019626 & 15.695 & 153.71 & 0.33578 & 0.24532 & 0.02450 & 0.28660  \\
%%l9bi.d.37c 
%\Mgsnineeight & 0.70696 & 0.018496 & 15.696 & 153.93 & 0.33542 & 0.24686 & 0.02307 & 0.28344  \\ % fixed 7/1/19
%%l9bi.d.17 
%\Mdifa    & 0.70911 & 0.019623 & 15.667 & 153.55 & 0.33847 & 0.24482 & 0.02450 & 0.28838  \\
%%l9bi.d.18 
%\Mdifb   & 0.70714 & 0.020054 & 15.718 & 154.68 & 0.33240 & 0.24165 & 0.02450 & 0.29064  \\
%\hline
%\end{tabular}
%\end{table}

\begin{table}
\caption{\clabel{tab:modelchar}
%\notecd [Model characteristics]
Characteristics of the models in Table~\ref{tab:modelpar}.
$X_0$ and $Z_0$ are the initial hydrogen and heavy-element abundances, 
$T_{\rm c}$, $\rho_{\rm c}$ and $X_c$ are the central temperature, density
and hydrogen abundance of the model of the present Sun, $\Ys$ is
the surface helium abundance, $\Zs/\Xs$ is the present ratio between
the surface heavy-element and hydrogen abundances and $d_{\rm cz}$
is the depth of the convective envelope.
}
\vskip 4mm
\centering
\begin{tabular}{l|ccccccccc}
\hline
Model   & $X_0$ & $Z_0$ & $T_{\rm c}$ & $\rho_{\rm c}$ & $X_{\rm c}$ & $Y_{\rm s}$ & $\Zs/\Xs$ & $d_{\rm cz}/R$ \\
        &       &       & $(10^6 \K)$ & $(\g \cm^{-3})$  &             &             &           &       &        \\
\hline
{\MS} & 0.70911 & 0.019631 & 15.667 & 153.86 & 0.33765 & 0.24464 & 0.02450 & 0.28844 \\
%{\MSprime} & 0.70810 & 0.019629 & 15.691 & 153.69 & 0.33636 & 0.24528 & 0.02450 & 0.28584 \\
{\Mage} & 0.70887 & 0.019617 & 15.659 & 153.37 & 0.33923 & 0.24496 & 0.02450 & 0.28805 \\
{\MRs} & 0.70914 & 0.019629 & 15.666 & 153.85 & 0.33769 & 0.24465 & 0.02450 & 0.28864 \\
{\MLs} & 0.70957 & 0.019639 & 15.642 & 153.25 & 0.33961 & 0.24425 & 0.02450 & 0.28839 \\
{\Meoszerofive} & 0.70872 & 0.019617 & 15.670 & 154.05 & 0.33717 & 0.24505 & 0.02450 & 0.28881 \\
{\Mopa} & 0.70852 & 0.019628 & 15.663 & 153.70 & 0.33794 & 0.24506 & 0.02450 & 0.28793 \\
{\Mopb} & 0.70906 & 0.019623 & 15.667 & 153.91 & 0.33755 & 0.24476 & 0.02450 & 0.28866 \\
{\Mopalninesix} & 0.70817 & 0.019625 & 15.692 & 153.70 & 0.33619 & 0.24530 & 0.02450 & 0.28658 \\
{\Msurfop} & 0.70808 & 0.019633 & 15.691 & 153.68 & 0.33636 & 0.24527 & 0.02450 & 0.28589 \\
{\Mgsnineeight} & 0.70696 & 0.018496 & 15.696 & 153.93 & 0.33542 & 0.24686 & 0.02307 & 0.28345 \\
{\Mopzerofive} & 0.71110 & 0.018540 & 15.647 & 153.63 & 0.33929 & 0.24365 & 0.02307 & 0.28693 \\
{\MCM} & 0.70905 & 0.019638 & 15.667 & 153.84 & 0.33768 & 0.24462 & 0.02450 & 0.28854 \\
{\Madelb} & 0.70933 & 0.019632 & 15.640 & 153.77 & 0.34209 & 0.24447 & 0.02450 & 0.28823 \\
{\Mnacre} & 0.70960 & 0.019661 & 15.661 & 154.37 & 0.34315 & 0.24397 & 0.02450 & 0.28771 \\
{\Mhethreeeq} & 0.70888 & 0.019621 & 15.661 & 153.63 & 0.33840 & 0.24488 & 0.02450 & 0.28835 \\
{\Mnoelscrn} & 0.71028 & 0.019734 & 15.752 & 157.09 & 0.34111 & 0.24269 & 0.02450 & 0.28599 \\
{\Mdifa} & 0.70913 & 0.019623 & 15.665 & 153.54 & 0.33886 & 0.24480 & 0.02450 & 0.28837 \\
{\Mdifb} & 0.70716 & 0.020054 & 15.715 & 154.67 & 0.33282 & 0.24163 & 0.02450 & 0.29061 \\
{\Mnodif} & 0.71798 & 0.017590 & 15.455 & 149.71 & 0.36024 & 0.26443 & 0.02450 & 0.27324 \\
\hline
\end{tabular}
\end{table}

\begin{table}
\caption{\clabel{tab:modeldiff}
%\notecd [Model differences]
Differences between the model quantities in Table~\ref{tab:modelchar}
and the corresponding properties of Model~{\MS}.
}
\vskip 4mm
\centering
\begin{tabular}{l|ccccccccc}
\hline
Model   & $\delta X_0$ & $\delta Z_0$ & $\delta T_{\rm c}/T_{\rm c}$ & $\delta \rho_{\rm c}/\rho_{\rm c}$ & $\delta X_{\rm c}$ & $\delta Y_{\rm s}$ & $\delta (\Zs/\Xs)$ & $\delta(d_{\rm cz}/R)$ \\
& $\times 10^3$ & $\times 10^3$ & $\times 10^3$ & $\times 10^3$ & $\times 10^3$ & $\times 10^3$ & $\times 10^3$ & $\times 10^3$ \\
\hline
{\MS} &    0.000 &    0.000 &    0.000 &    0.000 &    0.000 &    0.000 &    0.000 &    0.000 \\ 
%{\MSprime} & -1.003 &   -0.002 &    1.539 &   -1.145 &   -1.289 &    0.641 &   0.000 &   -2.607 \\ 
{\Mage} &  -0.234 &   -0.013 &   -0.499 &   -3.182 &    1.588 &    0.314 &   0.000 &   -0.399 \\ 
{\MRs} &    0.034 &   -0.002 &   -0.026 &   -0.074 &    0.044 &    0.008 &   0.000 &    0.195 \\ 
{\MLs} &    0.464 &    0.008 &   -1.572 &   -3.977 &    1.961 &   -0.391 &   0.000 &   -0.059 \\ 
{\Meoszerofive} &  -0.383 &   -0.014 &    0.237 &    1.236 &   -0.475 &    0.412 &   0.000 &    0.365 \\ 
{\Mopa} &  -0.591 &   -0.003 &   -0.234 &   -1.087 &    0.292 &    0.418 &    0.000 &   -0.518 \\ 
{\Mopb} &  -0.048 &   -0.007 &    0.040 &    0.266 &   -0.092 &    0.119 &   0.000 &    0.217 \\ 
{\Mopalninesix} &  -0.935 &   -0.006 &    1.637 &   -1.058 &   -1.453 &    0.663 &   0.000 &   -1.864 \\ 
{\Msurfop} &  -1.029 &    0.002 &    1.545 &   -1.190 &   -1.281 &    0.625 &    0.000 &   -2.553 \\ 
{\Mgsnineeight} &  -2.152 &   -1.134 &    1.857 &    0.440 &   -2.222 &    2.220 &   -1.430 &   -4.999 \\ 
{\Mopzerofive} &   1.988 &   -1.090 &   -1.265 &   -1.541 &    1.640 &   -0.991 &   -1.430 &   -1.511 \\ 
{\MCM} &  -0.056 &    0.007 &    0.005 &   -0.133 &    0.036 &   -0.021 &   0.000 &    0.094 \\ 
{\Madelb} &   0.218 &    0.001 &   -1.698 &   -0.642 &    4.444 &   -0.168 &   0.000 &   -0.214 \\ 
{\Mnacre} &   0.489 &    0.030 &   -0.339 &    3.279 &    5.502 &   -0.667 &   0.000 &   -0.733 \\ 
{\Mhethreeeq} &  -0.229 &   -0.009 &   -0.369 &   -1.495 &    0.755 &    0.243 &   0.000 &   -0.097 \\ 
{\Mnoelscrn} &   1.174 &    0.103 &    5.457 &   20.993 &    3.464 &   -1.953 &   0.000 &   -2.450 \\ 
{\Mdifa} &   0.022 &   -0.007 &   -0.122 &   -2.110 &    1.217 &    0.157 &   0.000 &   -0.078 \\ 
{\Mdifb} &  -1.951 &    0.423 &    3.112 &    5.214 &   -4.828 &   -3.009 &   0.000 &    2.168 \\ 
{\Mnodif} &   8.868 &   -2.040 &  -13.477 &  -27.007 &   22.592 &   19.793 &   0.000 &  -15.202 \\ 
\hline
\end{tabular}
\end{table}

%\notecd [Brief discussion of properties of differences. 
%Check \citet{Christ2005}, \citet{Christ1997a},
%papers for simple representations (although
%some of this may have to wait until next iteration).
%At least stress sound speed in convection zone].

Since the effects of the changes are subtle, some care is required in
specifying and computing the differences.%
\footnote{An example of the confusion that may be caused by even
quite subtle problems in the computation is provided by
\citet{Guenth1989}; see \citet{Christ1991a}.}
Here I consider differences (also denoted $\delta_r$) at fixed 
fractional radius $r/R$, where $R$ is the photospheric radius.
%\notecd [Need to check for consistency].
It should be noted, however, that \citet{Christ1997b} found differences 
$\delta_m$ at
fixed mass fraction $m/M$ more illuminating for studies of the effects 
on oscillation frequencies of near-surface modifications to the model.
Such differences are also more appropriate for studying evolutionary
effects on stellar models.
They showed that the two differences are related by
\begin{eqnarray}
\delta_m f &=& \delta_r f + \delta_m r {\dd f \over \dd r} \nonumber \\
\delta_r f &=& \delta_m f + \delta_r m {\dd f \over \dd m} \; ,
\label{eq:diffrel}
\end{eqnarray}
for any model quantity $f$.

A prerequisite for sensible studies of solar models and their dependence
on the physics is that adequate numerical precision is reached. 
I discuss this in Appendix~\ref{sec:numacc}.

\begin{figure}[htp]
\centerline{\includegraphics[width=\figwidth]{\fig/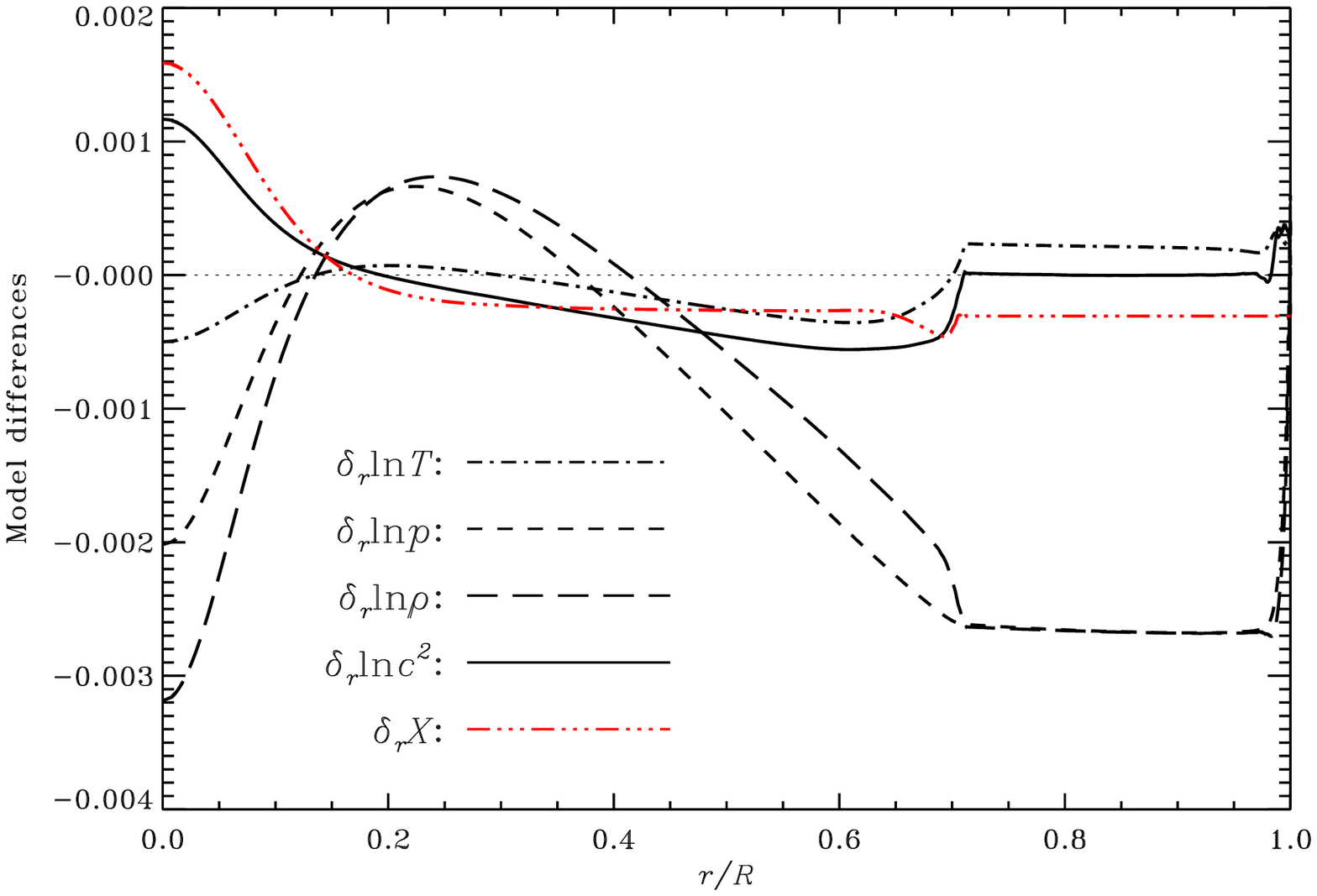}}
\caption{
Model changes at fixed fractional radius resulting from a change in age,
from the reference value of 4.6 Gyr used in Model~S to Model~{\Mage}
with an age of 4.57 Gyr (see Sect.~\ref{sec:basicpar}),
in the sense (Model~\Mage) -- (Model~S).
The line styles are defined in the figure.
The thin dotted line marks zero change.
%\notecd [change\_age.idl; dgr.l9bi.d.06c-d.02c]
}
\clabel{fig:changeage}
\end{figure}

\begin{figure}[htp]
\centerline{\includegraphics[width=\figwidth]{\fig/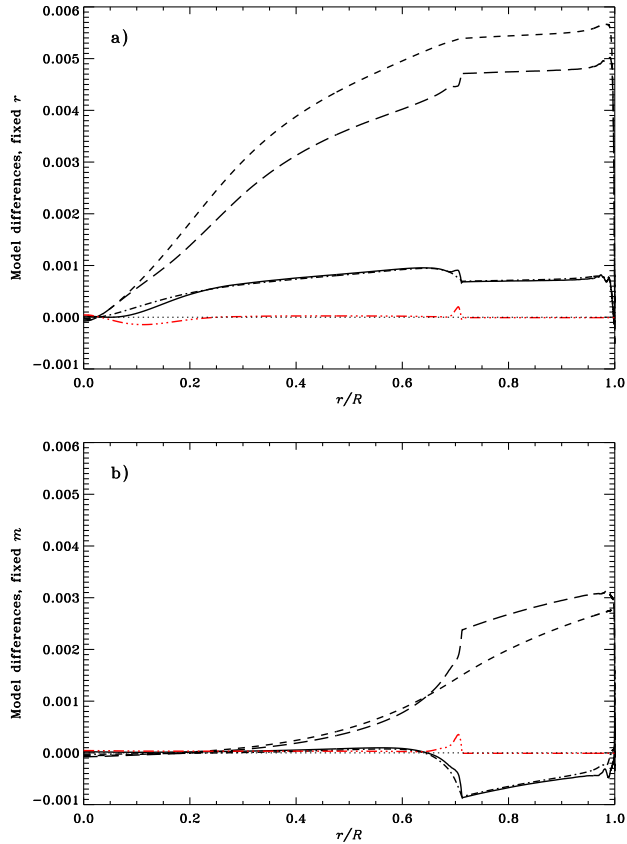}}
\caption{
Model changes at fixed fractional radius (panel a) and fixed mass (panel b),
resulting from a change
in photospheric radius, from the reference value of
$6.9599 \times 10^{10} \cm$ used in Model~S to the value of
$6.95508 \times 10^{10} \cm$ in the sense (Model~\MRs) -- (Model~S).
Line styles are as defined in Fig.~\ref{fig:changeage}.
%\notecd [change\_radius.idl; dgr.l9bi.d.05c-d.02c]
%\notecd [Add a, b.]
}
\clabel{fig:changeradius}
\end{figure}

\begin{figure}[htp]
\centerline{\includegraphics[width=\figwidth]{\fig/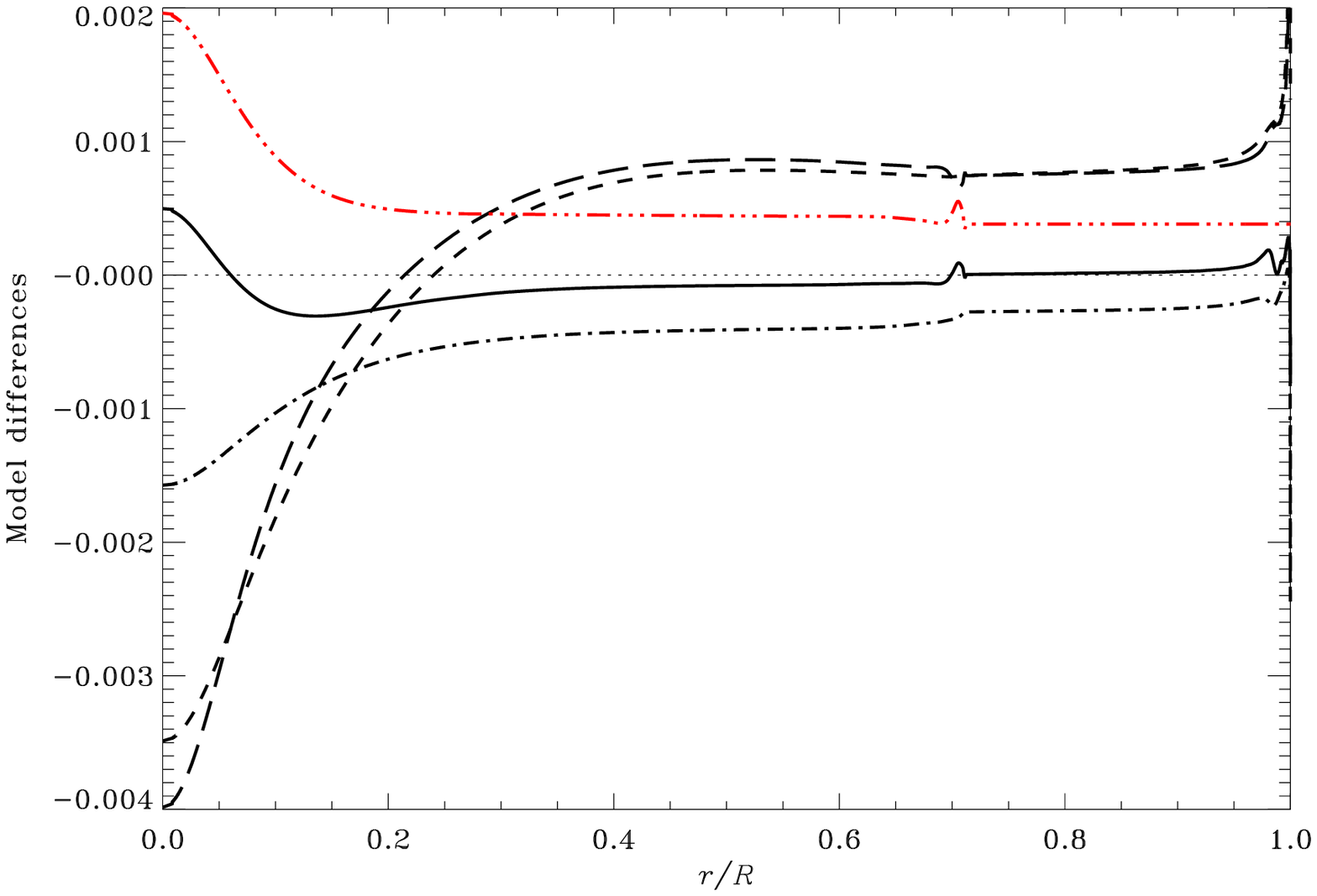}}
\caption{
Model changes at fixed fractional radius 
resulting from a change
in surface luminosity, from the reference value of
$3.846 \times 10^{33} \erg \s^{-1}$ used in Model~S to the value of
$3.828 \times 10^{33} \erg \s^{-1}$ adopted by \citet{Mamaje2015}
as the nominal solar luminosity, in the sense (Model~\MLs) -- (Model~S).
Line styles are as defined in Fig.~\ref{fig:changeage}.
%\notecd [change\_luminosity.idl; dgr.l9bi.d.41c-d.02c]
}
\clabel{fig:changelum}
\end{figure}

I first consider changes in the global parameters characterizing the model.
Figure~\ref{fig:changeage} shows the effect of decreasing the model age
to the now generally accepted value of 4.57 Gyr 
(see Sect.~\ref{sec:basicpar}),
compared with the reference value of 4.6 Gyr in Model~S.
To match the solar luminosity at this lower age, a slightly smaller initial 
hydrogen abundance is required, increasing $\mu$ ({\cf} Eq.~\ref{eq:homlum}); 
on the other hand, the increased central hydrogen abundance reflects
the shorter time spent in hydrogen burning. 
As predicted above, the sound-speed difference is virtually zero in
the convection zone, except in the ionization zones near the surface 
where the change results from the change in composition and the resulting
change in $\Gamma_1$.
Also, $\delta_r \ln p$ and $\delta_r \ln \rho$ are nearly constant and
nearly identical in the bulk of the convection zone 
({\cf} Eq.~\ref{eq:appdelp}) and the change in temperature reflects the
change in the mean molecular weight.

A related issue concerns the neglect of pre-main-sequence evolution in 
Model~S, where evolution starts from an essentially homogeneous 
zero-age main-sequence model.
This was investigated by \citet{Morel2000} who found that, with a shift in
the evolution by 25 Myr, the resulting calibrated solar models differed
by only a few parts in $10^4$.
Thus the assumption of an initial ZAMS model is adequate.

The effects of changing the radius, from the reference value of
$6.9599 \times 10^{10} \cm$ to the value of $6.95508 \times 10^{10} \cm$
found by \citet{Brown1998},
is illustrated in Fig.~\ref{fig:changeradius}a.
Here there is obviously a change in the sound speed in the convection zone, 
and consequently
$\delta_r \ln p$ and $\delta_r \ln \rho$, while still approximately constant
in the convection zone, differ.
Considering the changes in the radiative interior, 
the use of differences at fixed $r/R$
is in fact somewhat misleading in this case.
Much of the change shown in Fig.~\ref{fig:changeradius}a is essentially
a geometrical effect, corresponding to the gradient term in the 
second of Eqs~(\ref{eq:diffrel}); 
the corresponding differences at fixed $m$ (see Fig.~\ref{fig:changeradius}b)
become very small in the deep interior.
As a result, the value of $X_0$ required to calibrate the model is
virtually unchanged.

As illustrated in Fig.~\ref{fig:changelum} the change in luminosity from the
reference value of $3.846 \times 10^{33} \erg \s^{-1}$ to the value
$3.828 \times 10^{33} \erg \s^{-1}$ inferred from \citet{Kopp2016} has 
modest effects on the structure.
According to Eq.~(\ref{eq:homlum}) the calibration to lower luminosity 
requires a decrease in $\mu$ and hence an increase in $X$, 
accompanied by a decrease in temperature, which is evident in the figure.
In the central regions the lower luminosity also corresponds to a smaller
nuclear burning of hydrogen and hence a larger abundance.
The difference in sound speed is minute.

\begin{figure}[htp]
\centerline{\includegraphics[width=\figwidth]{\fig/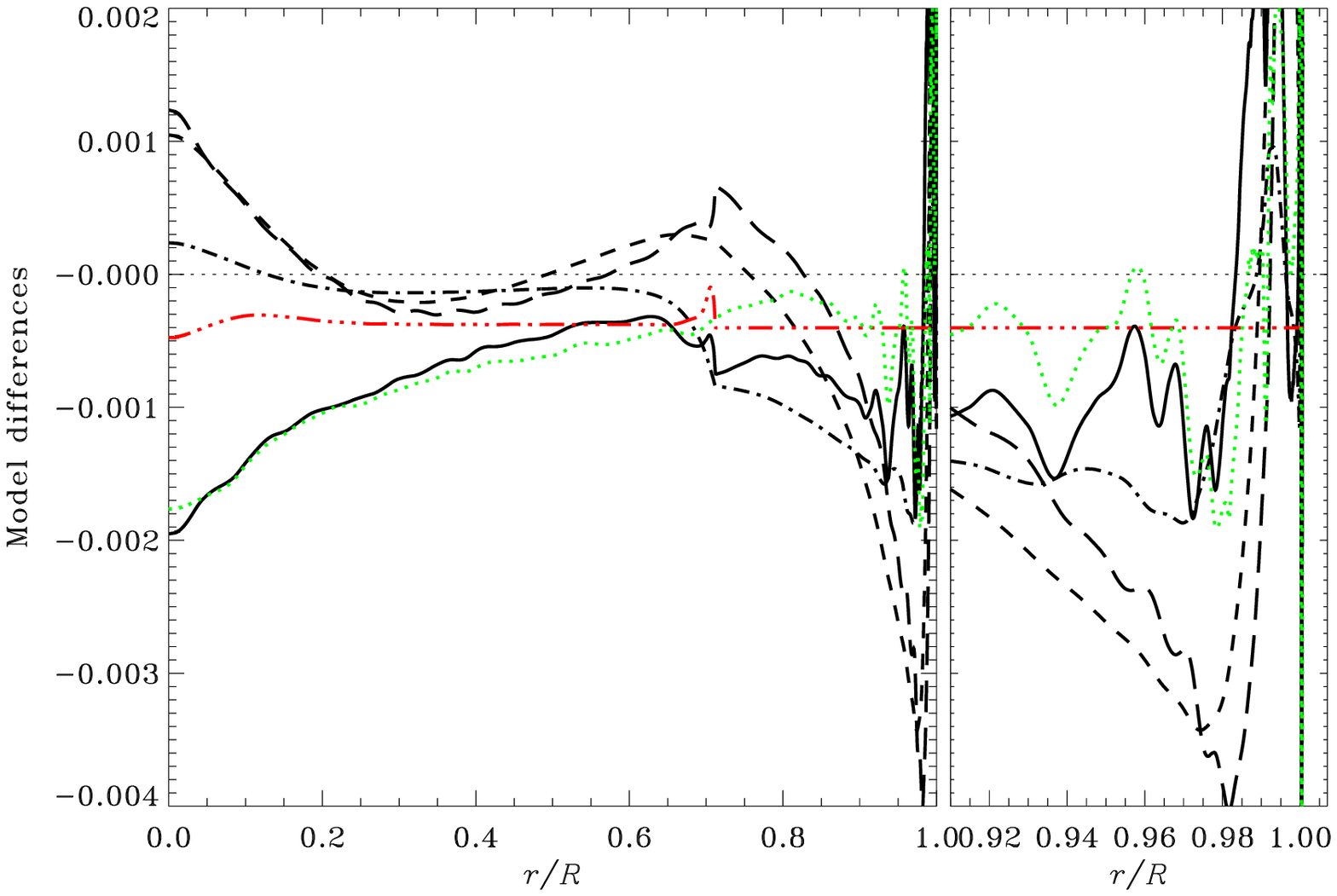}}
\caption{
Model changes at fixed fractional radius,
between Model~{\Meoszerofive} using the OPAL 2005 equation of state 
and Model~S, in the sense (Model~{\Meoszerofive}) --  (Model~S).
%\notecd [Updated OPAL EOS, with Z interpolation - Model~S.
%Note that in the core $\delta_r c^2/c^2 \simeq \delta_r \Gamma_1/\Gamma_1$.].
Line styles are as defined in Fig.~\ref{fig:changeage},
with the addition of the dotted green line showing $\delta_r \Gamma_1/\Gamma_1$.
%\notecd [change\_eos\_liv05.idl; dgr.l9bi.d.40c-d.02c]
}
\clabel{fig:changeeos}
\end{figure}

As discussed in Sect.~\ref{sec:eos} the OPAL equation of state has been
substantially updated since the computation of Model~S.
Figure~\ref{fig:changeeos} compares a model computed with the up-to-date 
OPAL\,2005 version with Model~S.
The effects in the bulk of the model are rather modest, with somewhat
larger changes in the near-surface layers.
%\notecd [which could be presented/discussed and might need a closer look].
A significant failing in the earlier tables was the neglect of relativistic
effects on the electrons in the central regions, which have a significant
effect on $\Gamma_1$ (see also Eq.~\ref{eq:elrelgam}).
This in fact dominates the sound-speed difference in the deeper parts of the
model in Fig.~\ref{fig:changeeos}.

\begin{figure}[htp]
\centerline{\includegraphics[width=\figwidth]{\fig/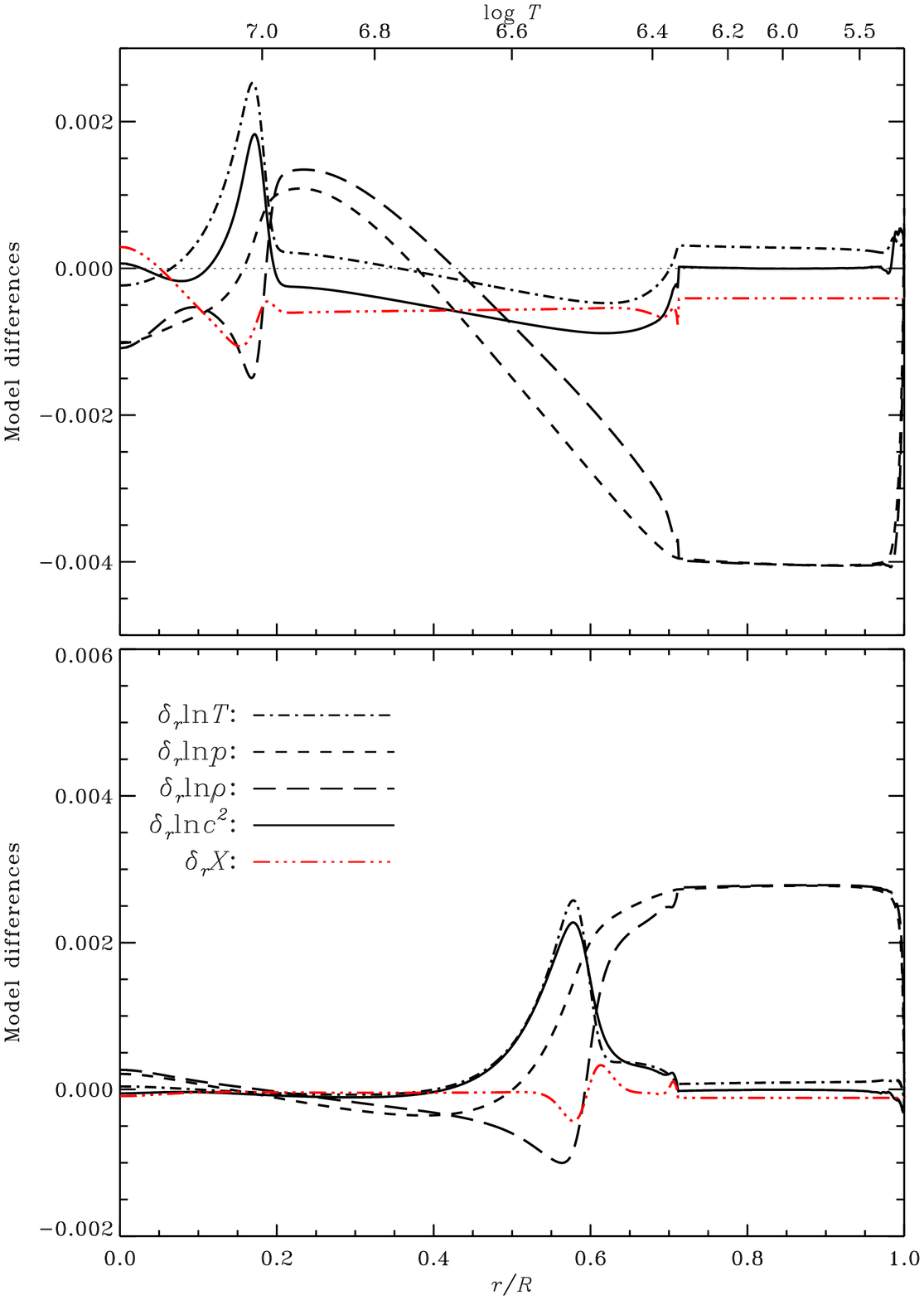}}
\caption{
Model changes at fixed fractional radius, resulting from localized
changes to the opacity described by Eq.~(\ref{eq:opmod})
with $A_\kappa = 0.02, \Delta_\kappa = 0.02$,
in the sense (modified model) -- (Model~S).
The top panel shows results for Model~{\Mopa}, with $\log T_\kappa = 7$,
and the bottom panel results for Model~{\Mopb}, with $\log T_\kappa = 6.5$.
Results are shown as a function of fractional radius (bottom
abscissa) and $\log T$ (top abscissa), 
and the line styles are defined in the figure. 
%and lines styles are as defined in Fig.~\ref{fig:changeage}.
%\notecd [change\_loc\_opacity.idl; dgr.l9bi.d.15c-d.02c, dgr.l9bi.d.16c-d.02c]
}
\clabel{fig:changelocopac}
\end{figure}

%\notecd [Also stress linearity; perhaps \citet{Tripat1998}].

%\notecd [Opacity modification:]

Perhaps the most uncertain aspect of the stellar internal microphysics is
the opacity (see also Sections \ref{sec:opac} and \ref{sec:modcorcomp}).
\citet{Tripat1998} made a detailed investigation of the effect on calibrated
solar models of localized modifications to the opacity.
They replaced $\log \kappa$, $\log$ being logarithm to base 10,
by $\log \kappa + \delta \log \kappa$, where
\be
\delta \log \kappa =
A_\kappa \exp[ -(\log T - \log T_\kappa)^2/\Delta_\kappa^2] \; ,
\eel{eq:opmod}
for a range of $\log T_\kappa$.
They also demonstrated a nearly linear response for even fairly large
modifications, by changing $A_\kappa$ from 0.1 to 0.2.
The response of solar models to opacity changes was also investigated
by \citet{Villan2010}.
As examples, Fig.~\ref{fig:changelocopac} shows the changes to the model
resulting from opacity changes of the form given in Eq.~(\ref{eq:opmod})
at $\log T_\kappa = 7$ and $6.5$. 
It is striking that the changes in temperature and hence sound speed are largely
localized in the vicinity of the opacity change, with a somewhat broader
response of pressure and density.
For the deeper opacity change a modest change in the hydrogen abundance
is required to calibrate the model to the correct luminosity:
the increase in opacity would tend to reduce the luminosity and
this is compensated by a decrease in $X$ and hence an increase in $\mu$,
in accordance with the homology scaling in Eq.~(\ref{eq:homlum}).

The behaviour of $\delta_r \ln T$ can be understood from the equation
for the temperature gradient (Eqs \ref{eq:nabla} and \ref{eq:nablarad})
which we write as
\be
{\dd \ln T \over \dd r} = - {3 \over 4 a \clight } {\kappa \rho \over T^4}
{L(r) \over 4 \pi r^2} \; ,
\eel{eq:tgrad}
or
\be
{\dd \over \dd r} \left( {\delta_r T \over T} \right) =
- {3 \over 4 a \clight } {\kappa \rho \over T^4} {L(r) \over 4 \pi r^2} 
\left( {\delta_r \kappa \over \kappa} + {\delta_r \rho \over \rho} 
+ 4 {\delta_r T \over T} \right) \; ,
\eel{eq:deltgrad}
where I neglected the perturbation to $L$.
We write $\delta_r \kappa / \kappa = (\delta \kappa/\kappa)_{\rm int}
+ \kappa_T \delta_r T/T$, where $ (\delta \kappa/\kappa)_{\rm int}$ is
the intrinsic opacity change given by Eq. (\ref{eq:opmod}),
$\kappa_T = (\partial \ln \kappa / \partial \ln T)_{\rho, X_i}$ and
I neglected the dependence of $\kappa$ on $\rho$ and composition.
Then Eq.~(\ref{eq:deltgrad}) can be written as
\be
{\dd \over \dd \ln T} \left( {\delta_r T \over T} \right) +
(4 - \kappa_T) {\delta_r T \over T} =
\left( {\delta \kappa \over \kappa} \right)_{\rm int} \; ,
\eel{eq:delteq}
neglecting again $\delta_r \rho / \rho$.
In the outer parts of the Sun the temperature is largely fixed, for small
changes in $X$, by Eq.~(\ref{eq:appdelt}).
Assuming that $\delta_r T/T \approx 0$ well outside the location 
$T = T_\kappa$ of the change in the opacity, and taking $\kappa_T$ as constant,
Eq.~(\ref{eq:delteq}) has the solution
\be
{\delta_r T \over T} \approx T^{-(4 - \kappa_T)} \int_{\ln T_{\rm s}}^{\ln T}
T'^{4 - \kappa_T}
\left( {\delta \kappa \over \kappa} \right)_{\rm int} \dd \ln T' \; ,
\eel{eq:deltsol}
where $T_{\rm s}$ is the surface temperature.
This explains the steep rise 
in the outer parts of the peak in $\delta_r T/T$ (and hence $\delta_r c^2/c^2$)
and, with $\kappa_T$ typically around $-2$ to $-3$,
the relatively rapid decay on the inner side.

To analyse the properties of $\delta_r p$ and $\delta_r \rho$
I assume the ideal gas law, Eq.~(\ref{eq:idealg}), and neglect
the change in the mean molecular weight, such that
$\delta_r \ln \rho \approx \delta_r \ln p - \delta_r \ln T$.
From the equation (\ref{eq:hydrostat}) of hydrostatic equilibrium,
neglecting the change in $m$, it then follows that
\be
{\dd \delta_r \ln p \over \dd \ln p} \approx - \delta_r \ln T \; .
\eel{eq:delpeq}
Below the location of the opacity and temperature change pressure and
density are relatively unaffected.
Thus the local change in pressure is dominated by the increase with
increasing $r$ in the peak of $\delta_r \ln T$, while $\delta_r \ln \rho$
has a negative dip in this region, but follows the increase in 
$\delta_r \ln p$ outside it.
The global behaviour of $\delta_r \ln p$ and $\delta_r \ln \rho$
is constrained by the conservation of total mass, such that
\be
\int_0^R \delta_r \rho r^2 \dd r = 0 \; .
\eel{eq:massconv}
For $\log T_\kappa = 7$ (top panel in Fig.~\ref{fig:changelocopac})
the region of positive $\delta_r \ln \rho$ just outside the peak in 
$\delta_r \ln T$ therefore forces a region of negative $\delta_r \ln \rho$
in the outer parts of the model, including the convection zone where
$\delta_r \ln p$ and $\delta_r \ln \rho$, according to Eq.~(\ref{eq:appdelp}),
are approximately constant.
For $\log T_\kappa = 6.5$ (bottom panel) the region of negative 
$\delta_r \ln \rho$ in the peak of $\delta_r \ln T$ results in 
positive $\delta_r \ln \rho$ and $\delta_r \ln p$ in the convection zone.
The effect on the hydrogen abundance is less clear in simple terms, although
it must be related to the calibration to keep the luminosity fixed.
Given that the changes in the deep interior are minute for
$\log T_\kappa = 6.5$, it is understandable that $\delta_r X$ is very small
in this case, except in the region just below the convection zone that
is directly affected by changes in diffusion and settling.

\begin{figure}[htp]
\centerline{\includegraphics[width=\figwidth]{\fig/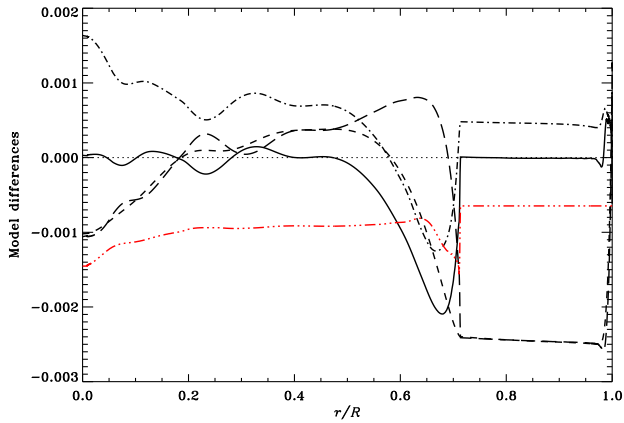}}
\caption{
Model changes at fixed fractional radius, between Model~{\Mopalninesix}
which uses the OPAL96 opacities and Model~S, where the OPAL92 tables were used,
in the sense ( Model~{\Mopalninesix}) -- (Model~S).
Line styles are as defined in Fig.~\ref{fig:changelocopac}.
%\notecd [change\_opacity.idl; dgr.l9bi.d.07c-d.02c]
}
\clabel{fig:changeopac}
\end{figure}

The OPAL opacity tables were updated by \citet{Iglesi1996}, relative to
the \citet{Rogers1992} tables used for Model~S.
As shown in Fig.~\ref{fig:changeopac}, comparing models that
both assumed the \citet{Greves1993} solar composition
but using respectively the OPAL96 and OPAL92 tables,
the revision of the opacity calculation has some effect on the structure,
including a relatively substantial change in the sound speed.
%\notecd [In fact, a surprising large change in $X$, to calibrate;
%might deserve a closer look, and perhaps further comments].

\begin{figure}[htp]
\centerline{\includegraphics[width=\figwidth]{\fig/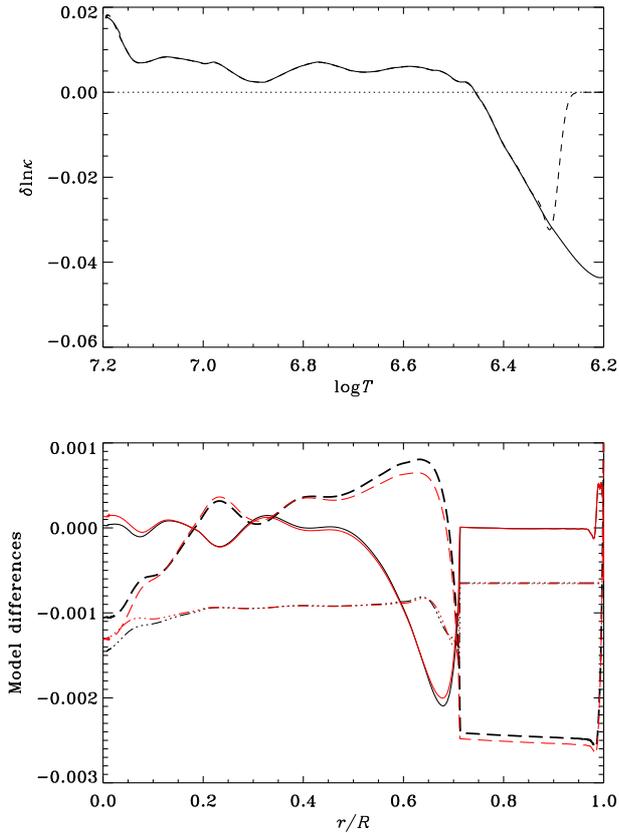}}
\caption{
Top panel: The solid curve shows logarithmic differences 
between the OPAL96 and the OPAL92 opacity, in the sense (OPAL96) - (OPAL92),
at fixed $\rho$, $T$ and composition in Model~\Mopalninesix.
The dashed curve shows a fit of functions of the form in Eq.~(\ref{eq:opmod}),
with $\Delta_\kappa = 0.02$ and on a grid in $\log T_\kappa$ between
$7.2$ and $6.2$ with a step of 0.01.
Bottom panel: differences $\delta_r \ln c^2$ (solid curves),
$\delta_r \ln \rho$ (long-dashed curve) and $\delta_r X$ 
(triple-dot-dashed curve).
The black curves show results from Fig.~\ref{fig:changeopac}, 
whereas the red curves show reconstructions based on `opacity kernels'
such as shown in Fig.~\ref{fig:changelocopac}, using the fit shown in
the top panel.
%\notecd [plot-delop.idl]
}
\clabel{fig:changefitopac}
\end{figure}

As noted by, for example, \citet{Tripat1998} and \citet{Vinyol2017}
responses to localized opacity changes such as shown in 
Fig.~\ref{fig:changelocopac} define `opacity kernels' that can be
used to reconstruct the effects of more general opacity changes.
An example is illustrated in Fig.~\ref{fig:changefitopac}.
Here the top panel shows a fit to the difference between the 
OPAL96 and OPAL92 tables in the radiative region, based on localized
opacity changes of the form in Eq.~(\ref{eq:opmod}) on a dense grid
in $\log T_\kappa$.
Applying the resulting amplitudes to the corresponding model differences
yields the red curves in the bottom panel, which are in excellent
agreement with the direct differences between the OPAL96 and OPAL92 models,
as illustrated in Fig.~\ref{fig:changeopac}.
The changes in $c^2$ and $\rho$ are dominated by the substantial
negative opacity difference at relatively low temperature, yielding 
a negative $\delta_r \ln c^2$ just below, and a negative $\delta_r \ln \rho$
within, the convection zone.
As noted above the change in $X$, on the other hand,
is insensitive to the opacity change in the outer parts 
of the radiative region, and hence the 
positive $\delta \ln \kappa$ in the deeper regions results in a 
negative $\delta_r X$.

\begin{figure}[htp]
\centerline{\includegraphics[width=\figwidth]{\fig/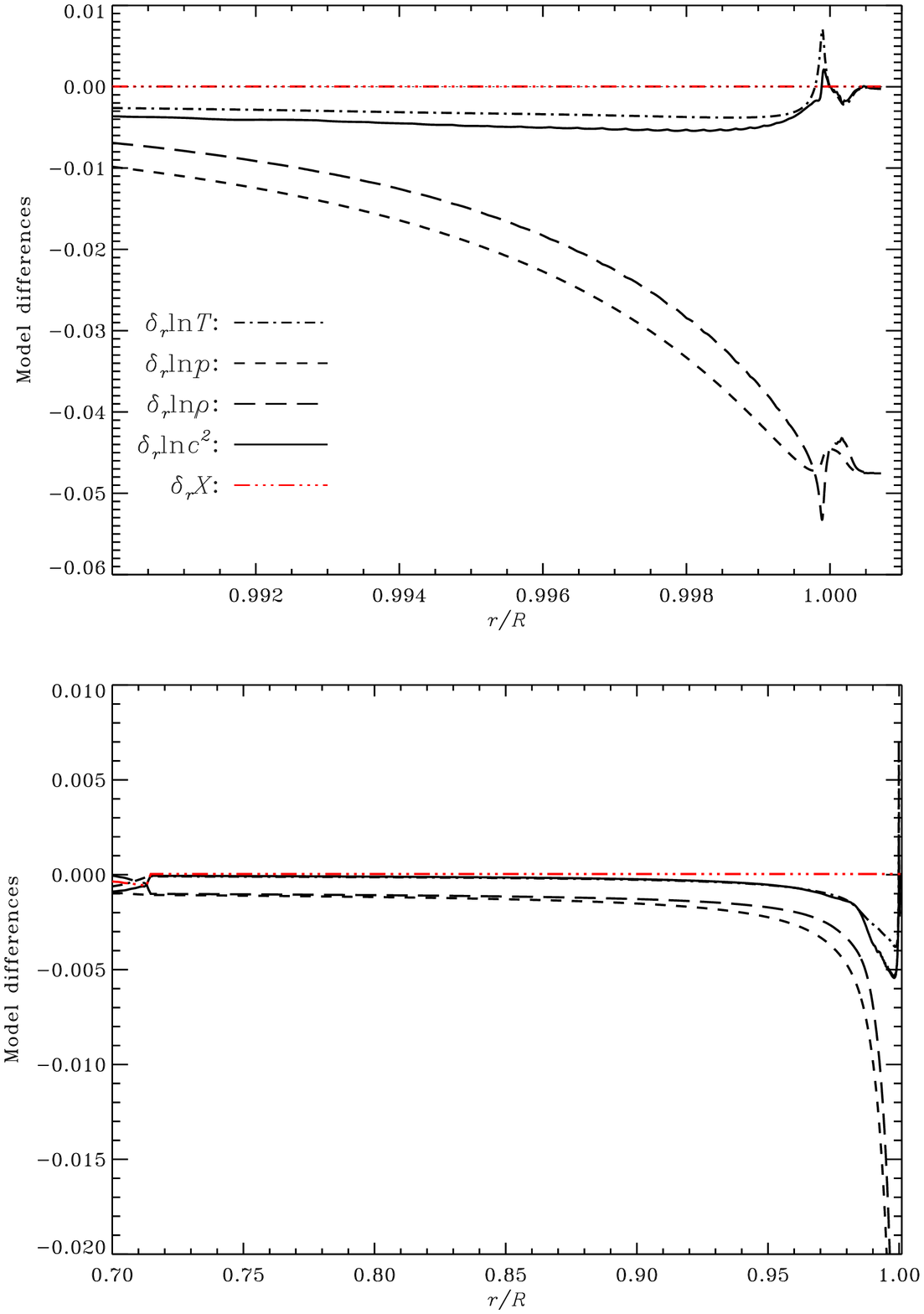}}
\caption{
Model changes at fixed fractional radius, between a model computed using
the Fergusson {\etal} (2005) low-temperature opacities
and Model~{\Mopalninesix}, which used the \citet{Kurucz1991} tables,
in the sense (Model~{\Msurfop}) - (Model~{\Mopalninesix});
in both cases the OPAL96 tables were used in the deeper parts of the model.
Line styles are defined in the top panel.
%as defined in Fig.~\ref{fig:changelocopac}.
%\notecd [change\_surf\_opacity.idl; dgr.l9bi.d.35c-d.07c]
}
\clabel{fig:changesurfopac}
\end{figure}

The effects of changing the atmospheric opacity are illustrated in
Fig.~\ref{fig:changesurfopac}, comparing the more recent tables
of \citet{Fergus2005} with the \citet{Kurucz1991} tables used in
the computation of Model~S.
There are significant changes in pressure and density in the atmosphere,
reflecting the integration of atmospheric structure at the given temperature
structure ({\cf} Sect.~\ref{sec:surface}, in particular Eqs \ref{eq:tau}
and \ref{eq:atmhyd}).
However, as discussed by \citet{Christ1997b} the effects of such
superficial changes in calibrated solar models
are very strongly confined to the near-surface layers; 
the differences in the bulk of the convection zone and in the radiative
interior are minute.%
\footnote{In the present comparison there are small contributions to the
differences in the radiative interior arising from the use of 
slightly different interpolation procedures.}

\begin{figure}[htp]
\centerline{\includegraphics[width=\figwidth]{\fig/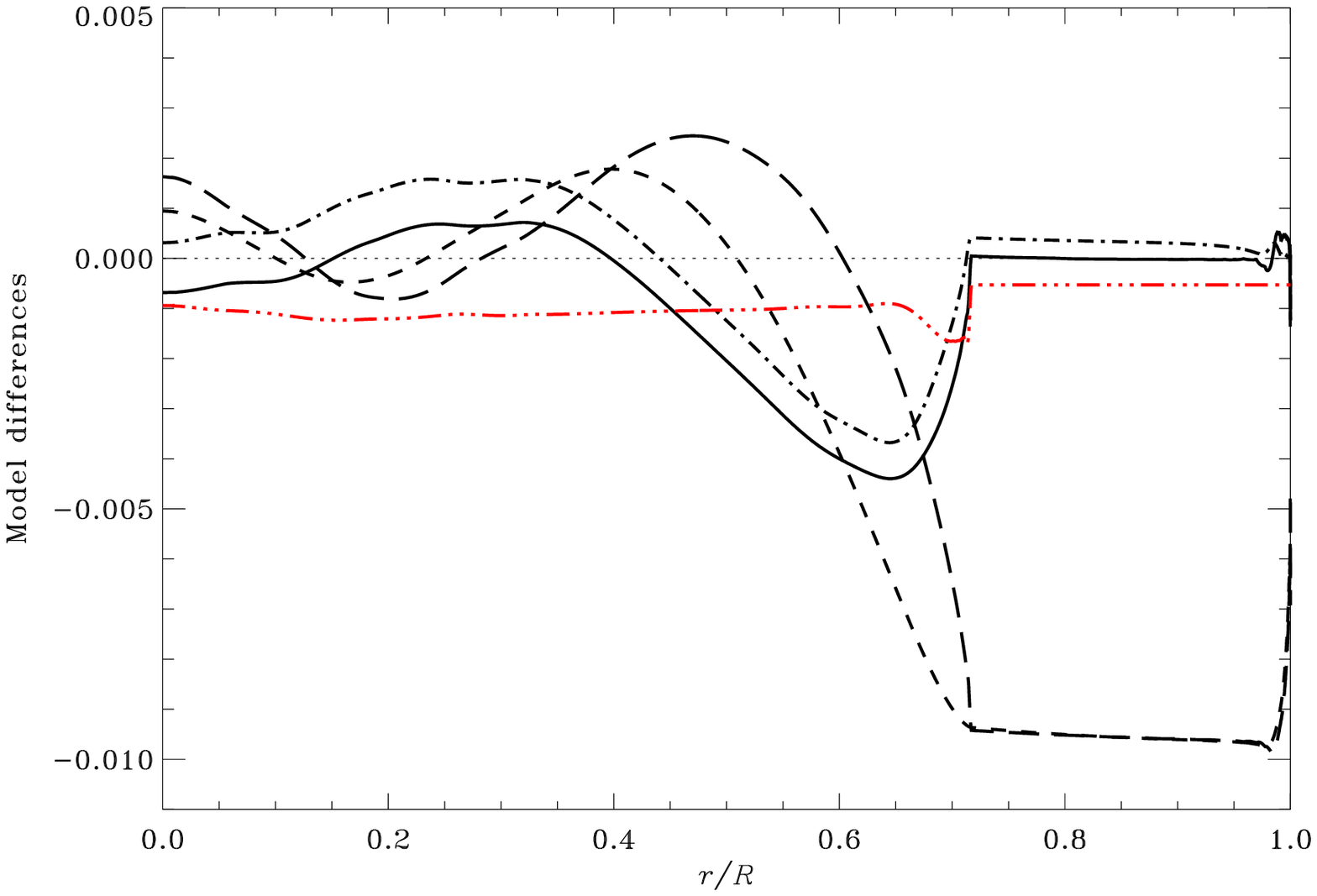}}
\caption{
Model changes at fixed fractional radius, between Model~{\Mgsnineeight}
using the \citet{Greves1998} composition and Model~{\Msurfop}
which used the \citet{Greves1993} composition (see Table~\ref{tab:abundances}),
in the sense (Model~{\Mgsnineeight}) -- (Model~{\Msurfop});
in both cases the \citet{Fergus2005} atmospheric and the OPAL96 interior
tables were used.
Line styles are as defined in Fig.~\ref{fig:changesurfopac}.
%\notecd [change\_opacity\_gs98.idl; dgr.l9bi.d.37c-d.35c]
}
\clabel{fig:changeopacgs98}
\end{figure}

Relative to the \citet{Greves1993} composition used in Model~S a modest
revision was proposed by \citet{Greves1998}; 
the compositions are compared in Table~\ref{tab:abundances} 
in Sect.~\ref{sec:newcomp} below.
This composition has seen extensive use in solar modelling.
The effects of this change on the model structure are illustrated in
Fig.~\ref{fig:changeopacgs98}, using for both compositions the
OPAL96 opacity tables.
There is evidently some change, at a level that is significant compared with
the helioseismic results in the
sound speed, as well as a modest change in the hydrogen abundance required
for luminosity calibration.
In particular, the 10 per~cent change in the oxygen abundance
({\cf} Table~\ref{tab:abundances}) and the general decrease in the
heavy-element abundance ({\cf} Table~\ref{tab:modeldiff}) cause 
a decrease in the opacity of up to 4 per cent just below the convection zone,
leading the a significant decrease in the sound speed in the outer parts
of the radiative region, as shown in Fig.~\ref{fig:changeopacgs98}.
As discussed in Sect.~\ref{sec:newcomp} the much greater revision 
since 2000 of the determination of the solar surface composition
has had very substantial effects on solar models.

\begin{figure}[htp]
\centerline{\includegraphics[width=\figwidth]{\fig/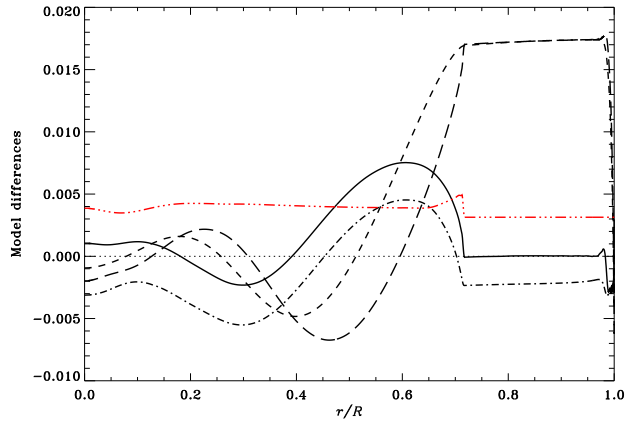}}
\caption{
Model changes at fixed fractional radius, between Model~{\Mopzerofive}
using the OP05 opacity tables \citep[e.g.,][]{Seaton2005}
and Model~{\Mgsnineeight} using the OPAL96 tables, in the sense
(Model~{\Mopzerofive}) -- (Model~{\Mgsnineeight});
in both cases the GS98 composition and the \citet{Fergus2005}
low-temperature opacities were used.
%\notecd [change\_opacity\_op05.idl; dgr.l9bi.d.36c-d.37c]
Line styles are as defined in Fig.~\ref{fig:changesurfopac}.
%\notecd [Could be interesting to look at inversion for this model].
}
\clabel{fig:changeopacop05}
\end{figure}

An indication of the effects of the uncertainties in the opacity computations
may be obtained by comparing the use of the OPAL tables with the results
of the independent OP project \citep[\eg,][]{Seaton1994, Badnel2005, 
Seaton2005};
the differences between the tables are illustrated in Fig.~\ref{fig:opcomp}
(note that this shows OPAL -- OP).
In Fig.~\ref{fig:changeopacop05} models computed with the OP and OPAL tables
are compared, in both cases using the \citet{Greves1998} composition.
The effect is clearly substantial, with an increase in the sound speed in
the bulk of the radiative interior and in the hydrogen abundance resulting
from the luminosity calibration.
The model differences can at least qualitatively be understood from
the opacity kernels discussed above. 
The differences in sound speed, pressure and density are probably dominated
by the positive table differences at temperatures just below the
convection zone, while the change in the hydrogen abundance is dominated
by the negative table differences in the deeper parts of the model.
Other comparisons of different opacity calculations were carried out, 
for example, by \citet{Neufor2001b}, who compared OPAL and the Los Alamos
LEDCOP tables, and \citet{LePenn2015b}, comparing OPAL and the recent 
OPAS tables \citep{Blanca2012, Mondet2015} developed at CEA, France.
%\notecd [Probably reference to Villante et al. 2018?].

\begin{figure}[htp]
\centerline{\includegraphics[width=\figwidth]{\fig/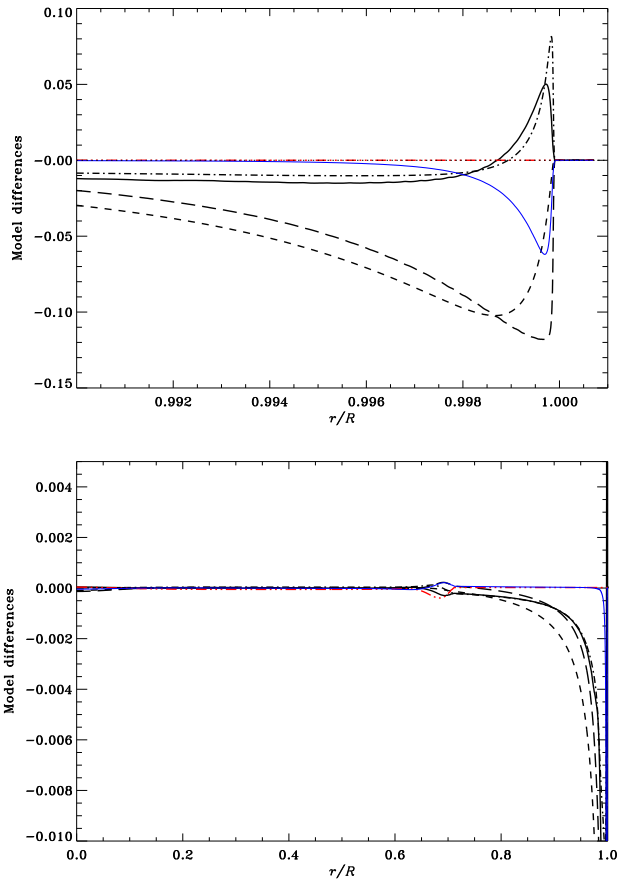}}
\caption{
Model changes at fixed fractional radius,
between Model~{\MCM} emulating the \citet{Canuto1991} treatment of
near-surface convection and Model~S,
in the sense (Model~\MCM) -- (Model~S).
Line styles are as defined in Fig.~\ref{fig:changesurfopac}, with the
addition of the solid blue line which shows the difference 
$\delta_q \ln c^2$ of squared sound speed at fixed mass fraction $q$.
%\notecd [change\_conv\_canuto.idl; dgr.l9bi.d.24c-d.02c]
}
\clabel{fig:canuto}
\end{figure}

As discussed in Sect.~\ref{sec:convection} there is considerable uncertainty
in the treatment of convection in the strongly super-adiabatic region
just below the photosphere (see Fig.~\ref{fig:gwcon}).
In calibrated solar models, however, this has little effect on the structure
of the bulk of the model.
To illustrate this Fig.~\ref{fig:canuto} shows differences between a model
computed using the \citet{Canuto1991} treatment, as implemented by
\citet{Montei1996}, and Model~S.
There are substantial differences in the near-surface region, but these
are very strongly confined, with the differences being extremely small in the
lower parts of the convection zone and the radiative interior
\citep[see also][]{Christ1997b}.
%\notecd [Might consider comments on the effects on frequencies, as required 
%later].
This effect is similar to the effect of modifying the atmospheric opacity,
shown in Fig.~\ref{fig:changesurfopac}.
As illustrated by the solid blue line, the difference in squared sound speed
at fixed mass fraction is much more strongly confined near the surface than
the difference at fixed fractional radius.
It was argued by \citet{Christ1997b} that, consequently, $\delta_q \ln c^2$
provides a better representation of the effects of the near-surface 
modification on the oscillation frequencies.
In fact, model differences such as these or those shown in 
Fig.~\ref{fig:changesurfopac} provide a model for the near-surface
errors in traditional structure and oscillation modelling which
have an important effect on helio- and asteroseismic investigation.
To illustrate this, Fig.~\ref{fig:canutofreq} below shows
frequency differences between the models illustrated in Fig.~\ref{fig:canuto}.

\begin{figure}[htp]
\centerline{\includegraphics[width=\figwidth]{\fig/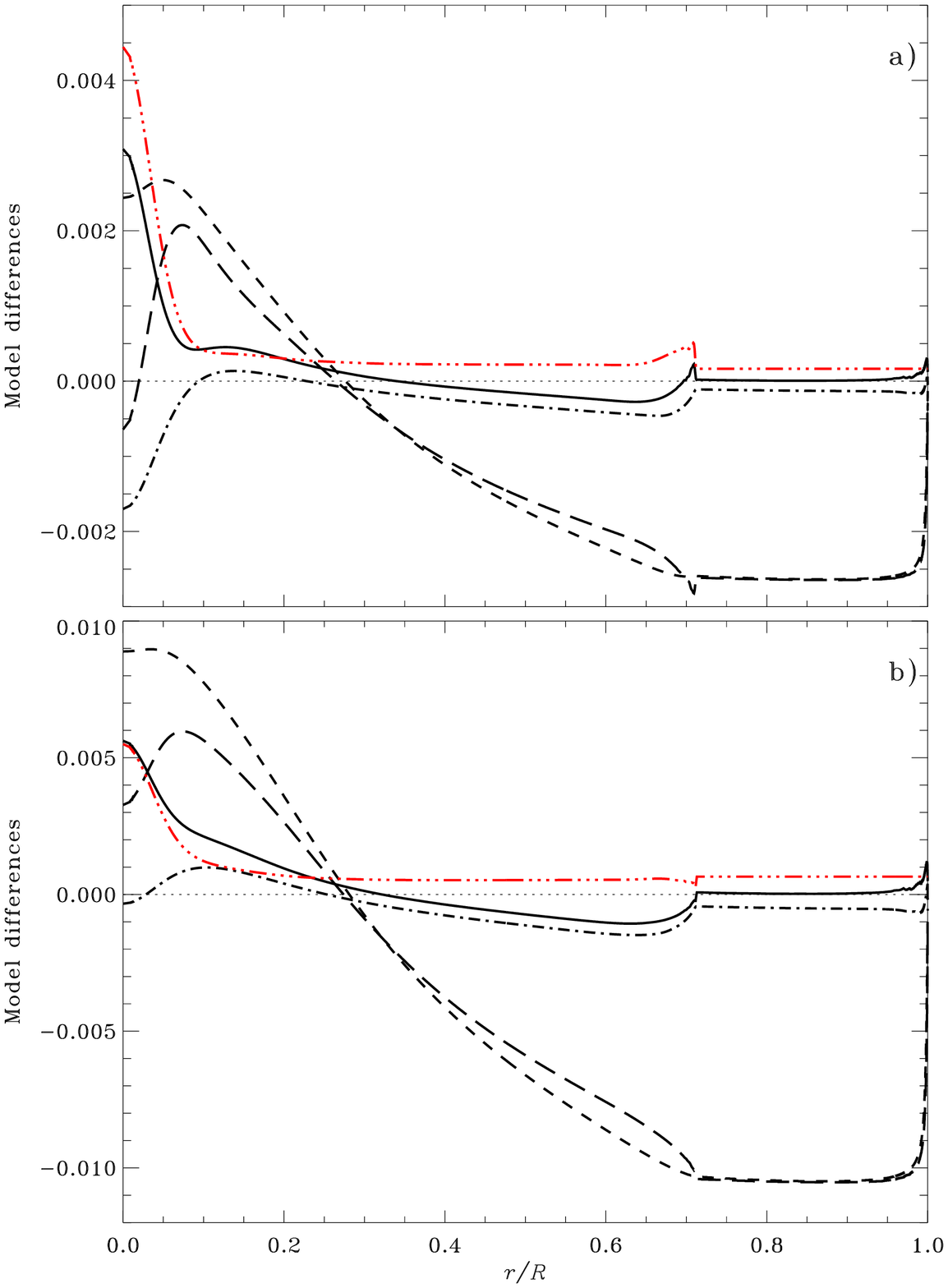}}
\caption{
Model changes at fixed fractional radius, corresponding to changes
in the nuclear reaction parameters, compared with Model~S which used
parameters largely from \citet{Bahcal1995}.
Panel a) shows differences for Model~{\Madelb}, using the \citet{Adelbe2011}
parameters, in the sense (Model~{\Madelb}) -- (Model~S),
and panel b) shows differences for Model~{\Mnacre} using the
\citet{Angulo1999} (NACRE) parameters, with the reaction $\nitfourteen + \hyd$
updated by \citet{Formic2004}, in the sense (Model~{\Mnacre} -- (Model~S).
Line styles are as defined in Fig.~\ref{fig:changesurfopac}.
%\notecd [change\_nucl\_comb.idl; dgr.l9bi.d.34c-d.02c, dgr.l9bi.d.39c-d.02c]
}
\clabel{fig:changenucl}
\end{figure}

The effects of the updates to the nuclear reaction parameters since Model~S 
are illustrated in Fig.~\ref{fig:changenucl}. 
Panel a) is based on a model computed with the \citet{Adelbe2011} parameters, 
while in panel b) the NACRE rates \citep{Angulo1999} reaction rates,
with the \citet{Formic2004} update of the $\nitfourteen$ rate, were used.
In both cases the dominant change to the overall reaction rate
was at the highest temperatures and is closely related to 
updated quantities for the CNO reactions;
at fixed conditions the energy generation decreased by 5 -- 8 per cent
relative to the formulation used in Model~S.
This is directly reflected in the higher hydrogen abundance 
(see also Table~\ref{tab:modeldiff}) and hence
higher sound speed in the core, in both cases.
Calibration to fixed luminosity caused modest changes in the structure in
the other parts of the models.
It should be noticed that while the differences in $\epsilon$ 
at fixed $\rho$, $T$ and composition for
the \citet{Adelbe2011} rates are largely confined to the region
where $\log T \ge 7.1$, the differences in the NACRE rates extend
more broadly, leading to the substantially larger model differences
in the NACRE case, Fig.~\ref{fig:changenucl}b.

%%\begin{figure}[htp]
%%\def\epsfsize#1#2{0.5#1}
%\centerline{\includegraphics[width=\figwidth]{\fig/change_nucl_NACRE28.eps}}
%\caption{
%Model changes at fixed fractional radius,
%\notecd [Use \citet{Angulo1999} reaction rates (NACRE),
%updated by \citet{Formic2004} - Model~S].
%Lines styles are as defined in Fig.~\ref{fig:changeage}.
%\notecd [change\_nucl\_NACRE28.idl; dgr.l9bi.d.39c-d.02c]
%}
%\clabel{fig:changeNACRE28}
%\end{figure}
%}

A potential simplification of the calculation is to assume that $\helthree$
is in nuclear equilibrium.
The region where this is satisfied approximately corresponds to the
rising part of the $\helthree$ abundance shown in Fig.~\ref{fig:he3}
and hence in fact covers most of the region of nuclear energy generation
in the present Sun.
However, the change in the hydrogen abundance over solar evolution
does depend on the details of the nuclear reactions.
As illustrated in Fig.~\ref{fig:change3He} assuming nuclear equilibrium
of $\helthree$ throughout the evolution indeed generally has a minute effect on
the resulting model of the present Sun.
The peak in $\delta_r X$ at $r/R \approx 0.27$ corresponds closely to the
peak in the $\helthree$ abundance ({\cf} Fig.~\ref{fig:he3}
and probably reflects the local conversion of hydrogen into $\helthree$.
%\notecd [Could try to understand the origin of the composition differences].

\begin{figure}[htp]
\centerline{\includegraphics[width=\figwidth]{\fig/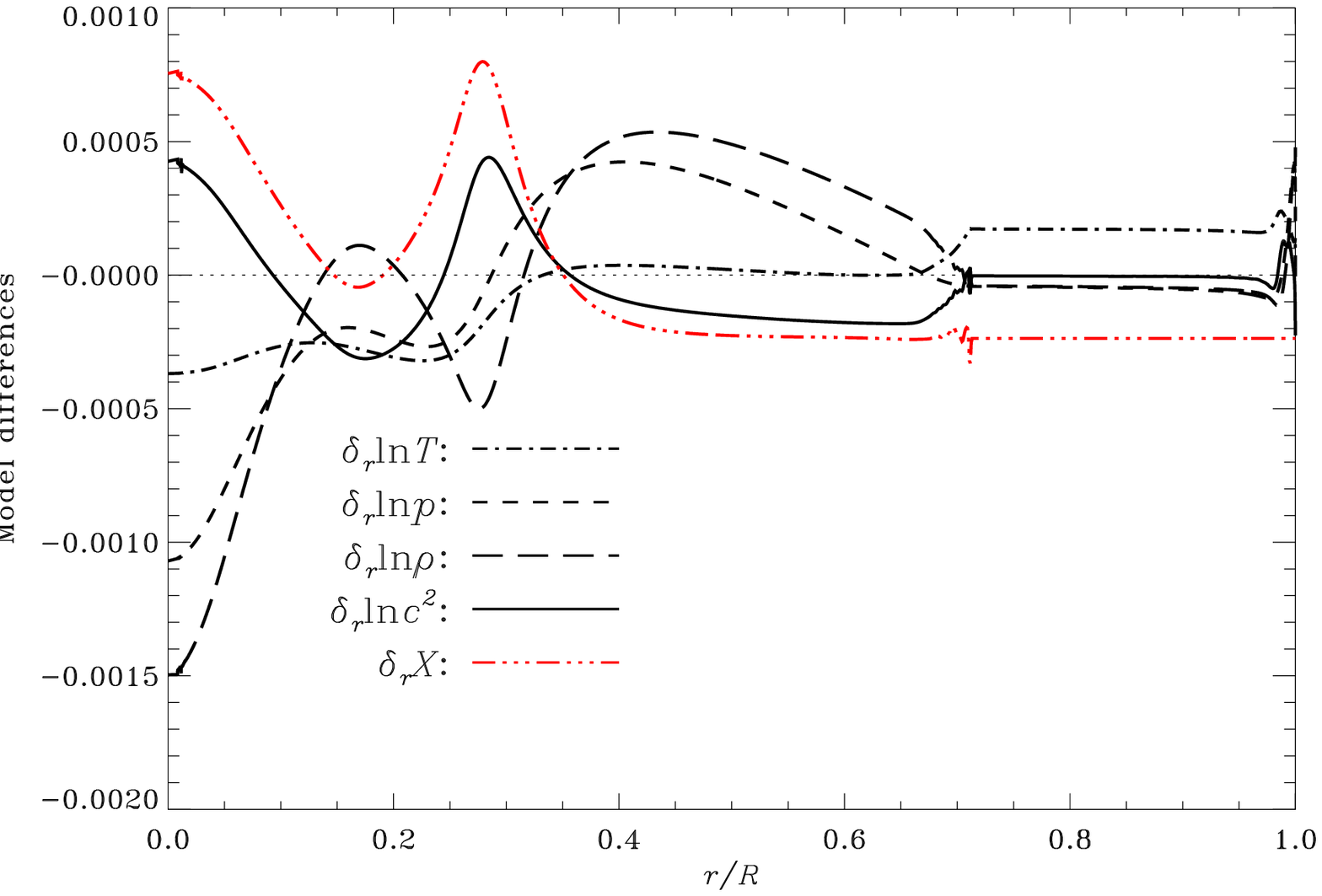}}
\caption{
Model changes at fixed fractional radius, between Model~{\Mhethreeeq}
where $\helthree$ is assumed to be in nuclear equilibrium and Model~S,
in the sense (Model~{\Mhethreeeq}) -- (Model~S).
Line styles are defined in the figure.
%as defined in Fig.~\ref{fig:changeage}.
%\notecd [change\_nucl\_3He.idl; dgr.l9bi.d.02c\_eq-d.02c]
}
\clabel{fig:change3He}
\end{figure}

\begin{figure}[htp]
\centerline{\includegraphics[width=\figwidth]{\fig/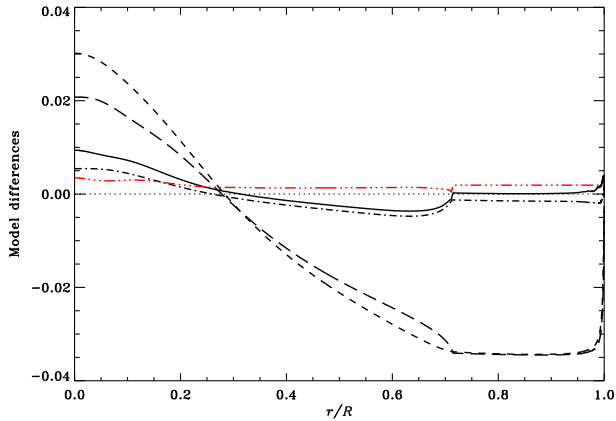}}
\caption{
Model changes at fixed fractional radius, between Model~{\Mnoelscrn}
where electron screening is switched off and Model~S, in the sense
(Model~{\Mnoelscrn}) -- (Model~S).
Line styles are as defined in Fig.~\ref{fig:change3He}.
%\notecd [change\_screen.idl; dgr.l9bi.d.20c-d.02c]
}
\clabel{fig:changescreen}
\end{figure}

As mentioned in Sect.~\ref{sec:engenr} there has been some discussion
about the validity of the classical \citet{Salpet1954} model of static 
screening of nuclear reactions, with dynamical simulations indicating
absence of screening \citep{Mussac2011}.
The effects of switching off all screening of nuclear reactions are 
illustrated in Fig.~\ref{fig:changescreen}.
At fixed conditions corresponding to Model~S this results in a 
reduction in the nuclear energy-generation rate of up to 9 per~cent
near the centre, where the CNO cycle plays some role
({\cf} Fig.~\ref{fig:cno}), and around
5 per~cent further out, where the PP chains dominate.
To achieve luminosity calibration this is compensated by increases
in temperature, hydrogen abundance and density, the latter increase
requiring a decrease in density in the outer parts of the model
to conserve the total mass ({\cf} Eq.~\ref{eq:massconv}).
%  Note that homology scaling (e.g., stel-struc, uges-09.5.pdf) shows
%  that an decrease in epsilon corresponds to an increase in the
%  luminosity, which then requires an decrease in mu, i.e., an
%  increase in X, for re-calibration.
%  This argument could be presented and made a little more quantitative?
The effects show some similarity to the effects of the revision of nuclear
parameters (Fig.~\ref{fig:changenucl}), probably reflecting also here
the larger reduction in the rates of the more temperature-sensitive reactions,
but the changes are clearly of a much larger magnitude.
Indeed, \citet{Weiss2001} pointed out that the resulting model is 
inconsistent with the constraints provided by the helioseismically
determined sound speed
\citep[{\cf} Sect.~\ref{sec:heliostruc}; see also][]{Christ2010}.

%\notecd [Modification of diffusion coefficient:
%multiply $D_i$ ({\cf} Eq.~\ref{eq:diffusion}) by
%a factor $\beta_D$.]
%
%\notecd [Modification of settling velocity
%multiply $V_i$ etc. ({\cf} Eq.~\ref{eq:diffusion}) by
%a factor $\beta_V$.]

\begin{figure}[htp]
\centerline{\includegraphics[width=\figwidth]{\fig/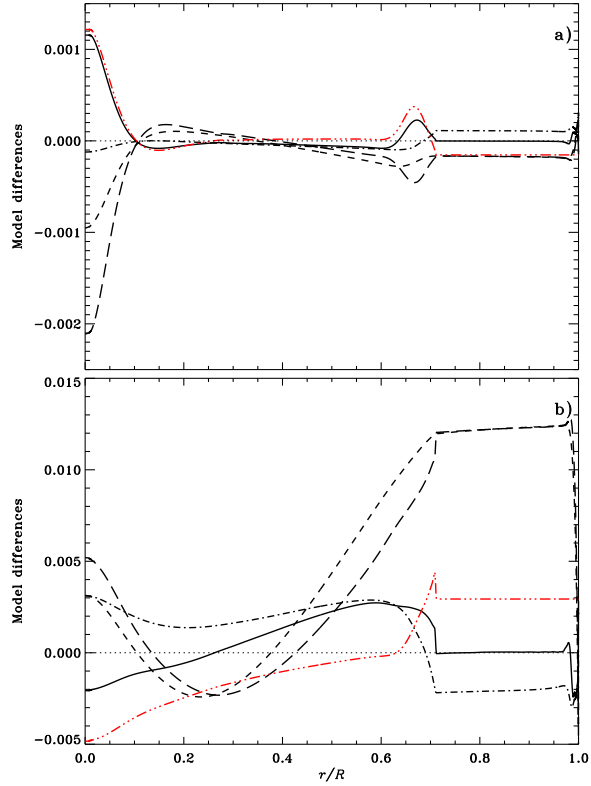}}
\caption{
Model changes at fixed fractional radius, resulting from changes
to the diffusion and settling coefficients, compared with Model~S.
Panel a) shows differences for Model~{\Mdifa} where the diffusion coefficient
$D_i$ ({\cf} Eq.~\ref{eq:diffusion}) was increased by a factor 1.2,
in the sense (Model~{\Mdifa}) -- (Model~S).
Panel b) shows differences for Model~{\Mdifb} where both $D_i$ and 
the settling velocity $V_i$ were increased by a factor 1.2,
in the sense (Model~{\Mdifb}) -- (Model~S).
Line styles are as defined in Fig.~\ref{fig:change3He}.
%\notecd [change\_diffus.idl; dgr.l9bi.d.17c-d.02c; dgr.l9bi.d.18c-d.02c]
}
\clabel{fig:changediffus}
\end{figure}

\begin{figure}[htp]
\centerline{\includegraphics[width=\figwidth]{\fig/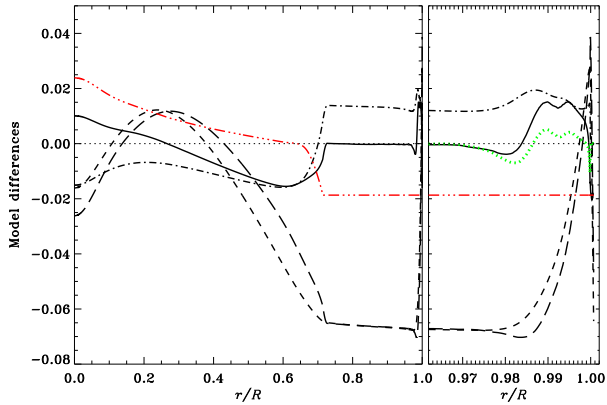}}
\caption{
Model changes at fixed fractional radius, comparing Model~{\Mnodif}
neglecting diffusion and settling with Model~S,
in the sense (Model~{\Mnodif}) - (Model~S).
Line styles are as defined in Fig.~\ref{fig:change3He};
in addition, in the right-hand expanded view of the outer helium and hydrogen
ionization zones the green dotted curve shows $\delta_r \ln \Gamma_1$.
%\notecd [change\_nodiffus\_nsur.idl; dgr.l9bi.03c-d.30c]
}
\clabel{fig:nodiffus}, 
\end{figure}

To illustrate the sensitivity of the models to the detailed treatment of
diffusion and settling Fig.~\ref{fig:changediffus} shows the effect of
increasing $D_i$ ({\cf} Eq.~\ref{eq:diffusion}) by a factor 1.2 (panel a)
or increasing both $D_i$ and $V_i$ by this factor (panel b).
In the former case the effects are small, the dominant changes being
confined to the core where the increased diffusion partly smoothes the
hydrogen profile, leading to an increase in the hydrogen abundance, with
a corresponding increase in the sound speed.
There are additional even smaller changes associated with the gradient
in hydrogen abundance caused by settling just below the convection zone.
When also the settling velocity is increased the changes are more substantial,
including a significant increase in the hydrogen abundance in the convection
zone and a noticeable increase in the sound speed below the
convection zone; note also the near-surface sound-speed changes,
of similar shape but opposite sign to the effects of neglecting 
diffusion and settling (see Fig.~\ref{fig:nodiffus} below) and, as in that case,
reflecting the thermodynamic response to the change in the helium abundance.

Finally, it should be recalled that early `standard solar models'
did not include effects of diffusion and settling.
It was shown by \citet{Christ1993} that including just diffusion and
settling of helium led to a substantial improvement 
in the comparison between the model
and helioseismic inferences of sound speed, and hence more recent solar models,
such as Model~S, include full treatment of diffusion.
To illustrate this effect Fig.~\ref{fig:nodiffus} compares a model 
ignoring diffusion but otherwise corresponding to Model~S, including the
calibration, with Model~S.
It is evident that the change in the hydrogen abundance (which obviously
to a large extent reflects the Model~S hydrogen profile illustrated in
Fig.~\ref{fig:Xmod}) has a substantial effect on the sound speed, hence
affecting the comparison with the helioseismic inference.
There are more subtle effects on the sound speed near the surface that in
part arises from the change in $\Gamma_1$ caused by the change of the helium
abundance in the helium ionization zones, and which affects the frequencies
of acoustic modes.
This effect illustrates the potential for helioseismic determination of the 
solar envelope helium abundance (see Sect.~\ref{sec:specasp}).
\section{Tests of solar models}
\clabel{sec:test}

%\notecd [Remind that `classical' observables do not provide a test.]
%
The models discussed so far have explicitly been computed to match the
`classical' observed quantities of the Sun: the initial composition 
$(X_0, Z_0)$ has been chosen to match the solar luminosity and present
surface composition and the choice of mixing length has been made to 
match an assumed solar radius, at the assumed present age of the Sun.
Since the model has thus been adjusted to match the observed 
$\Ls$, $R$ and $\Zs/\Xs$ these quantities provide no independent test
of the calculation, beyond the feeble constraint that apparently reasonable
values of the required parameters can be found which match the observables.

As discussed in the introduction,
very detailed independent testing of the model computation has become
possible through helioseismology, by means of extensive observations of
solar oscillations.
Additional information relevant to the structure of the
solar core results from the detection of neutrinos originating from
the nuclear reactions ({\cf} Eq.~\ref{eq:hydfus}).
Finally, I briefly consider the surface abundances of light elements or
isotopes which provide constraints on mixing processes in the solar
interior.

%\notecd [Briefly on the two types of tests available: helioseismology and
%neutrinos.]

\subsection{Helioseismic tests of solar structure}
\clabel{sec:helio}

%\notecd [Should check extent to which MJT paper will be available in parallel.
%Otherwise (and possibly even if) need a brief introduction to helioseismic
%techniques. In that case this probably expands to a full section,
%as does the neutrino subsection consequently.]

%\notecd [UPDATE: a little on excitation.  Note Jacoutot {\etal} 2008.]

%\notecd [General references, including C-D 2002 and Book!] 

Detailed reviews of the techniques and results of helioseismology
have been provided by \citet{Christ2002}, \citet{Basu2008} 
and \citet{Aerts2010};
An extensive review of solar oscillations and helioseismology was
provided by \citet{Basu2016} in \textit{Living Reviews of Solar Physics}.
A perhaps broader view, emphasizing also the limitations in the present results,
was provided by \citet{Gough2013a}.
None the less, it is appropriate here to provide a brief overview of
the techniques of helioseismology and to summarize the results on the
solar interior.

\subsubsection{Properties of solar oscillations}
\clabel{sec:basichelio}

Oscillations of the Sun are characterized by the degree $l$
and azimuthal order $m$,%
\footnote{This dual use of $m$, also denoting the mass inside a given point,
should cause little confusion.}
with $|m| \le l$,
of the spherical harmonic $Y_l^m(\theta, \phi)$ describing the mode,
where $\theta$ is co-latitude and $\phi$ is longitude,
and by its radial order $n$.
The degree provides a measure of the horizontal wave number $\kh$:
\be
\kh = {\sqrt{l(l+1)} \over r} \;,
\eel{eq:horwav}
at distance $r$ from the solar centre.
Thus, except for radial modes (with $l = 0$),
the average horizontal wavelength on the solar surface is
$\lambda_{\rm h,s} \simeq 2 \pi R/l$.
The azimuthal order measures the number of nodal lines crossing the equator.
The observed cyclic oscillation frequencies $\nu$, between roughly 
1 and 5 mHz, correspond to modes that predominantly have the character
of standing acoustic waves, or p modes, and, at high degree, surface
gravity waves, or f modes.
In the case of the p modes, the frequencies are predominantly determined by
the internal sound speed $c$,
with
\be
c^2 = {\Gamma_1 p \over \rho} \simeq {\Gamma_1 \kB T \over \mu \amu} \; ,
\eel{eq:soundspeed}
%\notecd [we might move this to the EOS section, to stress importance of
%$\Gamma_1$]
the latter expression assuming the ideal gas law 
({\cf} Eq.~\ref{eq:idealg}).

The f-mode frequencies are to a good approximation given 
by the deep-layer approximation for surface gravity waves,
determined by the surface gravitational acceleration.
Thus to leading order these modes provide little information about
the structure of the solar interior,
although a correction term, essentially reflecting the variation in
the appropriate gravitational acceleration with mode properties,
provides some sensitivity to the near-surface density profile
\citep{Gough1993, Chitre1998}.
The dependence on surface gravity has been used to determine, on the
basis of f-mode frequencies, the `seismic solar radius' \citep{Schou1997}
and its variation with solar cycle \citep[e.g.,][]{Kosovi2018a}.

Rotation (or other departures from spherical symmetry) induces a dependence
of the frequencies on the azimuthal order $m$.
To leading order the effect of rotation
simply corresponds to the advection of
the oscillation patterns by the angular velocity as averaged over
the region of the Sun sampled by a given mode.

From the dispersion relation for acoustic waves, and Eq.~\Eq{eq:horwav},
it is straightforward to show that the modes are
oscillatory as a function of $r$ in the region of the Sun which lies
outside an inner turning point, at distance $r = \rt$ from the
centre satisfying
\be
{c(\rt) \over \rt} = {\omega \over \sqrt{l(l+1)}} \; ,
\eel{eq:turn}
and evanescent interior to this point; here $\omega = 2 \pi \nu$
is the angular frequency of the mode.
Since the sound speed generally increases with decreasing $r$,
the turning point is close to the solar centre for very low degrees at
the observed frequencies, 
the modes becoming increasingly confined near the surface with increasing
degree.
From a physical point of view this behaviour of the modes {\rv corresponds to}
total internal {\rv reflection},
owing to the increase in the sound speed with depth,
of sound waves corresponding to the given degree:
the waves travel horizontally at the inner turning point.
With increasing degree the initial direction of the waves 
at the solar surface is more strongly inclined from the vertical and
the turning point is reached closer to the surface.

The frequency of a given acoustic mode reflects predominantly the structure
outside the turning point.
The observed modes have degree from 0 to more than 1000,
and hence turning points varying from very near the solar centre to 
immediately below the photosphere.
This variation in sensitivity allows the determination of the structure
with high resolution in the radial direction.
Very crudely, the high-degree modes give information about the near-surface
region of the Sun.
Given this, modes of slightly lower degree can be used to determine the
structure at slightly greater depth, and so on,
the analysis continuing to the solar core.
Similarly, modes of differing azimuthal order have different extent in 
latitude, those with $|m| \simeq l$ being confined near the equator and
modes with low $|m|$ extending over all latitudes;
thus observation of frequencies as a function of $m$ over a range
of degrees allows the determination of, for example, the angular
velocity as a function of both latitude and distance from the centre.

For completeness I note that there have also been claims of observed
solar oscillations with much longer periods.
Such modes would be internal gravity waves, or \emph{g modes},
with greater sensitivity to conditions in the solar core than the
acoustic modes.

%\notecd [Results on solar structure, emphasize sound-speed structure (and
%discuss briefly density, but also that things are dependent, given
%hydrostatic support; this deserves more work of course, but hardly
%at the present point).]

With the exception of the region just below the surface, and the atmosphere,
solar oscillations can be treated as adiabatic to a very high precision.
This approximation is generally used in computations of solar oscillation 
frequencies.
However, nonadiabatic effects in the oscillations are undoubtedly important 
in the near-surface region, as are the processes that excite the modes.
The physical treatment of these effects, involving the interaction between
convection and the oscillations, is uncertain, 
and so therefore are their effects on the oscillation frequencies
\citep[for a review, see][]{Houdek2015}.
Also, the structure of the near-surface region of the model is affected
by the uncertain effects of convection, including the general neglect of
turbulent pressure ({\cf} Sect.~\ref{sec:surface}).

\begin{figure}[htp]
\centerline{\includegraphics[width=\figwidth]{\fig/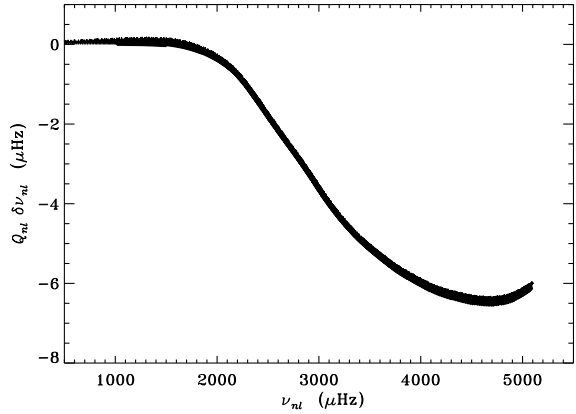}}
\caption{
Frequency differences for modes of degree $l \le 100$,
scaled by the inertia ratio $Q_{nl}$,
between a model emulating the \citet{Canuto1991} treatment of
near-surface convection and Model~S, in the sense (modified model) -- Model~S.
The corresponding model differences are shown in Fig.~\ref{fig:canuto}.
}
\clabel{fig:canutofreq}
\end{figure}

Such inadequacies in modelling the structure and the oscillations very near the
solar surface appear to dominate the differences between the observed 
frequencies and frequencies of solar models 
\citep[\eg,][]{Christ1984a, Dziemb1988, Christ1996}.
Fortunately, the effect of these near-surface uncertainties on the frequencies
in many cases has {\rv a relatively simple dependence on the mode frequency
and degree.
This follows from the fact that
the} physics of the modes, except at very high degree,
in the near-surface layers is insensitive to the degree and so, therefore,
is {\rv the direct effect of these layers} on the oscillation frequencies.
This, however, must be corrected for the fact that according to
Eq.~(\ref{eq:turn}) higher-degree modes involve a smaller fraction of the
star and hence are easier to perturb.
A quantitative measure of this effect is provided by {\it the mode inertia}
\be
E = {\int_V \rho |\bolddelr|^2 \dd V \over M |\bolddelr|_{\rm phot}^2 } \; ,
\eel{eq:inertia}
where the integral is over the volume of the star, $\bolddelr$ is the
displacement vector, and $|\bolddelr|_{\rm phot}$ is its norm at the
photosphere;
it may be shown that the frequency shift from a near-surface modification
is proportional to $E^{-1}$ \citep[e.g.,][]{Aerts2010}.
It is convenient to take out the frequency dependence of the inertia
by considering, instead of $E$, $Q = E/\bar E_0(\omega)$,
where $\bar E_0(\omega)$ is the inertia of a radial mode, interpolated
to the frequency $\omega$ of the mode considered,
effectively renormalizing the surface effect to the effect on radial modes.
The resulting functional form of the effect on the frequencies 
of the near-surface uncertainties is reflected by the last term in
Eq.~(\ref{eq:freqdif}) below \cite[\eg][]{Christ1988a, Aerts2010}.
Given the very extensive data available on solar oscillations
this property of the frequency differences caused by the near-surface effects
to a large extent allows their consequences to be
suppressed in the analysis of the observed oscillation frequencies, 
leading to reliable inferences of the internal structure
\citep[{\eg},][]{Dziemb1990, Dappen1991, Gough1996a}.
For distant stars, however, where only low-degree modes are observed,
the surface errors represent a significant source of uncertainty in the
analysis of the oscillation frequencies.
Various procedures have been developed to suppress these effects
in fits to the observed frequencies \citep[\eg][]{Kjelds2008, Ball2014},
or, alternatively, the fits can be based on frequency combinations defined
to be largely insensitive to them \citep{Roxbur2003, Oti2005}.

How errors in the near-surface region affect the oscillation frequencies
can be illustrated by the model differences shown in
Fig.~\ref{fig:canuto}, between a model using the \citet{Canuto1991}
treatment of convection and Model~S which used the \citet{Boehm1958}
mixing-length treatment.
Frequency differences between these two models are shown in
Fig.~\ref{fig:canutofreq}.
To compensate for the fact that with increasing degree the modes
involve a smaller part of the Sun ({\cf} Eq.~\ref{eq:turn})
the differences have been scaled by the normalized $Q_{nl}$, as discussed above.
%the so-called inertia of the mode,
%normalized to the interpolated inertia of radial modes at the same frequency
%\citep[see][]{Aerts2010}.
The figure clearly shows that with this scaling the frequency
differences are indeed largely independent of the degree.

\begin{figure}[htp]
\centerline{\includegraphics[width=\figwidth]{\fig/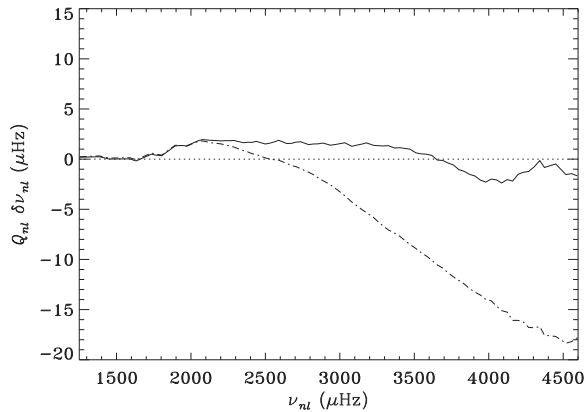}}
\caption{
Differences,
reduced to the case of radial modes (with $l = 0$),
between observed and modelled solar oscillation frequencies
against frequency, in the sense (Sun) -- (Model).
The dot-dashed curve uses adiabatic frequencies for a model
essentially corresponding to Model~S
\citep[][see Sect.~\ref{sec:models}]{Christ1996}.
The solid curve is based on a model where the outermost layers were 
replaced by a suitable average of a three-dimensional radiative-hydrodynamic
simulation of convection.
In addition, the frequencies were obtained from nonadiabatic calculations taking
the interaction with convection, including turbulent pressure, into account.
(Adopted from \citecl{Houdek2017}.)
}
\clabel{fig:hgsurf}
\end{figure}

Clearly an important goal is to understand the structure and
oscillation dynamics in the near-surface layers better and 
eventually model them consistently in the calculation of the oscillation
frequencies;
in this context the otherwise strongly constrained solar case will serve
as an important test.
A key aspect is the treatment of convection in the equilibrium model and
the oscillations (see also Sect.~\ref{sec:convection}).
\citet{Schlat1997} used a detailed atmospheric model and modelled the
outer layers of the convection zone by a variable mixing-length parameter
matched to a two-dimensional hydrodynamical simulation of convection;
they noted that the resulting model matched the observed solar oscillation
frequencies better than did the normal model.
A similar improvement of the frequencies was obtained by 
\citet{Rosent1995, Rosent1999} and \citet{Robins2003} 
by including suitable averages of
convection simulations in the modelling (see also Sect.~\ref{sec:convection}).
\citet{Sonoi2015} and \citet{Ball2016} studied the effect
on stellar oscillation frequencies
of using averaged simulations as the outer parts of stellar models, for
a range of stellar parameters.
\citet{Magic2016} also considered the patching of averaged simulations to
solar models and in addition devised corrections to the depth scale and density
in normal one-dimensional models that mimicked the effects on the 
frequencies of the patching.
In addition to normal simulations they carried out simulations with magnetic
fields, representing more active areas of the solar surface, determining
the effect of the resulting change in the structure of the solar layers
on the oscillation frequencies, although without considering the direct
effect of the field on the oscillations.
The analysis was extended to a broad range of stellar parameters,
ranging from the main sequence to the red-giant branch, by \citet{Trampe2017},
who emphasized
the importance of both the expansion of the near-photospheric layers by
the effect of turbulent pressure and the so-called `convective back-warming',
{\ie}, the effects of the convective fluctuations on the strongly
temperature-sensitive opacity.
In similar analyses, \citet{Sonoi2017} included also some
effects of the perturbation to the turbulent pressure,
based on a time-dependent convection formulation restricted to adiabatic
oscillations,
while \citet{Mancho2018} emphasized the sensitivity of the near-surface
frequency shifts to the metallicity of the stars.

An equally important contribution to the deficiencies in the model frequencies
is the physics of the oscillations in the near-surface region.
Here the energetics of the oscillations, including the perturbations to
the convective flux, must be taken into account in fully nonadiabatic 
calculations, and the perturbation to the turbulent pressure has a 
significant effect on the frequencies and the damping of the modes. 
To treat these effects requires a time-dependent modelling of convection
\citep[see][for a review]{Houdek2015}.
Time-dependent versions of the mixing-length theory were established by
\citet{Unno1967} and \citet{Gough1977c} and have been further developed since
then.
With a few exceptions the nonadiabatic calculations show that the modes
are intrinsically damped;
they are excited to the observed amplitudes by stochastic forcing from
convection, as confirmed by analysis of the observed amplitude distribution
\citep{Chapli1997}.
Consequently the observed linewidths in the frequency power spectra provide
a measure of the damping rates of the modes,
allowing calibration of parameters in the convection modelling such that
the computed damping rates match the observed linewidths.
Combining results from hydrodynamical simulations of
the outer layers with nonadiabatic computations using a non-local time-dependent
convection treatment including also the turbulent-pressure perturbation,
\citet{Houdek2017}, as illustrated in Fig.~\ref{fig:hgsurf},
obtained a much improved fit to the solar observed frequencies,
at the same time showing a reasonable fit to the observed damping rates.
Analyses of intrinsic or induced oscillations in hydrodynamical simulations
are providing further insight into the physics of the interaction between
convection and the oscillations \citep{Belkac2019, Zhou2019},
which may be used further to improve the simplified treatments based on
mixing-length formulations.
In an interesting analysis, \citet{Schou2020} determined the frequency
correction caused by the effect on the oscillations
of convection dynamics by matching eigenfunctions in standard 
oscillation calculations to eigenfunctions resulting from the convection
simulations. 

\subsubsection{Investigations of the structure and physics of 
the solar interior}
\clabel{sec:heliostruc}

%\notecd [A little here about data, before going into inversion results].
%
Very extensive helioseismic data have been acquired over the past decades,
from groundbased networks of observatories and from Space
\citep[for further details, see for example][]{Christ2002, Aerts2010}.
In most cases observations of radial velocity are carried out,
based on the Doppler effect, extending over months or years to achieve
sufficient frequency resolution, reduce the background noise and follow
possible temporal variations in the Sun.
Spatially resolved observations are analysed to isolate modes 
corresponding to a few combinations of $(l, m)$.%
\footnote{From the orthogonality of the spherical harmonics it follows that
a single pair $(l, m)$ could in principle be isolated if the entire surface
of the Sun were observed.}
From the resulting time series power spectra are constructed through
Fourier transform, and the frequencies of solar oscillations are determined
from the position of the peaks in the power spectra.
Low-degree modes have been studied in great detail through 
observations in disk-integrated light, observing the Sun as a star,
from the BiSON \citep{Chapli1996, Hale2016} and IRIS \citep{Fossat1991} 
networks, and with the GOLF instrument on the SOHO spacecraft
\citep{GabrieA1997}.
Modes of degree up to around 100 were studied for an extended
period of time with the LOWL instrument \citep{Tomczy1995}, 
extended to the two-station ECHO network, which has now stopped operation.
Also, the six-station GONG network \citep{Harvey1996}
has yielded nearly continuous data for modes of degree up to around 150
since late 1995, 
whereas modes including even higher degrees were studied with
the SOI/MDI instrument on SOHO \citep{Scherr1995, Rhodes1997}.
Since May 2010 these high-resolution observations have been taken over
by the HMI instrument on the Solar Dynamics Observatory \citep{Hoekse2018}, 
with regular MDI observations ending in April 2011.
Detailed analyses of the BiSON low-degree observations were carried out
by \citet{Broomh2009} and \citet{Davies2014}, while
\citet{Larson2015, Larson2018} analysed the MDI and HMI observations
for modes of degree up to $l \approx 300$.
At even higher degree the modes lose their individual nature owing to the
decreasing separation between adjacent modes and the increasing damping rates;
thus the analysis of these modes is affected by
systematic errors and interference between the modes
\citep{Korzen2004, Rabell2008}.
Here special techniques are required for the frequency determination as
discussed, {\eg}, by \citet{Reiter2015} and \citet{Reiter2020},
who analysed a 66-day high-resolution set of MDI observations.
It should be noticed that, according to Eq.~(\ref{eq:turn}),
these high-degree modes have their lower turning point quite 
close to the surface;
this makes them particularly interesting for the study 
of the near-surface layers \citep[e.g.,][]{DiMaur2002},
where thermodynamic effects associated with helium and hydrogen ionization
become relevant, and where, as discussed above, the properties of the structure
and the oscillations are somewhat uncertain.
Very extensive high-resolution data are being obtained with HMI, but these
have apparently so far not been analysed to determine properties
of high-degree modes.

Owing to their great potential for helioseismic investigations 
the g modes have been the target of major observational efforts.
\citet{Garcia2007} inferred the presence of g modes with 
the expected nearly uniform period spacing from periodicities in the
power spectrum of GOLF observations.
However, a review by \citet{Appour2010} found that the attempts up 
to that point to detect g modes were inconclusive.
Recently, \citet{Fossat2017} claimed evidence for g modes of degree $l = 1$ and
$2$ through an ingenious and complex analysis of the spacing between solar
acoustic low-degree modes observed with GOLF.
In a follow-up study \citet{Fossat2018} extended this to modes of degree
$3$ and $4$.
Interestingly, the results indicated a rapid rotation of the solar core,
possibly at variance with the results obtained from the analysis of solar
acoustic modes (see Fig.~\ref{fig:introt} below).
However, \citet{Schunk2018}, repeating the analysis, found that the results
were very sensitive to the details of the fits, including the assumed
starting time of the time series of observations.
{\rv A similar sensitivity to the details of the analysis was found by
\citet{Appour2019}, analysing a recalibrated version of the GOLF data
\citep{Appour2018}; on this basis they concluded that the results of
\citet{Fossat2017} and \citet{Fossat2018} were artefacts of the methodology.}
Also, the physical effects that might introduce the g-mode signal in the
acoustic-mode properties are so far unclear.
Indeed, although already \citet{Kenned1993} proposed this type of analysis they
noted that the coupling between the modes is such that to leading order
the p-mode frequencies
are insensitive to g modes of odd degree \citep[see also][]{Gough1993},
in conflict with the inferences of Fossat {\etal}
This was analysed in more detail by \citet{Boenin2019} and \citet{Scherr2019}.
Furthermore, Scherrer and Gough confirmed and extended the results of 
\citet{Schunk2018} and tried, and failed, to find a similar signal 
in the MDI and HMI data;
they also noted that the inferred rapid rotation of the solar core
is difficult to reconcile with the constraints obtained from extensive
analyses of well-observed solar acoustic modes (see Sect.~\ref{sec:heliorot}).
Thus the evidence for solar g modes remains uncertain, and I shall not
consider them further in this review.
%\notecd [\citet{Scherr2019, Boenin2019, Kenned1993}].

From Eq.~(\ref{eq:turn}) it follows that acoustic modes of low degree
penetrate to the stellar core.
This is particularly important for investigations of distant stars,
where only low-degree modes are observed (see Sect.~\ref{sec:stars}),
but low-degree acoustic modes have also been important for the study 
of the solar core, not least in connection with the solar neutrino problem
\citep[e.g.,][see also Sect.~\ref{sec:neutr}]{Elswor1990}.
The cyclic frequencies $\nu_{nl} = \omega_{nl}/2\pi$ of these modes
satisfy the asymptotic relation \citep{Tassou1980, Gough1993}
\be
\nu_{nl} \approx \Delta \nu \left(n + {l \over 2} + \varepsilon \right)
- d_{nl} \; ,
\eel{eq:pasymp}
where \emph{the large frequency separation}
\be
\Delta \nu = \left(2 \int_0^R {\dd r \over c} \right)^{-1}
\eel{eq:larsep}
is the inverse of the acoustic travel time across a stellar diameter
and $\varepsilon$ is a frequency-dependent phase 
related to the near-surface layers.
Thus to leading order the frequencies are uniformly spaced in radial order,
with degeneracy between modes with the same $n + l/2$.
This degeneracy is lifted by the small correction term $d_{nl}$,
leading to the \emph{small frequency separations}
\be
\delta \nu_{nl} = \nu_{nl} - \nu_{n-1\,l+2} \simeq
- (4 l + 6) {\Delta \nu \over 4 \pi^2 \nu_{nl}}
\int_0^R {\dd c \over \dd r} {\dd r \over r} \; .
\eel{eq:smlsep}
Since the integral is strongly weighted towards the stellar centre,
$\delta \nu_{nl}$ is a useful diagnostic for the properties of the stellar core,
including stellar age \citep[e.g.,][see also Sect.~\ref{sec:modrevcomp}]
{Christ1984b, Christ1988b, Ulrich1986}.

The extensive sets of observed solar oscillation frequencies 
make possible detailed inferences of the properties of solar structure,
through \emph{inverse analyses} of the observations.
Reviews of such inversion techniques were given by, for example,
\citet{Gough1991}, \citet{Gough1996a}, \citet{Basu2008}
and \citet{Basu2016}.
Assuming adiabatic oscillations, the frequencies are determined by
the dependence of pressure, density and gravity on $r$, as
well as on $\Gamma_1$ which relates the perturbations to pressure and density.
However, given that the solar model satisfies the equations of
hydrostatic support and mass, Eqs~\Eq{eq:hydrostat} and
\Eq{eq:mass}, the mass $m$ and $p$ can be computed once $\rho(r)$ is specified.
%\notecd [need to worry, as usual, about $m$ for azimuthal order and mass].
It follows that the adiabatic oscillation frequencies are fully defined
if $(\rho(r), \Gamma_1(r))$ is specified.
Alternatively equivalent pairs can be used; given that the frequencies
of acoustic modes are predominantly determined by the sound speed,
convenient choices are $(c^2, \rho)$ or $(u, \Gamma_1)$, 
$u = p/\rho$ being the squared isothermal sound speed.

It was demonstrated by \citet{Gough1984a} that a simple asymptotic relation
for the frequencies, first found by \citet{Duvall1982},
forms the basis for an approximate inversion for the solar sound speed;
this was used for the first inferences of the solar internal sound speed
by \citet{Christ1985}.
Such asymptotic techniques have been further developed by, for example,
\citet{Christ1989}, \citet{Voront1991a} and \citet{Marche2000}.

Alternatively, as originally noted by \citet{Gough1978a}
based on similar techniques in geophysics, 
a linearized relation that does not depend on the asymptotic properties
can be established between corrections to
the structure of a solar model, for example characterized by
differences at fixed $r$ $(\delta_r c^2, \delta_r \rho)$ between the Sun 
and the model, and the corresponding frequency differences.
This is based on the fact that the oscillation frequencies satisfy a
variational principle 
\citep[{\eg},][]{Chandr1964},
such that the frequency corrections are independent of corrections to
the eigenfunctions, to leading order.
However, the analysis must also take into account the inadequacies 
of the modelling of the near-surface layers discussed above.
As a result, the relative frequency differences can be written as
\be
{\delta \omega_{nl} \over \omega_{nl}} = 
\int_0^{\Rs} \left[ K_{c^2, \rho}^{nl}(r) {\delta_r c^2 \over c^2}(r)
+ K_{\rho,c^2}^{nl}(r) {\delta_r \rho \over \rho}(r) \right] \dd  r 
 + Q_{nl}^{-1} \CF(\omega_{nl}) \; ,
\eel{eq:freqdif}
where the kernels $K_{c^2, \rho}^{nl}$ and $K_{\rho, c^2}^{nl}$
are obtained from the reference solar model,
and $\Rs$ is the surface radius of the model.
The last term takes into account differences between the model and the Sun 
resulting from the inadequate modelling of the superficial layers and
their effects on the oscillations;
here $Q_{nl}$ is the mode inertia, normalized to the value for a radial mode
at the same frequency, and $\CF$ is a function of frequency characterizing
these near-surface effects.
To these relations must be added a
constraint on the density difference resulting from
the fact that the total mass of the model must be kept fixed; 
this can be expressed as
\be
0 = \int_0^{\Rs} {\delta_r \rho\over \rho} \rho r^2 \dd r \; ,
\eel{eq:denscons}
(as noted already in Eq.~\ref{eq:massconv}; see also the related
discussion of the model differences in Fig.~\ref{fig:changelocopac}),
which is formally of the same form as Eq.~\Eq{eq:freqdif}.
Thus this relation can be included directly in the analysis.
With a sufficiently extensive set of observed modes
the relations in Eq.~\Eq{eq:freqdif}
can be analysed to infer measures of the model differences.
Various techniques have been developed to carry out inversions
for the structure differences%
\footnote{\citet{Gough1978b} noted that
``[i]t remains to be seen whether the sun is likely to supply us with
enough data to make this [inverse analysis] possible''. 
The sun has indeed been generous.}
\citep[{\eg},][]{Gough1978b, Gough1985, Dziemb1990, GoughKos1990,
Dziemb1994, Antia1996, Basu1996}.
In all cases the techniques are characterized by \emph{trade-off parameters}
which determine the balance between the desired error and resolution 
of the inferences, as well as the weight given to the suppression of
unwanted contributions to the results;
in inferring the differences in sound speed, for example, the 
the so-called \emph{cross term}, {\ie}, the contribution
from the density differences, must be minimized.
The technical details of the various inversion techniques were
reviewed by \citet{Basu2016},
while \citet{Rabell1999} provided an
analysis of the commonly used technique of optimally localized
averages, including the appropriate choice of the required parameters.

Although the oscillation frequencies depend predominantly on the sound speed
it is also possible to carry out inversions to infer, for example, the
density difference between the Sun and the model.
It should be noticed, however, that the sound-speed and density inferences
are not independent: given the assumed constraints of hydrostatic 
equilibrium and mass equation,
determination of corrections to the hydrostatic structure
in terms of $\delta_r \rho$ and $\delta_r u$ are in principle equivalent;
indeed, \citet{Dziemb1990} pointed out how $\delta_r p$ and hence 
$\delta_r \rho$ can be determined directly from $\delta_r u$.
Since $c^2 = \Gamma_1 u$ and $\Gamma_1 \simeq 5/3$ in most of the Sun,
the inferences of $\delta_r u$ and $\delta_r c^2$ are also closely related.
%\notecd [Possibly some experiments to quantify the equivalence, using MP code;
%see $\sim$dimauro/EVOL/SUNG/dziemb/; $\sim$/papers/gchange/f.dimauro.030627c].

\begin{figure}[htp]
\centerline{\includegraphics[width=\figwidth]{\fig/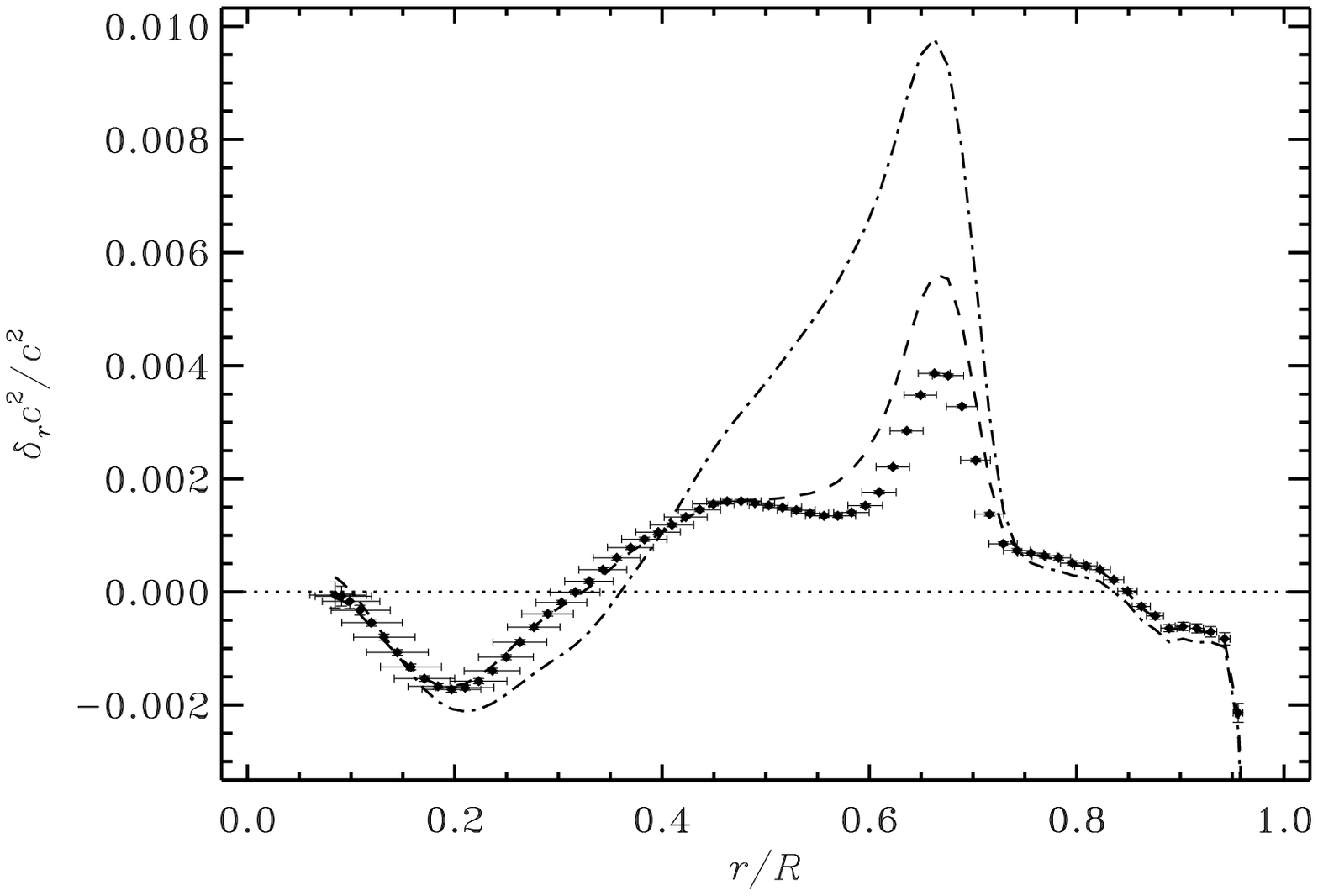}}
\centerline{\includegraphics[width=\figwidth]{\fig/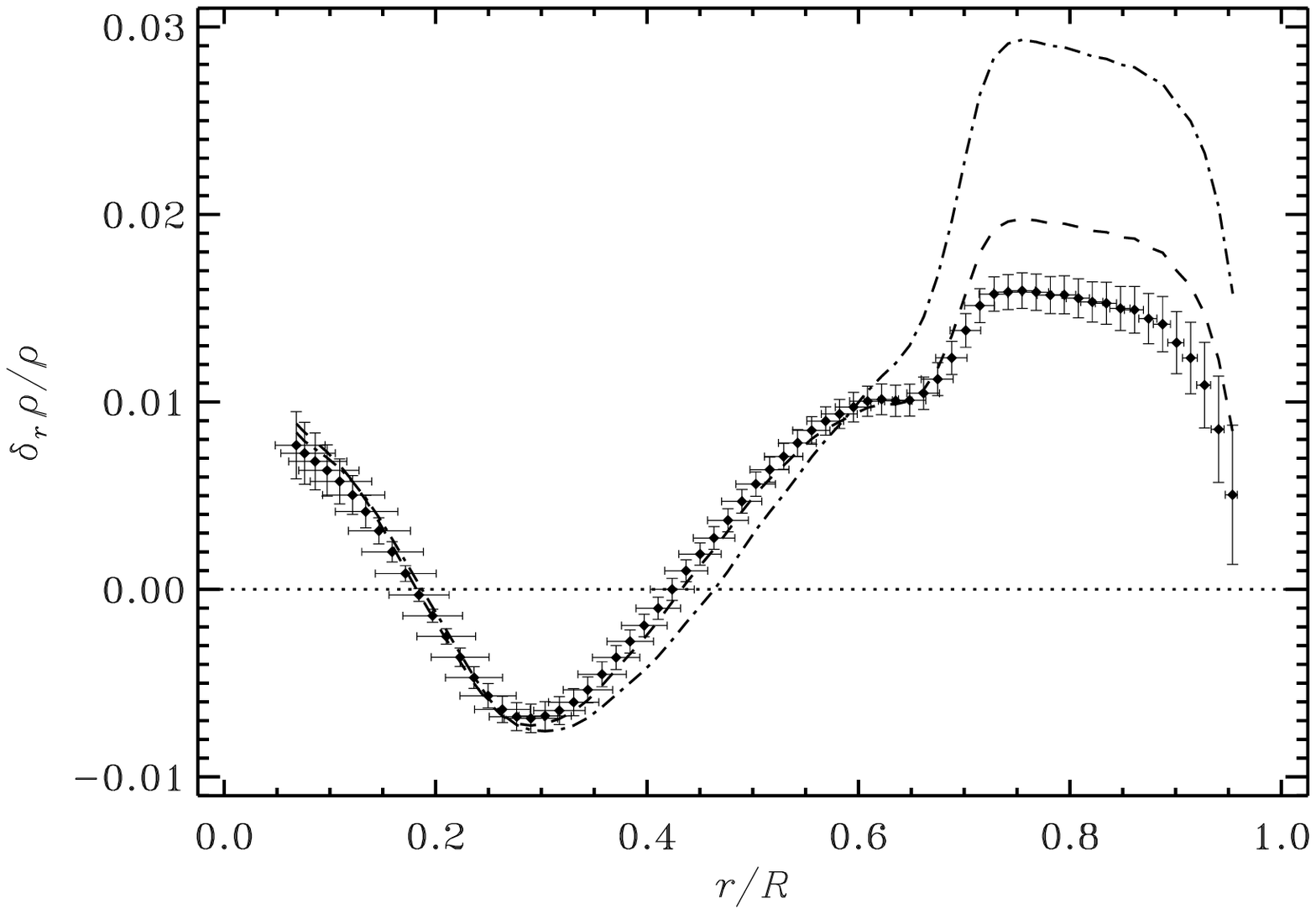}}
\caption{Results of helioseismic inversions.
The symbols show inferred relative differences in squared sound speed
(top) and density (bottom)
between the Sun and the original Model~S 
\citep[][see also Sect.~\ref{sec:models}]{Christ1996},
in the sense (Sun) -- (model); 
this uses the \citet{Greves1993} (GN93) heavy-element composition and the
\citet{Rogers1992} OPAL opacity tables.
The vertical bars show $1\,\sigma$ errors in the inferred values,
based on the errors, assumed statistically independent, in the
observed frequencies.
The horizontal bars extend from the first to the third quartile 
of the averaging kernels, to provide a measure of the resolution
of the inversion \citep[see][]{Basu2016}.
The dashed curves show results for the similar Model~\Mopalninesix, 
which used the \citet{Iglesi1996} tables,
whereas the dot-dashed curves are for Model~\Mgsnineeight, 
where the \citet{Greves1998} composition was used.
%(Adapted from \citecl{Basu1997a}.)
}
\clabel{fig:csqinv}
\end{figure}

As an example of the results on solar structure, Fig.~\ref{fig:csqinv}
shows inferred sound-speed and density differences between the Sun and 
Model~S \citep[see Sect.~\ref{sec:models} and][]{Christ1996}
with the \citet{Greves1993} heavy-element composition,
as well as for Model~\Mopalninesix, an updated version of this model,
with the same composition but using the \citet{Iglesi1996} opacity {\rv tables}
with a slightly reduced
opacity near the base of the convection zone and hence a reduced
sound speed and an increased sound-speed difference relative to the Sun.
This model was compared with Model~S in Fig.~\ref{fig:changeopac}.
In addition, results are shown for Model~{\Mgsnineeight}
with the \citet{Greves1998} composition
where the opacity is further somewhat reduced and the sound-speed difference
to the Sun increased;
the effects on the model of the change in composition were illustrated
in the comparison with Model~\Msurfop, very similar to Model~\Mopalninesix,
in Fig.~\ref{fig:changeopacgs98}.
The properties of the models are summarized in Tables~\ref{tab:modelpar} --
\ref{tab:modeldiff}.
%\notecd [All models should be defined in a table!].
The analysis \citep[see][]{Basu1997a} used a combination of LOWL and BiSON
frequencies.
Inversion was carried out by means of a technique of optimally localized
averages,
which explicitly characterizes the inferred quantities as 
averages of the differences with well-defined localized weight functions,
the so-called \emph{averaging kernels}; the widths of these provide a measure
of the resolution of the inversion
\citep[see][for details]{Basu2016}.
Also, the errors in the inferred differences are calculated
from the quoted errors in the observed frequencies.
The differences between the Sun and the models may be considered
as relatively small, although very significant
compared with the inferred errors, and highly systematic.
In particular, the observational errors are much smaller than the effects of
the relatively modest modification of the opacity tables illustrated by the
dashed curve.
In common with the model differences discussed in Sect.~\ref{sec:modsens}
the density differences are substantially larger
than the sound-speed differences.
Also, it is interesting that the differences are approximately constant in
the convection zone, in accordance with Eq.~\Eq{eq:appdelp}.
Independent analyses of other datasets
\citep[{\eg},][]{Goughetal1996, Kosovi1997a, Turck1997, Couvid2003}
have yielded very similar results, when applied to the same reference models.
As illustrated in Fig.~\ref{fig:compinv}a a model that does not include
diffusion and settling of helium and heavy elements results in
a much larger difference relative to the Sun 
\citep[see also][and Fig.~\ref{fig:nodiffus}]{Christ1993}.
%Also illustrated in the figure is that, as noted in Sect.~\ref{sec:modsens},
%models using
%the \citet{Greves1998} composition have a slightly smaller sound speed
%at the base of the convection zone and hence a somewhat higher 
%sound-speed difference in this region.
It is striking that the old Model~S, with some problems with the input
physics, fortuitously yields the best agreement with the inferred
solar structure.
Lest this relatively good agreement between the Sun and the models
leads to complacency,
I note that the revised abundances obtained by, \eg, \citet{Asplun2009}
cause much more dramatic effects on the comparison;
these are discussed in Sect.~\ref{sec:abundprob}.
%\notecd [UPDATE: Possibly around here also compare
%Models ${\rm S}_1$ and ${\rm S}_2$ with the helioseismic data].

\begin{figure}[htp]
\centerline{\includegraphics[width=\figwidth]{\fig/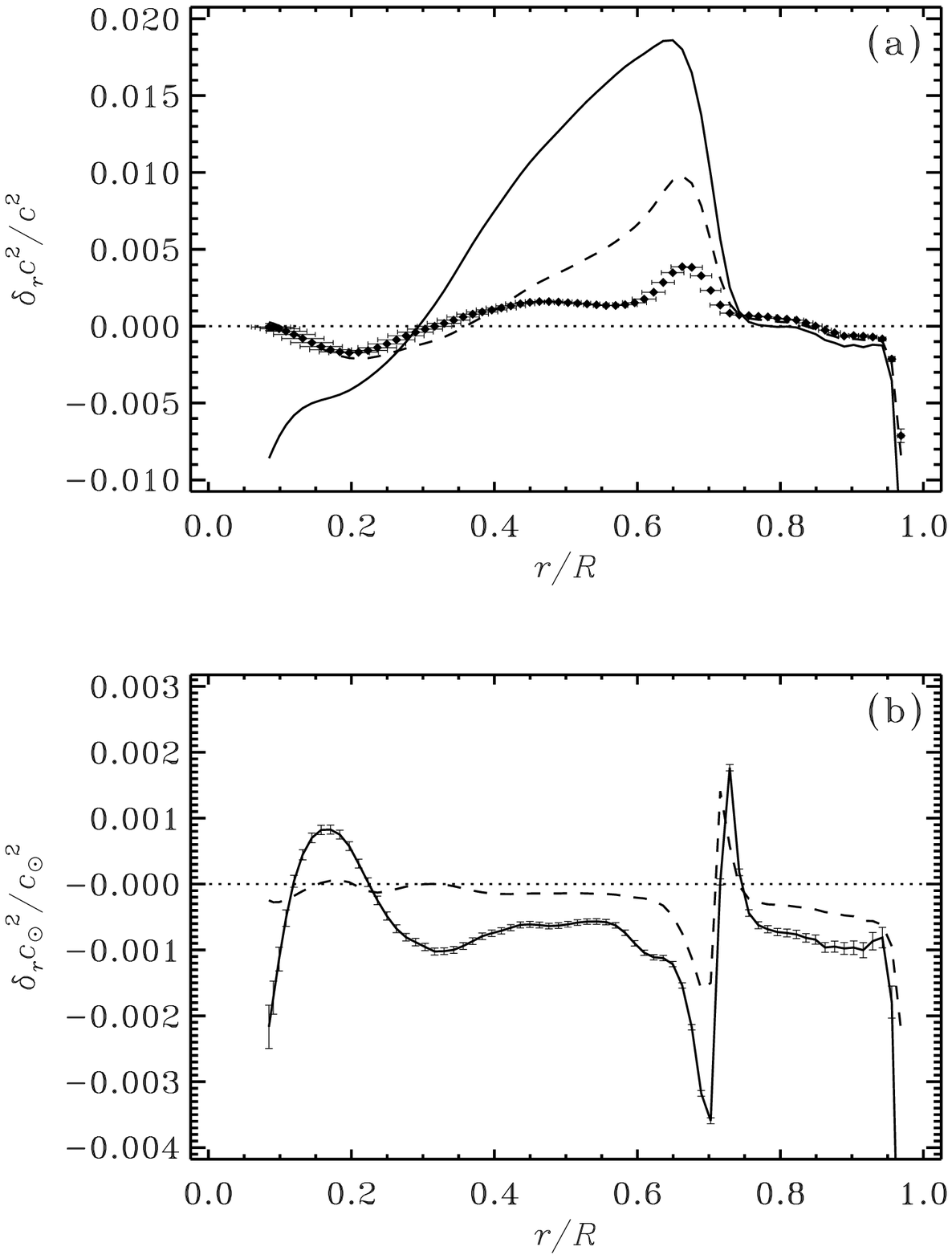}}
\caption{(a) As in Fig.~\ref{fig:csqinv} the symbols show
inferred difference in squared sound speed between the Sun
and the original Model~S,
%\citep[][see also Sect.~\ref{sec:models}]{Christ1996},
in the sense (Sun) -- (model). 
%The vertical bars show $1\,\sigma$ errors in the inferred values,
%based on the errors, assumed statistically independent, in the
%observed frequencies.
%The horizontal bars extend from the first to the third quartile 
%of the averaging kernels, to provide a measure of the resolution
%of the inversion \citep[see][]{Basu2016}.
The dashed curve shows results for Model~\Mgsnineeight, using the
\citet{Greves1998} composition, 
while the solid curve is for Model~\Mnodif, similar to Model~S
but neglecting diffusion and settling.
(b) Relative differences between inferred solar squared sound speed 
$c_\odot^2$ obtained
by correcting the reference model values with the helioseismically inferred 
differences ({\cf} Eq.~\ref{eq:csqrec}).
The solid curve compares the result obtained using 
the non-diffusive Model~{\Mnodif} as reference with the result of using Model~S,
in the sense (Model~\Mnodif) -- (Model~S); the standard deviations
were obtained by combining in quadrature the standard deviations
inferred in the two inversions.
The dashed curve similarly compares the 
Model~{\Mgsnineeight} inference with that from Model~S.
}
\clabel{fig:compinv}
\end{figure}

For high-degree modes the near-surface effects can no longer be regarded
as independent of degree.
Consequently, in Eq.~(\ref{eq:freqdif}) $\CF(\omega)$ must be
replaced by an expansion in $\tilde w = (l+ 1/2)/\omega$, with
$\CF$ as the leading $l$-independent term \citep{Gough1995}.
\citet{DiMaur2002} implemented inversion techniques to take the
expansion in $\tilde w$ into account and applied it to early 
high-degree observations from MDI.
In fact, \citet{Reiter2015} and \citet{Reiter2020} 
carried out inverse analyses including
high-degree modes, using Model~S as a reference, and noted a substantial
excess of the model sound speed within the upper five per cent of the model
(a tendency already hinted at in Fig.~\ref{fig:csqinv}).
However, since the analysis did not include
the $l$-dependent terms in the near-surface correction the result
should probably be regarded as preliminary.

\citet{Basu2000} carried out a detailed analysis of the various sources
of uncertainties in the helioseismic inferences of solar internal properties,
including the effects of different choices of observational data or
reference models. 
They found, for example, that the sound-speed structure resulting from applying
the inferred sound-speed difference to the reference model depended
relatively little on the 
assumed reference model, within a reasonable range of models.
Thus in this sense the analysis provides a robust determination of
the solar internal sound speed.
To illustrate this, Fig.~\ref{fig:compinv}b illustrates differences
between solar squared sound speeds, reconstructed from the model sound speed
and the helioseismically inferred sound-speed difference as
\be
c_\odot^2 = c_{\rm mod}^2\left(1 + {\delta_r c^2 \over c^2} \right) \; ,
\eel{eq:csqrec}
where $c_{\rm mod}^2$ is the squared sound speed in the reference model.
The figure shows differences of two such reconstructions relative to the
reconstruction based on Model~S, which is the model amongst those considered
so far that most closely resembles the Sun.%
\footnote{These differences are essentially equivalent to the
difference between $\delta_r c^2/c^2$ as inferred from inversion
of the frequency differences between the two models considered,
restricted to the observed mode set, and the actual $\delta_r c^2/c^2$
between the models
(the latter is illustrated relative to Model~S in Figs~\ref{fig:nodiffus} 
and \ref{fig:changeopacgs98} for, respectively, Model~{\Mnodif} and 
Model~\Mgsnineeight).
}
Even for the non-diffusive model (solid line), which shows a relatively 
substantial difference from the Sun, the departure from the Model~S 
reconstruction is less than 0.1 per~cent in most of the Sun, 
and for the \citet{Greves1998} model (dashed line) the departure is much 
smaller.
Apart from the uncertain central and near-surface regions the largest
departures are found {\rv just below the convection zone}, 
caused by the sharp gradients
in sound-speed differences between the models in this region,
which are not fully resolved owing to the finite resolution of the inversion.
%\notecd [UPDATE: Consider to include such a 'solar sound speed'
%in the online version].

It is fairly common \citep[{\eg},][]{DeglIn1997, Yang2016}
to compare solar models with an existing reconstructed solar
sound speed computed as in Eq.~(\ref{eq:csqrec}), based on some inversion;
in this case the choice of 
reference model in the original inversion clearly affects
the comparison and hence enters as a
component in the error in the inferred sound-speed difference.
It is then important that the selection of reference models included in the
estimate of that error is realistic 
(the error would obviously be overestimated by including, for example,
models without diffusion and settling).
On the other hand,
in the analyses in the present paper and in \citet{Vinyol2017}, for example,
the differences between the Sun and a model are
inferred directly by using the model as reference for a helioseismic inverse
analysis;
in this case the results provide a direct estimate of the difference
between the Sun and the specific model
subject to the observational error, the finite resolution of the inversion
and the success in suppressing the cross term and the surface contribution,
but without involving contributions to the error from the choice of reference
model.
Even so, Vinyoles {\etal} did include in their error analysis a
contribution obtained from the dispersion of inferences 
of the solar sound speed based on a
set of reference models from \citet{Bahcal2006},
varying the composition and other model parameters within the relevant errors.
Further investigations of the error estimates, in particular the effect of
error correlations, in solar structure inversions are certainly warranted.
%from the choice of reference model,
%although using a different set of reference models with varying physics,
%resulting in a combined error in the inferred
%sound-speed differences between the Sun and the model.
%This is the approach taken in the examples shown here.
%\notecd [But we need to worry about the importance of the cross term and the
%surface term; this certainly needs a little discussion here, and perhaps even
%a few experiments, at least in an update! What does Sarbani say about this
%point?].

\begin{figure}[htp]
\centerline{\includegraphics[width=\figwidth]{\fig/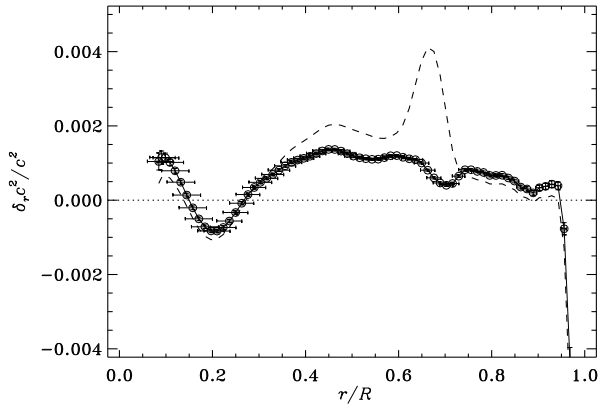}}
\caption{The symbols show the inferred difference in squared sound speed
between the Sun and a model with a suitably adjusted representation of
turbulent diffusion beneath the convection zone \citep{Christ2018a};
for comparison the dashed curve shows a model corresponding to the original
Model~S.
The models are based on the \citet{Asplun2009} composition, 
but with an opacity modification to restore the original Model~S
\citep[see][ and Fig.~\ref{fig:opaceff}a below]{Christ2010}.
(Adapted from \citecl{Christ2018a}.)
}
\clabel{fig:tachodif}
\end{figure}

The most noticeable feature of the sound-speed difference 
in Fig.~\ref{fig:csqinv} is the peak just below the convection zone.
This is a region of a strong gradient in the hydrogen abundance
caused by helium settling ({\cf} Fig.~\ref{fig:Xmod}).
Consequently, the difference can be reduced by partial mixing of the region;
this would increase the hydrogen abundance, and hence decrease the
mean molecular weight and increase the sound speed
({\cf} Eqs \ref{eq:meanmol} and \ref{eq:soundspeed}).
Such mixing might be induced by instabilities associated with the strong
gradient in the angular velocity in this so-called tachocline
({\cf} Fig.~\ref{fig:introt})
\citep[{\eg},][]{Brun1999, Elliot1999, Brun2002, Christ2007}.
Evidence for partial mixing has also been obtained from inverse analyses
designed to infer the composition structure of the solar interior
\citep{Antia1998, Takata2003a}.
\citet{Christ2018a} demonstrated that by imposing a combination of a suitable
modification of the opacity and suitable diffusive mixing the sound-speed
difference can be strongly reduced and the peak in the {\rv tachocline region}
essentially removed; 
the resulting inferred difference in the squared sound speed is illustrated
in Fig.~\ref{fig:tachodif}.
Such additional mixing is also implied by the partial destruction of lithium
(see Sect.~\ref{sec:lightcomp}).
The negative difference in the outer part of the core
could be similarly reduced by partly mixing of this region,
which would decrease the hydrogen abundance;
it is not obvious, however, that any realistic mechanism is available
which may cause such mixing.
%\notecd [And in any case need to refer to new results after revision of
%surface abundances].

\begin{figure}[htp]
\centerline{\includegraphics[width=\figwidth]{\fig/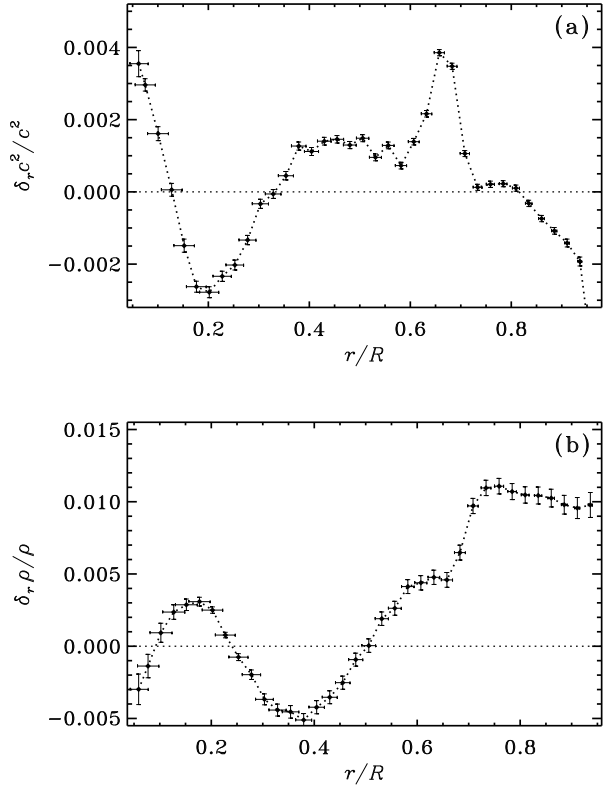}}
\caption{Inferred relative differences in squared sound speed (panel a)
and density (panel b) between the Sun and Model~S of \citet{Christ1996},
in the sense (Sun) -- (model);
the inversion is based on a combination of 14 years of low-degree data from
BiSON observations, corrected for solar-cycle frequency variations, and
data from the MDI experiment on the SOHO spacecraft.
The vertical bars show $1\,\sigma$ errors in the inferred values,
based on the errors, assumed statistically independent, in the
observed frequencies.
The horizontal bars provide a measure of the resolution of the inversion.
(Adapted from \citecl{Basu2009}.)
}
\clabel{fig:basucsqrho}
\end{figure}

%%\begin{figure}[htp]
%%\def\epsfsize#1#2{0.5#1}
%\centerline{\includegraphics[width=\figwidth]{\fig/rho-inv-basu.eps}}
%\caption{Inferred difference in density between
%the Sun and Model~S of \citet{Christ1996},
%in the sense (Sun) -- (model).
%See caption to Fig.~\ref{fig:basucsq}.
%(Adapted from \citecl{Basu2009}.)
%}
%\clabel{fig:basurho}
%\end{figure}
%}

A detailed investigation of the solar internal structure was carried
out by \citet{Basu2009}.
They used a combination of very extensive data on low-degree modes
from the BiSON network, carefully corrected for the variations caused by
the solar cycle, with data from the MDI experiment on SOHO;
to test the sensitivity of the result to the assumed data they also considered
other combinations of BiSON and MDI data.
The analysis was carried out relative to several reference models, including
Model~S considered also in Fig.~\ref{fig:csqinv}.
Figure \ref{fig:basucsqrho} shows the resulting sound-speed and 
density differences.
The results are clearly similar to those shown for Model~S
in Fig.~\ref{fig:csqinv}, obtained with a different dataset;
indeed, Basu {\etal} found that the inferred sound-speed differences showed
only minor dependence on the assumed dataset, although the detailed results
in the core depended slightly on the choice of low-degree data.
The main difference is in the convection zone where significant differences,
increasing in magnitude towards the surface, are found.
These are probably caused by residual effects in the treatment of
the unavoidable errors in the modelling of the near-surface layers; 
the data used in Fig.~\ref{fig:csqinv} did not include modes of degree above
100 and hence were less sensitive to the near-surface structure of the Sun.
I note that according to Eq.~\Eq{eq:approxcsq} we expect the sound speed 
in the bulk of the convection zone largely to match the solar sound speed,
assuming that the model has the
correct surface gravity, as was indeed found in the model comparisons
in Sect.~\ref{sec:modsens}.
Thus the inferred sound-speed differences between the Sun and the model
in the convection zone
further indicate problems with the inversion in the outer parts of the Sun,
including incomplete suppression of the near-surface inadequacies in the
modelling.

In the case of density Basu {\etal} found substantial sensitivity,
throughout the Sun, of the results to the choice of dataset.
They pointed out that this is related to the constraint in 
Eq.~\Eq{eq:denscons}.
The density difference in the core is relatively weakly constrained 
by the observations and hence differs substantially between the 
inversion results for the different datasets;
however, because of the constraint on the mass, any change in the density
difference in the core has to be compensated in the rest of the model.
This effect is further enhanced by the fact that the density is much higher
in the core, requiring a proportionally larger relative change in the outer
parts of the model.

The preceding discussion of helioseismic inversion implicitly assumed that
the solar radius $\Rsun$ was known, as indicated in Eq.~(\ref{eq:freqdif}).
In fact, as discussed in Sect.~\ref{sec:basicpar} there have been several
independent and not fully consistent determinations of $\Rsun$;
using an incorrect estimate could yield systematic errors in the 
inversion results.
A preliminary analysis of these issues was carried out 
by \citet{Takata2001a, Takata2003b}, including ways to improve the determination
of $\Rsun$ as part of the inversion process.
They found that the effects of errors in the assumed radius were small,
but not quite insignificant compared with the statistical error 
in the inferences.

Given the seismically determined sound speed and density in the Sun one
may construct a \emph{seismic model}, {\ie}, a model that is consistent with
the seismic results.
The first step is to reconstruct relevant aspects of solar structure
from seismically inferred differences,
as in Eq. (\ref{eq:csqrec}), with suitable extrapolation to the
regions of the Sun not covered by the inversions.
Additional properties, such as the mass distribution and pressure can be
obtained by invoking the equations of mass and hydrostatic equilibrium,
Eqs (\ref{eq:hydrostat}) and (\ref{eq:mass}).
This process may be iterated, by using the thus reconstructed model as reference
for a new inversion \citep[e.g.,][]{Antia1996, Buldge2020}.
Additional properties of the model can be constrained by
further equations of stellar structure combined with suitably chosen aspects
of the physics of the stellar interior \citep{Antia1997, Takata1998}.
Such a model, based on using Model~S as reference,
was presented by \citet{Gough2001}, with the underlying analysis and further
details discussed by \citet{Gough2004}.
The analysis constrained the composition and temperature structure
based on the nuclear energy-generation rate.
Given the inferred local luminosity and temperature, 
the opacity could be
estimated in the radiative region from Eqs~(\ref{eq:nabla}) and
(\ref{eq:nablarad}).
Interestingly, the differences between resulting
inferred opacity and the model opacity 
(based essentially on the model heavy-element abundance,
with the \citet{Greves1993} composition) were at most around 1.5 per cent.
%\notecd [Could include Gough or updated seismic model with the paper].

%\notecd [Here a paragraph on other tests of structure, still based on
%sound-speed inversion, such as depth of convection zone and overshoot].

\subsubsection{Specific aspects of the solar interior}
\clabel{sec:specasp}

In addition to the general behaviour of the internal solar sound speed
and density,
more specific aspects of the solar internal structure can be inferred.
Already the early asymptotic sound-speed inversion by \citet{Christ1985}
showed indications of the location of the base of the convection zone.
Further analyses have yielded tighter constraints on this
point, understood as the location where the thermal gradient becomes
substantially subadiabatic.
\citet{Christ1991b} determined the depth $d_{\rm cz}$ of the convection zone
as $d_{\rm cz} = (0.287 \pm 0.003) R$, a value confirmed by \citet{Kosovi1991}.
A very similar value, but with even higher precision, was determined
by \citet{Basu1997b} and \citet{Basu1998}.
Further information about conditions near the base of the convection zone
can be inferred from analysis of an oscillatory behaviour in the frequencies 
induced by the relatively rapid variations in solar structure in this region;
this corresponds to a so-called \emph{acoustic glitch},
where the structure varies on a scale small compared with the local wavelength
of the acoustic waves
\citep[{\eg},][]{Hill1986, Gough1988, Voront1988, Gough1990b}.
In particular, with the normal treatment of convection the second derivative
of sound speed is essentially discontinuous here;
also, simple models of convective overshoot \citep[{\eg},][]{Zahn1991}
predict a nearly adiabatic extension of the temperature gradient beneath
the unstable region, followed by an essentially discontinuous jump to
the radiative gradient, and hence a stronger variation in the sound speed.
From analysis of the oscillatory frequency variations associated with this
region the extent of such overshoot has been limited to a small
fraction of a pressure scale height
\citep[{\eg},][]{Basu1994a, Basu1994b, Montei1994, Roxbur1994, Christ1995a}.
More detailed modelling of overshoot 
\citep[{\eg},][]{Rempel2004, Rogers2006a, Xiong2001}
yields a smoother transition, possibly including a slightly subadiabatic
region in the lower parts of the convection zone.
Such overshoot models are not obviously constrained by the earlier helioseismic
analyses but might still be amenable to helioseismic investigation.
\citet{Christ2011} found that a model with somewhat smoothed overshoot 
was in fact in better agreement with the helioseismic data than was a
`standard' model such as Model~S;
in the latter helium settling causes a relatively sharp change in the sound
speed at the base of the convection zone and hence an oscillatory frequency
variation larger than observed.
%\notecd [Something from Houdek \& Gough???].

%\notecd [This short paragraph may be a little out of place].
%An interesting inverse analysis was presented by \citet{Buldge2017a} 
%in terms of the so-called Ledoux discriminant, {\ie}, the difference
%between the adiabatic and the actual density difference,
%which is sensitive to the detailed structure just below the convection zone.
%I return to this in Sect.~\ref{sec:modrevcomp}.

The departures of $\Gamma_1$ from the simple value for an ideal gas 
have very great interest as a diagnostics of the equation of state
and composition of the convection zone.
From the equation of state $\Gamma_1$ is given as a function
$\Gamma_1(p, \rho, \{X_i\})$ of the hydrostatic structure and composition,
with the abundance of helium having the strongest effect.  
The analysis of this dependence is simplified in the convection zone where
the structure is characterized by the approximately adiabatic gradient
and where composition can be assumed to be uniform due to the very
efficient convective mixing.
If the equation of state is assumed to be known, an inference of 
the sound speed can be used to infer the composition through its effect
on $\Gamma_1$.
In fact, it was noted by \citet{Gough1984b} and \citet{Dappen1986a} that the
second ionization of helium produces a signature in $\Gamma_1$,
acting as an acoustic glitch,
which is potentially a sensitive measure of the helium abundance.
This effect has been analysed in a variety of ways, using both asymptotic
and non-asymptotic techniques
\citep[{\eg},][]{Voront1991b, Dziemb1991, Kosovi1992, Antia1994, Perez1994,
Kosovi1996, Basu1998, Richar1998, DiMaur2002}.
The resulting values of the envelope helium abundance $Y_{\rm s}$
depend somewhat on the assumed equation of state, although
values around $Y_{\rm s} = 0.248$ are typically found,
with a formal uncertainty of as low as 0.001 and a somewhat larger
systematic uncertainty estimated from the use of different equations of state.
As an example, \citet{Basu2004a} obtained $\Ys = 0.2485 \pm 0.0034$,
taking into account also uncertainties in the equation of state.
It should be noted that this value is substantially lower than the
primordial value $Y_0 = 0.271$ required to calibrate solar models
({\cf} Sect.~\ref{sec:calib}), thus confirming the importance
of helium settling.
Indeed, the helioseismically inferred value is close to, and 
independent from, the value obtained in standard solar models
including settling;
in Model~S, for example, the envelope helium abundance is $Y_{\rm s} = 0.245$.
%\notecd [strong statement that should be tested and quantified].
Using the value $Y_{\rm s} = 0.2485 \pm 0.0034$ 
obtained by \citet{Basu2004a} and several different solar models,
\citet{Serene2010} estimated the primordial solar helium abundance as 
$Y_0 = 0.278 \pm 0.006$, which is essentially consistent with the 
value obtained from the Model~S calibration.
%\notecd [Somewhere should have tables of observed and model $Y_0$,
%$r_{\rm cz}$, with corresponding tables with the new composition later].
%\notecd [Check Gough (1996; Science) for determination of He abundance].

Observations of low-degree modes in distant stars 
provide a similar possibility for
determining the envelope helium abundance through the effect on
the acoustic-mode frequencies
\citep[{\eg},][]{Perez1998, Lopes2001, Houdek2004, Houdek2007a}.
As discussed in Sect.~\ref{sec:astero} this potential has 
been realized thanks to the very detailed asteroseismic
data obtained with the {\it Kepler} mission.

It is interesting, particularly in the light of the revisions
of solar surface abundances ({\cf} Sect.~\ref{sec:newcomp})
that it appears possible to constrain also abundances of the dominant
heavy elements through their effect on the equation of state.
I return to this in Sect.~\ref{sec:checkcomp}.
%\citep{Lin2005, Lin2007,Voront2013, Buldge2017b};
%the results appear to support abundances substantially lower than
%the value assumed in Model S,
%although, perhaps even more than for the determination of the 
%helium abundance,
%the results are sensitive to subtle details in the equation of state
%\notecd [could summarize all these helioseismic results in a table].
%\notecd [Heavy-element abundances should be discussed further,
%in connection with Asplund results].

\begin{figure}[htp]
\centerline{\includegraphics[width=\figwidth]{\fig/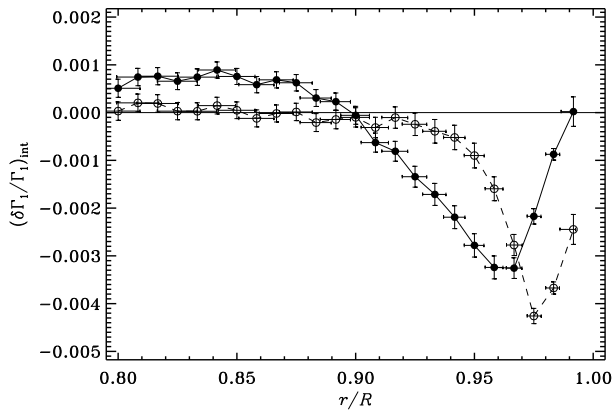}}
\caption{Relative intrinsic difference in $\Gamma_1$
in the sense (Sun) -- (model), inferred
from inversion of oscillation frequencies of degree up to $\sim 250$
%\notecd [quote range of degrees]
obtained with the MDI instrument.
The closed circles show results for a model using the MHD
equation of state while the open circles are for a model computed
with the OPAL equation of state.
As in Fig.~\ref{fig:csqinv} the vertical and horizontal bars measure
error and resolution, respectively.
(Adapted from \citecl{Basu1999}.)
}
\clabel{fig:gaminv}
\end{figure}

%\notecd [The following should be fixed taking into account DOG concerns on
%composition effects, and with suitable references].
The inverse analysis can be arranged to include also a contribution
from the \emph{intrinsic} difference $(\delta \Gamma_1)_{\rm int}$
between the solar equation of state and that assumed in the model,
{\ie}, the difference in $\Gamma_1$ at fixed $(p, \rho, \{X_i\})$,
allowing inferences to be made of $(\delta \Gamma_1)_{\rm int}$,
reflecting the errors in the equation of state used in the solar modelling
\citep{Basu1997c}.
Examples are shown in Fig.~\ref{fig:gaminv}, based on
analyses carried out by \citet{Basu1999}.
It is evident that the OPAL results are generally closer to the Sun,
although with some indications of the opposite tendency very
close to the surface.
It should be noted that a complete separation of the effects
of the intrinsic differences in $\Gamma_1$ and the differences in
composition is not possible; 
however, the compositional effects are strongly constrained by the
fact that the composition is uniform in the convection zone, 
and the effect of a reasonably uncertainty in, for example, the helium
abundance on the results shown in Fig.~\ref{fig:gaminv} is modest
\citep{Rabell2000}.
Further details about the equation-of-state differences, and hence 
also further constraints on the composition, could be obtained with
data on higher-degree modes \citep{DiMaur2002}.
%So far, however, the analysis of these modes has been affected by
%systematic errors and interference between the modes
%\citep{Korzen2004, Rabell2008},
%and it is not clear whether reliable frequencies are as yet available.
It should also be noted that specific details on the equation of state
can be investigated by comparing suitably parameterized formulations
of the equation of state with the helioseismically inferred properties;
an interesting example, involving a calibration of the size of hydrogen
and helium atoms and ions, was presented by \citet{Baturi2000}.

In the core of the Sun it might be expected that the equation of state
is relatively simple.
It was therefore somewhat surprising that \citet{Elliot1998a}
found significant differences in $\Gamma_1$ between the Sun and the model
close to the solar centre, in an inversion based on $(\rho, \Gamma_1)$.
They demonstrated that the differences arose
solely from the neglect of relativistic effects on the electrons
({\cf} Eq.~\ref{eq:elrelgam}) in the versions of the OPAL and MHD 
equations of state used at the time for the model calculation.
Taking these effects into account the inferred $\delta_r \Gamma_1$
was consistent with zero in the core to within errors.

\subsubsection{Investigations of solar internal rotation}
\clabel{sec:heliorot}

%\notecd [With reference to solar evolution should probably also summarize
%solar internal rotation as obtained from helioseismology;
%but check also what Miesch has done].

Solar rotation induces a splitting of the observed frequencies
according to their azimuthal order $m$.
\citet{Howe2009} provided an extensive review of the analysis of these data:
inversion of the rotational splittings
has provided detailed information about rotation in the solar interior
\citep[see also][]{Thomps2003}.
Already early results \citep{Duvall1984} indicated that the radiative
interior of the Sun rotates at a nearly uniform rate close to but
slightly below the surface equatorial rotation rate.
This was in striking contrast to models of solar evolution which
had led to the expectation of a possible relict rapidly rotating core
left over from an initial state of rapid rotation
(see Sect.~\ref{sec:pmsevol}).
An important consequence of the slow rotation of the solar interior is
that rotational oblateness causes no significant modification to the Sun's 
outer gravitational field, at a level which might affect tests of
Einstein's theory of general relativity on the basis of planetary motion
\citep[{\eg},][]{Pijper1998, Roxbur2001, Mecher2004, Antia2008}.

\begin{figure}[htp]
\centerline{\includegraphics[width=\figwidth]{\fig/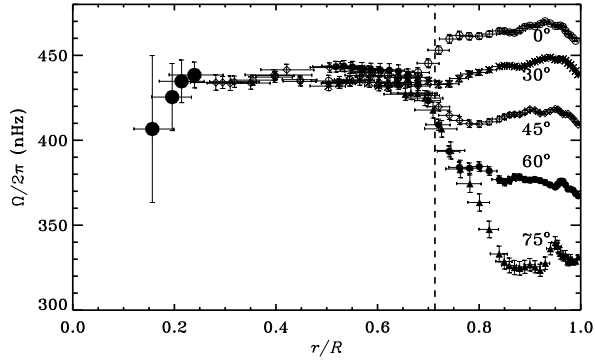}}
\caption{Inferred solar internal rotation rate $\Omega/2 \pi$ as a function of
fractional radius $r/R$, at the latitudes indicated.
As in Fig.~\ref{fig:csqinv} errors and resolution are indicated by the
vertical and horizontal bars.
Results in the outer parts of the Sun, including the convection zone
(the base of which is indicated by the dashed line) were obtained from
analysis of 144 days of SOI/MDI data \citep{Schou1998}.
Results below $r = 0.45 R$, with no latitude resolution, were
obtained by \citet{Chapli1999} from a combination of BiSON and LOWL data.
Note that the results become highly uncertain in the deep interior,
although with no indication of a rapidly rotating core.
}
\clabel{fig:introt}
\end{figure}

Results of the rotational inferences are summarized in 
Fig.~\ref{fig:introt}.
Throughout the convection zone rotation varies with latitude in a
manner similar to the directly observed surface differential rotation
({\cf} Eq.~\ref{eq:surfrot}),
although with subtler details such as the near-surface increase in
rotation rate with depth \citep[see also][]{Corbar2002, Bareka2014}.
Interestingly, indications of this near-surface variation 
were already found in the
early analysis of high-degree modes by \citet{Deubne1979}.
On the other hand, beneath the convection zone the rotation rate
is essentially independent of position.
The rotation rate of the solar core is highly uncertain, as indicated.
Only modes of the lowest degree reach the core ({\cf} Eq.~\ref{eq:turn}),
so that few azimuthal orders are available to determine the splitting;
also, even for these modes the contribution from the core to the
rotational splitting is small.
Several independent observations have confirmed the general 
near-uniformity of rotation in the deep solar interior
\citep[{\eg},][]{Chapli2001b, Fossat2003, Garcia2004}.
The slight indication in Fig.~\ref{fig:introt} of a downturn
in the core is interesting but obviously not significant.
Indeed, {\rv \citet{Corbar1998} found a strong dependence of the
inferred core rotation on the details of the data and inversion
techniques used. 
Also,} \citet{Chapli2004} demonstrated that with typical 
disk-integrated data, covering modes of degree $l = 1 - 3$, 
only a difference in the core rotation, for $r \le 0.2 R$
of at least of order $100 \nHz$
relative to the overall rate in the radiative interior would be
significantly detectable.

For completeness, I recall the claim made by \citet{Fossat2017} of rapid
rotation of the solar core, although based on a questionable analysis 
claiming detection of solar dipolar g modes (see Sect.~\ref{sec:heliostruc}).

Given the availability of independent datasets
it has been possible to test the consistency 
of the different observations and data-analysis techniques.
While there seems to be considerable consistency between 
inferences of the solar internal structure based on different datasets
\citep{Basu2003},
%\notecd [a little out of place, but probably o.k.]
systematic differences remain between results on solar rotation
\citep{Schou2002};
even in the latter case, however, the overall features in the inferred
rotation rates are largely consistent.
In a detailed analysis involving several different datasets and novel
inversion techniques \citet{EffDar2013} generally confirmed the results
shown in Fig.~\ref{fig:introt}, while emphasizing the uncertainty in
the inference of rotation of the solar core.

The transition between the different latitudinal variation of the rotation
in the convection zone and the radiative interior takes place in
a relatively thin region, \emph{the tachocline} \citep{Spiege1992}.
This region probably plays an important role in the presumed
dynamo action responsible for the solar magnetic field and its
cyclic 11-year variations \citep[{\eg},][]{Miesch2009, Charbo2020, Brun2017}.
%\notecd [Possibly more.]
Thus detailed studies of its properties have been carried out
\citep[see][and references therein]{Charbo1999}.
Charbonneau et al. determined the width, defined as the region over which
84 per cent of the variation in angular velocity takes place,
to be $w = (0.039 \pm 0.013) \Rsun$,%
\footnote{{\rv I note that this is consistent with the upper limit 
of $0.05 \Rsun$ inferred with a non-linear inversion technique by
\citet{Corbar1999}.}}
with the centre of the transition
being located at $r_{\rm c} = (0.693 \pm 0.002) \Rsun$.
The tachocline region was found to be slightly prolate, with $r_{\rm c}$ being
closer to the surface by $(0.024 \pm 0.004) \Rsun$ at a latitude of $60^\circ$
than at the equator.
No significant latitude variation in the width was found.
It should be noticed that the location of $r_{\rm c}$ places most of the
tachocline below the base of the convective envelope, at $r_{\rm bcz} =
(0.713 \pm 0.001) \Rsun$ (see Sect.~\ref{sec:specasp}).
\citet{Antia2011} confirmed the prolate nature of the tachocline and in
addition found a statistically significant increase with latitude in its width.

Rotation in the solar convection zone, including the latitudinal surface
differential rotation, is controlled by dynamical transport processes
within and just below the convection zone.
Simple arguments, and early numerical simulations of the convection zone,
indicate that the angular velocity should depend only on the distance to
the rotation axis, in what has been called `rotation on cylinders'
\citep[e.g.,][]{Gilman1976}, which is manifestly not shown by the 
helioseismic inferences in Fig.~\ref{fig:introt}.
More sophisticated simulations have produced rotation profiles quite 
similar to that of the Sun, with some suitable adjustment of parameters
\citep[see, for example][and references therein]{Miesch2006}.

Assuming that the Sun was born in a state of substantially more rapid 
rotation, some mechanism must obviously have been responsible for the
transport of angular momentum from the deep interior to the convection
zone, from which it has presumably been lost through coupling to
the solar wind.
Early models invoking angular-momentum transport through turbulent diffusion
\citep[{\eg},][]{Pinson1989, Chaboy1995} result in present internal rotation
rates several times the surface value, and hence are definitely inconsistent
with the helioseismic results.
A detailed analysis of the instabilities induced by rotation and 
angular-momentum transport was carried out by \citet{Mathis2018},
confirming that additional transport mechanisms would be required to
account for the observed solar rotation profile.
Angular-momentum transport by waves, originally 
proposed by \citet{Schatz1993} and further developed by
\citet{Kumar1997} and \citet{Talon1998},
may remain a possibility, although requiring a fairly elaborate
combination of effects \citep{Talon2002}.
As detailed by \citet{Talon2005} the model involves a so-called 
shear-layer oscillation just beneath the convection zone,
similar to the oscillation demonstrated in the laboratory experiment
of \citet{Plumb1978},
which filters the
gravity waves in such a way that in the deeper interior those waves dominate
that tend to slow down the radiative interior. 
Talon and Charbonnel also developed the model to provide 
lithium destruction consistent with
observations of open stellar clusters and of the Sun
\citep[see also][]{Charbl2005}.
The treatment of the gravity waves was based on a simplified model of
wave excitation by convective eddies within the convection zone
\citep[{\eg},][]{Kumar1997} which in particular ignored the likely
strong effects of the penetration of convective plumes into the stable layer
underneath \citep[{\eg},][]{Hurlbu1986}.
\citet{Rogers2006b} carried out detailed two-dimensional
numerical calculations of the
excitation of gravity waves in the solar convection zone and the resulting
transport of angular momentum.
They found strong effects of penetrating plumes just beneath the convection
zone, and hence pointed out
the need to consider the coupling between variations in
rotation in the convection zone, the tachocline and the deep radiative
interior.
Also, they noted that the gravity-wave spectrum resulting from the simulation
predicted a flux that was essentially independent of frequency, unlike the
spectrum used by \citet{Talon2005} which was strongly peaked at low frequency.
As a result, they questioned the viability of the gravity-wave mechanism to
slow down the solar core.
Further simulations by \citet{Rogers2008} of the properties of
gravity waves in the solar radiative interior,
driven at selected frequencies and wave numbers, 
confirmed the problems with the earlier models, particularly the effect
of a flat spectrum and the importance of previously neglected wave-wave
interactions, which are surely relevant under realistic circumstances where
a large number of waves are excited simultaneously
\citep[see also][]{Rogers2007}.
\citet{Deniss2008} showed that the gravity-wave transport would have a
tendency to produce large-scale oscillations in the angular velocity 
in the radiative interior which are not observed.
%\notecd [May include a little on further Rogers {\etal} paper here, on 
%quasi-linear approximation].
%\notecd [Need update on this; check with Tami].

A plausible alternative is that angular-momentum transport is
dominated by a weak primordial magnetic field 
\citep{Charbo1993, Gough1998}.
A more detailed analysis of such mechanisms revealed a strong sensitivity
to the assumed boundary conditions in the tachocline
\citep{Brun2006, Garaud2007a, Garaud2008a},
%\notecd [Garaud \& Rogers, Cambridge; 
%Garaud \& Brummell preprint??. Also possibly mention Brun \& Zahn]
in order to achieve the latitude-independent rotation in 
the radiative interior while satisfying Ferraro's law
of isorotation \citep{Ferrar1937},
according to which the angular velocity has a constant value along 
poloidal magnetic field lines.
This requires that the magnetic field is essentially
confined to the radiative interior;
were that not the case the latitude dependence of the angular velocity
in the convection zone would penetrate into the radiation zone,
which is not observed \citep[for a review, see][]{Garaud2007c}.
By imposing plausible boundary conditions at the base of the
convection zone on simulations of the dynamics of the tachocline,
\citet{Garaud2008b} obtained the required confinement.
The resulting angular velocity of the radiative interior was discussed
in terms of a simple model by \citet{Garaud2009}.
A recent analysis by \citet{Garaud2020}, based on simplified numerical
modelling and scaling of the relevant physical processes, 
suggested that the tachocline could be non-magnetic but
dominated by three-dimensional stratified
turbulence, possibly implying that it is much thinner than $0.01 \Rsun$;
given the finite resolution of the inverse analyses, this may well be 
consistent with the helioseismic inferences.

Weak dynamo action in the radiative interior driven by magnetic
instabilities \citep{Tayler1973, Spruit2002} has also been proposed
as a means of angular-momentum transport;
since the resulting transport is much more
efficient in the latitudinal than in the radial direction this could
account for the latitude-independent rotation below the convection zone.
\citet{Eggenb2005} demonstrated that this formulation could indeed
explain the evolution to the present solar internal rotation.
However, \citet{Deniss2007} questioned some aspects of the analysis
of \citet{Spruit2002} and hence the applicability of the mechanism
to the solar spin-down.
Also, \citet{Zahn2007},
using three-dimensional magneto-hydrodynamical simulations,
failed to find the required dynamo action under solar conditions,
{\rv although \citet{Braith2017} argued that this was caused by the assumed 
unrealistic high magnetic diffusivity.}
However, \citet{Fuller2019} revisited the Tayler-Spruit mechanism, 
including non-linear effects in the generation of the magnetic field;
on this basis they found that the resulting angular-momentum transport
plausibly could account for the nearly rigid rotation of the solar radiative
interior.
An overall analysis of the potential for magnetic effects to cause the
observed solar rotation profile was presented by \citet{Eggenb2019}.

It is probably fair to say that we do not yet have a full understanding
of the origin of the present solar internal rotation. 
One may hope that further constraints on the modelling may result from
asteroseismic results on the internal rotation of other solar-like stars
(see Sect.~\ref{sec:astero}).

An extensive review of the dynamics of the convection zone and tachocline 
was provided by \citet{Miesch2005},
while \citet{Gough2010} presented a detailed review of the 
angular-momentum coupling through the tachocline.
An overall discussion of the dynamics of the solar radiative interior,
in the light of the helioseismic results, was provided by \citet{Gough2015}.
%and a review of rotation in the solar interior is being provided
%by Sekii \& Shibahashi for {\it Living Reviews in Solar Physics}
%\notecd [probably replace by proper reference, when done].

\begin{figure}[htp]
\centerline{\includegraphics[width=\figwidthb]{\fig/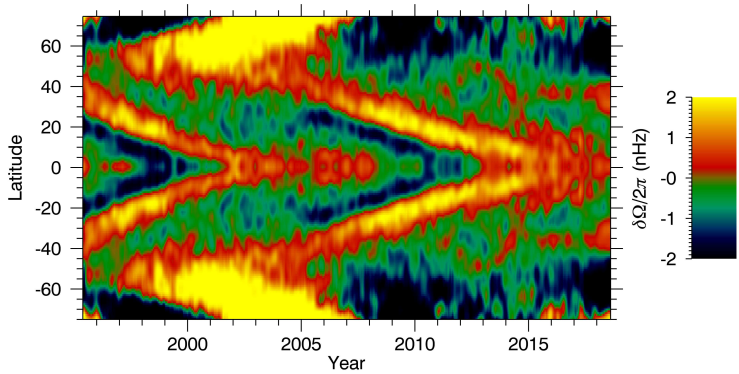}}
\caption{Rotation-rate residuals, 
relative to suitably defined average solar rotation rates,
inferred from inversions at a target depth of
$0.01 \, {\rm R}_\odot$ below the photosphere \citep{Howe2018}.
Results are shown as a function of time and latitude,
using the colour scale at the right. (Note that a rotation-rate variation
of 1\,nHz corresponds roughly to a flow speed of $4 \, {\rm m \, s^{-1}}$
at the solar equator.)
The analysis is based on data from the MDI, HMI and GONG experiments.
Figure courtesy of R. Howe, adapted from \citet{Howe2018}.
}
\clabel{fig:rhzonal}
\end{figure}

\subsubsection{Temporal changes of the solar interior}
\label{sec:tempchange}

The availability of the detailed helioseismic data
over an extended period has allowed studies of the time dependence
of solar internal structure and dynamics, 
potentially related to the solar magnetic cycle.
Changes in the oscillation frequencies, reflecting potential changes in
solar structure, {\rv were first detected by \citet{Woodar1985};}
the dominant variations are closely correlated with the surface magnetic field
\citep[{\eg},][]{Woodar1991, Bachma1993, Chapli2001a, Howe2002} and appear 
predominantly to be a near-surface effect.
In an analysis of the response of the Sun and its oscillation frequencies
to thermal perturbations \citet{Balmfo1996} concluded that deep-seated
perturbations of this nature were inconsistent with the lack of substantial
variations in the solar radius.
However, tentative but very interesting evidence was found by \citet{Gough1994}
of a possible solar-cycle related change in the frequency signature
of the acoustic glitch associated with the second helium ionization zone.
Similar effects were found in more detailed analyses of observed frequencies
by \citet{Basu2004b} and \citet{Verner2006a}, identified by Basu and Mandel
as possibly arising from a magnetic effect on the equation of state.
\citet{Gough2013b} carried out a detailed analysis of this effect, however,
raising some doubts about its reality.
%\notecd [And then Houdek \& Gough 201?].
%\notecd [UPDATE: a little more on the relevant physics; Goldreich+, Gough?].
Also, \citet{Baldne2008} found evidence for sound-speed changes
between solar activity minimum and maximum at the base of the convection zone,
although these have apparently so far not been further confirmed.

As reviewed by \citet{Howe2009}, the rotation throughout
and possibly below the solar convection zone shows clear changes with
solar cycle.
%largely in phase with the sunspot number.
The dominant variation has the form of \emph{zonal flows}, {\ie}, 
regions of slightly faster
and slower rotation, penetrating to substantial depth into the convection 
zone and shifting towards lower latitude with time.
Such variations in the surface rotation rate were previously detected
by \citet{Howard1980}, who identified them as \emph{torsional oscillations}
\citep[see also][]{Snodgr1985, Ulrich2001}.
As shown by, {\eg}, \citet{Howe2000} the behaviour is similar to the shift
towards the equator of the location of sunspots as the solar cycle advances,
in the so-called \emph{butterfly diagram} \citep{Hathaw2015}.
In addition, there is a band of more rapid flow moving towards the poles
\citep{Voront2002}.
Results covering the last full 22-year magnetic cycle are shown in
Fig.~\ref{fig:rhzonal}.
Interestingly, the slow decline of solar cycle 23 was matched by a slower
shift of the corresponding band of more rapid rotation \citep{Howeetal2009}.
Also, the first appearance of cycle 24 was visible in the zonal flows
well before the first appearance of active regions \citep{Howe2013}.
Recent analyses \citep{Howe2016, Howe2018} show a continuation of
this pattern;
as illustrated in Fig.~\ref{fig:rhzonal} the data now show the first
indications of the appearance of cycle 25.
The physical origin of these zonal flows is so far not understood,
although mean-field dynamo models have reproduced some
aspects of the flows, including the high-latitude branch
\citep[e.g.,][]{Rempel2007, Rempel2012, Pipin2019}.
\citet{Basu2019} also studied the time variation of solar rotation over
cycles 23 and 24. 
Interestingly, they found that the position and width of the tachocline 
did not vary, while there were significant variations with time in the
change in rotation rate across the tachocline.

These helioseismically inferred variations in the Sun provide a
potentially important diagnostics of the apparent changes in solar activity,
reflected by the delayed and unusually deep minimum between cycles 23 and 24
and the modest activity in cycle 24.
In fact, \citet{Basu2012} noted, after the fact, that the difference between
the frequency variations in the descending phases of cycle 22 and cycle 23
might have been used as a prediction of the unusual nature of the cycle 24
minimum,
while \citet{Howe2017} speculated that changes in the frequency response to
solar activity could reflect a more fundamental change in the solar dynamo.
\citet{Kosovi2018b} analysed solar f-mode frequency splittings from the
SOHO/MDI and SDO/HMI instruments extending over 21 years.
Much of the variation was clearly correlated with the solar-activity cycle,
while a coefficient related to the latitude variation of rotation showed
longer-term trends, which might also indicate changes in the solar internal
dynamics.
It is evident that a continuation of such observations over further solar
cycles is of great interest.

\subsection{Solar neutrino results}
\clabel{sec:neutr}

%\notecd [Briefly on history, resulting in neutrino `problem'; also a few remarks
%on attempts at solution.]
%
As discussed in Sect.~\ref{sec:engenr} the nuclear reactions generating
energy in the solar core unavoidably produce electron neutrinos.
Owing to their small cross section for interaction with matter the
neutrinos escape the Sun essentially unhindered.
From the total solar luminosity, and the energy generation of around
$26 \MeV$ for each produced $\helfour$,
it is easy to calculate that the neutrino flux at the distance of
the Earth from the Sun is around $6 \times 10^{10} \cm^{-2} \s^{-1}$.
It is evident that the detection of this neutrino flux would be a
strong confirmation of the importance of nuclear reactions in
the core of the present Sun, and potentially a very valuable
diagnostics of conditions in the solar core.
Here I provide a brief overview of some key aspects and results of the
study of solar neutrinos. 
More detailed
reviews of solar neutrino studies have been given by \citet{Bahcal1989},
\citet{Haxton1995}, \citet{Castel1997}, \citet{Kirste1999} and
\citet{Turck1999};
\citet{Bahcal2004a} and \citet{McDona2004} have provided reviews on
the theoretical and experimental situation, respectively,
and general overviews were given by \citet{Haxton2006}, \citet{Haxton2008a}
and \citet{Haxton2013}.
%\notecd [Need check of LRSP status, urgently.]
%\notecd [update].

\begin{figure}[htp]
%\def\epsfsize#1#2{0.6#1}
%\centerline{\epsfbox{\fig/colorbahcallserenellibs05OP.ps}}
%\centerline{
\begin{center}
%\hskip -20pt \includegraphics[scale=0.75]{\fig/neutrino_scale.eps}
%\vskip -50pt
%\rotatebox{-90}{\includegraphics[scale=0.5]
%{\fig/colorbahcallserenellibs05OP.ps}}
\includegraphics[scale=0.5]{\fig/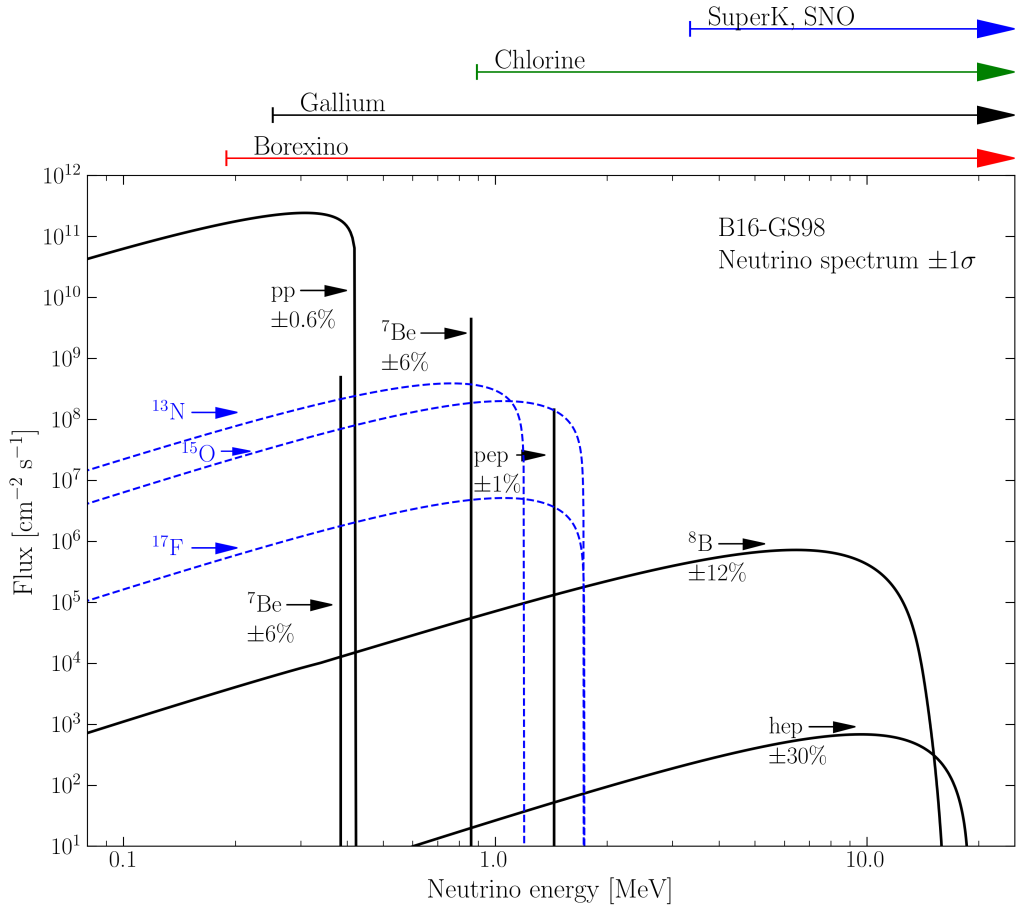}
%}
\end{center}
\caption{The energy spectrum of neutrinos predicted by a standard model
of the present Sun.
The neutrino fluxes from continuous sources are given in
units of $\cm^{-2} \s^{-1} \MeV^{-1}$
(despite the ordinate label) at one astronomical unit;
the line fluxes are in units of $\cm^{-2} \s^{-1}$.
The spectra from the PP chains are shown with continuous lines:
`pp' refers to the reaction
$\hyd(\hyd, \eplus \nu_{\rm e})\,\deut$,
`$\,\berseven$' to the reaction
$\berseven(\eminus, \nu_{\rm e})\,\litseven$,
and `$\,\boreight$' to the reaction
$\boreight(\eplus \nu_{\rm e})\,\bereight$.
In addition, two reactions are included which are
of no importance to the energy generation but 
of some significance to neutrino detections:
`pep' refers to the reaction
$\hyd(\hyd \,\, \eminus, \nu_{\rm e})\,\deut$,
and `hep' to the reaction
$\helthree(\hyd, \nu_{\rm e})\,\helfour$.
The spectra from the CNO cycle are shown with dashed lines:
`$\,\nitthirteen$' refers to the reaction
$\nitthirteen(\eplus \nu_{\rm e})\,\carthirteen$,
`$\,\oxyfifteen$' to the reaction
$\oxyfifteen (\eplus \nu_{\rm e})\,\nitfifteen$, and
`$\,\floseventeen$' to the reaction
$\floseventeen(\eplus \nu_{\rm e})\,\oxyseventeen$.
%The shadings and the bars at the top of the figure indicate the ranges
%of sensitivity of the different techniques for detecting the neutrinos.
%The figure was obtained from John Bahcall's homepage,
%at \url{http://www.sns.ias.edu/$\sim$jnb/}.
%(See \citecl{Bahcal2006a}.)
The neutrino spectra are based on Model B16-GS98, using the 
\citet{Greves1998} composition, from \citet{Vinyol2017}.
The arrows at the top schematically indicate the sensitivity ranges of the
various neutrino experiments (see text).
(Figure courtesy of A. {\rv Serenelli}.)
}
\clabel{fig:neutspect}
\end{figure}

The detection of solar neutrinos depends critically on the detailed
neutrino spectrum.
An example, computed for a standard solar model and referring to observations
at one astronomical unit, is shown in Fig.~\ref{fig:neutspect}.
Reactions involving $\eplus$ decay have continuous spectra,
reflecting the sharing of energy between the neutrino and the positron,
whereas the reactions involving $\eminus$ capture are characterized by
line spectra.
The spectrum is evidently dominated by the neutrinos 
from the $\hyd(\hyd, \eplus \nu_{\rm e})\,\deut$ reaction, which, however,
have a maximum energy of only $0.42 \MeV$.
In contrast, neutrinos from the
$\boreight(\eplus \nu_{\rm e})\,\bereight$ (a part of the PP-III chain;
{\cf} Eq.~\ref{eq:PPII}) and the
$\hyd(\hyd \,\, \eminus, \nu_{\rm e})\,\deut$ reaction
are relatively few in number
but have energies up to $15$ and $18.7 \MeV$, respectively.

%\notecd [UPDATE: Somewhere reference to sensitivity studies of neutrino fluxes
%and capture rates, including \citet{Bahcal2004b} and \citet{Bahcal2006a};
%also some Serenelli, no doubt.
%This is not the place, probably.]

%\notecd [Slightly more detailed discussion of various observation methods, 
%experimental results. For now, exclude SNO.]

\subsubsection{Problems with solar models?}
\clabel{sec:neutrprob}

The possibility of detecting high-energy neutrinos from the Sun was proposed
by \citet{Fowler1958}, following a revision of nuclear reaction rates
that indicated that the PP-III branch was more important than previously
thought, and further analysed by \citet{Bahcal1963}.
%\notecd [Need to check whether Bahcall did in fact suggest 37Cl.
%\citet{Bahcal1963} referred to an earlier Davis experiment with 37Cl.]
The first specific experiment was developed by
Raymond (Ray) Davis on the basis of the reaction
\be
\nu_{\rm e} + {}^{37}{\rm Cl} \rightarrow \eminus + {}^{37}{\rm Ar} \; 
\eel{eq:neutchlor}
\citep{Bahcal1964, Davis1964}.%
\footnote{A very personal presentation of Ray Davis's career
and the neutrino experiment was provided by \citet{Lande2009}.}
The detector consisted of a tank containing about $380\,000$ liter of
${\rm C}_2{\rm Cl}_4$ at a depth of $1480 \m$ in the 
Homestake mine in South Dakota;
the use of ${\rm C}_2{\rm Cl}_4$ provides a manageable way of handling the large
amount of chlorine, and the location helps reducing the background
from cosmic rays.
A discussion of the developments leading to this experiment and its
results has been provided by \citet{Davis2003}.
The reaction \Eq{eq:neutchlor} on average takes place 15 times a month 
in the tank;
the experiment is typically run for two months after which the argon produced
is flushed from the tank with helium and counted, utilizing the fact that
${}^{37}{\rm Ar}$ is radioactive, with a half-life of 35 days.
The neutrino flux is conventionally measured in units of Solar Neutrino
Units (SNU): $1 \SNU$ corresponds to $10^{-36}$ reactions per second
per target nuclei (in this case ${}^{37}{\rm Cl}$).
The initial results of the experiment \citep{Davis1968} found an upper limit
to the flux of $3 \SNU$, while the then predicted flux for a `standard'
model of the time \citep{Bahcal1968a}
was around 20 SNU \citep[{\eg},][]{Bahcal1968b}.
This was immediately recognized as a potentially serious problem for
our understanding of solar structure and energy generation;
an early review of the experimental and theoretical situation was
given by \citet{Bahcal1972}.
Despite continuing measurements and refinements of the modelling this
discrepancy persisted:
the final % \notecd [see \citet{Haxton2013}] 
average measured value is
$2.56\pm 0.16\,{\rm (statistical)}\,\pm 0.16\,{\rm (systematic)} \SNU$
\citep{Clevel1998}, 
while typical model predictions are around $8 \SNU$
\citep[{\eg},][]{Bahcal2001, Turck2001}.
The Homestake experiment has now ended.

\begin{figure}[htp]
%\def\epsfsize#1#2{0.6#1}
%\centerline{\epsfbox{\fig/as_theoryvsexpall_gs98.eps}}
%\centerline{\includegraphics[width=\textwidth]{\fig/as_theoryvsexpall_gs98.eps}}
\centerline{\includegraphics[width=\figwidthb]{\fig/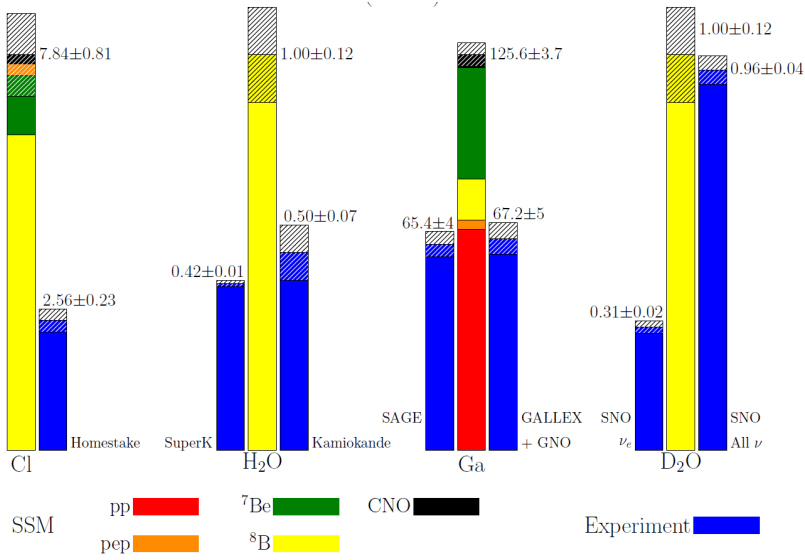}}
%\centerline{\epsfbox{\fig/colortheoryvsexp_snor.ps}}
%\centerline{\rotatebox{-90}{\includegraphics[scale=0.5]
%{\fig/colortheoryvsexp_05.ps}}}
\caption{Observed and computed neutrino capture rates, for a range of
neutrino experiments. 
In all cases the hatched regions indicate the $1\,\sigma$ uncertainties.
The dark blue bars show the observed values, in SNU for the
Cl and Ga experiments, and in terms of the ${}^8{\rm B}$ flux,
relative to the computed value of $5.46 \times 10^6 \cm^{-2} \s^{-1}$,
for the KamiokaNDE, SuperKamiokaNDE (SuperK) and SNO experiments.
The other bars show the computed values, the dominant contributions
being colour coded as indicated.
Theoretical results are for the so-called model B16-GS98 \citep{Vinyol2017}
For the observed values, see text.
%\notecd [Could need a little more detail].
Figure courtesy of A. Serenelli.
%  See http://www.sns.ias.edu/~jnb/ -> Solar Neutrinos -> viewgraphs
}
\clabel{fig:neutobs}
\end{figure}

An overview of this and other neutrino experiments, discussed in the following,
is provided by Fig.~\ref{fig:neutobs}.
The reaction \Eq{eq:neutchlor} is sensitive only to neutrinos with
energies exceeding $0.81 \MeV$ and, as indicated in the figure,
the predicted rate is dominated by $\boreight$ neutrinos.
Thus it provides no information about the neutrinos from the basic
$\hyd + \hyd$ reaction.

A second type of neutrino experiment uses neutrino scattering on 
electrons in water, causing {\v C}erenkov light from the resulting
energetic electrons.
Since the electrons are predominantly scattered in the forward direction
the detection is sensitive to the direction of the incoming neutrinos,
effectively producing a `neutrino image' of the Sun, 
albeit with low resolution.
Great care is taken to purify the water in the detector and shield it from
background radiation, including active background detection in a surrounding
volume of water.
Even so, owing to the dominant background at lower energies, these
experiments are limited to neutrino energies above a few MeV.
{\ie}, to the $\boreight$ and hep neutrinos.
Such experiments were initiated by Masatoshi Koshiba in Japan
\citep[see][]{Koshib2003}.
Early experiments carried out with the KamiokaNDE detector%
\footnote{The `NDE' in the name refers to `Nucleon Decay Experiment';
the experiment was originally designed to detect a possible 
proton decay.}
in the Kamioka Mine in the Japanese Alps,
with a detector volume of 2140 tons of water, found
a neutrino flux of less than half the predicted value \citep{Hirata1989};
the experiment also confirmed that the neutrinos
originated from the direction of the Sun.
The experiment was upgraded to Super-KamiokaNDE, with an inner detector
volume of $32\,000$ tons of water.
\citet{Fukuda2001} reported a measured flux of $\boreight$ neutrinos,
based on detection of around $18\,000$ neutrino events,
of $2.32 \pm 0.09 \times 10^6 \cm^{-2} \s^{-1}$,
{\ie}, 45 per cent of the value predicted by \citet{Bahcal2001}.
The most recent results, from the so-called Super-Kamiokande-IV phase
\citep{Abe2016}, yielded a measured flux of 
$2.31 \pm 0.05 \times 10^6 \cm^{-2} \s^{-1}$, and extending the
sensitivity down to neutrino energies of $3.5 \MeV$;
the resulting combined Super-Kamiokande result is
$2.35 \pm 0.04 \times 10^6 \cm^{-2} \s^{-1}$.
Interestingly, Abe {\etal} found a statistically significant 
day/night variation of around 4 per cent.

These measurements of solar neutrinos led to the award of the 2002 Nobel
prize in physics to Ray Davis and Masatoshi Koshiba.
% for establishing,
%respectively, the ${}^{37}{\rm Cl}$ and the Kamiokande experiments
\citep{Davis2003, Koshib2003}.
They shared the prize with Riccardo Giacconi, who got his part of the
prize for work in X-ray astronomy.

Detection of the neutrinos from the $\hyd + \hyd$ reaction can be made
with the reaction
\be
\nu_{\rm e} + {}^{71}{\rm Ga} \rightarrow \eminus + {}^{71}{\rm Ge} \; ,
\eel{eq:neutgal}
which is sensitive to neutrinos with energies exceeding $0.23 \MeV$.
The germanium must be extracted through chemical processing and counted.
Two independent experiments were established to use this technique:
the GALLEX experiment at the Laboratori Nazionali del Gran Sasso,
located under the Gran Sasso mountain, Italy,%
\footnote{The laboratory is very conveniently build adjacent to a 
motorway tunnel crossing the mountains, below around 1400\,m of rocks.}
and the SAGE%
\footnote{The acronym stands for {\bf S}oviet-{\bf A}merican
{\bf G}allium solar neutrino {\bf E}xperiment,
reflecting the period when the experiment was initiated.}
experiment at the Baksan Neutrino Observatory in Northern Kaukasus, Russia.
%\notecd [Would not hurt to get up to date on the status of these
%experiments].
The GALLEX experiment, using 30 tons of gallium, 
made the first detection of pp neutrinos \citep{Anselm1992},
at a capture rate of around $80 \SNU$.
This was confirmed by the SAGE experiment which in its full configuration
used 57 tons of gallium \citep{Abdura1994}.
This rate is essentially consistent with the flux of pp neutrinos 
(see Fig.~\ref{fig:neutobs}) but leaves
no room from contributions from neutrinos from the remaining reactions,
which would lead to a total predicted flux of around $130 \SNU$,
as illustrated in Fig.~\ref{fig:neutobs}.
The final result from GALLEX, based on data between 1992 and 1997,
was a capture rate of
$73.4\pm 6.0\,{\rm (statistical)}\,\pm 3.9\,{\rm (systematic)} \SNU$
\citep{Hampel1999, Kaethe2010};
the project continued under the name of GNO during the period
1998 -- 2003, with a combined Gallex/GNO rate of
$69.3\pm 4.1\,{\rm (statistical)}\,\pm 3.6\,{\rm (systematic)} \SNU$
\citep{Pandol2004, Altman2005}.
For SAGE a capture rate of
$65.4\pm 3.0\,{\rm (statistical)}\,\pm 2.7\,{\rm (systematic)} \SNU$
was found \citep{Abdura2009};
they also showed that the combined results of GALLEX, GNO and SAGE yielded
$66.1 \pm 3.1 \SNU$.%
\footnote{SAGE is still (2018) operating. 
At the 5th International Solar Neutrino Conference, TU-Dresden, June 2018,
V. Gavrin quoted the latest result of SAGE as $64.5^{+2.4}_{-2.3} \SNU$.}
% Fixed 3/2/18

%\notecd [Note that helioseismology all but excludes astrophysical solution
%to neutrino problem.]

The original discrepancy between the neutrino measurements and the model
predictions immediately led to attempts to modify solar models so as
to reduce the neutrino flux.
In most cases, this was done under the constraint that the total
solar nuclear energy generation rate was kept unchanged.
The initial detection with the ${}^{37}{\rm Cl}$ experiment was
predominantly sensitive to the $\boreight$ reactions.
Thus the predicted capture rate depended strongly on the branching
ratios between the PP-II and PP-I, and the PP-III and PP-II, chains
({\cf} Eqs \ref{eq:PPI} and \ref{eq:PPII}),
which in turn are very sensitive to temperature; 
at fixed nuclear luminosity the flux of $\boreight$ neutrinos scales
roughly as $T_{\rm c}^{18}$, where $T_{\rm c}$ is the central
temperature of the model.
% Exponent from \citet{Bahcal1989}, p. 151.
\citet{Sears1964} had already noticed a close relation between
the composition and the $\boreight$ neutrino flux:
decreasing the heavy-element abundance and hence, to maintain
the calibrated luminosity ({\cf} Eq.~\ref{eq:homlum}),
decreasing the helium abundance and hence the mean molecular weight,
reduced the central temperature and hence the neutrino flux.
Following the initial measurements, \citet{Iben1968, Iben1969}
made an extensive analysis of this sensitivity and concluded that
matching the observed upper limit would require an initial
solar helium abundance $Y_0$ of less than around 0.2;
Iben concluded that this would be inconsistent with the Galactic
helium abundance inferred from other objects, as well as with 
early estimates of the Big Bang helium production 
\citep[{\eg},][]{Peeble1966}.

Other attempts to reduce the capture rate through 
reducing the core temperature of the model were considered.
One possibility was substantial mixing of the core;
this would increase the central hydrogen abundance and hence allow
energy generation to take place at the required rate at a lower
temperature \citep[{\eg},][]{Bahcal1968c, Ezer1968}.
\citet{Dilke1972} proposed that recent core mixing, in what they called
`the solar spoon', might have reduced the nuclear energy generation
rate over a period of a few million years, such that the present 
neutrino capture rate would not be typical of a solar model in equilibrium;
owing to the solar thermal timescale of several million years such a
lack of equilibrium would not have immediately observable effects.
The mixing was supposed to have been initiated through instability to
oscillations \citep{Christ1974, Boury1975}.
An alternative mechanism to reduce the core temperature was to 
postulate a rapidly rotating core \citep{Barten1973, Demarq1973};%
\footnote{An earlier calculation by \citet{Ulrich1969} failed to
find a significant effect of even a very rapid rotation of the
central $0.3 \Msun$.}
this would reduce the gas pressure in the core required for hydrostatic
balance and hence the temperature, potentially leading to models
in agreement with the observed neutrino capture rate.
A reduction in $T_{\rm c}$ could also be accomplished by increasing
the efficiency of radiative energy transport in the radiative interior
or providing other, non-radiative, contributions to energy
transport \citep[{\eg},][]{Newman1976} and hence
decreasing the temperature gradient;
this was accomplished in 
the models of \citet{Iben1969} through a reduced heavy-element abundance
and hence reduced opacity.
\citet{Joss1974} proposed that this could be achieved,
maintaining the observed solar surface composition,
if the solar surface
had been contaminated by infalling material rich in heavy elements;
in that case the solar interior might have a much lower $Z$ and hence
a lower opacity.
The idea of stellar pollution was revived in connection
with the possibly detected high content of heavy elements in stars that
host planetary systems;
this could be the result of the accretion by the star of planets 
rich in heavy elements, which have migrated towards the star
\citep{Murray2002, Bazot2005}.
Also, as discussed in Sect.~\ref{sec:massloss}, 
accretion of metal-poor material has been invoked in the solar case
to account for the discrepancy between the present observed solar surface
abundance and helioseismic inferences of solar structure.
A more extreme proposal invoked the presence of the so-called
weakly interacting massive particles (`WIMPs').
Such particles had been proposed to account for
the `missing mass', {\eg}, in clusters of galaxies and galactic halos
\citep{Steigm1978, Steigm1985}; 
in fact, there is strong evidence that such non-baryonic dark matter 
dominates the matter content of the Universe 
\cite[see][for a review]{Sumner2002}.
If present in the solar interior they could contribute to the 
energy transport and hence reduce the temperature gradient required
for radiative transport \citep{Faulkn1985, Sperge1985, Gillil1986}.
This initially appeared to have some support from helioseismology,
models with WIMPs yielding improved agreement with early observations
of solar oscillation frequencies \citep{Dappen1986b, Faulkn1986};
however, improved observations \citep[{\eg},][]{Gelly1988}
and improved modelling \citep[{\eg},][]{Christ1992a}
have shown that this apparent agreement was in fact spurious.%
\footnote{However, studies of solar neutrinos and helioseismology may
still provide constraints on the subtler properties of dark matter
\citep[{\eg},][]{Turck2012, Lopes2014}.}

Given the improvements in the precision and extent of the solar oscillation
measurements, it became increasingly difficult to imagine 
that such modified solar models
could be found which were consistent both with the helioseismic inferences
and with the neutrino capture rate.
\citet{Elswor1990} pointed out that the measurements of 
the small frequency separations between low-degree modes, 
sensitive to the properties of the solar core
({\cf} Eq.~\ref{eq:smlsep}),
were consistent with normal solar models but inconsistent with
models proposed to reduce the neutrino flux
\citep[see also][]{Christ1991c}.
\citet{Dziemb1990} obtained lower limits on the solar neutrino flux in
models consistent with the results of helioseismic inversion and demonstrated
that these were inconsistent with the measured neutrino rates.
Admittedly, the helioseismic results are sensitive mainly to
the sound speed and not directly to the temperature upon which
the neutrino flux predominantly depends.
Thus, assuming the ideal-gas approximation (Eq.~\ref{eq:soundspeed})
helioseismology constrains $T/\mu$ but not $T$ and $\mu$ separately.
Even so, given the very small difference between the solar and
standard-model sound speed illustrated in Fig.~\ref{fig:csqinv},
a remarkable degree of fine tuning would be required to reduce
the temperature sufficiently to bring the neutrino predictions
in line with observations, while keeping the sound speed in accordance
with helioseismology \citep[{\eg},][]{Bahcal1997}.
Also, models modified to eliminate the remaining differences
between the model and the solar sound speed
produce neutrino fluxes very similar to those of standard models
\citep{Turck2001, Couvid2003}.
Thus the evidence was very strong that the structure of solar
models was basically correct, and that the solution to
the neutrino problem had to be found elsewhere.
It should be noted that this conclusion was also reached by,
for example, \citet{Castel1997} on the basis of analysis of apparent
inconsistencies between the results of the different neutrino experiments
which could not be resolved through modifications to the solar model.

%\notecd [Discuss introduction of neutrino oscillations, as a potential solution.
%Physical results on neutrino oscillations (need check for update).
%(Although order here is not quite obvious.)]
%
%\notecd [\citet{Patern1981} as early suggestion for physical solution.]
%
%\notecd [Fairly extensively on SNO results, now apparent consistency between
%observations and models.
%Usual spiel on the importance of interaction between physics and
%astrophysics in this test; moving beyond the standard (particle) model, etc.]

\subsubsection{Revision of neutrino physics: neutrino oscillations}
\clabel{sec:neutrosc}

Solutions to the neutrino discrepancy involving neutrino physics
were considered very early.
These are based on the existence of three different types, or
\emph{flavours}, of neutrinos: in addition to the electron neutrino
($\nu_{\rm e}$) produced in nuclear reactions in the Sun, 
muon ($\nu_\mu$) and tau ($\nu_\tau$) neutrinos also exist.
Although, in the Standard Model of particle physics, neutrinos are massless,
non-zero neutrino masses are possible in extensions of the model.
\citet{Pontec1967} and \citet{Gribov1969} noted that
in this case the three mass eigenstates of the neutrinos, which control their
propagation, would differ from the flavour eigenstates, causing an 
oscillation between the flavour states as the neutrinos propagate in vacuum.
%under certain
%assumptions about the properties of the neutrinos a transition is possible
%between the electron neutrinos and neutrinos of the other types.
If an appropriate fraction of the electron neutrinos were to be converted
into the other types, to which the ${}^{\rm 37}{\rm Cl}$ experiment is
not sensitive, the initial apparently anomalously low detection rate might be
explained.
A more detailed calculation of this effect, taking the neutrino spectrum
into account, was carried out by \citet{Bahcal1969b}.
Interestingly, in a brief note \citet{Patern1981} pointed out
that even the limited helioseismic data at that time provided support for
such a mechanism.
%Although such transitions can take place in vacuum, a more robust
%version of the process involves interactions with the electrons
In addition to the vacuum oscillations, transitions between the neutrino
flavours are mediated by the weak interaction between the neutrinos and the
electrons in solar matter \citep{Wolfen1978, Mikhey1985};
this is known as the \emph{MSW effect}.
The neutrino oscillations require that at least some of the neutrinos
have mass, the mass of $\nu_{\rm e}$ differing from that of the other types.
The transition rate depends on differences such as 
$\Delta m_{12}^2 = m_2^2 - m_1^2$
between the squared masses of the interacting neutrinos and
the so-called mixing angles, {\eg}, $\theta_{12}$.
As a result of the interaction with solar matter,
the \emph{survival probability}, {\ie}, the fraction of $\nu_{\rm e}$ that
reach terrestrial detectors, depends on the neutrino energy.
A concise summary of neutrino oscillations was provided by \citet{Haxton2013},
while \citet{Gonzal2003} gave a detailed review of the physics 
of neutrino mixing.

It was found possible to choose neutrino parameters such that the predictions
of neutrino oscillations
were consistent with the neutrino observations from the 
${}^{37}{\rm Cl}$, ${}^{71}{\rm Ga}$ and electron scattering experiments
\citep[for an overview, see][]{Bahcal1998}.
Some independent evidence for neutrino oscillations, involving the 
muon neutrinos, had been obtained from measurements of
neutrinos produced in the Earth's atmosphere by reactions
involving cosmic rays \citep[{\eg},][]{Fukuda1998};
this lent credence to the effect as an explanation of the solar
neutrino deficit.

Decisive tests of the mechanism came from
the Sudbury Neutrino Observatory (SNO) 
in Canada \citep[see][]{Boger2000, McDona2016}.
SNO measured solar high-energy (${}^8{\rm B}$)
neutrinos through reactions with deuterium
($\deut$) in heavy water as well as through electron scattering.
Thus the following neutrino reactions take place in the detector:
\bea
\nu_{\rm e} + \deut \rightarrow \hyd + \hyd + \eminus & \qquad \hbox{\rm (CC)} 
\; , \nonumber \\
\label{eq:sno}
\nu_x + \deut \rightarrow \hyd + \neut + \nu_x & \qquad \hbox{\rm (NC)} 
\; , \\
\nu_x + \eminus \rightarrow \nu_x + \eminus & \qquad \hbox{\rm (ES)} 
\; , \nonumber 
\eeanl{}
where $\nu_x$ are neutrinos of any flavour.
As indicated, the charged current (CC) reactions are sensitive only to
the electron neutrinos, while the neutral current (NC) reactions
are sensitive to all neutrino flavours;
electron scattering (ES) is mainly sensitive to $\nu_{\rm e}$ but
also has some, if reduced, sensitivity to $\nu_\mu$ and $\nu_\tau$.
In all cases the occurrence of a reaction is measured through 
the emission of {\v C}erenkov light.
In the case of electron scattering, as in the KamiokaNDE and 
SuperKamiokaNDE experiments, this is done for the electron on
which the neutrino scatters; this has a strong directionality around the
original direction of the neutrino.
In the case of the CC reactions the electron again is detected, although
with a different directional distribution.
Finally, for the NC interactions the neutrons are detected.
In the first phase of the experiment the neutrons reacted with $\deut$
to produce gamma-ray photon, which Compton scattered off electrons in the
water, resulting in emission of {\v C}erenkov light,
in this case essentially isotropically.
Thus from the angular distribution of the {\v C}erenkov light, as well as
from the energy spectra, the different reactions can be separated.
In the second phase the sensitivity was increased by dissolving NaCl 
in the heavy water,
the neutrons being detected 
through absorption in ${}^{35}{\rm Cl}$, gamma-ray emission,
Compton scattering on electrons and {\v C}erenkov-light emission. 
In the third and final phase the neutrons from the CC reactions were
detected by strings of proportional counters suspended in the heavy-water
container.

The initial analysis of the SNO results was based on comparing 
the rate measured with the charged-current reaction in 
Eq.~\Eq{eq:sno} with
the rate from previous electron-scattering KamiokaNDE and
SuperKamiokaNDE measurements
to deduce the number of $\nu_\mu$ and $\nu_\tau$, using the modest
sensitivity of the electron-scattering experiments to these flavours.
This provided a measure of the extent to which neutrino conversion
has taken place and therefore allowed an estimate of the original
neutrino production rate in the solar core.
The striking result was that the answer agreed, to within errors,
with the predictions of standard solar models \citep{Ahmad2001}.

The decisive demonstration of neutrino conversion was obtained
from measurements with SNO of the neutrino flux based on
the neutral-current reaction using neutron absorption in $\deut$
\citep{Ahmad2002},
which yielded a flux at the Earth of $\nu_{\rm e}$ from ${}^8 {\rm B}$ of
$(1.76 \pm 0.10)\times 10^6 \cm^{-2} \sec^{-1}$
and a flux of other neutrino types ($\nu_\mu$ and $\nu_\tau$) of
$(3.41 \pm 0.65)\times 10^6 \cm^{-2} \sec^{-1}$.
The total ${}^8 {\rm B}$ flux was found to be 
$(5.09 \pm 0.62)\times 10^6 \cm^{-2} \sec^{-1}$.
%\citep{Aharmi2007},
%with essentially consistent results from the salt-phase experiment
%\citep{Aharmi2005}.
which, as also indicated in Fig.~\ref{fig:neutobs},
is consistent with solar models.
The final combined results of the three phases of the SNO experiment
\citep{Aharmi2013}
yielded a total ${}^8 {\rm B}$ flux of
$(5.25 \pm 0.20)\times 10^6 \cm^{-2} \sec^{-1}$,
and a measured survival probability, at neutrino energy of $10 \MeV$,
of $0.317 \pm 0.018$,
giving a very strong confirmation of the presence of neutrino oscillations.
The 2015 Nobel Prize was awarded to Arthur B. McDonald 
for the detection of solar-neutrino oscillations \citep{McDona2016}.
He shared it with Takaaki Kajita who got the prize for the detection of
oscillations of muon neutrinos produced by cosmic-ray interactions in the 
upper atmosphere of the Earth \citep{Kajita2016}.
The heavy-water phase of the SNO experiment ended in 2006.
%  See press release at https://sno.phy.queensu.ca/sno/press_release/SNONewsReleaseNov15r.pdf

A broad range of neutrino results have been obtained over the last decade from 
the Borexino experiment \citep[][]{Alimon2009}.
This uses a 300-ton liquid scintillator for real-time
detection of solar neutrinos, established specifically to study the 
neutrinos resulting from the electron-capture decay of $\berseven$
({\cf} Eq.~\ref{eq:PPII});
furthermore, the background in the detector allows measurement 
of the ${}^8{\rm B}$~neutrinos down to an energy of $2.8 \MeV$.
This provides further constraints on the energy-dependence of
the oscillations between different neutrino flavours.
%An overview of the results from this experiment is provided in
%Figure~\ref{fig:neutborobs}.
\citet{Arpese2008} detected the signal from the $\berseven$ neutrino line
at 0.862~MeV,
and obtained a reduction relative to the model predictions which is
consistent with neutrino oscillations, given
the parameters determined from the earlier experiments.
Also, \citet{Bellin2010} considered the $\boreight$ spectrum at
energies at around 8.6 MeV;
comparing the results with the previous $\berseven$ results
demonstrated for the first time, using the same detector, an energy dependence 
of the reduction in the flux of $\nu_{\rm e}$ neutrinos 
that is consistent with
the matter-induced effects being important at the higher, and not the
lower, energy.

\begin{figure}[htp]
%\def\epsfsize#1#2{0.5#1}
%\centerline{\includegraphics[width=\figwidth]{\fig/ags05_opacity.eps}}
\centerline{\includegraphics[width=\figwidthb]{\fig/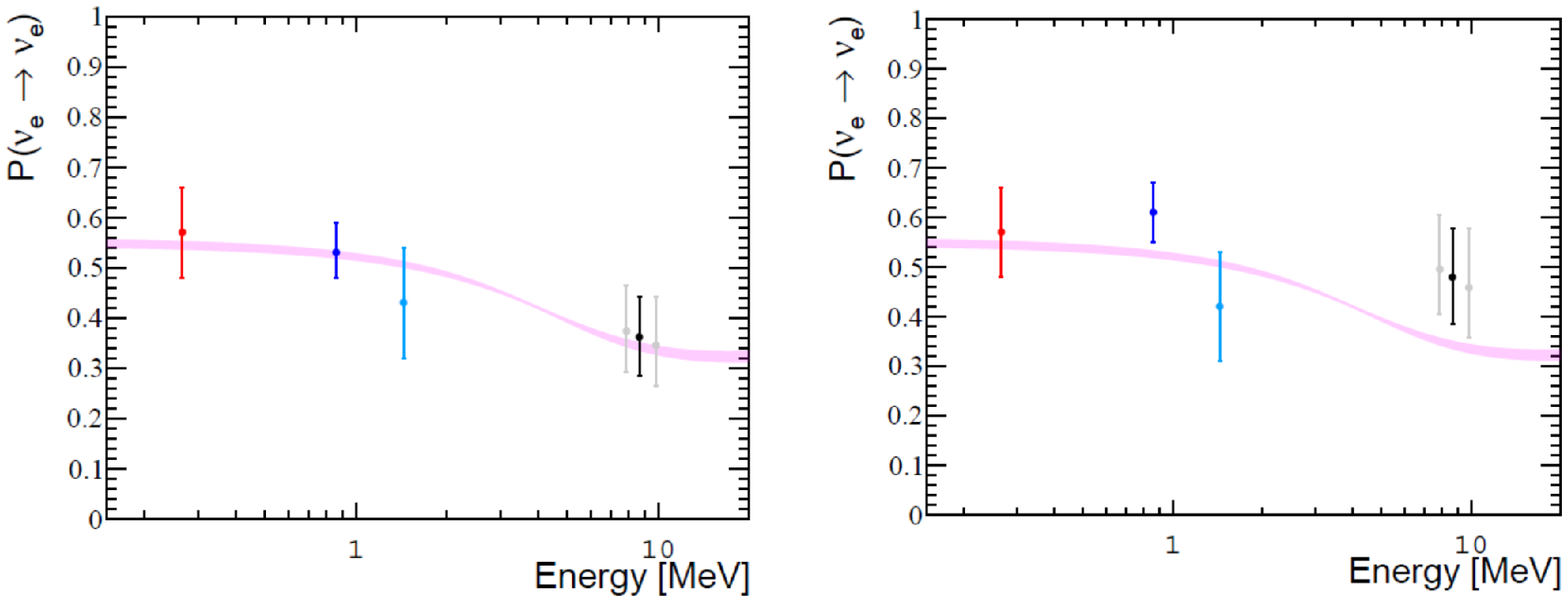}}
\caption{
Electron neutrino survival probability against neutrino energy,
based on comparing
Borexino measurements \citep{Agosti2020a} with solar models \citep{Vinyol2017}.
The red, blue and azure points show pp, ${}^7 {\rm Be}$ and pep
neutrinos \citep{Agosti2019}, and the black and grey points show
the combined and low-and high-energy results for ${}^8 {\rm B}$.
The model in the left panel used heavy-element abundances from
\citet{Greves1998}, while the model in the right panel is based on
the lower \citet{Asplun2009} abundances (see Sect.~\ref{sec:newcomp}).
The curves show computed survival probabilities based on 
neutrino-oscillation parameters from \citet{Esteba2017}.
\citep[From][]{Agosti2020a}.
}
\clabel{fig:neutsurv}
\end{figure}

\begin{figure}[htp]
\centerline{\includegraphics[width=\figwidthb]{\fig/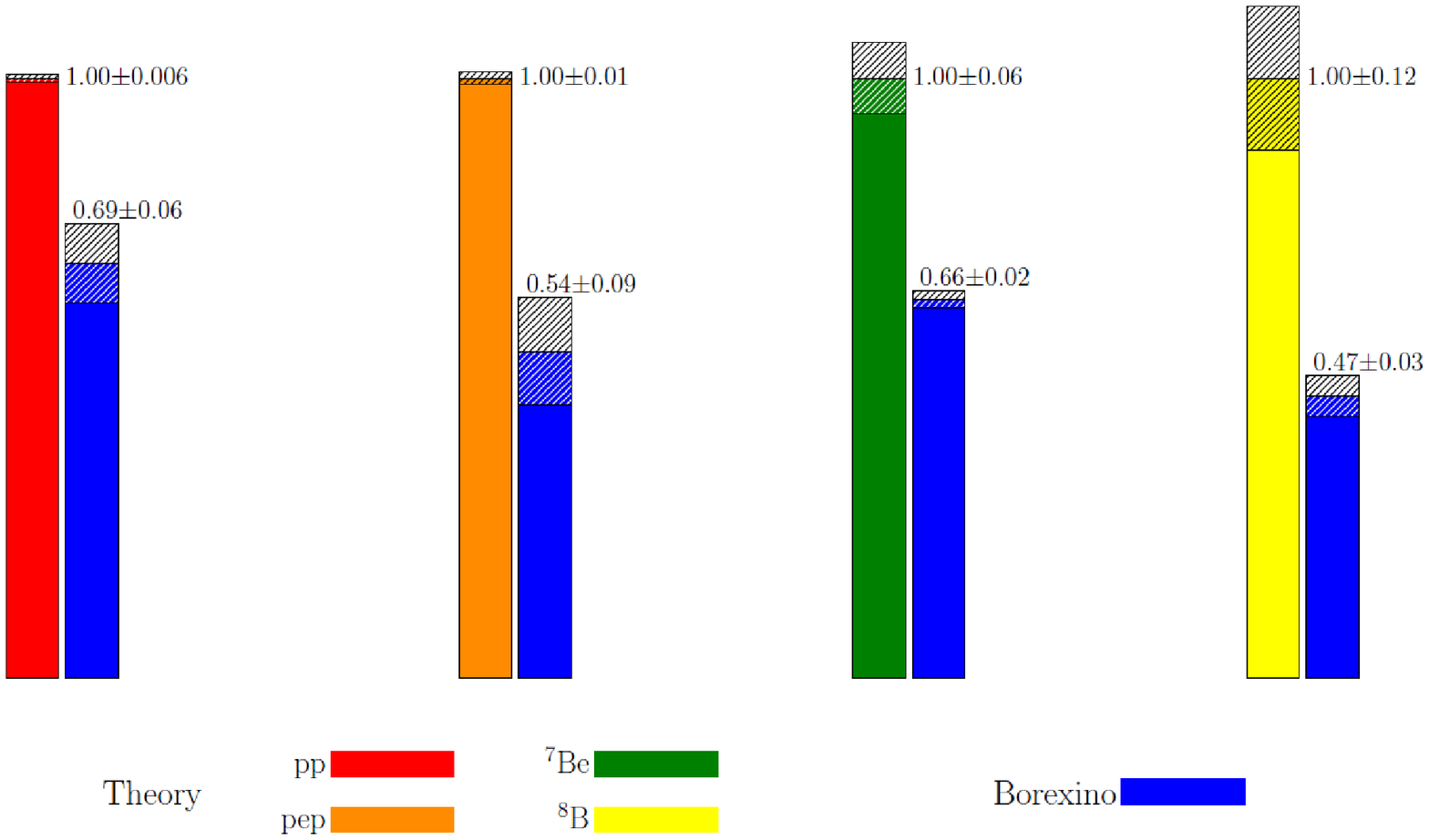}}
\caption{Observed and computed neutrino capture rates, for the Borexino
neutrino experiments \citep{Alimon2009}.
The observations, not corrected for neutrino oscillations,
were obtained from \citet{Agosti2018}.
The results are normalized by the computed values of the B16-GS98 model 
of \citet{Vinyol2017}:
$5.98 \times 10^{10} \cm^{-2} \s^{-1}$ for pp,
$1.44 \times 10^{8} \cm^{-2} \s^{-1}$ for pep,
$4.93 \times 10^{9} \cm^{-2} \s^{-1}$ for $\berseven$, and
$5.46 \times 10^{6} \cm^{-2} \s^{-1}$ for $\boreight$.
In all cases the hatched regions indicate the $1\,\sigma$ uncertainties.
%For the observed values, see text.
Figure courtesy of A. Serenelli and A. Ianni.
}
\clabel{fig:neutborobs}
\end{figure}

The sensitivity of the Borexino detector extends to energies much lower
than the energy cut-off for the pp neutrinos
({\cf} Fig.~\ref{fig:neutspect}).
Thus, after careful purification of the detector material 
\citet{Bellin2014} determined this basic flux of solar neutrinos,
taking neutrino oscillations into account,
to be $(6.6 \pm 0.7) \times 10^{10} \cm^{-2} \s^{-1}$, 
fully consistent with solar models. 
Also the $\nu_{\rm e}$ survival probability in this energy range
was found to be $0.64 \pm 0.12$.
\citet{Agosti2019} determined the fluxes of pp, pep and
${}^7 {\rm Be}$ neutrinos, whereas \citet{Agosti2020a} determined
the flux of ${}^8 {\rm B}$ neutrinos.
Combining these results with computed neutrino fluxes from solar models
provides a determination of the survival probability based on
data from a single experiment.
The results are illustrated in Fig.~\ref{fig:neutsurv} for two solar models,
compared with a prediction based on neutrino oscillations.
The results clearly follow the expected energy dependence fairly well, 
but with slight preference for the model with higher abundance $Z$ of
heavy elements (see also Sect.~\ref{sec:modrevcomp}).
The heavy-element abundance more directly affects the CNO neutrinos,
but {\rv \citet{Agosti2019} were only able to establish
an upper limit of a factor 1.5 - 2 higher than the model predictions}.
%\notecd [UPDATE after refereeing. See results from 2020 neutrino conference:
%res/serenelli/borexino\_cno\_neutrino2020.pdf.
%From
%https://indico.fnal.gov/event/43209/contributions/187871/attachments/129210/158592/borexino_cno_neutrino2020.pdf].

{\rv A combined analysis} of the Borexino results was presented by
\citet{Agosti2018} and is illustrated in Fig.~\ref{fig:neutborobs}.
Covering all reactions making substantial contributions to the
nuclear energy generation it also allowed an estimate of the total
solar nuclear luminosity, after taking flavour conversion into account. 
The result, $L_{\rm nucl} = (3.89^{+0.35}_{-0.42}) \times 10^{33} \erg \s^{-1}$,
is fully consistent with the observed solar luminosity and provides
a first demonstration of the instantaneous solar nuclear equilibrium
within a precision of 10 per cent.

{\rv Very recently \citet{Agosti2020b} announced a robust detection by
Borexino of CNO neutrinos, at a rate that this consistent with
both the low- and the high-metallicity models.}

In parallel with these efforts to study neutrino conversion
from solar observations, extensive terrestrial experiments have been
carried out, to obtain independent determinations of the neutrino-oscillation
parameters.
The KamLAND detector in Japan
\citep[{\eg},][]{Eguchi2003, Gando2011} measured the flux
of electron antineutrinos $\overline \nu_{\rm e}$ from
commercial nuclear reactors, with a clear signal of neutrino oscillations
which placed constraints on the oscillation parameters.
%Detailed analyses of data on solar neutrinos combined
%with the KamLAND results have been carried out
%%by \citet{Bahcal2004c} and 
%by \citet{Fogli2006}, to determine the
%parameters of neutrino oscillations;
%the latter analysis yielded 
%$\Delta m_{12}^2 = 7.92 \times 10^{-5} \eV^2$
%and $\sin^2 \theta_{12} = 0.31$ and strongly supported the significance of 
%matter effects in the interaction with electrons in the Sun, as in
%the MSW effect.
%
Other experiments have been developed
that direct beams of neutrinos from accelerators towards neutrino
detectors, over distances of several hundred kilometers.
A beam of muon antineutrinos $\bar \nu_\mu$ from the Fermilab accelerator
in Illinois was analysed
in the MINOS experiment \citep{Adamso2012}
with two detectors: a near detector one km from
the neutrino source and a far detector at the Soudan Underground Laboratory,
735 km away in Minnesota.
The OPERA experiment \citep{Agafon2015, Agafon2018} used a beam of neutrinos 
from CERN at Geneva to
search for conversions from muon to tau neutrinos at the Gran Sasso Laboratory,
730 km away.
In the NoVA experiment \citep[{\eg},][]{Adamso2016}
a beam of muon neutrinos was sent from the Fermilab accelerator
to a detector in Ash River, Minnesota, 810 km away.
The T2K experiment \citep[{\eg},][]{Abe2017} sends a neutrino beam from
the J-PARC accelerator at Tokai, Japan, to the SuperKamiokaNDE detector,
295 km away.
%  See http://t2k-experiment.org/t2k/
A review of such accelerator experiments was provided by \citet{Nakaya2016}.
However, the size of the Earth sets a natural limit to the
scale of terrestrial experiments.
Thus, observations of solar neutrinos remain
a very important possibility for studying 
the properties of the neutrino experimentally,
including the MSW effect and its consequences for the energy dependence
of the survival probability.
%provided that conditions in the solar core can be determined with
%sufficient accuracy, from observations of solar oscillations,
%that the Sun can be regarded as a well-calibrated neutrino source.
\citet{Bergst2016} analysed the solar and terrestrial neutrino data,
as a basis for a comparison with the predicted solar model results.
A comprehensive analysis of the available data, both solar and terrestrial,
was carried out by \citet{Esteba2017},
leading to the computed probability shown in Fig.~\ref{fig:neutsurv},
while \citet{Malton2016} discussed the importance of solar neutrinos in
investigations of neutrino physics and the resulting extensions 
beyond the Standard Model of particle physics.

With the improved understanding of the properties of the neutrinos,
and with further neutrino experiments,
we may increasingly use the observations of solar neutrinos as
constraints on the properties of the solar core, complementary to those
provided by helioseismology.
%An interesting example was provided by \citet{Gondol2009}.
An interesting example of such combined analysis of helioseismic and 
neutrino-based observations, to which I
return in Sect.~\ref{sec:modrevcomp}, was provided by \citet{Song2018}.
The present situation was summarized concisely and accurately
by \citet{Haxton2013}: 
``Effectively, the recent progress made on
neutrino mixing angles and mass differences has turned the neutrino 
into a well-understood probe of the Sun.
We now have two precise tools, helioseismology and neutrinos,
that can be used to see into the solar interior.
We have come full circle: The Homestake experiment was to have been 
a measurement of the solar core temperature, until the solar neutrino
problem intervened''.

\subsection{Abundances of light elements}

\clabel{sec:lightcomp}
The present solar surface abundances are the result of the composition of the
initial Sun as well as the result of processes which may have modified
the composition.
Thus in this sense they represent an archaeological record of solar
evolution.
Since the convective envelope is mixed on a timescale of a few months
its composition is uniform.
Thus the relevant aspects are the evolution of the composition beneath
the convection zone, as well as mixing processes which might link 
the composition in the deep solar interior to the solar surface.
Furthermore, substantial mass loss may expose material from deeper layers
at the surface (see also Sect.~\ref{sec:massloss}).
In this manner the surface composition provides a time integral over 
solar evolution of the processes in the solar interior.

For refractory elements, meteoritic abundances provide a measure of
the initial solar composition, {\eg}, measured relative to the abundance
of silicon which has presumably not been significantly affected 
by processes in the solar interior.
Very interesting cases are the light elements lithium, beryllium and boron
which are destroyed over the solar lifetime
by nuclear reactions at temperatures found in the solar interior.
Specifically, lithium is very substantially reduced over a period corresponding
to the solar age at temperatures above $2.5 \times 10^6 \K$,
while the corresponding critical temperatures for beryllium and boron are
$3.5 \times 10^6 \K$ and $5 \times 10^6 \K$, respectively.
Thus mixing down to these temperatures, or mass loss exposing material
that has been at such temperatures, should be reflected in reductions
in the abundances of these elements relative to the meteoritic values.

No significant depletion is found for boron \citep{Cunha1999}
or beryllium \citep{Balach1998, Asplun2004a},
limiting mixing to extend at most to temperatures 
less than $3.5 \times 10^6 \K$.
However, 
{\rv as mentioned in Section~\ref{sec:basicpar},}
the present solar surface abundance of lithium has been reduced
by a factor of around 150 \citep{Asplun2009}
relative to the meteoritic value, indicating mixing to temperatures exceeding
$2.5 \times 10^6 \K$. 
As noted by \citet{Schatz1969} this is substantially higher than the
temperature $T_{\rm bc}$ at the base of the solar convection zone,
during solar evolution.%
\footnote{In the case of Model~S \citep[][see Sect.~\ref{sec:models}]
{Christ1996}, for example,
$T_{\rm bc}$ decreases from $2.45 \times 10^6 \K$
to $2.2 \times 10^6 \K$ during evolution from the zero-age main sequence
to the present.}
Thus additional mixing, or mass loss, is required to account for the
lithium depletion.
{\rv It should be noted, however, that possible depletion 
in pre-main-sequence evolution, 
including a likely fully mixed convective phase, 
must be taken into account in using the lithium abundance as a diagnostic.
Theoretical \citep[{\eg}][]{Piau2002} and observational
\citep[{\eg}][]{Bouvie2016} studies show that this
depletion can be substantial, but strongly dependent on the details of
the evolution, including effects of rotation.
Also, the general uncertainty about the early evolution of stars
(see Section~\ref{sec:pmsevol}) must be taken into account.}

Additional constraints on mixing or mass-loss processes are provided
by the ratio $\helthree/\helfour$ between the abundances by
number of $\helthree$ and $\helfour$.
This has been measured in the solar wind and in lunar material, as
deposited from the solar wind.
As shown in Fig.~\ref{fig:he3} the nuclear reactions in the PP chains
cause a build-up of the $\helthree$ abundance with solar evolution.
This does not extend to the base of the convection zone;
however, mixing extending substantially deeper (or corresponding mass loss)
would evidently cause an increase in the isotope ratio at the solar surface
and hence in the solar wind.

The observational evidence was discussed by \citet{Bochsl1990}. 
They noted that the initial $\deut$ in the Sun has been converted
to $\helthree$ through the second reaction in the PP-I chain
({\cf} Eq.~\ref{eq:PPI}) which takes place at temperatures
as low as those found in the present solar convection zone.
% See Lebreton \& Maeder A\&A 175, 99.
From estimates of the primordial solar-system content of $\deut$ and
$\helthree$ they consequently estimated the initial ratio $\helthree/\helfour$
in the Sun as around $4.4 \times 10^{-4}$.
From solar-wind measurements, either from satellites or from foils exposed
in the Apollo missions, they found very similar values at present;
also, analyses of lunar material indicate that the ratio has not varied
much over the last few billion years \citep{Heber2003}.
An investigation of the composition of the solar wind
with the {\it Genesis} spacecraft yielded a value of 
$\helthree/\helfour = (4.64 \pm 0.09) \times 10^{-4}$ \citep{Heber2009},
consistent with the earlier results.%
\footnote{For completeness I note that \citet{Geiss1998} found 
that a correction is required to relate the solar-wind ratio to
the ratio at the solar surface; the magnitude is rather small, however,
compared with the uncertainty in the determination.}
The general conclusion, therefore, is that there has been little if any
enrichment of the solar convection zone with $\helthree$ 
during solar evolution.

Models including appropriately varying enhancements of the
diffusion coefficient assumed to be caused by turbulence,
can indeed account for the observed lithium depletion
\citep[{\eg},][]{Vaucla1978, Schatz1981, Lebret1987}.
In the latter two cases the $\helthree/\helfour$ ratio was also considered,
yielding a modest increase during solar evolution which may be inconsistent
with the present observational situation.
\citet{Christ1992b} presented a detailed analysis of simpler models,
assuming rapid mixing over a region below the convection zone and
taking into account the variation with time of the extent of the mixed region.
They found that to a good approximation the typical lithium-destruction
timescale, averaged over the mixed region and over solar age, 
could be approximated as twice the timescale at the base of the mixed
region in the present Sun.

The helioseismic investigations have provided further information
about conditions at the base of the convection zone.
As discussed in Sect.~\ref{sec:heliostruc}, 
the localized difference between the solar and model sound speed beneath
the convection zone (see Fig.~\ref{fig:csqinv}) may indicate
that the gradient in the hydrogen abundance,
caused by helium settling, is too steep in this part of the models,
suggesting the need for additional mixing.
Also, the sharp gradient in the angular velocity in the tachocline
({\cf} Sect.~\ref{sec:heliorot}) could give rise to 
dynamical processes leading to such mixing.
\citet{Richar1996} considered rotationally induced turbulent mixing,
following the description of \citet{Zahn1992}.
They obtained a reasonable sound-speed profile below the convection zone,
as well as the observed lithium depletion and the then assumed 
depletion of beryllium by a factor of two.
In a similar analysis, \citet{Brun1999} obtained a smoothed
sound-speed difference relative to the helioseismic results, 
together with the required lithium depletion, no depletion of
beryllium, and a $\helthree/\helfour$ ratio consistent with the 
inferred values.
Lithium destruction was also considered in the magnetically dominated
model by \citet{Gough1998} of the origin of the tachocline.
Clearly any modelling of these effects should aim for simultaneously reducing
the sound-speed difference just below the convection zone and obtain the
observed surface lithium abundance.
A recent analysis by \citet{Jorgen2018b} based on various forms of
imposed turbulent diffusion assumed to arise from convective overshoot
suggests that this may not be straightforward.
{\rv In Section~\ref{sec:massloss} I return to this issue, based on complex
modelling by \citet{Zhang2019} including early accretion, mass loss and
turbulent mixing.}

It is evident that the diagnostics of the solar internal structure provided
by these abundance determinations is less precise that those obtained
from helioseismology.
However, they must be kept in mind as constraints on any solar models.
In particular, they provide integral measures of the dynamics in the
solar interior over the solar lifetime, which is clearly closely 
related to the evolution of the solar internal rotation.
More generally, the observed dependence on stellar parameters
of lithium depletion in solar-like stars is an important diagnostics
of these processes, and the solar results must be understood in this
context \citep[see, for example][and Sect.~\ref{sec:stars}]{Charbl2005}.
Asteroseismic information about the internal rotation of stars is
extremely important in this connection.

%% \newpage

%===========================================================================

\section{The solar abundance problem}
\clabel{sec:abundprob}

%\notecd [At least need to discuss models with massloss, particularly Boothroy, 
%Sackmann detailed analysis. Interesting that this is not obviously
%inconsistent with helioseismology (and rather a nuisance).]
%
%\notecd [Are there more? New composition is hardly non-standard, by now.
%But we treat it as such, none the less, for now.]
%
The models presented in the preceding sections can be regarded as `classical' 
solar models of the late 20th century; 
they have been computed using well-established physics, including 
diffusion and settling, and are based on the observed parameters of the epoch.
Interestingly, as discussed in Sect.~\ref{sec:test}, they are 
in reasonable if not full agreement with the helioseismic inferences and
with the latest neutrino detections, taking into account flavour transitions.
In this sense it is perhaps reasonable to regard them as `standard' solar 
models.

Even so, the models obviously can, and should, be questioned.
The remaining differences in structure and physics between the Sun 
and the model, discussed in Sect.~\ref{sec:heliostruc}, 
obviously need to be understood.
More seriously, since around 2000 new determinations of the 
solar surface composition have led to substantial discrepancies between
the resulting solar models and the helioseismic results, forcing us to
reconsider the computation of solar models.
This is discussed below.
Indeed, we obviously need to question the simplified assumptions 
underlying the `standard' model computation.
One remaining serious uncertainty of potentially important consequences
for solar evolution is the treatment of the loss and redistribution of
angular momentum, briefly discussed in Sect.~\ref{sec:heliorot}.
Below I address a second issue, namely the assumption of no significant
mass loss during solar evolution, which was considered by
\citet{Sackma2003} in connection with the `faint early Sun' problem.

\subsection{Revisions to the inferred solar composition}

\clabel{sec:newcomp}
The `classical' modelling of the solar atmosphere in terms of a
static horizontally homogeneous layer is obviously oversimplified,
given the highly inhomogeneous and dynamic nature of the atmosphere.
This affects the profiles of the solar spectral lines and hence 
the determination of solar abundances.
In such analyses a semi-empirical mean structure of the atmosphere is
often used, based on observed properties such as the limb-darkening
function, {\ie}, the variation in intensity with position on the solar disk;
typical examples are \citet{Holweg1974, Vernaz1981}.
The dynamical aspects of the atmosphere are represented in terms of
parameterized `micro- and macro-turbulence', adjusted to match the observed
line profiles.
A second important issue are the departures from local thermodynamical
equilibrium (LTE) in the population of the
different states of ionization and excitation in the atoms in the atmosphere.
Proper treatment of such non-LTE (NLTE) effects requires detailed accounting
of the different radiative and collisional processes that affect the
population \citep[\eg,][]{Mihala1978}.

\begin{table}
\caption{\clabel{tab:abundances}
Selected solar photospheric abundances in terms of number densities,
on a logarithmic scale normalized such that the $\log N_{\rm H} = 12$,
where $N_{\rm H}$ is the number density of hydrogen and
$\log$ is to base 10; $\Zs/\Xs$ is the corresponding ratio between the
abundances by mass of heavy elements and hydrogen.
The following tabulations are included:
AG89: \citet{Anders1989}; GN93: \citet{Greves1993};
GS98: \citet{Greves1998}; AGS05: \citet{Asplunetal2005b}; 
AGSS09: \citet{Asplun2009}; and C11: \citet{Caffau2011}
(here elements not provided by Caffau {\etal}, indicated by italics,
were taken from \citet{Lodder2010}).
%\notecd [C11 supplemented from Lodders 2009].
}
\vskip 4mm

\centering
\begin{tabular}{l|cccccc}
\hline
& AG89 & GN93 & GS98 & AGS05 & AGSS09 & C11 \\
\hline
C   & 8.56 & 8.55 & 8.52 & 8.39  & 8.43 & 8.50 \\
N   & 8.05 & 7.97 & 7.92 & 7.78  & 7.83 & 7.86 \\
O   & 8.93 & 8.87 & 8.83 & 8.66  & 8.69 & 8.76 \\
Ne  & 8.09 & 8.07 & 8.08 & 7.84  & 7.93 & \emph{8.05} \\
Na  & 6.33 & 6.33 & 6.33 & 6.17  & 6.24 & \emph{6.29}  \\
Mg  & 7.58 & 7.58 & 7.58 & 7.53  & 7.60 & \emph{7.54} \\
Al  & 6.47 & 6.47 & 6.47 & 6.37  & 6.45 & \emph{6.46}   \\
Si  & 7.55 & 7.55 & 7.55 & 7.51  & 7.51 & \emph{7.53} \\
Fe  & 7.67 & 7.51 & 7.50 & 7.45  & 7.50 &  7.52 \\
\hline
$\Zs/\Xs$ & 0.0275 & 0.0245 & 0.0231 & 0.0165 & 0.0181 & 0.0209 \\
\hline
\end{tabular}
\end{table}

As discussed in Sect.~\ref{sec:convection},
hydrodynamical simulations of solar convection now yield a 
realistic representation of conditions in the uppermost parts of the
solar convection zone and in the solar atmosphere.
In particular, the spectral line profiles can be reproduced without the
use of additional parameters.
Application of these results to the determination of the solar 
abundance \citep{Allend2001, Allend2002, Asplun2004b, Asplunetal2005a}
provided increasingly strong evidence for a need to revise the
solar composition;
in particular, the inferred abundances of carbon, nitrogen and oxygen
were lower than previous determinations by more than 30 per cent,
resulting in $Z_{\rm s}/X_{\rm s} = 0.0165$ 
(compared, e.g., with the value 0.0245 obtained by \citet{Greves1993} and
used in Model~S).
%\notecd [Check when the term `Stagger code' was introduced].
Overviews of these initial results were provided by 
\citet{Asplun2005} and \citet{Asplunetal2005b} (in the following AGS05).

As reviewed in detail by \citet{Basu2008} and discussed extensively below
these changes in the composition assumed in solar modelling led to
substantial changes in the model structure and a drastic increase
in the helioseismically inferred difference between the Sun and the model
(see Fig.~\ref{fig:ags05csq}),
leading to questioning of the new abundance determinations.
For example, \citet{Ayres2006} criticized the atmospheric models obtained from
the hydrodynamical simulations, 
on the ground that they failed to match the observed centre--limb
variation over the solar disk in the continuum intensity;
a similar objection was raised by \citet{Pinson2009}.
%\footnote{However, the more recent simulations have corrected this
%\citep{Asplun2009, Pereir2013}.}
Also, Ayres {\etal} analysed weak CO features 
and obtained an abundance consistent with the old determinations.

Following this initial work, the Asplund {\etal} (AGS05) analysis was updated
by improved hydrodynamical models;
these did indeed, for the first time, succeed in reproducing 
the observed limb-darkening function, over a broad
range of wavelengths \citep{Pereir2013}. 
Furthermore, the analysis included careful consideration of NLTE effects,
whenever possible,
and of the choice of atomic input data and of spectral lines and
effects of line blending.
The resulting comprehensive composition results 
were presented by \citet{Asplun2009} (in the following AGSS09).
The revision led to a slight general increase in the abundances, although
still far from recovering the old values.
{\rv A recent update on these determinations was provided by \citet{Greves2019}.
}

%\notecd [Replace suitably by Lodders 2009, or return to this later].
Table~\ref{tab:abundances}
lists selected abundances from several determinations, 
including earlier results typically used in the computation 
of `standard' solar model;
I return to the \citet{Caffau2011} results below.
%It is clear that there has been a substantial reduction in the abundances
%of, in particular, oxygen, carbon and nitrogen and a corresponding reduction
%in the overall heavy-element abundance, reflected in $\Zs/\Xs$,
%although the effect is somewhat reduced in the latest results.

Interestingly, the revision to the solar abundances brings them more closely
in line with stars or other objects in the solar neighbourhood
\citep[{\eg},][]{Turck2004, Morel2009};
in contrast, the previous solar abundances tended to be substantially higher
than those of nearby hotter and therefore generally younger stars,
in conflict with the expectations of galactic chemical evolution.
This issue was further analysed by \citet{Nieva2012}, on the basis of
a characterization of the composition of matter in the present solar 
neighbourhood based on extensive observations of abundances of early B-type
stars.
They found that, even with the AGSS09 abundances, the Sun was substantially
over-abundant compared with the solar neighbourhood and concluded on this
basis that the Sun was formed in a region at a Galactocentric distance of
5 to 6\,kpc, where the heavy-element abundance was higher, and has 
subsequently migrated to its present distance of 8\,kpc.

Independent hydrodynamical modelling and abundance analysis is obviously
highly desirable.
\citet{Caffau2008} used the ${\rm CO^5BOLD}$ code%
\footnote{{\bf CO}nservative {\bf CO}de for the {\bf CO}mputation of
{\bf CO}mpressible {\bf CO}nvection in a {\bf BO}x of {\bf L} {\bf D}imensions,
L=2,3}
\citep{Freyta2002, Wedeme2004} to determine the oxygen abundance,
obtaining $8.76 \pm 0.07$ on the logarithmic scale used in 
Table~\ref{tab:abundances}.
It appears that the quite substantial increase relative to 
AGS05 in part was caused by a different assignment of the continuum
in the abundance analysis, resulting in increased equivalent widths
of the lines considered.
Also, \citet{Caffau2009} similarly determined the nitrogen abundance 
as $7.86 \pm 0.12$;
from these determinations they obtained $\Zs/\Xs = 0.0213$,
relatively close to the old determinations.
An overview of the results of these efforts are also included in 
Table~\ref{tab:abundances}, based on \citet{Caffau2011} and supplemented
by \citet{Lodder2010}.
A more careful comparison between the assumptions and 
results of these different abundance analyses is certainly needed,
to understand
the differences between these results and those of \citet{Asplun2009}.
Interestingly, a comparison carried out by \citet{Beeck2012} of different
hydrodynamical simulations of the solar near-surface layers,
including the so-called Stagger code \citep[{\eg}][]{Collet2011}%
\footnote{See also \url{https://starformation.hpc.ku.dk/?q=node/18}.
The code was originally described by Nordlund \& Galsgaard (1995),
in a document which unfortunately seems no longer to be available.}
that is closely related to the codes
used in the analyses by Asplund {\etal} and the ${\rm CO^5BOLD}$ code,
found good agreement between
the mean structure and turbulent behaviour between the codes.
This suggests that the differences between the AGSS09
and C11 compositions arise from more subtle aspects of the analysis including,
as also hinted above, the basic analysis of the observations.

The noble gases present particular problems for the abundance determination,
since they have no lines in the solar photospheric spectrum.
Particularly important is neon which, as shown in Fig.~\ref{fig:kapder},
makes a substantial contribution to the opacity.
Estimates of the abundances can be obtained from the solar wind or
solar energetic particles, or from lines formed in the higher layers of
the solar atmosphere, including the corona.
These determinations suffer from the uncertain effects of element separation
in the solar corona
which depends on the first ionization potential 
\citep[the so-called FIP effect, \eg,][]{Marsch1995, Laming2015}.
An alternative technique, used by AGSS09, is to determine the ratio, e.g.,
Ne/O between a noble gas and oxygen which may be expected to suffer 
approximately the same separation effect, and hence the neon abundance.
Recently \citet{Young2018} provided a re-assessment of data on the
transition region in the quiet Sun, on the basis of new atomic data,
to obtain a higher Ne/O ratio and a logarithmic neon abundance of 8.08.
Given the derivative in Fig.~\ref{fig:kapder}
this increase in the neon abundance of around 40 per~cent relative to
the assumed AGSS09 value would correspond
to an increase in the opacity of up to 5 per~cent, just below the
convection zone.

Detailed reviews on the solar and solar-system composition, 
%based on these and other sources and 
with emphasis on the abundance
determinations of refractory elements in meteorites,
were also given by \citet{Lodder2003, Lodder2010} and \citet{Lodder2009}.
Indeed, \citet{Vinyol2017} argued that the meteoritic abundances
of the refractory elements are likely more accurate, and certainly more
precise, than the photospheric abundances and hence should be used in
preference, when available.
In practice, the resulting differences for the elements listed in 
Table~\ref{tab:abundances} are very small.

\begin{figure}[htp]
%\def\epsfsize#1#2{0.5#1}
%\centerline{\includegraphics[width=\figwidth]{\fig/ags05_opacity.eps}}
\centerline{\includegraphics[width=\figwidth]{\fig/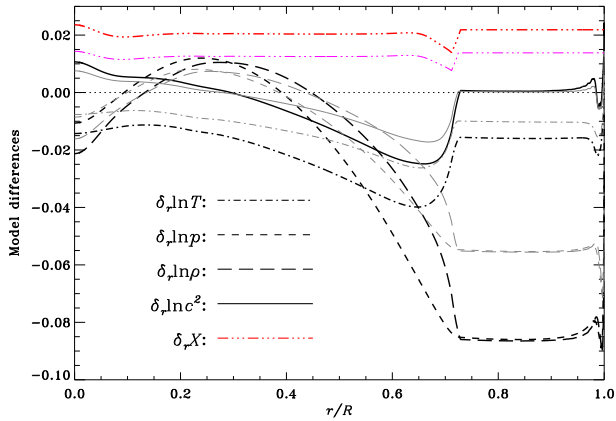}}
\caption{
Model changes at fixed fractional radius resulting from the
use of the \citet{Asplunetal2005b} abundances, relative to Model~S,
in the sense (Model~\Magszerofive) -- (Model~S).
The line styles are defined in the figure.
The thin dotted line marks zero change.
The thinner grey and magenta lines show the corresponding differences
for Model~{\Magsszeronine} using the AGSS09 composition.
}
\clabel{fig:ags05}
\end{figure}

\begin{table}
\caption{\clabel{tab:newmodelpar}
%\notecd [Model parameters]
Parameters of solar models. Age, $R$ and $L$ are for the model
of the present Sun.
OPAL92, OPAL96 and OP05 refer to the opacity tables by
\citet{Rogers1992}, \citet{Iglesi1996} and \citet{Badnel2005}, respectively,
while Kur91 and Fer05 indicate low-temperature opacities from
\citet{Kurucz1991} and \citet{Fergus2005}.
The heavy-element abundance used in the opacities are GN93 \citep{Greves1993},
GS98 \citep{Greves1998}, AGS05 \citep{Asplunetal2005b} or
AGSS09 \citep{Asplun2009}.
The first four models
use the same physics as Model~S, apart from the opacity tables and composition,
as indicated.
The final three models use the \citet{Adelbe2011} nuclear parameters.
Model~[AGSS09, mod. opac.] uses the AGSS09 opacity but modified as a function
of temperature in a manner to recover approximately the structure of Model~S
(see Sect.~\ref{sec:modcorcomp} and Fig.~\ref{fig:opaceff});
also, here the OPAL\,2005 \citep{Rogers2002} equation of state was used.
Model [Zhang19] refers to Model~TWA of \citet{Zhang2019}
(see Sect.~\ref{sec:massloss}), using the same equation of state and
nuclear parameters.
Model [Vinyoles17, AGSS09met] refers to the \citet{Vinyol2017} model
using the AGSS09 abundances, with some revision based on meteoritic
abundances of refractory elements;
here the equation of state was obtained from the FreeEOS formulation
\citep[see][]{Cassis2003a}.
For further details on model physics, see Sect.~\ref{sec:microphys}.
}
\vskip 4mm
\centering
{\small
\begin{tabular}{l|cccccc}
\hline
Model & Age & $R$  & $L$ & Opacity & Surface  & Surface \\
      & (Gyr) & $(10^{10} \cm)$ & $(10^{33} \erg \s^{-1})$ & tables &  opacity & comp. \\
\hline
{\MS}     & 4.60 & 6.9599 & 3.846 & OPAL92  & Kur91 & GN93  \\
\Mgsnineeight  & 4.60 & 6.9599  & 3.846 & OPAL96  & Fer05 &GS98  \\
\Magszerofive  & 4.60 & 6.9599  & 3.846 & OPAL96  & Fer05 &AGS05  \\
\Magsszeronine  & 4.60 & 6.9599  & 3.846 & OPAL96  & Fer05 &AGSS09  \\
{[}AGSS09,  & 4.60 & 6.9599  & 3.846 & OPAL96  & Fer05 & AGSS09  \\
mod. opac.{]} &  --  &  -- & -- & -- & -- & -- \\
{[}Zhang19{]}  & 4.57 & 6.9598  & 3.842 & OPAL96  & Fer05 & AGSS09  \\
{[}Vinyoles17,  & 4.57 & 6.9597  & 3.842 & OP05  & Fer05 & AGSS09  \\
	AGSS09met{]} &  --  &  -- & -- & -- & -- & (met) \\
\hline
\end{tabular}
}
\end{table}

\begin{table}
\caption{\clabel{tab:newmodelchar}
%\notecd [Model characteristics]
Characteristics of the models in Table~\ref{tab:newmodelpar}
with updated abundances.
$Y_0$ and $Z_0$ are the initial helium and heavy-element abundances, 
$T_{\rm c}$, $\rho_{\rm c}$ and $X_c$ are the central temperature, density
and hydrogen abundance of the model of the present Sun, 
$\Zs/\Xs$ is the present ratio between
the surface heavy-element and hydrogen abundances,
$\Zs$ and $\Ys$ are the surface heavy-element and helium abundances,
and $d_{\rm cz}$ is the depth of the convective envelope.
The last lines give helioseismically inferred solar values of $\Ys$
\citep{Basu2004a} and $d_{\rm cz}/R$ \citep{Christ1991b, Basu1997b}.
For details on the models, see the caption to Table~\ref{tab:newmodelpar}.
}
\vskip 4mm
\small
\centering
\hskip -1 cm
\begin{tabular}{l|ccccccccc}
\hline
Model   & $Y_0$ & $Z_0$ & $T_{\rm c}$ & $\rho_{\rm c}$ & $X_{\rm c}$ & $\Zs/\Xs$ & $\Zs$ & $Y_{\rm s}$ & $d_{\rm cz}/R$ \\
        &       &       & $(10^6 \K)$ & $(\g \cm^{-3})$  &             &             &           &       &        \\
\hline
{\MS} &           0.27126 & 0.019631 & 15.667 & 153.86 & 0.33765 & 0.02450 & 0.01806 & 0.24464 & 0.28844 \\
{\Mgsnineeight} & 0.27454 & 0.018496 & 15.696 & 153.93 & 0.33542 & 0.02307 & 0.01698 & 0.24686 & 0.28345 \\
{\Magszerofive} & 0.25664 & 0.013731 & 15.445 & 150.62 & 0.36126 & 0.01650 & 0.01253 & 0.22832 & 0.27098 \\
{\Magsszeronine} & 0.26342 & 0.014864 & 15.547 & 151.40 & 0.35204 & 0.01810 & 0.01360 & 0.23529 & 0.27558 \\
{[}AGSS09, & 0.27657 & 0.014514 & 15.639 & 154.18 & 0.34201 & 0.01810 & 0.01335 & 0.24901 & 0.28899 \\
mod. opac{{]}}        & -- & -- & -- & -- & -- & -- & -- & -- & -- \\
{[}Zhang19{{]}} & 0.27565 & 0.014717 & 15.579 & 153.35 & 0.34855 & 0.01880 & 0.01393 & 0.24505 & 0.28899 \\
{[}Vinyoles17,       & 0.26137  & 0.01485 & 15.439 & 148.91 & 0.36227 & 0.01781 & 0.01344 & 0.23167 & 0.27770 \\
AGSS09met{{]}}        & -- & -- & -- & -- & -- & -- & -- & -- & -- \\
\hline
Sun  &     &   --    &   --     &   --   &   --   &   --   &    --    &  $0.2485$     & $0.287 $ \\
     &     &   --    &   --     &   --   &   --   &   --        & -- &  $\pm 0.0034$ & $\pm 0.001$ \\
\hline
\end{tabular}
\end{table}

\subsection{Effects on solar models of the revised composition}
\clabel{sec:modrevcomp}

The effect on solar models of the change in the heavy-element abundance
arises predominantly from the resulting change in the opacity.
From Fig.~\ref{fig:kapder} it is obvious that oxygen makes a large
contribution to the opacity in much of the interior.
Thus the reduction in the oxygen abundance leads to a decrease in
the opacity; this reduces the depth of the convection zone,
as well as the temperature gradient in the radiative interior,
leading to the reduction in the sound speed in much of the interior.
An extensive review of the effects on solar models, and the broader
consequences for modelling other stars, was provided by \citet{Buldge2019a}.

\begin{figure}[htp]
\centerline{\includegraphics[width=\figwidth]{\fig/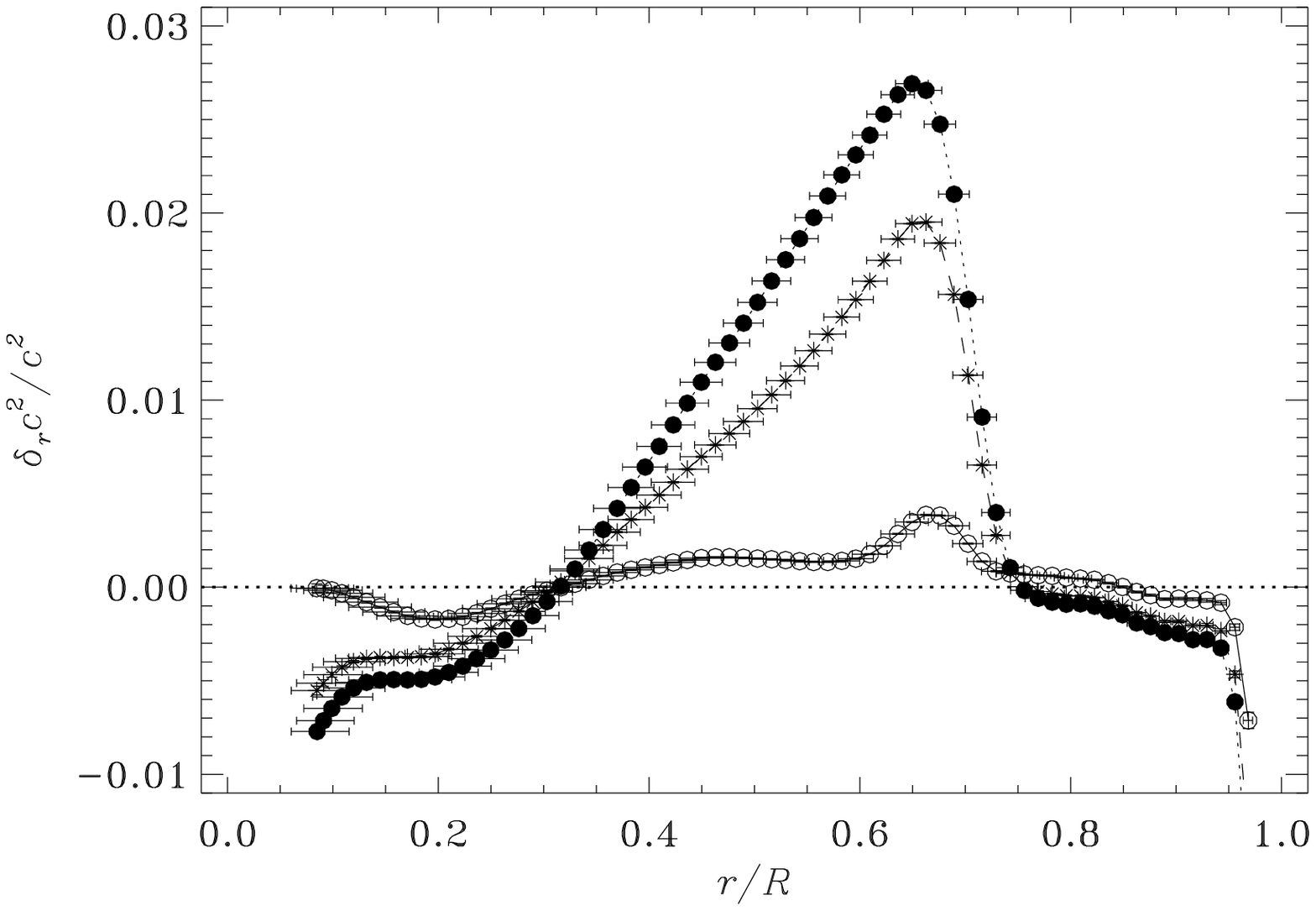}}
\caption{Inferred difference in squared sound speed between
the Sun and three solar models, in the sense (Sun) -- (model).
The open circles use Model~S ({\cf} Fig.~\ref{fig:csqinv}),
the filled circles the corresponding Model~{\Magszerofive}
based on the \citet{Asplunetal2005b} composition and the stars
Model~{\Magsszeronine} based on the \citet{Asplun2009} composition.
The vertical bars show $1\,\sigma$ errors in the inferred values,
based on the errors, assumed statistically independent, in the
observed frequencies.
The horizontal bars provide a measure of the resolution of the inversion.
}
\clabel{fig:ags05csq}
\end{figure}

To illustrate these effects
I consider Models~{\Magszerofive} and {\Magsszeronine}
together with Models~S and {\Mgsnineeight},
summarized in Table~\ref{tab:newmodelpar}.
The effects on the model of the present Sun of the changes in composition,
relative to Model~S,
are shown in Fig.~\ref{fig:ags05} and Table~\ref{tab:newmodelchar}.
The substantial decrease in the sound speed in the radiative interior 
is obvious, 
as is the reduction in the depth of the convection zone.
Also, according to Eq.~\Eq{eq:homlum} the reduction in the
heavy-element abundance must be balanced by a reduction in the
mean molecular weight, to keep the luminosity fixed,
and hence to an increase in $X$, as shown in Fig.~\ref{fig:ags05},
and a corresponding decrease of the helium abundance in the convection zone
(see Table~\ref{tab:newmodelchar}).

These changes in the solar model have had a drastic effect on the comparison
with the helioseismic results.
An extensive review of these consequences was provided by \citet{Basu2008},
while a more recent, but brief, review is in \citet{Serene2016a}.
The maximum relative change in $c^2$ resulting from using AGS05,
around 2 per cent, is substantially
larger than the difference between the Sun and Model~S illustrated in
Fig.~\ref{fig:csqinv} and of the opposite sign.
Thus the new abundances greatly increase the discrepancy between the 
model and helioseismically inferred solar sound speed.
This is illustrated in Fig.~\ref{fig:ags05csq}.
%If AGSS09 is used the effect is somewhat smaller, but still very substantial.
As expected from Fig.~\ref{fig:ags05},
the effect on the sound speed extends through much of the radiative interior;
in particular, it is not only a consequence of the error in the depth
of the convection zone of the model (see also Fig.~\ref{fig:modopcsq}).
Also, as illustrated in Table~\ref{tab:newmodelchar}
the envelope helium abundance and convection-zone depth of the model
differ strongly from the helioseismically inferred values.
Using the more recent AGSS09 composition reduces the discrepancies
with the helioseismic results somewhat (see Fig.~\ref{fig:ags05csq} and
Table~\ref{tab:newmodelchar}) although they remain substantial.

It was in fact immediately obvious that the revised composition
created problems in matching solar models to the helioseismic inferences.
\citet{Basu2004a} considered envelope models, demonstrating that 
a substantial increase in opacity would be needed to bring the models
in accordance with the seismic observations.
A similar conclusion was reached by \citet{Bahcal2004}, based
on the depth of the convection zone.
\citet{Guzik2004}, \citet{Montal2004}, \citet{Turck2004},
%Pijpers {\etal} \citep[see][]{Christ2004}
and \citet{Bahcal2005c}
showed that the sound speed in models with the revised composition
differed much more from the helioseismically determined behaviour than
for models with the old composition, as illustrated in Fig.~\ref{fig:ags05csq}.
In a detailed analysis based on the convection-zone depth and envelope
helium abundance \citet{Delaha2006} concluded that models with the AGS05
composition were inconsistent with the helioseismic inferences to a very
high degree of significance, while models with the old composition were
essentially consistent with the observations.

\begin{figure}[htp]
\centerline{\includegraphics[width=\figwidth]{\fig/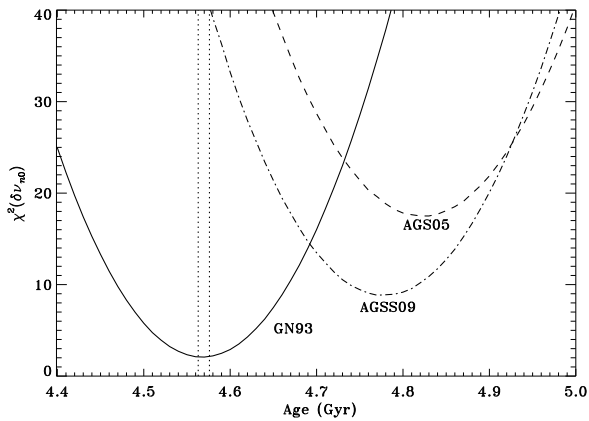}}
\caption{
Goodness of fit for the small frequency separation
$\delta \nu_{n0} = \nu_{n0} - \nu_{n-1\,2}$, fitting solar models of
varying age to the observations of \citet{Chapli2007}.
All models were calibrated to the observed surface luminosity and radius
and a specified value of $\Zs/\Xs$.
The solid curve shows results for the GN93 composition, the dashed curve
for the
AGS05 composition and the dot-dashed curve for the AGSS09 composition.
The vertical dotted lines indicate the interval of solar age obtained by
Wasserburg, in \citet{Bahcal1995}.
Adopted from \citet{Christ2009a}.
}
\clabel{fig:agefit}
\end{figure}

%\notecd [Somewhere here also include comparison in terms of determining
%solar age from small separations, with appropriate references to earlier work,
%of course.]

Other aspects of the solar oscillation frequencies show similarly large
inconsistencies for models computed with the revised abundances.
A convenient measure of conditions in the solar core is provided by
the small frequency separations
$\delta \nu_{nl} = \nu_{nl} - \nu_{n-1 \, l+2}$,
where $\nu_{nl}$ is the cyclic frequency of a mode of degree $l$ and
radial order $n$, which according to asymptotic theory
\citep[{\eg}][]{Tassou1980} is largely determined by the sound-speed
gradient in the stellar core ({\cf} Eq.~\ref{eq:smlsep}).
%\notecd [Could already be presented in the helioseismic section;
%worth using in the neutrino section that this early ruled out non-standard
%solar models, and becomes relevant also in the asteroseismic section].
\citet{Basu2007} considered a broad range of models with varying 
composition and opacity tables and found that models with the
GS98 composition were largely consistent with the observed values,
whereas the AGS05 composition resulted in a very significant
departure from the observations.
A detailed analysis of this nature was carried out by \citet{Chapli2007}
who carried out fits to the observations to constrain the heavy-element
abundance;
this resulted in a lower limit of $Z = 0.0187$, far higher than in the
models computed with AGS05 or AGSS09.
\citet{Zaatri2007} also found that the AGS05 composition resulted in small
frequency separations that were inconsistent with observations.
An illustration of the effect of the composition on the small 
frequency separations can be obtained by considering the calibration of the
solar age, based on fits to the small separations, following
\citet{Dziemb1999} and \citet{Bonann2002}.
Here the age is determined from $\chi^2$ fits to $\delta \nu_{n0}$,
for models of varying age but calibrated to the correct radius, luminosity
and assumed surface composition as characterized by $\Zs/\Xs$.
As illustrated in Fig.~\ref{fig:agefit}, using the old composition the 
best-fitting model has an age of 4.57\,Gyr,
very close to the age of $4.570 \pm 0.006 \Gyr$ obtained from meteorites
\citep[Wasserburg, in][]{Bahcal1995},
and the minimum $\chi^2$ is reasonable.
On the other hand, with the AGS05 composition the best-fitting model leads
to a high $\chi^2$ at an age of 4.83\,Gyr which is much higher than the
proper solar age, while using AGSS09 leads to a best-fitting age of 4.77\Gyr;
in both cases the fits are inconsistent with the meteoritic age, at a
high level of significance.
The inconsistencies involving properties sensitive to the solar core
clearly underline that the effects of the revised composition are
not confined to the vicinity of the convection zone.

\begin{figure}[htp]
\centerline{\includegraphics[width=\figwidth]{\fig/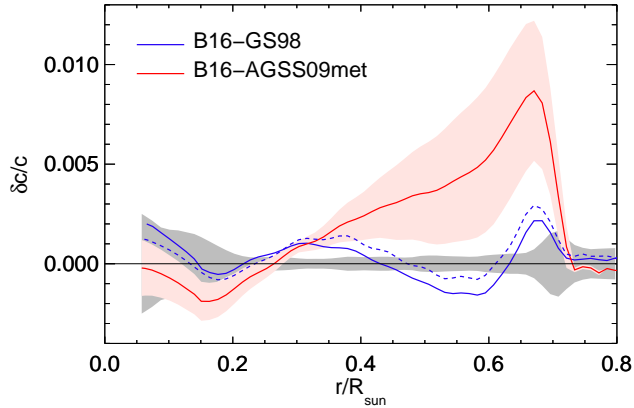}}
\caption{
Relative sound-speed differences, in the sense (Sun) -- (model),
at fixed fractional radius.
The red curve is based on a model using the AGSS09 abundances, updated
with meteoritic abundances for the refractory elements,
while the blue solid curve used the GS98 abundances (the dashed curve
corresponds to an older GS98 calculation). 
The red shaded region shows estimated effects of modelling uncertainties, 
while the grey shaded band is an estimate of the effects of errors in inferring 
the sound speed from results of an inversion.
(For comparison with, e.g., Fig.~\ref{fig:ags05csq}, note that the latter
shows differences in squared sound speed.)
Adopted from \citet{Vinyol2017}.
(Figure courtesy of Aldo Serenelli.)
}
\clabel{fig:vincsq}
\end{figure}

%\notecd [Needs revision, extension, more ....]
An early analysis based on the AGSS09 abundances was carried out by
\citet{Serene2009}
who in addition to the purely photospheric AGSS09 composition provided
in Table~\ref{tab:abundances} considered the effects of replacing
the abundances of certain elements,
such as magnesium and iron, with the probably more reliable meteoritic values.
This resulted in a slight decrease in the metallicity, relative to AGSS09
as analysed here, and a corresponding small increase in the sound-speed
discrepancy.
In a broad review of the solar interior
\citet{Basu2015} compared models computed with the different abundances
listed in Table~\ref{tab:abundances}
with the helioseismic results.
Interestingly, the C11 abundances gave results very similar to those for GS98,
despite the lower CNO abundances; these were compensated by other abundance
differences, leading to roughly similar opacities.

A very extensive analysis of the effects of the revised composition on solar
models was carried out by \citet{Vinyol2017}.
As did \citet{Serene2009} they included meteoritic abundances of
refractory elements, resulting in what they called the AGSS09met composition.
The modelling used up-to-date physics: the FreeEOS equation of state,
reaction rates based on an update of those provided by \citet{Adelbe2011},
OP opacities \citep{Badnel2005}, and diffusion using the formulation of
\citet{Thoul1994}.
A careful analysis was carried out of the errors in the model, based on the
errors in the input parameters, particularly the composition,
the nuclear reactions and the opacity; 
the opacity uncertainty was scaled according to the difference between
OP and OPAL opacities, as well as the results of the 
\citet{Bailey2015} experiments (discussed in Sect.~\ref{sec:modcorcomp})
and assumed to vary linearly with $\log T$.
Figure~\ref{fig:vincsq} shows the resulting relative sound-speed difference
using the AGSS09met composition,
compared with corresponding results using GS98, obtained from inversion
of a combination of BiSON and MDI frequencies.
The dominant modelling uncertainty, common to the two sets of results,
is shown as the red shaded region for the AGSS09met results.
The shaded grey area illustrates what the authors take to be the uncertainty
resulting from the inversion;
it includes a modest contribution from the choice of reference model which,
given that the inversion is carried out directly based on the model and
the observed frequencies, is essentially irrelevant 
(see Sect.~\ref{sec:heliostruc}).
They also compared with the helioseismically inferred values the
envelope helium abundance and location of the
base of the convection zone in the models;
results for their AGSS09 model are also shown in Table~\ref{tab:newmodelchar}
and are clearly similar to those for Model~{\Magsszeronine}.
The conclusion of the analysis was a statistically significant preference
for the GS98 composition;
this was particularly strong when excluding from the comparison
the bump in $\delta c/c$ just below the convection zone, 
which is likely associated with mixing processes missing 
in the modelling (see Fig.~\ref{fig:tachodif}).
The uncertainty in the inferred sound-speed difference in Fig.~\ref{fig:vincsq}
is dominated by the abundance uncertainties and the assumed range in the opacity
uncertainty.
A similar analysis, although with a more sophisticated analysis of the 
opacity uncertainty and involving a reconstruction of the solar
opacity profile, was carried out by \citet{Song2018}.
I return to the effects of opacity below.

\begin{figure}[htp]
\centerline{\includegraphics[width=\figwidth]{\fig/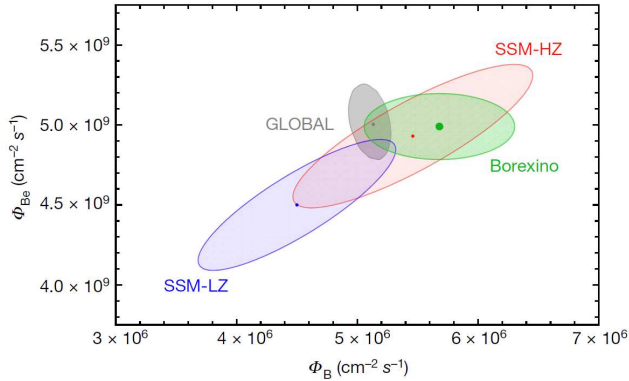}}
\caption{Comparison of the observed and computed
$\boreight \; (\Phi_{\rm B})$
and $\berseven \; (\Phi_{\rm Be})$ neutrino fluxes, indicated by
68 per cent confidence contours.
The red and blue areas show model results using the GS98 (SSM-HZ) and
AGSS09 (SSM-LZ) compositions \citep{Vinyol2017}.
The green area shows Borexino results, while the grey area was obtained
from a combined analysis of all solar, as well as the KAMLAND, data.
From \citet{Agosti2018}.
}
\clabel{fig:borcomp}
\end{figure}

Introducing the AGS05 abundances had a relatively
modest effect on the predicted neutrino production
compared with the uncertainties in the predictions and observations
\citep{Turck2004, Bahcal2005b, Bahcal2005c}.
This is a consequence of the small reduction in the temperature in the core
of the model, around 1 per cent (see Fig.~\ref{fig:ags05}),
leading to a reduction of around 20 per cent in the flux of
$\boreight$ neutrinos, with smaller changes in the other neutrino fluxes.
A detailed investigation of the uncertainties in the predicted neutrino flux
was carried out by \citet{Bahcal2005e};
they found that their so-called conservative uncertainties in
the surface composition, estimated from the differences between the 
compositions of individual elements in the emerging revised determinations
and GS98, still provided the largest contribution to the
total uncertainty in the computed neutrino fluxes.
As part of their detailed revised solar modelling, discussed above,
\citet{Vinyol2017} carried out a careful analysis of the impact of the AGSS09
composition on the predicted solar neutrinos, including a determination of
the uncertainties in the predictions taking into account other 
uncertainties in the modelling, and using updated measured neutrino fluxes,
including the Borexino $\berseven$ results.
They found reductions, although barely significant compared with the
model uncertainties, in the $\boreight$ and $\berseven$ fluxes,
as a result of the reduction in the core temperature (see Fig.~\ref{fig:ags05});
comparison with the observations showed a slight but insignificant
preference for the GS98 composition, as also hinted by Fig.~\ref{fig:neutsurv}.
\citet{Agosti2018} presented the effect of the solar composition on the
flux of $\boreight$ and $\berseven$ neutrino fluxes
({\cf} Fig.~\ref{fig:borcomp}) and concluded that the neutrino data
provide no significant distinction between the GS98 and AGS09 composition.
Similarly, \citet{Bergst2016} concluded that current neutrino data have
``\emph{absolutely} no preference for either [the GS98 or the AGSS09] model''.

\begin{figure}[htp]
\centerline{\hskip -6cm\includegraphics[width=\figwidthb]{\fig/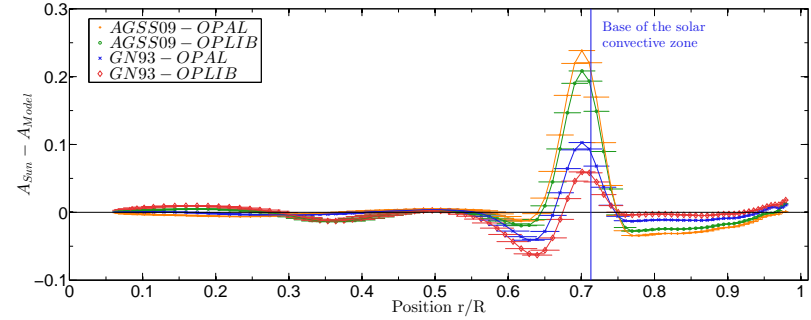}}
\vskip 0.5cm
\caption{
Inversion for differences, at fixed fractional radius,
in the Ledoux discriminant
({\cf} Eq.~\ref{eq:ledoux}) between the Sun and models with the OPAL and 
OPLIB opacities, using the GN93 and AGSS09 compositions (see legend).
Horizontal bars show the resolution and the (barely visible) vertical
error bars are propagated from the observational frequency errors.
Adopted from \citet{Buldge2017a}.
(Figure courtesy of Ga\"el Buldgen.)
}
\clabel{fig:buldledoux}
\end{figure}

\citet{Buldge2017a} applied a very interesting procedure to the analysis
of the effects of the composition updates.
This is based on inversion for differences in the \emph{Ledoux discriminant},
\be
A = {\dd \ln \rho \over \dd r} - {1 \over \Gamma_1} {\dd \ln p \over \dd r} \;
\eel{eq:ledoux}
\citep[see also][]{GoughKos1993},
using the structure pair $(A, \Gamma_1)$.
They applied the analysis to models computed with the FreeEOS equation of state
\citep[developed by A. Irwin, see][and Sect.~\ref{sec:eos}]{Cassis2003a},
with opacities from the OPAL and OPLIB tables 
(see Sect.~\ref{sec:opac}) and using both the GN93 and AGSS09 compositions.
Some results are shown in Fig.~\ref{fig:buldledoux}.
Interestingly, the differences are concentrated just below the convection 
zone, in the region of the bump in the sound-speed differences in
the GN93 models (e.g., Fig.~\ref{fig:csqinv}).
This suggests that the differences in $A$ are sensitive to this feature,
even in the AGSS09 models where it is hidden by the larger general 
sound-speed difference.
Just below the convection zone the GN93 models are closer to the Sun, while
around $r = 0.64 R$ the differences are smaller for the AGSS09 models.
In a second interesting analysis \citet{Buldge2017c} carried out inversion
for the same four models in terms of $S_{5/3} = p/\rho^{5/3}$, which
in the ideal-gas approximation is closely related to the specific entropy,
using again $\Gamma_1$ as the second variable.
Here substantial differences were found in the convection zone, essentially
corresponding to different values of the specific entropy in the adiabatic
part of the convection zone resulting from the model calibration, with some
preference for the GN93 models.
These analyses are potentially very valuable tools, as supplements to the
more common sound-speed and density inversion, particularly for the
investigation of the lower boundary of the convective envelope which
undoubtedly is the site of substantial uncertainties in the modelling,
related to possible overshooting or other types of mixing beyond 
the convection zone.

In the following I discuss possible solutions to the problems of solar models
with the revised abundances;
however I note already now that
these were discussed in more detail by \citet{Basu2008},
based on the AGS05 composition,
leading to the general conclusion that no definite satisfactory solution
had at that time been found.
This still holds.

\subsection{Are the revised abundances correct?}
\clabel{sec:checkcomp}

Given the difficulty in reconciling the AGS05 and AGSS09 abundances with the
helioseismic results, it has been natural to question these abundances.
In their favour is the fact that they bring the Sun into closer agreement
with the abundances of objects in the solar neighbourhood, as mentioned above,
although this is perhaps not decisive.

An independent determination of the envelope heavy-element abundances can
in principle be obtained from the effects of the heavy elements 
on the thermodynamic properties of the gas 
and the resulting influence on the solar oscillation frequencies
or the helioseismically inferred properties \citep{Gong2001b, Mussac2009}.
This is analogous to the determination of the envelope helium abundance
discussed in Sect.~\ref{sec:specasp},
although obviously far more demanding, given the lower abundance and the
correspondingly smaller effects.
{\rv \citet{Takata2001b} carried out an early inverse analysis targeting 
the heavy-element abundance and found an indication that it was lower by 
20 -- 30\,\% than in Model S in the convection zone.}
Early results by \citet{Lin2005} provided slight indications for a decrease
in the heavy-element abundance, relative to the \citet{Greves1993} value,
while \citet{Antia2006} and \citet{Lin2007} obtained results consistent
with the GN93 abundances \citep[for a review, see also][]{Basu2008}. 
A somewhat indirect determination was made by
\citet{Houdek2011} who used low-degree observations from BiSON,
combining analysis of the helium glitch with use of the asymptotic behaviour
of the acoustic modes,
to determine a seismic measure of solar age and the heavy-element abundance
through model calibration,
focusing on the structure of the core resulting from the hydrogen fusion.
The age was consistent with the value obtained from radioactive decay,
while the inferred heavy-element abundance, $Z_{\rm s} = 0.0142$,
was intermediate from the values obtained for the GS98 and AGSS09 compositions.
A potential problem with the analysis may be indicated by the fact that
the model fitting resulted in an envelope helium abundance 
of $Y_{\rm s} = 0.224$, substantially below values obtained from 
helioseismic analyses of just the effects of the helium glitch
(see Sect.~\ref{sec:specasp}).
It is evident that the use of $\Gamma_1$ as a composition diagnostics
depends critically on the assumed equation of state,
probably even more for the heavy-element abundance than in the case of 
the determination of the helium abundance.
{\rv Careful analyses were carried out by \citet{Voront2013, Voront2014}},
fitting helioseismic observations to solar convective-envelope models
based on a variety of
equations of state, including the so-called SAHA-S implementation
\citep{Baturi2013}.
They found that SAHA-S provided a substantially better fit to the observations
than other formulations, with a heavy-element abundance in the range
$Z = 0.008 - 0.013$, {\ie}, strongly supporting the 
revised low values of $Z$, while acknowledging that a complete solar
model with this abundance would be inconsistent with seismic inferences
of the radiative interior.
\citet{Buldge2017b} carried out numerical inversions based on corrections
to the Ledoux discriminant ({\cf} Eq.~\ref{eq:ledoux}), $Y_{\rm s}$ and
$Z_{\rm s}$;
the analysis was tailored to obtain determinations of $\delta Z_{\rm s}$,
suppressing the contributions from $A$ and $Y_{\rm s}$
\citep[see also Sect.~\ref{sec:heliostruc}, and][]{Basu2016}.
The results showed a substantial scatter, depending on the choice of
reference model and inversion details, but with a strong trend towards
a heavy-element abundance substantially below GS98, 
in accordance with the results of \citet{Voront2013}.
Thus several independent lines of investigation point towards the
lower abundance, in support of AGSS09. 

In principle the composition of the solar atmosphere can be directly 
sampled through analysis of the solar wind.
In practice, this is greatly affected by the fractionation of elements
taking place in the acceleration of the solar wind, particularly the FIP effect.
%
%\footnote{for First Ionization Potential; \eg, \citet{Marsch1995,
%Laming2015}}
However, it was argued by \citet{vonSte2016} that this effect is
largely absent in polar coronal holes.
Thus they used observations of the solar wind 
from the {\it Ulysses} spacecraft, with an
orbit passing repeatedly over the solar poles, to estimate the solar
photospheric composition.
Interestingly, the inferred heavy-element abundance, $Z = 0.0196 \pm 0.0014$,
is consistent with the older and helioseismically preferred composition.
On the other hand, these results were criticized by \citet{Serene2016b} who 
pointed out that using the detailed composition inferred by von Steiger
and Zurbuchen substantially increased the neutrino-flux discrepancy of the
models; they furthermore questioned the analysis of the FIP effect.
{\rv Similar results were reached in a detailed analysis by \citet{Vagnoz2017}.
}

%\notecd [More on CNO-abundances from neutrinos; to be revised and reorganized].

An interesting connection between the abundance issue and the 
neutrino observations was noted by \citet{Haxton2008b}.
They pointed out that future development in detector technology may allow
measurement of the flux of neutrinos from the $\nitthirteen$ and
$\oxyfifteen$ decays (see Eq.~\ref{eq:CNO}),
and hence of the rate of the CNO reactions;
given the present well-determined nuclear parameters this would provide
%an estimate of the primordial abundance of carbon and nitrogen.%
%\footnote{Since the $\nitfourteen(\hyd, \helfour)$ is the slowest reaction
%in the cycle almost all the initial carbon is converted into nitrogen by
%the reactions \notecd [a point that could be made earlier!].}
%Detection of neutrinos from the reactions in the CNO cycle would provide
an independent determination of the CNO abundances in the solar interior.
The resulting numerical relation between the flux of CNO neutrinos and
the central heavy-element abundance of the Sun was derived by 
\citet{Gough2019}.
{\rv Very recent Borexino results \citep{Agosti2020b}
provide a solid detection of the flux that, however, is consistent
with both the high- and low-metallicity compositions. 
New} detectors are being developed with the specific goal of
reaching a sufficiently low background to detect the CNO signals
\citep[for an overview, see][]{Bonven2018}.
A detailed analysis was carried out by \citet{Cerden2018} of the 
potential of the Borexino detector and planned new detectors 
for making a significant determination of the CNO composition of the solar core.
One example of {\rv a planned} detector potentially capable of distinguishing
between the low- and high-CNO models is the Jinping detector in China
\citep{Wan2019}, with a planned liquid-scintillator detector mass of up to
4\,kton.

\begin{figure}[htp]
\centerline{\includegraphics[width=\figwidth]{\fig/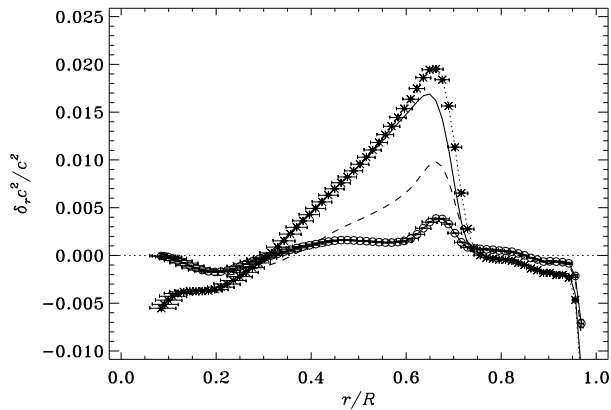}}
\caption{
Inferred differences in squared sound speed between the Sun and 
four solar models, in the sense (Sun) -- (model).
As in Fig.~\ref{fig:ags05csq} the open circles are for Model~S and
the stars (connected by a dotted line) for Model~{\Magsszeronine} using the AGSS09 composition;
the dashed curve shows the results for Model~{\Mgsnineeight} based on
the GS98 composition.
The solid curve shows results for a model corresponding 
to Model~{\Magsszeronine} but with a localized change in the opacity
near the base of the convection
zone to bring the depth of the convection zone into agreement with 
Model~{\Mgsnineeight}.
}
\clabel{fig:modopcsq}
\end{figure}

%\notecd [Also around here discuss very recent von Steiger and Zurbuchen
%2016 paper.
%And \citet{Voront2013}; somewhere this is already touched upon.]

\subsection{Possible corrections to the solar models}
\clabel{sec:modcorcomp}

The very serious discrepancies between the
models with the new composition and the helioseismic results
have led to many attempts to find modifications to the
models that will improve the agreement.
As reviewed by \citet{Guzik2006, Guzik2008} 
%\notecd [and possibly also Santa Fe paper. No, just on mass loss effects].
these attempts have met with limited success.

A perhaps not uncommon misconception is that the principal effect
of the revised composition is the decrease in the depth of the convection zone.
\citet{Basu2015}, for example, stated that
`[t]he most dramatic manifestation of the change of metallicities is
the change in the position of the convection-zone base, 
which changes the sound-speed difference between solar models and the Sun',
implying that the change in the sound speed is caused by the change in the
location of the base of the convection zone.
To test this I applied a localized change to the opacity in the
AGSS09 model in Fig.~\ref{fig:ags05csq}, of the form used in equation (1)
of \cite{Christ2018a} but calibrated to obtain the same depth of the
convection zone as in Model~{\Mgsnineeight} ({\cf} Fig.~\ref{fig:compinv}).
%\notecd [At some point it would help to assign names to these models!].
As indicated by Fig.~\ref{fig:changelocopac} such a local
opacity modification has a local effect on the sound speed;
a more detailed discussion of the effects
on the model structure was provided by \citet{Christ2018a}.
The helioseismically inferred differences in the squared sound speed
between this model and the Sun are compared in Fig.~\ref{fig:modopcsq}
with the corresponding results for Model~S, Model~{\Mgsnineeight}
and Model~{\Magsszeronine}.
The figure shows a small shift in the sound-speed difference in the 
opacity-modified model compared with the original Model~{\Magsszeronine},
corresponding to the shift in the base of the convection zone,
and a related modest decrease in the maximum of the sound-speed difference;
however, in the bulk of the radiative interior the difference for the
original and modified AGSS09 models are very similar.
Thus it is clear that the sound-speed difference is not just 
a consequence of the shift in the base of the convection zone
\citep[see also][]{Ayukov2017}.

\begin{figure}[htp]
\centerline{\includegraphics[width=\figwidth]{\fig/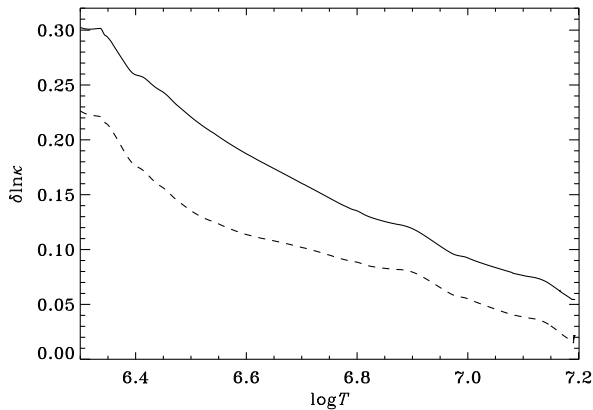}}
\caption{
Intrinsic opacity corrections, assumed to be functions of temperature alone,
required to bring models with the revised composition into agreement with 
Model~S.
The solid curve is for the AGS05 composition and the dashed curve is for
the AGSS09 composition.
Adopted from \citet{Christ2009b, Christ2010}.
}
\clabel{fig:opaccor}
\end{figure}

\begin{figure}[htp]
\centerline{\includegraphics[width=\figwidth]{\fig/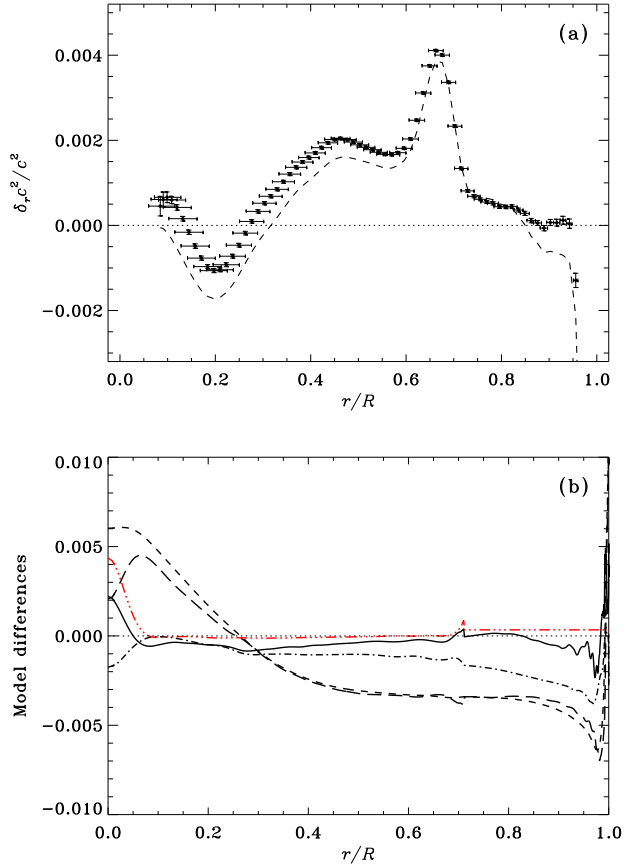}}
\caption{
(a) Result of sound-speed inversion using as reference a model based on
the AGSS09 opacities, but with the modification shown as a dashed curve in
Fig.~\ref{fig:opaccor}. 
The dashed curve shows the inversion result against Model~S, illustrated
in Fig.~\ref{fig:csqinv}, the caption of which also defines the error bars.
(b) Logarithmic differences between the model with the modified AGSS09
opacities and Model~S.
Line styles are defined in Fig.~\ref{fig:changeage}.
}
\clabel{fig:opaceff}
\end{figure}

As the heavy elements predominantly affect the structure through the 
opacity, an obvious correction to the model calculations is to increase the
opacity. 
This was noted by \citet{Basu2004a} and \citet{Montal2004} who estimated that
an opacity increase of 10 -- 20 per cent would be required. 
\citet{Bahcal2005c} found the opacity difference between models with the
old and new composition to be up to around 15 per cent, the largest values
being close to the base of the convection zone and reflecting the contribution
from the oxygen abundance illustrated in Fig.~\ref{fig:kapder}.
\citet{Christ2009b} evaluated the change in opacity,
assumed to be a function of temperature,
required to reproduce the structure of Model~S with the AGS05 composition.
The result is shown in Fig.~\ref{fig:opaccor}, including
also the similar analysis based on the AGSS09 composition \citep{Christ2010};
at the base of the convection zone the required increase 
is around 30 per cent when AGS05 is used,
while AGSS09 requires an opacity increase of up to around 23 per cent.
The effects of the latter increase on the results of sound-speed inversion
and model structure are shown in Fig.~\ref{fig:opaceff}.
It is evident that the opacity modification, applied to the AGSS09 opacities,
largely recovers the difference in squared sound speed
between the Sun and the model structure found with Model~S
(see also Fig.~\ref{fig:csqinv}).
Furthermore, comparing panel (b) with Fig.~\ref{fig:ags05} shows that most
of the difference in other properties of the model structure is also suppressed.
In particular, as shown in Table~\ref{tab:newmodelchar} the model is as 
successful as Model~S in matching the inferred solar envelope helium abundance
and depth of the convection zone.
A similar estimate of the required opacity change, but based on combining
intrinsic changes to the opacity with changes in the composition and taking
into account also the constraints of the observed neutrino fluxes,
was obtained by \citet{Villan2014}.
%\notecd [Might also include model in overall table of model quantities,
%to illustrate effect on depth of convection zone and envelope He].

%\notecd [New results on opacity, Le Pennec, Guzik, Buldgen, Serenelli ...]

\begin{figure}[htp]
\centerline{\includegraphics[width=\figwidth]{\fig/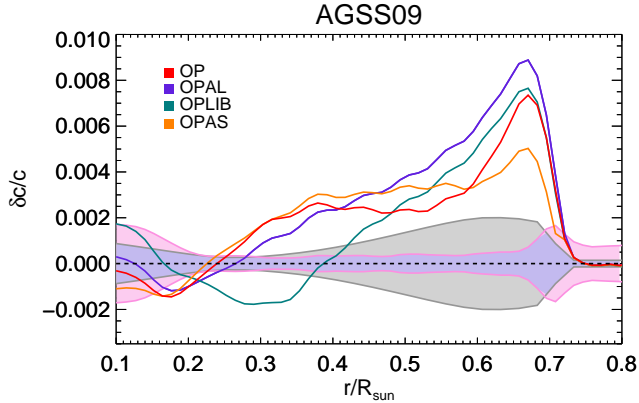}}
\caption{Inferred relative sound-speed differences,
at fixed fractional radius, between the Sun
and models with the AGSS09 composition and using the OP, OPAL, OPLIB and OPAS
opacity tables.
The pink shaded region indicates the uncertainty resulting from the inversion
procedure, whereas the grey area indicates uncertainties in the modelling
\citep[see][]{Vinyol2017}.
From Villante, Serenelli and Vinyoles (in preparation).
Figure courtesy of Aldo Serenelli.
}
\clabel{fig:csqopcomp}
\end{figure}

It is far from clear that such intrinsic increases in the opacity are
realistic.
A measure of the uncertainty in the opacities is perhaps provided
by differences between the totally independent calculations and their
effects on the results;
as discussed in Sect.~\ref{sec:opac} several such calculations are now
available.
Fig.~\ref{fig:changeopacop05} shows that replacing the OPAL tables
by OP increases the squared sound speed by up to about 0.7 per cent
below the convection zone, resulting in a modest reduction in the difference
between the Sun and the model.
Analyses using the more recent OPAS and 
OPLIB tables, with the AGSS09 composition,
have been carried out by {\rv \citet{Buldge2019a} and}
Villante, Serenelli and Vinyoles (in preparation).
The resulting sound-speed profiles are compared with the Sun
in Fig.~\ref{fig:csqopcomp}.
While OP and OPLIB yield results rather similar to those for OPAL,
the sound-speed difference for OPAS is generally lower than the rest,
probably reflecting the somewhat higher opacity just below the convection zone,
shown in Fig.~\ref{fig:opcomp}.
{\rv \citet{Buldge2019a} noted that the generally lower OPAS opacity in the 
bulk of the radiative interior requires a lower helium abundance 
for luminosity calibration and hence exacerbates the discrepancy between
the model and the helioseismically inferred surface helium abundance.}
Interestingly using OPLIB with the AGSS09 composition results in 
small frequency separations $\delta \nu_{nl}$ in good agreement with
the observations, while, as discussed above, using the OPAL opacities and
AGSS09 results in very significant differences between model and observations
\citep{Buldge2017c}.
On the other hand, the OPLIB opacities result in a substantial reduction in
the core temperature and hence in neutrino fluxes that are inconsistent
with the observations (A. Serenelli, private communication).
This is a strong demonstration of the complementary information available
from helioseismic and neutrino data, and makes the OPLIB less 
attractive for solar modelling.
In any case, the spread between different current opacity tables and its
dependence on temperature
in no way justify the opacity correction illustrated in Fig.~\ref{fig:opaccor}.

It cannot be excluded that effects ignored by current opacity calculations,
or contributions from other chemical elements not included in the calculations,
could have a substantial effect.
Thus it is very interesting that \citet{Bailey2015}, in an experiment
at conditions close to those corresponding
to the base of the solar convection zone obtained using the so-called
Z-pinch technique, measured absorption coefficients
for iron substantially higher than those resulting from atomic modelling
and used in opacity determinations.
Further experiments on chromium and nickel by \citet{Nagaya2019},
using the same facility, also found substantial discrepancies but of
a somewhat different nature, particularly for chromium, indicating
sensitivity to the details of atomic structure.
The origin of these differences between atomic modelling and
experiments is still not clear, and 
independent experiments now under way or being planned
\citep[\eg,][]{LePenn2015a, Perry2020} will be very valuable.
However, they indicate that there may be significant deficiencies 
in our understanding of the physics of the opacity.
\citet{Trampe2018} made an estimate of the consequences
for opacity calculations of the Bailey {\etal} results, 
indicating that it may correspond to increases not dissimilar to those shown 
in Fig.~\ref{fig:opaccor} to correct for the effects of the AGSS09
composition.
Also, \citet{Pradha2018} reviewed issues with current opacity calculations
that might account for the experimental and solar discrepancies.

Alternatively, the opacity could be increased by increasing the abundances
of other elements to compensate for the decrease in the abundances of oxygen
in the AGSS05 and AGSS09 composition tables.
Figure~\ref{fig:kapder} shows that neon contributes substantially to the
opacity.
As in the case of helium, the neon abundance cannot be determined directly
from photospheric spectral lines, and hence is highly uncertain.
The same is true of argon.
\citet{Antia2005} found that an increase by a factor of around 4 in the neon
abundance could bring their envelope models in agreement with helioseismology.
\citet{Bahcal2005d} considered increases of both neon and argon and found
models with abundance increases of around a factor of three
that approximately matched the helioseismically inferred sound speed.
Similar effects of increases in the neon abundance were found by
\citet{Zaatri2007}.
Possible support for such increases was provided by the determination
by \citet{Drake2005} of similarly high neon abundances in what was claimed
to be solar-like stars.
However, the relevance of these abundances for the solar case has
been seriously questioned \citep[{\eg},][]{Schmel2005, Robrad2008}.
Also, \citet{Morel2008} found no evidence in the neon abundances 
of near-by B stars for such a high neon content.
On the other hand,
\citet{Young2018} obtained
an increase by about 40 per cent in the chromospheric Ne/O ratio,
increasing the logarithmic normalized abundance
({\cf} Table~\ref{tab:abundances}) from the value 7.93 quoted by AGSS09
to 8.08.
As shown by \citet{Buldge2019b} the resulting increase in the opacity
(see also Fig.~\ref{fig:kapder}) results in a modest increase in the depth
of the convection zone and the envelope helium abundance, although still
far from enough to match the observed values.

%\notecd [Other proposed solutions; change to diffusion, accretion, mass loss
%(with references); get back to mass loss later.]

An obvious question is the extent to which the observed surface abundance
is representative of the abundance of the radiative interior and hence
of the opacity.
In normal solar models settling causes a significant difference
between the surface heavy-element abundance and the abundance
beneath the convection zone (see Fig.~\ref{fig:Xmod}).
Increasing the rates of diffusion and settling therefore
increases the heavy-element abundance in the interior
relative to the surface and hence compensates for the decrease in the
surface abundance.
This is indeed the case \citep[{\eg},][]{Basu2004a, Montal2004, Guzik2005,
Christ2007, Yang2007},
although to obtain a significant effect a considerable change
(by factors of 1.5 or more) have to be made;
this may be physically unrealistic.
Also, the resulting models typically have an envelope helium abundance
substantially below the helioseismically inferred value.
The effects of increasing diffusion are illustrated 
in Fig.~\ref{fig:changediffus}b which shows that
an increase by 20 per cent in the diffusion and settling coefficients
for both helium and heavy elements
leads to a relative increase in the squared sound speed by about 0.3 per cent,
with a similar increase in the surface hydrogen abundance.
This is also reflected in the decrease in the envelope helium abundance
for Model~{\Mdifb}
shown in Tables~\ref{tab:modelchar} and \ref{tab:modeldiff}.
Compensating for the effect on the interior sound speed 
of the revised abundances while maintaining the envelope helium abundance
would require a strong increase in the heavy-element settling with little
change in helium settling; 
this seems hard to justify physically.
%\notecd [Argument could be strengthened by a general table of general properties
%of the models illustrated in Sect.~\ref{sec:modsens}].

\citet{Ayukov2017} carried out an extensive analysis of solar models
with the various heavy-element compositions, based on the analysis of
solar envelope models by \citet{Voront2013}.
As a constraint on the properties of the solar convective envelope they
used the quantity $M_{75}$ defined as the mass, in units of $\Msun$,
inside a distance of $0.75 \Rsun$ from the solar centre.
This is determined by the density structure in the convection zone and hence
essentially characterizes the entropy in the adiabatic part of the convection
zone.
From the results of \citet{Voront2013} they chose $M_{\rm 75} = 0.9822$ as
a reference and aimed to fit this, together with the radius, luminosity and 
various seismic parameters of the model.
In addition to various forms of opacity changes they also included a possible
increase in the $\hyd + \hyd$ reaction rate. 
They did obtain a model providing a generally good fit, with essentially
the AGSS09 composition, but requiring an increase in the reaction rate
of around 5 per cent, much higher than its estimated uncertainty.
They noted that this could be accounted for by a major increase in the electron
screening of the reaction, although in fact molecular dynamics calculations
have indicated that electron screening could be be far less efficient than
normally assumed (see also Sect.~\ref{sec:engenr}).
Furthermore, the $\boreight$ neutrino flux was substantially lower than
observed.

A comprehensive analysis of solar modelling and helioseismic diagnostics
was carried out by \citet{Buldge2019b}.
The modelling used the GN93, GS98 and AGSS09 compositions, a range of different
opacity tables, and different equations of state.
In addition, a variety of modifications to the modelling, including opacity
modifications and convective overshoot or turbulent diffusion below the
convective envelope were considered.
The helioseismic analyses was carried out in terms of the sound speed,
{\rv the Ledoux discriminant ({\cf} Eq.~\ref{eq:ledoux}) and the entropy proxy
$S_{5/3} = p/\rho^{5/3}$}, as well as
the envelope helium abundance and the depth of the convective envelope.
Buldgen {\etal} concluded that obtaining a model in agreement with the
observations, given the revised surface composition, will require
addressing several different aspects of solar modelling.
As a very important point they noted that the often subtle issues involved
in the analysis of differences between models and observations require improved 
confidence in the modelling, which can only be achieved by careful
comparison of the results of independent modelling codes.

I finally note that the present surface heavy-element abundance could
be lower than the interior composition as a result of
later accretion of material less
rich in heavy elements; also, early solar mass loss has a significant effect on
the present internal sound speed \citep{Guzik2009}.
I return to the consequences of these effects,
in relation to the revised abundances, in the following section.

\subsection{Effects of accretion or mass loss}
\clabel{sec:massloss}

The solar models considered so far have all been evolved at constant mass,
neglecting any effects of mass loss or accretion.
The present rate of mass loss to the solar wind, around 
$2 \times 10^{-14} \Msun \yr^{-1}$ \citep[\eg,][]{Schrij2007}, is too
low to have a significant effect on solar evolution.
The same is true of the loss of mass resulting from the fusion of 
hydrogen to helium in the solar core.%
\footnote{Interestingly, detailed analysis of the orbit of the NASA MESSENGER
Mercury orbiter \citep{Genova2018} has determined a relative change of
$G {\rm M}_\odot$ of $(-6.13 \pm 1.47) \times 10^{-14} \yr^{-1}$,
which is consistent with the combined effect of the present solar wind and
the conversion of hydrogen into helium, thus constraining any
possible time variation of $G$.}
However, accretion or a much higher mass-loss rate in the past 
cannot {\it a priori} be excluded.

A simple way to obtain the observed surface composition,
maintaining a higher heavy-element abundance in the radiative interior
as apparently required by the helioseismic constraints,
is to postulate that the solar convection zone has been affected
by the accretion of material low in heavy elements \citep{Guzik2005};
this possibility has also been proposed in connection with 
detailed comparisons between the surface compositions of the Sun and
similar stars (see Sect.~\ref{sec:twins}).
However, it appears to be difficult to 
construct such models that satisfy all the helioseismic constraints
\citep{Guzik2006, Castro2007, Guzik2010}.
An extensive investigation of models with accretion, varying the timing
of the accretion during early solar evolution and the composition
and mass of the accreted material, was carried out by \citet{Serene2011},
comparing the GS98 and AGSS09 compositions;
the models were compared both with the helioseismic inferences and
the neutrino data.
The conclusion was that, over the extended set of parameters considered,
accretion was unable to achieve an agreement with the solar data
for models using the AGSS09 composition that matched the results for
the traditional model using the GS98 composition.

The possible effects of mass loss on the solar abundance problem
are less obvious although, as discussed below, significant.
An obvious consequence of a higher initial solar mass would be a higher
initial solar luminosity, as indicated by the luminosity scaling relation,
Eq.~(\ref{eq:homlum}); this has the potential to alleviate
the `faint early Sun problem' ({\cf} Sect.~\ref{sec:msevol}).
Also, by dragging material originally at greater depth
and hence at higher temperature into the convection zone, substantial
mass loss would change the composition of the solar surface;
in particular, it would lead to increased destruction of lithium
\citep{Weyman1965}
and increase the abundance ratio $\helthree/\helfour$
(see also Sect.~\ref{sec:lightcomp}).
\citet{Guzik1987} computed evolution sequences with exponentially
decreasing mass loss, starting at a mass of $2 \Msun$ and calibrated
to match solar properties at the present age.
They found that such high mass loss led to the complete destruction of 
lithium and beryllium, thus requiring additional processes in the near-surface
region to account for the observed abundances.
Apart from this, no obvious conflicts with the then known properties of 
the Sun were identified;
Guzik {\etal} did note that the $\helthree$ abundance on the solar surface
was strongly increased in the mass-losing models, but they did not 
consider the available observations sufficiently secure to rule
out such models.
\citet{Swenso1992} considered mass loss as an explanation of the
observed lithium abundances in the Sun and in the Hyades cluster.
In the solar case, they found that the observed present solar lithium abundance
could be accounted for with an initial solar mass of $1.1 \Msun$ and
either exponentially decreasing or constant mass loss.
A similar conclusion had been reached by \citet{Boothr1991}.

The availability of detailed helioseismic data obviously
provides further constraints on the mass-losing models.
\citet{Guzik1995} compared models with a total loss of $0.1 \Msun$, 
to match the lithium destruction, with observed frequencies from
\citet{Duvall1988}.
They concluded that such mass loss extending
over a timescale substantially exceeding $0.2 \Gyr$ was ruled out by
the observed frequencies.
Mass loss on a shorter timescale had little effect on the structure of the
present Sun;
%indeed, it is obvious that the effect on the present Sun
%arises almost entirely from the change in the composition profile,
%resulting from the mass loss, and the evolution of the composition profile
%during the first $0.2 \Gyr$ is modest.
indeed, it is obvious that early mass loss affects the structure of 
the present Sun
almost entirely from the resulting change in the composition profile,
and the evolution of the composition profile
during the first $0.2 \Gyr$ is modest.
Consequently, the computed frequencies were very similar to those
of a model without mass loss.
However, such rapid mass loss required an initial mass-loss rate of
around $5 \times 10^{-10} \Msun \yr^{-1}$, more than four orders of magnitude
higher than the present rate.

A detailed analysis of the helioseismic implications of early mass loss
was carried out by \citet{Sackma2003}.
This was motivated by the possible problem posed by the low initial
luminosity of the Sun, given evidence for liquid water on the Earth and
possibly Mars in the early phases of their evolution;
they noted that an early higher solar mass would increase the solar 
luminosity and decrease the distance between the Sun and the planets,
both leading to a higher solar flux at the Earth and Mars.
They considered three different mass-loss models, all calibrated to 
correspond to the present solar wind at solar age, and initial masses between
1.01 and $1.07 \Msun$.
The sound speed in the model of the present Sun was compared with the
helioseismic inference of \citet{Basu2000}.
\citet{Sackma2003} found that an initial mass of $1.07 \Msun$ would lead to
a flux at Mars high enough $3.8 \Gyr$ ago to be consistent with liquid water.
The effects in this case on the present solar sound-speed profile were
quite modest; in fact, mass loss slightly \emph{decreased} the difference
between the helioseismic and the model sound speed,
although the effect was not significant, given other uncertainties in
the modelling.
They noted that even with a mass loss of $0.07 \Msun$ additional mixing
would be required to account for the observed lithium depletion;
the helium isotope ratio was not discussed but is likely not significantly
affected by such a modest early mass loss.

\begin{figure}[htp]
\centerline{\includegraphics[width=\figwidth]{\fig/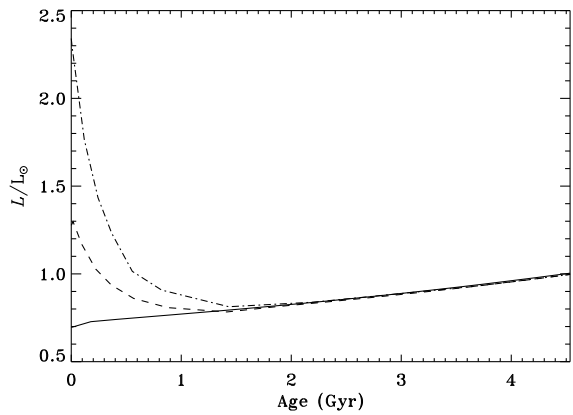}}
\caption{
Evolution with age in surface luminosity, in units of the present luminosity
of the Sun, for a model without mass loss (solid curve)
and mass-losing models with an initial mass of $1.15 \Msun$ (dashed curve)
and $1.3 \Msun$ (dot-dashed curve).
The models were calibrated to match solar properties at the present age
of the Sun. 
They were computed with the AGS05 composition \citep{Asplunetal2005b}.
Adapted from \citet{Guzik2010}; data courtesy of Joyce Guzik.
}
\clabel{fig:masslosslum}
\end{figure}

\begin{figure}[htp]
\centerline{\includegraphics[width=\figwidth]{\fig/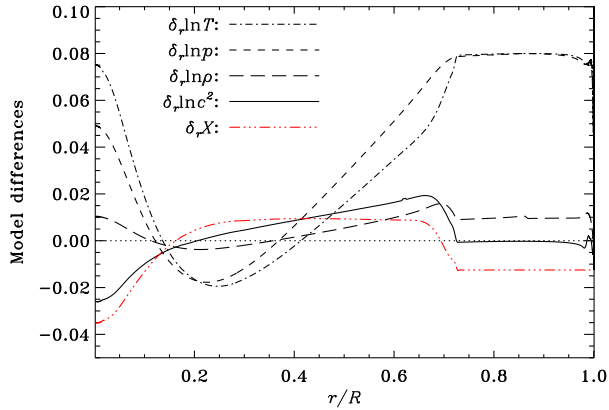}}
\caption{
Differences, at fixed fractional radius,
between models of the present Sun in a mass-losing
evolution sequence with initial mass $1.3 \Msun$ and a normal sequence, in the 
sense (mass-losing model) -- (normal model).
The line styles are defined in the figure; the dotted line marks 
zero difference.
The models were calibrated to match solar properties and the present age 
of the Sun. 
They were computed with the AGS05 composition \citep{Asplunetal2005b}.
Adapted from \citet{Guzik2010}; data courtesy of Joyce Guzik.
}
\clabel{fig:masslossdif}
\end{figure}

\citet{Minton2007} considered the mass loss required to ensure that the
average temperature of the Earth had been above freezing throughout the 
evolution of the solar system.
They found that this could be accomplished with an initial mass as
low as $1.026 \Msun$, with a resulting model at the present age which would
likely be consistent with helioseismic inferences.
However, they noted that the required mass-loss rate during the early stages
of solar evolution would have been substantially higher than the rates observed
in sun-like stars at similar stages in their evolution.
In addition, they found that solar mass loss would have had some effect on
the dynamics of the bodies in the solar system, although none with 
clear observable consequences at present.

Following \citet{Sackma2003}, \citet{Guzik2009} and \citet{Guzik2010}
investigated the effect
of mass loss on the comparison with the helioseismic sound-speed 
inferences, given the revision of the solar composition
(see Sect.~\ref{sec:modrevcomp}).
Interestingly, they found that a model with initial mass of $1.3 \Msun$
and an exponentially decreasing mass-loss rate with an e-folding time
of $0.45 \Gyr$, using the AGS05 composition, largely reproduced 
a model with no mass loss and the GN93 composition.
However, such a large amount of mass loss would bring to the solar
surface material that had been exposed to temperatures in excess of 
$5 \times 10^6 \K$, resulting in a complete destruction of lithium.
Also, the initial mass-loss rate of $6.6 \times 10^{-10} \Msun \yr^{-1}$
may be inconsistent with observations of other similar young stars.

To illustrate the effects of mass loss on solar evolution and the present
solar structure Fig.~\ref{fig:masslosslum} shows the evolution in
the surface luminosity for a normal model and two mass-losing models 
of initial mass $1.15$ and $1.3 \Msun$ \citep{Guzik2010}.
All models were calibrated to match the solar properties at the present
age of the Sun.
The mass-loss rate was assumed to decrease exponentially with age with an
e-folding time of $0.45 \Gyr$.
The initial luminosity is evidently well above the present solar luminosity
in both mass-losing models;
however, with the assumed rapid decrease in the mass loss the minimum 
luminosity is still only about 80\% of the present solar luminosity.
The effect on the structure of the model of the present Sun is illustrated
in Fig.~\ref{fig:masslossdif} for a starting mass of $1.3 \Msun$.
Comparison with Fig.~\ref{fig:ags05} confirms that such mass loss to
a large extent compensates for change in solar models caused by
the change in the surface composition,
from the \citet{Greves1993} to the \citet{Asplunetal2005b} values.

Lacking direct determinations of the early solar mass loss, constraints can
be sought from observations of young solar analogues.
Based on radio observations of such stars \citet{Fichti2017} concluded that
the total amount of mass lost by the Sun in the early phases of main-sequence
evolution was likely at most 0.4\%.
From the results discussed here this would clearly be insufficient
to compensate for the low early solar luminosity
or the change in the solar surface composition.

\begin{figure}[htp]
\centerline{\includegraphics[width=\figwidth]{\fig/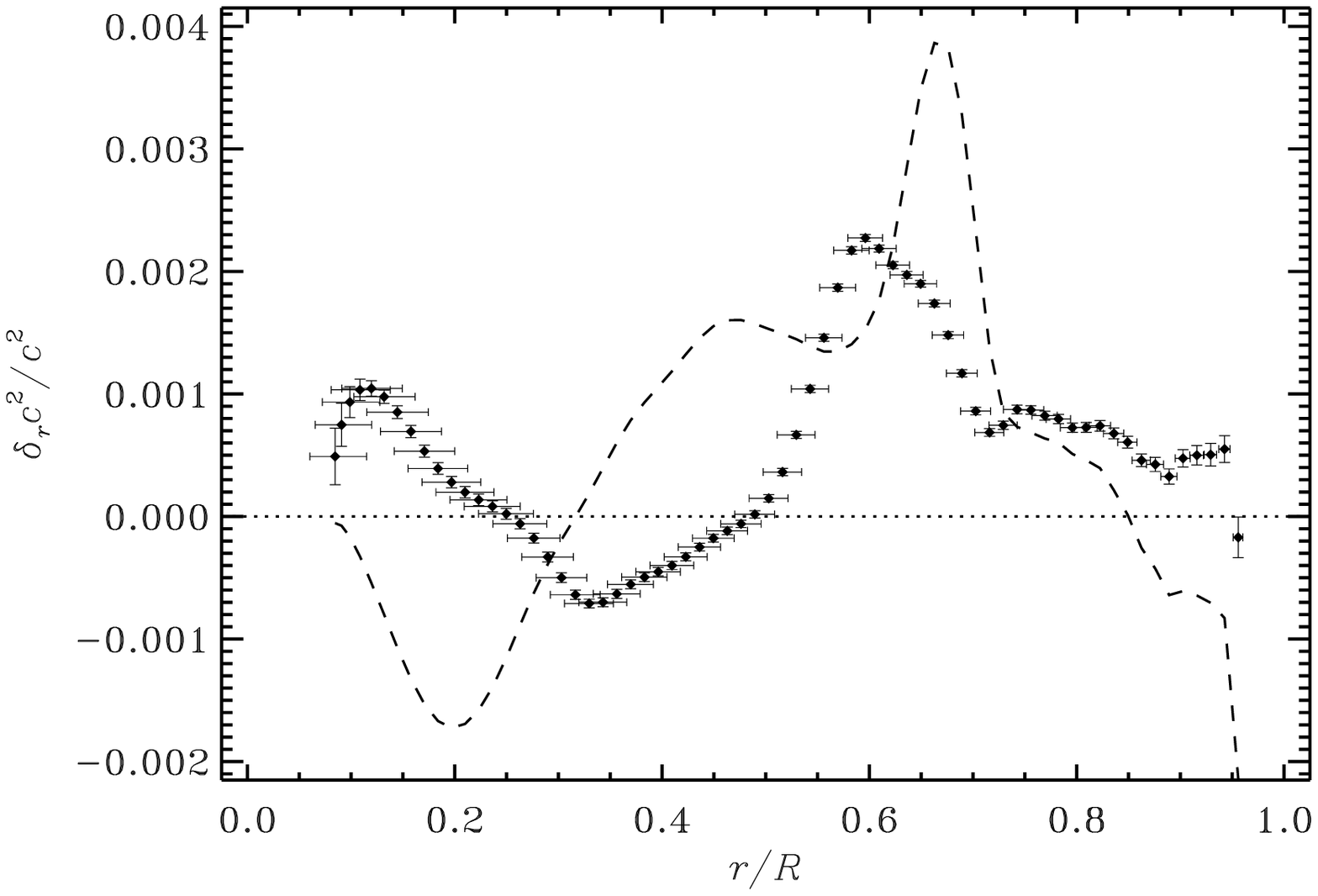}}
\centerline{\includegraphics[width=\figwidth]{\fig/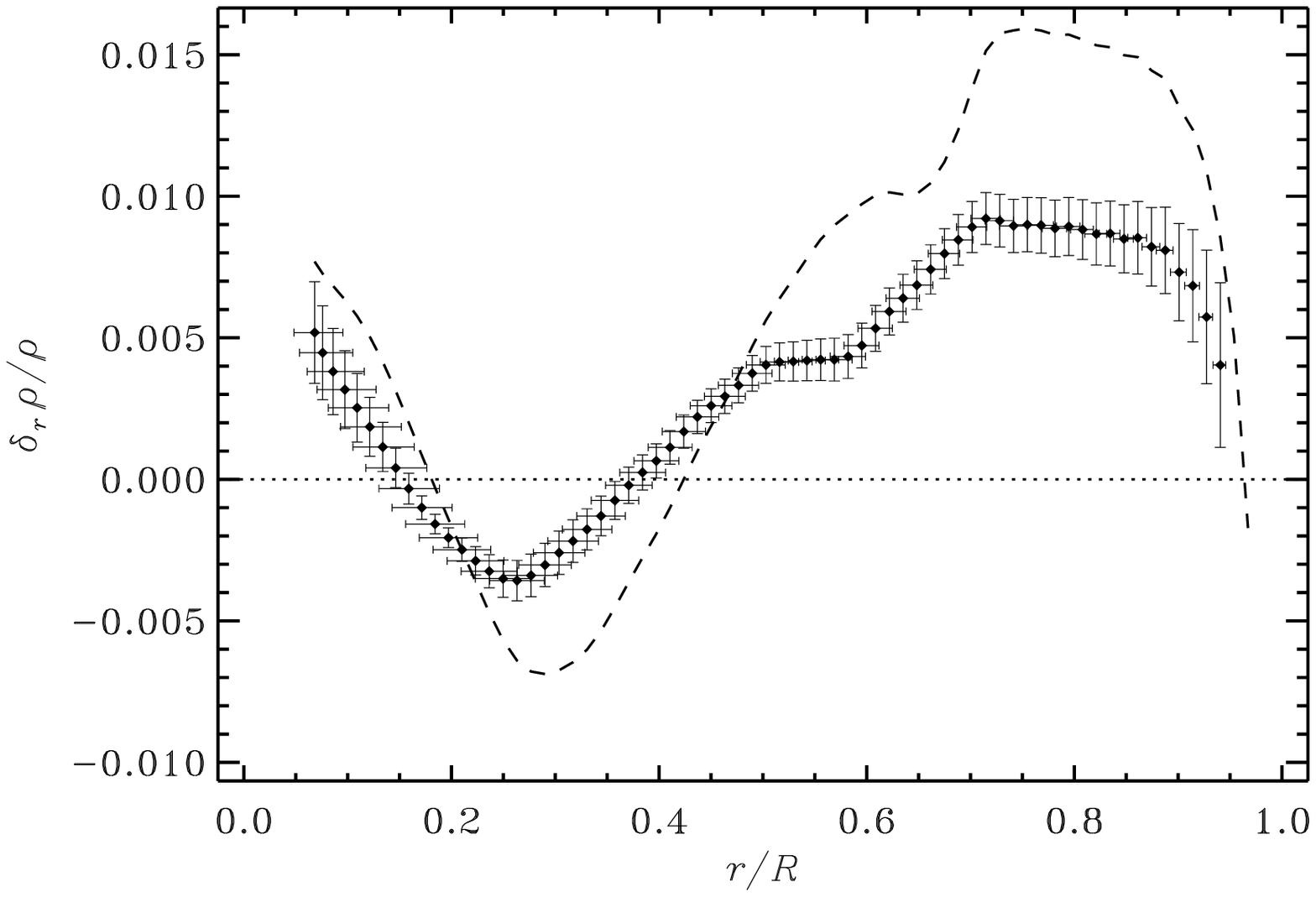}}
\caption{Results of helioseismic inversions,
for Model~TWA of \citet{Zhang2019},
including mixing below the convection zone and chemically differentiated
accretion and mass loss in early phases of stellar evolution.
The symbols show inferred relative differences in squared sound speed
(top) and density (bottom)
between the Sun and the model.
The vertical bars show $1\,\sigma$ errors in the inferred values,
based on the errors, assumed statistically independent, in the
observed frequencies.
The horizontal bars extend from the first to the third quartile 
of the averaging kernels, to provide a measure of the resolution
of the inversion \citep[see][]{Basu2016}.
For comparison, the dashed curves show results Model~S
(see also Fig.~\ref{fig:csqinv}).
}
\clabel{fig:zhanginv}
\end{figure}

A comprehensive effort to match observational data for the Sun,
given the revised solar composition, was carried out by \citet{Zhang2019},
involving both pre-main-sequence accretion and early mass loss.
In addition to the helioseismic data, the models were also fitted to the
observed lithium abundance (see also Sect.~\ref{sec:lightcomp}) and
tested against the observed neutrino data.
The models used the AGSS09 composition with the updated Neon abundance
following \citet{Young2018}.
Overshoot below the convection zone was treated using a model of the transport
of turbulent kinetic energy.
The most novel aspects of the model were the inclusion of selective and
somewhat heuristic
composition effects in the pre-main-sequence accretion and early mass loss, 
to match the detailed distribution of the helium abundance, as inferred from
the helioseismically determined sound speed.
Inferred sound-speed and density differences for the resulting so-called
Model~TWA are illustrated in Fig.~\ref{fig:zhanginv},
compared with the results for Model~S,
while overall model properties are included in Table~\ref{tab:newmodelchar}.
Even though largely using the AGSS09 composition the model clearly provides
a better match to solar sound speed and density than does Model~S,
while the convection-zone depth and envelope helium abundance are
in good agreement with the helioseismically inferred values.

%% \newpage

%===========================================================================

\section{Towards the distant stars}
\clabel{sec:stars}

Although the main focus here is the Sun, it is of course interesting to
consider broader aspects of stars, in relation to those of the Sun.
An important question in this regard is whether the Sun is in fact a typical
star.
\citet{Gustaf1998} addressed this in a paper with the title
``Is the Sun a sun-like star?''.
He answered this in the affirmative, find that the Sun is indeed typical
of stars with similar mass and age.
One important exception is that the Sun is a single {\rv star}, setting it apart
from the many stars that are in binary systems.
A second possible exception concerns the detailed mixture of heavy elements;
I return to this below.

\begin{figure}[htp]
\centerline{\includegraphics[width=\figwidthb]{\fig/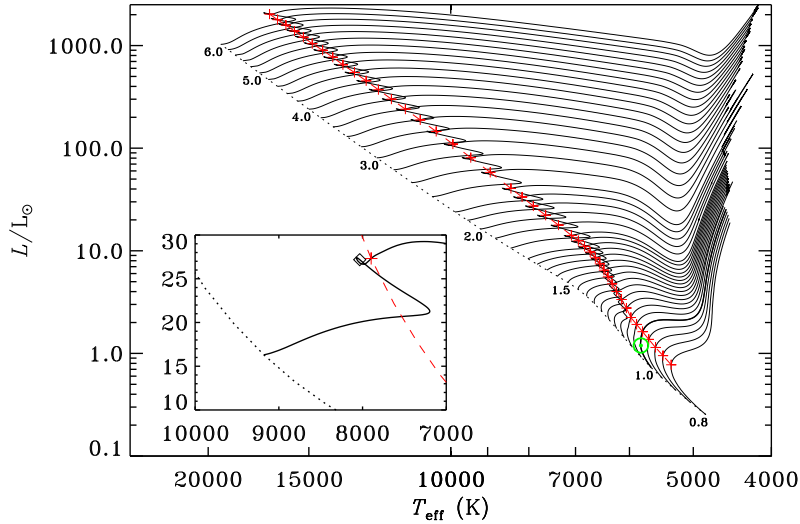}}
\caption{Evolution tracks during and just after central hydrogen burning for stellar
masses between $0.8$ and $6 \Msun$.
Selected masses are indicated in the figure. The track for $1 \Msun$ is
shown with a bolder curve, and the location of the Sun is marked by the green
sun symbol ($\odot$).
The models are characterized by an initial composition with
$X_0 = 0.7062$, $Z_0 = 0.01963$ and a mixing-length parameter
$\alphamlt = 1.8914$.
Evolution starts from chemically homogeneous models on the
\emph{Zero Age Main Sequence} (ZAMS), indicated by the dotted curve.
The red dashed curve and plusses mark the
\emph{Terminal Age Main Sequence} (TAMS),
where the central hydrogen abundance decreases below $10^{-5}$.
The inset shows the evolution track for $2 \Msun$ on an expanded scale.
Here the diamond marks the point where the convective core disappears.
}
\clabel{fig:HRMS}
\end{figure}

To place solar evolution into a broader context, Fig.~\ref{fig:HRMS} shows
evolutionary tracks for a broad range of stars, on the 
main sequence and just beyond.
%\notecd [Use simplified physics for now.
%Point out striking transition to convective core, and the uncertainties
%in where that happens].
To avoid problems with excessively rapid settling for masses only slightly
higher than solar (see Sect.~\ref{sec:diffus}),
diffusion and settling were neglected in these calculations.
Otherwise the physics corresponds essentially to what was used in
Sect.~\ref{sec:standard}, and the mixing-length parameter and initial
abundance were calibrated to obtain a model at the present age of the Sun
matching the observed properties.
In accordance with Eq.~\Eq{eq:homlum}
the luminosity generally increases with evolution during central hydrogen
burning.
However, it is evident that the qualitative behaviour of the evolution
tracks changes at a mass of around $1.15 \Msun$,
with the appearance of a `hook', where the effective temperature increases
with age for a brief period.
This reflects that such more massive stars, unlike the Sun,
have a convective core.
The convective instability 
is a result of the increasing central temperature and hence increasing 
importance of the highly temperature-sensitive CNO cycle
(see Eq.\ \ref{eq:tdep}).
This causes the energy generation to be strongly concentrated near the
stellar centre, leading to a high value of $L(r)/m(r)$ near the centre
and hence, according to Eqs~\Eq{eq:nablarad} and \Eq{eq:schwarz},
to a tendency for convective instability.
Since the convective core is fully mixed, the hydrogen abundance decreases
uniformly in the core up to the point where hydrogen disappears in all,
or much of, the region where the temperature is high enough to allow
nuclear burning.
In the last phases of central hydrogen burning this causes a contraction of
the entire star to drive up the central temperature in order to maintain
the luminosity, hence leading to the increase in the effective temperature.
This behaviour stops when the energy generation is taken over by hydrogen
burning in a shell around the hydrogen-depleted core;
as in the lower-mass stars the surface radius of the star increases with 
evolution and the effective temperature therefore drops.

%\notecd [Briefly on modelling (very briefly!). Main point to emphasize here
%is the potential to study more general aspects of stellar structure
%and evolution, particularly phenomena not seen in the Sun, and with
%emphasis, of course, on asteroseismology.]

It is evident that much of the detailed discussion of solar modelling and
evolution presented in this paper is immediately relevant to other stars.
Indeed, a key aspect of the helioseismic investigations of the solar
interior is the ability to test the theory of stellar structure and evolution
in very considerable detail.
Also, the Sun is in many ways an ideal case for such tests,
even apart from the obvious advantage of its proximity.
Compared with most other stars its properties are relatively simple.
It has had no convective core during the bulk of its main-sequence evolution.%
\footnote{A convective core briefly appears in the final stages of the
pre-main-sequence evolution as the CNO cycle ({\cf} Eq.~\ref{eq:CNO})
reaches equilibrium \citep[{\eg},][]{Morel2000}.}
It is slowly rotating, so that rotation has no obvious immediate consequences
for the structure of the present Sun.
The physical conditions of matter in the Sun are relatively benign,
the departures from the ideal-gas equation of state being modest although still
large enough to be investigated with helioseismology.
Thus it is perhaps not unreasonable to hope that even our simple models
can give a reasonable representation of the properties of the solar 
interior, and this indeed seemed to be the case, as discussed in
Sect.~\ref{sec:heliostruc}, at least until the revision of solar
abundances (see Sect.~\ref{sec:modrevcomp}).

%\notecd [Discuss obvious need for more complex modelling; this could be
%inspired by GONG06 paper].

Such complacency is clearly naive, however, given the potential of
the solar interior for complexities far beyond our simple models.
As discussed in Sect.~\ref{sec:heliorot} the origin of the present
internal solar rotation is not understood.
It is very likely that the phenomena leading to the present near-uniform
rotation of the solar radiative interior has had some effect also on
solar structure, for example through associated mixing processes.
Also, it should be kept in mind that even the relatively successful
models, such as Model~S discussed extensively here, show a highly significant
departure from the helioseismic inferences ({\cf} Fig.~\ref{fig:csqinv}).
However, it is perhaps mainly the consequences for solar models
of the revision of the solar
composition that has served as a wake-up call for reconsidering the basics
of solar modelling.
As discussed by \citet{Guzik2006} there seems to be no 
straightforward way to reconcile normal models computed with this composition
with the helioseismic inferences.
This should motivate looking for more serious flaws in our understanding
of stellar structure and evolution.

Abundances of solar-like stars are often measured relative to those of the Sun. 
Thus, the modifications to the inferred solar abundances
discussed in Sect.~\ref{sec:abundprob} affect
also the modelling of other stars.
As an example, \citet{Vanden2007} noted that isochrones for the
open cluster M67, computed based on the AGS05 solar composition,
provided a worse match to the observed colour-magnitude diagram
than did models based on the GS98 composition.
Specifically, the best-fitting isochrone lacked the hook near
the end of central hydrogen burning.
%which reflects the presence
%of a convective core (see Sect.~\ref{sec:stars}).
Such a hook is found with the GS98 composition and appears to
be reflected in the observations.
In this case the dominant consequence of the change in the composition
is the decrease in the importance of the CNO cycle in hydrogen burning 
resulting from the reduced abundances, and hence a reduced tendency
for convective instability in the core.
It was pointed out, however, by \citet{Magic2010} that favouring GS98
on this basis depended critically on other assumptions in the modelling. 
Including, for example, diffusion and settling (which was not taken into
account by VandenBerg {\etal}) the GS98 and AGS05 models were equally
successful in reproducing the hook, while other aspects of the modelling
similarly had substantial effects on the morphology of the isochrones;
effects on the properties of convective cores of composition and 
other aspects of the model physics were also investigated 
by \citet{Christ2010}.
Thus, although the details of the morphology is an interesting diagnostics 
of the model physics, it does not provide a definite constraint on any one
feature such as the composition.

One obvious failing of standard modelling is that rotation is ignored.
The dynamical effect, resulting from the centrifugal force, is relatively 
straightforward to include, assuming that the rotation rate is given,
at least for relatively slow rotation allowing a spherical approximation
with a modified equation of hydrostatic equilibrium.
For more rapid rotation departures from spherical symmetry must be
modelled explicitly.
This is the goal of the ESTER project 
\citep[Evolution STEllaire en Rotation;][]{Espino2013, Rieuto2016},
to carry out fully self-consistent two-dimensional calculations 
of stellar structure.
A recent example is the modelling of the rapidly rotating star Altair 
\citep{Boucha2020},
for which detailed interferometric observations are available on the surface
distortion and temperature variations induced by rotation.

Even more difficult is the treatment of circulation and instabilities
associated with rotation, and of the evolution of the internal 
angular velocity and associated transport processes, 
which is still far from fully understood.
\citet{Zahn1992} developed a simplified, if hardly simple, treatment of these
processes which has seen extensive use in computations of the evolution of
massive stars 
\citep[for a review, see][]{Maeder2000}
and has been further developed
by, for example, \citet{Maeder1998} and \citet{Mathis2004}.
Effects on these processes from diffusion-induced gradients in the
mean molecular weight were considered by \citet{Theado2003a, Theado2003b},
while \citet{Talon2005} developed a combined treatment of the effects
of rotation, internal gravity waves and atomic diffusion.
\citet{Maeder2009} provided a comprehensive discussion of
the effects of rotation on stellar evolution.
Transport by gravity waves was proposed by \citet{Schatz1993, Schatz1996}
and has been extensively discussed in connection with solar internal rotation
({\cf} Sect.~\ref{sec:heliorot}).
As discussed there, effects of magnetic fields are also likely to be relevant.
Ambitious efforts to include all these effects in stellar modelling
were discussed by \citet{Mathis2006} and \citet{Palaci2006}
\citep[for a recent overview, see][]{Aerts2019}.

%\notecd [\citet{Charbl2005} on rotation and Li abundances of solar-type stars.
%\citet{Theado2003c} on abundances of Li and Be in main-sequence stars].

Observational tests of these models obviously require considerations of
stars other than the Sun.
An important constraint comes from the dependence of stellar surface
rotation on the mass and age of the star,
which may provide additional constraints on the, so far somewhat
uncertain, processes responsible for the evolution of the solar internal 
rotation (see Sect.~\ref{sec:heliorot}). 
Additional information comes from the stellar surface abundances and
their dependence on stellar types which reflect the mixing processes
in the stellar interiors, possibly associated with the evolution of rotation.
Particularly important are the abundances of lithium and beryllium
(see also Sect.~\ref{sec:lightcomp});
since these elements are destroyed by nuclear reactions at relatively
modest temperature,
their abundances provide stringent constraints on
the depth to which significant mixing has occurred
(see also Sections~\ref{sec:basicpar} and \ref{sec:lightcomp}).
\citet{Theado2003c} showed that the dependence on effective temperature
of the lithium and beryllium abundances in stars in the Hyades cluster 
could be well explained in a model combining rotationally induced mixing
with an appropriate treatment of the gradient in the mean molecular weight
resulting from helium settling.
Also, \citet{Charbl2005} showed that modelling the evolution of rotation
by gravity-wave transport could account for the dependence of 
lithium depletion on stellar age.

\citet{Israel2009} found an interesting possible relation
between enhanced lithium depletion and the presence of planets
around Sun-like stars, including the Sun.
\citet{Bouvie2008} related this to the rotational history of the stars;
he suggested that the planet formation could be related to locking 
to a long-lived proto-planetary disk which would lead to slow rotation 
of the outer layers of the star and hence a strong internal rotation gradient,
causing mixing and lithium destruction.
In a careful study of solar-twin stars, however, \citet{Carlos2016}
found a strong correlation between lithium abundance and age but no
indication of enhanced depletion in planet hosts.
Even so, a close connection was found by \citet{Bouvie2018} between
rotation rates, determined photometrically, of stars in the Pleiades cluster
and their Li abundances, with slowly rotating stars showing a stronger Li
destruction;
this provides some support to the relation inferred by \citet{Bouvie2008}
between long-lived disk locking, rotation and the Li destruction and points to
the importance of such abundance studies in investigations of stellar
evolution.
It should be noted that the general issue of lithium destruction
has an important relation to cosmology,
given the observed nearly uniform deficiency of lithium in halo stars
compared with the predictions of Big Bang nucleosynthesis
\citep[for a review, see][]{Cyburt2016}, perhaps raising questions
about the cosmological models.
However,
a detailed analysis by \citet{Korn2006, Korn2007} of abundances in the globular
cluster NGC\,6397 demonstrated the importance of settling and 
turbulent mixing for the lithium abundance in old metal-poor stars;
they concluded that these processes can account for a previously 
inferred discrepancy between the observed abundances in such stars and
the predictions of Big Bang nucleosynthesis.

\begin{figure}[htp]
\centerline{\includegraphics[width=\figwidth]{\fig/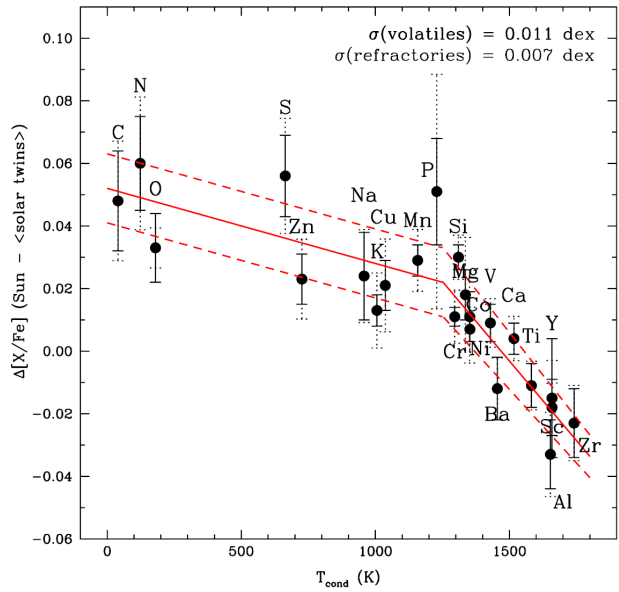}}
\caption{Logarithmic differences ({\cf} Eq.~\ref{eq:feh}) 
between the solar surface abundances, normalized to
iron, and the averages of stars identified as being `solar twins'.
The abscissa shows the condensation temperature \citep{Lodder2003}
of the corresponding element in the proto-solar nebula.
\citep[From][]{Melend2009}.
%\notecd [Copyright issue here.]
}
\clabel{fig:solarabund}
\end{figure}

\subsection{Solar twins}
\clabel{sec:twins}

%\notecd [This paragraph needs updates, following Bengt Gustafssons talk, etc;
%{\cf} Bengt's ISSI paper: is the Sun a solar-like star?
%Check Nissen papers].
A very interesting particular class of stars are the so-called `solar twins',
{\ie} stars with properties very similar to those of the Sun.
Very interesting analyses have been carried out comparing 
the solar surface composition with such stars,
benefitting from the development of very precise techniques for stellar
abundance determinations \citep[see, {\eg},][for a review]{Nissen2018}.
%\citep[\eg,][]{Melend2006, Melend2007}.
Specifically, a solar twin is defined by requiring that
the effective temperature, gravity and metallicity,
characterized by [Fe/H], should agree with the Sun to within one standard
deviation.
Here the logarithmic abundance difference is defined by
\be
{\rm [A/B]} = \log(N_{\rm A}/N_{\rm B})_* - 
\log(N_{\rm A}/N_{\rm B})_\odot \; ,
\eel{eq:feh}
where $N_{\rm A}$ and $N_{\rm B}$ are the abundances of elements A and B,
$\log$ is logarithm to base 10 and the difference is between the stellar and
solar values.
Fixing thus the iron abundance relative to hydrogen to the solar value,
for a set solar twins 
\citet{Melend2009} and \citet{Ramire2009} compared abundances for
other elements, relative to iron,
with the corresponding solar abundances.
As illustrated in the example in Fig.~\ref{fig:solarabund}
this showed a highly systematic dependence on the condensation temperature
of the element.
%strongly indicating a relation between the solar composition
%and the formation of the solar system.
Mel\'endez {\etal} related this to the formation of the solar system.
Specifically, if planetary systems are not generally found in the solar twins,
condensation leading to
the formation of the solar-system planets may have depleted the
material accreting on the proto-sun of refractory elements, leading to
the observed dependence of the solar abundance deficit on condensation
temperature.
The effect of accretion on the final solar composition depends
critically on the mass contained in the convectively mixed region
during the relevant accretion phase.
In most models of pre-main-sequence evolution the star goes through a
fully convective phase (see also Sect.~\ref{sec:pmsevol})
which would require an unrealistically large 
amount of material condensated in the form of rocky planets or planet cores
to account for the observed solar composition depletion.
This led \citet{Melend2009} to propose a rather late accretion,
at a point in the evolution of the proto-sun where the convective
envelope had reached approximately the present extent.
Alternatively, \citet{Nordlu2010} recalled the detailed modelling
of pre-main-sequence evolution by \citet{Wuchte2003}
which indicated that
the convective envelope did not involve a large fraction of the stellar
mass during the accretion phase,
as also found by \citet{Baraff2010} in models with episodic infall
(see also Sect.~\ref{sec:pmsevol}).
This might lead to a sufficient depletion of the convection-zone 
abundance with realistic condensation.
Nordlund also noted that the resulting difference between the solar surface
composition and the composition of the radiative interior might resolve
the conflict between the effect on solar models
of the composition revision by \citet{Asplun2005, Asplun2009}
and the helioseismically inferred solar structure
(see Sect.~\ref{sec:modrevcomp}; 
I recall, however, that such models apparently do not provide a fully
satisfactory solution to the discrepancy).
%\citep[{\eg},][]{Guzik2006, Castro2007}).
%however, this certainly deserves further detailed investigation.
The results of \citet{Melend2009} were confirmed by the analysis
of a much larger sample of stars by \citet{Bedell2018}, 
who also pointed to a possible connection with the existence of
the solar system.
I note that this argument is somewhat weakened by the ubiquitous presence
of planets inferred by the {\it Kepler} mission
\citep[\eg,][]{Batalh2014}, although, as pointed out by Bedell {\etal}, 
a planetary system matching the properties of the solar system
has yet {\rv to be} found.
As an alternative explanation \citet{Gustaf2018} suggested that 
material accreted in the later phases of solar formation could have
been depleted in refractory material through cleansing of dust by the
effect of solar radiation on the dust grains.
A more detailed discussion of these composition differences and 
the proposed explanations was provided by \citet{Nissen2018}.

%\notecd [End with the obvious point that studying other stars will in the
%end inform our modelling of the Sun, by breaking parameter degeneracies,
%or whatever].

%\notecd [Also, possibly return to hope of some constraints on stellar internal
%rotation, although limited (until Stellar Imager!)].

\subsection{Asteroseismology}
\clabel{sec:astero}

Despite the importance of the abundance studies it is evident that 
observations with more direct sensitivity to stellar interiors would be
very valuable.
As in the case of helioseismology, the study, known as \emph{asteroseismology},%
\footnote{For the etymology of this nomenclature, see \citet{Gough1996b}.} 
of stellar interiors from observations of oscillations provides such
a possibility.
Extensive reviews of asteroseismology were provided by, for example,
%\citet{Kjelds2004}, 
\citet{Cunha2007a} and \citet{Aerts2010},
while \citet{Chapli2013} discussed solar-like oscillators and
\citet{Hekker2017} considered the very interesting seismology
of red giants.
A review of the field was provided recently by \citet{Garcia2019}.

%\notecd [The following clearly needs substantial updates].
Although stellar properties have been investigated by means of observations
of stellar oscillations for several decades
\citep[{\eg},][]{Peters1973, Bradle1994},
the field has developed rapidly in recent years owing to large-scale
observational projects and new observing techniques.
Particularly dramatic has been the development of observations of
solar-like oscillations.
A major breakthrough came with new spectroscopic techniques that enabled
the analysis of oscillations in radial velocity with amplitudes 
of a few $\cm \s^{-1}$ \citep[{\eg},][]{Kjelds2005}.
%\notecd [Possibly be more precise, optimistic; a reference or two].
Missions for space-based high-precision photometry combining the search
for extra-solar planets (exoplanets) using the transit technique
with asteroseismology have revolutionized stellar astrophysics.
The {\it CoRoT}\,%
\footnote{{\bf Co}nvection, {\bf Ro}tation and {\bf T}ransits}
satellite \citep{Baglin2009, Baglin2012},
launched in 2006 and operating until 2012,
yielded asteroseismic data for a substantial number of stars.
Much more extensive data were obtained from the NASA {\it Kepler mission}
\citep{Gillil2010, Boruck2016}
%\citep[{\eg},][]{Christ2007b, Christ2008b}
which was launched in March 2009 into an Earth-trailing heliocentric orbit.
It operated in the nominal mission observing one field in the Cygnus-Lyra
region until May 2013, when the second of four reaction wheels failed;
since then it was repurposed to the K2 mission \citep{Howell2014},
observing a large number of fields along the ecliptic for around 80 days each.
The mission was finally stopped in October 2018, when the spacecraft ran
out of fuel.
The TESS%
\footnote{{\bf T}ransiting {\bf E}xoplanet {\bf S}urvey {\bf S}atellite}
mission \citep{Ricker2014}, launched in April 2018,
is surveying about 80 per cent of the sky,
emphasizing relatively bright and nearby stars in a search for exoplanets and
carrying out asteroseismology of a large number of stars.
In the slightly more distant future very extensive studies, coordinating 
investigations of extra-solar planetary systems and asteroseismic studies
of stellar properties, will be carried out with the ESA PLATO%
\footnote{{\bf PLA}netary {\bf T}ransits and {\bf O}scillations of stars.}
mission \citep{Rauer2014}, which was adopted in 2017 
for a planned launch in 2026.
%analyses of the initial asteroseismic observations from the mission
%exceed expectations \citep[for a review, see][]{Gillil2010}.
%Further improvements, particularly in the data quality, will result from
%the planned SIAMOIS%
%\footnote{{\bf S}eismic {\bf I}nterferometer {\bf A}iming to
%{\bf M}easure {\bf O}scillations in the {\bf I}nterior of {\bf S}tars}
%project to carry out observations from Antarctica
%\citep{Mosser2008} and from

Even given the huge advances provided by the space-based photometric
observations, ground-based radial-velocity observations still offer
important advantages, particularly in terms of the ratio between the 
oscillation signal and the stellar background noise, which is much higher
for radial velocity than for photometric observations of solar-like
oscillations \citep{Harvey1988, Grunda2007}.
Also, with a dedicated network of telescopes extended observations can be
obtained for particularly interesting stars.
This is the goal of the planned
8-station SONG%
\footnote{{\bf S}tellar {\bf O}bservations {\bf N}etwork {\bf G}roup}
global network dedicated to asteroseismology
\citep{Grunda2014} which is under development.
Currently (2020) one node of the network, 
at Observatorio del Teide on Tenerife, in collaboration with
Instituto de Astrof\'{i}sica de Canarias has been in operation since 2014;
one remarkable result is several hundred nights of observations of the
subgiant $\mu$ Her \citep{Grunda2017}.
Two additional nodes are under development in China and at
University of Southern Queensland, Australia,
while collaboration is sought for additional nodes.

In the foreseeable future the lack of spatial resolution in general
limits observations of stellar oscillations to modes of spherical harmonic
degree of at most 3.%
\footnote{An exception is observation of rapidly rotating stars,
where \emph{Doppler imaging} allows study of modes of higher degree
\citep[see][]{Aerts2010}.}
At a very basic level the oscillation frequencies scale as $t_{\rm dyn}^{-1}$
({\cf} Eq.~\ref{eq:tdyn}), {\ie}, as $\bar \rho^{1/2} \propto M^{1/2} R^{-3/2}$,
where $\bar \rho$ is the mean density of the star.
In particular, for solar-like oscillations,
{\ie}, acoustic modes of high radial order,
the large frequency separation
({\cf} Eq.~\ref{eq:larsep}) satisfies $\Delta \nu \propto \bar \rho^{1/2}$.
Also,
Eq.~\Eq{eq:turn} shows that these are the modes which penetrate most deeply
and hence provide information about the stellar core.
The change in sound speed resulting from the fusion of hydrogen to helium
affects the frequencies in a manner that provides information about the
evolutionary state of the star and hence its age
\citep{Christ1984b, Christ1988b, Ulrich1986};
the sensitivity to the central composition
is reflected in the dependence of the small frequency separation
on an integral of the sound-speed gradient, weighted towards the centre
({\cf} Eq.~\ref{eq:smlsep}),
although the determination is obviously affected by other 
uncertainties in the stellar modelling \citep{Gough1987}.
These properties make solar-like oscillations powerful tools for
determining the global properties of stars, {\ie}, mass, radius and age,
which are very important for the characterization of exoplanetary systems
\citep[for recent reviews, see][]{Christ2018b, Lundkv2018}.
\citet{Lebret2014} made a careful analysis of asteroseismic data for
a star observed by CoRoT and demonstrated that, combining these with 
`classical' observations, precise estimates of the mass and age of the star
could be obtained.
\citet{SilvaA2015} carried out a comprehensive asteroseismic analysis 
of stars detected as exoplanet hosts by {\it Kepler},
demonstrating that precise stellar parameters could be obtained.
Also, the so-called LEGACY sample of {\it Kepler} stars, selected as
being particularly well-observed for asteroseismology, was the basis of
extensive analyses by \citet{Lund2017} and \citet{SilvaA2017}.
This sample will undoubtedly form the basis for further investigations of
the detailed properties of these stars.

The sharp gradient in composition and hence sound speed at the
edge of a convective core has distinctive effects on the frequencies
%The latter effects will be within reach of the space observations that are now
%under way
\citep[{\eg}][]{Mazumd2006, Cunha2007b}.
As has already been found in solar data (see Sect.~\ref{sec:heliostruc})
sufficiently extensive observations will also be sensitive to effects of
other such acoustic glitches,
{\ie}, aspects of the structure of the star which vary on a scale short 
compared with the wavelength of the oscillations;
examples are effects of helium ionization on $\Gamma_1$ and the
change in the sound-speed gradient at the base of a convective envelope
\citep[{\eg},][]{Perez1998, Montei2000, Ballot2004, Verner2006b}.
A careful analysis of this type of investigation was provided by
\citet{Houdek2007a},
with particular emphasis on the determination of the envelope helium abundance.
Through constraining other aspects of the star,
such analyses may also help reducing the systematic errors in the
determination of stellar age
\citep{Montei2002, Mazumd2005, Houdek2007b, Houdek2011}.
\citet{Mazumd2014} identified acoustic glitches associated with both
the base of the convective envelope and the second helium ionization zones
in a number of stars observed by {\it Kepler}, while
\citet{Verma2014} used the helium glitch to determine the helium abundance
in the solar analog binary 16 Cyg, observed by {\it Kepler}.
In a remarkable analysis \citet{Verma2017} used the acoustic glitches to
determine the depth of the convective envelope and the helium abundance
in the {\it Kepler} LEGACY stars.
Such a largely independent determination % of the helium abundance
is very valuable in breaking the degeneracy in fits to asteroseismic data
between the mass and the helium abundance,
implicit in the relation (\ref{eq:homlum}) for luminosity in terms of
mass and mean molecular weight.
In an interesting application, \citet{Verma2019} used determinations of
helium abundances in three stars with masses around $1.4 \Msun$ to constrain
the extent of extra mixing below the convective envelope required to
counteract helium settling (see also Sect.~\ref{sec:diffus}).

Investigation of internal rotation based on just low-degree modes is 
restricted by the small number of $m$ values available and the limited 
sensitivity of the frequencies to rotation in the deep interior
\citep[{\eg},][]{Lund2014b}.
However, determination of the rotational splitting provides an average
of the rotation rate of the stellar interior; combined with measurement 
of the surface rotation rate, {\eg}, from photometric variations induced by
spots, this can give some information on the variation of rotation with position
in the star and hence on the effects of the evolution of internal rotation.
Also, the relative amplitudes of the different $m$ components provide 
information about the inclination $i$ of the rotation axis, if the average
intrinsic amplitude is assumed to be independent of $m$
\citep{Gizon2003, Ballot2006}.
This has been used to study the inclination between the rotation axis and 
the orbital plane for exoplanets detected using the transit technique
\citep[{\eg}][]{Huber2013, Lund2014a, Campan2016}.
\citet{Benoma2015} determined the mean interior rotation rate from
observations of rotational splitting and combined that with spectroscopic
measurements of $v \sin i$, for 22 main-sequence stars observed by
{\it Kepler}, with $i$ determined from the asteroseismic data.
In most cases the results were consistent with no variation of angular
velocity between the surface and the interior.
Interestingly, this is essentially consistent with the properties of solar
rotation, as inferred from helioseismology ({\cf} Sect.~\ref{sec:heliorot}).
For completeness I note that in more evolved stars, such as subgiants and 
red giants, modes with a mixed character between p and g modes allow
detailed determination of the rotation of the deep interiors of the stars,
showing an increasing ratio between the core and envelope rotation rate,
although far less drastic than predicted by models of the evolution of
stellar rotation \citep[see][for reviews]{Chapli2013, Hekker2017}.

In some cases the {\it Kepler} data were sensitive to the
dependence of the rotational splitting on $m$, leading to constraints on the
variation of angular velocity with latitude.
In this way \citet{Benoma2018} showed the presence of latitudinal 
differential rotation in some stars, in the same sense as in the Sun,
{\ie}, with the equator rotating more rapidly than the poles.
Combining asteroseismic measurement of the differential rotation with 
rotation periods from photometric variations induced by starspots
\citet{Bazot2018} inferred the presence in a {\it Kepler} star of
cyclic activity variation, including the shift of the preferred
latitude of starspots, qualitatively similar to the solar butterfly
diagram (see Sect.~\ref{sec:tempchange}), although with a much shorter period.
{\rv These investigations of stellar rotation are clearly important also
for a broader understanding of how the rotation depends on stellar 
properties, with a mutual interplay between the deep probing of solar
rotation and the broad investigation of stars of different mass and
evolutionary stage.}

More detailed information about the variation of rotation with depth and
latitude {\rv in distant stars} 
will require observations with spatial resolution.
Such observations are planned with the interferometric {\it Stellar Imager}
\citep{Schrij2007, Carpen2009}, now under concept study as a NASA project.
This would allow observation of modes of degree as high as 60 in selected
stars and hence inference of the rotation rate in the entire radiative
interior of a star as the Sun, including a possible tachocline
(see Fig.~\ref{fig:introt}).
Such observations are crucial for the study of the effect of the dynamics
of the base of the convection zone on the dynamo mechanism likely responsible
for stellar cyclic activity.
Needless to say, such observations of modes of relatively high degree would
also revolutionize investigations of stellar internal structure.

%\notecd [Here need some famous last words, possibly as a separate section].

%% \newpage

\section{Concluding remarks}
\clabel{sec:concluding}

When I started my PhD-studies in 1973 in Cambridge
under the supervision of Douglas Gough very little was known about the
solar interior.
The apparent deficit of solar neutrinos in the Davis experiment was 
a serious concern, leading to a range of proposals for possible changes
to solar and stellar modelling, with potentially important consequences
for our general understanding of stellar evolution.
The initial goal of my project was to carry out more reliable calculations
of the stability of the Sun towards g-mode oscillations,
which might affect solar structure and decrease the computed neutrino flux.
Part of this was to develop a more accurate code to calculate solar
evolution.

The direction of the project changed drastically with the first announcements
of possible global solar oscillations, 
into what became part of the early development of helioseismology.
As will be clear from this review, the results of this development have
fundamentally changed our investigations of the solar interior and,
as a result, our general knowledge about stellar evolution.
We now know the structure of most of the present Sun, as characterized by,
for example, the sound speed and density to a remarkable accuracy.
In parallel, the improved understanding of the properties of neutrinos
and new experiments to detect them have advanced the study of solar 
neutrinos to a point where the measurements of the neutrino flux 
provide additional very valuable information about the properties 
of the solar core.
Strikingly, at the level of precision often applied in astrophysics, 
the agreement between models and the observationally inferred properties
is reasonable, typically within a few per cent.
This applies to models where no direct attempts have been made to 
adjust parameters to match the observations, 
apart from the classical calibration of initial composition and treatment
of convection to obtain the correct radius, luminosity and overall
surface composition of the model of the present Sun.

A fascinating aspect of these solar investigations is that the
accuracy and information content of the data on solar oscillations is
far higher than most astrophysical data.
This makes it meaningful to use the observations as deep probes of the 
physics of the solar interior.
This sensitivity also makes the Sun a potential detector of more
esoteric physical effects, such as the effects of dark matter.
In fact, the accuracy and agreement with the observations that have been
reached in solar modelling is very far from matching the accuracy of
the observations. 
A striking example is the thermodynamical properties of solar matter,
where models based on the current sophisticated treatments
still do not match the observations.
An additional open issue that has emerged in the last two decades is
the revision of determinations of the solar surface composition, which
has increased the discrepancy between the models and the helioseismic
inferences and cast doubt on the calculation of opacities
or other aspects of solar modelling.

Indeed, it should be no surprise that current simple solar models
are inadequate;
the surprise is perhaps rather that they work as comparatively well as they do.
The models neglect a number of physical processes that must have been
active in the Sun during its evolution and still affect it.
This includes the evolution of solar rotation, involving redistribution
of angular momentum and likely flows that would change the composition
structure of the Sun.
Also, magnetic fields are typically neglected, 
yet they could have a significant effect on the structure and dynamics
of the solar interior.
The next steps in the investigation of the Sun will surely involve models
that take such effects into account, considering the interplay between the
structure and dynamics of the solar interior.
Here insight into the relevant physical processes can be sought in
increasingly, but far from fully, realistic detailed hydrodynamical
simulations.
An important issue is to understand the origin and effects of the solar
cyclic magnetic activity and the extent to which it involves larger parts
of the Sun.
Also, the helioseismic data accumulated over the preceding more than two
decades have very far from been fully exploited and offer excellent 
opportunities for tests of such refined modelling.
New data-analysis techniques are required to make full use of the data,
including understanding their statistical properties and the extent to which
the resulting conclusions are significant.

From the understanding of solar oscillations as caused by stochastic 
excitation by near-surface convection follows that all stars with outer
convection zones are expected to show similar oscillations;
the question is whether or not they are detectable.
Early detections of such oscillations were made with ground-based
spectroscopic observations starting in the 1990s, but the real breakthrough
and a revolution in asteroseismology based on solar-like oscillations,
starting in 2007,
came with the photometric space missions CoRoT and {\it Kepler}.
We now have extensive data on stellar oscillations for a very broad range
of stars in all evolutionary phases, and hence an excellent possibility
to study stellar evolution over a broad range of parameters in mass and
chemical composition, as a complement to the detailed investigations in
the solar case.
This will undoubtedly also improve our understanding of the Sun,
perhaps providing pointers towards the resolution of the discrepancies
caused by the revised determinations of the solar surface composition.
As for the Sun, the detailed exploitation of these data is just beginning,
and there is a huge potential for continuing investigations, in parallel
with the improvements in the techniques for modelling, and leading
to a continued development in our understanding of solar and stellar
structure and evolution.

The fields of solar and stellar astrophysics are very much alive;
and so, therefore, should this review be.

%% \newpage

%===========================================================================

\begin{acknowledgements}
\clabel{sec:acknowledgements}

I am very grateful to Douglas Gough for introducing me, as his PhD student,
to stellar structure and evolution and involving me in the development
of helioseismology, thus forming the basis for this review.
Many colleagues have contributed through discussions or specific
contributions to the writing of the review, and the following will
unavoidably be an incomplete list.
I am grateful to Werner D\"appen and Regner Trampedach
for discussions on the equation of state and the opacity, 
and to H.~M. Antia for the data illustrated in Fig.~\ref{fig:kapder}.
G\"unter Houdek is thanked for extensive discussions on the effects
of convection on pulsations and Nick Featherstone for contributing the
result on large-scale hydrodynamical simulations in Fig.~\ref{fig:gwcon}.
I thank Sarbani Basu for fruitful collaboration on solar structure inversions
and for contributing the data shown in Fig.~\ref{fig:basucsqrho}.
Rachel Howe is thanked for collaboration over many years 
on helioseismic analysis of solar rotation, and for providing
Fig.~\ref{fig:rhzonal},
and Roger Ulrich is thanked for a useful correspondence on solar
torsional oscillations.
I thank Tami Rogers for interesting discussions on the transport
of angular momentum in the solar interior,
{\rv and Travis Metcalfe and Jen van Saders for discussions on
gyrochronology}.
Remo Collet and Poul Erik Nissen are thanked for useful discussions on the
solar composition, and Bengt Gustafsson is thanked for valuable insight
into the issues of the composition of solar twins.
I thank Kai Zuber for organizing
the 5th International Solar Neutrino Conference, Dresden,
and for inviting {\rv me} to give a talk there; 
this provided
a great deal of information and inspiration on the study of solar neutrinos.
I am grateful to G\"unter Houdek and Werner D\"appen for providing the
equation of state and opacity packages used in the solar-model calculations
and to Maria Pia Di Mauro for providing the 
helioseismic structure inversion code that has been used extensively here.
Aldo Serenelli and Francesco Villante are thanked for helpful discussions
on solar modelling, particularly the neutrino aspects, and for
contributing Figs~\ref{fig:opcomp}, \ref{fig:neutobs}, \ref{fig:vincsq} and
\ref{fig:csqopcomp}, and I thank Aldo Ianni and Serenelli for providing
Fig.~\ref{fig:neutborobs} and Serenelli for providing Fig.~\ref{fig:neutspect}.
I thank Ga\"el Buldgen for providing Fig.~\ref{fig:buldledoux}
and Joyce Guzik and Katie Mussack for discussions on solar mass loss,
and for contributing the information used in Figs~\ref{fig:masslosslum}
and \ref{fig:masslossdif}.
I am grateful to Achim Weiss for discussions 
on general stellar evolution, in particular the helium flash,
and to Andrea Miglio for contributing the results shown in Fig.~\ref{fig:pms}.
Finally, Achim Weiss, G\"unter Houdek, {\rv Ga\"el Buldgen, 
Vladimir Baturin and the anonymous referee} are thanked for reading and
providing very valuable comments on {\rv earlier versions} of the manuscript.
Funding for the Stellar Astrophysics Centre is provided by the Danish National
Research Foundation (Grant DNRF106).
%\notecd [Here the acknowledgements can be included.]
%\notecd [Thank Antia for opacity derivatives in plot, Fig.~\ref{fig:kapder}].
%\notecd [Thank D\"appen and Trampedach for discussions of EOS and opacity].
%\notecd [Thank Miglio for data used in Fig.~\ref{fig:pms}].
%\notecd [Thank Tami Rogers for discussions of solar rotation].
%\notecd [Thank Sacha Brun for mean gradient in convective simulations].
%\notecd [Thank Achim Weiss for help with He flash].
%\notecd [Thank Joyce Guzik and Katie Mussack for help with mass loss].
%\notecd [Thank Aldo Serenelli and Francesco Villante 
%for general discussions, including neutrinos].
%\notecd [Thanks Aldo Ianni for Borexino figure].
%\notecd [Thank Rachel Howe for helioseismic rotation].
%\notecd [Thank Remo and Poul Erik for discussions on abundances].
%\notecd [Thanks Regner for discussions about opacity].
%\notecd [Thank Kai Zuber for organizing, and inviting to talk at, 
%the 5th International Solar Neutrino Conference, Dresden,
%with a great deal of information and inspiration on solar neutrinos].
%\notecd [Thank Roger Ulrich for references on torsional oscillations].
%\notecd [Thank Q.S. Zhang for discussion of model TWA (or whatever)].
%\notecd [Thank Bengt Gustafsson for discussion of selective accretion].
\end{acknowledgements}

%% \newpage

\appendix

\normalsize

\section{Numerical accuracy}
\clabel{sec:numacc}

To make full use of the accuracy provided by, {\eg}, the helioseismic
observations, and to carry out reliable analyses of the sensitivity 
of the models to the assumed parameters and physics, 
the models should be computed with adequate numerical precision.
A simple, if not complete, test is to compare models
with different numbers of spatial meshpoints or timesteps.
The models computed for the present investigation used the
Aarhus Stellar Evolution Code \citep[ASTEC][]{Christ2008a}.
The calculations were started from an essentially chemically homogeneous 
zeo-age main-sequence (ZAMS) model (see also Sect.~\ref{sec:models}).
The calculations used 2400 meshpoints distributed according to the
variation with position of a number of key variables, 
using a form of the first-derivative stretching introduced by
\citet{Gough1975}.
The step in time between successive models was determined by 
constraining the maximum change amongst suitably scaled
changes in a number of variables to be below a quantity $\Delta y_{\rm max}$,
estimating the change from the preceding timestep;
this typically results in 23\,--\,24 timesteps from the ZAMS 
to the present solar age.

\begin{figure}[htp]
\centerline{\includegraphics[width=\figwidth]{\fig/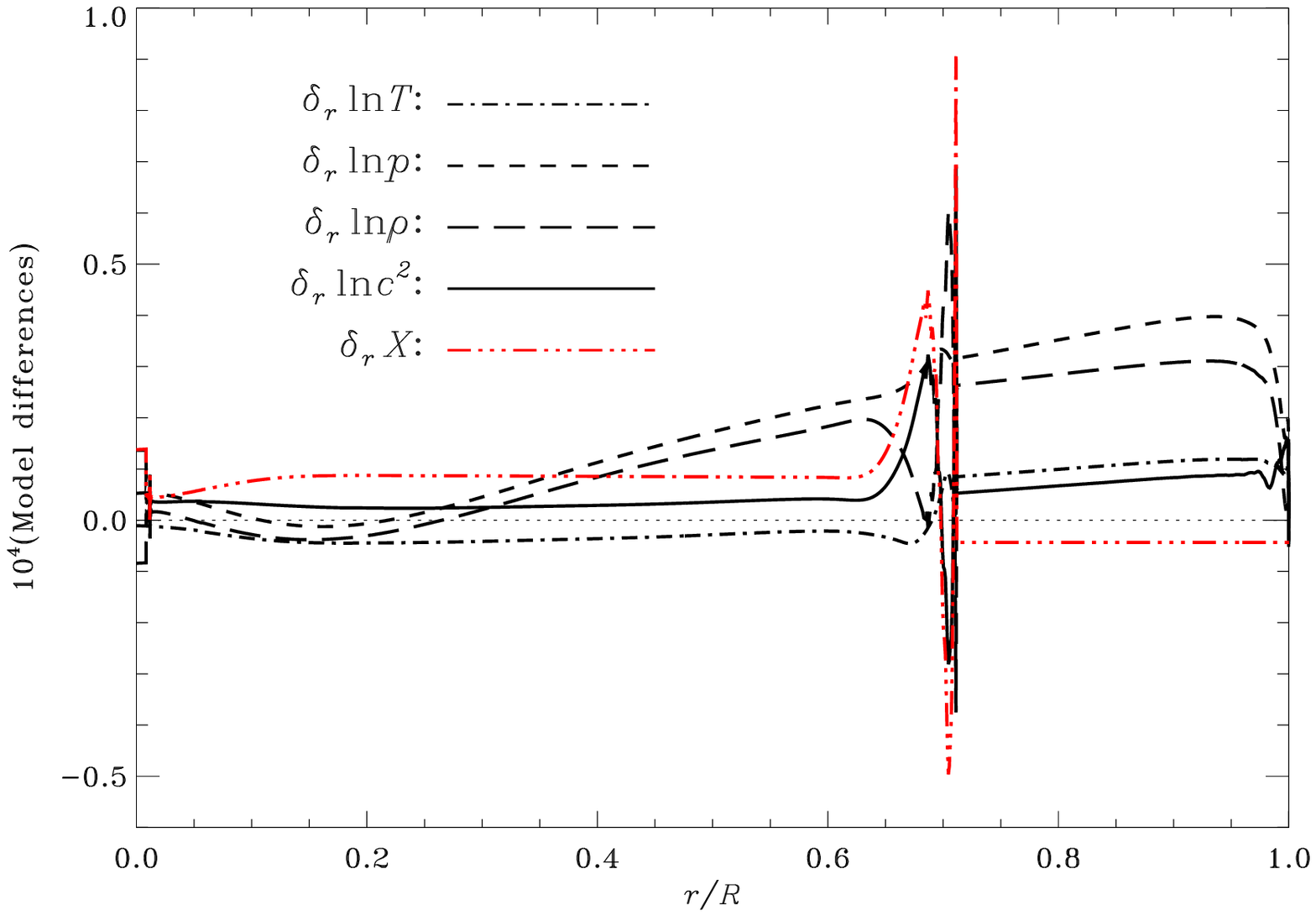}}
\centerline{\includegraphics[width=\figwidth]{\fig/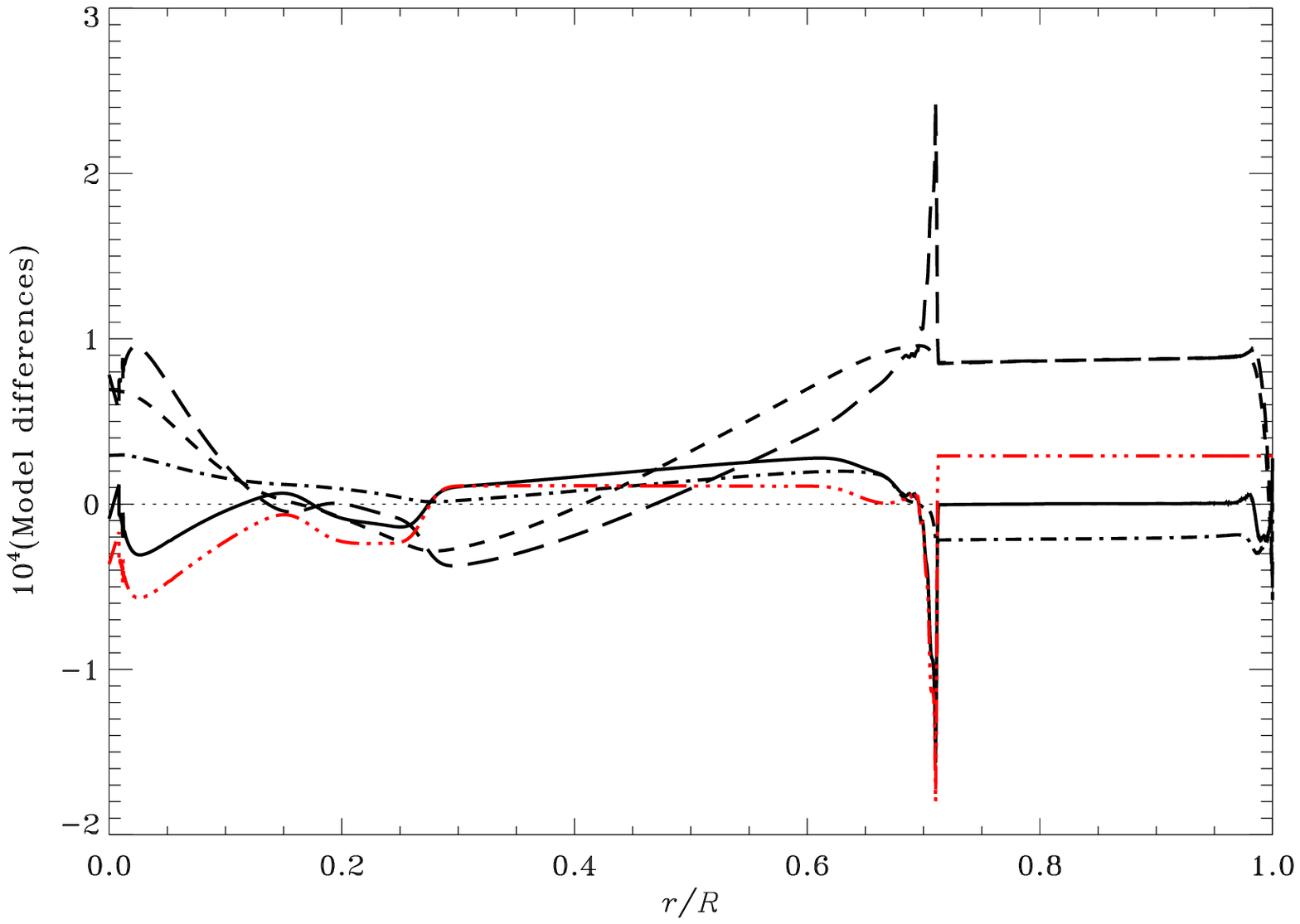}}
\caption{Differences between Model~{\MS} and (top panel) Model~[Mesh]
with increased number of mesh points and (bottom panel)
Model~[Mts] with increased number of timesteps, in the sense
(Model~S) -- (modified model).
%\notecd [dgr.l9bi.d.02c-d.03c, dgr.l9bi.d.02c-d.04c]
}
\clabel{fig:meshtime}
\end{figure}

To test the sensitivity of the model to these numerical procedures,
two additional models of the present Sun were computed,
both otherwise corresponding precisely
to Model~S in terms of parameters and physics, and calibrated to the
appropriate radius, luminosity and $\Zs/\Xs$:

\begin{itemize}

\item Model [Mesh]: Doubling the number of mesh points to 2400, 23 timesteps

\item Model [Tstp]: With 2400 mesh points, halving $\Delta y_{\rm max}$, 
resulting in 43 timesteps.

\end{itemize}

\noindent
Differences between these modified models and Model~S are plotted in
Fig.~\ref{fig:meshtime},
and numerical differences in selected quantities, relative to Model~S,
are shown in Table~\ref{tab:meshtime}.
Given that the numerical method predominantly uses second-order 
spatially and temporally centred schemes \citep[see][]{Christ2008a}
the numerical errors roughly scale as the inverse square of the number
of mesh points or timesteps.
Accordingly we can estimate the actual numerical errors of Model~{\MS}
resulting from the spatial and temporal discretization to be approximately
4/3 times the differences illustrated.

It is evident that the differences resulting from the improved numerical
treatment are small,
certainly substantially smaller than the differences resulting
from changes to the model physics (see Sect.~\ref{sec:modsens},
in particular Table~\ref{tab:modeldiff}) and
the differences between the Sun and Model~S (see Fig.~\ref{fig:csqinv}).
A dominant feature in both panels in Fig.~\ref{fig:meshtime} are the spikes
at the base of the convection zone, undoubtedly related to the change
in the depth of the convection zone.
Thus from this point of view some trust may be appropriate in the results
presented here; I also note that in the model comparisons presented in
Sect.~\ref{sec:modsens} the models are probably similarly affected by
these numerical errors, so that they largely cancel in the differences
illustrated.

\begin{table}
\caption{\clabel{tab:meshtime}
%\notecd [Model differences]
Differences between Model~{\MS} and Model~[Mesh] with an increased
number of mesh points and Model~[Tstp] with increased number of
timesteps, in the sense (Model~{\MS}) -- (Modified model).
}
\vskip 4mm
\centering
\begin{tabular}{l|ccccccc}
\hline
Model   & $\delta X_0$ & $\delta Z_0$ & $\delta T_{\rm c}/T_{\rm c}$ & $\delta \rho_{\rm c}/\rho_{\rm c}$ & $\delta X_{\rm c}$ & $\delta Y_{\rm s}$ &  $\delta(d_{\rm cz}/R)$ \\
\hline
& $\times 10^5$ & $\times 10^5$ & $\times 10^5$ & $\times 10^5$ & $\times 10^5$ & $\times 10^5$ & $\times 10^5$  \\
\hline
[Mesh] &  0.877 & -0.111 & -0.106 & -0.842 & 1.378 & 0.444 & 0.624 \\
{[}Tstp] & 1.133 & 0.147 & 2.945 & 7.816 & -3.623 & -2.982 & 5.766 \\
\hline
\end{tabular}
\end{table}

It should be noted that this simple analysis is only a first step in verifying
the numerical reliability of the solar models.
Additional potential numerical problems can arise from the interpolations
required in the tables of the equation of state and opacity.
A probably more serious issue are potential errors in the coding of the
stellar evolution programme used in the calculation.
The most reliable way to detect such errors is to compare results from
independent codes,
keeping identical, as far as possible,
the assumptions and physics of the calculation.
For solar modelling, early comparisons of this nature were carried out
by \citet{ChristReit1995} and \citet{Reiter1995}.
Comparisons involving several codes were organized for main-sequence stars
in \citet{Montei2008} and for red-giant stars by \citet{SilvaA2020}.
A detailed comparison of independent solar-modelling calculations is
planned for an update to the present paper.

%% \newpage

%===================================================================================

%\bibliography{LivRevSolar}
\bibliographystyle{spbasic}      % basic style, author-year citations
%%\bibliography{refs}   % name your BibTeX data base

\begin{thebibliography}{}



%\bibitem[Abdurashitov {\it et~al.\/}(1999)]{Abdura1999}
%\biblab{Abdura1999}
%Abdurashitov, J. N., Gavrin, V. N., Girin, S. V., {\etal}, 1999.
%%Gorbachev, V. V.,
%%Ibragimova, T. V., Kalikhov, A. V., Khairnasov, N. G.,
%%Knodel, T. V., Mirmov, I. N., Shikhin, A. A., Veretenkin, E. P.,
%%Vermul, V. M., Yants, V. E., Zatsepin, G. T., Bowles, T. J.,
%%Teasdale, W. A., Wark, D. L., Cherry, M. L., Nico, J. S.,
%%Cleveland, B. T., Davis, R., Lande, K., Wildenhain, P. S.,
%%Elliott, S. R. \& Wikerson, J. F., 1999.
%[Measurement of the solar neutrino capture rate with gallium metal].
%{\it Phys. Rev. C}, {\bf 60}, 055801(1-32).

\bibitem[Abdurashitov {\it et~al.\/}(1994)]{Abdura1994}
\biblab{Abdura1994}
Abdurashitov, J.~N., Faizov, E.~L., Gavrin, V.~N., {\etal}, 1994.
%Gusev, A.~O.,
%Kalikhov, A.~V., Knodel, T.~V., Knyshenko, I.~I., Kornoukhov, V.~N.,
%Mirmov, I.~N., Pshukov, A.~M., Shalagin, A.~M., Shikhin, A.~A.,
%Timofeyev, P.~V., Veretenkin, E.~P., Vermul, V.~M., Zatsepin, G.~T.,
%Bowles, T.~J., Nico, J.~S., Teasdale, W.~A., Wark, D.~L., Wilkerson, J.~F.,
%Cleveland, B.~T., Daily, T., Davis~Jr. R., Lande, K., Lee, C.~K.,
%Wildenhain, P.~W., Elliott, S.~R., Cherry, M.~L. \& Kouzes, R.~T., 1994.
[Results from SAGE (the Russian-American solar neutrino Experiment)].
{\it Phys. Lett. B}, {\bf 328}, 234--248.

\bibitem[Abdurashitov {\it et~al.\/}(2009)]{Abdura2009}
\biblab{Abdura2009}
Abdurashitov, J. N., Gavrin, V. N., Gorbachev, V. V., Gurkina, P. P.,
Ibragimova, T. V., Kalikhov, A. V., Khairnasov, N. G.,
Knodel, T. V., Mirmov, I. N., Shikhin, A. A., Veretenkin, E. P.,
Yants, V. E. \& Zatsepin, G. T., 2009.
[Measurement of the solar neutrino capture rate with gallium metal. III.
Results from the 2002--2007 data-taking period].
{\it Phys. Rev. C}, {\bf 80}, 015807(1--16).

\bibitem[Abe {\it et~al.\/}(2016)]{Abe2016}
\biblab{Abe2016}
Abe, K., Haga, Y., Hayato, Y., Ikeda, M., Iyogi, K., Kameda, J., 
Kishimoto, Y., Marti, L., Miura M., 
Moriyama, S., Nakahata, M., {\etal} (Super-Kamiokande Collaboration), 2016.
[Solar neutrino measurements in Super-Kamiokande-IV].
{\it Phys. rev. D.}, {\bf 94}, 052010-(1--32).

\bibitem[Abe {\it et~al.\/}(2017)]{Abe2017}
\biblab{Abe2017}
Abe, K., Amey, J., Andreopoulos, C., Antonova, M., Aoki, S., Ariga, A., 
Ashida, Y., Ban, S., Barbi, M., Barker, G. J., Barr, G., Barry, C., 
Batkiewicz, M., Berardi, V., Berkman, S., Bhadra, S., Bienstock, S., 
Blondel, A., Bolognesi, S., Bordoni, S., Boyd, S. B., Brailsford, D., 
Bravar, A., Bronner, C., Buizza Avanzini, M., {\etal}
(the T2K Collaboration), 2017.
[Measurement of neutrino and antineutrino oscillations by the T2K experiment 
including a new additional sample of $\nu_{\rm e}$ interactions 
at the far detector].
{\it Phys. Rev. D}, {\bf 96}, 092006-(1--49).

\bibitem[Adams(2010)]{Adams2010}
\biblab{Adams2010}
Adams, F. C., 2010.
[The birth environment of the solar system].
{\it Annu. Rev. Astron. Astrophys.}, {\bf 48}, 47--85.

\bibitem[Adamson {\it et~al.\/}(2012)]{Adamso2012}
\biblab{Adamso2012}
Adamson, P., Ayres, D. S., Backhouse, C., Barr, G., Bishai, M., Blake, A., 
Bock, G. J., Boehnlein, D. J., Bogert, D., Cao, S. V., Childress, S., 
Coelho, J. A. B., Corwin, L., Cronin-Hennessy, D., Danko, I. Z., 
de Jong, J. K., Devenish, N. E., Diwan, M. V., Escobar, C. O., Evans, J. J., 
Falk, E., Feldman, G. J., Frohne, M. V., Gallagher, H. R., Gomes, R. A., {\etal}
(the MINOS Collaboration), 2012.
[Improved Measurement of Muon Antineutrino Disappearance in MINOS].
{\it Phys. Rev. Lett.}, {\bf 108}, 191801-(1--5).

\bibitem[Adamson {\it et~al.\/}(2016)]{Adamso2016}
\biblab{Adamso2016}
Adamson, P., Ader, C., Andrews, M., Anfimov, N., Anghel, I., Arms, K., 
Arrieta-Diaz, E., Aurisano, A., Ayres, D. S., Backhouse, C., Baird, M., 
Bambah, B. A., Bays, K., Bernstein, R., Betancourt, M., Bhatnagar, V., 
Bhuyan, B., Bian, J., Biery, K., Blackburn, T., Bocean, V., Bogert, D., 
Bolshakova, A., Bowden, M., Bower, C., Broemmelsiek, D., Bromberg, C., 
Brunetti, G., {\etal} (NOvA Collaboration), 2016.
[First Measurement of Electron Neutrino Appearance in NOvA].
{\it Phys. Rev. Lett.}, {\bf 116}, 151806-(1--7).

\bibitem[Adelberger {\it et~al.\/}(1998)]{Adelbe1998}
\biblab{Adelbe1998}
Adelberger, E. G., Austin, S. M., Bahcall, J. N., {\etal}, 1998.
%Balantekin, A. B.,
%Bogaert, G., Brown, L. S., Buchmann, L., Cecil, F. E., Champagne, A. E,
%de Braeckeleer, L., Duba, C. A., Elliott, S. R.,
%Freedman, S. J., Gai, M., Goldring, G.,
%Gould, C. R., Gruzinov, A., Haxton, W. C., Heeger, K. M.,
%Henley, E., Johnson, C. W., 
%Kamionkowski, M., Kavanagh, R. W., Koonin, S. E., Kubodera, K.,
%Langanke, K., Motobayashi, T., Pandharipande, V., Parker, P.,
%Robertson, R. G. H.,
%Rolfs, C., Sawyer, R. F., Shaviv, N., Shoppa, T. D., Snover, K. A., 
%Swanson, E., Tribble, R. E., Turck-Chi{\`e}ze, S. \& Wilkerson, J. F., 1998.
[Solar fusion cross sections].
{\it Rev. Mod. Phys.}, {\bf 70}, 1265--1291.

\bibitem[Adelberger {\it et~al.\/}(2011)]{Adelbe2011}
\biblab{Adelbe2011}
Adelberger, E. G., Garc\'{\i}a, A., Robertson, R. G. H., {\etal}, 2011.
%Snover, K. A.,
%Balantekin, A. B., Heeger, K., Ramsey-Musolf, M. J., Bemmerer, D.,
%Junghans, A., Bertulani, C. A., Chen, J.-W., Costantini, H., Prati, P.,
%Couder, M., Uberseder, E., Wiescher, M., Cyburt, R., Davids, B.,
%Freedman, S. J., Gai, M., Gazit, D., Gialanella, L., Imbriani, G.,
%Greife, U., Hass, M., Haxton, W. C., Itahashi, T., Kubodera, K.,
%Langanke, K., Leitner, D., Leitner, M., Vetter, P., Winslow, L.,
%Marcucci, L. E., Motobayashi, T., Mukhamedzhanov, A., Tribble, R. E.,
%Nollett, K. M., Nunes, F. M., Park, T.-S., Parker, P. D.,
%Schiavilla, R., Simpson, E. C., Spitaleri, C., Strieder, F.,
%Trautvetter, H.-P., Suemmerer, K. \& Typel, S., 2011.
[Solar fusion cross sections. II. The {\it pp} chain and CNO cycles].
{\it Rev. Mod. Phys.}, {\bf 83}, 195--245.

\bibitem[Aerts {\it et~al.\/}(2010)]{Aerts2010}
\biblab{Aerts2010}
Aerts, C., Christensen-Dalsgaard, J. \& Kurtz, D. W., 2010.
{\it Asteroseismology},
Springer, Heidelberg.

\bibitem[Aerts {\it et~al.\/}(2018)]{Aerts2019}
\biblab{Aerts2019}
Aerts, C., Mathis, S. \& Rogers, T. M., 2019.
[Angular momentum transport in stellar interiors].
{\it Annu. Rev. Astron. Astrophys.}, {\bf 57}, 35--78.

\bibitem[Agafonova {\it et~al.\/}(2015)]{Agafon2015}
\biblab{Agafon2015}
Agafonova, N., Aleksandrov, A., Anokhina, A., Aoki, S., Ariga, A., Ariga, T., 
Bender, D., Bertolin, A., Bodnarchuk, I., Bozza, C., Brugnera, R., 
Buonaura, A., Buontempo, S., B\"uttner, B., Chernyavsky, M., Chukanov, A., 
Consiglio, L., D'Ambrosio, N., de Lellis, G., de Serio, M., 
Del Amo Sanchez, P., di Crescenzo, A., di Ferdinando, D., di Marco, N., 
Dmitrievski, S., Dracos, M., Duchesneau, D., Dusini, S., Dzhatdoev, T., {\etal}
(OPERA Collaboration), 2015.
[Discovery of $\tau$ Neutrino Appearance in the CNGS Neutrino Beam 
with the OPERA Experiment].
{\it Phys. Rev. Lett.}, {\bf 115}, 121802-(1--7).

\bibitem[Agafonova {\it et~al.\/}(2018)]{Agafon2018}
\biblab{Agafon2018}
Agafonova, N., Alexandrov, A., Anokhina, A., Aoki, S., Ariga, A., Ariga, T.,
Bertolin, A., Bozza, C., Brugnera, R., Buonaura, A., Buontempo, S.,
Chernyavskiy, M., Chukanov, A., Consiglio, L., D'Ambrosio, N.,
De Lellis, G., De Serio, M., del Amo Sanchez, P., Di Crescenzo, A.,
Di Ferdinando, D., Di Marco, N., Dmitrievsky, S., Dracos, M.,
Duchesneau, D., Dusini, S., Dzhatdoev, T., Ebert, J., Ereditato, A.,
Favier, J., Fini, R. A., Fornari, F., T. Fukuda, {\etal} (OPERA Collaboration),
2018.
[Final results of the OPERA experiment on $\nu_\tau$ appearance
in the CNGS neutrino beam].
{\it Phys. Rev. Lett.}, {\bf 120}, 211801-(1--7).
% Note: 'Dmitrievski' in Agafon2015 and 'Dmitrievsky' in Agafon2018 is o.k. :-(

\bibitem[Agostini {\it et~al.\/}(2018)]{Agosti2018}
\biblab{Agosti2018}
Agostini, M., Altenm{\"u}ller, K., Appel, S., Atroshchenko, V., 
Bagdasarian, Z., 
Basilico, D., Bellini, G., Benziger, J., Bick, D., Bonfini, G., Bravo, D., 
Caccianiga, B., Calaprice, F., Caminata, A., Caprioli, S., Carlini, M., 
Cavalcante, P., Chepurnov, A., Choi, K., Collica, L., {\etal},
(Borexino Collaboration), 2018.
[Comprehensive measurement of $pp$--chain solar neutrinos].
{\it Nature}, {\bf 562}, 505--510.

%\bibitem[Aharmim {\it et~al.\/}(2007)]{Aharmi2007}
%\biblab{Aharmi2007}
%Aharmim, B., Ahmad, Q.~R., Ahmed, S.~N., Allen, R.~C., Andersen, T.~C.,
%Anglin, J.~D., B{\"u}hler, G., Barton, J.~C., Beier, E.~W., Bercovitch, M.,
%{\etal}, 2007.
%[Determination of the $\nu_e$ and total ${}^8{\rm B}$ solar neutrino flux
%using the Sudbury Neutrino Observatory Phase I data set].
%{\it Phys. Rev. C}, {\bf 75}, 045502-(1--69).
%
%\bibitem[Aharmim {\it et~al.\/}(2005)]{Aharmi2005}
%\biblab{Aharmi2005}
%Aharmim, B., Ahmed, S.~N., Anthony, A.~E., Beier, E.~W., Bellerive, A.,
%Bergevin, M., Biller, S.~D., Boger, J., Boulay, M.~G., Bowler, M.~G.,
%{\etal}, 2005.
%[Electron energy spectra, fluxes, and day-night asymmetries of ${}^8{\rm B}$
%solar neutrinos from measurements with NaCl dissolved in the heavy-water
%detector at the Sudbury Neutrino Observatory].
%{\it Phys. Rev. C}, {\bf 72}, 055502-(1--47).

{\rv
\bibitem[Agostini {\it et~al.\/}(2019)]{Agosti2019}
\biblab{Agosti2019}
Agostini, M., Altenm{\"u}ller, K., Appel, S., Atroshchenko, V., 
Bagdasarian, Z., 
Basilico, D., Bellini, G., Benziger, J., Bonfini, G., Bravo, D., 
Caccianiga, B., Calaprice, F., Caminata, A., Cappelli, L., Caprioli, S.,
Carlini, M., 
Cavalcante, P., Cavanna, F., Chepurnov, A., Choi, K., Collica, L., {\etal}
(Borexino Collaboration), 2019.
[First simultaneous precision spectroscopy of $pp$, $^7$Be, and $pep$ 
solar neutrinos with Borexino Phase-II].
{\it Phys. Rev. D.}, {\bf 100}, 082004-(1--17).
{\tt [arXiv:hep-ex/1707.09279]}
}

{\rv
\bibitem[Agostini {\it et~al.\/}(2020a)]{Agosti2020a}
\biblab{Agosti2020a}
Agostini, M., Altenm{\"u}ller, K., Appel, S., Atroshchenko, V., 
Bagdasarian, Z., 
Basilico, D., Bellini, G., Benziger, J., Bick, D., Bravo, D., 
Caccianiga, B., Calaprice, F., Caminata, A., 
Cavalcante, P., Chepurnov, A., {\etal},
(Borexino Collaboration), 2020a.
[Improved measurement of ${}^8 {\rm B}$ solar neutrinos with 
1.5 kt$\,\cdot\,$y of Borexino exposure].
{\it Phys. Rev. D.}, {\bf 101}, 062001-(1--14).
{\tt [arXiv:hep-ex/1709.00756]}
}

{\rv
\bibitem[Agostini {\it et~al.\/}(2020b)]{Agosti2020b}
\biblab{Agosti2020b}
Agostini, M., Altenm{\"u}ller, K., Appel, S., Atroshchenko, V., Bagdasarian, Z.,
Basilico, D., Bellini, G., Benziger, J., Biondi, R., Bravo, D.,
Caccianiga, B., Calaprice, F., Caminata, A., Cavalcante, P., Chepurnov, A.,
D'Angelo, D., Davini, S., Derbin, A., Di Giacinto, A., Di Marcello, V.,
Ding, X.F., Di Ludovico, A., Di Noto, L., Drachnev, I., Formozov, A.,
Franco, D., Galbiati, C., Ghiano, C., Giammarchi, M., Goretti, A.,
G{\"o}ttel, A.S., Gromov, M., Guffanti, D., Ianni, Aldo, Ianni, Andrea,
Jany, A., Jeschke, D., Kobychev, V., Korga, G., Kumaran, S., Laubenstein, M.,
Litvinovich, E., Lombardi, P., Lomskaya, I., Ludhova, L., Lukyanchenko, G.,
Lukyanchenko, L., Machulin, I., Martyn, J., Meroni, E., Meyer, M.,
Miramonti, L., Misiaszek, M., Muratova, V., Neumair, B., Nieslony, M.,
Nugmanov, R., Oberauer, L., Orekhov, V., Ortica, F., Pallavicini, M.,
Papp, L., Pelicci, L., Penek, \"O., Pietrofaccia, L., Pilipenko, N.,
Pocar, A., Raikov, G., Ranalli, M.T., Ranucci, G., Razeto, A., Re, A.,
Redchuk, M., Romani, A., Rossi, N., Sch{\"o}nert, S., Semenov, D.,
Settanta, G., Skorokhvatov, M., Singhal, A., Smirnov, O., Sotnikov, A.,
Suvorov, Y., Tartaglia, R., Testera, G., Thurn, J., Unzhakov, E.,
Villante, F.L., Vishneva, A., Vogelaar, R.B., von Feilitzsch, F.,
Wojcik, M., Wurm, M., Zavatarelli, S., Zuber, K., Zuzel, G.,
The Borexino Collaboration, 2020b.
[Experimental evidence of neutrinos produced in the CNO fusion cycle
in the Sun].
{\it Nature}, {\bf 587}, 577 -- 582.
%{\tt [arXiv:2006.15115 [astro-ph.SR]]}
}

\bibitem[Aharmim {\it et~al.\/}(2013)]{Aharmi2013}
\biblab{Aharmi2013}
Aharmim, B., Ahmed, S. N., Anthony, A. E., Barros, N., Beier, E. W.,
Bellerive, A., Beltran, B., Bergevin, M., Biller, S. D., Boudjemline, K., 
{\rv Boulay}, M. G., Cai, B., Chan, Y. D., Chauhan, D., Chen, M.,
Cleveland, B. T., Cox, G. A., {\etal} (SNO Collaboration), 2013.
[Combined analysis of all three phases of solar neutrino data from
the Sudbury Neutrino Observatory].
{\it Phys. Rev. C}, {\bf 88}, {\rv 025501}-(1--27).

\bibitem[Ahmad {\it et~al.\/}(2001)]{Ahmad2001}
\biblab{Ahmad2001}
Ahmad, Q. R., Allen, R. C., Andersen, T. C., {\etal}, 2001.
%Anglin, J. D.,
%B{\"u}hler, G., Barton, J. C., Beier, E. W.,
%Bercovitch, M., Bigu, J., Biller, S., Black, R. A., Blevis, I.,
%Boardman, R. J., Boger, J., Bonvin, E., 
%Boulay, M. G., Bowler, M. G., Bowles, T. J., Brice, S. J.,
%Browne, M. C., Bullard, T. V., Burritt, T. H.,
%Cameron, K., Cameron, J., Chan, Y. D., Chen, M., Chen, H. H., Chen, X.,
%Chon, M. C., Cleveland, B. T.,
%Clifford, E. T. H., Cowan, J. H. M., Cowen, D. F., Cox, G. A., Dai, Y.,
%Dai, X., Dalnoki-Veress, F.,
%Davidson, W. F., Doe, P. J., Doucas, G., Dragowsky, M. R., Duba, C. A.,
%Duncan, F. A., Dunmore, J.,
%Earle, E. D., Elliott, S. R., Evans, H. C., Ewan, G. T., Farine, J., 
%Fergani, H., Ferraris, A. P., Ford, R. J.,
%Fowler, M. M., Frame, K., Frank, E. D., Frati, W., Germani, J. V.,
%Gil, S., Goldschmidt, A., Grant, D. R.,
%Hahn, R. L., Hallin, A. L., Hallman, E. D., Hamer, A., Hamian, A. A.,
%Haq, R. U., Hargrove, C. K., Harvey, P. J.,
%Hazama, R., Heaton, R., Heeger, K. M., Heintzelman, W. J., Heise, J.,
%Helmer, R. L., Hepburn, J. D.,
%Heron, H., Hewett, J., Hime, A., Howe, M., Hykawy, J. G., Isaac, M. C. P.,
%Jagam, P., Jelley, N. A., Jillings, C.,
%Jonkmans, G., Karn, J., Keener, P. T., Kirch, K., Klein, J. R., 
%Knox, A. B., Komar, R. J., Kouzes, R.,
%Kutter, T., Kyba, C. C. M., Law, J., Lawson, I. T., Lay, M.,
%Lee, H. W., Lesko, K. T., Leslie, J. R., Levine, I.,
%Locke, W., Lowry, M. M., Luoma, S., Lyon, J., Majerus, S., Mak, H. B., 
%Marino, A. D., McCauley, N., 
%McDonald, A. B., McDonald, D. S., McFarlane, K., McGregor, G.,
%McLatchie, W., Meijer Drees, R., Mes, H.,
%Mifflin, C., Miller, G. G., Milton, G., Moffat, B. A., Moorhead, M.,
%Nally, C. W., Neubauer, M. S., 
%Newcomer, F. M., Ng, H. S., Noble, A. J., Norman, E. B., Novikov, V. M.,
%O'Neill, M., Okada, C. E., 
%Ollerhead, R. W., Omori, M., Orrell, J. L., Oser, S. M., Poon, A. W. P.,
%Radcliffe, T. J., Roberge, A., 
%Robertson, B. C., Robertson, R. G. H., Rowley, J. K., Rusu, V. L.,
%Saettler, E., Schaffer, K. K., Schuelke, A.,
%Schwendener, M. H., Seifert, H., Shatkay, M., Simpson, J. J.,
%Sinclair, D., Skensved, P., Smith, A. R.,
%Smith, M. W. E., Starinsky, N., Steiger, T. D., Stokstad, R. G.,
%Storey, R. S., Sur, B., Tafirout, R., Tagg, N.,
%Tanner, N. W., Taplin, R. K., Thorman, M., Thornewell, P., Trent, P. T.,
%Tserkovnyak, Y. I.,
%Van Berg, R., Van de Water, R. G., Virtue, C. J., Waltham, C. E., 
%Wang, J.-X., Wark, D. L., West, N.,
%Wilhelmy, J. B., Wilkerson, J. F., Wilson, J., Wittich, P.,
%Wouters, J. M. \& Yeh, M., 2001.
[Measurement of the rate of $\nu_e + d \rightarrow p + p + e^-$
interactions produced by ${}^8{\rm B}$
solar neutrinos at the Sudbury Neutrino Observatory].
{\it Phys. Rev. Lett.}, {\bf 87}, 071301(1-6).

\bibitem[Ahmad {\it et~al.\/}(2002)]{Ahmad2002}
\biblab{Ahmad2002}
Ahmad, Q. R., Allen, R. C., Andersen, T. C., {\etal}, 2002.
%Anglin, J. D., Barton, J. C.,
%Beier, E. W., Bercovitch, M., Bigu, J., Biller, S. D., Black, R. A.,
%Blevis, I., Boardman, R. J., Boger, J., Bonvin, E., Boulay, M. G.,
%Bowler, M. G., Bowles, T. J., Brice, S. J., Browne, M. C., Bullard, T. V.,
%B{\"u}hler, G., Cameron, J., Chan, Y. D., Chen, H. H., Chen, M., Chen, X.,
%Cleveland, B. T., Clifford, E. T. H., Cowan, J. H. M., Cowen, D. F.,
%Cox, G. A., Dai, X., Dalnoki-Veress, F., Davidson, W. F., Doe, P. J.,
%Doucas, G., Dragowsky, M. R., Duba, C. A., Duncan, F. A., Dunford, M.,
%Dunmore, J. A., Earle, E. D., Elliott, S. R., Evans, H. C., Ewan, G. T.,
%Farine, J., Fergani, H., Ferraris, A. P., Ford, R. J., Formaggio, J. A.,
%Fowler, M. M., Frame, K., Frank, E. D., Frati, W., Gagnon, N., Germani, J. V.,
%Gil, S., Graham, K., Grant, D. R., Hahn, R. L., Hallin, A. L., Hallman, E. D.,
%Hamer, A. S., Hamian, A. A., Handler, W. B., Haq, R. U., Hargrove, C. K.,
%Harvey, P. J., Hazama, R., Heeger, K. M., Heintzelman, W. J., Heise, J.,
%Helmer, R. L., Hepburn, J. D., Heron, H., Hewett, J., Hime, A., Howe, M.,
%Hykawy, J. G., Isaac, M. C. P., Jagam, P., Jelley, N. A., Jillings, C.,
%Jonkmans, G., Kazkaz, K., Keener, P. T., Klein, J. R., Knox, A. B.,
%Komar, R. J., Kouzes, R., Kutter, T., Kyba, C. C. M., Law, J., Lawson, I. T.,
%Lay, M., Lee, H. W., Lesko, K. T., Leslie, J. R., Levine, I., Locke, W.,
%Luoma, S., Lyon, J., Majerus, S., Mak, H. B., Maneira, J., Manor, J.,
%Marino, A. D., McCauley, N., McDonald, A. B., McDonald, D. S., McFarlane, K.,
%McGregor, G., Meijer Drees, R., Mifflin, C., Miller, G. G., Milton, G.,
%Moffat, B. A., Moorhead, M., Nally, C. W., Neubauer, M. S., Newcomer, F. M.,
%Ng, H. S., Noble, A. J., Norman, E. B., Novikov, V. M., O'Neill, M.,
%Okada, C. E., Ollerhead, R. W., Omori, M., Orrell, J. L., Oser, S. M.,
%Poon, A. W. P., Radcliffe, T. J., Roberge, A., Robertson, B. C.,
%Robertson, R. G. H., Rosendahl, S. S. E., Rowley, J. K., Rusu, V. L.,
%Saettler, E., Schaffer, K. K., Schwendener, M. H., Sch{\"u}lke, A., 
%Seifert, H.,
%Shatkay, M., Simpson, J. J., Sims, C. J., Sinclair, D., Skensved, P.,
%Smith, A. R., Smith, M. W. E., Spreitzer, T., Starinsky, N., Steiger, T. D.,
%Stokstad, R. G., Stonehill, L. C., Storey, R. S., Sur, B., Tafirout, R.,
%Tagg, N., Tanner, N. W., Taplin, R. K., Thorman, M., Thornewell, P. M.,
%Trent, P. T., Tserkovnyak, Y. I., Van Berg, R., Van de Water, R. G.,
%Virtue, C. J., Waltham, C. E., Wang, J.-X., Wark, D. L.,
%West, N., Wilhelmy, J. B., Wilkerson, J. F., Wilson, J. R., Wittich, P.,
%Wouters, J. M. \& Yeh, M., 2002.
[Direct evidence for neutrino flavor transformation from neutral-current
interactions in the Sudbury Neutrino Observatory].
{\it Phys. Rev. Lett.}, {\bf 89}, 011301-(1--6).

%\bibitem[Ahmed {\it et~al.\/}(2004)]{Ahmed2004}
%\biblab{Ahmed2004}
%Ahmed, S. N., Anthony, A. E., Beier, E. W., {\etal}, 2004.
%Bellerive, A., Biller, S. D.,
%Boger, J., Boulay, M. G., Bowler, M. G., Bowles, T. J., Brice, S. J.,
%Bullard, T. V., Chan, Y. D., Chen, M., Chen, X., Cleveland, B. T., 
%Cox, G. A., Dai, X., Dalnoki-Veress, F., Doe, P. J., Dosanjh, R. S.,
%Doucas, G., Dragowsky, M. R., Duba, C. A., Duncan, F. A., Dunford, M.,
%Dunmore, J. A., Earle, E. D., Elliott, S. R., Evans, H. C., Ewan, G. T., 
%Farine, J., Fergani, H., Fleurot, F., Formaggio, J. A., Fowler, M. M.,
%Frame, K., Fulsom, B. G., Gagnon, N., Graham, K., Grant, D. R., Hahn, R. L., 
%Hall, J. C., Hallin, A. L., Hallman, E. D., Hamer, A. S., 
%Handler, W. B., Hargrove, C. K., Harvey, P. J., Hazama, R., Heeger, K. M.,
%Heintzelman, W. J., Heise, J., Helmer, R. L., Hemingway, R. J., Hime, A., 
%Howe, M. A., Jagam, P., Jelley, N. A., Klein, J. R., Kos, M. S., 
%Krumins, A. V., Kutter, T., Kyba, C. C. M., Labranche, H., Lange, R.,
%Law, J., Lawson, I. T., Lesko, K. T., Leslie, J. R., Levine, I., Luoma, S., 
%MacLellan, R., Majerus, S., Mak, H. B., Maneira, J., Marino, A. D.,
%McCauley, N., McDonald, A. B., McGee, S., McGregor, G., Mifflin, C.,
%Miknaitis, K. K. S., Miller, G. G., Moffat, B. A., Nally, C. W., 
%Nickel, B. G., 
%Noble, A. J., Norman, E. B., Oblath, N. S., Okada, C. E., Ollerhead, R. W.,
%Orrell, J. L., Oser, S. M., Ouellet, C., Peeters, S. J. M., Poon, A. W. P., 
%Robertson, B. C., Robertson, R. G. H., Rollin, E., Rosendahl, S. S., 
%Rusu, V. L., Schwendener, M. H., Simard, O., Simpson, J. J., Sims, C. J.,
%Sinclair, D., Skensved, P., Smith, M. W. E., Starinsky, N., Stokstad, R. G., 
%Stonehill, L. C., Tafirout, R., Takeuchi, Y., Te{\v s}i{\'c}, G., Thomson, M.,
%Thorman, M., Van Berg, R., Van de Water, R. G., Virtue, C. J., 
%Wall, B. L., Waller, D., Waltham, C. E., Wan Chan Tseung, H., Wark, D. L.,
%West, N., Wilhelmy, J. B., Wilkerson, J. F., Wilson, J. R., Wouters, J. M.,
%Yeh, M. \& Zuber, K., 2004.
%[Measurement of the total active ${}^8{\rm B}$ solar neutrino flux at 
%the Sudbury Neutrino Observatory with enhanced neutral current sensitivity].
%{\it Phys. Rev. Lett.}, {\bf 92}, 181301-(1--6).

\bibitem[Alexander and Ferguson(1994)]{Alexan1994}
\biblab{Alexan1994}
Alexander, D. R. \& Ferguson, J. W., 1994.
[Low-temperature Rosseland opacities].
{\it Astrophys. J.}, {\bf 437}, 879--891.

\bibitem[Alibert {\it et~al.\/}(2005)]{Aliber2005}
\biblab{Aliber2005}
Alibert, Y., Mordasini, C., Benz, W. \& Winisdoeffer, C., 2005.
[Models of giant planet formation with migration and disc evolution].
{\it Astron. Astrophys.}, {\bf 434}, 343--353.

\bibitem[Alimonti {\it et~al.\/}(2009)]{Alimon2009}
\biblab{Alimon2009}
Alimonti, G., Arpesella, C., Back, H., Balata, M., Bartolomei, D.,
de~Bellefon, A., Bellini, G., Benziger, J., Bevilacqua, A., Bondi, D.,
Bonetti, S., Brigatti, A., Caccianiga, B., Cadonati, L., Calaprice, F.,
Carraro, C., Cecchet, G., Cereseto, R., Chavarria, A., Chen, M.,
Chepurnov, A., Cubaiu, A., Czech, W., D'Angelo, D., Dalnoki-Veress, F.,
De~Bari, A., De~Haas, E., Derbin, A., Deutsch, M., Di~Credico, A.,
Di~Ludovico, A., Di~Pietro, G., Eisenstein, R., Elisei, F., Etenko, A.,
von~Feilitzsch, F., Fernholz, R., Fomenko, K., Ford, R., Franco, D.,
Freudiger, B., Gaertner, N., Galbiati, C., Gatti, F., Gazzana, S.,
Gehman, V., Giammarchi, M., Giugni, D., Goeger-Neff, M., Goldbrunner, T.,
Golubchikov, A., Goretti, A., Grieb, C., Hagner, C., Hagner, T., Hampel, W.,
Harding, E., Hardy, S., Hartmann, F.~X., von~Hentig, R., Hertrich, T.,
Heusser, G., Hult, M., Ianni, A., Ianni, An., Ioannucci, L., Jaenner, K.,
Joyce, M., de~Kerret, H., Kidner, S., Kiko, J., Kirsten, T., Kobychev, V.,
Korga, G., Korschinek, G., Kozlov, Yu., Kryn, D., La Marche, P.,
Lagomarsino, V.,
Laubenstein, M., Lendvai, C., Leung, M., Lewke, T., Litvinovich, E., Loer, B.,
Loeser, F., Lombardi, P., Ludhova, L., Machulin, I., Malvezzi, S., Manco, A.,
Maneira, J., Maneschg, W., Manno, I., Manuzio, D., Manuzio, G., Marchelli, M.,
Martemianov, A., Masetti, F., Mazzucato, U., McCarty, K., McKinsey, D.,
Meindl, Q., Meroni, E., Miramonti, L., Misiaszek, M., Montanari, D.,
Monzani, M.~E., Muratova, V., Musico, P., Neder, H., Nelson, A.,
Niedermeier, L., Nisi, S., Oberauer, L., Obolensky, M., Orsini, M.,
Ortica, F., Pallavicini, M., Papp, L., Parcells, R., Parmeggiano, S.,
Parodi, M., Pelliccia, N., Perasso, L., Pocar, A., Raghavan, R.,
Ranucci, G., Rau, W., Razeto, A., Resconi, E., Risso, P., Romani, A.,
Rountree, D., Sabelnikov, A., Saggese, P., Saldhana, R., Salvo, C.,
Scardaoni, R., Schimizzi, D., Sch\"onert, S., Schubeck, K.~H.,
Shutt, T., Siccardi, F., Simgen, H., Skorokhvatov, M., Smirnov, O.,
Sonnenschein, A., Soricelli, F., Sotnikov, A., Sukhotin, S., Sule, C.,
Suvorov, Y., Tarasenkov, V., Tartaglia, R., Testera, G., Vignaud, D.,
Vitale, S., Vogelaar, R.~B., Vyrodov, V., Williams, B., Wojcik, M.,
Wordel, R., Wurm, M., Zaimidoroga, O., Zavatarelli, S. \& Zuzel, G., 2009.
[The Borexino detector at the Laboratori Nazionali del Gran Sasso].
{\it Nucl. Instr. and Meth. A}, {\bf 600}, 568--593.

\bibitem[Allen(1973)]{Allen1973}
\biblab{Allen1973}
Allen, C. W., 1973.
{\it Astrophysical Quantities}, 3rd edition, 
Athlone Press, London.

\bibitem[Allende Prieto(2016)]{Allend2016}
\biblab{Allend2016}
Allende Prieto, C., 2016.
[Solar and stellar photospheric abundances].
{\it Living Rev. Solar Phys.}, {\bf 13}, 1.

\bibitem[Allende Prieto {\it et~al.\/}(2001)]{Allend2001}
\biblab{Allend2001}
Allende Prieto, C., Lambert, D. L. \& Asplund, M., 2001.
[The {\it forbidden} abundance of oxygen in the Sun].
{\it Astrophys. J.}, {\bf 556}, L63--L66.

\bibitem[Allende Prieto {\it et~al.\/}(2002)]{Allend2002}
\biblab{Allend2002}
Allende Prieto, C., Lambert, D. L. \& Asplund, M., 2002.
[A reappraisal of the solar photospheric C/O ratio].
{\it Astrophys. J.}, {\bf 573}, L137--L140.

\bibitem[ALMA Partnership(2015)]{ALMA2015}
\biblab{ALMA2015}
ALMA Partnership, Brogan, C. L., P{\'e}rez, L. M., Hunter, T. R., Dent, W. R. F.,
Hales, A. S., Hills, R. E., Corder, S., Fomalont, E. B., Vlahakis, C.,
Asaki, Y.,
Barkats, D., Hirota, A., Hodge, J. A., Impellizzeri, C. M. V., Kneissl, R.,
Liuzzo, E., Lucas, R., Marcelino, N., Matsushita, S., Nakanishi, K.,
Phillips, N., Richards, A. M. S., Toledo, I., Aladro, R., Broguiere, D.,
Cortes, J. R., Cortes, P. C., Espada, D., Galarza, F., Garcia-Appadoo, D.,
Guzman-Ramirez, L., Humphreys, E. M., Jung, T., Kameno, S., Laing, R. A.,
Leon, S., Marconi, G., Mignano, A., Nikolic, B., Nyman, L.-A., Radiszcz, M.,
Remijan, A., Rod{\'o}n, J. A., Sawada, T., Takahashi, S., Tilanus, R. P. J.,
Vila Vilaro, B., Watson, L. C., Wiklind, T., Akiyama, E., Chapillon, E.,
de Gregorio-Monsalvo, I., Di Francesco, J., Gueth, F., Kawamura, A.,
Lee, C.-F., Nguyen Luong, Q., Mangum, J., Pietu, V., Sanhueza, P., Saigo, K.,
Takakuwa, S., Ubach, C., van Kempen, T., Wootten, A., Castro-Carrizo, A.,
Francke, H., Gallardo, J., Garcia, J., Gonzalez, S., Hill, T., Kaminski, T.,
Kurono, Y., Liu, H.-Y., Lopez, C., Morales, F., Plarre, K., Schieven, G.,
Testi, L., Videla, L., Villard, E., Andreani, P., Hibbard, J. E. \&
Tatematsu, K., 2015.
[The 2014 ALMA Long Baseline Campaign: First results from
high angular resolution observations toward the HL Tau Region].
{\it Astrophys. J.}, {\bf 808}, L3-(1--10).

\bibitem[Altmann {\it et~al.\/}(2005)]{Altman2005}
\biblab{Altman2005}
Altmann, M., Balata, B., Belli, P., Bellotti, E., Bernabei, R., Burkert, E.,
Cattadori, C., Cerulli, R., Chiarini, M., Cribier, M., d'Angelo, S.,
Del Re, G., Ebert, K. H., von Feilitzsch, F., Ferrari, N., Hampel, W.,
Hartmann, F. X., Henrich, E., Heusser, G., Kaether, F., Kiko, J., 
Kirsten, T., Lachenmaier, T., Lanfranchi, J., Laubenstein, M.,
L\"utzenkirchen, K., Mayer, K., Moegel, P., Motta, D., Nisi, S.,
Oehm, J., Pandola, L., Petricca, F., Potzel, W., Richter, H.,
Schoenert, S., Wallenius, M., Wojcik, M. \& Zanotti, L., 2005.
[Complete results for five years of GNO solar neutrino observations].
{\it Phys. Lett. B}, {\bf 616}, 174--190.

\bibitem[Anders and Grevesse(1989)]{Anders1989}
\biblab{Anders1989}
Anders, E. \& Grevesse, N., 1989.
[Abundances of the elements: meteoritic and solar].
{\it Geochim. Cosmochim. Acta}, {\bf 53}, 197--214.

\bibitem[Andreasen and Petersen(1988)]{Andrea1988}
\biblab{Andrea1988}
Andreasen, G. K. \& Petersen, J. O., 1988.
[Double mode pulsating stars and opacity changes].
{\it Astron. Astrophys.}, {\bf 192}, L4--L6.

\bibitem[Angelou {\it et~al.\/}(2011)]{Angelo2011}
\biblab{Angelo2011}
Angelou, G. C., Church, R. P., Stancliffe, R. J., Lattanzio, J. C. \&
Smith, G. H., 2011.
[Thermohaline mixing and its role in the evolution of carbon and nitrogen
abundances in globular cluster red giants: the test case of Messier 3].
{\it Astrophys. J.}, {\bf 728}, 729-(1--12).

\bibitem[Angulo {\it et~al.\/}(1999)]{Angulo1999}
\biblab{Angulo1999}
Angulo, C., Arnould, M., Rayet, M., {\etal}, 1999.
%Descouvemont, P., Baye, D.,
%Leclercq-Willain, C., Coc, A., Barhoumi, S., Aguer, P., Rolfs, C.,
%Kunz, R., Hammer, J. W., Mayer, A., Paradellis, T., Kossionides, S.,
%Chronidou, C., Spyrou, K., Degl'Innocenti, S., Fiorentini, G.,
%Ricci, B., Zavatarelli, S., Providencia, C., Wolters, H., Soares, J.,
%Grama, C., Rahighi, J., Shotter, A. \& Lamehi Rachti, M., 1999.
[A compilation of charged-particle induced thermonuclear reaction rates].
{\it Nucl. Phys. A}, {\bf 656}, 3--183.

\bibitem[Angulo {\it et~al.\/}(2005)]{Angulo2005}
\biblab{Angulo2005}
Angulo, C., Champagne, A. E. \& Trautvetter, H.-P., 2005.
[$R$-matrix analysis of the ${}^{14}{\rm N}({\rm p}, \gamma){}^{15}{\rm O}$
astrophysical $S$-factor].
{\it Nucl. Phys. A}, {\bf 758}, 391c--394c.

\bibitem[Anselmann {\it et~al.\/}(1992)]{Anselm1992}
\biblab{Anselm1992}
Anselmann, P., Hampel, W., Heusser, G., {\etal}, 1992.
%Kiko, J., Kirsten, T.,
%Pernicka, E., Plaga, R., R{\"o}nn, U., Sann, M., Schlosser, C., Wink, R.,
%W{\'o}jcik, M., Ammon, R. v., Ebert, K. H., Fritsch, T., Hellriegel, K.,
%Henrich, E., Stieglitz, L., Weyrich, F.,
%Balata, M., Bellotti, E., Ferrari, N., Lalla, H., Stolarczyk, T.,
%Cattadori, C., Cremonesi, O., Fiorini, E., Pezzoni, S., Zanotti, L.,
%Feilitzsch, F. v., M{\"o}ssbauer, R., Schanda, U.,
%Berthomieu, G., Schatzman, E., Carmi, I., Dostrovsky, I.,
%Bacci, C., Belli, P., Bernabei, R., d'Angelo, S., Paoluzi, L.,
%Charbit, S., Cribier, M., Dupont, G., Gosset, L., Rich, J., Spiro, M.,
%Tao, C., Vignaud, D., Hahn, R. L., Hartmann, F. X., Rowley, J. K.,
%Stoenner, R. W. \& Wesener, J., 1992.
[Solar neutrinos observed by GALLEX at Gran Sasso].
{\it Phys. Lett. B}, {\bf 285}, 376--389.

\bibitem[Antia(1996)]{Antia1996}
\biblab{Antia1996}
Antia, H. M., 1996.
[Nonasymptotic helioseismic inversion: iterated seismic solar model].
{\it Astron. Astrophys.}, {\bf 307}, 609--623.

\bibitem[Antia and Basu(1994)]{Antia1994}
\biblab{Antia1994}
Antia, H. M. \& Basu, S., 1994.
[Measuring the helium abundance in the solar envelope:
the role of the equation of state].
{\it Astrophys. J.}, {\bf 426}, 801--811.

\bibitem[Antia and Basu(2005)]{Antia2005}
\biblab{Antia2005}
Antia, H. M. \& Basu, S., 2005.
[The discrepancy between solar abundances and helioseismology].
{\it Astrophys. J.}, {\bf 620}, L129--L132.

\bibitem[Antia and Basu(2006)]{Antia2006}
\biblab{Antia2006}
Antia, H. M. \& Basu, S., 2006.
[Determining solar abundances using helioseismology].
{\it Astrophys. J.}, {\bf 644}, 1292--1298.

\bibitem[Antia and  Basu(2011)]{Antia2011}
\biblab{Antia2011}
Antia, H. M. \& Basu, S., 2011.
[Revisiting the solar tachocline: average properties and temporal variations].
{\it Astrophys. J.}, {\bf 735}, L45-(1--6).

\bibitem[Antia and  Chitre(1997)]{Antia1997}
\biblab{Antia1997}
Antia, H. M. \& Chitre, S. M., 1997.
[Helioseismic models and solar neutrino fluxes].
{\it Mon. Not. R. Astron. Soc.}, {\bf 289}, L1--L4.

\bibitem[Antia and Chitre(1998)]{Antia1998}
\biblab{Antia1998}
Antia, H. M. \& Chitre, S. M., 1998.
[Determination of temperature and chemical composition profiles
in the solar interior from seismic models].
{\it Astron. Astrophys.}, {\bf 339}, 239--251.

\bibitem[Antia {\it et~al.\/}(2008)]{Antia2008}
\biblab{Antia2008}
Antia, H. M., Chitre, S. M. \& Gough, D. O., 2008.
[Temporal variations in the Sun's rotational kinetic energy].
{\it Astron. Astrophys.}, {\bf 477}, 657--663.

{\rv
\bibitem[Appourchaux and Corbard(2019)]{Appour2019}
\biblab{Appour2019}
Appourchaux, T. \& Corbard, T., 2019.
[Searching for $g$ modes. II. Unconfirmed $g$-mode detection in the
power spectrum of the time series of round-trip travel time].
{\it Astron. Astrophys.}, {\bf 624}, A106-(1--11).
}

\bibitem[Appourchaux {\it et~al.\/}(2010)]{Appour2010}
\biblab{Appour2010}
Appourchaux, T., Belkacem, K., Broomhall, A.-M., Chaplin, W. J.,
Gough, D. O., Houdek, G., Provost, J., Baudin, F., Boumier, P.,
Elsworth, Y., Garc\'{\i}a, R. A., Andersen, B., Finsterle, W.,
Fr\"ohlich, C., Gabriel, A., Grec, G., Jim\'enez, A., Kosovichev, A.,
Sekii, T., Toutain, T. \& Turck-Chi\`eze, S., 2010.
[The quest for the solar g modes].
{\it Astron. Astrophys. Rev.}, {\bf 18}, 197--277.
%{\tt [arXiv:0910.0848v2 [astro-ph]]}

{\rv
\bibitem[Appourchaux {\it et~al.\/}(2018)]{Appour2018}
\biblab{Appour2018}
Appourchaux, T., Boumier, P., Leibacher, J. W. \& Corbard, T., 2018.
[Searching for $g$ modes. I. A new calibration of the GOLF instrument].
{\it Astron. Astrophys.}, {\bf 617}, A108-(1--7).
}

\bibitem[Armitage(2011)]{Armita2011}
\biblab{Armita2011}
Armitage, P. J., 2011.
[Dynamics of protoplanetary disks].
{\it Annu. Rev. Astron. Astrophys.}, {\bf 49}, 195--236.

\bibitem[Armitage(2017)]{Armita2017}
\biblab{Armita2017}
Armitage, P. J., 2017.
[Lecture notes on the formation and early evolution of planetary systems].
{\tt [arXiv:astro-ph/0701485v6]}

\bibitem[Arpesella {\it et~al.}(2008)]{Arpese2008}
\biblab{Arpese2008}
Arpesella, C., Bellini, G., Benziger, J., {\etal}, 2008.
%Bonetti, S., Caccianiga, B.,
%Calaprice, F., Dalnoki-Veress, F., D'Angelo, D., de Kerret, H., Derbin, A.,
%Deutsch, M., Etenko, A., Fomenko, K., Ford, R., Franco, D., Freudiger, B.,
%Galbiati, C., Gazzana, S., Giammarchi, M., Goeger-Neff, M., Goretti, A.,
%Grieb, C., Hardy, S., Heusser, G., Ianni, A., Ianni, A., Joyce, M.,
%Korga, G., Kryn, D., Laubenstein, M., Leung, M., Litvinovich, E., Lombardi, P.,
%Ludhova, L., Machulin, I., Manuzio, G., Martemianov, A., Masetti, F.,
%McCarty, K., Meroni, E., Miramonti, L., Misiaszek, M., Montanari, D.,
%Monzani, M.~E., Muratova, V., Niedermeier, L., Oberauer, L., Obolensky, M.,
%Ortica, F., Pallavicini, M., Papp, L., Perasso, L., Pocar, A., Raghavan, R.~S.,
%Ranucci, G., Razeto, A., Sabelnikov, A., Salvo, C., Sch\"onert, S.,
%Simgen, H., Smirnov, O., Skorokhvatov, M., Sonnenschein, A., Sotnikov, A.,
%Sukhotin, S., Suvorov, Y., Tarasenkov, V., Tartaglia, R., Testera, G.,
%Vignaud, D., Vitale, S., Vogelaar, R.~B., von Feilitzsch, F., Wojcik, M.,
%Zaimidoroga, O., Zavatarelli, S. \& Zuzel, G., 2008.
[First real-time detection of ${}^7{\rm Be}$ solar neutrinos by Borexino].
{\it Phys. Lett. B}, {\bf 658}, 101--108.

\bibitem[Ashbrook(1968)]{Ashbro1968}
\biblab{Ashbro1968}
Ashbrook, J., 1968.
[Astronomical scrapbook. The gradual recognition of helium].
{\it Sky and Telescope}, {\bf 36}, 87.

\bibitem[Asplund(2004)]{Asplun2004a}
\biblab{Asplun2004a}
Asplund, M., 2004.
[Line formation in solar granulation. V.
Missing UV-opacity and the photospheric Be abundance].
{\it Astron. Astrophys.}, {\bf 417}, 769--774.

\bibitem[Asplund(2005)]{Asplun2005}
\biblab{Asplun2005}
Asplund, M., 2005.
[New light on stellar abundance analysis: departures from LTE and 
homogeneity].
{\it Annu. Rev. Astron. Astrophys.}, {\bf 43}, 481--540.

\bibitem[Asplund {\it et~al.\/}(2000)]{Asplun2000}
\biblab{Asplun2000}
Asplund, M., Nordlund, {\AA}., Trampedach, R., Allende Prieto, C. \& 
Stein, R. F., 2000.
[Line formation in solar granulation. I.
Fe line shapes, shifts and asymmetries].
{\it Astron. Astrophys.}, {\bf 359}, 729--742.

\bibitem[Asplund {\it et~al.\/}(2004)]{Asplun2004b}
\biblab{Asplun2004b}
Asplund, M., Grevesse, N., Sauval, A. J., Allende Prieto, C. \&
Kiselman, D., 2004.
[Line formation in solar granulation. IV.
[O I], O I and OH lines and the photospheric O abundance].
{\it Astron. Astrophys.}, {\bf 417}, 751--768
(Erratum: {\it Astron. Astrophys.}, {\bf 435}, 339--340).

\bibitem[Asplund {\it et~al.\/}(2005a)]{Asplunetal2005a}
\biblab{Asplunetal2005a}
Asplund, M., Grevesse, N., Sauval, A. J., Allende Prieto, C. \&
Blomme, R., 2005a.
[Line formation in solar granulation. VI. [CI], CI, CH and ${\rm C}_2$
lines and the photospheric C abundance].
{\it Astron. Astrophys.}, {\bf 431}, 693--705.

\bibitem[Asplund {\it et~al.\/}(2005b)]{Asplunetal2005b}
\biblab{Asplunetal2005b}
Asplund, M., Grevesse, N. \& Sauval, A. J., 2005b.
[The solar chemical composition].
In {\it Cosmic Abundances as Records of Stellar Evolution and Nucleosynthesis},
eds T. G. Barnes III \& F. N. Bash, {\it ASP Conf. Ser.}, {\bf 336},
{\rv ASP, San Francisco,} 
p. 25--38.  % [{\tt astro-ph/0410214}]

\bibitem[Asplund {\it et~al.\/}(2009)]{Asplun2009}
\biblab{Asplun2009}
Asplund, M., Grevesse, N., Sauval, A. J. \& Scott, P., 2009.
[The chemical composition of the Sun].
{\it Annu. Rev. Astron. Astrophys.}, {\bf 47}, 481--522.

\bibitem[Auwers(1891)]{Auwers1891}
\biblab{Auwers1891}
Auwers, A., 1891.
[Der Sonnendurchmesser und der Venusdurchmesser nach den Beobachtungen
an den Heliometern der deutschen Venus-Expedition].
{\it Astron. Nachr.}, {\bf 128}, 361--375.

\bibitem[Ayres {\it et~al.\/}(2006)]{Ayres2006}
\biblab{Ayres2006}
Ayres, T. R., Plymate, C. \& Keller, C. U., 2006.
[Solar carbon monoxide, thermal profiling and the abundances of C, O,
and their isotopes].
{\it Astrophys. J. Suppl.}, {\bf 165}, 618--651.

\bibitem[Ayukov and  Baturin(2017)]{Ayukov2017}
\biblab{Ayukov2017}
Ayukov, S. V. \& Baturin, V. A., 2017.
[Helioseismic models of the Sun with a low heavy element abundance].
{\it Astron. Zh.}, {\bf 94}, 894--906
(English translation: {\it Astronomy Reports}, {\bf 61}, 901--913).

\bibitem[Bachmann and Brown(1993)]{Bachma1993}
\biblab{Bachma1993}
Bachmann, K. T. \& Brown, T. M., 1993.
[$p$-mode frequency variation in relation to global solar activity].
{\it Astrophys. J.}, {\bf 411}, L45--L48.

\bibitem[Badnell {\it et~al.\/}(2005)]{Badnel2005}
\biblab{Badnel2005}
Badnell, N. R., Bautista, M. A., Butler, K., Delahaye, F., Mendoza, C.,
Palmeri, P., Zeippen, C. J. \& Seaton, M. J., 2005.
[Updated opacities from the Opacity Project].
{\it Mon. Not. R. Astron. Soc.}, {\bf 360}, 458--464.

%\bibitem[Baglin {\it et~al.\/}(2006)]{Baglin2006}
%\biblab{Baglin2006}
%Baglin, A., Michel, E., Auvergne, M. and the CoRoT team, 2006.
%[The seismology programme of the CoRoT space mission].
%In {\it Proc. SOHO 18 / GONG 2006 / HELAS I Conf.
%Beyond the spherical Sun},
%ed. K. Fletcher, ESA SP-624, ESA Publications Division,
%Noordwijk, The Netherlands.

\bibitem[Baglin {\it et~al.\/}(2009)]{Baglin2009}
\biblab{Baglin2009}
Baglin, A., Auvergne, M., Barge, P., Deleuil, M., Michel, E. and
the CoRoT Exoplanet Science Team, 2009.
[CoRoT: Description of the mission and early results].
In {\it Proc. IAU Symp. 253, Transiting Planets},
eds F. Pont, D. Sasselov \& M. Holman,
IAU and Cambridge University Press, 71--81.

\bibitem[Baglin {\it et~al.\/}(2012)]{Baglin2012}
\biblab{Baglin2012}
Baglin, A., Michel, E., and the CoRoT Team, 2012.
[CoRoT: A few highlights and their impact on understanding stellar structure].
In {\it Proceedings of the 61st Fujihara Seminar: Progress in
solar/stellar physics with helio- and asteroseismology}.
H. Shibahashi, M. Takata \& A. E. Lynas-Gray, eds,
{\it ASP Conf. Ser.}, {\bf 462}, {\rv ASP, San Francisco}, p. 492--504.

\bibitem[Bahcall(1964)]{Bahcal1964}
\biblab{Bahcal1964}
Bahcall, J. N., 1964.
[Solar neutrinos. I. Theoretical].
{\it Phys. Rev. Lett.}, {\bf 12}, 300--302.

\bibitem[Bahcall(1989)]{Bahcal1989}
\biblab{Bahcal1989}
Bahcall, J. N., 1989.
{\it Neutrino astrophysics},
Cambridge University Press, Cambridge.

\bibitem[Bahcall(2006)]{Bahcal2006a}
\biblab{Bahcal2006a}
Bahcall, J. N., 2006.
[Solar models and solar neutrinos].
In {\it Proc. Nobel Symposium 129,  2004: Neutrino Physics},
eds L. Bergstr{\"o}m, O. Botner, P. Carlson, P.O. Hulth, \&T. Ohlsson,
The Royal Swedish Academy of Sciences,
{\it Physica Scripta}, {\bf T121}, 46--50.
%{\tt [arXiv:hep-ph/0412068~v1]}.

\bibitem[Bahcall and Frautschi(1969)]{Bahcal1969b}
\biblab{Bahcal1969b}
Bahcall, J. N. \& Frautschi, S. C., 1969.
[Lepton non-conservation and solar neutrinos].
{\it Phys. Lett. B.}, {\bf 29}, 623--625.

\bibitem[Bahcall and Pe\~na-Garay(2004)]{Bahcal2004a}
\biblab{Bahcal2004a}
Bahcall, J. N. \& Pe\~na-Garay, C., 2004.
[Solar models and solar neutrino oscillations].
{\it New Journal of Physics}, {\bf 6}-63, 1--19.

\bibitem[Bahcall and Pinsonneault(1992a)]{Bahcal1992a}
\biblab{Bahcal1992a}
Bahcall, J. N. \& Pinsonneault, M. H., 1992a.
[Helium diffusion in the Sun].
{\it Astrophys. J.}, {\bf 395}, L119--L122.

\bibitem[Bahcall and Pinsonneault(1992b)]{Bahcal1992b}
\biblab{Bahcal1992b}
Bahcall, J. N. \& Pinsonneault, M. H., 1992b.
[Standard solar models, with and without helium diffusion
and the solar neutrino problem].
{\it Rev. Mod. Phys.}, {\bf 64}, 885--926.

\bibitem[Bahcall and Pinsonneault(1995)]{Bahcal1995}
\biblab{Bahcal1995}
Bahcall, J. N. \& Pinsonneault, M. H., 1995.
(With an appendix by G. J. Wasserburg).
[Solar models with helium and heavy-element diffusion].
{\it Rev. Mod. Phys.}, {\bf 67}, 781--808.

\bibitem[Bahcall and Pinsonneault(2004)]{Bahcal2004b}
\biblab{Bahcal2004b}
Bahcall, J. N. \& Pinsonneault, M. H., 2004.
[What do we (not) know theoretically about solar neutrino fluxes?].
{\it Phys. Rev. Lett.}, {\bf 92}, 121301-(1--4).

\bibitem[Bahcall and Sears(1972)]{Bahcal1972}
\biblab{Bahcal1972}
Bahcall, J. N. \& Sears, R. L., 1972.
[Solar neutrinos].
{\it Annu. Rev. Astron. Astrophys.}m {\bf 10}, 25--44.

\bibitem[Bahcall and Serenelli(2005)]{Bahcal2005e}
\biblab{Bahcal2005e}
Bahcall, J. N. \& Serenelli, A. M., 2005.
[How do uncertainties in the surface chemical composition of the Sun affect
the predicted solar neutrino fluxes?].
{\it Astrophys. J.}, {\bf 626}, 530--542.

\bibitem[Bahcall and Shaviv(1968)]{Bahcal1968a}
\biblab{Bahcal1968a}
Bahcall, J. N. \& Shaviv, G., 1968.
[Solar models and neutrino fluxes].
{\it Astrophys. J.}, {\bf 153}, 113--125.

\bibitem[Bahcall and Ulrich(1988)]{Bahcal1988}
\biblab{Bahcal1988}
Bahcall, J. N. \& Ulrich, R. K., 1988.
[Solar models, neutrino experiments and helioseismology].
{\it Rev. Mod. Phys.}, {\bf 60}, 297--372.

\bibitem[Bahcall {\it et~al.\/}(1963)]{Bahcal1963}
\biblab{Bahcal1963}
Bahcall, J. N., Fowler, W. A., Iben, I. \& Sears, R. L., 1963.
[Solar neutrino flux].
{\it Astrophys. J.}, {\bf 137}, 344--346.

\bibitem[Bahcall {\it et~al.\/}(1968)]{Bahcal1968b}
\biblab{Bahcal1968b}
Bahcall, J. N., Bahcall, N. A. \& Shaviv, G., 1968.
[Present status of the theoretical predictions for the ${}^{36}{\rm Cl}$
solar-neutrino experiment%
\footnote{SIC!}%
].
{\it Phys. Rev. Lett.}, {\bf 20}, 1209--1212.
%  SIC: 36Cl rather than 37Cl!

\bibitem[Bahcall {\it et~al.\/}(1968)]{Bahcal1968c}
\biblab{Bahcal1968c}
Bahcall, J. N., Bahcall, N. A. \& Ulrich, R. K., 1968.
[Mixing in the Sun and neutrino fluxes].
{\it Astrophys. Lett.}, {\bf 2}, 91--95.

\bibitem[Bahcall {\it et~al.\/}(1969)]{Bahcal1969a}
\biblab{Bahcal1969a}
Bahcall, J. N., Bahcall, N. A. \& Ulrich, R. K., 1969.
[Sensitivity of the solar-neutrino fluxes].
{\it Astrophys. J.}, {\bf 156}, 559--568.

\bibitem[Bahcall {\it et~al.\/}(1997)]{Bahcal1997}
\biblab{Bahcal1997}
Bahcall, J. N., Pinsonneault, M. H., Basu, S. \& Christensen-Dalsgaard, J.,
1997.
[Are standard solar models reliable?].
{\it Phys. Rev. Lett.}, {\bf 78}, 171--174.

\bibitem[Bahcall {\it et~al.\/}(1998)]{Bahcal1998}
\biblab{Bahcal1998}
Bahcall, J. N., Krastev, P. I. \& Smirnov, A. Yu., 1998.
[Where do we stand with neutrino oscillations?].
{\it Phys. Rev. D}, {\bf 59}, 096016-(1-22).

\bibitem[Bahcall {\it et~al.\/}(2001)]{Bahcal2001}
\biblab{Bahcal2001}
Bahcall, J. N., Pinsonneault, M. H. \& Basu, S., 2001.
[Solar models: current epoch and time dependences, neutrinos, and
helioseismological properties].
{\it Astrophys. J.}, {\bf 555}, 990--1012.
% 'dependences' sic!

\bibitem[Bahcall {\it et~al.\/}(2002)]{Bahcal2002}
\biblab{Bahcal2002}
Bahcall, J. N., Brown, L. S., Gruzinov, A. \& Sawyer, R. F., 2002.
[The Salpeter plasma correction for solar fusion reactions].
{\it Astron. Astrophys.}, {\bf 383}, 291--295.

%\bibitem[Bahcall {\it et~al.\/}(2004)]{Bahcal2004c}
%\biblab{Bahcal2004c}
%Bahcall, J. N., Gonzalez-Garcia, M. C. \& Pe\~na-Garay, C., 2004.
%[Solar neutrinos before and after Neutrino 2004].
%{\it J. High Energy Phys.}, {\bf 08(2004)}, 016-(1--26).

\bibitem[Bahcall {\it et~al.\/}(2004)]{Bahcal2004}
\biblab{Bahcal2004}
Bahcall, J. N., Serenelli, A. M. \& Pinsonneault, M., 2004.
[How accurately can we calculate the depth of the solar convective zone?].
{\it Astrophys. J.}, {\bf 614}, 464--471. % [astro-ph/0403604v1].

\bibitem[Bahcall {\it et~al.\/}(2005b)]{Bahcal2005c}
\biblab{Bahcal2005c}
Bahcall, J. N., Basu, S., Pinsonneault, M. \& Serenelli, A. M., 2005b.
[Helioseismological implications of recent solar abundance determinations].
{\it Astrophys. J.}, {\bf 618}, 1049--1056. % {\tt [astro-ph/0407060v1]}.

\bibitem[Bahcall {\it et~al.\/}(2005c)]{Bahcal2005d}
\biblab{Bahcal2005d}
Bahcall, J. N., Basu, S. \& Serenelli, A. M., 2005c.
[What is the neon abundance of the Sun?].
{\it Astrophys. J.}, {\bf 631}, 1281--1285. 

\bibitem[Bahcall {\it et~al.\/}(2005a)]{Bahcal2005b}
\biblab{Bahcal2005b}
Bahcall, J. N., Serenelli, A. M. \& Basu, S., 2005a.
[New solar opacities, abundances, helioseismology, and neutrino fluxes].
{\it Astrophys. J.}, {\bf 621}, L85--L88.

\bibitem[Bahcall {\it et~al.\/}(2006)]{Bahcal2006}
\biblab{Bahcal2006}
Bahcall, J. N., Serenelli, A. M. \& Basu, S., 2006.
[10,000 standard solar models: a Monte Carlo simulation].
{\it Astrophys. J. Suppl.}, {\bf 165}, 400--431.

\bibitem[Bailey {\it et~al.\/}(2015)]{Bailey2015}
\biblab{Bailey2015}
Bailey, J. E., Nagayama, T., Loisel, G. P., Rochau, G. A., Blancard, C.,
Colgan, J., Cosse, P., Faussurier, G., Fontes, C. J.,
Gilleron, F., Golovkin, I., Hansen, S. B., Iglesias, C. A., Kilcrease, D. P.,
MacFarlane, J. J., Mancini, R. C., Nahar, S. N., Orban, C.,
Pain, J.-C., Pradhan, A. K., Sherrill, M. \& Wilson, B. G., 2015.
[A higher-than-predicted measurement of iron opacity at solar interior
temperatures].
{\it Nature}, {\bf 517}, 56--59.

\bibitem[Baker and  Gough(1979)]{Baker1979}
\biblab{Baker1979}
Baker, N. H. \& Gough, D. O., 1979.
[Pulsations of model RR Lyrae stars].
{\it Astrophys. J.}, {\bf 234}, 232--244.

\bibitem[Balachandran and Bell(1998)]{Balach1998}
\biblab{Balach1998}
Balachandran, S. C. \& Bell, R. A., 1998.
[Shallow mixing in the solar photosphere inferred from revised beryllium
abundances].
{\it Nature}, {\bf 392}, 791--793.

\bibitem[Baldner and Basu(2008)]{Baldne2008}
\biblab{Baldne2008}
Baldner, C. S. \& Basu, S., 2008.
[Solar cycle related changes at the base of the convection zone].
{\it Astrophys. J.}, {\bf 686}, 1349--1361.

\bibitem[Ball and  Gizon(2014)]{Ball2014}
\biblab{Ball2014}
Ball, W. H. \& Gizon, L., 2014.
[A new correction of stellar oscillation frequencies for near-surface
effects].
{\it Astron. Astrophys.}, {\bf 568}, A123-(1--10).
(Erratum: {\it Astron. Astrophys.}, {\bf 569}, C2.)

\bibitem[Ball {\it et~al.\/}(2016)]{Ball2016}
\biblab{Ball2016}
Ball, W. H., Beeck, B., Cameron, R. H. \& Gizon, L., 2016.
[MESA meets MURaM. Surface effects in main-sequence solar-like oscillators 
computed using three-dimensional radiation hydrodynamics simulations].
{\it Astron. Astrophys.}, {\bf 592}, A159-(1--8).

\bibitem[Ballot {\it et~al.\/}(2004)]{Ballot2004}
\biblab{Ballot2004}
Ballot, J., Turck-Chi\`eze, S. \& Garc\'{\i}a, R. A., 2004.
[Seismic extraction of the convective extent in solar-like stars. The
observational point of view].
{\it Astron. Astrophys.}, {\bf 423}, 1051--1061.

\bibitem[Ballot {\it et~al.\/}(2006)]{Ballot2006}
\biblab{Ballot2006}
Ballot, J., Garc\'{\i}a, R. A. \& Lambert, P., 2006.
[Rotation speed and stellar axis inclination from modes: how
{\it CoRoT\/} would see other suns].
{\it Mon. Not. R. Astron. Soc.}, {\bf 369}, 1281--1286.

%\bibitem[Balmforth and Gough(1990)]{Balmfo1990}
%\biblab{Balmfo1990}
%Balmforth, N. J. \& Gough, D. O., 1990.
%[Mixing-length theory and the excitation of solar acoustic oscillations].
%{\it Solar Phys.}, {\bf 128}, 161--193.

\bibitem[Balmforth(1992)]{Balmfo1992}
\biblab{Balmfo1992}
Balmforth, N. J., 1992.
[Solar pulsational stability. I: Pulsation-mode thermodynamics].
{\it Mon. Not. R. astr. Soc.}, {\bf 255}, 603--631.

\bibitem[Balmforth and  Gough(1991)]{Balmfo1991}
\biblab{Balmfo1991}
Balmforth, N. J. \& Gough, D. O., 1991.
[The vibrational stability of the sun].
In 
{\it Challenges to theories of the structure of moderate-mass stars},
{\it Lecture Notes in Physics}, vol. {\bf 388}, 
eds Gough, D. O. \& Toomre, J., Springer, Heidelberg, p. 221--224.

\bibitem[Balmforth {\it et~al.\/}(1996)]{Balmfo1996}
\biblab{Balmfo1996}
Balmforth, N. J., Gough, D. O. \& Merryfield, W. J., 1996.
[Structural changes to the Sun through the solar cycle].
{\it Mon. Not. R. Astron. Soc.}, {\bf 278}, 437--448.


\bibitem[Baraffe and  Chabrier(2010)]{Baraff2010}
\biblab{Baraff2010}
Baraffe, I. \& Chabrier, G., 2010.
[Effect of episodic accretion on the structure and the lithium depletion
of low-mass stars and planet-hosting stars].
{\it Astron. Astrophys.}, {\bf 521}, A44-(1--8).

\bibitem[Baraffe {\it et~al.\/}(2009)]{Baraff2009}
\biblab{Baraff2009}
Baraffe, I., Chabrier, G. \& Gallardo, J., 2009.
[Episodic accretion at early stages of evolution of low-mass stars and
brown dwarfs: a solution for the observed luminosity spread in H-R diagrams?].
{\it Astrophys. J.}, {\bf 702}, L27--L31.

\bibitem[Barekat {\it et~al.\/}(2014)]{Bareka2014}
\biblab{Bareka2014}
Barekat, A., Schou, J. \& Gizon, L., 2014.
[The radial gradient of the near-surface shear layer of the Sun].
{\it Astron. Astrophys.}, {\bf 570}, L12-(1--4).

\bibitem[Barnes(2003)]{Barnes2003}
\biblab{Barnes2003}
Barnes, S. A., 2003.
[On the rotational evolution of solar- and late-type stars, its magnetic
origins and the possibility of stellar gyrochronology].
{\it Astrophys. J.}, {\bf 586}, 464--479.

\bibitem[Barnes(2010)]{Barnes2010}
\biblab{Barnes2010}
Barnes, S. A., 2010.
[A simple nonlinear model for the rotation of main-sequence cool stars. I.
Introduction, implications for gyrochronology, and color-period diagrams].
{\it Astrophys. J.}, {\bf 722}, 222--234.

{\rv
\bibitem[Barnes {\it et~al.\/}(2016)]{Barnes2016}
\biblab{Barnes2016}
Barnes, S. A., Spada, F. \& Weingrill, J., 2016.
[Some aspects of cool main sequence star ages derived from stellar rotation
(gyrochronology)].
{\it Astron. Nach.}, {\bf 337}, 810--814.
}

\bibitem[Bartenwerfer(1973)]{Barten1973}
\biblab{Barten1973}
Bartenwerfer, D., 1973.
[Differential rotation, magnetic fields and the solar neutrino flux].
{\it Astron. Astrophys.}, {\bf 25}, 455--456.

\bibitem[Basu(1998)]{Basu1998}
\biblab{Basu1998}
Basu, S., 1998.
[Effects of errors in the solar radius on helioseismic inferences].
{\it Mon. Not. R. Astron. Soc.}, {\bf 298}, 719--728.

\bibitem[Basu(2016)]{Basu2016}
\biblab{Basu2016}
Basu, S., 2016.
[Global seismology of the Sun].
{\it Living Rev. Sol. Phys.}, {\bf 13}, 2 (pp 1--110).
\url{https://doi.org/10.1007/s41116-016-0003-4}.

\bibitem[Basu and Antia(1994)]{Basu1994b}
\biblab{Basu1994b}
Basu, S. \& Antia, H. M., 1994.
[Effects of diffusion on the extent of overshoot below the solar convection
zone].
{\it Mon. Not. R. Astron. Soc.}, {\bf 269}, 1137--1144.

\bibitem[Basu and Antia(1997)]{Basu1997b}
\biblab{Basu1997b}
Basu, S. \& Antia, H. M., 1997.
[Seismic measurement of the depth of the solar convection zone].
{\it Mon. Not. R. Astron. Soc.}, {\bf 287}, 189--198.

\bibitem[Basu and Antia(2004)]{Basu2004a}
\biblab{Basu2004a}
Basu, S. \& Antia, H. M., 2004.
[Constraining solar abundances using helioseismology].
{\it Astrophys. J.}, {\bf 606}, L85--L88.

\bibitem[Basu and Antia(2008)]{Basu2008}
\biblab{Basu2008}
Basu, S. \& Antia, H. M., 2008.
[Helioseismology and solar abundances].
{\it Phys. Rep.}, {\bf 457}, 217--283.

\bibitem[Basu and  Antia(2019)]{Basu2019}
\biblab{Basu2019}
Basu, S. \& Antia, H. M., 2019.
[Changes in solar rotation over two cycles].
{\it Astrophys. J.}, {\bf 883}, 93-(1--10).

\bibitem[Basu and Christensen-Dalsgaard(1997)]{Basu1997c}
\biblab{Basu1997c}
Basu, S. \& Christensen-Dalsgaard, J., 1997.
[Equation of state and helioseismic inversions].
{\it Astron. Astrophys.}, {\bf 322}, L5--L8.

\bibitem[Basu and  Mandel(2004)]{Basu2004b}
\biblab{Basu2004b}
Basu, S. \& Mandel, A., 2004.
[Does solar structure vary with solar magnetic activity?]
{\it Astrophys. J.}, {\bf 617}, L155--L158.

\bibitem[Basu and Thompson(1996)]{Basu1996}
\biblab{Basu1996}
Basu, S. \& Thompson, M. J., 1996.
[On constructing seismic models of the Sun].
{\it Astron. Astrophys.}, {\bf 305}, 631--642.

\bibitem[Basu {\it et~al.\/}(1994)]{Basu1994a}
\biblab{Basu1994a}
Basu, S., Antia, H. M. \& Narasimha, D., 1994.
[Helioseismic measurement of the extent of overshoot below
the solar convection zone].
{\it Mon. Not. R. Astron. Soc.}, {\bf 267}, 209--224.

\bibitem[Basu {\it et~al.\/}(1997)]{Basu1997a}
\biblab{Basu1997a}
Basu, S., Chaplin, W. J., Christensen-Dalsgaard, J.,
Elsworth, Y., Isaak, G.~R., New, R., Schou, J.,
Thompson, M. J. \& Tomczyk, S., 1997.
[Solar internal sound speed as inferred from combined BiSON and LOWL
oscillation frequencies].
{\it Mon. Not. R. Astron. Soc.}, {\bf 292}, 243--251. 

\bibitem[Basu {\it et~al.\/}(1999)]{Basu1999}
\biblab{Basu1999}
Basu, S., D{\"a}ppen, W. \& Nayfonov, A., 1999.
[Helioseismic analysis of the hydrogen partition function in the solar
interior].
{\it Astrophys. J.}, {\bf 518}, 985--993.

\bibitem[Basu {\it et~al.\/}(2000)]{Basu2000}
\biblab{Basu2000}
Basu, S., Pinsonneault, M. H. \& Bahcall, J. N., 2000.
[How much do helioseismological inferences depend on the assumed reference
model?].
{\it Astrophys. J.}, {\bf 529}, 1084--1100.

\bibitem[Basu {\it et~al.\/}(2003)]{Basu2003}
\biblab{Basu2003}
Basu, S., Christensen-Dalsgaard, J., Howe, R., Schou, J.,
Thompson, M. J., Hill, F. \& Komm, R., 2003.
[A comparison of solar p-mode parameters from MDI and GONG: mode frequencies
and structure inversions].
{\it Astrophys. J.}, {\bf 591}, 432--445.

\bibitem[Basu {\it et~al.\/}(2007)]{Basu2007}
\biblab{Basu2007}
Basu, S., Chaplin, W. J., Elsworth, Y., New, R., Serenelli, A. M. \&
Verner, G. A., 2007.
[Solar abundances and helioseismology: fine-structure spacings and
separation ratios of low-degree $p$-modes].
{\it Astrophys. J.}, {\bf 655}, 660--671.

\bibitem[Basu {\it et~al.\/}(2009)]{Basu2009}
\biblab{Basu2009}
Basu, S., Chaplin, W. J., Elsworth, Y., New, R. \& Serenelli, A. M., 2009.
[Fresh insights on the structure of the solar core].
{\it Astrophys. J.}, {\bf 699}, 1403--1417.
%{\tt [arXiv:0905.0651v2 [astro-ph]]}.

\bibitem[Basu {\it et~al.\/}(2012)]{Basu2012}
\biblab{Basu2012}
Basu, S., Broomhall, A.-M., Chaplin, W. J. \& Elsworth, Y., 2012.
[Thinning of the Sun's magnetic layer: the peculiar solar minimum 
could have been predicted].
{\it Astrophys. J.}, {\bf 758}, 43-(1--6).

\bibitem[Basu {\it et~al.\/}(2015)]{Basu2015}
\biblab{Basu2015}
Basu, S., Grevesse, N., Mathis, S. \& Turck-Chi{\`e}ze, S., 2015.
[Understanding the internal chemical composition and physical processes 
of the solar interior].
{\it Space Sci. Rev.}, {\bf 196}, 49--77.

\bibitem[Batalha(2014)]{Batalh2014}
\biblab{Batalh2014}
Batalha, N., 2014.
[Exploring exoplanet populations with NASA's Kepler Mission].
{\it PNAS}, {\bf 111}, 12647--12654.

\bibitem[Baturin {\it et~al.\/}(2000)]{Baturi2000}
\biblab{Baturi2000}
Baturin, V. A., D{\"a}ppen, W., Gough, D. O. \& Vorontsov, S. V., 2000.
[Seismology of the solar envelope: sound-speed gradient in the 
convection zone and its diagnosis of the equation of state].
{\it Mon. Not. R. Astron. Soc.}, {\bf 316}, 71--83.

\bibitem[Baturin {\it et~al.\/}(2013)]{Baturi2013}
\biblab{Baturi2013}
Baturin, V. A., Ayukov, S. V., Gryaznov, V. K., Iosilevskiy, I. L.,
Fortov, V. E. \& Starostin, A. N., 2013.
[The current version of the SAHA-S equation of state: improvement and
perspective].
In {\it Progress in physics of the Sun and stars: 
a new era in helio- and asteroseismology}.
H. Shibahashi \& A. E. Lynas-Gray, eds,
{\it ASP Conf. Ser.}, {\bf 479}, {\rv ASP, San Francisco}, p. 11--18.

\bibitem[Baturin {\it et~al.\/}(2017)]{Baturi2017}
\biblab{Baturi2017}
Baturin, V. A., D{\"a}ppen, W., Morel, P., Oreshina, A. V., Th{\'e}venin, F.,
Gryaznov, V. K., Iosilevskiy, I. L., Starostin, A. N. \& Fortov, V. E.,
2017.
[Equation of state SAHA-S meets stellar evolution code CESAM2k].
{\it Astron. Astrophys.}, {\bf 606}, A129-(1--8).

\bibitem[Baturin {\it et~al.\/}(2019)]{Baturi2019}
\biblab{Baturi2019}
Baturin, V. A., D{\"a}ppen, W., Oreshina, A. V., Ayukov, S. V. \&
Gorshkov, A. B., 2019.
[Interpolation of equation-of-state data].
{\it Astron. Astrophys.}, {\bf 626}, A108-(1--11).

\bibitem[Bazot {\it et~al.\/}(2005)]{Bazot2005}
\biblab{Bazot2005}
Bazot, M., Vauclair, S., Bouchy, F. \& Santos, N. C., 2005.
[Seismic analysis of the planet-hosting star $\mu$ Arae].
{\it Astron. Astrophys.}, {\bf 440}, 615--621.

\bibitem[Bazot {\it et~al.\/}(2018)]{Bazot2018}
\biblab{Bazot2018}
Bazot, M., Nielsen, M. B., Mary, D., Christensen-Dalsgaard, J., Benomar, O.,
Petit, P., Gizon, L., Sreenivasan, K. R. \& White, T. R., 2018.
[Butterfly diagram of a Sun-like star observed using asteroseismology].
{\it Astron. Astrophys.}, {\bf 619}, L9-(1--8).

\bibitem[Bedell {\it et~al.\/}(2018)]{Bedell2018}
\biblab{Bedell2018}
Bedell, M., Bean, J. L., Mel{\'e}ndez, J., Spina, L., Ram\'{\i}rez, I.,
Asplund, M., Alves-Brito, A., dos Santos, L., Dreizler, S., Yong, D., 
Monroe, T. \& Casagrande, L., 2018.
[The chemical homogeneity of Sun-like stars in the solar neighborhood].
{\it Astrophys. J.}, {\bf 865}, 68-(1--13).

\bibitem[Beeck {\it et~al.\/}(2012)]{Beeck2012}
\biblab{Beeck2012}
Beeck, B., Collet, R., Steffen, M., Asplund, M., Cameron, R. H., Freytag, B., 
Hayek, W., Ludwig, H.-G. \& Sch{\"u}ssler, M., 2012.
[Simulations of the solar near-surface layers with the CO5BOLD, MURaM,
and Stagger codes].
{\it Astron. Astrophys.}, {\bf 539}, A121-(1--11).

\bibitem[Belkacem {\it et~al.\/}(2019)]{Belkac2019}
\biblab{Belkac2019}
Belkacem, K., Kupka, F., Samadi, R. \& Grimm-Strele, H., 2019.
[Solar $p$-mode damping rates: insight from a 3D hydrodynamical simulation].
{\it Astrophys. J.}, {\bf 625}, A20-(1--15).

\bibitem[Bellini {\it et~al.\/}(2010)]{Bellin2010}
\biblab{Bellin2010}
Bellini, G., Benziger, J., Bonetti, S., Buizza~Avanzini, M., Caccianiga, B.,
Cadonati, L., Calaprice, F., Carraro, C., Chavarria, A., Dalnoki-Veress, F.,
D'Angelo, D., Davini, S., de~Kerret, H., Derbin, A., Etenko, A., Fomenko, K.,
Franco, D., Galbiati, C., Gazzana, S., Ghiano, C., Giammarchi, M.,
Goeger-Neff, M., Goretti, A., Guardincerri, E., Hardy, S., Ianni, Aldo,
Ianni, Andrea, Joyce, M., Korga, G., Kryn, D., Laubenstein, M., Leung, M.,
Lewke, T., Litvinovich, E., Loer, B., Lombardi, P., Ludhova, L., Machulin, I.,
Manecki, S., Maneschg, W., Manuzio, G., Meindl, Q., Meroni, E., Miramonti, L.,
Misiaszek, M., Montanari, D., Muratova, V., Oberauer, L., Obolensky, M.,
Ortica, F., Pallavicini, M., Papp, L., Perasso, L., Perasso, S., Pocar, A.,
Raghavan, R.~S., Ranucci, G., Razeto, A., Re, A., Risso, P., Romani, A.,
Rountree, D., Sabelnikov, A., Saldanha, R., Salvo, C., Sch\"onert, S.,
Simgen, H., Skorokhvatov, M., Smirnov, O., Sotnikov, A., Sukhotin, S.,
Suvorov, Y., Tartaglia, R., Testera, G., Vignaud, D., Vogelaar, R.~B.,
von~Feilitzsch, F., Winter, J., Wojcik, M., Wright, A., Wurm, M., Xu, J.,
Zaimidoroga, O., Zavatarelli, S. \& Zuzel, G., 2010.
[Measurement of the solar ${}^8 {\rm Be}$ neutrino rate with a 
liquid scintillator target and 3 MeV energy threshold in the Borexino detector].
{\it Phys. Rev. D.}, {\bf 82}, 033006-(1--10).
{\tt [arXiv:0808.2868v1 [astro-ph]]} 

\bibitem[Bellini {\it et~al.\/}(2014)]{Bellin2014}
\biblab{Bellin2014}
Bellini, G., Benziger, J., Bick, D., Bonfini, G., Bravo, D., Caccianiga, B., 
Cadonati, L., Calaprice, F., Caminata, A., Cavalcante, P., Chavarria, A., 
Chepurnov, A., D'Angelo, D., Davini, S., Derbin, A., Empl, A., Etenko, A., 
Fomenko, K., Franco, D. {\etal}, (Borexino Collaboration), 2014.
[Neutrinos from the primary proton--proton fusion process in the Sun].
{\it Nature}, {\bf 512}, 383--386.

\bibitem[Benomar {\it et~al.\/}(2015)]{Benoma2015}
\biblab{Benoma2015}
Benomar, O., Takata, M., Shibahashi, H., Ceillier, T. \& Garc\'{\i}a, R. A.,
2015.
[Nearly uniform internal rotation of solar-like main-sequence stars
revealed by space-based asteroseismology and spectroscopic measurements].
{\it Mon. Not. R. Astron. Soc.}, {\bf 452}, 2654--2674.

\bibitem[Benomar {\it et~al.\/}(2018)]{Benoma2018}
\biblab{Benoma2018}
Benomar, O., Bazot, M., Nielsen, M. B., Gizon, L., Sekii, T.,
Takata, M., Hotta, H., Hanasoge, S., Sreenivasan, K. R. \&
Christensen-Dalsgaard, J., 2018.
[Asteroseismic detection of latitudinal differential rotation in
13 Sun-like stars].
{\it Science}, {\bf 361}, 1231--1234. 

\bibitem[Bergstr{\"o}m {\it et~al.\/}(2016)]{Bergst2016}
\biblab{Bergst2016}
Bergstr{\"o}m, J., Gonzales-Garcia, M. C., Maltoni, M., Pe\~na-Garay, C.,
Serenelli, A. \& Song, N., 2016.
[Updated determination of solar neutrino fluxes from solar neutrino data].
{\it J. High Energ. Phys.}, {\bf 3}, 132-(1--18). 

\bibitem[Berthomieu {\it et~al.\/}(1993)]{Bertho1993}
\biblab{Bertho1993}
Berthomieu, G., Provost, J., Morel, P. \& Lebreton, Y., 1993.
[Standard solar models with CESAM code: neutrinos and helioseismology].
{\it Astron. Astrophys.}, {\bf 268}, 775--791.

\bibitem[Bethe(1939)]{Bethe1939}
\biblab{Bethe1939}
Bethe, H. A., 1939.
[Energy production in stars].
{\it Phys. Rev.}, {\bf 55}. 434--456.

\bibitem[Bethe and Chritchfield(1938)]{Bethe1938}
\biblab{Bethe1938}
Bethe, H. A. \& Chritchfield, C. L., 1938.
[The formation of deuterons by proton combination].
{\it Phys. Rev.}, {\bf 54}, 248--254.

\bibitem[Biermann(1932)]{Bierma1932}
\biblab{Bierma1932}
Biermann, L., 1932.
[Untersuchungen \"uber der inneren Aufbau der Sterne. IV. Konvektionszonen
im Innern der Sterne].
{\it Z. Astrophys.}, {\bf 5}, 117--139.

\bibitem[Biermann(1942)]{Bierma1942}
\biblab{Bierma1942}
Biermann, L., 1942.
[{\"U}ber das Ionisationsgleichgewicht und den Aufbau der
Wasserstoffkonvektionszone].
{\it Z. Astrophys.}, {\bf 21}, 320--346.

\bibitem[Blancard {\it et~al.\/}(2012)]{Blanca2012}
\biblab{Blanca2012}
Blancard, C., Coss{\'e}, P. \& Faussurier, G., 2012.
[Solar mixture opacity calculations using detailed configuration and level
accounting treatments].
{\it Astrophys. J.}, {\bf 745}, 10-(1--7).

\bibitem[Bochsler {\it et~al.\/}(1990)]{Bochsl1990}
\biblab{Bochsl1990}
Bochsler, P., Geiss, J. \& Maeder, A., 1990.
[The abundance of $^3 He$ in the solar wind -
a constraint for models of solar evolution].
{\it Solar Phys.}, {\bf 128}, 203--215.

\bibitem[Bodenheimer(1995)]{Bodenh1995}
\biblab{Bodenh1995}
Bodenheimer, P., 1995.
[Angular momentum evolution of young stars and disks].
{\it Annu. Rev. Astron. Astrophys.}, {\bf 33}, 199--238.

\bibitem[Boger {\it et~al.\/}(2000)]{Boger2000}
\biblab{Boger2000}
Boger, J., Hahn, R.~L., Rowley, J.~K., {\etal}
%Carter, A.~L., Hollebone, B.,
%Kessler, D., Blevis, I., Dalnoki-Veress, F., DeKok, A., Farine, J.,
%Grant, D.~R., Hargrove, C.~K., Laberge, G., Levine, I., McFarlane, K.,
%Mes, H., Noble, A.~T., Novikov, V.~M., O'Neill, M., Shatkay, M., Shewchuk, C.,
%Sinclair, D., Clifford, E.~T.~H., Deal, R., Earle, E.~D., Gaudette, E.,
%Milton, G., Sur, B., Bigu, J., Cowan, J.~H.~M., Cluff, D.~L., Hallman, E.~D.,
%Haq, R.~U., Hewett, J., Hykawy, J.~G., Jonkmans, G., Michaud, R., Roberge, A.,
%Roberts, J., Saettler, E., Schwendener, M.~H., Seifert, H., Sweezey, D.,
%Tafirout, R., Virtue, C.~J., Beck, D.~N., Chan, Y.~D., Chen, X.,
%Dragowsky, M.~R., Dycus, F.~W., Gonzalez, J., Isaac, M.~C.~P., Kajiyama, Y.,
%Koehler, G.~W., Lesko, K.~T., Moebus, M.~C., Norman, E.~B., Okada, C.~E.,
%Poon, A.~W.~P., Purgalis, P., Schuelke, A., Smith, A.~R., Stokstad, R.~G.,
%Turner, S., Zlimen, I., Anaya, J.~M., Bowles, T.~J., Brice, S.~J., Esch, E.-I.,
%Fowler, M.~M., Goldschmidt, A., Hime, A., McGirt, A.~F., Miller, G.~G.,
%Teasdale, W.~A., Wilhelmy, J.~B., Wouters, J.~M., Anglin, J.~D.,
%Bercovitch, M., Davidson, W.~F., Storey, R.~S., Biller, S., Black, R.~A.,
%Boardman, R.~J., Bowler, M.~G., Cameron, J., Cleveland, B., Ferraris, A.~P.,
%Doucas, G., Heron, H., Howard, C., Jelley, N.~A., Knox, A.~B., Lay, M.,
%Locke, W., Lyon, J., Majerus, S., Moorhead, M., Omori, M., Tanner, N.~W.,
%Taplin, R.~K., Thorman, M., Wark, D.~L., West, N., Barton, J.~C., Trent, P.~T.,
%Kouzes, R., Lowry, M.~M., Bell, A.~L., Bonvin, E., Boulay, M., Dayon, M.,
%Duncan, F., Erhardt, L.~S., Evans, H.~C., Ewan, G.~T., Ford, R.,
%Hallin, A., Hamer, A., Hart, P.~M., Harvey, P.~J., Haslip, D.,
%Hearns, C.~A.~W., Heaton, R., Hepburn, J.~D., Jillings, C.~J., Korpach, E.~P.,
%Lee, H.~W., Leslie, J.~R., Liu, M.-Q., Mak, H.~B., McDonald, A.~B.,
%MacArthur, J.~D., McLatchie, W., Moffat, B.~A., Noel, S., Radcliffe, T.~J.,
%Robertson, B.~C., Skensved, P., Stevenson, R.~L., Zhu, X., Gil, S.,
%Heise, J., Helmer, R.~L., Komar, R.~J., Nally, C.~W., Ng, H.~S.,
%Waltham, C.~E., Allen, R.~C., B{\"u}hler, G., Chen, H.~H., Aardsma, G.,
%Andersen, T., Cameron, K., Chon, M.~C., Hanson, R.~H., Jagam, P., Karn, J.,
%Law, J., Ollerhead, R.~W., Simpson, J.~J., Tagg, N., Wang, J.-X., 
%Alexander, C., Beier, E.~W., Cook, J.~C., Cowen, D.~F., Frank, E.~D., 
%Frati, W., Keener, P.~T.,
%Klein, J.~R., Mayers, G., McDonald, D.~S., Neubauer, M.~S., Newcomer, F.~M.,
%Pearce, R.~J., Van~de~Water, R.~G., Van~Berg, R., Wittich, P.,
%Ahmad, Q.~R., Beck, J.~M., Browne, M.~C., Burritt, T.~H., Doe, P.~J.,
%Duba, C.~A., Elliott, S.~R., Franklin, J.~E., Germani, J.~V., Green, P.,
%Hamian, A.~A., Heeger, K.~M., Howe, M., Meijer~Drees, R., Myers, A.,
%Robertson, R.~G.~H., Smith, M.~W.~E., Steiger, T.~D., Van~Wechel, T.,
%Wilkerson, J.~F., 2000.
(the SNO collaboration), 2000.
[The Sudbury Neutrino Observatory].
{\it Nucl. Inst. Meth.}, {\bf A449}, 172--207.
%  See also http://www.sno.phy.queensu.ca/

\bibitem[B{\"o}hm-Vitense(1958)]{Boehm1958}
\biblab{Boehm1958}
B{\"o}hm-Vitense, E., 1958.
[{\"U}ber die Wasserstoffkonvektionszone in Sternen
verschiedener Effektivtemperaturen und Leuchtkr{\"a}fte].
{\it Z. Astrophys.}, {\bf 46}, 108--143.

\bibitem[Bonanno {\it et~al.\/}(2002)]{Bonann2002}
\biblab{Bonann2002}
Bonanno, A., Schlattl, H. \& Patern\`o, L., 2002.
[The age of the Sun and the relativistic corrections in the EOS].
{\it Astron. Astrophys.}, {\bf 390}, 1115--1118.

\bibitem[B{\"o}ning {\it et~al.\/}(2019)]{Boenin2019}
\biblab{Boenin2019}
B{\"o}ning, V. C. A., Hu, H. \& Gizon, L., 2019.
[Signature of solar $g$ modes in first-order $p$-mode frequency shifts].
{\it Astron. Astrophys.}, {\bf 629}, A26-(1--16).

\bibitem[Bonventre {\it et~al.\/}(2018)]{Bonven2018}
\biblab{Bonven2018}
Bonventre, R. \& Orebi Gann, G. D., 2018.
[Sensitivity of a low threshold directional detector to CNO-cycle
solar neutrinos].
{\it Eur. Phys. J. C}, {\bf 78}, 435-(1--11).

\bibitem[Boothroyd and Sackmann(2003)]{Boothr2003}
\biblab{Boothr2003}
Boothroyd, A. I. \& Sackmann, I.-Juliana, 2003.
[Our Sun. IV. The standard model and helioseismology: consequences of
uncertainties in input physics and in observed solar parameters].
{\it Astrophys. J.}, {\bf 583}, 1004--1023.

\bibitem[Boothroyd {\it et~al.\/}(1991)]{Boothr1991}
\biblab{Boothr1991}
Boothroyd, A. I., Sackmann, I.-J. \& Fowler, W. A., 1991.
[Our Sun. II. Early mass loss of 0.1 $M_{\odot}$ and the case
of the missing lithium].
{\it Astrophys. J.}, {\bf 377}, 318--329.

\bibitem[Borucki(2016)]{Boruck2016}
\biblab{Boruck2016}
Borucki, W. J., 2016.
[{\it Kepler} Mission: development and overview].
{\it Rep. Prog. Phys.}, {\bf 79}, 036901-(1--49).

\bibitem[Bouchaud {\it et~al.\/}(2020)]{Boucha2020}
\biblab{Boucha2020}
Bouchaud, K., Domiciano de Souza, A., Rieutord, M., Reese, D. R. \& 
Kervella, P., 2020.
[A realistic two-dimensional model of Altair].
{\it Astron. Astrophys.}, {\bf 633}, A78-(1--20).

\bibitem[Boury {\it et~al.\/}(1975)]{Boury1975}
\biblab{Boury1975}
Boury, A., Gabriel, M., Noels, A., Scuflaire, R. \& Ledoux, P., 1975.
[Vibrational instability of a 1 $M_{\odot}$ star towards non-radial
oscillations].
{\it Astron. Astrophys.}, {\bf 41}, 279--285.

\bibitem[Bouvier(2008)]{Bouvie2008}
\biblab{Bouvie2008}
Bouvier, J., 2008.
[Lithium depletion and the rotational history of exoplanet host stars].
{\it Astron. Astrophys.}, {\bf 489}, L53--L56.

{\rv
\bibitem[Bouvier {\it et~al.\/}(2016)]{Bouvie2016}
\biblab{Bouvie2016}
Bouvier, J., Lanzafame, A. C., Venuti, L., Klutsch, A., Jeffries, R., 
Frasca, A., Moraux, E., Biazzo, K., Messina, S., Micela, G., Randich, S., 
Stauffer, J., Cody, A. M., Flaccomio, E., Gilmore, G., Bayo, A., 
Bensby, T., Bragaglia, A., Carraro, G., Casey, A. Costado, M. T., 
Damiani, F., Delgado Mena, E., Donati, P., Franciosini, E., 
Hourihane, A., Koposov, S., Lardo, C., Lewis, J., Magrini, L., Monaco, L., 
Morbidelli, L., Prisinzano, L., Sacco, G., Sbordone, L., Sousa, S. G., 
Vallenari, A., Worley, C. C., Zaggia, S., Zwitter, T., 2016.
[The {\it Gaia}-ESO Survey: A lithium-rotation connection at 5 Myr?].
{\it Astron. Astrophys.}, {\bf 590}, A78-(1--9).
}

\bibitem[Bouvier {\it et~al.\/}(2018)]{Bouvie2018}
\biblab{Bouvie2018}
Bouvier, J., Barrado, D., Moraux, E., Stauffer, J., Rebull, L., 
Hillenbrand, L., Bayo, A., Boisse, I., Bouy, H., DiFolco, E.,
Lillo-Box, J. \& Morales Calder{\'o}n, M., 2018.
[The lithium-rotation connection in the 125 Myr-old Pleiades cluster].
{\it Astron. Astrophys.}, {\bf 613}, A63-(1--9).

\bibitem[Bradley and Winget(1994)]{Bradle1994}
\biblab{Bradle1994}
Bradley, P. A. \& Winget, D. E., 1994.
[An asteroseismological determination of the structure of the DBV
white dwarf GD 358].
{\it Astrophys. J.}, {\bf 430}, 850--857.

{\rv
\bibitem[Braithwaite and Spruit(2017)]{Braith2017}
\biblab{Braith2017}
Braithwaite, J. \& Spruit, H.C., 2017.
[Magnetic fields in non-convective regions of stars].
{\it R. Soc. Open Sci.}, {\bf 4}, 160271-(1--45). 
}

\bibitem[Brennecka {\it et~al.\/}(2020)]{Brenne2020}
\biblab{Brenne2020}
Brennecka, G. A., Burkhardt, C., Budde, G., Kruijer, T. S., Nimmo, F. \&
Kleine, T., 2020.
[Astronomical context of Solar System formation from molybdenum isotopes 
in meteorite inclusions].
{\it Science}, {\bf 370}, 837 -- 840.

\bibitem[Brown and Christensen-Dalsgaard(1998)]{Brown1998}
\biblab{Brown1998}
Brown, T. M. \& Christensen-Dalsgaard, J., 1998.
[Accurate determination of the solar photospheric radius].
{\it Astrophys. J.}, {\bf 500}, L195--L198.

\bibitem[Broomhall {\it et~al.\/}(2009)]{Broomh2009}
\biblab{Broomh2009}
Broomhall, A. M., Chaplin, W. J., Davies, G. R., Elsworth, Y., Fletcher, S. T.,
Hale, S. J., Miller, B. \& New, R., 2009.
[Definitive Sun-as-a-star p-mode frequencies: 23 years of BiSON observations].
{\it Mon. Not. R. Astron. Soc.}, {\bf 396}, L100--L104.

\bibitem[Brun and  Browning(2017)]{Brun2017}
\biblab{Brun2017}
Brun, A. S. \& Browning, M. K., 2017.
[Magnetism, dynamo action and the solar-stellar connection].
{\it Living Rev. Sol. Phys.}, {\bf 14}, 4. 
\url{https://doi.org/10.1007/s41116-017-0007-8}

\bibitem[Brun and Zahn(2006)]{Brun2006}
\biblab{Brun2006}
Brun, A. S. \& Zahn, J.-P., 2006.
[Magnetic confinement of the solar tachocline].
{\it Astron. Astrophys.}, {\bf 457}, 665--674.

\bibitem[Brun {\it et~al.\/}(1998)]{Brun1998}
\biblab{Brun1998}
Brun, A. S., Turck-Chi\`eze, S. \& Morel, P., 1998.
[Standard solar models in the light of new helioseismic constraints.
I. The solar core].
{\it Astrophys. J.}, {\bf 506}, 913--925.

\bibitem[Brun {\it et~al.\/}(1999)]{Brun1999}
\biblab{Brun1999}
Brun, A. S., Turck-Chi{\`e}ze, S. \& Zahn, J. P., 1999.
[Standard solar models in the light of new helioseismic constraints. II.
Mixing below the convection zone].
{\it Astrophys. J.}, {\bf 525}, 1032--1041.
(Erratum: {\it Astrophys. J.}, {\bf 536}, 1005).

\bibitem[Brun {\it et~al.\/}(2002)]{Brun2002}
\biblab{Brun2002}
Brun, A. S., Antia, H. M., Chitre, S. M. \& Zahn, J.-P., 2002.
[Seismic tests for solar models with tachocline mixing].
{\it Astron. Astrophys.}, {\bf 391}, 725--739.

\bibitem[Br{\"u}ggen and  Gough(1997)]{Brugge1997}
\biblab{Brugge1997}
Br{\"u}ggen, M. \& Gough, D. O., 1997.
[On electrostatic screening of ions in astrophysical plasmas].
{\it Astrophys. J.}, {\bf 488}, 867--871.

\bibitem[Br{\"u}ggen and  Gough(2000)]{Brugge2000}
\biblab{Brugge2000}
Br{\"u}ggen, M. \& Gough, D. O., 2000.
[Free energy of a screened ion pair].
{\it J. Math. Phys.}, {\bf 41}, 260--283.

\bibitem[Buldgen {\it et~al.\/}(2017a)]{Buldge2017a}
\biblab{Buldge2017a}
Buldgen, G., Salmon, S. J. A. J., Godart, M., Noels, A., Scuflaire, R.,
Dupret, M. A., Reese, D. R., Colgan, J., Fontes, C. J., Eggenberger, P.,
Hakel, P., Kilcrease, D. P. \& Richard, O., 2017.
[Inversions of the Ledoux discriminant: a closer look at the tachocline].
{\it Mon. Not. R. Astron. Soc.}, {\bf 472}, L70--L74.

\bibitem[Buldgen {\it et~al.\/}(2017b)]{Buldge2017b}
\biblab{Buldge2017b}
Buldgen, G., Salmon, S. J. A. J., Noels, A., Scuflaire, R., Dupret, M. A. \&
Reese, D. R., 2017.
[Determining the metallicity of the solar envelope using seismic inversion
techniques].
{\it Mon. Not. R. Astron. Soc.}, {\bf 472}, 751--764.

\bibitem[Buldgen {\it et~al.\/}(2017c)]{Buldge2017c}
\biblab{Buldge2017c}
Buldgen, G., Salmon, S. J. A. J., Noels, A., Scuflaire, R.,
Reese, D. R., Dupret, M.-A., Colgan, J., Fontes, C. J., Eggenberger, P.,
Hakel, P., Kilcrease, D. P. \& Turck-Chi{\`e}ze, S., 2017.
[Seismic inversion for the solar entropy. A case for improving the standard 
solar model].
{\it Astron. Astrophys.}, {\bf 607}, A58-(1--5).

\bibitem[Buldgen {\it et~al.\/}(2019b)]{Buldge2019b}
\biblab{Buldge2019b}
Buldgen, G., Salmon, S. J. A. J., Noels, A., Scuflaire, R.,
Montalban, J., Baturin, V. A., Eggenberger, P., Gryaznov, V. K.,
Iosilevskiy, I. L.,
Meynet, G., Chaplin, W. J., Miglio, A., Oreshina, A. V., Richard, O. \&
Starostin, A. N., 2019b.
[Combining multiple structural inversions to constrain the solar 
modelling problem].
{\it Astron. Astrophys.}, {\bf 621}, A33-(1--16).
%{\tt [arXiv:1809.08958v1 [astro-ph.SR]]}.

\bibitem[Buldgen {\it et~al.\/}(2019a)]{Buldge2019a}
\biblab{Buldge2019a}
Buldgen, G., Salmon, S. \& Noels, A., 2019a.
[Progress in global helioseismology: a new light on the solar modeling
problem and its implications for solar-like stars].
{\it Frontiers in Astronomy and Space Sciences}, {\bf 6}, 42-(1--28).

\bibitem[Buldgen {\it et~al.\/}(2020)]{Buldge2020}
\biblab{Buldge2020}
Buldgen, G., Eggenberger, P., Baturin, V. A., Corbard, T.,
Christensen-Dalsgaard, J., Salmon, S. J. A. J., Noels, A., Oreshina, A. V. \&
Scuflaire, R., 2020.
[Seismic solar models from Ledoux discriminant inversions].
{\it Astron. Astrophys.}, {\bf 642}, A36-(1 -- 14).

\bibitem[Burgers(1969)]{Burger1969}
\biblab{Burger1969}
Burgers, J. M., 1969.
{\it Flow equations for composite gases},
Academic Press, New York.

\bibitem[Caffau {\it et~al.\/}(2008)]{Caffau2008}
\biblab{Caffau2008}
Caffau, E., Ludwig, H.-G., Steffen, M., Ayres, T. R., Bonifacio, P., 
Cayrel, R., Freytag, B. \& Plez, B., 2008.
[The photospheric solar oxygen project. I. Abundance analysis of atomic lines
and influence of atmospheric models].
{\it Astron. Astrophys.}, {\bf 488}, 1031--1046.

\bibitem[Caffau {\it et~al.\/}(2009)]{Caffau2009}
\biblab{Caffau2009}
Caffau, E., Maiorca, E., Bonifacio, P., Faraggiana, R., Steffen, M.,
Ludwig, H.-G., Kamp, I. \& Busso, M., 2009.
[The solar photospheric nitrogen abundance. Analysis of atomic transitions
with 3D and 1D model atmospheres].
{\it Astron. Astrophys.}, {\bf 498}, 877--884.
%{\tt [arXiv:0903.3406v1 [astro-ph.SR]]}

\bibitem[Caffau {\it et~al.\/}(2011)]{Caffau2011}
\biblab{Caffau2011}
Caffau, E., Ludwig, H.-G., Steffen, M., Freytag, B. \& Bonifacio, P., 2011.
[Solar chemical abundances determined with a CO5BOLD 3D model atmosphere].
{\it Solar Phys.}, {\bf 268}, 255--269.

\bibitem[Campante {\it et~al.\/}(2016)]{Campan2016}
\biblab{Campan2016}
Campante, T. L., Lund, M. N., Kuszlewicz, J. S., Davies, G. R.,
Chaplin, W. J., Albrecht, S., Winn, J. N., Bedding, T. R., Benomar, O.,
Bossini, D., Handberg, R., Santos, A. R. G., Van Eylen, V., Basu, S.,
Christensen-Dalsgaard, J., Elsworth, Y. P., Hekker, S., Hirano, T.,
Huber, D., Karoff, C., Kjeldsen, H., Lundkvist, M. S., North, T. S. H.,
Silva Aguirre, V., Stello, D. \& White, T. R., 2016.
[Spin-orbit alignment of exoplanet systems: ensemble analysis using 
asteroseismology].
{\it Astrophys. J.}, {\bf 819}, 85-(1--28).

\bibitem[Canuto and Mazzitelli(1991)]{Canuto1991}
\biblab{Canuto1991}
Canuto, V. M. \& Mazzitelli, I., 1991.
[Stellar turbulent convection: a new model and applications].
{\it Astrophys. J.}, {\bf 370}, 295--311.

\bibitem[Canuto {\it et~al.\/}(1996)]{Canuto1996}
\biblab{Canuto1996}
Canuto, V. M., Goldman, I. \& Mazzitelli, I., 1996.
[Stellar turbulent convection: a self-consistent model].
{\bf Astrophys. J.}, {\it 473}, 550--559.

\bibitem[Carlos {\it et~al.\/}(2016)]{Carlos2016}
\biblab{Carlos2016}
Carlos, M., Nissen, P. E. \& Mel{\'e}ndez, J., 2016.
[Correlation between lithium abundance and ages of solar twin stars].
{\it Astron. Astrophys.}, {\bf 587}, A100-(1--6).

\bibitem[Carrington(1863)]{Carrin1863}
\biblab{Carrin1863}
Carrington, R. C., 1863. 
{\it Spots on the Sun, from November 9, 1853, to March 24, 1861,
made at Redhill},
Williams and Norgate, London.

\bibitem[Carpenter {\it et~al.\/}(2009)]{Carpen2009}
\biblab{Carpen2009}
Carpenter, K. G., Schrijver, C. J. \& Karovska, M., 2009.
[The Stellar Imager (SI) project: a deep space UV/Optical Interferometer
(UVOI) to observe the Universe at 0.1 milli-arcsec angular resolution].
{\it Astrophys. Space Sci.}, {\bf 320}, 217--223.

\bibitem[Cassisi {\it et~al.\/}(2003a)]{Cassis2003a}
\biblab{Cassis2003a}
Cassisi, S., Salaris, M. \& Irwin, A. W., 2003a.
[The initial helium content of galactic globular cluster stars from
the $R$-parameter: comparison with the cosmic microwave background
constraint].
{\it Astrophys. J.}, {\bf 588}, 862--870.

\bibitem[Cassisi {\it et~al.\/}(2003b)]{Cassis2003b}
\biblab{Cassis2003b}
Cassisi, S., Schlattl, H., Salaris, M. \& Weiss, A., 2003b.
[First full evolutionary computation of the helium flash-induced mixing
in Population II stars].
{\it Astrophys. J.}, {\bf 582}, L43--L46.

\bibitem[Castellani {\it et~al.\/}(1997)]{Castel1997}
\biblab{Castel1997}
Castellani, V., Degl'Innocenti, S., Fiorentini, G.,
Lissia, M. \& Ricci, B., 1997.
[Solar neutrinos: Beyond standard solar models].
{\it Phys. Rep.}, {\bf 281}, 309--398.

\bibitem[Castro {\it et~al.\/}(2007)]{Castro2007}
\biblab{Castro2007}
Castro, M., Vauclair, S. \& Richard, O., 2007.
[Low abundances of heavy elements in the solar outer layers: comparisons
of solar models with helioseismic inversions].
{\it Astron. Astrophys.}, {\bf 463}, 755--758.

%\bibitem[Catala(2008)]{Catala2008}
%\biblab{Catala2008}
%Catala, C., and the PLATO consortium, 2008.
%[PLATO: PLAnetary Transits and Oscillations of stars].
%In {\it Proc. HELAS II International Conference: Helioseismology,
%Asteroseismology and the MHD Connections, G\"ottingen, August 2007},
%eds L. Gizon \& M. Roth,
%{\it J. Phys.: Conf. Ser.}, {\bf 118}, 012040(1--11).

\bibitem[{\v C}elebonovi{\'c} {\it et~al.\/}(2004)]{Celebo2004}
\biblab{Celebo2004}
{\v C}elebonovi{\'c}, V., D{\"a}ppen, W. \& Gough, D., (eds), 2004.
{\it Equation-of-State and Phase-Transition Issues in
Models of Ordinary Astrophysical Matter},
AIP Conf. Proc., AIP, Melville, New York, pp. i--x, 1--312.

\bibitem[Cerde{\~n}o {\it et~al.\/}(2018)]{Cerden2018}
\biblab{Cerden2018}
Cerde{\~n}o, D. G., Davis, J. H., Fairbairn, M. \& Vincent, A. C., 2018.
[CNO neutrino Grand Prix: the race to solve the solar metallicity problem].
{\it J. Cosmology and Astroparticle Physics}, {\bf 04}, 037-(i, 1--22).

\bibitem[Chaboyer {\it et~al.\/}(1995)]{Chaboy1995}
\biblab{Chaboy1995}
Chaboyer, B., Demarque, P., Guenther, D. B. \& Pinsonneault, M. H., 1995.
[Rotation, diffusion and overshoot in the Sun:
effects on the oscillation frequencies and the neutrino flux].
{\it Astrophys. J.}, {\bf 446}, 435--444.

\bibitem[Chandrasekhar(1939)]{Chandr1939}
\biblab{Chandr1939}
Chandrasekhar, S., 1939.
{\it An introduction to the study of stellar structure},
University of Chicago Press, Chicago (reissued 1957, by Dover Publ.)

\bibitem[Chandrasekhar(1964)]{Chandr1964}
\biblab{Chandr1964}
Chandrasekhar, S., 1964.
[A general Variational Principle Governing the Radial and the
Non-radial Oscillations of Gaseous Masses].
{\it Astrophys. J.}, {\bf 139}, 664--674.

\bibitem[Chaplin and  Miglio(2013)]{Chapli2013}
\biblab{Chapli2013}
Chaplin, W. J. \& Miglio, A., 2013.
[Asteroseismology of solar-type and red-giant stars].
{\it Annu. Rev. Astron. Astrophys.}, {\bf 51}, 353--392.

\bibitem[Chaplin {\it et~al.\/}(1996)]{Chapli1996}
\biblab{Chapli1996}
Chaplin, W. J., Elsworth, Y., Howe, R., Isaak, G. R., McLeod, C. P.,
Miller, B. A., van der Raay, H. B., Wheeler, S. J. \& New, R., 1996.
[BiSON performance].
{\it Solar Phys.}, {\bf 168}, 1--18.

\bibitem[Chaplin {\it et~al.\/}(1997)]{Chapli1997}
\biblab{Chapli1997}
Chaplin, W. J., Elsworth, Y., Howe, R., Isaak, G. R., McLeod, C. P.,
Miller, B. A. \& New, R., 1997.
[The observation and simulation of stochastically excited solar p modes].
{\it Mon. Not. R. Astron. Soc.}, {\bf 287}, 51--56.

\bibitem[Chaplin {\it et~al.\/}(1999)]{Chapli1999}
\biblab{Chapli1999}
Chaplin, W. J., Christensen-Dalsgaard, J.,
Elsworth, Y., Howe, R., Isaak, G. R., Larsen, R. M., New, R., Schou, J.,
Thompson, M. J. \& Tomczyk, S., 1999.
[Rotation of the solar core from BiSON and LOWL frequency observations].
{\it Mon. Not. R. Astron. Soc.}, {\bf 308}, 405--414.

\bibitem[Chaplin {\it et~al.\/}(2001a)]{Chapli2001a}
\biblab{Chapli2001a}
Chaplin, W. J., Elsworth, Y., Isaak, G. R., Marchenkov, K. I., 
Miller, B. A. \& New, R., 2001a.
[Changes to low-$\ell$ solar p-mode frequencies over the solar cycle:
correlations on different time-scales].
{\it Mon. Not. R. Astron. Soc.}, {\bf 322}, 22--30.

\bibitem[Chaplin {\it et~al.\/}(2001b)]{Chapli2001b}
\biblab{Chapli2001b}
Chaplin, W. J., Elsworth, Y., Isaak, G. R., Marchenkov, K. I.,
Miller, B. A. \& New, R., 2001b.
[Rigid rotation of the solar core? On the reliable extraction of
low-$\ell$ rotational p-mode splittings from full-disc observations of the Sun].
{\it Mon. Not. R. Astron. Soc.}, {\bf 327}, 1127--1136.

\bibitem[Chaplin {\it et~al.\/}(2004)]{Chapli2004}
\biblab{Chapli2004}
Chaplin, W. J., Sekii, T., Elsworth, Y. \& Gough, D. O., 2004.
[On the detectability of a rotation-rate gradient in the solar core].
{\it Mon. Not. R. Astron. Soc.}, {\bf 355}, 535--542.

\bibitem[Chaplin {\it et~al.\/}(2007)]{Chapli2007}
\biblab{Chapli2007}
Chaplin, W. J., Serenelli, A. M., Basu, S., Elsworth, Y., New, R. \&
Verner, G. A., 2007.
[Solar heavy-element abundance: constraints from frequency separation ratios
of low-degree $p$-modes].
{\it Astrophys. J.}, {\bf 670}, 872--884.

{\rv
\bibitem[Charbonneau(2020)]{Charbo2020}
\biblab{Charbo2020}
Charbonneau, P., 2020.
[Dynamo models of the solar cycle].
{\it Living Rev. Solar Phys.}, {\bf 17},  4. URL (cited on 4/10/20): 
\url{https://doi.org/10.1007/s41116-020-00025-6}
}

\bibitem[Charbonneau and MacGregor(1993)]{Charbo1993}
\biblab{Charbo1993}
Charbonneau, P. \& MacGregor, K. B., 1993.
[Angular momentum transport in magnetized stellar radiative zones.
II. The solar spin-down].
{\it Astrophys. J.}, {\bf 417}, 762--780.

\bibitem[Charbonneau {\it et~al.\/}(1999)]{Charbo1999}
\biblab{Charbo1999}
Charbonneau, P., Christensen-Dalsgaard, J., Henning, R., 
Larsen, R. M., Schou, J., Thompson, M. J. \& Tomczyk, S., 1999.
[Helioseismic constraints on the structure of the solar tachocline].
{\it Astrophys. J.}, {\bf 527}, 445--460.

\bibitem[Charbonnel and Talon(2005)]{Charbl2005}
\biblab{Charbl2005}
Charbonnel, C. \& Talon, S., 2005.
[Influence of gravity waves on the internal rotation and Li abundance of
solar-type stars].
{\it Science}, {\bf 309}, 2189--2191.

\bibitem[Charbonnel and  Zahn(2007)]{Charbl2007}
\biblab{Charbl2007}
Charbonnel, C. \& Zahn, J.-P., 2007.
[Thermohaline mixing: a physical mechanism governing the photospheric
composition of low-mass stars].
{\it Astron. Astrophys.}, {\bf 467}, L15--L18.

\bibitem[Chitre {\it et~al.\/}(1998)]{Chitre1998}
\biblab{Chitre1998}
Chitre, S. M., Christensen-Dalsgaard, J. \& Thompson, M. J., 1998.
[Diagnostic potential of the solar f modes].
In {\it Structure and dynamics of the
interior of the Sun and Sun-like stars; Proc. SOHO 6/GONG 98 Workshop},
eds S.G. Korzennik \& A. Wilson, ESA SP-418, ESA Publications Division,
Noordwijk, The Netherlands, p. 141--145.

\bibitem[Christensen-Dalsgaard(1982)]{Christ1982}
\biblab{Christ1982}
Christensen-Dalsgaard, J., 1982.
[On solar models and their periods of oscillation].
{\it Mon. Not. R. astr. Soc.}, {\bf 199}, 735--761.

\bibitem[Christensen-Dalsgaard(1984)]{Christ1984a}
\biblab{Christ1984a}
Christensen-Dalsgaard, J., 1984a.
[Solar Oscillations].
{\it The Hydromagnetics of the Sun}, p. 3--12,
ESA SP-220, ESTEC, Noordwijk.

\bibitem[Christensen-Dalsgaard(1984b)]{Christ1984b}
\biblab{Christ1984b}
Christensen-Dalsgaard, J., 1984b.
[What will asteroseismology teach us?].
{\it Space Research Prospects in Stellar Activity and Variability},
p. 11--45,
eds Mangeney, A. \& Praderie, F., Paris Observatory Press.

\bibitem[Christensen-Dalsgaard(1988a)]{Christ1988a}
\biblab{Christ1988a}
Christensen-Dalsgaard, J., 1988a.
[Study of solar structure based on p-mode helioseismology].
{\it Seismology of the Sun \& Sun-like Stars}, p. 431--450,
eds Domingo, V. \& Rolfe, E. J., 
ESA SP-286, ESA Publications Division,
Noordwijk, The Netherlands.

\bibitem[Christensen-Dalsgaard(1988b)]{Christ1988b}
\biblab{Christ1988b}
Christensen-Dalsgaard, J., 1988b.
[A Hertzsprung-Russell diagram for stellar oscillations].
{\it Proc. IAU Symposium No 123, Advances in helio- and asteroseismology},
p. 295--298,
eds Christensen-Dalsgaard, J. \& Frandsen, S., 
Reidel, Dordrecht.

\bibitem[Christensen-Dalsgaard(1991)]{Christ1991c}
\biblab{Christ1991c}
Christensen-Dalsgaard, J., 1991.
[Some aspects of the theory of solar oscillations].
{\it Geophys. Astrophys. Fluid Dynamics}, {\bf 62}, 123--152.

\bibitem[Christensen-Dalsgaard(1992)]{Christ1992a}
\biblab{Christ1992a}
Christensen-Dalsgaard, J., 1992a.
[Solar models with enhanced energy transport in the core].
{\it Astrophys. J.}, {\bf 385}, 354--362.

\bibitem[Christensen-Dalsgaard(1997)]{Christ1997a}
\biblab{Christ1997a}
Christensen-Dalsgaard, J., 1997.
[Effects of convection on the mean solar structure].
In {\it SCORe'96: Solar Convection and Oscillations and their
Relationship},
eds Pijpers, F. P., Christensen-Dalsgaard, J. \& Rosenthal, C. S., 
Kluwer, Dordrecht, p. 3--22.

\bibitem[Christen\-sen-Dalsgaard(2002)]{Christ2002}
\biblab{Christ2002}
Christensen-Dalsgaard, J., 2002.
[Helioseismology].
{\it Rev. Mod. Phys.}, {\bf 74}, 1073--1129. % Oct. 2002, [astro-ph/0207403].

\bibitem[Christensen-Dalsgaard(2004)]{Christ2004}
\biblab{Christ2004}
Christensen-Dalsgaard, J., 2004.
[An overview of helio- and asteroseismology].
In {\it Proc. SOHO 14 - GONG 2004: ``Helio- and Asteroseismology: Towards
a golden future''; Yale, July 12--16 2004}, ed. Danesy, D., ESA SP-559,
ESA Publication Division, Noordwijk, The Netherlands, p. 1--33.

\bibitem[Christensen-Dalsgaard(2008)]{Christ2008a}
\biblab{Christ2008a}
Christensen-Dalsgaard, J., 2008.
[ASTEC--the Aarhus STellar Evolution Code].
{\it Astrophys. Space Sci.},  {\bf 316}, 13--24.

\bibitem[Christensen-Dalsgaard(2009)]{Christ2009a}
\biblab{Christ2009a}
Christensen-Dalsgaard, J., 2009.
[The Sun as a fundamental calibrator of stellar evolution].
In {\it Proc. IAU Symp. 258, The Ages of Stars},
eds E. E. Mamajek, D. R. Soderblom \& R. F. G. Wyse,
IAU and Cambridge University Press, p. 431--442.
%{\tt [arXiv:0904.0358v1 [astro-ph.SR]]}

\bibitem[Christensen-Dalsgaard and Di Mauro(2007)]{Christ2007}
\biblab{Christ2007}
Christensen-Dalsgaard, J. \& Di Mauro, M. P., 2007.
[Diffusion and helioseismology].
In {\it Stellar Evolution and Seismic Tools for Asteroseismology: Diffusive
Processes in Stars and Seismic Analysis}, eds C. W. Straka, Y. Lebreton \&
M. J. P. F. G. Monteiro, EAS Publ. Ser., {\bf 26},
EDP Sciences, Les Ulis, France, p. 3--16.

\bibitem[Christensen-Dalsgaard and D{\"a}ppen(1992)]{Christ1992}
\biblab{Christ1992}
Christensen-Dalsgaard, J. \& D{\"a}ppen, W., 1992.
[Solar oscillations and the equation of state].
{\it Astron. Astrophys. Rev.}, {\bf 4}, 267--361.

\bibitem[Christensen-Dalsgaard and Gough(1976)]{Christ1976}
\biblab{Christ1976}
Christensen-Dalsgaard, J. \& Gough, D. O., 1976.
[Towards a heliological inverse problem].
{\it Nature}, {\bf 259}, 89--92.

\bibitem[Christensen-Dalsgaard and  Houdek(2010)]{Christ2010}
\biblab{Christ2010}
Christensen-Dalsgaard, J. \& Houdek, G., 2010.
[Prospects for asteroseismology].
In {\it Proc. HELAS Workshop on `Synergies between solar and
stellar modelling', Rome 22--26 June 2009},
eds M. Marconi, D. Cardini \& M. P. Di Mauro,
{\it Astrophys. Space Sci.}, {\bf 328}, 51--66.
%{\tt [arXiv:0911.4629 [astro-ph]]}.

\bibitem[Christensen-Dalsgaard and  Reiter(1995)]{ChristReit1995}
\biblab{ChristReit1995}
Christensen-Dalsgaard, J. \& Reiter, J., 1995.
[A comparison of precise solar models with simplified physics].
In {\it Proc. GONG'94: Helio- and Astero-seismology from Earth and Space},
eds Ulrich, R. K., Rhodes Jr, E. J. \& D{\"a}ppen, W.,
Astronomical Society of the Pacific Conference Series, San Francisco,
{\bf 76}, 136--139.

\bibitem[Christensen-Dalsgaard and Silva Aguirre(2018)]{Christ2018b}
\biblab{Christ2018b}
Christensen-Dalsgaard, J. \& Silva Aguirre, V., 2018.
[Ages for exoplanet host stars].
In {\it Handbook of Exoplanets}, eds Deeg, H.J. \& Belmonte, J.A,
Springer, Cham, pp 1--18.
\url{https://doi.org/10.1007/978-3-319-30648-3_184-1}
%{\tt [arXiv:1803.03125 [astro-ph.SR]]}

\bibitem[Christensen-Dalsgaard and Thompson(1991)]{Christ1991a}
\biblab{Christ1991a}
Christensen-Dalsgaard, J. \& Thompson, M. J., 1991.
[The response of the adiabatic exponent $\Gamma_1$
to modifications of solar models].
{\it Astrophys. J.}, {\bf 367}, 666--670.

\bibitem[Christensen-Dalsgaard and Thompson(1997)]{Christ1997b}
\biblab{Christ1997}
Christensen-Dalsgaard, J. \& Thompson, M. J., 1997.
[On solar p-mode frequency shifts caused by near-surface model changes].
{\it Mon. Not. R. Astron. Soc.}, {\bf 284}, 527--540.

\bibitem[Christensen-Dalsgaard {\it et~al.\/}(1974)]{Christ1974}
\biblab{Christ1974}
Christensen-Dalsgaard, J., Dilke, F. W. W. \& Gough, D. O., 1974.
[The stability of a solar model to non-radial oscillations].
{\it Mon. Not. R. astr. Soc.}, {\bf 169}, 429--445.

\bibitem[Christensen-Dalsgaard {\it et~al.\/}(1985)]{Christ1985}
\biblab{Christ1985}
Christensen-Dalsgaard, J., Duvall, T. L., Gough, D. O., Harvey, J. W. \& 
Rhodes Jr, E. J., 1985.
[Speed of sound in the solar interior].
{\it Nature}, {\bf 315}, 378--382.

\bibitem[Christensen-Dalsgaard {\it et~al.\/}(1988)]{ChristDL1988}
\biblab{ChristDL1988}
Christensen-Dalsgaard, J., D{\"a}ppen, W. \& Lebreton, Y., 1988.
[Solar oscillation frequencies and the equation of state].
{\it Nature}, {\bf 336}, 634--638.

\bibitem[Christensen-Dalsgaard {\it et~al.\/}(1989)]{Christ1989}
\biblab{Christ1989}
Christensen-Dalsgaard, J., Gough, D. O. \& Thompson, M. J., 1989.
[Differential asymptotic sound-speed inversions].
{\it Mon. Not. R. astr. Soc.}, {\bf 238}, 481--502.

\bibitem[Christensen-Dalsgaard {\it et~al.\/}(1991)]{Christ1991b}
\biblab{Christ1991b}
Christensen-Dalsgaard, J., Gough, D. O. \& Thompson, M. J., 1991.
[The depth of the solar convection zone].
{\it Astrophys. J.}, {\bf 378}, 413--437.

\bibitem[Christensen-Dalsgaard {\it et~al.\/}(1992)]{Christ1992b}
\biblab{Christ1992b}
Christensen-Dalsgaard, J., Gough, D. O. \& Thompson, M. J., 1992.
[On the rate of destruction of lithium in late-type main-sequence stars].
{\it Astron. Astrophys.}, {\bf 264}, 518--528.

\bibitem[Christensen-Dalsgaard {\it et~al.\/}(1993)]{Christ1993}
\biblab{Christ1993}
Christensen-Dalsgaard, J., Proffitt, C. R. \& Thompson, M. J., 1993.
[Effects of diffusion on solar models and their oscillation frequencies].
{\it Astrophys. J.}, {\bf 403}, L75--L78.

\bibitem[Christensen-Dalsgaard {\it et~al.\/}(1995)]{Christ1995a}
\biblab{Christ1995a}
Christensen-Dalsgaard, J., Monteiro, M. J. P. F. G. \& Thompson, M. J., 1995.
[Helioseismic estimation of convective overshoot in the Sun].
{\it Mon. Not. R. Astron. Soc.}, {\bf 276}, 283--292.

\bibitem[Christensen-Dalsgaard {\it et~al.\/}(1996)]{Christ1996}
\biblab{Christ1996}
Christensen-Dalsgaard, J., D{\"a}ppen, W., Ajukov, S. V., Anderson, E. R.,
Antia, H. M., Basu, S., Baturin, V. A., Berthomieu, G., Chaboyer, B.,
Chitre, S. M., Cox, A. N., Demarque, P., Donatowicz, J., Dziembowski, W. A.,
Gabriel, M., Gough, D. O., Guenther, D. B., Guzik, J. A., Harvey, J. W.,
Hill, F., Houdek, G., Iglesias, C. A., Kosovichev, A. G., Leibacher, J. W.,
Morel, P., Proffitt, C. R., Provost, J., Reiter, J., Rhodes Jr., E. J.,
Rogers, F. J., Roxburgh, I. W., Thompson, M. J., Ulrich, R. K., 1996.
[The current state of solar modeling].
{\it Science}, {\bf 272}, 1286--1292.

\bibitem[Christensen-Dalsgaard {\it et~al.\/}(2005)]{Christ2005}
\biblab{Christ2005}
Christensen-Dalsgaard, J., Di Mauro, M. P., Schlattl, H. \& Weiss, A., 2005.
[On helioseismic tests of basic physics].
{\it Mon. Not. R. Astron. Soc.}, {\bf 356},  587--595.

%\bibitem[Christensen-Dalsgaard {\it et~al.\/}(2007)]{Christ2007b}
%\biblab{Christ2007b}
%Christensen-Dalsgaard, J., Arentoft, T., Brown, T. M., Gilliland, R. L.,
%Kjeldsen, H., Borucki, W. J. \& Koch, D., 2007.
%[The {\it Kepler\/} Asteroseismic Investigation].
%In {\it Proc. HELAS II International Conference: Helioseismology,
%Asteroseismology and the MHD Connections, G\"ottingen, August 2007},
%eds L. Gizon \& M. Roth,
%{\it J. Phys.: Conf. Ser.}, in the press.

%\bibitem[Christensen-Dalsgaard {\it et~al.\/}(2008)]{Christ2008b}
%\biblab{Christ2008b}
%Christensen-Dalsgaard, J., Arentoft, T., Brown, T. M., Gilliland, R. L.,
%Kjeldsen, H., Borucki, W. J. \& Koch, D., 2008.
%[The {\it Kepler\/} Asteroseismic Investigation].
%In {\it Proc. HELAS II International Conference: Helioseismology,
%Asteroseismology and the MHD Connections, G\"ottingen, August 2007},
%eds L. Gizon \& M. Roth,
%{\it J. Phys.: Conf. Ser.}, in the press.

\bibitem[Christensen-Dalsgaard {\it et~al.\/}(2009)]{Christ2009b}
\biblab{Christ2009b}
Christensen-Dalsgaard, J., Di Mauro, M. P., Houdek, G. \& Pijpers, F., 2009.
[On the opacity change required to compensate for the revised solar
composition].
{\it Astron. Astrophys.}, {\bf 494}, 205--208.

\bibitem[Christensen-Dalsgaard {\it et~al.\/}(2011)]{Christ2011}
\biblab{Christ2011}
Christensen-Dalsgaard, J., Monteiro, M. J. P. F. G., Rempel, M. \&
Thompson, M. J., 2011.
[A more realistic representation of overshoot at the base
of the solar convective envelope as seen by helioseismology].
{\it Mon. Not. R. Astron. Soc.}, {\bf 414}, 1158--1174.

\bibitem[Christensen-Dalsgaard {\it et~al.\/}(2018)]{Christ2018a}
\biblab{Christ2018a}
Christensen-Dalsgaard, J., Gough, D. O. \& Knudstrup, E., 2018a.
[On the hydrostatic stratification of the solar tachocline].
{\it Mon. Not. R. Astron. Soc.}, {\bf 477}, 3845--3852.
%{\tt [arXiv:1803.08675 [astro-ph.SR]]}

\bibitem[Claverie {\it et~al.\/}(1979)]{Claver1979}
\biblab{Claver1979}
Claverie, A., Isaak, G. R., McLeod, C. P., van der Raay, H. B. \&
Roca Cortes, T., 1979.
[Solar structure from global studies of the 5-minute oscillation].
{\it Nature}, {\bf 282}, 591--594.

\bibitem[Clayton(1968)]{Clayto1968}
\biblab{Clayto1968}
Clayton, D. D., 1968.
{\it Principles of Stellar Evolution and Nucleosynthesis}, 
McGraw-Hill, New York.

\bibitem[Cleveland {\it et~al.\/}(1998)]{Clevel1998}
\biblab{Clevel1998}
Cleveland, B. T., Daily, T., Davis Jr, R., Distel, J. R., Lande, K.,
Lee, C. K., Wildenhain, P. S. \& Ullman, J., 1998.
[Measurement of the solar electron neutrino flux with the Homestake 
chlorine detector].
{\it Astrophys. J.}, {\bf 496}, 505--526.

\bibitem[Cohen and  Taylor(1987)]{Cohen1987}
\biblab{Cohen1987}
Cohen, E. R. \& Taylor, B. N., 1987.
[The 1986 adjustment of the fundamental physical constants].
{\it Rev. Mod. Phys.}, {\bf 59}, 1121--1148.

\bibitem[Colgan {\it et~al.\/}(2016)]{Colgan2016}
\biblab{Colgan2016}
Colgan, J., Kilcrease, D. P., Magee, N. H., Sherrill, M. E., Abdallah, J., 
Hakel, P., Fontes, C. J., Guzik, J. A. \& Mussack, K. A., 2016.
[A new generation of Los Alamos Opacity tables].
{\it Astrophys. J.}, {\bf 817}, 116-(1--10).

\bibitem[Collet {\it et~al.\/}(2011)]{Collet2011}
\biblab{Collet2011}
Collet, R., Hayek, W., Asplund, M., Nordlund, {\AA}., Trampedach, R. \&
Gudiksen, B., 2011.
[Three-dimensional surface convection simulations of metal-poor stars.
The effect of scattering on the photospheric temperature stratification].
{\it Astron. Astrophys.}, {\bf 528}, A32-(1--12).

\bibitem[Connelly {\it et~al.\/}(2012)]{Connel2012}
\biblab{Connel2012}
Connelly, J. N., Bizzarro, M., Krot, A. N., Nordlund, {\AA}.,
Wielandt, D. \& Ivanova, M. A., 2012.
[The absolute chronology and thermal processing of solids in the
solar protoplanetary disk].
{\it Science}, {\bf 338}, 651--655.

%\bibitem[Collins {\it et~al.\/}(2007)]{Collin2007}
%\biblab{Collin2007}
%Collins, W., Colman, R., Haywood, J., Manning, M. R. \& Mote, P., 2007.
%[The physical science behind climate change].
%{\it Scientific American}, {\bf 297} (August 48--55.

\bibitem[Corbard and  Thompson(2002)]{Corbar2002}
\biblab{Corbar2002}
Corbard, T. \& Thompson, M. J., 2002.
[The subsurface radial gradient of solar angular velocity from MDI
$f$-mode observations].
{\it Solar Phys.}, {\bf 205}, 211--229.

{\rv
\bibitem[Corbard {\it et~al.\/}(1998)]{Corbar1998}
\biblab{Corbar1998}
Corbard, T., Di Mauro, M. P., Sekii, T. and the GOLF team, 1998.
[The solar internal rotation from the GOLF splittings].
In {\it Structure and dynamics of the
interior of the Sun and Sun-like stars; Proc. SOHO 6/GONG 98 Workshop}, 
eds S.G. Korzennik \& A. Wilson, ESA SP-418, ESA Publications Division,
Noordwijk, The Netherlands, p. 741--746.
}

{\rv
\bibitem[Corbard {\it et~al.\/}(1999)]{Corbar1999}
\biblab{Corbar1999}
Corbard, T., Blanc-F{\'e}raud, L., Berthomieu, G. \& Provost, J., 1999.
[Non linear regularization for helioseismic inversions.
Application for the study of the solar tachocline].
{\it Astron. Astrophys.}, {\bf 344}, 696--708.
}

\bibitem[Cossette and  Rast(2016)]{Cosset2016}
\biblab{Cosset2016}
Cossette, J.-F. \& Rast, M. P., 2016.
[Supergranulation as the largest buoyantly driven convective scale of the Sun].
{\it Astrophys. J.}, {\bf 829}, L17-(1--5).

\bibitem[Couvidat {\it et~al.\/}(2003)]{Couvid2003}
\biblab{Couvid2003}
Couvidat, S., Turck-Chi{\`e}ze, S. \& Kosovichev, A. G., 2003.
[Solar seismic models and the neutrino predictions].
{\it Astrophys. J.}, {\bf 599}, 1434--1448.

\bibitem[Cox(1991)]{Cox1991}
\biblab{Cox1991}
Cox, A. N., 1991.
[Masses of RR{\it d} variables using Livermore OPAL opacities].
{\it Astrophys. J.}, {\bf 381}, L71--L74.

\bibitem[Cox(2000)]{Cox2000}
\biblab{Cox2000}
Cox, A. N., (ed.), 2000.
{\it Allen's Astrophysical Quantities, Fourth Ed.}, 
Springer, New York.

\bibitem[Cox and Stewart(1970)]{Cox1970}
\biblab{Cox1970}
Cox, A. N. \& Stewart, J. N., 1970.
[Rosseland opacity tables for Population I compositions].
{\it Astrophys. J. Suppl.}, {\bf 19}, 243--279.

\bibitem[Cox and Tabor(1976)]{Cox1976}
\biblab{Cox1976}
Cox, A. N. \& Tabor, J. E., 1976.
[Radiative opacity tables for 40 stellar mixtures].
{\it Astrophys. J. Suppl.}, {\bf 31}, 271--312.

\bibitem[Cox {\it et~al.\/}(1989)]{Cox1989}
\biblab{Cox1989}
Cox, A. N., Guzik, J. A. \& Kidman, R. B., 1989.
[Oscillations of solar models with internal element diffusion].
{\it Astrophys. J.}, {\bf 342}, 1187--1206.

\bibitem[Cox {\it et~al.\/}(1992)]{Cox1992}
\biblab{Cox1992}
Cox, A. N., Morgan, S. M., Rogers, F. J. \& Iglesias, C. A., 1992.
[An opacity mechanism for the pulsations of OB stars].
{\it Astrophys. J.}, {\bf 393}, 272--277.

\bibitem[Cox and Giuli(1968)]{Cox1968}
\biblab{Cox1968}
Cox, J. P. \& Giuli, R. T., 1968.
{\it Principles of Stellar Structure}, Gordon and Breach, New York.

\bibitem[Crowley(2000)]{Crowle2000}
\biblab{Crowle2000}
Crowley, T. J., 2000.
[Causes of climate change over the past 1000 years].
{\it Science}, {\bf 289}, 270--277.

\bibitem[Cunha and Smith(1999)]{Cunha1999}
\biblab{Cunha1999}
Cunha, K. \& Smith, V. V., 1999.
[A determination of the solar photospheric boron abundance].
{\it Astrophys. J.}, {\bf 512}, 1006--1013.

\bibitem[Cunha and  Metcalfe(2007)]{Cunha2007b}
\biblab{Cunha2007b}
Cunha, M. S. \& Metcalfe, T. S., 2007.
[Asteroseismic signatures of small convective cores].
{\it Astrophys. J.}, {\bf 666}, 413--422.

\bibitem[Cunha {\it et~al.\/}(2007)]{Cunha2007a}
\biblab{Cunha2007a}
Cunha, M. S., Aerts, C., Christensen-Dalsgaard, J., Baglin, A., Bigot, L.,
Brown, T. M., Catala, C., Creevey, O. L., Domiciano de Souza, A.,
Eggenberger, P.,
Garcia, P. J. V., Grundahl, F., Kervella, P., Kurtz, D. W., Mathias, P.,
Miglio, A., Monteiro, M. J. P. F. G., Perrin, G., Pijpers, F. P., Pourbaix, D.,
Quirrenbach, A., Rousselet-Perraut, K., Teixeira, T. C., Th\'evenin, F. \&
Thompson, M. J., 2007.
[Asteroseismology and interferometry].
{\it Astron. Astrophys. Rev.},  {\bf 14}, 217--360.
%{\tt [arXiv:0709.4613v1 [astro-ph]]}

\bibitem[Cubasch {\it et~al.\/}(2013)]{Cubasc2013}
\biblab{Cubasc2013}
Cubasch, U., Wuebbles, D., Chen, D., Facchini, M.~C., Frame, D.,
Mahowald, N. \& Winther, J.-G., 2013.
[Introduction].
In: {\it Climate Change 2013: The Physical Science Basis.
Contribution of Working Group I to the Fifth Assessment Report of
the Intergovernmental Panel on Climate Change},
eds Stocker, T.~F., Qin, D., Plattner, G.-K., Tignor, M., Allen, S.~K.,
Boschung, J., Nauels, A., Xia, Y., Bex, V. \&  Midgley, P.~M. 
Cambridge University Press, Cambridge, United Kingdom.

\bibitem[Cyburt {\it et~al.\/}(2016)]{Cyburt2016}
\biblab{Cyburt2016}
Cyburt, R. H., Fields, B. D., Olive, K. A. \& Yeh, T.-H., 2016.
[Big bang nucleosynthesis: present status].
{\it Rev. Mod. Phys.}, {\bf 88}, 015004-(1--22).

\bibitem[D{\"a}ppen(1993)]{Dappen1993}
\biblab{Dappen1993}
D{\"a}ppen, W., 1993.
[The equation of state].
In {\it Proc. IAU Colloq. 137: Inside the stars},
eds Baglin, A. \& Weiss, W. W., 
Astronomical Society of the Pacific Conference Series, San Francisco,
{\bf 40}, 208--221.

\bibitem[D{\"a}ppen(2004)]{Dappen2004}
\biblab{Dappen2004}
D{\"a}ppen, W., 2004.
[Equations of state for solar and stellar modelling].
In {\it Equation-of-State and Phase-Transition Issues in
Models of Ordinary Astrophysical Matter},
eds V. {\v C}elebonovi{\'c},  W. D{\"a}ppen \& D. Gough,
AIP Conf. Proc. vol. 731, AIP, Melville, New York, p. 3--17.

\bibitem[D{\"a}ppen(2007)]{Dappen2007}
\biblab{Dappen2007}
D{\"a}ppen, W., 2007.
[Seismic abundance determination in the Sun and in stars].
In Stancliffe R. J., Dewi J., Houdek G., Martin R. G., Tout C.A., eds,
AIP Conf. Proc. vol. 948, {\it Unsolved Problems in Stellar Physics}.
American Institute of Physics, Melville, p. 179--190.

\bibitem[D{\"a}ppen(2010)]{Dappen2010}
\biblab{Dappen2010}
D{\"a}ppen, W., 2010.
[Accurate and versatile equations of state for the Sun and Sun-like stars].
{\it Astrophys. Space Sci.}, {\bf 328}, 139--146.

\bibitem[D{\"a}ppen and Gough(1986)]{Dappen1986a}
\biblab{Dappen1986a}
D{\"a}ppen, W. \& Gough, D. O., 1986.
[Progress report on helium abundance determination].
In {\it Seismology of the Sun and the distant Stars}, p. 275--280,
ed. Gough, D. O., Reidel, Dordrecht.

\bibitem[D{\"a}ppen and Guzik(2000)]{Dappen2000}
\biblab{Dappen2000}
D{\"a}ppen, W. \& Guzik, J. A., 2000.
[Astrophysical equation of state and opacity].
In {\it Variable Stars as Essential Astrophysical Tools},
ed. C. \.{I}bano\u{g}lu, Kluwer Academic Publishers, p. 177--212.

\bibitem[D{\"a}ppen and Mao(2009)]{Dappen2009}
\biblab{Dappen2009}
D{\"a}ppen, W. \& Mao, D., 2009.
[A smooth equation of state for solar and stellar abundance determinations].
{\it J. Phys. A: Math. Theor.}, {\bf 42}, 214006-(1--5).

\bibitem[D{\"a}ppen {\it et~al.\/}(1986)]{Dappen1986b}
\biblab{Dappen1986b}
D{\"a}ppen, W., Gilliland, R. L. \& Christensen-Dalsgaard, J., 1986.
[Weakly interacting massive particles, solar neutrinos and solar oscillations].
{\it Nature}, {\bf 321}, 229--231.

\bibitem[D{\"a}ppen {\it et~al.\/}(1988)]{Dappen1988}
\biblab{Dappen1988}
D{\"a}ppen, W., Mihalas, D., Hummer, D. G. \& Mihalas, B. W., 1988.
[The equation of state for stellar envelopes. III. Thermodynamic quantities].
{\it Astrophys. J.}, {\bf 332}, 261--270.

\bibitem[D{\"a}ppen {\it et~al.\/}(1991)]{Dappen1991}
\biblab{Dappen1991}
D{\"a}ppen, W., Gough, D. O., Kosovichev, A. G. \& Thompson, M. J., 1991.
[A new inversion for the hydrostatic stratification of the Sun].
In 
{\it Challenges to theories of the structure of moderate-mass stars},
{\it Lecture Notes in Physics}, vol. {\bf 388}, p. 111--120,
eds Gough, D. O. \& Toomre, J., Springer, Heidelberg.

\bibitem[Davies {\it et~al.\/}(2014)]{Davies2014}
\biblab{Davies2014}
Davies, G. R., Broomhall, A. M., Chaplin, W. J., Elsworth, Y. \& Hale, S. J.,
2014.
[Low-frequency low-degree solar p-mode properties from 22 years of
Birmingham Solar Oscillations Network data].
{\it Mon. Not. R. Astron. Soc.}, {\bf 439}, 2025--2032.

\bibitem[Davis(1964)]{Davis1964}
\biblab{Davis1964}
Davis, R., 1964.
[Solar neutrinos. I. Experimental].
{\it Phys. Rev. Lett.}, {\bf 12}, 303--305.

\bibitem[Davis(2003)]{Davis2003}
\biblab{Davis2003}
Davis, R., 2003.
[Nobel lecture: A half-century with solar neutrinos].
{\it Rev. Mod. Phys.}, {\bf 75}, 985--994.

\bibitem[Davis {\it et~al.\/}(1968)]{Davis1968}
\biblab{Davis1968}
Davis, R., Harmer, D. S. \& Hoffman, K. C., 1968.
[Search for neutrinos from the Sun].
{\it Phys. Rev. Lett.}, {\bf 20}, 1205--1209.

\bibitem[Deal {\it et~al.\/}(2020)]{Deal2020}
\biblab{Deal2020}
Deal, M., Goupil, M.-J., Marques, J. P., Reese, D. R. \& Lebreton, Y., 2020.
[Chemical mixing in low mass stars. I. Rotation against atomic diffusion 
including radiative acceleration].
{\it Astron. Astrophys.}, {\bf 633}, A23-(1--15).

\bibitem[Degl'Innocenti {\it et~al.\/}(1997)]{DeglIn1997}
\biblab{DeglIn1997}
Degl'Innocenti, S., Dziembowski, W. A., Fiorentini, G. \& Ricci, B., 1997.
[Helioseismology and standard solar models].
{\it Astroparticle Phys.}, {\bf 7}, 77--95.

\bibitem[Delahaye and Pinsonneault(2006)]{Delaha2006}
\biblab{Delaha2006}
Delahaye, F. \& Pinsonneault, M., 2006.
[The solar heavy-element abundances. I. Constraints from stellar interiors].
{\it Astrophys. J.}, {\bf 649}, 529--540.

\bibitem[Demarque {\it et~al.\/}(1973)]{Demarq1973}
\biblab{Demarq1973}
Demarque, P., Mengel, J. G. \& Sweigart, A. V., 1973.
[Rotating solar models with low neutrino flux].
{\it Astrophys. J.}, {\bf 183}, 997--1004.
(Erratum: {\it Nature}, {\bf 252}, 368; 1974).

{\rv
\bibitem[Denissenkov and Merryfield(2011)]{Deniss2011}
\biblab{Deniss2011}
Denissenkov, P. A. \& Merryfield, W. J., 2011.
[Thermohaline mixing: does it really govern the atmospheric chemical
composition of low-mass red giants?].
{\it Astrophys. J.}, {\bf 727}, L8-(1--4).
}

\bibitem[Denissenkov and Pinsonneault(2007)]{Deniss2007}
\biblab{Deniss2007}
Denissenkov, P. A. \& Pinsonneault, M., 2007.
[A revised prescription for the Tayler-Spruit dynamo: magnetic angular momentum transport in stars].
{\it Astrophys. J.}, {\bf 655}, 1157--1165.

\bibitem[Denissenkov {\it et~al.\/}(2008)]{Deniss2008}
\biblab{Deniss2008}
Denissenkov, P. A., Pinsonneault, M. \& MacGregor, K. B., 2008.
[What prevents internal gravity waves from disturbing the solar uniform
rotation?].
{\it Astrophys. J.}, {\bf 684}, 757--769.

\bibitem[Deubner(1975)]{Deubne1975}
\biblab{Deubne1975}
Deubner, F.-L., 1975.
[Observations of low wavenumber nonradial eigenmodes of the Sun].
{\it Astron. Astrophys.}, {\bf 44}, 371--375.

\bibitem[Deubner {\it et~al.\/}(1979)]{Deubne1979}
\biblab{Deubne1979}
Deubner, F.-L., Ulrich, R. K. \& Rhodes, E. J., 1979.
[Solar $p$-mode oscillations as a tracer of radial differential rotation].
{\it Astron. Astrophys.}, {\bf 72}, 177--185.

\bibitem[Dilke and Gough(1972)]{Dilke1972}
\biblab{Dilke1972}
Dilke, F. W. W. \& Gough, D. O., 1972.
[The solar spoon].
{\it Nature}, {\bf 240}, 262--264 \& 293--294.

\bibitem[Di Mauro(2003)]{DiMaur2003}
\biblab{DiMaur2003}
Di Mauro, M. P., 2003.
[Helioseismology: a fantastic tool to probe the interior of the Sun].
{\it The Sun's Surface and Subsurface},
{\it Lecture Notes in Physics}, vol. {\bf 599}, p. 31--67,
ed. Rozelot, J.-P., Springer, Heidelberg.
{\tt [arXiv:1212.5077 [astro-ph.SR]]}

\bibitem[Di Mauro {\it et~al.\/}(2002)]{DiMaur2002}
\biblab{DiMaur2002}
Di Mauro, M. P., Christensen-Dalsgaard, J., Rabello-Soares, M. C. \&
Basu, S., 2002.
[Inferences on the solar envelope with high-degree modes].
{\it Astron. Astrophys.}, {\bf 384}, 666--677.

\bibitem[Dipierro {\it et~al.\/}(2015)]{Dipier2015}
\biblab{Dipier2015}
Dipierro, G., Price, D., Laibe, G., Hirsh, K., Cerioli, A. \& Lodato, G.,
2015.
[On planet formation in HL Tau].
{\it Mon. Not. R. Astron. Soc.}, {\bf 453}, L73--L77.

\bibitem[Drake and Testa(2005)]{Drake2005}
\biblab{Drake2005}
Drake, J. J. \& Testa, P., 2005.
[The solar model problem solved by the abundance of neon in stars of
the local cosmos].
{\it Nature}, {\bf 436}, 525--528. %  [{\tt astro-ph/0506182 v1}]

\bibitem[Duvall(1982)]{Duvall1982}
\biblab{Duvall1982}
Duvall, T. L., 1982.
[A dispersion law for solar oscillations].
{\it Nature}, {\bf 300}, 242--243.

\bibitem[Duvall and Harvey(1983)]{Duvall1983}
\biblab{Duvall1983}
Duvall, T. L. \& Harvey, J. W., 1983.
[Observations of solar oscillations of low and intermediate degree].
{\it Nature}, {\bf 302}, 24--27.

\bibitem[Duvall {\it et~al.\/}(1984)]{Duvall1984}
\biblab{Duvall1984}
Duvall, T. L., Dziembowski, W. A., Goode, P. R.,
Gough, D. O., Harvey, J. W. \& Leibacher, J. W., 1984.
[The internal rotation of the Sun].
{\it Nature}, {\bf 310}, 22--25.

\bibitem[Duvall {\it et~al.\/}(1988)]{Duvall1988}
\biblab{Duvall1988}
Duvall, T. L., Harvey, J. W., Libbrecht, K. G., Popp, B. D. \&
Pomerantz, M. A., 1988.
[Frequencies of solar $p$-mode oscillations].
{\it Astrophys. J.}, {\bf 324}, 1158--1171.

\bibitem[Dziembowski {\it et~al.\/}(1988)]{Dziemb1988}
\biblab{Dziemb1988}
Dziembowski, W., Patern{\'o}, L. \& Ventura, R., 1988.
[How comparison between observed and calculated $p$-mode eigenfrequencies
can give information on the internal structure of the Sun].
{\it Astron. Astrophys.}, {\bf 200}, 213--217.

\bibitem[Dziembowski {\it et~al.\/}(1990)]{Dziemb1990}
\biblab{Dziemb1990}
Dziembowski, W. A., Pamyatnykh, A. A. \& Sienkiewicz, R., 1990.
[Solar model from helioseismology and the neutrino flux problem].
{\it Mon. Not. R. astr. Soc.}, {\bf 244}, 542--550.

\bibitem[Dziembowski {\it et~al.\/}(1991)]{Dziemb1991}
\biblab{Dziemb1991}
Dziembowski, W. A., Pamyatnykh, A. A. \& Sienkiewicz, R., 1991.
[Helium content in the solar convective envelope from helioseismology].
{\it Mon. Not. R. astr. Soc.}, {\bf 249}, 602--605.

\bibitem[Dziembowski {\it et~al.\/}(1994)]{Dziemb1994}
\biblab{Dziemb1994}
Dziembowski, W. A., Goode, P. R., Pamyatnykh, A. A. \& Sienkiewicz, R., 1994.
[A seismic model of the Sun's interior].
{\it Astrophys. J.}, {\bf 432}, 417--426.

\bibitem[Dziembowski {\it et~al.\/}(1999)]{Dziemb1999}
\biblab{Dziemb1999}
Dziembowski, W. A., Fiorentini, G., Ricci, B. \& Sienkiewicz, R., 1999.
[Helioseismology and the solar age].
{\it Astron. Astrophys.}, {\bf 343}, 990--996.

\bibitem[Eddington(1920)]{Edding1920}
\biblab{Edding1920}
Eddington, A. S., 1920.
[The internal constitution of the stars].
{\it Nature}, {\bf 106}, 14--20.

\bibitem[Eddington(1926)]{Edding1926}
\biblab{Edding1926}
Eddington, A. S., 1926.
{\it The internal constitution of the stars},
Cambridge University Press, Cambridge.

\bibitem[Eff-Darwich and  Korzennik(2013)]{EffDar2013}
\biblab{EffDar2013}
Eff-Darwich, A. \& Korzennik, S. G., 2013.
[The dynamics of the solar radiative zone].
{\it Solar Phys.}, {\bf 287}, 43--56.

\bibitem[Eggenberger {\it et~al.\/}(2005)]{Eggenb2005}
\biblab{Eggenb2005}
Eggenberger, P., Maeder, A. \& Meynet, G., 2005.
[Stellar evolution with rotation and magnetic fields. IV. The solar rotation
profile].
{\it Astron. Astrophys.}, {\bf 440}, L9--L12.

\bibitem[Eggenberger {\it et~al.\/}(2019)]{Eggenb2019}
\biblab{Eggenb2019}
Eggenberger, P., Buldgen, G. \& Salmon, S. J. A. J., 2019.
[Rotation rate of the solar core as a key constraint to magnetic 
angular momentum transport in stellar interiors].
{\it Astron. Astrophys.}, {\bf 626}, L1-(1--5).

\bibitem[Eggleton(1971)]{Egglet1971}
\biblab{Egglet1971}
Eggleton, P. P., 1971.
[The evolution of low mass stars].
{\it Mon. Not. R. astr. Soc.}, {\bf 151}, 351--364.

\bibitem[Eggleton {\it et~al.\/}(1973)]{Egglet1973}
\biblab{Egglet1973}
Eggleton, P. P., Faulkner, J. \& Flannery, B. P., 1973.
[An approximate equation of state for stellar material].
{\it Astron. Astrophys.}, {\bf 23}, 325--330.

\bibitem[Eggleton {\it et~al.\/}(2006)]{Egglet2006}
\biblab{Egglet2006}
Eggleton, P. P., Dearborn, D. S. P. \& Lattanzio, J. C., 2006.
[Deep mixing of ${}^3{\rm He}$: reconciling Big Bang and stellar 
nucleosynthesis].
{\it Science}, {\bf 314}, 1580--1583.

\bibitem[Eguchi {\it et~al.\/}(2003)]{Eguchi2003}
\biblab{Eguchi2003}
Eguchi, K., Enomoto, S., Furuno, K., {\etal}, 2003.
%Goldman, J., Hanada, H., Ikeda, H.,
%Ikeda, K., Inoue, K., Ishihara, K., Itoh, W., Iwamoto, T., Kawaguchi, T.,
%Kawashima, T., Kinoshita, H., Kishimoto, Y., Koga, M., Koseki, Y., Maeda, T.,
%Mitsui, T., Motoki, M., Nakajima, K., Nakajima, M., Nakajima, T., Ogawa, H.,
%Owada, K., Sakabe, T., Shimizu, I., Shirai, J., Suekane, F., Suzuki, A.,
%Tada, K., Tajima, O., Takayama, T., Tamae, K., Watanabe, H., Busenitz, J.,
%Djurcic, Z., McKinny, K., Mei, D.-M., Piepke, A., Yakushev, E., Berger, B.~E.,
%Chan, Y.~D., Decowski, M.~P., Dwyer, D.~A., Freedman, S.~J., Fu, Y.,
%Fujikawa, B.~K., Heeger, K.~M., Lesko, K.~T., Luk, K.-B., Murayama, H.,
%Nygren, D.~R., Okada, C.~E., Poon, A.~W.~P., Steiner, H.~M., Winslow, L.~A.,
%Horton-Smith, G.~A., McKeown, R.~D., Ritter, J., Tipton, B.,
%Vogel, P., Lane, C.~E., Miletic, T., Gorham, P.~W.,
%Guillian, G., Learned, J.~G., Maricic, J., Matsuno, S.,
%Pakvasa, S., Dazeley, S., Hatakeyama, S., Murakami, M., Svoboda, R.~C.,
%Dieterle, B.~D., DiMauro, M., Detwiler, J., Gratta, G., Ishii, K.,
%Tolich, N., Uchida, Y., Batygov, M., Bugg, W., Cohn, H., Efremenko, Y.,
%Kamyshkov, Y., Kozlov, A., Nakamura, Y., De~Braeckeleer, L., Gould, C.~R.,
%Karwowski, H.~J., Markoff, D.~M., Messimore, J.~A., Nakamura, K.,
%Rohm, R.~M., Tornow, W., Young, A.~R. \& Wang, Y.-F., 2003.
[First results from KamLAND: Evidence for reactor antineutrino disappearance].
{\it Phys. Rev. Lett.}, {\bf 90}, 021802-(1--6).

\bibitem[Elliott(1998)]{Elliot1998b}
\biblab{Elliot1998b}
Elliott, J. R., 1998.
[Helioseismic constraints on new solar models from the MoSEC code].
{\it Astron. Astrophys.}, {\bf 334}, 703--712.

\bibitem[Elliott and Gough(1999)]{Elliot1999}
\biblab{Elliot1999}
Elliott, J. R. \& Gough, D. O., 1999.
[Calibration of the thickness of the solar tachocline].
{\it Astrophys. J.}, {\bf 516}, 475--481.

\bibitem[Elliott and Kosovichev(1998)]{Elliot1998a}
\biblab{Elliot1998a}
Elliott, J. R. \& Kosovichev, A. G., 1998.
[The adiabatic exponent in the solar core].
{\it Astrophys. J.}, {\bf 500}, L199--L202.

\bibitem[Elsworth {\it et~al.\/}(1990)]{Elswor1990}
\biblab{Elswor1990}
Elsworth, Y., Howe, R., Isaak, G. R., McLeod, C. P. \& New, R., 1990.
[Evidence from solar seismology against non-standard solar-core models].
{\it Nature}, {\bf 347}, 536--539.

\bibitem[Emden(1907)]{Emden1907}
\biblab{Emden1907}
Emden, R., 1907.
{\it Gaskugeln},
B. G. Teubner, Leibzig.

\bibitem[Epstein and  Pinsonneault(2014)]{Epstei2014}
\biblab{Epstei2014}
Epstein, C. R. \& Pinsonneault, M. H., 2014.
[How good a clock is rotation? The stellar rotation-mass-age relationship
for old field stars].
{\it Astrophys. J.}, {\bf 780}, 159-(1--24).

\bibitem[Espinosa Lara {\it et~al.\/}(2013)]{Espino2013}
\biblab{Espino2013}
Espinosa Lara, F. \& Rieutord, M., 2013.
[Self-consistent 2D models of fast-rotating early-type stars].
{\it Astron. Astrophys.}, {\bf 552}, A35-(1--16).

\bibitem[Esteban {\it et~al.\/}(2017)]{Esteba2017}
\biblab{Esteba2017}
Esteban, I., Gonzalez-Garcia, M., C., Maltoni, M., Martinez-Soler, I. \&
Schwetz, T., 2017.
[Updated fit to three neutrino mixing: exploring the accelerator-reactor 
complementarity].
{\it J. High Energ. Phys.}, 2017:87-(1--31). 

\bibitem[Ezer and Cameron(1968)]{Ezer1968}
\biblab{Ezer1968}
Ezer, D. \& Cameron, A. G. W., 1968.
[Solar spin-down and neutrino fluxes].
{\it Astrophys. Lett.}, {\bf 1}, 177--179.

\bibitem[Faulkner(2004)]{Faulkn2004}
\biblab{Faulkn2004}
Faulkner, J., 2004.
[Red giants: then and now].
In {\it The Scientific Legacy of Fred Hoyle}, 
ed. D. Gough, Cambridge University Press, p. 149--226.

\bibitem[Faulkner and Gilliland(1985)]{Faulkn1985}
\biblab{Faulkn1985}
Faulkner, J. \& Gilliland, R. L., 1985.
[Weakly interacting, massive particles and the solar neutrino flux].
{\it Astrophys. J.}, {\bf 299}, 994--1000.

\bibitem[Faulkner {\it et~al.\/}(1986)]{Faulkn1986}
\biblab{Faulkn1986}
Faulkner, J., Gough, D. O. \& Vahia, M. N., 1986.
[Weakly interacting massive particles and solar oscillations].
{\it Nature}, {\bf 321}, 226--229.

\bibitem[Featherstone and  Hindman(2016)]{Feathe2016}
\biblab{Feathe2016}
Featherstone, N. A. \& Hindman, B. W., 2016.
[The emergence of solar supergranulation as a natural consequence of
rotationally constrained interior convection].
{\it Astrophys. J.}, {\bf 830}, L15-(1--6).

\bibitem[Ferguson {\it et~al.\/}(2005)]{Fergus2005}
\biblab{Fergus2005}
Ferguson, J. W., Alexander, D. R., Allard, F., Barman, T., Bodnarik, J. G.,
Hauschildt, P. H., Heffner-Wong, A. \& Tamanai, A., 2005.
[Low-temperature opacities].
{\it Astrophys. J.}, {\bf 623}, 585--596.

\bibitem[Ferraro(1937)]{Ferrar1937}
\biblab{Ferrar1937}
Ferraro, V. C. A., 1937.
[The non-uniform rotation of the Sun and its magnetic field].
{\it Mon. Not. R. astr. Soc.}, {\bf 97}, 458--472.

\bibitem[Feulner(2012)]{Feulne2012}
\biblab{Feulne2012}
Feulner, G., 2012.
[The faint early Sun problem].
{\it Rev. Geophys.}, {\bf 50}, RG2006-(1--29).

\bibitem[Fichtinger {\it et~al.\/}(2017)]{Fichti2017}
\biblab{Fichti2017}
Fichtinger, B., G{\"u}del, M., Mutel, R. L., Hallinan, G., Gaidos, E.,
Skinner, S. L., Lynch, C. \& Gayley, K. G., 2017.
[Radio emission and mass loss rate limits of four young solar-type stars].
{\it Astron. Astrophys.}, {\bf 599}, A127-(1--11).

%\bibitem[Fogli {\it et~al.\/}(2006)]{Fogli2006}
%\biblab{Fogli2006}
%Fogli, G. L., Lisi, E., Marrone, A. \& Palazzo, A., 2006.
%[Global analysis of three-flavor neutrino masses and mixings].
%{\it Prog. Particle Nuclear Phys.}, {\bf 57}, 742--795.

\bibitem[Formicola {\it et~al.\/}(2004)]{Formic2004}
\biblab{Formic2004}
Formicola, A., Imbriani, G., Costantini, H., Angulo, C., Bemmerer, D.,
Bonetti, R., Broggini, C., Corvisiero, P., Cruz, J., Descouvemont, P.,
F\"ul\"op, Z., Gervino, G., Guglielmetti, A., Gustavino, C., Gy\"urky, G.,
Jesus, A.~P., Junker, M., Lemut, A., Menegazzo, R., Prati, P.,
Roca, V., Rolfs, C., Romano, M., Rossi~Alvarez, C., Sch\"umann, F.,
Somorjai, E., Straniero, O., Strieder, F., Terrasi, F., Trautvetter, H.~P.,
Vomiero, A. \& Zavatarelli, S., 2004.
[Astrophysical $S$-factor of ${}^{14}{\rm N}(p, \gamma){}^{15}{\rm O}$].
{\it Phys. Lett. B}, {\bf 591}, 61--68.

\bibitem[Fossat(1991)]{Fossat1991}
\biblab{Fossat1991}
Fossat, E., 1991.
[The IRIS network for full disk helioseismology:
Present status of the programme].
{\it Solar Phys.}, {\bf 133}, 1--12.

\bibitem[Fossat and  Schmider(2018)]{Fossat2018}
\biblab{Fossat2018}
Fossat, E. \& Schmider, F. X., 2018.
[More about solar g modes].
{\it Astron. Astrophys.}, {\bf 612}, L1-(1--8).

\bibitem[Fossat {\it et~al.\/}(2003)]{Fossat2003}
\biblab{Fossat2003}
Fossat, E., Salabert, D., Cacciani, A., Ehgamberdiev, S., Gelly, B.,
Grec, G., Hoeksema, J. T., Kholikov, S., Lazrek, M., 
Palle, P., Schmider, F. X. \& Tomczyk, S., 2003.
[Eleven years of IRIS frequencies and splittings].
In {\it Proc. SOHO 12/ GONG+ 2002. Local and Global Helioseismology:
The Present and Future},
ed. A. Wilson, ESA SP-517, ESA Publications Division,
Noordwijk, The Netherlands, p. 139--144.

\bibitem[Fossat {\it et~al.\/}(2017)]{Fossat2017}
\biblab{Fossat2017}
Fossat, E., Boumier, P., Corbard, T., Provost, J., Salabert, D.,
Schmider, F. X., Gabriel, A. H., Grec, G., Renaud, C., Robillot, J. M.,
Roca-Cort{\'e}s, T., Turck-Chi{\`e}ze, S., Ulrich, R. K. \& Lazrek, M., 2017.
[Asymptotic $g$ modes: Evidence for rapid rotation of the solar core].
{\it Astron. Astrophys.}, {\bf 604}, A40-(1--17).

\bibitem[Fowler(1958)]{Fowler1958}
\biblab{Fowler1958}
Fowler, W. A., 1958.
[Completion of the proton-proton reaction chain and the possibility of
energetic neutrino emission by hot stars].
{\it Astrophys. J.}, {\bf 127}, 551--556.

\bibitem[Fowler {\it et~al.\/}(1967)]{Fowler1967}
\biblab{Fowler1967}
Fowler, W. A., Caughlan, G. R. \& Zimmerman, B. A., 1967.
[Thermonuclear reaction rates].
{\it Annu. Rev. Astron. Astrophys.}, {\bf 5}, 525--570.

\bibitem[Fowler {\it et~al.\/}(1975)]{Fowler1975}
\biblab{Fowler1975}
Fowler, W. A., Caughlan, G. R. \& Zimmerman, B. A., 1975.
[Thermonuclear reaction rates, II].
{\it Annu. Rev. Astron. Astrophys.}, {\bf 13}, 69--112.

\bibitem[Fr\"ohlich and Lean(2004)]{Frohli2004}
\biblab{Frohli2004}
Fr\"ohlich, C. \& Lean, J., 2004.
[Solar radiative output and its variability: evidence and mechanisms].
{\it Astron. Astrophys. Rev.}, {\bf 12}, 273--320.

\bibitem[Freytag {\it et~al.\/}(2002)]{Freyta2002}
\biblab{Freyta2002}
Freytag, B., Steffen, M. \& Dorch, B., 2002.
[Spots on the surface of Betelgeuse --- results from new 3D stellar
convection models].
{\it Astron. Nachr.}, {\bf 323}, 213--219.

\bibitem[Freytag {\it et~al.\/}(2012)]{Freyta2012}
\biblab{Freyta2012}
Freytag, B., Steffen, M., Ludwig, H.-G. \& Wedemeyer-B{\"o}hm, S., 2012.
[Simulations of stellar convection with CO5BOLD].
{\it J. Comp. Phys.}, {\bf 231}, 919--959.

\bibitem[Fukuda {\it et~al.\/}(1998)]{Fukuda1998}
\biblab{Fukuda1998}
Fukuda, Y., Hayakawa, T., Ichihara, E., {\etal}, 1998.
%Inoue, K., Ishihara, K., Ishino, H.,
%Itow, Y., Kajita, T., Kameda, J.,
%Kasuga, S., Kobayashi, K., Kobayashi, Y., Koshio, Y., Martens, K., Miura, M.,
%Nakahata, M., Nakayama, S.,
%Okada, A., Oketa, M., Okumura, K., Ota, M., Sakurai, N., Shiozawa, M.,
%Suzuki, Y., Takeuchi, Y., Totsuka, Y.,
%Yamada, S., Earl, M., Habig, A., Hong, J. T., Kearns, E., Kim, S. B.,
%Masuzawa, M., Messier, M. D.,
%Scholberg, K., Stone, J. L., Sulak, L. R., Walter, C. W., Goldhaber, M.,
%Barszczak, T., Gajewski, W.,
%Halverson, P. G., Hsu, J., Kropp, W. R., Price, L. R., Reines, F.,
%Sobel, H. W., Vagins, M. R., Ganezer, K. S.,
%Keig, W. E., Ellsworth, R. W., Tasaka, S., Flanagan, J. W., 
%Kibayashi, A., Learned, J. G., Matsuno, S., Stenger, V.,
%Takemori, D., Ishii, T., Kanzaki, J., Kobayashi, T., Nakamura, K.,
%Nishikawa, K., Oyama, Y., Sakai, A.,
%Sakuda, M., Sasaki, O., Echigo, S., Kohama, M., Suzuki, A. T.,
%Haines, T. J., Blaufuss, E., Sanford, R.,
%Svoboda, R., Chen, M. L., Conner, Z., Goodman, J. A., Sullivan, G. W.,
%Mori, M., Goebel, F., Hill, J., Jung, C. K.,
%Mauger, C., McGrew, C., Sharkey, E., Viren, B., Yanagisawa, C., Doki, W.,
%Ishizuka, T., Kitaguchi, Y.,
%Koga, H., Miyano, K., Okazawa, H., Saji, C., Takahata, M.,
%Kusano, A., Nagashima, Y., Takita, M.,
%Yamaguchi, T., Yoshida, M., Etoh, M., Fujita, K., Hasegawa, A., Hasegawa, T.,
%Hatakeyama, S.,
%Iwamoto, T., Kinebuchi, T., Koga, M., Maruyama, T., Ogawa, H.,
%Saito, M., Suzuki, A., Tsushima, F., Koshiba, M.,
%Nemoto, M., Nishijima, K., Futagami, T., Hayato, Y., Kanaya, Y., 
%Kaneyuki, K., Watanabe, Y.,
%Kielczewska, D., Doyle, R., George, J.,
%Stachyra, A., Wai, L., Wilkes, J. \& Young, K., 1998.
[Measurement of a small atmospheric $\nu_\mu/\nu_e$ ratio].
{\it Phys. Lett. B}, {\bf 433}, 9--18.

\bibitem[Fukuda {\it et~al.\/}(2001)]{Fukuda2001}
\biblab{Fukuda2001}
Fukuda, S., Fukuda, Y., Ishitsuka, M., {\etal}, 2001.
%Itow, Y., Kajita, T.,
%Kameda, J., Kaneyuki, K., Kobayashi, K., Koshio, Y.,
%Miura, M., Moriyama, S., Nakahata, M., Nakayama, S., Okada, A.,
%Sakurai, N., Shiozawa, M., Suzuki, Y.,
%Takeuchi, H., Takeuchi, Y., Toshito, T., Totsuka, Y., Yamada, S.,
%Desai, S., Earl, M., Kearns, E., Messier, M. D.,
%Scholberg, K., Stone, J. L., Sulak, L. R., Walter, C. W., Goldhaber, M.,
%Barszczak, T., Casper, D., Gajewski, W.,
%Kropp, W. R., Mine, S., Liu, D. W., Price, L. R., Smy, M. B.,
%Sobel, H. W., Vagins, M. R., Ganezer, K. S.,
%Keig, W. E., Ellsworth, R. W., Tasaka, S., Kibayashi, A., Learned, J. G.,
%Matsuno, S., Takemori, D., Hayato, Y.,
%Ishii, T., Kobayashi, T., Nakamura, K., Obayashi, Y., Oyama, Y.,
%Sakai, A., Sakuda, M., Kohama, M.,
%Suzuki, A. T., Inagaki, T., Nakaya, T., Nishikawa, K., Haines, T. J.,
%Blaufuss, E., Dazeley, S., Lee, K. B.,
%Svoboda, R., Goodman, J. A., Guillian, G., Sullivan, G. W., Turcan, D.,
%Habig, A., Hill, J., Jung, C. K.,
%Martens, K., Malek, M., Mauger, C., McGrew, C., Sharkey, E., Viren, B.,
%Yanagisawa, C., Mitsuda, C.,
%Miyano, K., Saji, C., Shibata, T., Kajiyama, Y., Nagashima, Y.,
%Nitta, K., Takita, M., Yoshida, M.,
%Kim, H. I., Kim, S. B., Yoo, J., Okazawa, H., Ishizuka, T.,
%Etoh, M., Gando, Y., Hasegawa, T., Inoue, K.,
%Ishihara, K., Maruyama, T., Shirai, J., Suzuki, A., Koshiba, M.,
%Hatakeyama, Y., Ichikawa, Y., Koike, M.,
%Nishijima, K., Fujiyasu, H., Ishino, H., Morii, M., Watanabe, Y.,
%Golebiewska, U., Kielczewska, D.,
%Boyd, S. C., Stachyra, A. L., Wilkes, R. J. \& Young, K. K., 2001.
[Solar ${}^8{\rm B}$ and hep neutrino measurements from 1258 days
of Super-Kamiokande data].
{\it Phys. Rev. Lett.}, {\bf 86}, 5651--5655.

\bibitem[Fuller {\it et~al.\/}(2019)]{Fuller2019}
\biblab{Fuller2019}
Fuller, J., Piro, A. L. \& Jermyn, A. S., 2019.
[Slowing the spins of stellar cores].
{\it Mon. Not. R. Astron. Soc.}, {\bf 485}, 3661--3680.

\bibitem[Gabriel {\it et~al.\/}(1997)]{GabrieA1997}
\biblab{GabrieA1997}
Gabriel, A. H., Charra, J., Grec, G., Robillot, J.-M., Roca Cort{\'e}s, T.,
Turck-Chi{\`e}ze, S., Ulrich, R., Basu, S., Baudin, F., Bertello, L.,
Boumier, P., Charra, M., Christensen-Dalsgaard, J., Decaudin, M.,
Dzitko, H., Foglizzo, T., Fossat, E., Garc\'{\i}a, R. A., 
Herreros, J. M., Lazrek, M., Pall{\'e}, P. L., P{\'e}trou, N.,
Renaud, C. \& R{\'e}gulo, C., 1997.
[Performance and early results from the GOLF instrument flown
on the SOHO mission].
{\it Solar Phys.}, {\bf 175}, 207--226.

\bibitem[Gabriel(1991)]{Gabrie1991}
\biblab{Gabrie1991}
Gabriel, M., 1991.
[Accuracy tests for the computation of solar models].
In {\it Challenges to theories of the structure of moderate-mass stars},
{\it Lecture Notes in Physics}, vol. {\bf 388}, p. 51--55,
eds Gough, D. O. \& Toomre, J., Springer, Heidelberg.

\bibitem[Gabriel(1994)]{Gabrie1994}
\biblab{Gabrie1994}
Gabriel, M., 1994.
[On the p-mode spectrum of solar models].
{\it Astron. Astrophys.}, {\bf 281}, 551--560.

\bibitem[Gabriel(1997)]{GabrieM1997}
\biblab{GabrieM1997}
Gabriel, M., 1997.
[Influence of heavy element and rotationally induced diffusions
on solar models].
{\it Astron. Astrophys.}, {\bf 327}, 771--778.

\bibitem[Gabriel {\it et~al.\/}(1984)]{Gabrie1984}
\biblab{Gabrie1984}
Gabriel, M., Noels, A. \& Scuflaire, R., 1984.
[Influence of opacities, partition function and hydrogen diffusion
on the 5 min. solar oscillations].
{\it Mem. Soc. Astron. Ital.}, {\bf 55}, 169--174.

\bibitem[Gaidos {\it et~al.\/}(2009)]{Gaidos2009}
\biblab{Gaidos2009}
Gaidos, E., Krot, A. N. \& Hus, G. R., 2009.
[On the oxygen isotopic composition of the solar system].
{\it Astrophys. J.}, {\bf 705}, L163--L167.

\bibitem[Gallet and  Bouvier(2013)]{Gallet2013}
\biblab{Gallet2013}
Gallet, F. \& Bouvier, J., 2013.
[Improved angular momentum evolution for solar-like stars].
{\it Astron. Astrophys.}, {\bf 556}, A36-(1--15).

\bibitem[Gando {\it et~al.\/}(2011)]{Gando2011}
\biblab{Gando2011}
Gando, A., Gando, Y., Ichimura, K., Ikeda, H., Inoue, K., Kibe, Y., 
Kishimoto, Y., Koga, M., Minekawa, Y., Mitsui, T., Morikawa, T., 
Nagai, N., Nakajima, K., Nakamura, K., Narita, K., {\etal}
(KamLAND Collaboration), 2011.
[Constraints on $\theta_{13}$ from a three-flavor oscillation analysis 
of reactor antineutrinos at KamLAND].
{\it Phys. Rev. D}, {\bf 83}, 052002-(1--11).

\bibitem[Garaud(2007)]{Garaud2007c}
\biblab{Garaud2007c}
Garaud, P., 2007.
[Magnetic confinement of the solar tachocline].
In {\it The solar tachocline},
eds D. W. Hughes, R. Rosner \& N. O. Weiss,
Cambridge University Press, p. 147--181.

\bibitem[Garaud(2020)]{Garaud2020}
\biblab{Garaud2020}
Garaud, P., 2020.
[The tachocline revisited].
In {\it Dynamics of the Sun and Stars--Honoring the Life and Work of
Michael J. Thompson},
eds Monteiro, M. J. P. F. G., Garc\'{\i}a, R., Christensen-Dalsgaard, J. \&
McIntosh, S. W., Astrophysics and Space Science Proceedings, vol 57.
Springer, Cham, pp. 207 -- 220.

\bibitem[Garaud and Brummell(2008)]{Garaud2008a}
\biblab{Garaud2008a}
Garaud, P. \& Brummell, N. H., 2008.
[On the penetration of meridional circulation below the solar convection zone].
{\it Astrophys. J.}, {\bf 674}, 498--510.
%{\tt [arXiv:0708.0258v1 [astro-ph]]}

\bibitem[Garaud and Garaud(2008)]{Garaud2008b}
\biblab{Garaud2008b}
Garaud, P. \& Garaud, J.-D., 2008.
[Dynamics of the solar tachocline--II: the stratified case].
{\it Mon. Not. R. Astron. Soc.}, {\bf 391}, 1239--1258.

\bibitem[Garaud and Guervilly(2009)]{Garaud2009}
\biblab{Garaud2009}
Garaud, P. \& Guervilly, C., 2009.
[The rotation rate of the solar radiative zone].
{\it Astrophys. J.}, {\bf 695}, 799--808.

\bibitem[Garaud and Rogers(2007)]{Garaud2007a}
\biblab{Garaud2007a}
Garaud, P. \& Rogers, T., 2007.
[Solar rotation].
In Stancliffe R.J., Dewi J., Houdek G., Martin R.G., Tout C.A., eds,
AIP Conf. Proc. vol. 948, {\it Unsolved Problems in Stellar Physics}.
American Institute of Physics, Melville, p. 237--248.

\bibitem[Garc\'{\i}a and  Ballot(2019)]{Garcia2019}
\biblab{Garcia2019}
Garc\'{\i}a, R. A. \& Ballot, J., 2019.
[Asteroseismology of solar-type stars].
{\it Living Rev. Solar Phys.}, {\bf 16}, 4-(1--99).

\bibitem[Garc\'{\i}a {\it et~al.\/}(2004)]{Garcia2004}
\biblab{Garcia2004}
Garc\'{\i}a, R. A., Corbard, T., Chaplin, W. J., Couvidat, S.,
Eff-Darwich, A., Jim{\'e}nez-Reyes, S. J., Korzennik, S. G., Ballot, J.,
Boumier, P., Fossat, E., Henney, C. J., Howe, R., Lazrek, M.,
Lochard, J., Pall{\'e}, P. L. \& Turck-Chi{\`e}ze, S., 2004.
[About the rotation of the solar radiative interior].
{\it Solar Phys.}, {\bf 220}, 269--285.

\bibitem[Garc\'{\i}a {\it et~al.\/}(2007)]{Garcia2007}
\biblab{Garcia2007}
Garc\'{\i}a, R. A., Turck-Chi{\`e}ze, S., 
Jim{\'e}nez-Reyes, S. J., Ballot, J., Pall{\'e}, P., Eff-Darwich, A.,
Mathur, S. \& Provost, J., 2007.
[Tracking solar gravity modes: the dynamics of the solar core].
{\it Science}, {\bf 316}, 1591--1593.

%\bibitem[Gavrin(2001)]{Gavrin2001}
%\biblab{Gavrin2001}
%Gavrin, V. N., 2001.
%[Solar neutrino results from SAGE].
%{\it Nucl. Phys. B (Proc. Suppl)}, {\bf 91}, 36--43.

\bibitem[Geiss and Gloeckler(1998)]{Geiss1998}
\biblab{Geiss1998}
Geiss, J. \& Gloeckler, G., 1998.
[Abundances of deuterium and helium-3 in the protosolar cloud].
{\it Space Sci. Rev.}, {\bf 84}, 239--250.

\bibitem[Geiss and  Gloeckler(2007)]{Geiss2007}
\biblab{Geiss2007}
Geiss, J. \& Gloeckler, G., 2007.
[Linking primordial to solar and Galactic composition].
{\it Space Sci. Rev.}, {\bf 130}, 5--26.

\bibitem[Gelly {\it et~al.\/}(1988)]{Gelly1988}
\biblab{Gelly1988}
Gelly, B., Fossat, E., Grec, G. \& Schmider, F.-X., 1988.
[Solar calibration of asteroseismology].
{\it Astron. Astrophys.}, {\bf 200}, 207--212.

\bibitem[Genova {\it et~al.\/}(2018)]{Genova2018}
\biblab{Genova2018}
Genova, A., Mazarico, E., Goossens, S., Lemoine, F. G., Neumann, G. A.,
Smith, D. E. \& Zuber, M. T., 2018.
[Solar system expansion and strong equivalence principle as seen by the
NASA MESSENGER mission].
{\it Nature Comm.}, {\bf 9}, 289-(1--9).

\bibitem[Gesicki {\it et~al.\/}(2018)]{Gesick2018}
\biblab{Gesick2018}
Gesicki, K., Zijlstra, A. A. \& Miller Bertolami, M. M., 2018.
[The mysterious age invariance of the planetary nebula luminosity function
bright cut-off].
{\it Nature Astron.}, {\bf 2}, 580--584.

\bibitem[Gilliland {\it et~al.\/}(1986)]{Gillil1986}
\biblab{Gillil1986}
Gilliland, R. L., Faulkner, J., Press, W. H. \& Spergel, D. N., 1986.
[Solar models with energy transport by weakly interacting particles].
{\it Astrophys. J.}, {\bf 306}, 703--709.

\bibitem[Gilliland {\it et~al.\/}(2010)]{Gillil2010}
\biblab{Gillil2010}
Gilliland, R. L., Brown, T. M., Christensen-Dalsgaard, J., Kjeldsen, H.,
Aerts, C., Appourchaux, T., Basu, S., Bedding, T. R., Chaplin, W. J.,
Cunha, M. S., De Cat, P., De Ridder, J., Guzik, J. A.,
Handler, G., Kawaler, S., Kiss, L., Kolenberg, K., Kurtz, D. W.,
Metcalfe, T. S., Monteiro, M. J. P. F. G., Szab{\'o}, R.,
Arentoft, T., Balona, L., Debosscher, J., Elsworth, Y. P., Quirion, P.-O.,
Stello, D., Su{\'a}rez, J. C., Borucki, W. J., Jenkins, J. M.,
Koch, D., Kondo, Y., Latham, D. W., Rowe, J. F. \& Steffen, J. H., 2010.
[{\it Kepler} asteroseismology program: Introduction and first results].
{\it Publ. Astron. Soc. Pacific}, {\bf 122}, 131--143.

\bibitem[Gilman(1976)]{Gilman1976}
\biblab{Gilman1976}
Gilman, P., 1976.
[Theory of rotation in a deep rotating spherical shell, and its application
to the Sun].
In {\it Proc. IAU Symposium no. 74, Basic Mechanisms of Solar Activity}, 
eds V. Bumba \& J. Kleczek, Reidel, Dordrecht, p. 207--228.

\bibitem[Gingerich {\it et~al.\/}(1971)]{Ginger1971}
\biblab{Ginger1971}
Gingerich, O., Noyes, R. W., Kalkofen, W. \& Cuny, Y., 1971.
[The Harvard-Smithsonian Reference Atmosphere].
{\it Sol. Phys.}, {\bf 18}, 347--365.

\bibitem[Gizon and  Birch(2005)]{Gizon2005}
\biblab{Gizon2005}
Gizon, L. \& Birch, A. C., 2005.
[Local helioseismology].
{\it Living Rev. Solar Phys.}, {\bf 2},  6. URL (cited on 14/10/06):
\url{https://doi.org/10.12942/lrsp-2005-6}

\bibitem[Gizon and Solanki(2003)]{Gizon2003}
\biblab{Gizon2003}
Gizon, L. \& Solanki, S., 2003.
[Determining the inclination of the rotation axis of a Sun-like star].
{\it Astrophys. J.}, {\bf 589}, 1009--1019.

%\bibitem[Gondolo and Raffelt(2009)]{Gondol2009}
%\biblab{Gondol2009}
%Gondolo, P. \& Raffelt, G. G., 2009.
%[Solar neutrino limit on axions and keV-mass bosons].
%{\it Phys. Rev. D.}, {\bf 79}, 107301-(1--4).

\bibitem[Gong {\it et~al.\/}(2001a)]{Gong2001a}
\biblab{Gong2001a}
Gong, Z., D\"appen, W. \& Zejda, L., 2001a.
[MHD equation of state with relativistic electrons].
{\it Astrophys. J.}, {\bf 546}, 1178--1182.

\bibitem[Gong {\it et~al.\/}(2001b)]{Gong2001b}
\biblab{Gong2001b}
Gong, Z., D\"appen, W. \& Nayfonov, A., 2001b.
[Effects of heavy elements and excited states in the equation of state
of the solar interior].
{\it Astrophys. J.}, {\bf 563}, 419--433.

\bibitem[Gonzalez-Garcia and Nir(2003)]{Gonzal2003}
\biblab{Gonzal2003}
Gonzalez-Garcia, M. C. \& Nir, Y., 2003.
[Neutrino masses and mixing: evidence and implications].
{\it Rev. Mod. Phys.}, {\bf 75}, 345--402.

\bibitem[Goodson {\it et~al.\/}(2016)]{Goodso2016}
\biblab{Goodso2016}
Goodson, M. D., Luebbers, I., Heitsch, F. \& Frazer, C. C., 2016.
[Chemical enrichment of the pre-solar cloud by supernova dust grains].
{\it Mon. Not. R. Astron. Soc.}, {\bf 462}, 2777--2791.

\bibitem[Goswami and Vanhala(2000)]{Goswam2000}
\biblab{Goswam2000}
Goswami, J. N. \& Vanhala, H. A. T., 2000.
[Extinct radionuclides and the origin of the solar system].
In {\it Protostars and Planets IV}, 
eds V. Mannings, A. P. Boss \& S. S. Russell, University of Arizona Press,

\bibitem[Gough(1977a)]{Gough1977a}
\biblab{Gough1977a}
Gough, D. O., 1977a.
[Random remarks on solar hydrodynamics].
{\it Proc. IAU Colloq. No. 36:
The energy balance and hydrodynamics of the solar chromosphere and corona},
p. 3--36,
eds Bonnet, R. M. \& Delache, P., G. de Bussac, Clairmont-Ferrand.

\bibitem[Gough(1977b)]{Gough1977b}
\biblab{Gough1977b}
Gough, D. O., 1977b. % Note: year corrected 1/5/02
[The current state of stellar mixing-length theory].
In {\it Problems of stellar convection, IAU Colloq. No. 38},
{\it Lecture Notes in Physics, vol. 71}.
eds Spiegel, E. A. \& Zahn, J.-P., Springer-Verlag, Berlin, p. 15--56.

\bibitem[Gough(1977c)]{Gough1977c}
\biblab{Gough1977c}
Gough, D. O., 1977c.
[Mixing-length theory for pulsating stars].
{\it Astrophys. J.}, {\bf 214}, 196--213.

\bibitem[Gough(1978a)]{Gough1978a}
\biblab{Gough1978a}
Gough, D. O., 1978a.
[On the power of the five minute oscillations to resolve the
solar angular velocity].
{\it Proc. Workshop on solar rotation},
eds Belvedere, G. \& Paterno, L., University of Catania Press, p. 255--268.

\bibitem[Gough(1978b)]{Gough1978b}
\biblab{Gough1978b}
Gough, D. O., 1978b. 
[The significance of solar oscillations].
In {\it Pleins feux sur la physique solaire. Proc. 2me Assembl{\'e}e 
Europ{\'e}enne de Physique Solaire}, CNRS, Paris, p. 81--103.

\bibitem[Gough(1984a)]{Gough1984a}
\biblab{Gough1984a}
Gough, D. O., 1984a.
[On the rotation of the Sun].
{\it Phil. Trans. R. Soc. London, Ser. A}, {\bf 313}, 27--38.

\bibitem[Gough(1984b)]{Gough1984b}
\biblab{Gough1984b}
Gough, D. O., 1984b.
[Towards a solar model].
{\it Mem. Soc. Astron. Ital.}, {\bf 55}, 13--35.

\bibitem[Gough(1985)]{Gough1985}
\biblab{Gough1985}
Gough, D. O., 1985.
[Inverting helioseismic data].
{\it Solar Phys.}, {\bf 100}, 65--99.

\bibitem[Gough(1987)]{Gough1987}
\biblab{Gough1987}
Gough, D. O., 1987.
[Seismological measurement of stellar ages].
{\it Nature}, {\bf 326}, 257--259.

\bibitem[Gough(1990a)]{Gough1990a}
\biblab{Gough1990a}
Gough, D. O., 1990a.
[The internal structure of late-type main-sequence stars].
In {\it Astrophysics. Recent progress and future possibilities},
eds B. Gustafsson \& P. E. Nissen,
{\it Mat.-fys. Meddel. Kgl. Danske Vidensk. Selsk.}, {\bf vol. 42, No. 4}, 13--50.

\bibitem[Gough(1990b)]{Gough1990b}
\biblab{Gough1990b}
Gough, D. O., 1990b.
[Comments on helioseismic inference].
{\it Progress of seismology of the sun and stars},
{\it Lecture Notes in Physics}, vol. {\bf 367}, 283--318,
eds Osaki, Y. \& Shibahashi, H., Springer, Berlin.

\bibitem[Gough(1993)]{Gough1993}
\biblab{Gough1993}
Gough, D. O., 1993.
[Course 7. Linear adiabatic stellar pulsation].
In {\it Astrophysical fluid dynamics, Les Houches Session XLVII},
eds Zahn, J.-P. \& Zinn-Justin, J., Elsevier, Amsterdam, p. 399--560.

\bibitem[Gough(1994)]{Gough1994}
\biblab{Gough1994}
Gough, D., 1994.
[What can we learn from oscillation studies about irradiance and 
radius changes?].
In {\it Proc. IAU Colloq. 143: The Sun as a variable star:
Solar and stellar irradiance variations}, 
eds J. M. Pap, C. Fr{\"o}hlich, H. S. Hudson \& S. Solanki,
Cambridge University Press, p. 252--263.
%\notecd [To be checked!]

\bibitem[Gough(1996b)]{Gough1996b}
\biblab{Gough1996b}
Gough, D. O., 1996b.
[Astereoasteroseismology (correspondence to the Editors)].
{\it Observatory}, {\bf 116}, 313--315.

\bibitem[Gough(1996a)]{Gough1996a}
\biblab{Gough1996a}
Gough, D., 1996a.
[Testing solar models: the inverse problem].
In {\it Proc. VI IAC Winter School ``The structure of the Sun"},
eds T. Roca Cort{\'e}s \& F. S{\'a}nchez, Cambridge University Press,
p. 141--228.

\bibitem[Gough(2004)]{Gough2004}
\biblab{Gough2004}
Gough, D., 2004.
[The power of helioseismology to address issues of fundamental physics].
In {\it Equation-of-State and Phase-Transition Issues in
Models of Ordinary Astrophysical Matter},
eds V. {\v C}elebonovi{\'c}, W. D{\"a}ppen \& D. Gough,
AIP Conf. Proc. Vol. 731, AIP, Melville, New York, p. 119--138.

\bibitem[Gough(2010)]{Gough2010}
\biblab{Gough2010}
Gough, D. O., 2010.
[Angular-momentum coupling through the tachocline].
In {\it Magnetic coupling between the interior and the atmosphere
of the Sun}, eds S. S. Hasan \& R. J. Rutten, 
Astrophysics and Space Science Proceedings,
Springer, Berlin, Heidelberg, p. 68--85.
%{\tt [arXiv:0905.4924v1 [astro-ph.SR]]}

\bibitem[Gough(2013a)]{Gough2013a}
\biblab{Gough2013a}
Gough, D. O., 2013a.
[What have we learned from helioseismology, what have we really learned, 
and what do we aspire to learn?].
{\it Solar Phys.}, {\bf 287}, 9--41.

\bibitem[Gough(2013b)]{Gough2013b}
\biblab{Gough2013b}
Gough, D. O., 2013b.
[Commentary on a putative magnetic field variation in the solar convection
zone].
{\it Mon. Not. R. Astron. Soc.}, {\bf 435}, 3148--3158.

\bibitem[Gough(2015)]{Gough2015}
\biblab{Gough2015}
Gough, D. O., 2015.
[Some glimpses from helioseismology at the dynamics of the deep solar
interior].
{\it Space Sci. Rev.}, {\bf 196}, 15--47.

\bibitem[Gough(2019)]{Gough2019}
\biblab{Gough2019}
Gough, D. O., 2019.
[Anticipating the Sun's heavy-element abundance].
{\it Mon. Not. R. Astron. Soc.}, {\bf 485}, L114--L115.

\bibitem[Gough and Kosovichev(1990)]{GoughKos1990}
\biblab{GoughKos1990}
Gough, D. O. \& Kosovichev, A. G., 1990.
[Using helioseismic data to probe the hydrogen abundance in the solar core].
{\it Proc. IAU Colloquium No 121, Inside the Sun},
p. 327--340,
eds Berthomieu G. \& Cribier M.,

\bibitem[Gough and  Kosovichev(1993)]{GoughKos1993}
\biblab{GoughKos1993}
Gough, D. O. \& Kosovichev, A. G., 1993.
[Seismic analysis of stellar p-mode spectra].
In {\it Proc. GONG 1992: Seismic investigation of the Sun and stars},
ed. Brown, T. M.,
Astronomical Society of the Pacific Conference Series, San Francisco,
{\bf 42}, 351--354.

\bibitem[Gough and McIntyre(1998)]{Gough1998}
\biblab{Gough1998}
Gough, D. O. \& McIntyre, M. E., 1998.
[Inevitability of a magnetic field in the Sun's radiative interior].
{\it Nature}, {\bf 394}, 755--757.

\bibitem[Gough and Scherrer(2001)]{Gough2001}
\biblab{Gough2001}
Gough, D. O. \& Scherrer, P. H., 2001.
[The solar interior].
In {\it The Century of Space Science},
eds Bleeker, J. A. M., Geiss, J. \& Huber, M. C. E., Kluwer, Dordrecht,
p. 1035--1063.

\bibitem[Gough and  Thompson(1988)]{Gough1988}
\biblab{Gough1988}
Gough, D. O. \& Thompson, M. J., 1988.
[Magnetic perturbations to stellar oscillation eigenfrequencies].
{\it Proc. IAU Symposium No 123, Advances in helio- and asteroseismology},
eds Christensen-Dalsgaard, J. \& Frandsen, S., 
Reidel, Dordrecht, p. 155--160.

\bibitem[Gough and Thompson(1991)]{Gough1991}
\biblab{Gough1991}
Gough, D. O. \& Thompson, M. J., 1991.
[The inversion problem].
In {\it Solar interior and atmosphere},
eds Cox, A. N., Livingston, W. C. \& Matthews, M.,
Space Science Series, University of Arizona Press, p. 519--561.

\bibitem[Gough and  Vorontsov(1995)]{Gough1995}
\biblab{Gough1995}
Gough, D. O. \& Vorontsov, S. V., 1995.
[Seismology of the solar envelope: measuring the acoustic phase shift
generated in the outer layers].
{\it Mon. Not. R. Astron. Soc.}, {\bf 273}, 573--582.

\bibitem[Gough and Weiss(1976)]{Gough1976}
\biblab{Gough1976}
Gough, D. O. \& Weiss, N. O., 1976.
[The calibration of stellar convection theories].
{\it Mon. Not. R. astr. Soc.}, {\bf 176}, 589--607.

\bibitem[Gough {\it et~al.\/}(1975)]{Gough1975}
\biblab{Gough1975}
Gough, D. O., Spiegel, E. A. \& Toomre, J., 1975.
[Highly stretched meshes as functionals of solutions].
{\it Lecture Notes in Physics}, 
{\bf 35}, 191--196,
ed. Richtmyer, R. D., Springer, Heidelberg.

\bibitem[Gough {\it et~al.\/}(1996)]{Goughetal1996}
\biblab{Goughetal1996}
Gough, D. O., Kosovichev, A. G., Toomre, J., Anderson, E. R., Antia, H. M.,
Basu, S., Chaboyer, B., Chitre, S. M., Christensen-Dalsgaard, J.,
Dziembowski, W. A., Eff-Darwich, A., Elliott, J. R., Giles, P. M.,
Goode, P. R., Guzik, J. A., Harvey, J. W., Hill, F., Leibacher, J. W.,
Monteiro, M. J. P. F. G., Richard, O., Sekii, T., Shibahashi, H.,
Takata, M., Thompson, M. J., Vauclair, S., Vorontsov, S. V., 1996.
[The seismic structure of the Sun].
{\it Science}, {\bf 272}, 1296--1300.

\bibitem[Gounelle and Meibom(2007)]{Gounel2007}
\biblab{Gounel2007}
Gounelle, M. \& Meibom, A., 2007.
[The oxygen isotopic composition of the Sun as a test of the supernova origin
of ${}^{26}{\rm Al}$ and ${}^{41}{\rm Ca}$].
{\it Astrophys. J.}, {\bf 664}, L123--L125.

\bibitem[Greaves(2005)]{Greave2005}
\biblab{Greave2005}
Greaves, J. S., 2005.
[Disks around stars and the growth of planetary systems].
{\it Science}, {\bf 307}, 68--71. % Special section, review

\bibitem[Grec {\it et~al.\/}(1980)]{Grec1980}
\biblab{Grec1980}
Grec, G., Fossat, E. \& Pomerantz, M., 1980.
[Solar oscillations: full disk observations from the geographic South Pole].
{\it Nature}, {\bf 288}, 541--544.

\bibitem[Greer {\it et~al.\/}(2015)]{Greer2015}
\biblab{Greer2015}
Greer, B. J., Hindman, B. W., Featherstone, N. A. \& Toomre, J., 2015.
[Helioseismic imaging of fast convective flows throughout the 
near-surface shear layer].
{\it Astrophys. J.}, {\bf 803}, L17-(1--5).

{\rv
\bibitem[Grevesse(2019)]{Greves2019}
\biblab{Greves2019}
Grevesse, N., 2019.
[The solar chemical composition: past and present].
In {\it Proc. Workshop ``How much do we trust stellar models?''}, 
{\it Bull. Soc. Roy. Sci. de Li{\`e}ge}, {\bf 88}, Actes de Colloques, 5--14.
}

\bibitem[Grevesse and Noels(1993)]{Greves1993}
\biblab{Greves1993}
Grevesse, N. \& Noels, A., 1993.
[Cosmic abundances of the elements].
In {\it Origin and evolution of the Elements},
eds N. Prantzos, E. Vangioni-Flam \& M. Cass{\'e} 
(Cambridge: Cambridge Univ. Press), 15--25.

\bibitem[Grevesse and Sauval(1998)]{Greves1998}
\biblab{Greves1998}
Grevesse, N. \& Sauval, A. J., 1998.
[Standard solar composition].
{\it Proc. ISSI Workshop on Solar Composition and its
Evolution--from Core to Corona},
eds C. Fr\"ohlich, M. C. E. Huber, S. Solanki \& R. von Steiger,
{\it Space Science Reviews}, {\bf 85}, Kluwer, Dordrecht, 161--174.

\bibitem[Gribov and Pontecorvo(1969)]{Gribov1969}
\biblab{Gribov1969}
Gribov, V. \& Pontecorvo, B., 1969.
[Neutrino astronomy and lepton charge].
{\it Phys. Lett. B}, {\bf 28}, 493--496.

\bibitem[Grundahl {\it et~al.\/}(2007)]{Grunda2007}
\biblab{Grunda2007}
Grundahl, F., Kjeldsen, H., Christensen-Dalsgaard, J., Arentoft, T. \&
Frandsen, S., 2007.
[Stellar Oscillations Network Group].
{\it Proc. Vienna Workshop on the Future of Asteroseismology},
eds G. Handler \& G. Houdek, {\it Comm. in Asteroseismology}, 
{\bf 150}, 300--306.

\bibitem[Grundahl {\it et~al.\/}(2014)]{Grunda2014}
\biblab{Grunda2014}
Grundahl, F., Christensen-Dalsgaard, J., Pall{\'e}, P. L., Andersen, M. F.,
Frandsen, S., Harps{\o}e, K., J{\o}rgensen, U. G., Kjeldsen, H.,
Rasmussen, P. K., Skottfelt, J., S{\o}rensen, A. N. \& Hage, A. T., 2014.
[Stellar Observations Network Group: the prototype is nearly ready].
In {\it Proc. IAU symp. 301, Precision Asteroseismology},
eds J. A. Guzik, W. J. Chaplin, G. Handler \& A. Pigulski,
IAU and Cambridge University Press, 69--75.

\bibitem[Grundahl {\it et~al.\/}(2017)]{Grunda2017}
\biblab{Grunda2017}
Grundahl, F., Fredslund Andersen, M., Christensen-Dalsgaard, J., Antoci, V.,
Kjeldsen, H., Handberg, R., Houdek, G., Bedding, T. R., Pall{\'e}, P. L.,
Jessen-Hansen, J., Silva Aguirre, V., White, T. R., Frandsen, S.,
Albrecht, S., Andersen, M. I., Arentoft, T., Brogaard, K., Chaplin, W. J.,
Harps{\o}e, K., J{\o}rgensen, U. G., Karovicova, I., Karoff, C.,
Kj{\ae}rgaard Rasmussen, P., Lund, M. N., Sloth Lundkvist, M., Skottfelt, J.,
Norup S{\o}rensen, A., Tronsgaard, R. \& Weiss, E., 2017.
[First results from the Hertzsprung SONG Telescope: asteroseismology of the
G5 subgiant $\mu$ Herculis].
{\it Astrophys. J.}, {\bf 836}, 142-(1--12).

\bibitem[Gryaznov {\it et~al.\/}(2004)]{Gryazn2004}
\biblab{Gryazn2004}
Gryaznov, V. K., Ayukov, S. V., Baturin, V. A., Iosilevskiy, I. L.,
Starotsin, A. N. \& Fortov, V. E., 2004.
[SAHA-S model: equation of state and thermodynamic functions of solar plasma].
In {\it Equation-of-State and Phase-Transition Issues in
Models of Ordinary Astrophysical Matter},
eds V. {\v C}elebonovi{\'c}, W. D{\"a}ppen \& D. Gough,
AIP Conf. Proc. Vol. 731, AIP, Melville, New York, p. 147--161.

\bibitem[G\"udel(2007)]{Gudel2007}
\biblab{Gudel2007}
G\"udel, M., 2007.
[The Sun in time: activity and environment].
{\it Living Rev. Solar Phys.}, {\bf 4},  3.  
\url{http://www.livingreviews.org/lrsp-2007-3}

\bibitem[Guenther {\it et~al.\/}(1992)]{Guenth1992}
\biblab{Guenth1992}
Guenther, D. B., Demarque, P., Kim, Y.-C. \& Pinsonneault, M. H., 1992.
[Standard solar model].
{\it Astrophys. J.}, {\bf 387}, 372--393.

\bibitem[Guenther {\it et~al.\/}(1989)]{Guenth1989}
\biblab{Guenth1989}
Guenther, D. B., Jaffe, A. \& Demarque, P., 1989.
[The standard solar model: composition, opacities and seismology].
{\it Astrophys. J.}, {\bf 345}, 1022--1033.

\bibitem[Guenther {\it et~al.\/}(1996)]{Guenth1996}
\biblab{Guenth1996}
Guenther, D. B., Kim, Y.-C. \& Demarque, P., 1996.
[Seismology of the standard solar model:
tests of diffusion and the OPAL and MHD equations of state].
{\it Astrophys. J.}, {\bf 463}, 382--390.

\bibitem[Gustafsson(1998)]{Gustaf1998}
\biblab{Gustaf1998}
Gustafsson, B., 1998.
[Is the Sun a sun-like star?].
{\it Proc. ISSI Workshop on Solar Composition and its
Evolution--from Core to Corona},
eds C. Fr{\"o}hlich, M. C. E. Huber, S. Solanki \& R. von Steiger,
{\it Space Science Reviews}, {\bf 85}, 419--428, Kluwer, Dordrecht.

\bibitem[Gustafsson(2018)]{Gustaf2018}
\biblab{Gustaf2018}
Gustafsson, B., 2018.
[Dust cleansing of star-forming gas. II. Did late accretion flows change
the chemical composition of the solar atmosphere?].
{\it Astron. Astrophys.}, {\bf 620}, A53-(1--10).

\bibitem[Gustafsson {\it et~al.\/}(2008)]{Gustaf2008}
\biblab{Gustaf2008}
Gustafsson, B., Edvardsson, B., Eriksson, K., J{\o}rgensen, U. G.,
Nordlund, {\AA}. \& Plez, B., 2008.
[A grid of MARCS model atmospheres for late-type stars. I. Methods and
general properties].
{\it Astron. Astrophys.}, {\bf 486}, 951--970.

%\bibitem[Gustafsson(2008)]{Gustaf2008}
%\biblab{Gustaf2008}
%Gustafsson, B., 2008.
%[Is the Sun unique as a star---and if so, why?].
%{\it Phys. Scr.}, {\bf T130}, 014036-(1 --6).

\bibitem[Guzik(2006)]{Guzik2006}
\biblab{Guzik2006}
Guzik, J. A., 2006.
[Reconciling the revised solar abundances with helioseismic constraints].
In {\it Proc. SOHO 18 / GONG 2006 / HELAS I Conf.
Beyond the spherical Sun},
ed. K. Fletcher, ESA SP-624, ESA Publications Division,
Noordwijk, The Netherlands.

\bibitem[Guzik(2008)]{Guzik2008}
\biblab{Guzik2008}
Guzik, J. A., 2008.
[Problems for the standard solar model arising from the new solar mixture].
{\it Mem. della Societa Astronomica Italiana}, {\bf 79}, 481--489.

\bibitem[Guzik and Cox(1995)]{Guzik1995}
\biblab{Guzik1995}
Guzik, J. A. \& Cox, A. N., 1995.
[Early solar mass loss, element diffusion, and solar oscillation frequencies].
{\it Astrophys. J.}, {\bf 448}, 905--914.

\bibitem[Guzik and Mussack(2010)]{Guzik2010}
\biblab{Guzik2010}
Guzik, J. A. \& Mussack, K., 2010.
[Exploring mass loss, low-Z accretion, and convective overshoot
in solar models to mitigate the solar abundance problem].
{\it Astrophys. J.}, {\bf 713}, 1108--1119.
%{\tt [arXiv:1001.0648v1 [astro-ph]]}

\bibitem[Guzik and Watson(2004)]{Guzik2004}
\biblab{Guzik2004}
Guzik, J. A. \& Watson, L. C., 2004.
[Can recently derived solar photospheric abundances be consistent with
helioseismology?].
In {\it Proc. SOHO 14 - GONG 2004: ``Helio- and Asteroseismology: Towards
a golden future''; Yale, July 12--16 2004}, ed. Danesy, D., ESA SP-559,
ESA Publication Division, Noordwijk, The Netherlands, p. 456--459.

\bibitem[Guzik {\it et~al.\/}(1987)]{Guzik1987}
\biblab{Guzik1987}
Guzik, J. A., Willson, L. A. \& Brunish, W. M., 1987.
[A comparison between mass-losing and standard solar models].
{\it Astrophys. J.}, {\bf 319}, 957--965.

\bibitem[Guzik {\it et~al.\/}(2005)]{Guzik2005}
\biblab{Guzik2005}
Guzik, J. A., Watson, L. S. \& Cox, A. N., 2005.
[Can enhanced diffusion improve helioseismic agreement for solar models
with revised abundances?].
{\it Astrophys. J.}, {\bf 627}, 1049--1056.

\bibitem[Guzik {\it et~al.\/}(2009)]{Guzik2009}
\biblab{Guzik2009}
Guzik, J. A., Keady, J. J. \& Kilcrease, D. P., 2009.
[Early solar mass loss, opacity uncertainties, and the solar abundance problem].
In {\it Stellar pulsation: Challenges for theory and observation},
eds J. A. Guzik and P. A. Bradley, AIP,
AIP Conf. Proc. Ser., Vol. 1170, p. 577--581.

\bibitem[Haberreiter {\it et~al.\/}(2008)]{Haberr2008}
\biblab{Haberr2008}
Haberreiter, M., Schmutz, W. \& Kosovichev, A. G., 2008.
[Solving the discrepancy between the seismic and photospheric solar radius].
{\it Astrophys. J.}, {\bf 675}, L53--L56.

\bibitem[Hale {\it et~al.\/}(2016)]{Hale2016}
\biblab{Hale2016}
Hale, S. J., Howe, R., Chaplin, W. J., Davies, G. R. \& Elsworth, Y. P., 2016.
[Performance of the {\it Birmingham Solar-Oscillations Network}].
{\it Solar Phys.}, {\bf 291}, 1--28.

\bibitem[Hampel {\it et~al.\/}(1999)]{Hampel1999}
\biblab{Hampel1999}
Hampel, W., Handt, J., Heusser, G., Kiko, J., {\etal}, 1999.
%Kirsten, T., 
%Laubenstein, M., Pernicka, E., Rau, W., Wojcik, M., Zakharov, Y., v. Ammon, R.,
%Ebert, K. H., Fritsch, T., Heidt, D., Henrich, E., Stielglitz, L.,
%Weirich, F., Balata, M., Sann, M.,
%Hartmann, F. X., Bellotti, E., Cattadori, C., Cremonesi, O., 
%Ferrari, N., Fiorini, E., Zanotti, L., Altmann, M., v. Feilitzsch, F., 
%M{\"o}{\ss}bauer, R., W{\"a}nninger, S., Berthomieu, G., Schatzman, E.,
%Carmi, I., Dostrovsky, I., Bacci, C., Belli, P., Bernabei, R., d'Angelo, S.,
%Paoluzi, L., Cribier, M., Rich, J., Spiro, M., Tao, C.,
%Vignaud, D., Boger, J., Hahn, R. L., 
%Rowley, J. K., Stoenner, R. W. \& Weneser, J., 1999.
[GALLEX solar neutrino observations: results from GALLEX IV].
{\it Phys. Lett. B}, {\bf 447}, 127--133.

\bibitem[Hanasoge {\it et~al.\/}(2012)]{Hanaso2012}
\biblab{Hanaso2012}
Hanasoge, S. M., Duvall, T. L. \& Sreenivasan, K. R., 2012.
[Anomalously weak solar convection].
{\it PNAS}, {\bf 109}, 11928--11932.

\bibitem[Hanasoge {\it et~al.\/}(2016)]{Hanaso2016}
\biblab{Hanaso2016}
Hanasoge, S., Gizon, L. \& Sreenivasan, K. R., 2016.
[Seismic sounding of convection in the Sun].
{\it Ann. Rev. Fluid. Mech.}, {\bf 48}, 191--217.

{\rv
\bibitem[Hanasoge {\it et~al.\/}(2020)]{Hanaso2020}
\biblab{Hanaso2020}
Hanasoge, S. M., Hotta, H. \& Sreenivasan, K. R., 2020.
[Turbulence in the Sun is suppressed on large scales and confined to
equatorial regions].
{\it Science Advances}, {\bf 6}, eaba9639-(1--8).}

\bibitem[Harsano {\it et~al.\/}(2018)]{Harson2018}
\biblab{Harson2018}
Harsono, D., Bjerkeli, P., van der Wiel, M. H. D., Ramsey, J. P., 
Maud, L. T., Kristensen, L. E. \& J{\o}rgensen, J. K., 2018.
[Evidence for the start of planet formation in a young circumstellar disk].
{\it Nat. Astron.}, {\rv {\bf 2}, 646--651.}
\url{https://doi.org/10.1038/s41550-018-0497-x}

\bibitem[Harvey(1988)]{Harvey1988}
\biblab{Harvey1988}
Harvey, J. W., 1988.
[Techniques for observing stellar oscillations].
In {\it Proc. IAU Symposium No 123, Advances in helio- and asteroseismology},
eds Christensen-Dalsgaard, J. \& Frandsen, S.,
Reidel, Dordrecht, p. 497--511.

\bibitem[Harvey {\it et~al.\/}(1996)]{Harvey1996}
\biblab{Harvey1996}
Harvey, J. W., Hill, F., Hubbard, R. P., Kennedy, J. R., Leibacher, J. W.,
Pintar, J. A., Gilman, P. A., Noyes, R. W., Title, A. M., Toomre, J.,
Ulrich, R. K., Bhatnagar, A., Kennewell, J. A., Marquette, W.,
Partr{\'o}n, J., Sa{\'a}, O. \& Yasukawa, E., 1996.
[The Global Oscillation Network Group (GONG) project].
{\it Science}, {\bf 272}, 1284--1286.

\bibitem[Hathaway(2015)]{Hathaw2015}
\biblab{Hathaw2015}
Hathaway, D. H., 2015.
[The solar cycle].
{\it Living Rev. Solar Phys.}, {\bf 12},  4.  
\url{https://doi.org/10.1007/lrsp-2015-4}

\bibitem[Haxton(1995)]{Haxton1995}
\biblab{Haxton1995}
Haxton, W. C., 1995.
[The solar neutrino problem].
{\it Annu. Rev. Astron. Astrophys.}, {\bf 33}, 459--503.

\bibitem[Haxton(2008)]{Haxton2008a}
\biblab{Haxton2008a}
Haxton, W. C., 2008.
[Solar neutrinos: models, observations and new opportunities].
{\it Publ. Astron. Soc. Australia}, {\bf 25}, 44--51.

\bibitem[Haxton and Serenelli(2008)]{Haxton2008b}
\biblab{Haxton2008b}
Haxton, W. C. \& Serenelli, A. M., 2008.
[CN cycle solar neutrinos and the Sun's primordial core metallicity].
{\it Astrophys. J.}, {\bf 687}, 678--691.

\bibitem[Haxton {\it et~al.\/}(2006)]{Haxton2006}
\biblab{Haxton2006}
Haxton, W. C., Parker, P. D. \& Rolfs, C. E., 2006.
[Solar hydrogen burning and neutrinos].
{\it Nucl. Phys. A.}, {\bf 777}, 226--253.

\bibitem[Haxton {\it et~al.\/}(2013)]{Haxton2013}
\biblab{Haxton2013}
Haxton, W. C., Hamish-Robertson, R. G. \& Serenelli, A. M., 2013.
[Solar neutrinos: status and prospects].
{\it Annu. Rev. Astron. Astrophys.}, {\bf 51}, 21--61.

\bibitem[Hayashi and  Hoshi(1961)]{Hayash1961}
\biblab{Hayash1961}
Hayashi, C. \& Hoshi, R., 1961.
[The outer envelope of giant stars with surface convection zone].
{\it Publ. Astron. Soc. Japan}, {\bf 13}, 442--449.

\bibitem[Heber {\it et~al.\/}(2003)]{Heber2003}
\biblab{Heber2003}
Heber, V. S., Baur, H. \& Wieler, R., 2003.
[Helium in lunar samples analyzed by high-resolution stepwise etching:
implications for the temporal constancy of solar wind isotope composition].
{\it Astrophys. J.}, {\bf 597}, 602--614.

\bibitem[Heber {\it et~al.\/}(2009)]{Heber2009}
\biblab{Heber2009}
Heber, V. S., Wieler, R., Baur, H., Olinger, C., Friedmann, T. A. \&
Burnett, D. S., 2009.
[Noble gas composition of the solar wind as collected by the Genesis mission].
{\it Geochim. Cosmochim Acta}, {\bf 73}, 7414--7432.

\bibitem[Hekker and  Christensen-Dalsgaard(2017)]{Hekker2017}
\biblab{Hekker2017}
Hekker, S. \& Christensen-Dalsgaard, J., 2017.
[Giant star seismology].
{\it Astron. Astrophys. Rev.}, {\bf 25}, 1-(1--122). 

\bibitem[Henrici(1962)]{Henric1962}
\biblab{Henric1962}
Henrici, P., 1962.
{\it Discrete variable methods in ordinary differential equations},
John Wiley \& Sons, New York.

\bibitem[Henyey {\it et~al.\/}(1955)]{Henyey1955}
\biblab{Henyey1955}
Henyey, L. G., LeLevier, R. \& Lev'ee, R. D., 1955.
[The early phases of stellar evolution].
{\it Publ. Astron. Soc. Pacific}, {\bf 67}, 154--160.

\bibitem[Henyey {\it et~al.\/}(1959)]{Henyey1959}
\biblab{Henyey1959}
Henyey, L. G., Wilets, L., B{\"o}hm, K. H., LeLevier, R. \& Levee, R. D., 1959.
[A method for automatic computation of stellar evolution].
{\it Astrophys. J.}, {\bf 129}, 628--636.

\bibitem[Henyey {\it et~al.\/}(1964)]{Henyey1964}
\biblab{Henyey1964}
Henyey, L. G., Forbes, J. E. \& Gould, N. L., 1964.
[A new method of automatic computation of stellar evolution].
{\it Astrophys. J.}, {\bf 139}, 306--317.

\bibitem[Henyey {\it et~al.\/}(1965)]{Henyey1965}
\biblab{Henyey1965}
Henyey, L. G., Vardya, M. S. \& Bodenheimer, P., 1965.
[Studies in stellar evolution. III. The calculation of model envelopes].
{\it Astrophys. J.}, {\bf 142}, 841--854.

\bibitem[Herwig(2005)]{Herwig2005}
\biblab{Herwig2005}
Herwig, F., 2005.
[Evolution of asymptotic giant branch stars].
{\it Annu. Rev. Astron. Astrophys.}, {\bf 43}, 435--479.

\bibitem[Hill and Rosenwald(1986)]{Hill1986}
\biblab{Hill1986}
Hill, H. A. \& Rosenwald, R. D., 1986.
[Deviations from the normal mode spectrum of asymptotic theory.
I. Identification of quasi-periodic departures in the low-degree spectrum of
the solar 5-min oscillations].
{\it Astrophys. Space Sci.}, {\bf 126}, 335--356.

\bibitem[Hill {\it et~al.\/}(1976)]{Hill1976}
\biblab{Hill1976}
Hill, H. A., Stebbins, R. T. \& Brown, T. M., 1976.
[Recent oblateness observations: Data, interpretation and
significance for earlier work].
{\it Atomic Masses and Fundamental Constants}, {\bf 5}
(ed. Sanders, J. H. \& Wapstra, A. H.) p. 622--628 (Plenum Press).

\bibitem[Hirata {\it et~al.\/}(1989)]{Hirata1989}
\biblab{Hirata1989}
Hirata, K.~S., Kajita, T., Kifune, T., {\etal}, 1989.
%Kihara, K., Nakahata, M.,
%Nakamura, K., Ohara, S., Oyama, Y., Sato, N., Takita, M., Totsuka, Y.,
%Yaginuma, Y., Mori, M., Suzuki, A., Takahashi, K., Tanimori, T., Yamada, M.,
%Koshiba, M., Suda, T., Miyano, K., Miyata, H., Takei, H., Kaneyuki, K.,
%Nagashima, Y., Suzuki, Y., Beier, E.~W., Feldscher, L.~R., Frank, E.~D.,
%Frati, W., Kim, S.~B., Mann, A.~K., Newcomer, F.~M., Van Berg, R. \&
%Zhang, W., 1989.
[Observation of ${}^8{\rm B}$ solar neutrinos in the Kamiokande-II
detector].
{\it Phys. Rev. Lett.}, {\bf 63}, 16--19.

\bibitem[Hoeksema {\it et~al.\/}(2018)]{Hoekse2018}
\biblab{Hoekse2018}
Hoeksema, J. T., Baldner, C. S., Bush, R. I., Schou, J. \& Scherrer, P. H.,
2018.
[On-orbit performance of the {\it Helioseismic and Magnetic Imager} instrument
onboard the {\it Solar Dynamics Observatory}].
{\it Solar Phys.}, {\bf 293}, 45-(1--49).

\bibitem[Holweger and  M{\"u}ller(1974)]{Holweg1974}
\biblab{Holweg1974}
Holweger, H. \& M{\"u}ller, E. A., 1974.
[The photospheric barium spectrum: solar abundance and collision broadening
of Ba II lines by hydrogen].
{\it Solar Phys.}, {\bf 39}, 19--30.

\bibitem[Houdek(2000)]{Houdek2000}
\biblab{Houdek2000}
Houdek, G., 2000.
[Convective effects on p-mode stability in Delta Scuti stars].
In {\it Delta Scuti and related stars},
eds M. Breger \& M. H. Montgomery,
{\it ASP Conf. Ser.}, {\bf 210}, ASP, San Francisco, p. 454--463.

\bibitem[Houdek(2004)]{Houdek2004}
\biblab{Houdek2004}
Houdek, G., 2004.
[Asteroseismic helium abundance determination].
In {\it Equation-of-State and Phase-Transition Issues in
Models of Ordinary Astrophysical Matter},
eds V. {\v C}elebonovi{\'c}, W. D{\"a}ppen \& D. Gough,
AIP Conf. Proc. Vol. 731, AIP, Melville, New York, p. 193--207.

\bibitem[Houdek and  Dupret(2015)]{Houdek2015}
\biblab{Houdek2015}
Houdek, G. \& Dupret, M.-A., 2015.
[Interaction between convection and pulsation].
{\it Living Rev. Solar Phys.}, {\bf 12},  8. 
\url{https://doi.org/10.1007/lrsp-2015-8}

\bibitem[Houdek and Gough(2007a)]{Houdek2007a}
\biblab{Houdek2007a}
Houdek, G. \& Gough, D. O., 2007a.
[An asteroseismic signature of helium ionization].
{\it Mon. Not. R. Astron. Soc.}, {\bf 375}, 861--880.

\bibitem[Houdek and  Gough(2007)]{Houdek2007b}
\biblab{Houdek2007b}
Houdek, G. \& Gough, D. O., 2007b.  
[On the seismic age of the Sun].
In Stancliffe R.J., Dewi J., Houdek G., Martin R.G., Tout C.A., eds,
AIP Conf. Proc., {\it Unsolved Problems in Stellar Physics}.
American Institute of Physics, Melville, p. 219--224.

\bibitem[Houdek and  Gough(2011)]{Houdek2011}
\biblab{Houdek2011}
Houdek, G. \& Gough, D. O., 2011.  
[On the seismic age and heavy-element abundance of the Sun].
{\it Mon. Not. R. Astron. Soc.}, {\bf 418}, 1217--1230.

\bibitem[Houdek {\it et~al.\/}(1999)]{Houdek1999}
\biblab{Houdek1999}
Houdek, G., Balmforth, N. J., Christensen-Dalsgaard, J. \& Gough, D. O., 1999.
[Amplitudes of stochastically excited oscillations in main-sequence stars].
{\it Astron. Astrophys.}, {\bf 351}, 582--596.

\bibitem[Houdek {\it et~al.\/}(2017)]{Houdek2017}
\biblab{Houdek2017}
Houdek, G., Trampedach, R., Aarslev, M. J. \& Christensen-Dalsgaard, J., 2017.
[On the surface physics affecting solar oscillation frequencies].
{\it Mon. Not. R. Astron. Soc.}, {\bf 464}, L124--L128. % refereed
%{\tt [arXiv:1609.06129 [astro-ph.SR]]}

\bibitem[Howard and  LaBonte(1980)]{Howard1980}
\biblab{Howard1980}
Howard, R. \& LaBonte, B. J., 1980.
[The Sun is observed to be a torsional oscillator with a period of 11 years].
{\it Astrophys. J.}, {\bf 239}, L33--L36.

\bibitem[Howe(2009)]{Howe2009}
\biblab{Howe2009}
Howe, R., 2009.
[Solar interior rotation and its variation].
{\it Living Rev. Solar Phys.}, {\bf 6},  1. URL (cited on 16/4/09): 
\url{http://www.livingreviews.org/lrsp-2009-1}

\bibitem[Howe(2016)]{Howe2016}
\biblab{Howe2016}
Howe, R., 2016.
[Solar interior structure and dynamics].
{\it Asian J. Phys.}, {\bf 25}, 311--324.

\bibitem[Howe {\it et~al.\/}(2000)]{Howe2000}
\biblab{Howe2000}
Howe, R., Christensen-Dalsgaard, J., Hill, F., Komm, R. W.,
Larsen, R. M., Schou, J., Thompson, M. J. \& Toomre, J., 2000.
[Deeply penetrating banded zonal flows in the solar convection zone].
{\it Astrophys. J.}, {\bf 533}, L163--L166.

\bibitem[Howe {\it et~al.\/}(2002)]{Howe2002}
\biblab{Howe2002}
Howe, R., Komm, R. W. \& Hill, F., 2002.
[Localizing the solar cycle frequency shifts in global $p$-modes].
{\it Astrophys. J.}, {\bf 580}, 1172--1187.

\bibitem[Howe {\it et~al.\/}(2009)]{Howeetal2009}
\biblab{Howeetal2009}
Howe, R., Christensen-Dalsgaard, J., Hill, F., Komm, R., Schou, J. \&
Thompson, M. J., 2009.
[A note on the torsional oscillation at solar minimum].
{\it Astrophys. J.}, {\bf 701}, L87--L90.

\bibitem[Howe {\it et~al.\/}(2013)]{Howe2013}
\biblab{Howe2013}
Howe, R., Christensen-Dalsgaard, J., Hill, F., Komm, R., Larson, T. P.,
Rempel, M., Schou, J.  \& Thompson, M. J., 2013.
[The high-latitude branch of the solar torsional oscillation in the rising
phase of cycle 24].
{\it Astrophys. J.}, {\bf 767}, L20-(1--4).

\bibitem[Howe {\it et~al.\/}(2017)]{Howe2017}
\biblab{Howe2017}
Howe, R., Davies, G. R., Chaplin, W. J., Elsworth, Y., Basu, S., Hale, S. J.,
Ball, W. H. \& Komm, R. W., 2017.
[The Sun in transition? Persistence of near-surface structural changes
through Cycle 24].
{\it Mon. Not. R. Astron. Soc.}, {\bf 470}, 1935--1942.

\bibitem[Howe {\it et~al.\/}(2018)]{Howe2018}
\biblab{Howe2018}
Howe, R., Hill, F., Komm, R., Chaplin, W. J., Elsworth, Y., Davies, G. R.,
Schou, J. \& Thompson, M. J., 2018.
[Signatures of Cycle 25 in subsurface zonal flows].
{\it Astrophys. J.}, {\bf 862}, L5-(1--6).

\bibitem[Howell {\it et~al.\/}(2014)]{Howell2014}
\biblab{Howell2014}
Howell, S. B., Sobeck, C., Haas, M., Still, M., Barclay, T., Mullally, F.,
Troeltzsch, J., Aigrain, S., Bryson, S. T., Caldwell, D., Chaplin, W. J.,
Cochran, W. D., Huber, D., Marcy, G. W., Miglio, A., Najita, J. R.,
Smith, M., Twicken, J. D. \& Fortney, J. J., 2014.
[The K2 Mission: Characterization and Early results].
{\it Publ. Astron. Soc. Pacific}, {\bf 126}, 398--408.

\bibitem[Huber {\it et~al.\/}(2013)]{Huber2013}
\biblab{Huber2013}
Huber, D., Carter, J. A., Barbieri, M., Miglio, A., Deck, K. M.,
Fabrycky, D. C., Montet, B. T., Buchhave, L. A., Chaplin, W. J.,
Hekker, S., Montalb{\'a}n, J., Sanchis-Ojeda, R., Basu, S., Bedding, T. R.,
Campante, T. L., Christensen-Dalsgaard, J., Elsworth, Y. P., Stello, D.,
Arentoft, T., Ford, E. B., Gilliland, R. L., Handberg, R., Howard, A. W.,
Isaacson, H., Asher Johnson, J., Karoff, C., Kawaler, S. D.,
Kjeldsen, H., Latham, D. W., Lund, M. N., Lundkvist, M., Marcy, G. W.,
Metcalfe, T. S., Silva Aguirre, V. \& J. N. Winn, 2013.
[Stellar spin-orbit misalignment in a multi-planet system].
{\it Science}, {\bf 342}, 331--334.

\bibitem[Hummer and Mihalas(1988)]{Hummer1988}
\biblab{Hummer1988}
Hummer, D. G. \& Mihalas, D., 1988.
[The equation of state for stellar envelopes. I. An occupation probability
formalism for the truncation of internal partition functions].
{\it Astrophys. J.}, {\bf 331}, 794--814.

\bibitem[Hurlburt {\it et~al.\/}(1986)]{Hurlbu1986}
\biblab{Hurlbu1986}
Hurlburt, N. E., Toomre, J. \& Massaguer, J. M., 1986.
[Nonlinear compressible convection penetrating into stable layers
and producing internal gravity waves].
{\it Astrophys. J.}, {\bf 311}, 563--577.

\bibitem[Iben(1968)]{Iben1968}
\biblab{Iben1968}
Iben, I., 1968.
[Solar neutrinos and the solar helium abundance].
{\it Phys. Rev. Lett.}, {\bf 21}, 1208--1212.

\bibitem[Iben(1969)]{Iben1969}
\biblab{Iben1969}
Iben, I., 1969.
[The ${\rm Cl}^{37}$ solar neutrino experiment and the solar helium abundance].
{\it Ann. Phys. New York}, {\bf 54}, 164--203.

\bibitem[Iglesias and Rogers(1991)]{Iglesi1991}
\biblab{Iglesi1991}
Iglesias, C. A., \& Rogers, F. J., 1991.
[Opacities for the solar radiative interior].
{\it Astrophys. J.}, {\bf 371}, 408--417.

\bibitem[Iglesias and Rogers(1995)]{Iglesi1995}
\biblab{Iglesi1995}
Iglesias, C. A. \& Rogers, F. J., 1995.
[Discrepancies between OPAL and OP opacities at high densities 
and temperatures].
{\it Astrophys. J.}, {\bf 443}, 460--463.

\bibitem[Iglesias and Rogers(1996)]{Iglesi1996}
\biblab{Iglesi1996}
Iglesias, C. A. \& Rogers, F. J., 1996.
[Updated OPAL opacities].
{\it Astrophys. J.}, {\bf 464}, 943--953.

\bibitem[Iglesias {\it et~al.\/}(1987)]{Iglesi1987}
\biblab{Iglesi1987}
Iglesias, C. A., Rogers, F. J. \& Wilson, B. G., 1987.
[Reexamination of the metal contribution to astrophysical opacities].
{\it Astrophys. J.}, {\bf 322}, L45--L48.

\bibitem[Iglesias {\it et~al.\/}(1992)]{Iglesi1992}
\biblab{Iglesi1992}
Iglesias, C. A., Rogers, F. J. \& Wilson, B. G., 1992.
[Spin-orbit interaction effects on the Rosseland mean opacity].
{\it Astrophys. J.}, {\bf 397}, 717--728.

\bibitem[Isella {\it et~al.\/}(2016)]{Isella2016}
\biblab{Isella2016}
Isella, A., Guidi, G., Testi, L., Liu, S., Li, H., Li, S., Weaver, E.,
Boehler, Y., Carperter, J. M., De Gregorio-Monsalvo, I., Manara, C. F.,
Natta, A.,
P{\'e}rez, L. M., Ricci, L., Sargent, A., Tazzari, M. \& Turner, N., 2016.
[Ringed structures of the HD 163296 protoplanetary disk revealed by ALMA]
{\it Phys. Rev. Lett.}, {\bf 117}, 251101-(1--8).

\bibitem[Israelian {\it et~al.\/}(2009)]{Israel2009}
\biblab{Israel2009}
Israelian, G., Delgado Mena, E., Santos, N. C., Sousa, S. G., Mayor, M.,
Udry, S., Dom\'{\i}nguez Cerde{\~n}a, C., Rebolo, R. \& Randich, S., 2009.
[Enhanced lithium depletion in Sun-like stars with orbiting planets].
{\it Nature}, {\bf 462}, 189--191.

\bibitem[Jensen and  Haugb{\o}lle(2018)]{Jensen2018}
\biblab{Jensen2018}
Jensen, S. S. \& Haugb{\o}lle, T., 2018.
[Explaining the luminosity spread in young clusters: proto and pre-main
sequence stellar evolution in a molecular cloud environment].
{\it Mon. Not. R. Astron. Soc.}, {\bf 474}, 1176--1193.

\bibitem[Johansen and  Lambrechts(2017)]{Johans2017}
\biblab{Johans2017}
Johansen, A. \& Lambrechts, M., 2017.
[Forming planets via pebble accretion].
{\it Annu. Rev. Earth Planet. Sci.}, {\bf 45}, 359--387.

\bibitem[J{\o}rgensen {\it et~al.\/}(2017a)]{Jorgen2017a}
\biblab{Jorgen2017a}
J{\o}rgensen, A. C. S., Weiss, A., Mosumgaard, J. R., Silva Aguirre, V. \&
Sahlholdt, C. L., 2017a.
[Theoretical oscillation frequencies for solar-type dwarfs 
from stellar models with $\langle 3 {\rm D} \rangle$-atmospheres].
{\it Mon. Not. R. Astron. Soc.}, {\bf 472}, 3264--3276.

\bibitem[J{\o}rgensen and  Christensen-Dalsgaard(2017b)]{Jorgen2017b}
\biblab{Jorgen2017b}
J{\o}rgensen, A. C. S. \& Christensen-Dalsgaard, J., 2017b.
[A semi-analytical computation of the theoretical uncertainties of the solar
neutrino flux].
{\it Mon. Not. R. Astron. Soc.}, {\bf 471}, 4802--4805.

\bibitem[J{\o}rgensen and  Weiss(2018)]{Jorgen2018b}
\biblab{Jorgen2018b}
J{\o}rgensen, A. C. S. \& Weiss, A., 2018.
[Addressing the acoustic tachocline anomaly and the lithium depletion
problem at the same time].
{\it Mon. Not. R. Astron. Soc.}, {\bf 481}, 4389--4396.

\bibitem[J{\o}rgensen {\it et~al.\/}(2018)]{Jorgen2018a}
\biblab{Jorgen2018a}
J{\o}rgensen, A. C. S., Mosumgaard, J. R., Weiss, A., Silva Aguirre, V. \&
Christensen-Dalsgaard, J., 2018.
[Coupling 1D stellar evolution with 3D-hydrodynamical simulations 
on-the-fly--I: a new standard solar model].
{\it Mon. Not. R. Astron. Soc.}, {\bf 481}, L35--L39.

\bibitem[J{\o}rgensen(1991)]{Joerge1991}
\biblab{Joerge1991}
J{\o}rgensen, U. G., 1991.
[Advanced stages in the evolution of the Sun].
{\it Astron. Astrophys.}, {\bf 246}, 118--136.

\bibitem[Joss(1974)]{Joss1974}
\biblab{Joss1974}
Joss, P. C., 1974.
[Are stellar surface heavy-element abundances systematically enhanced?].
{\it Astrophys. J.}, {\bf 191}, 771--774.

\bibitem[Kaether {\it et~al.\/}(2010)]{Kaethe2010}
\biblab{Kaethe2010}
Kaether, E., Hampel, W., Heusser, G., Kiko, J. \& Kirsten, T., 2010.
[Reanalysis of the GALLEX solar neutrino flux and source experiments].
{\it Phys. Lett. B}, {\bf 685}, 47--54.

\bibitem[Kajita(2016)]{Kajita2016}
\biblab{Kajita2016}
Kajita, T.,2016.
[Nobel lecture: Discovery of atmospheric neutrino oscillations].
{\it Rev. Mod. Phys.}, {\bf 88}, 030501-(1--7).

\bibitem[Kanbur and Simon(1994)]{Kanbur1994}
\biblab{Kanbur1994}
Kanbur, S. M. \& Simon, N. R., 1994.
[Comparative pulsation calculations with OP and OPAL opacities].
{\it Astrophys. J.}, {\bf 420}, 880--883.

{\rv
\bibitem[Kawaler(1988)]{Kawale1988}
\biblab{Kawale1988}
Kawaler, S. D., 1988.
[Angular momentum loss in low-mass stars].
{\it Astrophys. J.}, {\bf 333}, 236--247.
}

%\bibitem[Kippenhahn and Weigert(1990)]{Kippen1990}
%\biblab{Kippen1990}
%Kippenhahn, R. \& Weigert, A., 1990.
%{\it Stellar structure and evolution},
%Springer-Verlag, Berlin.

\bibitem[Kennedy {\it et~al.\/}(1993)]{Kenned1993}
\biblab{Kenned1993}
Kennedy, J. R., Jefferies, S. M. \& Hill, F., 1993.
[Solar g-mode signatures in p-mode signals].
In {\it Proc. GONG 1992: Seismic investigation of the Sun and stars},
ed. Brown, T. M.,
Astronomical Society of the Pacific Conference Series, San Francisco,
{\bf 42}, 273--276.

\bibitem[Kippenhahn {\it et~al.\/}(2012)]{Kippen2012}
\biblab{Kippen2012}
Kippenhahn, R., Weigert, A. \& Weiss, A., 2012.
{\it Stellar structure and evolution, Second Edition},
Springer-Verlag, Berlin.

\bibitem[Kiriakidis {\it et~al.\/}(1992)]{Kiriak1992}
\biblab{Kiriak1992}
Kiriakidis, M., El Eid, M. F. \& Glatzel, W., 1992.
[Heavy element opacities and the pulsations of $\beta$ Cephei stars].
{\it Mon. Not. R. astr. Soc.}, {\bf 255}, 1P--5P.

\bibitem[Kirsten(1999)]{Kirste1999}
\biblab{Kirste1999}
Kirsten, A., 1999.
[Solar neutrino experiments: results and implications].
{\it Rev. Mod. Phys.}, {\bf 71}, 1213--1232.

%\bibitem[Kjeldsen and Bedding(2004)]{Kjelds2004}
%\biblab{Kjelds2004}
%Kjeldsen, H. \& Bedding, T. R., 2004.
%[Latest observational results of solar-like oscillations in other stars].
%In {\it Proc. SOHO 14 - GONG 2004: ``Helio- and Asteroseismology: Towards
%a golden future''; Yale, July 12--16 2004}, ed. Danesy, D., ESA SP-559,
%ESA Publication Division, Noordwijk, The Netherlands, p. 101--112.

\bibitem[Kjeldsen {\it et~al.\/}(2005)]{Kjelds2005}
\biblab{Kjelds2005}
Kjeldsen, H., Bedding, T. R., Butler, R. P., Christensen-Dalsgaard, J.,
Kiss, L. L., McCarthy, C., Marcy, G. W., Tinney, C. G. \& Wright, J. T., 2005.
[Solar-like oscillations in $\alpha$ Centauri B].
{\it Astrophys. J.}, {\bf 635}, 1281--1290.

\bibitem[Kjeldsen {\it et~al.\/}(2008)]{Kjelds2008}
\biblab{Kjelds2008}
Kjeldsen, H., Bedding, T. R. \& Christensen-Dalsgaard, J., 2008.
[Correcting stellar oscillation frequencies for near-surface effects].
{\it Astrophys. J.}, {\bf 683}, L175--L178. 
%{\tt [arXiv:0807.1769v1 [astro-ph]]}

\bibitem[Kopp {\it et~al.\/}(2016)]{Kopp2016}
\biblab{Kopp2016}
Kopp, G., Krivova, N., Wu, C. J. \& Lean, J., 2016.
[The impact of the revised sunspot record on solar irradiance reconstructions].
{\it Solar Phys.}, {\bf 291}, 2951--2965.
%{\tt [arXiv:1601.05397 [astro-ph]]}.

\bibitem[Korn {\it et~al.\/}(2006)]{Korn2006}
\biblab{Korn2006}
Korn, A. J., Grundahl, F., Richard, O., Barklem, P. S., Mashonkina, L.,
Collet, R., Piskunov, N. \& Gustafsson, B., 2006.
[A probable stellar solution to the cosmological lithium discrepancy].
{\it Nature}, {\bf 442}, 657--659.

\bibitem[Korn {\it et~al.\/}(2007)]{Korn2007}
\biblab{Korn2007}
Korn, A. J., Grundahl, F., Richard, O., Mashonkina, L., Barklem, P. S.,
Collet, R., Gustafsson, B. \& Piskunov, N., 2007.
[Atomic diffusion and mixing in old stars. I. Very Large Telescope 
FLAMES-UVES observations of stars in NGC 6397].
{\it Astrophys. J.}, {\bf 671}, 402--419.

\bibitem[Korycansky {\it et~al.\/}(2001)]{Koryca2001}
\biblab{Koryca2001}
Korycansky, D. G., Laughlin, G. \& Adams, F. C., 2001.
[Astronomical engineering: a strategy for modifying planetary orbits].
{\it Astrophys. Space Sci.}, {\bf 275}, 349--366.

\bibitem[Korzennik {\it et~al.\/}(2004)]{Korzen2004}
\biblab{Korzen2004}
Korzennik, S. G., Rabello-Soares, M. C. \& Schou, J., 2004.
[On the determination of Michelson Doppler Imager high-degree mode frequencies].
{\it Astrophys. J.}, {\bf 602}, 481--515.

\bibitem[Koshiba(2003)]{Koshib2003}
\biblab{Koshib2003}
Koshiba, M., 2003.
[Nobel lecture: Birth of neutrino astrophysics].
{\it Rev. Mod. Phys.}, {\bf 75}, 1011--1020.

\bibitem[Kosovichev(1995)]{Kosovi1995}
\biblab{Kosovi1995}
Kosovichev, A. G., 1995.
[The upper convective boundary layer].
In: {\it Proc. Fourth SOHO Workshop: Helioseismology},
eds Hoeksema, J. T., Domingo, V., Fleck, B. \& Battrick, B., 
ESA SP-376, Vol. 1, ESTEC, Noordwijk, p. 165--176.

\bibitem[Kosovichev(1996)]{Kosovi1996}
\biblab{Kosovi1996}
Kosovichev, A. G., 1996.
[Helioseismic measurements of elemental abundances in the Sun's interior].
{\it Proc. Conf. on ``Windows on the Sun's interior'', Bombay, Oct. 1995;
Bull. Astron. Soc. India}, {\bf 24}, 355--358.

\bibitem[Kosovichev and Fedorova(1991)]{Kosovi1991}
\biblab{Kosovi1991}
Kosovichev, A. G. \& Fedorova, A. V., 1991.
[Construction of a seismic model of the sun].
{\it Astron. Zh.}, {\bf 68}, 1015--1029
(English translation: {\it Sov. Astron.}, {\bf 35}, 507--513).

\bibitem[Kosovichev and  Rozelot(2018a)]{Kosovi2018a}
\biblab{Kosovi2018a}
Kosovichev, A. \& Rozelot, J.-P., 2018a.
[Cyclic changes of the Sun's seismic radius].
{\it Astrophys. J.}, {\bf 861}, 90-(1--5).
%{\tt [arXiv:1805.09385v1 [astro-ph.SR]]}.

\bibitem[Kosovichev and  Rozelot(2018b)]{Kosovi2018b}
\biblab{Kosovi2018b}
Kosovichev, A. \& Rozelot, J.-P., 2018b.
[Solar cycle variations of rotation and asphericity in the near-surface
shear layer].
{\it J. Atmospheric and Solar-Terrestrial Physics}, {\bf 176}, 21--25.

%\bibitem[Kosovichev(1997)]{Kosovi1997b}
%\biblab{Kosovi1997b}
%Kosovichev, A. G., 1997.
%[Inferences of element abundances from helioseismic data].
%In "Robotic Exploration Close to the Sun: Scientific Basis",
%ed. S.R. Habbal, AIP Conf. Proc. 385, Amer. Inst. Phys.,
%Woodbury, NY, p. 159--166.

\bibitem[Kosovichev {\it et~al.\/}(1992)]{Kosovi1992}
\biblab{Kosovi1992}
Kosovichev, A. G., Christensen-Dalsgaard, J., D{\"a}ppen, W., 
Dziembowski, W. A., Gough, D. O. \& Thompson, M. J., 1992.
[Sources of uncertainty in direct seismological measurements 
of the solar helium abundance].
{\it Mon. Not. R. Astron. Soc.}, {\bf 259}, 536--558.

\bibitem[Kosovichev {\it et~al.\/}(1997)]{Kosovi1997a}
\biblab{Kosovi1997a}
Kosovichev, A. G., Schou, J., Scherrer, P. H., Bogart, R. S.,
Bush, R. I., Hoeksema, J. T., Aloise, J., Bacon, L., Burnette, A.,
de Forest, C., Giles, P. M., Leibrand, K., Nigam, R., Rubin, M., 
Scott, K., Williams, S. D., Basu, S., Christensen-Dalsgaard, J.,
D\"appen, W., Rhodes Jr, E. J., Duvall Jr, T. L., Howe, R., Thompson, M. J.,
Gough, D. O., Sekii, T., Toomre, J., Tarbell, T. D., Title, A. M.,
Mathur, D., Morrison, M., Saba, J. L. R., Wolfson, C. J., Zayer, I. \&
Milford, P. N., 1997.
[Structure and rotation of the solar interior: initial results from the
MDI medium-l program].
{\it Solar Phys.}, {\bf 170}, 43--61.

\bibitem[Krishna Swamy(1966)]{Krishn1966}
\biblab{Krishn1966}
Krishna Swamy, K. S., 1966.
[Profiles of strong lines in K-dwarfs].
{\it Astrophys. J.}, {\bf 145}, 174--194.

\bibitem[Kuffmeier {\it et~al.\/}(2018)]{Kuffme2018}
\biblab{Kuffme2018}
Kuffmeier, M., Frimann, S., Jensen, S. S. \& Haugb{\o}lle, T., 2018.
[Episodic accretion: the interplay of infall and disk instabilities].
{\it Mon. Not. R. Astron. Soc.}, {\bf 475}, 2642--2658.

\bibitem[Kumar and Quataert(1997)]{Kumar1997}
\biblab{Kumar1997}
Kumar, P. \& Quataert, E. J., 1997.
[Angular momentum transport by gravity waves and its effect 
on the rotation of the solar interior].
{\it Astrophys. J.}, {\bf 475}, L143--L146.

\bibitem[Kump {\it et~al.\/}(2000)]{Kump2000}
\biblab{Kump2000}
Kump, L. R., Brantley, S. L. \& Arthur, M. A., 2000.
[Chemical weathering, atmospheric ${\rm CO_2}$, and climate].
{\it Annu. Rev. Earth Planet. Sci.}, {\bf 28}, 611--667.

\bibitem[Kurucz(1991)]{Kurucz1991}
\biblab{Kurucz1991}
Kurucz, R. L., 1991.
[New opacity calculations].
In {\it Stellar atmospheres: beyond classical models},
eds Crivellari, L., Hubeny, I. \& Hummer, D. G.,
NATO ASI Series, Kluwer, Dordrecht, p. 441--448.

\bibitem[Kurucz(1996)]{Kurucz1996}
\biblab{Kurucz1996}
Kurucz, R. L., 1996.
[A new opacity-sampling model atmosphere program for arbitrary abundances].
In {Proc. IAU Symp. 176, Stellar surface structure},
eds K. G. Strassmeier \& J. L. Linsky, Kluwer, Dordrecht, p. 523--526.

\bibitem[Lada and Shu(1990)]{Lada1990}
\biblab{Lada1990}
Lada, C. J. \& Shu, F. H., 1990.
[The formation of sunlike stars].
{\it Science}, {\bf 248}, 564--572.

\bibitem[Laming(2015)]{Laming2015}
\biblab{Laming2015}
Laming, J. M., 2015.
[The FIP and the inverse FIP effects in solar and stellar coronae].
{\it Living Rev. Solar Phys.}, {\bf 12}, 2.

\bibitem[Lande(2009)]{Lande2009}
\biblab{Lande2009}
Lande, K., 2009.
[The life of Raymond Davis, Jr. and the beginning of neutrino astronomy].
{\it Annu. Rev. Nucl. Part. Sci.}, {\bf 59}, 21--39.

\bibitem[Lane(1870)]{Lane1870}
\biblab{Lane1870}
Lane, J. H., 1870.
[On the theoretical temperature of the Sun; under the hypothesis of a 
gaseous mass maintaining its volume by its internal heat, and depending
on the laws of gases as known to terrestrial experiments].
{\it American Journal of Science, 2nd ser.}, {\bf 50}, 57--74.

\bibitem[Larson and  Schou(2015)]{Larson2015}
\biblab{Larson2015}
Larson, T. P. \& Schou, J., 2015.
[Improved helioseismic analysis of medium-$\ell$ data from the
{\it Michelson Doppler Imager}].
{\it Solar Phys.}, {\bf 290}, 3221--3256.

\bibitem[Larson and  Schou(2018)]{Larson2018}
\biblab{Larson2018}
Larson, T. P. \& Schou, J., 2018.
[Global-mode analysis of full-disk data from the
{\it Michelson Doppler Imager} and the {\it Helioseismic and Magnetic Imager}].
{\it Solar Phys.}, {\bf 293}, 29-(1--28).

\bibitem[Lebreton and Maeder(1987)]{Lebret1987}
\biblab{Lebret1987}
Lebreton, Y. \& Maeder, A., 1987.
[Stellar evolution with turbulent diffusion mixing.
VI. The solar model, surface $^7$Li and $^3$He
abundances, solar neutrinos and oscillations].
{\it Astron. Astrophys.}, {\bf 175}, 99--112.

\bibitem[Lebreton and  Goupil(2014)]{Lebret2014}
\biblab{Lebret2014}
Lebreton, Y. \& Goupil, M. J., 2014.
[Asteroseismology for ``\`a la carte'' stellar age-dating and weighing.
Age and mass of the CoRoT exoplanet host HD 52265].
{\it Astron. Astrophys.}, {\bf 569}, A21-(1--24).

\bibitem[Lebreton {\it et~al.\/}(2008)]{Lebret2008}
\biblab{Lebret2008}
Lebreton, Y., Monteiro, M. J. P. F. G., Montalb{\'a}n, J., Moya, A.,
Baglin, A., Christensen-Dalsgaard, J., Goupil, M.-J., Michel, E., Provost, J.,
Roxburgh, I. W., Scuflaire, R., and ESTA team, 2008.
[The CoRoT Evolution and Seismic Tool Activity].
{\it Astrophys. Space Sci.}, {\bf 316}, 1--12.

\bibitem[Leibacher and Stein(1971)]{Leibac1971}
\biblab{Leibac1971}
Leibacher, J. \& Stein, R. F., 1971.
[A new description of the solar five-minute oscillation].
{\it Astrophys. Lett.}, {\bf 7}, 191--192.

\bibitem[Leighton {\it et~al.\/}(1962)]{Leight1962}
\biblab{Leight1962}
Leighton, R. B., Noyes, R. W. \& Simon, G. W., 1962.
[Velocity fields in the solar atmosphere I. Preliminary report].
{\it Astrophys. J.}, {\bf 135}, 474--499.

\bibitem[Le~Pennec {\it et~al.\/}(2015a)]{LePenn2015a}
\biblab{LePenn2015a}
Le Pennec, M., Ribeyre, X., Ducret, J.-E. \& Turck-Chi{\`e}ze, S., 2015a.
[New opacity measurement principle for LMJ-PETAL laser facility].
{\it High Energy Density Physics}, {\bf 17}, 162--167.

\bibitem[Le~Pennec {\it et~al.\/}(2015b)]{LePenn2015b}
\biblab{LePenn2015b}
Le Pennec, M., Turck-Chi\`eze, S., Salmon, S., Blancard, C.,
Coss\'e, P., Faussurier, G. \& Mondet, G., 2015b.
[First new solar models with OPAS opacity tables].
{\it Astrophys. J.}, {\bf 813}, L42-(1--6).

\bibitem[Li {\it et~al.\/}(2018)]{Li2018}
\biblab{Li2018}
Li, P. S., Klein, R. I. \& McKee, C.F., 2018.
[Formation of stellar clusters in magnetized, filamentary infrared
dark clouds].
{\it Mon. Not. R. Astron. Soc.}, {\bf 473}, 4220--4241.

\bibitem[Liang(2004)]{Liang2004}
\biblab{Liang2004}
Liang, A., 2004.
[Emulating the OPAL equation of state in the chemical-picture formulation].
In {\it Equation-of-State and Phase-Transition Issues in
Models of Ordinary Astrophysical Matter},
eds V. {\v C}elebonovi{\'c}, W. D{\"a}ppen \& D. Gough,
AIP Conf. Proc. Vol. 731, AIP, Melville, New York, p. 106--116.

\bibitem[Lin and D{\"a}ppen(2004)]{Lin2005}
\biblab{Lin2005}
Lin, C.-H. \& D{\"a}ppen, W., 2005.
[The chemical composition and equation of state of the Sun inferred
from seismic models through an inversion procedure].
In {\it Equation-of-State and Phase-Transition Issues in
Models of Ordinary Astrophysical Matter},
eds V. {\v C}elebonovi{\'c}, W. D{\"a}ppen \& D. Gough,
AIP Conf. Proc. Vol. 731, AIP, Melville, New York, p. 230--236.

\bibitem[Lin {\it et~al.\/}(2007)]{Lin2007}
\biblab{Lin2007}
Lin, C.-H., Antia, H. M. \& Basu, S., 2007.
[Seismic study of the chemical composition of the solar convection zone].
{\it Astrophys. J.}, {\bf 668}, 603--610.

\bibitem[Lin and  D{\"a}ppen(2010)]{LinDap2010}
\biblab{LinDap2010}
Lin, H.-H. \& D{\"a}ppen, W., 2010.
[Emulating the OPAL equation of state].
{\it Astrophys. Space Sci.}, {\bf 328}, 175--178.

\bibitem[Lissauer(1993)]{Lissau1993}
\biblab{Lissau1993}
Lissauer, J. J., 1993.
[Planet formation].
{\it Annu Rev. Astron. Astrophys.}, {\bf 31}, 129--174.

\bibitem[Lodders(2003)]{Lodder2003}
\biblab{Lodder2003}
Lodders, K., 2003.
[Solar system abundances and condensation temperatures of the elements].
{\it Astrophys. J.}, {\bf 591}, 1220--1247.

\bibitem[Lodders(2010)]{Lodder2010}
\biblab{Lodder2010}
Lodders, K., 2010.
[Solar system abundances of the elements].
In {\it Principles and perspectives in cosmochemistry},
{\it Astrophys. Space Sci. Proc.}, Springer, Berlin Heidelberg, p. 379--417.
{\tt [arXiv:1010.27464 [astro-ph.SR]]}.

\bibitem[Lodders {\it et~al.\/}(2009)]{Lodder2009}
\biblab{Lodder2009}
Lodders, K., Palme, H. \& Gail, H.-P., 2009.
[Abundances of the elements in the solar system].
In {Landolt-B{\"o}rnstein, New Series}, Vol. VI/4B, ed. J. E. Tr{\"u}mper,
New York, Springer, p. 560--630.

\bibitem[Lopes and Gough(2001)]{Lopes2001}
\biblab{Lopes2001}
Lopes, I. P. \& Gough, D. O., 2001.
[Seismology of stellar envelopes: probing the outer layers of a star
through the scattering of acoustic waves].
{\it Mon. Not. R. Astron. Soc.}, {\bf 322}, 473--485.

\bibitem[Lopes {\it et~al.\/}(2014)]{Lopes2014}
\biblab{Lopes2014}
Lopes, I., Kadota, K. \& Silk, J., 2014.
[Constraint on light dipole dark matter from helioseismology].
{\it Astrophys. J.}, {\bf 780}, L15-(1--4).

{\rv
\bibitem[Lorenzo-Oliveira {\it et~al.\/}(2020)]{Lorenz2020}
\biblab{Lorenz2020}
Lorenzo-Oliveira, D., Mel{\'e}ndez, J., Ponte, G. \& Galarza, J. Y., 2020.
[The ancient main-sequence solar proxy HIP\,102152 unveils the activity
and rotational fate of our Sun].
{\it Mon. Not. R. Astron. Soc.}, {\bf 495}, L61 -- L65.
}

\bibitem[Lovelock and  Whitfield(1982)]{Lovelo1982}
\biblab{Lovelo1982}
Lovelock, J. E. \& Whitfield, M., 1982.
[Life span of the biosphere].
{\it Nature}, {\bf 296}, 561--563.

\bibitem[Ludwig {\it et~al.\/}(1999)]{Ludwig1999}
\biblab{Ludwig1999}
Ludwig, H.-G., Freytag, B. \& Steffen, M., 1999.
[A calibration of the mixing-length for solar-type stars
based on hydrodynamical simulations. I. Methodological aspects and
results for solar metallicity].
{\it Astron. Astrophys.}, {\bf 346}, 111--124.

\bibitem[Ludwig {\it et~al.\/}(2008)]{Ludwig2008}
\biblab{Ludwig2008}
Ludwig, H.-G., Caffau, E. \& Ku{\v c}inskas, A., 2008.
[Radiation-hydrodynamics simulations of surface convection in low-mass stars:
connections to stellar structure and asteroseismology].
In {\it Proc. IAU Symp. 252: The Art of Modelling Stars in the
$21^{\rm st}$ Century},
eds L. Deng \& K. L. Chan,
IAU and Cambridge University Press, p. 75--81.

\bibitem[Lund {\it et~al.\/}(2014a)]{Lund2014a}
\biblab{Lund2014a}
Lund, M. N., Lundkvist, M., Silva Aguirre, V., Houdek, G., Casagrande, L.,
Van Eylen, V., Campante, T. L., Karoff, C., Kjeldsen, H., Albrecht, S.,
Chaplin, W. J., Nielsen, M. B., Degroote, P., Davies, G. R. \& Handberg, R.,
2014a.
[Asteroseismic inference on the spin-orbit misalignment and stellar parameters
of HAT-P-7].
{\it Astron. Astrophys.}, {\bf 570}, A54-(1--16).

\bibitem[Lund {\it et~al.\/}(2014b)]{Lund2014b}
\biblab{Lund2014b}
Lund, M. L., Miesch, M. S. \& Christensen-Dalsgaard, J., 2014b.
[Differential rotation in main-sequence solar-like stars: qualitative
inference from asteroseismic data].
{\it Astrophys. J.}, {\bf 790}, 121-(1--28). 
(Erratum: {\it Astrophys. J.}, {\bf 794}, 96-(1--2).)

\bibitem[Lund {\it et~al.\/}(2017)]{Lund2017}
\biblab{Lund2017}
Lund, M. N., Silva Aguirre, V., Davies, G. R., Chaplin, W. J.,
Christensen-Dalsgaard, J., Houdek, G., White, T. R., Bedding, T. R.,
Ball, W. H., Huber, D., Antia, H. M., Lebreton, Y., Latham, D. W.,
Handberg, R., Verma, K., Basu, S., Casagrande, L., Justesen, A. B.,
Kjeldsen, H. \& Mosumgaard, J. R., 2017.
[Standing on the shoulders of dwarfs: the {\it Kepler} Asteroseismic
LEGACY Sample. I.  Oscillation mode parameters].
{\it Astrophys. J.}, {\bf 835}, 172-(1--31).
Erratum: {\it Astrophys. J.}, {\bf 850}, 110-(1--7).

\bibitem[Lundkvist {\it et~al.\/}(2018)]{Lundkv2018}
\biblab{Lundkv2018}
Lundkvist, M. S., Huber, D., Silva Aguirre, V. \& Chaplin, W. J., 2018.
[Characterizing host stars using asteroseismology].
In {\it Handbook of Exoplanets}, eds Deeg, H.J. \& Belmonte, J.A,
Springer, Cham, pp 1--24.
\url{https://doi.org/10.1007/978-3-319-30648-3_177-1}
%{\tt [arXiv:1804.02214 [astro-ph.SR]]}

\bibitem[Maeder(2009)]{Maeder2009}
\biblab{Maeder2009}
Maeder, A., 2009.
{\it Physics, formation and evolution of rotating stars},
Springer, Berlin.

\bibitem[Maeder and Meynet(2000)]{Maeder2000}
\biblab{Maeder2000}
Maeder, A. \& Meynet, G., 2000.
[The evolution of rotating stars].
{\it Annu. Rev. Astron. Astrophys.}, {\bf 38}, 143--190.

\bibitem[Maeder and Zahn(1998)]{Maeder1998}
\biblab{Maeder1998}
Maeder, A. \& Zahn, J.-P., 1998.
[Stellar evolution with rotation. III. Meridional circulation with
$\mu$-gradients and non-stationarity].
{\it Astron. Astrophys.}, {\bf 334}, 1000--1006.

{\rv
\bibitem[Maeder {\it et~al.\/}(2013)]{Maeder2013}
\biblab{Maeder2013}
Maeder, A., Meynet, G., Lagarde, M. \& Charbonnel, C., 2013.
[The thermohaline, Richardson, Rayleigh-Taylor, Solberg-H{\o}iland,
and GSF criteria in rotating stars].
{\it Astron. Astrophys.}, {\bf 553}, A1-(1--7).
}

\bibitem[Magic and  Weiss(2016)]{Magic2016}
\biblab{Magic2016}
Magic, Z. \& Weiss, A., 2016.
[Surface-effect corrections for the solar model].
{\it Astron. Astrophys.}, {\bf 592}, A24-(1--10).

\bibitem[Magic {\it et~al.\/}(2010)]{Magic2010}
\biblab{Magic2010}
Magic, Z., Serenelli, A., Weiss, A. \& Chaboyer, B., 2010.
[On using the color-magnitude diagram morphology of M67 to test
solar abundances].
{\it Astrophys. J.}, {\bf 718}, 1378--1387.

\bibitem[Magic {\it et~al.\/}(2013)]{Magic2013}
\biblab{Magic2013}
Magic, Z., Collet, R., Asplund, M., Trampedach, R., Hayek, W., Chiavassa, A.,
Stein, R. F. \& Nordlund, {\AA}, 2013.
[The STAGGER-grid: A grid of 3D stellar atmosphere models. I. Methods and
general properties].
{\it Astron. Astrophys.}, {\bf 557}, A26-(1--30).

\bibitem[Magic {\it et~al.\/}(2015)]{Magic2015}
\biblab{Magic2015}
Magic, Z., Weiss, A. \& Asplund, M., 2015.
[The STAGGER-grid: A grid of 3D stellar atmosphere models. III The relation to
mixing length convection theory].
{\it Astron. Astrophys.}, {\bf 573}, A89-(1--17).

\bibitem[Maltoni and  Smirnov(2016)]{Malton2016}
\biblab{Malton2016}
Maltoni, M. \& Smirnov, A. Y., 2016.
[Solar neutrinos and neutrino physics].
{\it Eur. Phys. J. A}, {\bf 52}, 87-(1--16).

{\rv
\bibitem[Mamajek(2009)]{Mamaje2009}
\biblab{Mamaje2009}
Mamajek, E. E., 2009.
[Initial conditions of plane formation: lifetimes of primordial disks].
{\it AIP Conf. Proc.}, {\bf 1158}, 3-(1--8).
{\tt [arXiv:0906.5011 [astro-ph.SR]]}
}

\bibitem[Mamajek {\it et~al.\/}(2015)]{Mamaje2015}
\biblab{Mamaje2015}
Mamajek, E. E., Prsa, A., Torres, G., Harmanec, P., Asplund, M., Bennett, P. D.,
Capitaine, N., Christensen-Dalsgaard, J., Depagne, E., Folkner, W. M.,
Haberreiter, M., Hekker, S., Hilton, J. L., Kostov, V., Kurtz, D. W., 
Laskar, J.,
Mason, B. D., Milone, E. F., Montgomery, M. M., Richards, M. T., Schou, J. \&
Stewart, S. G.
(IAU Inter-Division A-G Working Group on Nominal Units for Stellar \& Planetary
Astronomy), 2015.
[Resolution B3 on recommended nominal conversion constants for selected solar
and planetary properties].
{\tt [arXiv:1510.07674v1 [astro-ph.SR]]}.

\bibitem[Manchon {\it et~al.\/}(2018)]{Mancho2018}
\biblab{Mancho2018}
Manchon, L., Belkacem, K., Samadi, R., Sonoi, T., Marques, J. P. C.,
Ludwig, H.-G. \& Caffau,  E., 2018.
[Influence of metallicity on the near-surface effect on oscillation
frequencies].
{\it Astron. Astrophys.}, {\bf 620}, A107-(1--14).

\bibitem[Mao {\it et~al.\/}(2009)]{Mao2009}
\biblab{Mao2009}
Mao, D., Mussack, K. \& D\"appen, W., 2009.
[Dynamic screening in solar plasma].
{\it Astrophys. J.}, {\bf 701}, 1204--1208.
%{\tt [arXiv:0906.3406v1 [astro-ph]]}.

\bibitem[Marchenkov {\it et~al.\/}(2000)]{Marche2000}
\biblab{Marche2000}
Marchenkov, K., Roxburgh, I. \& Vorontsov, S., 2000.
[Non-linear inversion for the hydrostatic structure of the solar interior].
{\it Mon. Not. R. Astron. Soc.}, {\bf 312}, 39--50.

\bibitem[Margulis and Lovelock(1974)]{Margul1974}
\biblab{Margul1974}
Margulis, L. \& Lovelock, J. E., 1974.
[Biological modulation of the Earth's atmosphere].
{\it Icarus}, {\bf 21}, 471--489.

\bibitem[Mark(1947)]{Mark1947}
\biblab{Mark1947}
Mark, C., 1947.
[The neutron density near a plane surface].
{\it Phys. Rev.}, {\bf 72}, 558--564.

\bibitem[Marsch {\it et~al.\/}(1995)]{Marsch1995}
\biblab{Marsch1995}
Marsch, E., von Steiger, R. \& Bochsler, P., 1995.
[Element fractionation by diffusion in the solar chromosphere].
{\it Astron. Astrophys.}, {\bf 301}, 261--276.

\bibitem[Mathis and Zahn(2004)]{Mathis2004}
\biblab{Mathis2004}
Mathis, S. \& Zahn, J.-P., 2004.
[Transport and mixing in the radiation zones of rotating stars. I. 
Hydrodynamical processes].
{\it Astron. Astrophys.}, {\bf 425}, 229--242.

%\bibitem[Mathis {\it et~al.\/}(2007)]{Mathis2007}
%\biblab{Mathis2007}
%Mathis, S., Decressin, T., Palacios, A., Eggenberger, P., Siess, L.,
%Talon, S., Charbonnel, C., Turck-Chi\`eze, S. \& Zahn, J.-P., 2007.
%[Meridional circulation in the radiation zone of rotating stars: origins,
%behaviors and consequences on stellar evolution].
%{\it Astron. Nachr.}, {\bf 328}, 1062--1065.

\bibitem[Mathis {\it et~al.\/}(2006)]{Mathis2006}
\biblab{Mathis2006}
Mathis, S., Decressin, T., Palacios, A., Siess, L., Charbonnel, C.,
Turck-Chi\`eze, S. \& Zahn, J.-P., 2006.
[Dynamical processes in stellar radiation zones: secular magnetohydrodynamics
of rotating stars].
In {\it Proc. SOHO 18 / GONG 2006 / HELAS I Conf.
Beyond the spherical Sun},
ed. K. Fletcher, ESA SP-624, ESA Publications Division,
Noordwijk, The Netherlands.

\bibitem[Mathis {\it et~al.\/}(2018)]{Mathis2018}
\biblab{Mathis2018}
Mathis, S., Prat, V., Amard, L., Charbonnel, C., Palacios, A.,
Lagarde, N. \& Eggenberger, P., 2018.
[Anisotropic turbulent transport in stably stratified rotating stellar
radiation zones].
{\it Astron. Astrophys.}, {\bf 620}, A22-(1--16).
%{\tt [arXiv:1808.01814v1 [astro-ph.SR]]}

{\rv
\bibitem[Matt {\it et~al.\/}(2015)]{Matt2015}
\biblab{Matt2015}
Matt, S. P., Brun, A. S., Baraffe, I., Bouvier, J. \& Chabrier, G., 2015.
[The mass-dependence of angular momentum evolution in Sun-like stars].
{\it Astrophys. J.}, {\bf 799}, L23-(1--6).
(Erratum: {\it Astrophys. J.}, {\bf 870}, L27).
}

\bibitem[Mazumdar(2005)]{Mazumd2005}
\biblab{Mazumd2005}
Mazumdar, A., 2005.
[Asteroseismic diagrams for solar-type stars].
{\it Astron. Astrophys.}, {\bf 441}, 1079--1086.

\bibitem[Mazumdar {\it et~al.\/}(2006)]{Mazumd2006}
\biblab{Mazumd2006}
Mazumdar, A., Basu, S., Collier, B. L. \& Demarque, P., 2006.
[Asteroseismic diagnostics of stellar convective cores].
{\it Mon. Not. R. Astron. Soc.}, {\bf 372}, 949--958.

\bibitem[Mazumdar {\it et~al.\/}(2014)]{Mazumd2014}
\biblab{Mazumd2014}
Mazumdar, A., Monteiro, M. J. P. F. G., Ballot, J., Antia, H. M., Basu, S.,
Houdek, G., Mathur, S., Cunha, M. S., Silva Aguirre, V., Garc\'{\i}a, R. A.,
Salabert, D., Christensen-Dalsgaard, J., Metcalfe, T. S.,
Sanderfer, D. T., Seader, S. E., Smith, J. C. \& Chaplin, W. J., 2014.
[Measurement of acoustic glitches in solar-type stars from oscillation
frequencies observed by {\it Kepler}].
{\it Astrophys. J.}, {\bf 782}, 18-(1--17).
%{\tt [arXiv:1312.4907 [astro-ph]]}

\bibitem[McDonald(2004)]{McDona2004}
\biblab{McDona2004}
McDonald, A. B., 2004.
[Solar neutrinos].
{\it New Journal of Physics}, {\bf 6}-121, 1--17.

\bibitem[McDonald(2016)]{McDona2016}
\biblab{McDona2016}
McDonald, A. B., 2016.
[Nobel lecture: The Sudbury Neutrino Observatory: Observation of flavor
change for solar neutrinos].
{\it Rev. Mod. Phys.}, {\bf 88}, 030502-(1--9).

\bibitem[McKee and Ostriker(2007)]{McKee2007}
\biblab{McKee2007}
McKee, C. F. \& Ostriker, E. C., 2007.
[Theory of star formation].
{\it Annu. Rev. Astron. Astrophys.}, {\bf 45}, 565--687.

%\bibitem[Mel\'endez and Ram\'{\i}rez(2007)]{Melend2007}
%\biblab{Melend2007}
%Mel\'endez, J. \& Ram\'{\i}rez, I., 2007.
%[HIP 56948: a solar twin with a low lithium abundance].
%{\it Astrophys. J.}, {\bf 669}, L89--L92.

\bibitem[Mecheri {\it et~al.\/}(2004)]{Mecher2004}
\biblab{Mecher2004}
Mecheri, R., Abdelatif, T., Irbah, A., Provost, J. \& Berthomieu, G., 2004.
[New values of gravitational moments $J_2$ and $J_4$ deduced from
helioseismology].
{\it Solar Phys.}, {\bf 222}, 191--197.

\bibitem[Mel\'endez {\it et~al.\/}(2009)]{Melend2009}
\biblab{Melend2009}
Mel\'endez, J., Asplund, M., Gustafsson, B. \& Yong, 2009.
[The peculiar solar composition and its possible relation to planet formation].
{\it Astrophys. J.}, {\bf 704}, L66--L70.

{\rv
\bibitem[Metcalfe and Egeland(2019)]{Metcal2019}
\biblab{Metcal2019}
Metcalfe, T. S. \& Egeland, R., 2019.
[Understanding the limitations of gyrochronology for old field stars].
{\it Astrophys. J.}, {\bf 871}, 39-(1--6).
}

%\bibitem[Mel\'endez {\it et~al.\/}(2006)]{Melend2006}
%\biblab{Melend2006}
%Mel\'endez, J., Dodds-Eden, K. \& Robles, J. A., 2006.
%[HD 98618: a star closely resembling our Sun].
%{\it Astrophys. J.}, {\bf 641}, L133--L136.

\bibitem[Michaud(1970)]{Michau1970}
\biblab{Michau1970}
Michaud, G., 1970.
[Diffusion processes in peculiar A stars].
{\it Astrophys. J.}, {\bf 160}, 641--658.

\bibitem[Michaud and Proffitt(1993)]{Michau1993}
\biblab{Michau1993}
Michaud, G. \& Proffitt, C. R., 1993.
[Particle transport processes].
In {\it Proc. IAU Colloq. 137: Inside the stars},
eds Baglin, A. \& Weiss, W. W., 
Astronomical Society of the Pacific Conference Series, San Francisco,
{\bf 40}, 246--259.

\bibitem[Michaud {\it et~al.\/}(2011)]{Michau2011}
\biblab{Michau2011}
Michaud, G., Richer, J. \& Vick, M., 2011.
[Sirius A: turbulence or mass loss?].
{\it Astron. Astrophys.}, {\bf 534}, A18-(1--10).

\bibitem[Miesch(2005)]{Miesch2005}
\biblab{Miesch2005}
Miesch, M. S., 2005.
[Large-Scale Dynamics of the Convection Zone and Tachocline]. 
{\it Living Rev. Solar Phys.}, {\bf 2},  1. URL (cited on 30/5/05): 
\url{http://www.livingreviews.org/lrsp-2005-1}

\bibitem[Miesch and Toomre(2009)]{Miesch2009}
\biblab{Miesch2009}
Miesch, M. S. \& Toomre, J., 2009.
[Turbulence, magnetism, and shear in stellar interiors].
{\it Annu. Rev. Astron. Astrophys.}, {\bf 41}, 317--345.

\bibitem[Miesch {\it et~al.\/}(2006)]{Miesch2006}
\biblab{Miesch2006}
Miesch, M. S., Brun, A. S. \& Toomre, J., 2006.
[Solar differential rotation influenced by latitudinal entropy variations
in the tachocline].
{\it Astrophys. J.}, {\bf 641}, 618--625.

\bibitem[Mihalas(1970)]{Mihala1970}
\biblab{Mihala1970}
Mihalas, D., 1970.
{\it Stellar Atmospheres}, 1st ed.,
W. H. Freeman, San Francisco.

\bibitem[Mihalas(1978)]{Mihala1978}
\biblab{Mihala1978}
Mihalas, D., 1978.
{\it Stellar Atmospheres}, 2nd ed.,
W. H. Freeman, San Francisco.

\bibitem[Mihalas {\it et~al.\/}(1988)]{Mihala1988}
\biblab{Mihala1988}
Mihalas, D., D{\"a}ppen, W. \& Hummer, D. G., 1988.
[The equation of state for stellar envelopes. II. Algorithm and
selected results].
{\it Astrophys. J.}, {\bf 331}, 815--825.

\bibitem[Mihalas {\it et~al.\/}(1990)]{Mihala1990}
\biblab{Mihala1990}
Mihalas, D., Hummer, D. G., Mihalas, B. W. \& D{\"a}ppen, W., 1990.
[The equation of state for stellar envelopes. IV. Thermodynamic
quantities and selected ionization fractions for six elemental mixes].
{\it Astrophys. J.}, {\bf 350}, 300--308.

\bibitem[Mikheyev and Smirnov(1985)]{Mikhey1985}
\biblab{Mikhey1985}
Mikheyev, S. P. \& Smirnov, A. Yu., 1985.
[Resonance enhancement of oscillations in matter and solar neutrino
spectroscopy].
{\it Yad. Fiz.}, {\bf 42}, 1441--1448
(English translation: {\it Sov. J. Nucl. Phys.}, {\bf 42}, 913--917).

\bibitem[Miller Bertolami(2016)]{Miller2016}
\biblab{Miller2016}
Miller Bertolami, M. M., 2016. 
[New models for the evolution of post-asymptotic giant branch stars
and central stars of planetary nebulae].
{\it Astron. Astrophys.}, {\bf 588}, A25-(1--21).

\bibitem[Minton and Malhotra(2007)]{Minton2007}
\biblab{Minton2007}
Minton, D. A. \& Malhotra, R., 2007.
[Assessing the massive young Sun hypothesis to solve the warm young Earth 
puzzle].
{\it Astrophys. J.}, {\bf 660}, 1700--1706. % {\tt [astro-ph/0612321v2]}.

\bibitem[Moc\'ak {\it et~al.\/}(2008)]{Mocak2008}
\biblab{Mocak2008}
Moc\'ak, M., M\"uller, E., Weiss, A. \& Kifonidis, K., 2008.
[The core helium flash revisited. I. One and two-dimensional hydrodynamical
simulations].
{\it Astron. Astrophys.}, {\bf 490}, 265--277.

\bibitem[Moc\'ak {\it et~al.\/}(2009)]{Mocak2009}
\biblab{Mocak2009}
Moc\'ak, M., M\"uller, E., Weiss, A. \& Kifonidis, K., 2009.
[The core helium flash revisited. II. Two and three-dimensional hydrodynamic
simulations].
{\it Astron. Astrophys.}, {\bf 501}, 659--677.
%{\tt [arXiv:0811.4083v1 [astro-ph]]}.

%\bibitem[Mohr and Taylor(2005)]{Mohr2005}
%\biblab{Mohr2005}
%Mohr, P. J. \& Taylor, B. N., 2005.
%[CODATA recommended values of the fundamental physical constants: 2002].
%{\it Rev. Mod. Phys.}, {\bf 77}, 1--107.

\bibitem[Mohr {\it et~al.\/}(2015)]{Mohr2016}
\biblab{Mohr2016}
Mohr, P. J., Newell, D. B. \& Taylor, B. N., 2016.
[CODATA recommended values of the fundamental physical constants: 2014].
{\it J. Phys. Chem. Reference Data} {\bf 45}, 043102-(1--11).
\url{https://doi.org/10.1063/1.4954402},
{\tt [arXiv:1507.07956v1 [physics.atom-ph]]}.

\bibitem[Mojzsis {\it et~al.\/}(2001)]{Mojzsi2001}
\biblab{Mojzsi2001}
Mojzsis, S. J., Harrison, T. M. \& Pidgeon, R. T., 2001.
[Oxygen-isotope evidence from ancient zircons for liquid water at the
Earth's surface 4,300 Myr ago].
{\it Nature}, {\bf 409}, 178--181.

\bibitem[Mondet {\it et~al.\/}(2015)]{Mondet2015}
\biblab{Mondet2015}
Mondet, G., Blancard, C., Coss{\'e}, P. \& Faussurier, G., 2015.
[Opacity calculations for solar mixtures].
{\it Astrophys. J. Suppl.}, {\bf 220}, 2-(1--7).

\bibitem[Montalb\'an {\it et~al.\/}(2004)]{Montal2004}
\biblab{Montal2004}
Montalb\'an, J., Miglio, A., Noels, A., Grevesse, N. \&
Di Mauro, M. P., 2004.
[Solar model with CNO revised abundances].
In {\it Proc. SOHO 14 - GONG 2004: ``Helio- and Asteroseismology: Towards
a golden future''; Yale, July 12--16 2004}, ed. Danesy, D., ESA SP-559,
ESA Publication Division, Noordwijk, The Netherlands, p. 574--576.

\bibitem[Monteiro(2008)]{Montei2008}
\biblab{Montei2008}
Monteiro, M. J. P. F. G. (ed.), 2008.
{\it Evolution and Seismic Tools for Stellar Astrophysics}.
{\it Astrophys. Space Sci.}, {\bf 316}.

\bibitem[Monteiro {\it et~al.\/}(1994)]{Montei1994}
\biblab{Montei1994}
Monteiro, M. J. P. F. G., Christensen-Dalsgaard, J. \& Thompson, M. J., 1994.
[Seismic study of overshoot at the base of the solar convective envelope].
{\it Astron. Astrophys.}, {\bf 283}, 247--262.

\bibitem[Monteiro {\it et~al.\/}(1996)]{Montei1996}
\biblab{Montei1996}
Monteiro, M. J. P. F. G., Christensen-Dalsgaard, J. \& Thompson, M. J., 1996.
[Seismic properties of the Sun's superadiabatic layer.
I. Theoretical modelling and parametrization of the uncertainties].
{\it Astron. Astrophys.}, {\bf 307}, 624--634.

\bibitem[Monteiro {\it et~al.\/}(2000)]{Montei2000}
\biblab{Montei2000}
Monteiro, M. J. P. F. G., Christensen-Dalsgaard, J. \& Thompson, M. J., 2000.
[Seismic study of stellar convective regions: the base of convective
envelopes in low-mass stars].
{\it Mon. Not. R. Astron. Soc.}, {\bf 316}, 165--172.

\bibitem[Monteiro {\it et~al.\/}(2002)]{Montei2002}
\biblab{Montei2002}
Monteiro, M. J. P. F. G., Christensen-Dalsgaard, J. \&
Thompson, M. J., 2002.
[Asteroseismic Inference for Solar-Type Stars].
In {\it Proc. 1st Eddington Workshop, `Stellar Structure
and Habitable Planet Finding'},
eds F. Favata, I. W. Roxburgh and D. Galad\'{\i}-Enr\'{\i}quez,
ESA SP-485, ESA Publications Division, Noordwijk, The Netherlands,
p. 291--298.

\bibitem[Montmerle {\it et~al.\/}(2006)]{Montme2006}
\biblab{Montme2006}
Montmerle, T., Augereau, J.-C., Chaussidon, M., Gounelle, M., Marty, B. \&
Morbidelli, A., 2006.
[Solar system formation and early evolution: the first 100 million years].
{\it Earth, Moon and Planets}, {\bf 98}, 39--95.

\bibitem[Morel {\it et~al.\/}(1994)]{Morel1994}
\biblab{Morel1994}
Morel, P., van 't Veer, C., Provost, J., Berthomieu, G., Castelli, F.,
Cayrel, R., Goupil, M. J. \& Lebreton, Y., 1994.
[Incorporating the atmosphere in stellar structure models: the solar case].
{\it Astron. Astrophys.}, {\bf 286}, 91--102.

\bibitem[Morel {\it et~al.\/}(1999)]{Morel1999}
\biblab{Morel1999}
Morel, P., Pichon, B., Provost, J. \& Berthomieu, G., 1999.
[Solar models and NACRE thermonuclear reaction rates].
{\it Astron. Astrophys.}, {\bf 350}, 275--285.

\bibitem[Morel {\it et~al.\/}(2000)]{Morel2000}
\biblab{Morel2000}
Morel, P., Provost, J. \& Berthomieu, G., 2000.
[About the time of evolution of a solar model].
{\it Astron. Astrophys.}, {\bf 353}, 771--774.

\bibitem[Morel(2009)]{Morel2009}
\biblab{Morel2009}
Morel, T., 2009.
[Abundances of massive stars: some recent developments].
In {\it Proc. 38$^{\rm th}$ Li\`ege International Astrophysical Colloquium:
Evolution and Pulsation of Massive Stars on the Main Sequence and Close to it,
Li\`ege, July 7--11 2008},
{\it Comm. in Asteroseismology}, {\bf 158}, 122--130.
{\tt [arXiv:0811.4114v1 [astro-ph]]}

\bibitem[Morel and Butler(2008)]{Morel2008}
\biblab{Morel2008}
Morel, T. \& Butler, K., 2008.
[The neon content of nearby B-type stars and its implications for the 
solar model problem].
{\it Astron. Astrophys.}, {\bf 487}, 307--315.

\bibitem[Moskalik and Dziembowski(1992)]{Moskal1992b}
\biblab{Moskal1992b}
Moskalik, P. \& Dziembowski, W. A., 1992.
[New opacities and the origin of the $\beta$ Cephei pulsation].
{\it Astron. Astrophys.}, {\bf 256}, L5--L8.

\bibitem[Moskalik {\it et~al.\/}(1992)]{Moskal1992a}
\biblab{Moskal1992a}
Moskalik, P., Buchler, J. R. \& Marom, A., 1992.
[Toward a resolution of the bump and beat Cepheid mass discrepancies].
{\it Astrophys. J.}, {\bf 385}, 685--693.

%\bibitem[Mosser {\it et~al.\/}(2008)]{Mosser2008}
%\biblab{Mosser2008}
%Mosser, B., Appourchaux, T., Catala, C., Buey, J.-T. and the 
%SIAMOIS team, 2008.
%[SIAMOIS: Seismic Interferometer to Measure Oscillations in the
%Interior of Stars].
%In {\it Proc. HELAS II International Conference: Helioseismology,
%Asteroseismology and the MHD Connections, G\"ottingen, August 2007},
%eds L. Gizon \& M. Roth,
%{\it J. Phys.: Conf. Ser.}, in the press.

\bibitem[Mosumgaard {\it et~al.\/}(2017)]{Mosumg2017}
\biblab{Mosumg2017}
Mosumgaard, J. R., Silva Aguirre, V., Weiss, A., Christensen-Dalsgaard, J. \&
Trampedach, R., 2017.
[Improving 1D stellar models with 3D atmospheres].
In {\it Seismology of the Sun and the Distant stars 2016},
eds M. J. P. F. G. Monteiro, M. S. Cunha \& J. M. Ferreira,
{\it EPJ Web of Conferences}, {\bf 160}, 03009-(1--4).
%{\tt [arXiv:1610.07323v1 [astro-ph]]}

\bibitem[Mosumgaard {\it et~al.\/}(2018)]{Mosumg2018}
\biblab{Mosumg2018}
Mosumgaard, J. R., Ball, W.H., Silva Aguirre, V., Weiss, A. \&
Christensen-Dalsgaard, J., 2018.
[Stellar models with calibrated convection and temperature stratification
from 3D hydrodynamic simulations].
{\it Mon. Not. R. Astron. Soc.}, {\bf 478}, 5650--5659.

\bibitem[Mosumgaard {\it et~al.\/}(2020)]{Mosumg2020}
\biblab{Mosumg2020}
Mosumgaard, J. R., J{\o}rgensen, A. C. S., Weiss, A.,
Silva Aguirre, V. \& Christensen-Dalsgaard, J., 2020.
[Coupling 1D stellar evolution with 3D-hydrodynamical simulations
on-the-fly II: Stellar evolution and asteroseismic applications].
{\it Mon. Not. R. Astron. Soc.}, {\bf 491}, 1160--1173. % refereed

\bibitem[Murray and Chaboyer(2002)]{Murray2002}
\biblab{Murray2002}
Murray, N. \& Chaboyer, B., 2002.
[Are stars with planets polluted?].
{\it Astrophys. J.}, {\bf 566}, 442--451.

\bibitem[Mussack and  D{\"a}ppen(2011)]{Mussac2011}
\biblab{Mussac2011}
Mussack, K. \& D{\"a}ppen, W., 2011.
[Dynamic screening correction for solar $p{-}p$ reaction rates].
{\it Astrophys. J.}, {\bf 729}, 96-(1--6).

\bibitem[Mussack and Gough(2009)]{Mussac2009}
\biblab{Mussac2009}
Mussack, K. \& Gough, D., 2009.
[Measuring solar abundances with seismology].
In {\it Proc. GONG2008/SOHO21 meeting:
Solar-stellar Dynamos as revealed by Helio- and Asteroseismology},
eds M. Dikpati, T. Arentoft, I. Gonz\'alez Hern\'andez, C. Lindsey \& F. Hill,
{\it ASP Conf. Ser.}, {\bf 416}, {\rv ASP, San Francisco}, p. 203--207.
%{\tt [arXiv:0810.2722 [astro-ph.SR]]}

\bibitem[Mussack {\it et~al.\/}(2007)]{Mussac2007}
\biblab{Mussac2007}
Mussack, K., Mao, D. \& D\"appen, W., 2007.
[Evaluation of molecular-dynamic simulations for the study of hot dense
Coulomb systems].
In Stancliffe R. J., Dewi J., Houdek G., Martin R. G., Tout C.A., eds,
AIP Conf. Proc. vol. 948, {\it Unsolved Problems in Stellar Physics}.
American Institute of Physics, Melville, p. 207--211.

\bibitem[Nagayama {\it et~al.\/}(2019)]{Nagaya2019}
\biblab{Nagaya2019}
Nagayama, T., Bailey, J. E., Loisel, G. P., Dunham, G. S., Rochau, G. A.,
Blancard, C., Colgan, J., Coss{\'e}, P., Faussurier, G., Fontes, C. J.,
Gilleron, F., Hansen, S. B., Iglesias, C. A., Golovkin, I. E.,
Kilcrease, D. P., MacFarlane, J. J., Mancini, R. C., More, R. M.,
Orban, C., Pain, J.-C., Sherrill, M. E. \& Wilson, B. G., 2019.
[Systematic studies of $L$-shell opacity at stellar interior temperatures].
{\it Phys. Rev. Lett.}, {\bf 122}, 235001-(1--7).

\bibitem[Nakaya and  Plunkett(2016)]{Nakaya2016}
\biblab{Nakaya2016}
Nakaya, T. \& Plunkett, R. K., 2016.
[Neutrino oscillations with the MINOS, MINOS+, T2K, and NOvA experiments].
{\it New J. Phys.}, {\bf 18}, 15009-(1--28).

\bibitem[Nayfonov {\it et~al.\/}(1999)]{Nayfon1999}
\biblab{Nayfon1999}
Nayfonov, A., D{\"a}ppen, W., Hummer, D. G. \& Mihalas, D., 1999.
[The MHD equation of state with post-Holtsmark microfield distributions].
{\it Astrophys. J.}, {\bf 526}, 451--464.

\bibitem[Neuforge-Verheecke {\it et~al.\/}(2001a)]{Neufor2001a}
\biblab{Neufor2001a}
Neuforge-Verheecke, C., Goriely, S., Guzik, J. A., Swenson, F. J. \&
Bradley, P. A., 2001a.
[Seismological tests of solar models calculated with the NACRE reaction rates
and the Grevesse and Sauval 1998 mixture].
{\it Astrophys. J.}, {\bf 550}, 493--502.

\bibitem[Neuforge-Verheecke {\it et~al.\/}(2001b)]{Neufor2001b}
\biblab{Neufor2001b}
Neuforge-Verheecke, C., Guzik, J. A., Keady, J. J., Magee, N. H.,
Bradley, P. A. \& Noels, A., 2001b.
[Helioseismic tests of the new Los Alamos LEDCOP opacities].
{\it Astrophys. J.}, {\bf 561}, 450--454.

\bibitem[Newman and Fowler(1976)]{Newman1976}
\biblab{Newman1976}
Newman, M. J. \& Fowler, W. A., 1976.
[Solar models of low neutrino counting rate: energy transport by processes
other than radiative transfer].
{\it Astrophys. J.}, {\bf 207}, 601--604.

\bibitem[Nieva and  Przybilla(2012)]{Nieva2012}
\biblab{Nieva2012}
Nieva, M.-F. \& Przybilla, N., 2012.
[Present-day cosmic abundances. A comprehensive study of nearby early B-type 
stars and implications for stellar and Galactic evolution and interstellar
dust models].
{\it Astron. Astrophys.}, {\bf 539}, A143-(1--57).

\bibitem[Nissen and  Gustafsson(2018)]{Nissen2018}
\biblab{Nissen2018}
Nissen, P. E. \& Gustafsson, B., 2018.
[High-precision stellar abundances of elements: methods and applications].
{\it Astron. Astrophys. Rev.}, {\bf 26}, 6-(1--70).

\bibitem[Noerdlinger(1977)]{Noerdl1977}
\biblab{Noerdl1977}
Noerdlinger, P. D., 1977.
[Diffusion of helium in the Sun].
{\it Astron. Astrophys.}, {\bf 57}, 407--415.

\bibitem[Nordlund(2010)]{Nordlu2010}
\biblab{Nordlu2010}
Nordlund, {\AA}, 2010.
[Solar twins and possible solutions of the solar and Jupiter abundance
problems].
Submitted to {\it Astrophys. J. Lett.}.
{\tt [arXiv:0908.3479v2 [astro-ph]]}.

\bibitem[Nordlund and Stein(2009)]{Nordlu2009b}
\biblab{Nordlu2009b}
Nordlund, {\AA}. \& Stein, R. F., 2009.
[Accurate radiation hydrodynamics and MHD modeling of 3-D stellar 
atmospheres].
In {\it Recent Directions in Astrophysical Quantitative Spectroscopy and
Radiation Hydrodynamics}, 
eds I. Hubeny, J. M. Stone, K. MacGregor \& K. Werner,
AIP Conf. Proc. vol. 1171, AIP, Melville, New York, p. 242--259.

\bibitem[Nordlund {\it et~al.\/}(2009)]{Nordlu2009a}
\biblab{Nordlu2009a}
Nordlund, {\AA}., Stein, R. F. \& Asplund, M., 2009.
[Solar surface convection].
{\it Living Rev. Solar Phys.}, {\bf 6},  2. URL (cited on 5/5/09): 
\url{http://www.livingreviews.org/lrsp-2009-2}

\bibitem[Ot\'{\i} Floranes {\it et~al.\/}(2005)]{Oti2005}
\biblab{Oti2005}
Ot\'{\i} Floranes, H., Christensen-Dalsgaard, J. \& Thompson, M. J., 2005.
[The use of frequency-separation ratios for asteroseismology].
{\it Mon. Not. R. Astron. Soc.}, {\bf 356}, 671--679.

\bibitem[Padoan {\it et~al.\/}(2016)]{Padoan2016}
\biblab{Padoan2016}
Padoan, P., Pan, L., Haugb{\o}lle, T. \& Nordlund, {\AA}., 2016.
[Supernova driving. I. The origin of molecular cloud turbulence].
{\it Astrophys. J.}, {\bf 822}, 11-(1--28).

\bibitem[Pain {\it et~al.\/}(2017)]{Pain2017}
\biblab{Pain2017}
Pain, J.-C., Gilleron, F. \& Comet, M., 2017.
[Detailed opacity calculations for astrophysical applications].
{\it Atoms}, {\bf 5}, 22-(1--29).

\bibitem[Palacios {\it et~al.\/}(2006)]{Palaci2006}
\biblab{Palaci2006}
Palacios, A., Talon, S., Turck-Chi\`eze, S. \& Charbonnel, C., 2006.
[Dynamical processes in the solar radiative interior].
In {\it Proc. SOHO 18 / GONG 2006 / HELAS I Conf.
Beyond the spherical Sun},
ed. K. Fletcher, ESA SP-624, ESA Publications Division,
Noordwijk, The Netherlands.

\bibitem[Palla and  Stahler(1993)]{Palla1993}
\biblab{Palla1993}
Palla, F. \& Stahler, S. W., 1993.
[The pre-main-sequence evolution of intermediate-mass stars].
{\it Astrophys. J.}, {\bf 418}, 414--425.

\bibitem[Pandola(2004)]{Pandol2004}
\biblab{Pandol2004}
Pandola, L., 2004.
[Search for time modulations in the Gallex/GNO solar neutrino data].
{\it Astroparticle Phys.}, {\bf 22}, 219--226.

\bibitem[Patern{\`o}(1981)]{Patern1981}
\biblab{Patern1981}
Patern{\`o}, L., 1981.
[Solar oscillations as evidence for neutrino mass].
{\it Mem. Soc. Astron. Ital.}, {\bf 52}, 471--473.

\bibitem[Paxton {\it et~al.\/}(2011)]{Paxton2011}
\biblab{Paxton2011}
Paxton, B., Bildsten, L., Dotter, A., Herwig, F., Lesaffre, P. \&
Timmes, F., 2011.
[Modules for Experiments in Stellar Astrophysics (MESA)].
{\it Astrophys. J. Suppl.}, {\bf 192}, 3-(1--35).

\bibitem[Peebles(1966)]{Peeble1966}
\biblab{Peeble1966}
Peebles, P. J. E., 1966.
[Primordial helium abundance and the primordial fireball. II].
{\it Astrophys. J.}, {\bf 146}, 542--552.

\bibitem[Pereira {\it et~al.\/}(2013)]{Pereir2013}
\biblab{Pereir2013}
Pereira, T. M. D., Asplund, M., Collet, R., Thaler, I., Trampedach, R. \&
Leenaarts, J., 2013.
[How realistic are solar model atmospheres?].
{\it Astron. Astrophys.}, {\bf 554}, A118-(1--16).

\bibitem[P{\'e}rez Hern{\'a}ndez and Christensen-Dalsgaard(1994)]{Perez1994}
\biblab{Perez1994}
P{\'e}rez Hern{\'a}ndez, F. \& Christensen-Dalsgaard, J., 1994.
[The phase function for stellar acoustic oscillations--III.
The solar case].
{\it Mon. Not. R. Astron. Soc.}, {\bf 269}, 475--492.

\bibitem[P{\'e}rez Hern{\'a}ndez and Christensen-Dalsgaard(1998)]{Perez1998}
\biblab{Perez1998}
P{\'e}rez Hern{\'a}ndez, F. \& Christensen-Dalsgaard, J., 1998.
[The phase function for stellar acoustic oscillations - IV. Solar-like stars].
{\it Mon. Not. R. Astron. Soc.}, {\bf 295}, 344--352.

\bibitem[Perry {\it et~al.\/}(2020)]{Perry2020}
\biblab{Perry2020}
Perry, T. S., Heeter, R. F., Opachich, Y. P., Johns, H. M., King, J. A.,
Dodd, E. S., DeVolder, B. G., Sherrill, M. E., Wilson, B. G.,
Iglesias, C. A., Kline, J. L., Flippo, K. A.,
Cardenas, T., Schneider, M. B., Liedahl, D. A., Urbatsch, T. J.,
Douglas, M. R., Bailey, J. E. \& Rochau, G. A., 2020.
[Progress toward NIF opacity measurements].
{\it High Energy Density Physics}, {\bf 35}, 100728-(1--3).

\bibitem[Petersen(1973)]{Peters1973}
\biblab{Peters1973}
Petersen, J. O., 1973.
[Masses of double mode Cepheid variables determined by analysis of
period ratios].
{\it Astron. Astrophys.}, {\bf 27}, 89--93.

{\rv
\bibitem[Piau and Turck-Chi{\`e}ze(2002)]{Piau2002}
\biblab{Piau2002}
Piau, L. \& Turck-Chi{\`e}ze, S., 2002.
[Lithium depletion in pre-main-sequence solar-like stars].
{\it Astrophys.J.}, {\bf 566}, 419--434.
}

\bibitem[Pijpers(1998)]{Pijper1998}
\biblab{Pijper1998}
Pijpers, F. P., 1998.
[Helioseismic determination of the solar gravitational quadrupole moment].
{\it Mon. Not. R. Astron. Soc.}, {\bf 297}, L76--L80.

\bibitem[Pinsonneault and  Delahaye(2009)]{Pinson2009}
\biblab{Pinson2009}
Pinsonneault, M. \& Delahaye, F., 2009.
[The solar heavy element abundances. II. Constraints from stellar atmospheres].
{\it Astrophys. J.}, {\bf 704}, 1174--1188.

\bibitem[Pinsonneault {\it et~al.\/}(1989)]{Pinson1989}
\biblab{Pinson1989}
Pinsonneault, M. H., Kawaler, S. D., Sofia, S. \& Demarque, P., 1989.
[Evolutionary models of the rotating Sun].
{\it Astrophys. J.}, {\bf 338}, 424--452.

\bibitem[Pipin and  Kosovichev(2019)]{Pipin2019}
\biblab{Pipin2019}
Pipin, V. V. \& Kosovichev, A. G., 2019.
[On the origin of solar torsional oscillations and extended solar cycle].
{\it Astrophys. J.}, {\bf 887}, 215-(1--16).

\bibitem[Plumb and McEwan(1978)]{Plumb1978}
\biblab{Plumb1978}
Plumb, R. A. \& McEwan, A. D., 1978.
[The instability of a forced standing wave in a viscous stratified fluid:
a laboratory analogue of the quasi-biennial oscillation].
{\it J. Atmos. Sci.}, {\bf 35}, 1827--1839.

\bibitem[Pontecorvo(1967)]{Pontec1967}
\biblab{Pontec1967}
Pontecorvo, B., 1967.
[Neutrino experiments and the problem of conservation of leptonic charge].
{\it Zh. Eksp. Teor. Fiz.}, {\bf 53}, 1717--1725
(English translation: {\it Sov. Phys. JETP}, {\bf 26}, 984--988; 1968).

\bibitem[Pradhan and Nahar(2018)]{Pradha2018}
\biblab{Pradha2018}
Pradhan, A. K. \& Nahar, S. N., 2018.
[Recalculation of astrophysical opacities: overview, methodology and
atomic calculations].
In {\it Workshop on Astrophysical Opacities}, eds Mendoza, C., 
Turck-Chi{\`e}ze, S. \& Colgan, J., {\it ASP Conf. Ser.}, {\bf 515}, 
ASP, San Francisco, p. 79--88.

\bibitem[Proffitt(1994)]{Proffi1994}
\biblab{Proffi1994}
Proffitt, C. R., 1994.
[Effects of heavy-element settling on solar neutrino fluxes and interior
structure].
{\it Astrophys. J.}, {\bf 425}, 849--855.

\bibitem[Proffitt and Michaud(1991)]{Proffi1991}
\biblab{Proffi1991}
Proffitt, C. R. \& Michaud, G., 1991.
[Gravitational settling in solar models].
{\it Astrophys. J.}, {\bf 380}, 238--250.

\bibitem[Pr{\v s}a {\it et~al.\/}(2016)]{Prsa2016}
\biblab{Prsa2016}
Pr{\v s}a, A., Harmanec, P., Torres, G., Mamajek, E., Asplund, M.,
Capitaine, N., Christensen-Dalsgaard, J., Depagne, \'E., Haberreiter, M.,
Hekker, S., Hilton, J., Kopp, G., Kostov, V., Kurtz, D. W., Laskar, J.,
Mason, B. D., Milone, E. F., Montgomery, M., Richards, M., Schmutz, W.,
Schou, J. \& Stewart, S. G., 2016.
[Nominal values for selected solar and planetary quantities:
IAU 2015 Resolution B3].
{\it Astron. J.}, {\bf 152}, 41-(1--7).

\bibitem[Rabello-Soares {\it et~al.\/}(1999)]{Rabell1999}
\biblab{Rabell1999}
Rabello-Soares, M. C., Basu, S. \& Christensen-Dalsgaard, J., 1999.
[On the choice of parameters in solar structure inversion].
{\it Mon. Not. R. Astron. Soc.}, {\bf 309}, 35--47.

\bibitem[Rabello-Soares {\it et~al.\/}(2000)]{Rabell2000}
\biblab{Rabell2000}
Rabello-Soares, M. C., Basu, S., Christensen-Dalsgaard, J. \&
Di Mauro, M. P., 2000.
[The potential of solar high-degree modes for structure inversion].
%In {\it Proc. Ninth SOHO Workshop: Helioseismic Diagnostics
%of Solar Convection and Activity},
{\it Solar Phys.}, {\bf 193}, 345--356.

\bibitem[Rabello-Soares {\it et~al.\/}(2008)]{Rabell2008}
\biblab{Rabell2008}
Rabello-Soares, M. C., Korzennik, S. G. \& Schou, J., 2008.
[Analysis of MDI high-degree mode frequencies and their rotational splittings].
{\it Solar Phys.}, {\bf 251}, 197--224.

\bibitem[Ram\'{\i}rez {\it et~al.\/}(2009)]{Ramire2009}
\biblab{Ramire2009}
Ram\'{\i}rez, I., Mel\'endez, J. \& Asplund, M., 2009.
[Accurate abundance patterns of solar twins and analogs. Does the anomalous
solar chemical composition come from planet formation?].
{\it Astron. Astrophys.}, {\bf 508}, L17--L20.

\bibitem[Rast(2020)]{Rast2020}
\biblab{Rast2020}
Rast, M. P., 2020.
[Deciphering solar convection].
In {\it Dynamics of the Sun and Stars--Honoring the Life and Work of
Michael J. Thompson},
eds Monteiro, M. J. P. F. G., Garc\'{\i}a, R., Christensen-Dalsgaard, J. \&
McIntosh, S. W., Astrophysics and Space Science Proceedings, vol 57.
Springer, Cham, pp. 149 -- 161.

\bibitem[Rauer {\it et~al.\/}(2014)]{Rauer2014}
\biblab{Rauer2014}
Rauer, H., Catala, C., Aerts, C., Appourchaux, T., Benz, W.,
Brandeker, A., Christensen-Dalsgaard, J., Deleuil, M., Gizon, L.,
et al., 2014.
[The PLATO 2.0 mission].
{\it Exp. Astron.}, {\bf 38}, 249--330.
%{\tt [arXiv:1310.0696 [astro-ph.EP]]}

\bibitem[Reinhold {\it et~al.\/}(2020)]{Reinho2020}
\biblab{Reinho2020}
Reinhold, T., Shapiro, A. I., Solanki, S. K., Montet, B. T., Krivova, N. A.,
Cameron, R. H. \& Amazo-G{\'o}mez, E. M., 2020.
[The Sun is less active than other solar-like stars].
{\it Science}, {\bf 368}, 518--521.

\bibitem[Reipurth and Bally(2001)]{Reipur2001}
\biblab{Reipur2001}
Reipurth, B. \& Bally, J., 2001.
[Herbig-Haro flows: probes of early stellar evolution].
{\it Annu. Rev. Astron. Astrophys.}, {\bf 39}, 403--455.

\bibitem[Reiter {\it et~al.\/}(1995)]{Reiter1995}
\biblab{Reiter1995}
Reiter, J., Walsh, L. \& Weiss, A., 1995.
[Solar models: a comparative study of two stellar evolution codes].
{\it Mon. Not. R. Astron. Soc.}, {\bf 274}, 899--908.

\bibitem[Reiter {\it et~al.\/}(2015)]{Reiter2015}
\biblab{Reiter2015}
Reiter, J., Rhodes, E. J., Kosovichev, A. G., Schou, J., Scherrer, P. H. \&
Larson, T. P., 2015.
[A method for the estimation of $p$-mode parameters from averaged 
solar oscillation power spectra].
{\it Astrophys. J.}, {\bf 803}, 92-(1--42).

\bibitem[Reiter {\it et~al.\/}(2020)]{Reiter2020}
\biblab{Reiter2020}
Reiter, J., Rhodes, E. J., Kosovichev, A. G., Scherrer, P. H.,
Larson, T. P. \& Pinkerton, S. F., 2020.
[A method for the estimation of f- and p-mode parameters and rotational 
splitting coefficients from un-averaged solar oscillation power spectra].
{\it Astrophys. J.}, {\bf 894}, 80-(1--41).

\bibitem[Rempel(2004)]{Rempel2004}
\biblab{Rempel2004}
Rempel, M., 2004.
[Overshoot at the base of the solar convection zone: a semianalytical 
approach].
{\it Astrophys. J.}, {\bf 607}, 1046--1064.

\bibitem[Rempel(2007)]{Rempel2007}
\biblab{Rempel2007}
Rempel, M., 2007.
[Origin of solar torsional oscillations].
{\it Astrophys. J.}, {\bf 655}, 651--659.

\bibitem[Rempel(2012)]{Rempel2012}
\biblab{Rempel2012}
Rempel, M., 2012.
[High-latitude solar torsional oscillations during phases of changing
magnetic cycle amplitude].
{\it Astrophys. J.}, {\bf 750}, L8-(1--4).

\bibitem[Rhodes {\it et~al.\/}(1997)]{Rhodes1997}
\biblab{Rhodes1997}
Rhodes, E. J., Kosovichev, A. G., Schou, J., Scherrer, P. H. \&
Reiter, J., 1997.
[Measurements of frequencies of solar oscillations from the
MDI medium-$l$ program].
{\it Solar Phys.}, {\bf 175}, 287--310.

\bibitem[Richard {\it et~al.\/}(1996)]{Richar1996}
\biblab{Richar1996}
Richard, O., Vauclair, S., Charbonnel, C. \& Dziembowski, W. A., 1996.
[New solar models including helioseismological constraints
and light-element depletion].
{\it Astron. Astrophys.}, {\bf 312}, 1000--1011.

\bibitem[Richard {\it et~al.\/}(1998)]{Richar1998}
\biblab{Richar1998}
Richard, O., Dziembowski, W. A., Sienkiewicz, R. \& Goode, P. R., 1998.
[On the accuracy of helioseismic determination of solar helium abundance].
{\it Astron. Astrophys.}, {\bf 338}, 756--760.

\bibitem[Richard {\it et~al.\/}(2001)]{Richar2001}
\biblab{Richar2001}
Richard, O., Michaud, G. \& Richer, J., 2001.
[Iron convection zones in B, A and F stars].
{\it Astrophys. J.}, {\bf 558}, 377--391.

\bibitem[Richer {\it et~al.\/}(1998)]{Richer1998}
\biblab{Richer1998}
Richer, J., Michaud, G., Rogers, F., Iglesias, C., Turcotte, S. \&
LeBlanc, F., 1998.
[Radiative accelerations for evolutionary model calculations].
{\it Astrophys. J.}, {\bf 492}, 833--842.

\bibitem[Richer {\it et~al.\/}(2000)]{Richer2000}
\biblab{Richer2000}
Richer, J., Michaud, G. and Turcotte, S., 2000.
[The evolution of AmFm stars, abundance anomalies, and turbulent transport].
{\it Astrophys. J.}, {\bf 529}, 338--356.

\bibitem[Richtmyer(1957)]{Richtm1957}
\biblab{Richtm1957}
Richtmyer, R. D., 1957.
{\it Difference methods for initial-value problems}.
Interscience Publishers, New York.

\bibitem[Ricker {\it et~al.\/}(2014)]{Ricker2014}
\biblab{Ricker2014}
Ricker, G. R., Winn, J. N., Vanderspek, R., Latham, D. W., Bakos, G. {\`A}.,
Bean, J. L., Berta-Thompson, Z. K., Brown, T. M., Buchhave, L., 
Butler, N. R., Butler, R. P., Chaplin, W. J., Charbonneau, D.,
Christensen-Dalsgaard, J., Clampin, M., Deming, D., Doty, J., Lee, N. D.,
Dressing, C., Dunham, E. W., Endl, M., Fressin, F., Ge, J., Henning, T.,
Holman, M. J., Howard, A. W., Ida, S., Jenkins, J., Jernigan, G.,
Johnson, J. A., Kaltenegger, L., Kawai, N., Kjeldsen, H., Laughlin, G.,
Levine, A. M., Lin, D., Lissauer, J. J., MacQueen, P., Marcy, G.,
McCullough, P. R., Morton, T. D., Narita, N., Paegert, M., Palle, E.,
Pepe, F., Pepper, J., Quirrenbach, A., Rinehart, S. A., Sasselov, D.,
Sato, B., Seager, S., Sozzetti, A., Stassun, K. G., Sullivan, P.,
Szentgyorgyi, A., Torres, G., Udry, S. \& Villasenor, J., 2014.
[The Transiting Exoplanet Survey Satellite]. % TESS
Proc. SPIE, Astronomical Telescopes + Instrumentation, {\bf 9143},
914320-(1--15).
{\tt [arXiv:1406.0151v1 [astro-ph]]}

\bibitem[Rieutord {\it et~al.\/}(2016)]{Rieuto2016}
\biblab{Rieuto2016}
Rieutord, M., Espinosa Lara, F. \& Pitigny, B., 2016.
[An algorithm for computing the 2D structure of fast rotating stars].
{\it J. Comp. Phys.}, {\bf 318}, 277--304.

{\rv
\bibitem[Rincon and Rieutord(2018)]{Rincon2018}
\biblab{Rincon2018}
Rincon, F. \& Rieutord, M., 2018.
[The Sun's supergranulation].
{\it Living Rev. Solar Phys.}, {\bf 15},  6. URL (cited on 9/10/20): 
\url{https://doi.org/10.1007/s41116-018-0013-5}
}

\bibitem[Robinson {\it et~al.\/}(2003)]{Robins2003}
\biblab{Robins2003}
Robinson, F. J., Demarque, P., Li, L. H., Kim, Y.-C., Chan, K. L. \&
Guenther, D. B., 2003.
[Three-dimensional convection simulations of the outer layers of the Sun
using realistic physics].
{\it Mon. Not. R. Astron. Soc.}, {\bf 340}, 923--936.

\bibitem[Robles {\it et~al.\/}(2008)]{Robles2008}
\biblab{Robles2008}
Robles, J. A., Lineweaver, C. H., Grether, D., Flynn, C., Egan, C. A.,
Pracy, M. B., Holmberg, J. \& Gardner, E., 2008.
[A comprehensive comparison of the Sun to other stars: search for
self-selection effects].
{\it Astrophys. J.}, {\bf 684}, 691--706.
(Erratum: {\it Astrophys. J.}, {\bf 689}, 1457).

\bibitem[Robrade {\it et~al.\/}(2008)]{Robrad2008}
\biblab{Robrad2008}
Robrade, J., Schmitt, J. H. M. M. \& Favata, F., 2008.
[Neon and oxygen in low activity stars: towards a coronal unification with
the Sun].
{\it Astron. Astrophys.}, {\bf 486}, 995--1002.

\bibitem[Rogers(1981)]{Rogers1981}
\biblab{Rogers1981}
Rogers, F., 1981.
[Equation of state of dense, partially degenerate, reacting plasmas].
{\it Phys. Rev.}, {\bf A24}, 1531--1543.

\bibitem[Rogers and Iglesias(1992)]{Rogers1992}
\biblab{Rogers1992}
Rogers, F. J. \& Iglesias, C. A., 1992.
[Radiative atomic Rosseland mean opacity tables].
{\it Astrophys. J. Suppl.}, {\bf 79}, 507--568.

\bibitem[Rogers and Iglesias(1994)]{Rogers1994}
\biblab{Rogers1994}
Rogers, F. J. \& Iglesias, C. A., 1994.
[Astrophysical opacities].
{\it Science}, {\bf 263}, 50--55.

\bibitem[Rogers and Nayfonov(2002)]{Rogers2002}
\biblab{Rogers2002}
Rogers, F. J. \& Nayfonov, A., 2002.
[Updated and expanded OPAL equation-of-state tables: implications 
for helioseismology].
{\it Astrophys. J.}, {\bf 576}, 1064--1074.

\bibitem[Rogers {\it et~al.\/}(1996)]{Rogers1996}
\biblab{Rogers1996}
Rogers, F. J., Swenson, F. J. \& Iglesias, C. A., 1996.
[OPAL Equation-of-State Tables for Astrophysical Applications].
{\it Astrophys. J.}, {\bf 456}, 902--908.

\bibitem[Rogers(2007)]{Rogers2007}
\biblab{Rogers2007}
Rogers, T. M., 2007.
[Numerical simulations of gravity wave driven shear flows in the solar
tachocline].
In Stancliffe R.J., Dewi J., Houdek G., Martin R.G., Tout C.A., eds,
AIP Conf. Proc. vol. 948, {\it Unsolved Problems in Stellar Physics}.
American Institute of Physics, Melville, p. 65--72.

\bibitem[Rogers and Glatzmaier(2006)]{Rogers2006b}
\biblab{Rogers2006b}
Rogers, T. M. \& Glatzmaier, G. A., 2006.
[Angular momentum transport by gravity waves in the solar interior].
{\it Astrophys. J.}, {\bf 653}, 756--764.

\bibitem[Rogers {\it et~al.\/}(2006)]{Rogers2006a}
\biblab{Rogers2006a}
Rogers, T. M., Glatzmaier, G. A. \& Jones, C. A., 2006.
[Numerical simulation of penetration and overshoot in the Sun].
{\it Astrophys. J.}, {\bf 653}, 765--773.

\bibitem[Rogers {\it et~al.\/}(2008)]{Rogers2008}
\biblab{Rogers2008}
Rogers, T. M., MacGregor, K. B. \& Glatzmaier, G. A., 2008.
[Gravity wave driven flows in the solar radiative interior].
{\it Mon. Not. R. Astron. Soc.}, {\bf 387}, 616--630.

\bibitem[Rosenthal {\it et~al.\/}(1995)]{Rosent1995}
\biblab{Rosent1995}
Rosenthal, C. S., Christensen-Dalsgaard, J., Houdek, G., Monteiro, M.J.P.F.G.,
Nordlund, {\AA}. \& Trampedach, R., 1995.
[Seismology of the solar surface regions].
In: {\it Proc. Fourth SOHO Workshop: Helioseismology},
eds Hoeksema, J. T., Domingo, V., Fleck, B. \& Battrick, B., 
ESA SP-376, vol. 2, ESTEC, Noordwijk, p. 459--464.

\bibitem[Rosenthal {\it et~al.\/}(1999)]{Rosent1999}
\biblab{Rosent1999}
Rosenthal, C. S., Christensen-Dalsgaard, J., Nordlund, {\AA}.,
Stein, R. F. \& Trampedach, R., 1999.
[Convective contributions to the frequencies of solar oscillations].
{\it Astron. Astrophys.}, {\bf 351}, 689--700.

\bibitem[Rosing and Frei(2004)]{Rosing2004}
\biblab{Rosing2004}
Rosing, M. T. \& Frei, R., 2004.
[U-rich Archaean sea-floor sediments from Greenland--indications of
$> 3700$ Ma oxygenic photosynthesis].
{\it Earth Planet. Sci. Lett.}, {\bf 217}, 237--244.

\bibitem[Rosing {\it et~al.\/}(2010)]{Rosing2010}
\biblab{Rosing2010}
Rosing, M. T., Bird, D. K., Sleep, N. H. \& Bjerrum, C. J., 2010.
[No climate paradox under the faint early Sun].
{\it Nature}, {\bf 464}, 744--747.

\bibitem[Rosseland(1924)]{Rossel1924}
\biblab{Rossel1924}
Rosseland, S., 1924.
[Note on the absorption of radiation within a star].
{\it Mon. Not. R. astr. Soc.}, {\bf 84}, 525--528.

\bibitem[Roxburgh(2001)]{Roxbur2001}
\biblab{Roxbur2001}
Roxburgh, I. W., 2001.
[Gravitational multipole moments of the Sun determined from helioseismic
estimates of the internal structure and rotation].
{\it Astron. Astrophys.}, {\bf 377}, 688--690.

\bibitem[Roxburgh and Vorontsov(1994)]{Roxbur1994}
\biblab{Roxbur1994}
Roxburgh, I. W. \& Vorontsov, S. V., 1994.
[Seismology of the solar envelope: the base of the convection zone
as seen in the phase shift of acoustic waves].
{\it Mon. Not. R. Astron. Soc.}, {\bf 268}, 880--888.

\bibitem[Roxburgh and  Vorontsov(2003)]{Roxbur2003}
\biblab{Roxbur2003}
Roxburgh, I. W. \& Vorontsov, S. V., 2003.
[The ratio of small to large separations of acoustic oscillations as 
a diagnostic of the interior of solar-like stars].
{\it Astron. Astrophys.}, {\bf 411}, 215--220.

\bibitem[Rudkj{\o}bing(1942)]{Rudkjo1942}
\biblab{Rudkjo1942}
Rudkj{\o}bing, M., 1942.
[{\"U}ber Konvektion in Sternatmosf{\"a}ren].
{\it Z. Astrophys.}, {\bf 21}, 254--268.
\bibitem[Russell(1929)]{Russel1929}
\biblab{Russel1929}
Russell, H. N., 1929.
[On the composition of the Sun's atmosphere].
{\it Astrophys. J.}, {\bf 70}, 11--82.

\bibitem[Rybicki and Denis(2001)]{Rybick2001}
\biblab{Rybick2001}
Rybicki, K. R. \& Denis, C., 2001.
[On the final destiny of the Earth and the Solar System].
{\it Icarus}, {\bf 151}, 130--137.

\bibitem[Sackmann {\it et~al.\/}(1993)]{Sackma1993}
\biblab{Sackma1993}
Sackmann, I.-Juliana, Boothroyd, A. I. \& Kraemer, K. E., 1993.
[Our Sun. III. Present and future].
{\it Astrophys. J.}, {\bf 418}, 457--468.

\bibitem[Sackmann and Boothroyd(2003)]{Sackma2003}
\biblab{Sackma2003}
Sackmann, I.-Juliana \& Boothroyd, A. I., 2003.
[Our Sun. V. A bright young Sun consistent with helioseismology and
warm temperatures on ancient Earth and Mars].
{\it Astrophys. J.}, {\bf 583}, 1024--1039.

\bibitem[Sagan and Mullen(1972)]{Sagan1972}
\biblab{Sagan1972}
Sagan, C. \& Mullen, G., 1972.
[Earth and Mars: evolution of atmospheres and surface temperatures].
{\it Science}, {\bf 177}, 52--56.

\bibitem[Salaris and  Cassisi(2015)]{Salari2015}
\biblab{Salari2015}
Salaris, M. \& Cassisi, S., 2015.
[Stellar models with mixing length and $T(\tau)$ relations calibrated
on 3D convection simulations].
{\it Astron. Astrophys.}, {\bf 577}, A60-(1--6).

\bibitem[Salaris {\it et~al.\/}(2002)]{Salari2002}
\biblab{Salari2002}
Salaris, M., Cassisi, S. \& Weiss, A., 2002.
[Red giant branch stars: the theoretical framework].
{\it Publ. Astron. Soc. Pacific}, {\bf 114}, 375--402.

{\rv
\bibitem[Salaris {\it et~al.\/}(2018)]{Salari2018}
\biblab{Salari2018}
Salaris, M., Cassisi, S., Schiavon, R. P. \& Pietrinferni, A., 2018.
[Effective temperatures of red giants in the APOKASC catalogue
and the mixing length calibration in stellar models].
{\it Astron. Astrophys.}, {\bf 612}, A68-(1--8).
}

\bibitem[Salpeter(1954)]{Salpet1954}
\biblab{Salpet1954}
Salpeter, E. E., 1954.
[Electron screening and thermonuclear reactions].
{\it Austr. J. Phys.}, {\bf 7}, 373--388.

\bibitem[Schatzman(1969)]{Schatz1969}
\biblab{Schatz1969}
Schatzman, E., 1969.
[Turbulent transport, solar lithium and solar neutrinos].
{\it Astrophys. Lett.}, {\bf 3}, 139--140.

\bibitem[Schatzman(1993)]{Schatz1993}
\biblab{Schatz1993}
Schatzman, E., 1993.
[Transport of angular momentum and diffusion by the action of
internal waves].
{\it Astron. Astrophys.}, {\bf 279}, 431--446.

\bibitem[Schatzman(1996)]{Schatz1996}
\biblab{Schatz1996}
Schatzman, E., 1996.
[Diffusion process produced by random internal waves].
{\it J. Fluid Mech.}, {\bf 322}, 355--382.

\bibitem[Schatzman {\it et~al.\/}(1981)]{Schatz1981}
\biblab{Schatz1981}
Schatzman, E., Maeder, A., Angrand, F. \& Glowinski, R., 1981.
[Stellar evolution with turbulent diffusion mixing. III. The solar
model and the neutrino problem].
{\it Astron. Astrophys.}, {\bf 96}, 1--16.

\bibitem[Scherrer and  Gough(2019)]{Scherr2019}
\biblab{Scherr2019}
Scherrer, P. H. \& Gough, D. O., 2019.
[A critical evaluation of recent claims concerning solar rotation].
{\it Astrophys. J.}, {\bf 877}, 42-(1--12).

\bibitem[Scherrer {\it et~al.\/}(1995)]{Scherr1995}
\biblab{Scherr1995}
Scherrer, P. H., Bogart, R. S., Bush, R. I., Hoeksema, J. T., 
Kosovichev, A. G., Schou, J., Rosenberg, W., Springer, L., Tarbell, T. D.,
Title, A., Wolfson, C. J., Zayer, I., and the MDI engineering team, 1995.
[The Solar Oscillation Investigation--Michelson Doppler Imager].
{\it Solar Phys.}, {\bf 162}, 129--188.

\bibitem[Schlattl {\it et~al.\/}(1997)]{Schlat1997}
\biblab{Schlat1997}
Schlattl, H., Weiss, A. \& Ludwig, H.-G., 1997.
[A solar model with improved subatmospheric stratification].
{\it Astron. Astrophys.}, {\bf 322}, 646--652.

\bibitem[Schlattl {\it et~al.\/}(2001)]{Schlat2001}
\biblab{Schlat2001}
Schlattl, H., Cassisi, S., Salaris, M. \& Weiss, A., 2001.
[On the helium flash in low-mass population III red giant stars].
{\it Astrophys. J.}, {\bf 559}, 1082--1093.

\bibitem[Schmelz {\it et~al.\/}(2005)]{Schmel2005}
\biblab{Schmel2005}
Schmelz, J. T., Nasraoui, K., Roames, J. K., Lippner, L. A. \& Garst, J. W.,
2005.
[Neon lights up a controversy: the solar Ne/O abundance].
{\it Astrophys. J.}, {\bf 634}, L197--L200.

\bibitem[Schou and Birch(2020)]{Schou2020}
\biblab{Schou2020}
Schou, J. \& Birch, A. C., 2020.
[Estimating the nonstructural component of the helioseismic surface term
using hydrodynamic simulations].
{\rv {\it Astron. Astrophys.}, {\bf 638}, A51-(1--9)}
%{\tt (arXiv:2004.13548v1 [astro-ph.SR])}

\bibitem[Schou {\it et~al.\/}(1997)]{Schou1997}
\biblab{Schou1997}
Schou, J., Kosovichev, A. G., Goode, P. R. \& Dziembowski, W. A., 1997.
[Determination of the Sun's seismic radius from the 
{\sl SOHO} Michelson Doppler Imager].
{\it Astrophys. J.}, {\bf 489}, L197--L200.

\bibitem[Schou {\it et~al.\/}(1998)]{Schou1998}
\biblab{Schou1998}
Schou, J., Antia, H. M., Basu, S., Bogart, R. S., Bush, R. I.,
Chitre, S. M., Christensen-Dalsgaard, J., Di Mauro, M. P.,
Dziembowski, W. A., Eff-Darwich, A.,
Gough, D. O., Haber, D. A., Hoeksema, J. T., Howe, R.,
Korzennik, S. G., Kosovichev, A. G., Larsen, R. M., Pijpers, F. P.,
Scherrer, P. H., Sekii, T., Tarbell, T. D., Title, A. M.,
Thompson, M. J., Toomre, J., 1998.
[Helioseismic studies of differential rotation in the solar
envelope by the Solar Oscillations Investigation using the
Michelson Doppler Imager].
{\it Astrophys. J.}, {\bf 505}, 390--417.

\bibitem[Schou {\it et~al.\/}(2002)]{Schou2002}
\biblab{Schou2002}
Schou, J., Howe, R., Basu, S., Christensen-Dalsgaard, J., Corbard, T.,
Hill, F., Larsen, R. M., Rabello-Soares, M. C. \& Thompson, M. J., 2002.
[A comparison of solar $p$-mode parameters from the Michelson Doppler Imager
and the Global Oscillation Network Group: splitting coefficients and rotation
inversions].
{\it Astrophys. J.}, {\bf 567}, 1234--1249.

\bibitem[Schrijver {\it et~al.\/}(2007)]{Schrij2007}
\biblab{Schrij2007}
Schrijver, C. J., Carpenter, K. G. \& Karovska, M., 2007.
[Dynamos, asteroseismology, and the Stellar Imager].
{\it Comm. in Asteroseismology}, {\bf 150}, 364--370.

\bibitem[Schr{\"o}der and  Smith(2008)]{Schrod2008}
\biblab{Schrod2008}
Schr{\"o}der, K.-O. \& Smith, R. C., 2008.
[Distant future of the Sun and Earth revisited].
{\it Mon. Not. R. Astron. Soc.}, {\bf 386}, 155--163.
%{\tt [arXiv:0801.4031v1 [astro-ph]]}.

\bibitem[Schumacher and  Sreenivasan(2020)]{Schuma2020}
\biblab{Schuma2020}
Schumacher, J. \& Sreenivasan, K. R., 2020.
[{\it Colloquium}: Unusual dynamics of convection in the Sun].
{\it Rev. Mod. Phys.}, {\bf 92}, 041001-(1 -- 25).

\bibitem[Schunker {\it et~al.\/}(2018)]{Schunk2018}
\biblab{Schunk2018}
Schunker, H., Schou, J., Gaulme, P. \& Gizon, L., 2018.
[Fragile detection of solar g modes by Fossat {\etal}].
{\it Solar Phys.}, {\bf 293}, 95-(1 -- 12).
{\tt [arXiv:1804.04407v1 [astro-ph.SR]]}

\bibitem[Schwarzschild(1906)]{Schwar1906}
\biblab{Schwar1906}
Schwarzschild, K., 1906.
[\"Uber das Gelichwicht der Sonnenatmosph{\"a}re].
{\it Nachr. Kgl. Ges. d. Wiss. zu G{\"o}tt., Math. Phys. Klasse 1906}, 41--53.

\bibitem[Schwarzschild(1958)]{Schwar1958}
\biblab{Schwar1958}
Schwarzschild, M., 1958.
{\it Structure and evolution of the stars},
Princeton University Press, Princeton, New Jersey.

\bibitem[Schwarzschild {\it et~al.\/}(1957)]{Schwar1957}
\biblab{Schwar1957}
Schwarzschild, M., Howard, R. \& H{\"a}rm, R., 1957.
[Inhomogeneous stellar models. V. A solar model with convective envelope 
and inhomogeneous interior].
{\it Astrophys. J.}, {\bf 125}, 233--241.

\bibitem[Scuflaire {\it et~al.\/}(2008)]{Scufla2008}
\biblab{Scufla2008}
Scuflaire, R., Th{\'e}ado, S., Montalb{\'a}n, J., Miglio, A., Bourge, P.-O.,
Thoul, A. \& Noels, A., 2008.
[CL\'ES, Code Li{\`e}geois d'\'Evolution Stellaire],
{\it Astrophys. Space Sci.}, {\bf 316}, 83--91.

%\bibitem[Sears(1959)]{Sears1959}
%\biblab{Sears1959}
%Sears, R. L., 1959.
%[An evolutionary sequence of solar models].
%{\it Astrophys. J.}, {\bf 129}, 489--495.

\bibitem[Sears(1964)]{Sears1964}
\biblab{Sears1964}
Sears, R. L., 1964.
[Helium content and neutrino fluxes in solar models].
{\it Astrophys. J.}, {\bf 140}, 477--484.

\bibitem[Seaton(2005)]{Seaton2005}
\biblab{Seaton2005}
Seaton, M. J., 2005.
[Opacity Project data on CD for mean opacities and radiative accelerations].
{\it Mon. Not. R. Astron. Soc.}, {\bf 362}, L1--L3.

\bibitem[Seaton and Badnell(2004)]{Seaton2004}
\biblab{Seaton2004}
Seaton, M. J. \& Badnell, N. R., 2004.
[A comparison of Rosseland-mean opacities from OP and OPAL].
{\it Mon. Not. R. Astron. Soc.}, {\bf 354}, 457--465. 

\bibitem[Seaton {\it et~al.\/}(1994)]{Seaton1994}
\biblab{Seaton1994}
Seaton, M. J., Yan, Y., Mihalas, D. \& Pradhan, A. K., 1994.
[Opacities for stellar envelopes].
{\it Mon. Not. R. Astron. Soc.}, {\bf 266}, 805--828.

\bibitem[Serenelli(2016)]{Serene2016a}
\biblab{Serene2016a}
Serenelli, A., 2016.
[Alive and well: a short review about standard solar models].
{\it Eur. Phys. J. A}, {\bf 52}, 78-(1--13).

\bibitem[Serenelli and  Basu(2010)]{Serene2010}
\biblab{Serene2010}
Serenelli, A. \& Basu, S., 2010.
[Determining the initial helium abundance of the Sun].
{\it Astrophys. J.}, {\bf 719}, 865--872.

\bibitem[Serenelli {\it et~al.\/}(2009)]{Serene2009}
\biblab{Serene2009}
Serenelli, A. M., Basu, S., Ferguson, J. W. \& Asplund, M., 2009.
[New solar composition: the problem with solar models revisited].
{\it Astrophys. J.}, {\bf 705}, L123--L127.

\bibitem[Serenelli {\it et~al.\/}(2011)]{Serene2011}
\biblab{Serene2011}
Serenelli, A. M., Haxton, W. C. \& Pe\~na-Garay, C., 2011.
[Solar models with accretion. I. Application to the solar abundance problem].
{\it Astrophys. J.}, {\bf 743}, 24-(1--20).

\bibitem[Serenelli {\it et~al.\/}(2016)]{Serene2016b}
\biblab{Serene2016b}
Serenelli, A., Scott, P., Villante, F. L., Vincent, A. C., Asplund, M.,
Basu, S., Grevesse, N. \& Pe\~nas-Garay, C., 2016.
[Implications of solar wind measurements for solar models and composition].
{\it Mon. Not. R. Astron. Soc.}, {\bf 463}, 2--9.

\bibitem[Severny {\it et~al.\/}(1979)]{Severn1979}
\biblab{Severn1979}
Severny, A. B., Kotov, V. A. \& Tsap, T. T., 1979.
[Solar oscillations and the problem of the internal structure of the sun].
{\it Astron. Zh.}, {\bf 56}, 1137--1148
(English translation: {\it Sov. Astron.}, {\bf 23}, 641--647).
%  First use, apparently, of the term `helioseismology'; {\cf} Chaplin

\bibitem[Shaviv(2004a)]{Shaviv2004a}
\biblab{Shaviv2004a}
Shaviv, G., 2004a.
[Numerical experiments in screening theory].
{\it Astron. Astrophys.}, {\bf 418}, 801--811.

\bibitem[Shaviv(2004b)]{Shaviv2004b}
\biblab{Shaviv2004b}
Shaviv, G., 2004b.
[The limit on mean field theories and nuclear screening in stellar plasmas].
In {\it Equation-of-State and Phase-Transition Issues in
Models of Ordinary Astrophysical Matter},
eds V. {\v C}elebonovi{\'c},  W. D{\"a}ppen \& D. Gough,
AIP Conf. Proc. vol. 731, AIP, Melville, New York, p. 67--82.

\bibitem[Shaviv(2009)]{Shaviv2009}
\biblab{Shaviv2009}
Shaviv, G., 2009.
{\it The life of stars. The controversial inception and emergence of the
theory of stellar structure}.
The Hebrew University Magnes Press and Springer-Verlag, Berlin Heidelberg.

\bibitem[Shaviv(2003)]{Shaviv2003}
\biblab{Shaviv2003}
Shaviv, N. J., 2003.
[Toward a solution to the early faint Sun paradox: a lower cosmic ray
flux from a stronger solar wind].
{\it J. Geophys. Res.}, {\bf 108}, 1437(1--8).

\bibitem[Shaviv and  Shaviv(1996)]{Shaviv1996}
\biblab{Shaviv1996}
Shaviv, N. J. \& Shaviv, G., 1996.
[The electrostatic screening of thermonuclear reactions in 
astrophysical plasmas. I].
{\it Astrophys. J.}, {\bf 468}, 433--444.

\bibitem[Shaviv and  Shaviv(2001)]{Shaviv2001}
\biblab{Shaviv2001}
Shaviv, N. J. \& Shaviv, G., 2001.
[The electrostatic screening of nuclear reactions in the Sun].
{\it Astrophys. J.}, {\bf 558}, 925--942.

\bibitem[Shu {\it et~al.\/}(1994)]{Shu1994}
\biblab{Shu1994}
Shu, F., Najita, J., Ostriker, E., Wilkin, F., Ruden, S. \&
Lizano, S., 1994.
[Magnetocentrifugally driven flows from young stars and disks.
1: A generalized model].
{\it Astrophys. J.}, {\bf 429}, 781--796.

\bibitem[Shu {\it et~al.\/}(2000)]{Shu2000}
\biblab{Shu2000}
Shu, F. H., Najita, J. R., Shang, H. \& Li, Z.-Y., 2000.
[X-winds: theory and observations].
In {\it Protostars and Planets IV}, 
eds V. Mannings, A. P. Boss \& S. S. Russell, University of Arizona Press,
p. 789--813.

\bibitem[Silva Aguirre {\it et~al.\/}(2015)]{SilvaA2015}
\biblab{SilvaA2015}
Silva Aguirre, V., Davies, G. R., Basu, S., Christensen-Dalsgaard, J.,
Creevey, O., Metcalfe, T. S., Bedding, T. R., Casagrande, L.,
Handberg, R., Lund, M. N., Nissen, P. E., Chaplin, W. J., Huber, D.,
Serenelli, A. M., Stello, D., Van Eylen, V., Campante, T. L., Elsworth, Y.,
Gilliland, R. L., Hekker, S., Karoff, C., Kawaler, S. D., Kjeldsen, H. \&
Lundkvist, M. S., 2015.
[Ages and fundamental properties of {\it Kepler} exoplanet host stars from
asteroseismology].
{\it Mon. Not. R. Astron. Soc.}, {\bf 452}, 2127--2148.
%{\tt [arXiv:1504.07992 [astro-ph.SR]]}

\bibitem[Silva Aguirre {\it et~al.\/}(2017)]{SilvaA2017}
\biblab{SilvaA2017}
Silva Aguirre, V., Lund, M. N., Antia, H. M., Ball, W. H., Basu, S.,
Christensen-Dalsgaard, J., Lebreton, Y., Reese, D. R., Verma, K.,
Casagrande, L., Justesen, A. B., Mosumgaard, J. R., Chaplin, W. J.,
Bedding, T. R., Davies, G. R., Handberg, R., Houdek, G., Huber, D.,
Kjeldsen, H., Latham, D. W., White, T. R., Coelho, H. R., Miglio, A. \&
Rendle, B., 2017.
[Standing on the shoulders of dwarfs: the {\it Kepler} Asteroseismic
LEGACY Sample. II. Radii, masses and ages].
{\it Astrophys. J.}, {\bf 835}, 173-(1--18).

\bibitem[Silva Aguirre {\it et~al.\/}(2020)]{SilvaA2020}
\biblab{SilvaA2020}
Silva Aguirre, V., Christensen-Dalsgaard, J., Cassisi, S.,
Miller Bertolami, M., Serenelli, A., Stello, D.,
Weiss, A., Angelou, G., Jiang, C., Lebreton, Y., Spada, F.,
Bellinger, E. P., Deheuvels, S., Ouazzani, R. M., Pietrinferni, A., 
Mosumgaard, J. R., Townsend, R. H. D., Battich, T., Bossini, D., 
Constantino, T., Eggenberger, P., Hekker, S., Mazumdar, A., Miglio, A.,
Nielsen, K. B. \& Salaris, M., 2020.  
[The Aarhus red giants challenge I.
Stellar structures in the red giant branch phase].
{\it Astron. Astrophys.} {\bf 635}, A164-(1--13).
%{\tt [arXiv:1912.04909v1 [astro-ph.SR]]}

\bibitem[Simon(1982)]{Simon1982}
\biblab{Simon1982}
Simon, N. R., 1982.
[A plea for reexamining heavy element opacities].
{\it Astrophys. J.}, {\bf 260}, L87--L90.

\bibitem[Skumanich(1972)]{Skuman1972}
\biblab{Skuman1972}
Skumanich, A., 1972.
[Time scales for Ca II emission decay, rotational braking, and lithium
depletion].
{\it Astrophys. J.}, {\bf 171}, 565--567.

\bibitem[Snodgrass and  Howard(1985)]{Snodgr1985}
\biblab{Snodgr1985}
Snodgrass, H. B. \& Howard, R., 1985.
[Torsional oscillations of the Sun].
{\it Science}, {\bf 228}, 945--952.

\bibitem[Soderblom {\it et~al.\/}(2001)]{Soderb2001}
\biblab{Soderb2001}
Soderblom, D. R., Jones, B. F. \& Fischer, D., 2001.
[Rotational studies of late-type stars. VII. M34 (NGC 1039) and the
evolution of angular momentum and activity in young solar-type stars].
{\it Astrophys. J.}, {\bf 563}, 334--340.

\bibitem[Solomon {\it et~al.\/}(2009)]{Solomo2009}
\biblab{Solomo2009}
Solomon, S., Plattner, G.-K., Knutti, R. \& Friedlingstein, P., 2009.
[Irreversible climate change due to carbon dioxide emissions].
{\it PNAS}, {\bf 106}, 1704--1709.

\bibitem[Song {\it et~al.\/}(2018)]{Song2018}
\biblab{Song2018}
Song, N., Gonzales-Garcia, M. C., Villante, F. L., Vinyoles, N. \&
Serenelli, A., 2018.
[Helioseismic and neutrino data-driven reconstruction of solar properties].
{\it Mon. Not. R. Astron. Soc.}, {\bf 477}, 1397--1413.
%{\tt [arXiv:1710.02147v1 [astro-ph.SR]]}

{\rv
\bibitem[Sonoi and Shibahashi(2012)]{Sonoi2012}
\biblab{Sonoi2012}
Sonoi, T. \& Shibahashi, H., 2012.
[Dipole low-order g-mode instability of metal-poor low-mass main-sequence
stars due to the $\varepsilon$ mechanism].
{\it Mon. Not. R. Astron. Soc.}, {\bf 422}, 2642--2647.
}

\bibitem[Sonoi {\it et~al.\/}(2015)]{Sonoi2015}
\biblab{Sonoi2015}
Sonoi, T., Samadi, R., Belkacem, K., Ludwig, H.-G., Caffau, E. \& Mosser, B.,
2015.
[Surface-effect corrections for solar-like oscillations using 
3D hydrodynamical simulations. I. Adiabatic oscillations].
{\it Astron. Astrophys.}, {\bf 583}, A112-(1--11).

\bibitem[Sonoi {\it et~al.\/}(2017)]{Sonoi2017}
\biblab{Sonoi2017}
Sonoi, T., Belkacem, K., Dupret, M.-A., Samadi, R., Ludwig, H.-G.,
Caffau, E. \& Mosser, B., 2017.
[Computation of eigenfrequencies for equilibrium models including 
turbulent pressure].
{\it Astron. Astrophys.}, {\bf 600}, A31-(1--11).

\bibitem[Spada {\it et~al.\/}(2018)]{Spada2018}
\biblab{Spada2018}
Spada, F., Demarque, P., Basu, S. \& Tanner, J. D., 2018.
[Improved calibration of the radii of cool stars based on 3D simulations of
convection: implications for the solar model].
{\it Astrophys. J.}, {\bf 869}, 135-(1--14).

\bibitem[Spergel and Press(1985)]{Sperge1985}
\biblab{Sperge1985}
Spergel, D. N. \& Press, W. H., 1985.
[Effect of hypothetical, weakly interacting, massive particles on
energy transport in the solar interior].
{\it Astrophys. J.}, {\bf 294}, 663--673.

\bibitem[Spiegel(1963)]{Spiege1963}
\biblab{Spiege1963}
Spiegel, E. A., 1963.
[A generalization of the mixing-length theory of turbulent convection].
{\it Astrophys. J.}, {\bf 138}, 216--225.

\bibitem[Spiegel and Zahn(1992)]{Spiege1992}
\biblab{Spiege1992}
Spiegel, E. A. \& Zahn, J.-P., 1992.
[The solar tachocline].
{\it Astron. Astrophys.}, {\bf 265}, 106--114.

\bibitem[Spruit(2002)]{Spruit2002}
\biblab{Spruit2002}
Spruit, H. C., 2002.
[Dynamo action by differential rotation in a stably stratified 
stellar interior].
{\it Astron. Astrophys.}, {\bf 381}, 923--932.

%\bibitem[Spruit {\it et~al.\/}(1990)]{Spruit1990}
%\biblab{Spruit1990}
%Spruit, H. C., Nordlund, {\AA}. \& Title, A. M., 1990.
%[Solar convection].
%{\it Annu. Rev. Astron. Astrophys.}, {\bf 28}, 263--301

\bibitem[Stahler(1994)]{Stahle1994}
\biblab{Stahle1994}
Stahler, S. W., 1994.
[Early stellar evolution].
{\it Pub. Astron. Soc. Pacific}, {\bf 106}, 337--343.

\bibitem[Stahler and  Palla(2004)]{Stahle2004}
\biblab{Stahle2004}
Stahler, S. W. \& Palla, F., 2004.
{\it The formation of stars}.
Wiley-VCH, Weinham, Germany

\bibitem[Steigman and Turner(1985)]{Steigm1985}
\biblab{Steigm1985}
Steigman, G. \& Turner, M. S., 1985.
[Cosmological constraints on the properties of weakly interacting massive
particles].
{\it Nucl. Phys. B}, {\bf 253}, 375--386.

\bibitem[Steigman {\it et~al.\/}(1978)]{Steigm1978}
\biblab{Steigm1978}
Steigman, G., Sarazin, C. L., Quintana, H. \& Faulkner, J., 1978.
[Dynamical interactions and astrophysical effects of stable heavy neutrinos].
{\it Astron. J.}, {\bf 83}, 1050--1061.

\bibitem[Stein and Nordlund(1989)]{Stein1989}
\biblab{Stein1989}
Stein, R. F. \& Nordlund, {\AA}., 1989.
[Topology of convection beneath the solar surface].
{\it Astrophys. J.}, {\bf 342}, L95--L98.

\bibitem[Stein and Nordlund(1998)]{Stein1998}
\biblab{Stein1998}
Stein, R. F. \& Nordlund, {\AA}., 1998.
[Simulations of solar granulation. I. General properties].
{\it Astrophys. J.}, {\bf 499}, 914--933.

\bibitem[Stein {\it et~al.\/}(2006)]{Stein2006}
\biblab{Stein2006}
Stein, R. F., Benson, D., Georgobiani, D. \& Nordlund, {\AA}., 2006.
[Supergranule scale convection simulations[.
In {\it Proc. SOHO 18 / GONG 2006 / HELAS I Conf.
Beyond the spherical Sun},
ed. K. Fletcher, ESA SP-624, ESA Publications Division,
Noordwijk, The Netherlands.

\bibitem[Stein {\it et~al.\/}(2009)]{Stein2009}
\biblab{Stein2009}
Stein, R. F., Georgobiani, D., Schafenberger, W., Nordlund, {\AA}. \&
Benson, D., 2009.
[Supergranulation scale convection simulations].
In {\it Proc. 15th Cambridge Workshop on Cool Stars, Stellar Systems and
the Sun}, 
AIP Conf. Proc. vol. 1094, AIP, Melville, New York, p. 764--767.

\bibitem[Stellingwerf(1976)]{Stelli1976}
\biblab{Stelli1976}
Stellingwerf, R. F., 1976.
[The role of turbulent pressure in mixing length convection].
{\it Astrophys. J.}, {\bf 206}, 543--547.

\bibitem[Str{\"o}mgren(1932)]{Stromg1932}
\biblab{Stromg1932}
Str{\"o}mgren, B., 1932.
[The opacity of stellar matter and the hydrogen content of the stars].
{\it Z. Astrophys.}, {\bf 4}, 118--152.

\bibitem[Str{\"o}mgren(1933)]{Stromg1933}
\biblab{Stromg1933}
Str{\"o}mgren, B., 1933.
[On the interpretation of the Hertzsprung-Russell-diagram].
{\it Z. Astrophys.}, {\bf 7}, 222--248.

\bibitem[Str{\"o}mgren(1950)]{Stromg1950}
\biblab{Stromg1950}
Str{\"o}mgren, B., 1950.
[On the extent of the convection zones in the solar interior].
{\it Matematisk Tidsskrift B, Festskrift til Jakob Nielsen}, 96--100.

\bibitem[Strugarek {\it et~al.\/}(2017)]{Struga2017}
\biblab{Struga2017}
Strugarek, A., Beaudoin, P., Charbonneau, P., Brun, A. S. \&
do Nascimento, J.-D., 2017.
[Reconciling solar and stellar magnetic cycles with nonlinear 
dynamo simulations].
{\it Science}, {\bf 357}, 185--187.

\bibitem[Sumner(2002)]{Sumner2002}
\biblab{Sumner2002}
Sumner, T.J., 2002.
[Experimental searches for dark matter].
{\it Living Rev. Relativ.}, {\bf 5}, 4 
\url{https://doi.org/10.12942/lrr-2002-4}.

\bibitem[Svensmark(2006)]{Svensm2006}
\biblab{Svensm2006}
Svensmark, H., 2006.
[Cosmic rays and the biosphere over 4 billion years].
{\it Astron. Nachr.}, {\bf 327}, 871--875.

\bibitem[Swenson and Faulkner(1992)]{Swenso1992}
\biblab{Swenso1992}
Swenson, F. J. \& Faulkner, J., 1992.
[Lithium dilution through main-sequence mass loss].
{\it Astrophys. J.}, {\bf 395}, 654--674.

\bibitem[Takata and  Gough(2001)]{Takata2001a}
\biblab{Takata2001a}
Takata, M. \& Gough, D. O., 2001.
[The influence of uncertainties in the Sun's radius on inversions for
the solar structure].
In {\it Helio- and Asteroseismology at the Dawn of the Millennium:
Proc. SOHO 10 / GONG 2000 Workshop}
ESA SP-464, ESA Publications Division,
Noordwijk, The Netherlands, p. 543--546.

\bibitem[Takata and  Gough(2003)]{Takata2003b}
\biblab{Takata2003b}
Takata, M. \& Gough, D. O., 2003.
[The seismic radius of the Sun, and structure inversions].
In {\it Proc. SOHO 12 / GONG+ 2002. Local and Global Helioseismology:
The Present and Future},
ed. A. Wilson, ESA SP-517, ESA Publications Division,
Noordwijk, The Netherlands, p. 397--400.

\bibitem[Takata and  Shibahashi(1998)]{Takata1998}
\biblab{Takata1998}
Takata, M. \& Shibahashi, H., 1998.
[Solar models based on helioseismology and the solar neutrino problem].
{\it Astrophys. J.}, {\bf 504}, 1035--1050.

{\rv
\bibitem[Takata and Shibahashi(2001)]{Takata2001b}
\biblab{Takata2001b}
Takata, M. \& Shibahashi, H., 2001.
[Solar metal abundance inferred from helioseismology].
In {\it Proc. IAU Symp. No 203: Recent insights into the physics
of the Sun and heliosphere: Highlights from SOHO and other space missions},
eds Brekke, P., Fleck, B. \& Gurman, J. B., 
ASP, San Francisco, p. 43--45.
}

\bibitem[Takata and Shibahashi(2003)]{Takata2003a}
\biblab{Takata2003a}
Takata, M. \& Shibahashi, H., 2003.
[Hydrogen abundance in the tachocline layer of the Sun].
{\it Publ. Astron. Soc. Japan}, {\bf 55}, 1015--1023.

\bibitem[Talon and Charbonnel(2003)]{Talon2003}
\biblab{Talon2003}
Talon, S. \& Charbonnel, C., 2003.
[Angular momentum transport by internal gravity waves. I. - Pop. I
main sequence stars].
{\it Astron. Astrophys.}, {\bf 405}, 1025--1032.

\bibitem[Talon and Charbonnel(2005)]{Talon2005}
\biblab{Talon2005}
Talon, S. \& Charbonnel, C., 2005.
[Hydrodynamical stellar models including rotation, internal gravity waves,
and atomic diffusion. I. Formalism and tests on Pop I dwarfs].
{\it Astron. Astrophys.}, {\bf 440}, 981--994.

\bibitem[Talon and Zahn(1998)]{Talon1998}
\biblab{Talon1998}
Talon, S. \& Zahn, J.-P., 1998.
[Towards a hydrodynamical model predicting the observed solar rotation
profile].
{\it Astron. Astrophys.}, {\bf 329}, 315--318.

\bibitem[Talon {\it et~al.\/}(2002)]{Talon2002}
\biblab{Talon2002}
Talon, S., Kumar, P. \& Zahn, J.-P., 2002.
[Angular momentum extraction by gravity waves in the Sun].
{\it Astrophys. J.}, {\bf 574}, L175--L178.

\bibitem[Tassoul and Tassoul(2004)]{Tassou2004}
\biblab{Tassou2004}
Tassoul, J.-L. \& Tassoul, M., 2004.
{\it A concise history of solar and stellar physics},
Princeton University Press, Princeton, New Jersey.

\bibitem[Tassoul(1980)]{Tassou1980}
\biblab{Tassou1980}
Tassoul, M., 1980.
[Asymptotic approximations for stellar nonradial pulsations].
{\it Astrophys. J. Suppl.}, {\bf 43}, 469--490.

\bibitem[Tayar {\it et~al.\/}(2017)]{Tayar2017}
\biblab{Tayar2017}
Tayar, J., Somers, G., Pinsonneault, M. H., Stello, D., Mints, A.,
Johnson, J. A., Zamora, O., Garc\'{\i}a-Hern\'andez, D. A., Maraston, C.,
Serenelli, A., Allende Prieto, C., Bastien, F. A., Basu, S., Bird, J. C.,
Cohen, R. E., Cunha, K., Elsworth, Y., Garc\'{\i}a, R. A., Girardi, L.,
Hekker, S., Holtzman, J., Huber, D., Mathur, S., M\'esz\'aros, S.,
Mosser, B., Shetrone, M., Silva Aguirre, V., Stassun, K., Stringfellow, G. S.,
Zasowski, G. \& Roman-Lopes, A., 2017.
[The correlation between mixing length and metallicity on the giant branch:
implications for ages in the {\it Gaia} era].
{\it Astrophys. J.}, {\bf 840}, 17-(1--12).

\bibitem[Tayler(1973)]{Tayler1973}
\biblab{Tayler1973}
Tayler, R. J., 1973.
[The adiabatic stability of stars containing magnetic fields---I].
{\it Mon. Not. R. astr. Soc.}, {\bf 161}, 365--380.

\bibitem[Th\'eado and Vauclair(2003a)]{Theado2003a}
\biblab{Theado2003a}
Th\'eado, S. \& Vauclair, S., 2003a.
[On the coupling between helium settling and rotation-induced mixing in
stellar radiative zones. I. Analytical approach].
{\it Astrophys. J.}, {\bf 587}, 777--783.

\bibitem[Th\'eado and Vauclair(2003b)]{Theado2003b}
\biblab{Theado2003b}
Th\'eado, S. \& Vauclair, S., 2003b.
[On the coupling between helium settling and rotation-induced mixing in
stellar radiative zones. II. Numerical approach].
{\it Astrophys. J.}, {\bf 587}, 784--794.

\bibitem[Th\'eado and Vauclair(2003c)]{Theado2003c}
\biblab{Theado2003c}
Th\'eado, S. \& Vauclair, S., 2003c.
[On the coupling between helium settling and rotation-induced mixing in
stellar radiative zones. III. Applications to light elements in 
Population I main-sequence stars].
{\it Astrophys. J.}, {\bf 587}, 795--805.

\bibitem[Thompson {\it et~al.\/}(2003)]{Thomps2003}
\biblab{Thomps2003}
Thompson, M. J., Christensen-Dalsgaard, J., Miesch, M. S. \& Toomre, J., 2003.
[The internal rotation of the Sun].
{\it Annu. Rev. Astron. Astrophys.}, {\bf 41}, 599--643.

\bibitem[Thoul and  Montalb{\'a}n(2007)]{Thoul2007}
\biblab{Thoul2007}
Thoul, A. \& Montalb{\'a}n, J., 2007.
[Microscopic diffusion in stellar plasmas].
In {\it Stellar Evolution and Seismic Tools for Asteroseismology: Diffusive
Processes in Stars and Seismic Analysis}, eds C. W. Straka, Y. Lebreton \&
M. J. P. F. G. Monteiro, EAS Publ. Ser., {\bf 26},
EDP Sciences, Les Ulis, France, p. 25--36.

\bibitem[Thoul {\it et~al.\/}(1994)]{Thoul1994}
\biblab{Thoul1994}
Thoul, A. A., Bahcall, J. N. \& Loeb, A., 1994.
[Element diffusion in the solar interior].
{\it Astrophys. J.}, {\bf 421}, 828--842.

\bibitem[Tomczyk {\it et~al.\/}(1995)]{Tomczy1995}
\biblab{Tomczy1995}
Tomczyk, S., Streander, K., Card, G., Elmore, D., Hull, H. \&
Cacciani, A., 1995.
[An instrument to observe low-degree solar oscillations].
{\it Solar Phys.}, {\bf 159}, 1--21.

\bibitem[Toomre {\it et~al.\/}(1977)]{Toomre1977}
\biblab{Toomre1977}
Toomre, J., Gough, D. O. \& Spiegel, E. A., 1977.
[Numerical solutions of single-mode convection equations].
{\it J. Fluid Mech.}, {\bf 79}, 1--31.

%\bibitem[Trampedach(2007)]{Trampe2007}
%\biblab{Trampe2007}
%Trampedach, R., 2007.
%[A new stellar atmosphere grid - in 3D].
%In Stancliffe R. J., Dewi J., Houdek G., Martin R. G., Tout C.A., eds,
%AIP Conf. Proc. vol. 948, {\it Unsolved Problems in Stellar Physics}.
%American Institute of Physics, Melville, p. 141--148.

\bibitem[Trampedach(2018)]{Trampe2018}
\biblab{Trampe2018}
Trampedach, R., 2018.
[The dark side of the Sun].
In {\it Workshop on Astrophysical Opacities},
eds C. Mendoza, S. Turck-Chi\`eze \& J. Colgan,
{\it ASP Conf. Ser.}, {\bf 515}, {\rv ASP, San Francisco,}
p. 29--32.
%[The dark side of the Sun. A plea for a next-generation opacity calculation].
%Submitted to
%{\it Proc. Second Workshop on Astrophysical Opacities (Kalamazoo, MI, 2017)}
%{\tt [arXiv:1804.04123 [astro-ph.SR]]}.

\bibitem[Trampedach {\it et~al.\/}(1999)]{Trampe1999}
\biblab{Trampe1999}
Trampedach, R., Stein, R. F., Christensen-Dalsgaard, J. \&
Nordlund, {\AA}., 1999.
[Stellar evolution with a variable mixing-length parameter].
In {\it Theory and Tests of Convection in Stellar Structure},
eds A. Gim{\'e}nez, E.F. Guinan \& B. Montesinos,
{\it ASP Conf. Ser.}, {\bf 173}, {\rv ASP, San Francisco}, p. 233--236.

\bibitem[Trampedach {\it et~al.\/}(2006)]{Trampe2006}
\biblab{Trampe2006}
Trampedach, R., D\"appen, W. \& Baturin, V. A., 2006.
[A synoptic comparison of the Mihalas-Hummer-D\"appen and OPAL equations of
state].
{\it Astrophys. J.}, {\bf 646}, 560--578.

\bibitem[Trampedach {\it et~al.\/}(2013)]{Trampe2013}
\biblab{Trampe2013}
Trampedach, R., Asplund, M., Collet, R., Nordlund, {\AA}. \& Stein, R. F., 2013.
[A grid of three-dimensional stellar atmosphere models of solar metallicity. 
I. General properties, granulation, and atmospheric expansion].
{\it Astrophys. J.}, {\bf 769}, 18-(1--15).

\bibitem[Trampedach {\it et~al.\/}(2014a)]{Trampe2014a}
\biblab{Trampe2014a}
Trampedach, R., Stein, R. F., Christensen-Dalsgaard, J., Nordlund, {\AA}. \&
Asplund, M., 2014a.
[Improvements to stellar structure models, based on a grid of
3D convection simulations--I. $T(\tau)$ relations].
{\it Mon. Not. R. Astron. Soc.}, {\bf 442}, 805--820.

\bibitem[Trampedach {\it et~al.\/}(2014b)]{Trampe2014b}
\biblab{Trampe2014b}
Trampedach, R., Stein, R., Christensen-Dalsgaard, J., Nordlund, {\AA} \&
Asplund, M., 2014b.
[Improvements to stellar structure models, based on a grid of
3D convection simulations--II. Calibrating the mixing-length formulation].
{\it Mon. Not. R. Astron. Soc.}, {\bf 445}, 4366--4384.

\bibitem[Trampedach {\it et~al.\/}(2017)]{Trampe2017}
\biblab{Trampe2017}
Trampedach, R., Aarslev, M. J., Houdek, G., Collet, R.,
Christensen-Dalsgaard, J., Stein, R. F. \& Asplund, M., 2017.
[The asteroseismic surface effect from a grid of 3D convection simulations. -
I. Frequency shifts from convective expansion of stellar atmospheres].
{\it Mon. Not. R. Astron. Soc.}, {\bf 466}, L43--L47.

\bibitem[Tripathy and  Christensen-Dalsgaard(1998)]{Tripat1998}
\biblab{Tripat1998}
Tripathy, S. C. \& Christensen-Dalsgaard, J., 1998.
[Opacity effects on the solar interior. I. Solar structure].
{\it Astron. Astrophys.}, {\bf 337}, 579--590.

\bibitem[Turck-Chi{\`e}ze(1999)]{Turck1999}
\biblab{Turck1999}
Turck-Chi{\`e}ze, S., 1999.
[The solar neutrino puzzle: the way ahead].
{\it New Astronomy}, {\bf 4}, 325--332.

\bibitem[Turck-Chi\`eze and Lopes(1993)]{Turck1993}
\biblab{Turck1993}
Turck-Chi\`eze, S. \& Lopes, I., 1993.
[Toward a unified classical model of the Sun: On the sensitivity
of neutrinos and helioseismology to the microscopic physics].
{\it Astrophys. J.}, {\bf 408}, 347--367.

\bibitem[Turck-Chi{\`e}ze and  Lopes(2012)]{Turck2012}
\biblab{Turck2012}
Turck-Chi{\`e}ze, S. \& Lopes, I., 2012.
[Solar-stellar astrophysics and dark matter].
{\it Research in Astron.Astrophys.}, {\bf 12}, 1107--1138.

\bibitem[Turck-Chi\`eze {\it et~al.\/}(1988)]{Turck1988}
\biblab{Turck1988}
Turck-Chi\`eze, S., Cahen, S., Cass\'e, M. \& Doom, C., 1988.
[Revisiting the standard solar model].
{\it Astrophys. J.}, {\bf 335}, 415--424.

\bibitem[Turck-Chi\`eze {\it et~al.\/}(1997)]{Turck1997}
\biblab{Turck1997}
Turck-Chi\`eze, S., Basu, S., Brun, A. S., Christensen-Dalsgaard, J.,
Eff-Darwich, A., Lopes, I., P\'erez Hern\'andez, F., Berthomieu, G.,
Provost, J., Ulrich, R. K., Baudin, F., Boumier, P., Charra, J.,
Gabriel, A. H., Garcia, R. A., Grec, G., Renaud, C., Robillot, J. M. \&
Roca Cort\'es, T., 1997.
[First view of the solar core from GOLF acoustic modes].
{\it Solar Phys.}, {\bf 175}, 247--265.

\bibitem[Turck-Chi{\`e}ze {\it et~al.\/}(2001)]{Turck2001}
\biblab{Turck2001}
Turck-Chi{\`e}ze, S., Couvidat, S., Kosovichev, A. G.,
Gabriel, A. H., Berthomieu, G., Brun, A. S.,
Christensen-Dalsgaard, J., Garc\'{\i}a, R. A., Gough, D. O.,
Provost, J., Roca-Cortes, T., Roxburgh, I. W. \& Ulrich, R. K., 2001.
[Solar neutrino emission deduced from a seismic model].
{\it Astrophys. J.}, {\bf 555}, L69--L73.

\bibitem[Turck-Chi\`eze {\it et~al.\/}(2004)]{Turck2004}
\biblab{Turck2004}
Turck-Chi\`eze, S., Couvidat, S., Piau, L., Ferguson, J., 
Lambert, P., Ballot, J., Garc\'{\i}a, R. A. \& Nghiem, P., 2004.
[Surprising Sun: a new step towards a complete picture?]
{\it Phys. Rev. Lett.}, {\bf 93}, 211102-(1--4).

\bibitem[Turck-Chi{\`e}ze {\it et~al.\/}(2016)]{Turck2016}
\biblab{Turck2016}
Turck-Chi{\`e}ze, S., Le Pennec, M., Ducret, J. E., Colgan, J., Kilcrease, D. P.,
Fontes, C. J., Magee, N., Gilleron, F. \& Pain, J. C., 2016.
[Detailed opacity comparison for an improved stellar modeling of the
envelopes of massive stars].
{\it Astrophys. J.}, {\bf 823}, 78-(1--15).

\bibitem[Turcotte and Christensen-Dalsgaard(1998)]{Turcot1998b}
\biblab{Turcot1998b}
Turcotte, S. \& Christensen-Dalsgaard, J., 1998.
[Solar models with non-standard chemical composition].
{\it Proc. ISSI Workshop on Solar Composition and its
Evolution--from Core to Corona},
eds C. Fr\"ohlich, M. C. E. Huber, S. Solanki \& R. von Steiger,
{\it Space Science Reviews}, {\bf 85}, 133--140. Kluwer, Dordrecht.

\bibitem[Turcotte {\it et~al.\/}(1998)]{Turcot1998a}
\biblab{Turcot1998a}
Turcotte, S., Richer, J., Michaud, G., Iglesias, C. A. \& Rogers, F. J., 1998.
[Consistent solar evolution model including diffusion and radiative
acceleration effects].
{\it Astrophys. J.}, {\bf 504}, 539--558.

\bibitem[Turcotte {\it et~al.\/}(2000)]{Turcot2000}
\biblab{Turcot2000}
Turcotte, S., Richer, J., Michaud, G. \& Christensen-Dalsgaard, J., 2000.
[The effect of diffusion on pulsations of stars on the upper main sequence.
$\delta$ Scuti and metallic A stars].
{\it Astron. Astrophys.}, {\bf 360}, 603--616.

\bibitem[Ulrich and Rhodes(1977)]{Ulrich1977}
\biblab{Ulrich1977}
Ulrich, R. K. \& Rhodes, E. J., 1977.
[The sensitivity of nonradial $p$ mode eigenfrequencies to solar
envelope structure].
{\it Astrophys. J.}, {\bf 218}, 521--529.

\bibitem[Ulrich(1969)]{Ulrich1969}
\biblab{Ulrich1969}
Ulrich, R. K., 1969.
[A rapidly rotating core and solar neutrinos].
{\it Astrophys. J.}, {\bf 158}, 427. % 1 page

\bibitem[Ulrich(1970)]{Ulrich1970}
\biblab{Ulrich1970}
Ulrich, R. K., 1970.
[The five-minute oscillations on the solar surface].
{\it Astrophys. J.}, {\bf 162}, 993--1001.

\bibitem[Ulrich(1972)]{Ulrich1972}
\biblab{Ulrich1972}
Ulrich, R. K., 1972.
[Thermohaline convection in stellar interiors].
{\it Astrophys. J.}, {\bf 172}, 165--177.

\bibitem[Ulrich(1986)]{Ulrich1986}
\biblab{Ulrich1986}
Ulrich, R. K., 1986.
[Determination of stellar ages from asteroseismology].
{\it Astrophys. J.}, {\bf 306}, L37--L40.

\bibitem[Ulrich(2001)]{Ulrich2001}
\biblab{Ulrich2001}
Ulrich, R. K., 2001.
[Very long lived wave patterns detected in the solar surface velocity signal].
{\it Astrophys. J.}, {\bf 560}, 466--475.

\bibitem[Ulrich {\it et~al.\/}(1988)]{Ulrich1988}
\biblab{Ulrich1988}
Ulrich, R. K., Boyden, J. E., Webster, L., Snodgrass, H. B.,
Padilla, S. P., Gilman, P. \& Shieber, T., 1988.
[Solar rotation measurements at Mount Wilson. V.
Reanalysis of 21 years of data].
{\it Solar Phys.}, {\bf 117}, 291--328.

\bibitem[Unno(1967)]{Unno1967}
\biblab{Unno1967}
Unno, W., 1967.
[The stellar radial pulsation coupled with the convection].
{\it Publ. Astron. Soc. Japan}, {\bf 19}, 140--153.

\bibitem[Uns{\"o}ld(1930)]{Unsold1930}
\biblab{Unsold1930}
Uns{\"o}ld, A., 1930.
[{\rv Konvektion in der Sonnenatmosph{\"a}re (nebst einer Bemerkung zur Deutung
der Novae)}].
{\it Z. Astrophys.}, {\bf 1}, 138--148.

\bibitem[Uns{\"o}ld(1931)]{Unsold1931}
\biblab{Unsold1931}
{\rv Uns{\"o}ld, A., 1931.
[Wasserstoff und Helium in Sternatmosph\"aren].}
{\it Z. Astrophys.}, {\bf 3}, 81--104.

{\rv
\bibitem[Vagnozzi {\it et~al.\/}(2017)]{Vagnoz2017}
\biblab{Vagnoz2017}
Vagnozzi, S., Freese, K. \& Zurbuchen, T. H., 2017.
[Solar models in light of new high metallicity measurements from solar wind
data].
{\it Astrophys. J.}, {\bf 839}, 55-(1--10).
}

\bibitem[VandenBerg {\it et~al.\/}(2007)]{Vanden2007}
\biblab{Vanden2007}
VandenBerg, D. A., Gustafsson, B., Edvardsson, B., Eriksson, K. \&
Ferguson, J., 2007.
[A constraint on $Z_\odot$ from fits of isochrones to the color-magnitude
diagram of M67].
{\it Astrophys. J.}, {\bf 666}, L105--L108.

\bibitem[van Saders {\it et~al.\/}(2012)]{vanSad2012}
\biblab{vanSad2012}
van Saders, J. \& Pinsonneault, M. H., 2012.
[The sensitivity of convection zone depth to stellar abundances: an absolute
stellar abundance scale from asteroseismology].
{\it Astrophys. J.}, {\bf 746}, 16-(1--16).

\bibitem[van Saders {\it et~al.\/}(2016)]{vanSad2016}
\biblab{vanSad2016}
van Saders, J. L., Ceillier, T., Metcalfe, T. S., Silva Aguirre, V., 
Pinsonneault, M. H., Garc\'{\i}a, R. A., Mathur, S. \& Davies, G. R., 2016.
[Weakened magnetic braking as the origin of anomalously rapid rotation
in old field stars].
{\it Nature}, {\bf 529}, 181--184.

\bibitem[Vauclair {\it et~al.\/}(1974)]{Vaucla1974}
\biblab{Vaucla1974}
Vauclair, G., Vauclair, S. \& Pamjatnikh, A., 1974.
[Diffusion processes in the envelopes of main-sequence A stars:
model variations due to helium depletion].
{\it Astron. Astrophys.}, {\bf 31}, 63--70.

\bibitem[Vauclair and  Richard(1998)]{Vaucla1998}
\biblab{Vaucla1998}
Vauclair, S. \& Richard, O., 1998.
[Consistent solar models including the $^7$Li and $^3$He constraints].
In {\it Structure and dynamics of the
interior of the Sun and Sun-like stars; Proc. SOHO 6/GONG 98 Workshop}, 
eds S.G. Korzennik \& A. Wilson, ESA SP-418, ESA Publications Division,
Noordwijk, The Netherlands, p. 427--429.

\bibitem[Vauclair {\it et~al.\/}(1978)]{Vaucla1978}
\biblab{Vaucla1978}
Vauclair, S., Vauclair, G., Schatzman, E. \& Michaud, G., 1978.
[Hydrodynamical instabilities in the envelopes of main-sequence stars:
constraints implied by the lithium, beryllium, and boron observations].
{\it Astrophys. J.}, {\bf 223}, 567--582.

\bibitem[Verma and Silva Aguirre(2019)]{Verma2019}
\biblab{Verma2019}
Verma, K. \& Silva Aguirre, V., 2019.
[Helium settling in F stars: constraining turbulent mixing using observed
helium glitch signature].
{\it Mon. Not. R. Astron. Soc.}, {\bf 489}, 1850--1858.

\bibitem[Verma {\it et~al.\/}(2014)]{Verma2014}
\biblab{Verma2014}
Verma, K., Faria, J. P., Antia, H. M., Basu, S., Mazumdar, A., 
Monteiro, M. J. P. F. G., Appourchaux, T., Chaplin, W. J., 
Garc\'{\i}a, R. A. \& Metcalfe, T. S., 2014.
[Asteroseismic estimate of helium abundance of a solar analog binary system].
{\it Astrophys. J.}, {\bf 790}, 138-(1--13).

\bibitem[Verma {\it et~al.\/}(2017)]{Verma2017}
\biblab{Verma2017}
Verma, K, Raodeo, K., Antia, H. M., Mazumdar, A., Basu, S.,
Lund, M. N. \& Silva Aguirre, V., 2017.
[Seismic measurement of the locations of the base of convection zone and helium
ionization zone for stars in the {\it Kepler} seismic LEGACY sample].
{\it Astrophys. J.},{\bf 837}, 47-(1--16).

\bibitem[Vernazza {\it et~al.\/}(1981)]{Vernaz1981}
\biblab{Vernaz1981}
Vernazza, J. E., Avrett, E. H. \& Loeser, R., 1981.
[Structure of the solar chromosphere. III. Models of the EUV
brightness components of the quiet Sun].
{\it Astrophys. J. Suppl.}, {\bf 45}, 635--725.

\bibitem[Verner {\it et~al.\/}(2006a)]{Verner2006a}
\biblab{Verner2006a}
Verner, G. A., Chaplin, W. J. \& Elsworth, Y., 2006.
[BiSON data show change in solar structure with magnetic activity].
{\it Astrophys. J.}, {\bf 640}, L95--L98.

\bibitem[Verner {\it et~al.\/}(2006b)]{Verner2006b}
\biblab{Verner2006b}
Verner, G. A., Chaplin, W. J. \& Elsworth, Y., 2006.
[The detectability of signatures of rapid variation in low-degree stellar
$p$-mode oscillation frequencies].
{\it Astrophys. J.}, {\bf 638}, 440--445.

\bibitem[Villante and  Ricci(2010)]{Villan2010}
\biblab{Villan2010}
Villante, F. L. \& Ricci, B., 2010.
[Linear solar models].
{\it Astrophys. J.}, {\bf 714}, 944--959.

\bibitem[Villante {\it et~al.\/}(2014)]{Villan2014}
\biblab{Villan2014}
Villante, F. L., Serenelli, A. M., Delahaye, F. \& Pinsonneault, M., 2014.
[The chemical composition of the Sun from helioseismic and solar neutrino data].
{\it Astrophys. J.}, {\bf 787}, 13-(1--14).

\bibitem[Vinyoles {\it et~al.\/}(2017)]{Vinyol2017}
\biblab{Vinyol2017}
Vinyoles, N., Serenelli, A. M., Villante, F. L., Basu, S., Bergstr{\"o}m, J.,
Gonzalez-Garcia, M. C., Maltoni, M., Pe\~na-Garay, C. \& Song, N., 2017.
[A new generation of standard solar models].
{\it Astrophys. J.}, {\bf 835}, 202-(1--16).

\bibitem[Vitense(1953)]{Vitens1953}
\biblab{Vitens1953}
Vitense, E., 1953.
[Die Wasserstoffkonvektionszone der Sonne].
{\it Z. Astrophys.}, {\bf 32}, 135--164.

\bibitem[von Paris {\it et~al.\/}(2008)]{vonPar2008}
\biblab{vonPar2008}
von Paris, P., Rauer, H., Grenfell, J. L., Patzer, B.,
Hedelt, P., Stracke, B., Trautmann, T. \& Schreier, F., 2008.
[Warming the early earth---CO$_2$ reconsidered].
{\it Planet. Space Sci.}, {\bf 56}, 1244--1259.

\bibitem[von Steiger and Zurbuchen(2016)]{vonSte2016}
\biblab{vonSte2016}
von Steiger, R. \& Zurbuchen, T. H., 2016.
[Solar metallicity derived from in situ solar wind composition].
{\it Astrophys. J.}, {\bf 816}, 13-(1--8).

\bibitem[von Weizs{\"a}cker(1937)]{vonWei1937}
\biblab{vonWei1937}
von Weizs{\"a}cker, C. F., 1937.
[\"Uber Elementumwandlungen im Innern der Sterne. I.].
{\it Physik. Zeitschr.}, {\bf 38}, 176--191.

\bibitem[von Weizs{\"a}cker(1938)]{vonWei1938}
\biblab{vonWei1938}
von Weizs{\"a}cker, C. F., 1938.
[\"Uber Elementumwandlungen im Innern der Sterne. II.].
{\it Physik. Zeitschr.}, {\bf 39}, 633--646.

\bibitem[Vorontsov(1988)]{Voront1988}
\biblab{Voront1988}
Vorontsov, S. V., 1988.
[A search of the effects of magnetic field in the solar five-minute 
oscillations].
{\it Proc. IAU Symposium No 123, Advances in helio- and asteroseismology},
p. 151--154,
eds Christensen-Dalsgaard, J. \& Frandsen, S.,
Reidel, Dordrecht.

\bibitem[Vorontsov and Shibahashi(1991)]{Voront1991a}
\biblab{Voront1991a}
Vorontsov, S. V. \& Shibahashi, H., 1991.
[Asymptotic inversion of the solar oscillation frequencies:
sound speed in the solar interior].
{\it Publ. Astron. Soc. Japan}, {\bf 43}, 739--753.

\bibitem[Vorontsov {\it et~al.\/}(1991)]{Voront1991b}
\biblab{Voront1991b}
Vorontsov, S. V., Baturin, V. A. \& Pamyatnykh, A. A., 1991.
[Seismological measurement of solar helium abundance].
{\it Nature}, {\bf 349}, 49--51.

\bibitem[Vorontsov {\it et~al.\/}(2002)]{Voront2002}
\biblab{Voront2002}
Vorontsov, S. V., Christensen-Dalsgaard, J., Schou, J.,
Strakhov, V. N. \& Thompson, M. J., 2002.
[Helioseismic measurement of solar torsional oscillations].
{\it Science}, {\bf 296}, 101--103.

\bibitem[Vorontsov {\it et~al.\/}(2013)]{Voront2013}
\biblab{Voront2013}
Vorontsov, S. V., Baturin, V. A., Ayukov, S. V. \& Gryaznov, V. K., 2013.
[Helioseismic calibration of the equation of state and chemical composition
in the solar envelope].
{\it Mon. Not. R. Astron. Soc.}, {\bf 430}, 1636--1652.

{\rv
\bibitem[Vorontsov {\it et~al.\/}(2014)]{Voront2014}
\biblab{Voront2014}
Vorontsov, S. V., Baturin, V. A., Ajukov, S. V. \& Gryaznov, V. K., 2014.
[Helioseismic measurements in the solar envelope using group velocities
of surface waves].
{\it Mon. Not. R. Astron. Soc.}, {\bf 441}, 3296--3305.
}

\bibitem[Walker {\it et~al.\/}(1981)]{Walker1981}
\biblab{Walker1981}
Walker, J. C. G., Hays, P. B. \& Kasting, J. F., 1981.
[A negative feedback mechanism for the long-term stabilization of Earth's
surface temperature].
{\it J. Geophys. Res.}, {\bf 86}, 9776--9782.

\bibitem[Wambsganss(1988)]{Wambsg1988}
\biblab{Wambsg1988}
Wambsganss, J., 1988.
[Hydrogen-helium-diffusion in solar models].
{\it Astron. Astrophys.}, {\bf 205}, 125--128.

\bibitem[Wan(2019)]{Wan2019}
\biblab{Wan2019}
Wan, L., 2019.
[Simulation and sensitivity studies for solar neutrinos at Jinping].
In {\it Solar Neutrinos. Proc. 5th International Solar Neutrino Conference},
eds M. Meyer \& K. Zuber, World Scientific, 381--389.

\bibitem[Wedemeyer {\it et~al.\/}(2004)]{Wedeme2004}
\biblab{Wedeme2004}
Wedemeyer, S., Freytag, B., Steffen, M., Ludwig, H.-G. \& Holweger, H., 2004.
[Numerical simulation of the three-dimensional structure and dynamics of
the non-magnetic solar chromosphere].
{\it Astron. Astrophys.}, {\bf 414}, 1121--1137.

\bibitem[Weiss and  Schlattl(2008)]{Weiss2008}
\biblab{Weiss2008}
Weiss, A. \& Schlattl, H., 2008.
[GARSTEC---the Garching Stellar Evolution Code.
The direct descendant of the legendary Kippenhahn code].
{\it Astrophys. Space Sci}, {\bf 316}, 99--106.

\bibitem[Weiss {\it et~al.\/}(2001)]{Weiss2001}
\biblab{Weiss2001}
Weiss, A., Flaskamp, M. \& Tsytovich, V. N., 2001.
[Solar models and electron screening].
{\it Astron. Astrophys.}, {\bf 371}, 1123--1127.

\bibitem[Weiss {\it et~al.\/}(2004)]{Weiss2004}
\biblab{Weiss2004}
Weiss, A., Hillebrandt, W., Thomas, H.-C. \& Ritter, H., 2004.
{\it Cox \& Giuli's Principles of Stellar Structure},
Cambridge Scientific Publishers, Cambridge, UK.

\bibitem[Weymann(1957)]{Weyman1957}
\biblab{Weyman1957}
Weymann, R., 1957.
[Inhomogeneous stellar models. VI. An improved solar model with the
carbon cycle included].
{\it Astrophys. J.}, {\bf 126}, 208--212.

\bibitem[Weymann and Sears(1965)]{Weyman1965}
\biblab{Weyman1965}
Weymann, R. \& Sears, R. L., 1965.
[The depth of the convective envelope on the lower main sequence and the
depletion of lithium].
{\it Astrophys. J.}, {\bf 142}, 174--181.

\bibitem[Wilde {\it et~al.\/}(2001)]{Wilde2001}
\biblab{Wilde2001}
Wilde, S. A., Valley, J. W., Peck, W. H. \& Graham, C. M., 2001.
[Evidence from detrital zircons for the existence of continental crusts
and oceans on the Earth 4.4 Gyr ago].
{\it Nature}, {\bf 409}, 175--178.

\bibitem[Williams and  Cieza(2011)]{Willia2011}
\biblab{Willia2011}
Williams, J. P. \& Cieza, L. A., 2011.
[Protoplanetary disks and their evolution].
{\it Annu. Rev. Astron. Astrophys.}, {\bf 49}, 67--117.

\bibitem[Willson(2000)]{Willso2000}
\biblab{Willso2000}
Willson, L. A., 2000.
[Mass loss from cool stars: impact on the evolution of stars and stellar 
populations].
{\it Annu. Rev. Astron. Astrophys.}, {\bf 38}, 573--611.

\bibitem[Willson(1997)]{Willso1997}
\biblab{Willso1997}
Willson, R. C., 1997.
[Total solar irradiance trend during solar cycles 21 and 22].
{\it Science}, {\bf 277}, 1963--1965.

%\bibitem[Willson {\it et~al.\/}(1988)]{Willso1988}
%\biblab{Willso1988}
%Willson, R. C., Hudson, H. S., 1988.
%[Solar luminosity variations in solar cycle 21].
%{\it Nature}, {\bf 332}, 810--812.

\bibitem[Winn and  Fabrycky(2015)]{Winn2015}
\biblab{Winn2015}
Winn, J. N. \& Fabrycky, D. C., 2015.
[The occurrence and architecture of exoplanetary systems].
{\it Annu. Rev. Astron. Astrophys.}, {\bf 53}, 409--447.

\bibitem[Wolfenstein(1978)]{Wolfen1978}
\biblab{Wolfen1978}
Wolfenstein, L., 1978.
[Neutrino oscillations in matter].
{\it Phys. Rev. D}, {\bf 17}, 2369--2374.

{\rv 
\bibitem[Woodard and Noyes(1985)]{Woodar1985}
\biblab{Woodar1985}
Woodard, M. F. \& Noyes, R. W., 1985.
[Change of solar oscillation eigenfrequencies with the solar cycle].
{\it Nature}, {\bf 318}, 449--450.
}

\bibitem[Woodard {\it et~al.\/}(1991)]{Woodar1991}
\biblab{Woodar1991}
Woodard, M. F., Kuhn, J. R., Murray, N. \& Libbrecht, K. G., 1991.
[Short-term changes in solar oscillation frequencies and solar activity].
{\it Astrophys. J.}, {\bf 373}, L81--L84.

\bibitem[Wuchterl and  Klessen(2001)]{Wuchte2001}
\biblab{Wuchte2001}
Wuchterl, G. \& Klessen, R. S., 2001.
[The first million years of the Sun: a calculation of the formation 
and early evolution of a solar mass star].
{\it Astrophys. J.}, {\bf 560}, L185--L188.

\bibitem[Wuchterl and Tscharnuter(2003)]{Wuchte2003}
\biblab{Wuchte2003}
Wuchterl, G. \& Tscharnuter, W. M., 2003.
[From clouds to stars. Protostellar collapse and the evolution to the
pre-main sequence. I. Equations and evolution in the Hertzsprung-Russell
diagram].
{\it Astron. Astrophys.}, {\bf 398}, 1081--1090.

\bibitem[Xiong(1977)]{Xiong1977}
\biblab{Xiong1977}
Xiong, D. R., 1977.
[Statistical theory of turbulent convection in pulsating variables].
{\it Acta Astron. Sinica}, {\bf 18}, 86--104.

\bibitem[Xiong(1989)]{Xiong1989}
\biblab{Xiong1989}
Xiong, D. R., 1989.
[Radiation-hydrodynamic equations for stellar oscillations].
{\it Astron. Astrophys.}, {\bf 209}, 126--134.

\bibitem[Xiong and Deng(2001)]{Xiong2001}
\biblab{Xiong2001}
Xiong, D R. \& Deng, L., 2001.
[The structure of the solar convective overshooting region].
{\it Mon. Not. R. Astron. Soc.}, {\bf 327}, 1137--1144.

\bibitem[Yang(2016)]{Yang2016}
\biblab{Yang2016}
Yang, W., 2016.
[Solar models with new low metal abundances].
{\it Astrophys. J.}, {\bf 821}, 108-(1--10).

\bibitem[Yang and Bi(2007)]{Yang2007}
\biblab{Yang2007}
Yang, W. M. \& Bi, S. L., 2007.
[Solar models with revised abundances and opacities].
{\it Astrophys. J.}, {\bf 658}, L67--L70.

\bibitem[Young(2018)]{Young2018}
\biblab{Young2018}
Young, P. R., 2018.
[Element abundance ratios in the quiet Sun transition region].
{\it Astrophys. J.}, {\bf 855}, 15-(1--7).

\bibitem[Zaatri {\it et~al.\/}(2007)]{Zaatri2007}
\biblab{Zaatri2007}
Zaatri, A., Provost, J., Berthomieu, G., Morel, P. \& Corbard, T., 2007.
[Sensitivity of low degree oscillations to the change in solar abundances].
{\it Astron. Astrophys.}, {\bf 469}, 1145--1149.

\bibitem[Zahn(1991)]{Zahn1991}
\biblab{Zahn1991}
Zahn, J.-P., 1991.
[Convective penetration in stellar interiors].
{\it Astron. Astrophys.}, {\bf 252}, 179--188.

\bibitem[Zahn(1992)]{Zahn1992}
\biblab{Zahn1992}
Zahn, J.-P., 1992.
[Circulation and turbulence in rotating stars].
{\it Astron. Astrophys.}, {\bf 265}, 115--132.

\bibitem[Zahn {\it et~al.\/}(2007)]{Zahn2007}
\biblab{Zahn2007}
Zahn, J.-P., Brun, A. S. \& Mathis, S., 2007.
[On magnetic instabilities and dynamo action in stellar radiation zones].
{\it Astron. Astrophys.}, {\bf 474}, 145--154.
%{\tt (arXiv:0707.3287v1 [astro-ph])}

\bibitem[Zhang {\it et~al.\/}(2019)]{Zhang2019}
\biblab{Zhang2019}
Zhang, Q.-S., Li, Y., Christensen-Dalsgaard, J., 2019.  
[Solar models with convective overshoot, solar-wind mass loss, 
and PMS disk accretion: helioseismic quantities, Li depletion 
and neutrino fluxes].
{\it Astrophys. J.}, {\bf 881}, 103-(1--26).

\bibitem[Zhou {\it et~al.\/}(2019)]{Zhou2019}
\biblab{Zhou2019}
Zhou, Y., Asplund, M. \& Collet, R., 2019.
[The amplitude of solar p-mode oscillations from three-dimensional
convection simulations].
{\it Astrophys. J.}, {\bf 880}, 13-(1--10).

\end{thebibliography}

\end{document}